\documentclass[letterpaper,12pt,oneside]{book}
\usepackage{amssymb}
\usepackage{amsfonts}
\usepackage[dvips]{graphicx}
\usepackage[spanish]{babel}

\date{}
\usepackage[latin1]{inputenc}

\def\thebibliography#1{\chapter*{Referencias
   }\list
  {[\arabic{enumi}]}{\settowidth\labelwidth{[#1]}\leftmargin\labelwidth
    \advance\leftmargin\labelsep
\usecounter{enumi}}
    \def\newblock{\hskip .11em plus .33em minus .07em}
    \sloppy\clubpenalty4000\widowpenalty4000
    \sfcode`\.=1000\relax}

\begin{document}

\pagestyle{plain}

\begin{center}
{\bf \large UNIVERSIDAD AUT\'ONOMA METROPOLITANA\\ 
UNIDAD CUAJIMALPA\\
DEPARTAMENTO DE MATEM\'ATICAS APLICADAS Y SISTEMAS}
\end{center}
\vskip 2cm


\vskip 2cm

\begin{center}
{\bf FUNCIONES ESPECIALES Y TRANSFORMADAS INTEGRALES\\
CON APLICACIONES A LA MEC\'ANICA CU\'ANTICA Y  ELECTRODIN\'AMICA}
\end{center}
\vskip 4.5cm

\begin{center}
{\bf JUAN \ MANUEL \ ROMERO \  SANPEDRO\\
jromero@correo.cua.uam.mx\\

}
\end{center}
\vskip 2cm


\tableofcontents

\newpage

\section{Introducci\'on }

Dec\'ia Galileo que los secretos de la naturaleza est\'an escritos en el lenguaje de las matem\'aticas.
I. Newton tambi\'en se dio cuenta de este hecho e invent\'o el c\'alculo infinitesimal para entender el movimiento de los planetas.
Al ser estudiados los fen\'omenos el\'ectricos y magn\'eticos surgieron nuevas matem\'aticas. Ahora esta
parte de la naturaleza se expres\'o  en t\'erminos de campos vectoriales y ecuaciones diferenciales parciales,
las cuales se resumen en las ecuaciones de Maxwell. Al resolver las ecuaciones de Maxwell surgieron funciones
con caracter\'isticas especiales, por ello se les llama funciones especiales. Para la electrodin\'amia, dentro de esas funciones especiales, de notable  importancia son las funciones de Bessel, los polinomios de Legendre, as\'i como las llamadas series de Fourier. Las cantidades importantes de la electrodin\'amica se expresan en t\'erminos de estas funciones. Despu\'es de mucho esfuerzo los matem\'aticos se dieron cuentas que estas funciones forman espacios vectoriales   con dimensi\'on infinita, lo que hoy se conoce como  espacios de Hilbert.  Textos cl\'asicos sobre electrodin\'anica se pueden ver en \cite{jackson:gnus,landau-medios:gnus} y referencias sobre espacios de Hilbert se pueden ver en \cite{reed:gnus,kolmogorov:gnus}. \\

Por otra parte, en un inicio nadie entend\'ia los fen\'omenos cu\'anticos y como expresarlos matem\'aticamente. Sin embargo, al proponer 
Schr\"odinger su ecuaci\'on de onda las cosas se entendieron un poco m\'as. Sorprendentemente al  resolver la ecuaci\'on de Schr\"odinger  surgieron funciones especiales, como los polinomios de Hermite y los polinomios de Laguerre. As\'i, los matem\'aticos de la \'epoca se dieron cuenta que estaban frente a una nueva aplicaci\'on de los espacios de Hilbert. Dos excelentes referencias sobre mec\'anica cu\'antica se pueden ver \cite{q1:gnus,q2:gnus}. Adem\'as,  se encontr\'o que una generalizaci\'on de las series Fourier, la transformada de Fourier, es de vital importancia para entender diversos fen\'omenos cu\'anticos. La transformada de Fourier es un caso particular de las transformadas integrales, un texto sobre este tema se puede ver en \cite{lokenath:gnus}.\\

Cabe se\~nalar que en un inicio las ecuaciones diferenciales que surg\'ian en la electrodin\'amica y en la mec\'anica cu\'antica se resolv\'ian mediante series de potencial, ese es el m\'etodo tradicional \cite{whittaker:gnus,morse:gnus,Samarskii:gnus}.   Sin embargo, el matem\'aticos franc\'es Jean Gaston Darboux se dio cuenta que muchas de esas ecuaciones se pueden resolver con lo que hoy se llama el m\'etodo de Factorizaci\'on. Al ser aplicado este m\'etodo  en mec\'anica cu\'antica mostr\'o su gran potencia. Referencias sobre este \'ultimo m\'etodo se pueden ver en \cite{infel:gnus,facto:gnus}. Otra aportaci\'on importante fue dada por el matem\'atico franc\'es   Sophus Lie, quien mostr\'o que la teor\'ia de grupos es de gran ayuda para encontrar soluciones de las ecuaciones diferenciales.  Posteriormente, Pauli mostr\'o que usando teor\'ia de grupos pod\'ia obtener los estados cu\'anticos de varios sistemas f\'isicos. As\'i, pocos se sorprendieron cuando se encontr\'o una relaci\'on entre algunos grupos, como el grupo de rotaciones, y algunas funciones especiales, como los polinomios de Legendre. Con el tiempo la teor\'ia de grupos se convirt\'io en una herramienta fundamental en varias \'areas de la f\'isica te\'orica. Textos sobre aplicaciones de la teor\'ia de grupos a las ecuaciones diferenciales se pueden ver  \cite{vilenki:gnus,copias:gnus}.\\

Notablemente, recientemente se ha encontrado que herramientas de la mec\'anica cu\'antica se pueden aplicar para abordar problemas de otras disciplinas. 
Por ejemplo, en estudios de las Finanzas \cite{baaquie:gnus} .\\
  
En este libro, se busca introducir al estudiante en las funciones especiales y las transformadas integrales junto con
sus aplicaciones en la electrodin\'amica y en la mec\'anica cu\'antica. El objetivo principal de este texto es dar al estudiante las herramientas b\'asicas para que aborde sin dificultad problemas avanzados a nivel licenciatura. Para que el lector tenga una visi\'on de los objetos matem\'aticos que se estudian, se da una introducci\'on a los espacios de Hilbert. En el libro se usan tres diferentes m\'etodos para obtener las funciones especiales. Se usan la series de potencia para obtener las funciones de Bessel. 
Se usa el m\'etodo de factorizaci\'on para  obtener los polinomios de Hermite y se usa el grupo de rotaciones para obtener los polinomios de Legendre y los arm\'onicos esf\'ericos.\\

A lo largo del texto, se realizan varios ejercicios para mostrar como se usan las funciones especiales. As\'i, se encuentran soluciones a ecuaciones como la ecuaci\'on de Laplace, de onda, de Helmholtz, de calor, de Schr\"odinger. 
En particular se obtienen los estados para el oscilador arm\'onico y del \'atomo de Hidr\'ogeno. Tambi\'en se obtienen
las funciones de Green de la ecuaci\'on de Laplace, de la ecuaci\'on de Helmholtz y la ecuaci\'on de onda.\\

El material de este texto ha sido utilizado en cursos para estudiantes de las carreras de F\'isica y Matem\'aticas de la UNAM y de  Matem\'aticas Aplicadas de la UAM-Cuajimalpa. Normalmente los f\'isicos se muestran m\'as interesados en los aspectos formales, mientras que los matem\'aticos se interesan m\'as en las aplicaciones. Se trat\'o de tener un punto de equilibrio para que las dos  clases  de estudiantes logren obtener fundamentos s\'olidos y al mismo tiempo sean capaces de abordar problemas sofisticados.

\chapter{ La Convenci\'on de Suma de Einstein, el Tensor de Levi-Civita y las Ecuaciones de Maxwell}

{\it Le calcul tensoriel sait mieux la physique que le physicien lui-même \\
( El c\'alculo tensorial sabe m\'as f\'{\i}sica que los mismos f\'{\i}sicos)\\
Paul Langevin  1964.\\}

\section{Introducci\'on}

En este cap\'{\i}tulo veremos algunas herramientas matem\'aticas
que facilitan la manipulaci\'on de operaciones vectoriales. Para ver
la eficacia de estas herramientas las aplicaremos a las ecuaciones de Maxwell
y a la fuerza de Lorentz. En particular, ocupando estas herramientas, obtendremos las cantidades 
conservadas que implican las ecuaciones de Maxwell.\\

Las ecuaciones de Maxwell son
\begin{eqnarray}
\vec\nabla\cdot \vec E&=&4\pi \rho \label{eq:m1},\\
\vec\nabla\times \vec E&=&-\frac{1}{c}\frac{\partial \vec B}{\partial t}
\label{eq:m2},\\
\vec \nabla\cdot \vec B&=&0,\label{eq:m3}\\
\vec\nabla\times \vec B&=&\frac{4\pi}{c}\vec J+
\frac{1}{c}\frac{\partial \vec E}{\partial t}.\label{eq:m4}
\end{eqnarray}
Donde $\rho$ es la densidad volum\'etrica de carga y $\vec J$ 
es la densidad de corriente el\'ectrica. De forma gen\'erica 
$\vec J$ se puede escribir como $\vec J=\rho(\vec x,t) \vec v(\vec x,t),$ 
con $\vec v(\vec x,t)$ la velocidad de las part\'{\i}culas 
cargadas en el punto $\vec x$ al tiempo $t.$\\

La ley de Gauss Eq. (\ref{eq:m1}) relaciona la densidad de carga el\'ectrica con 
el campo el\'ectrico. La ley de Faraday Eq. (\ref{eq:m2}) nos dice que un campo
magn\'etico que varia en el tiempo produce un campo el\'ectrico. La ecuaci\'on 
Eq. (\ref{eq:m3}) nos dice que no existen cargas magn\'eticas. La ley de Amp\`ere
Eq. (\ref{eq:m4}) nos dice dos cosas, la primera es que la corriente el\'ectrica produce 
campo magn\'etico y la segunda es que la variaci\'on temporal del campo el\'ectrico 
produce campo magn\'etico. Como se puede ver todas estas ecuaciones 
son lineales.\\

Adem\'as, la fuerza de Lorentz  nos dice que una part\'{\i}cula de masa $m$ y carga $q$ en un 
campo el\'ectrico $\vec E$ y magn\'etico $\vec B$ siente la fuerza
\begin{eqnarray}
m\vec a=\vec F=q\vec E+q\frac{\vec v}{c}\times \vec B.
\end{eqnarray}
Esta fuerza se puede obtener de las ecuaciones de Maxwell, sin embargo para 
interpretar los resultados que obtendremos la supondremos independiente.\\

Las ecuaciones de Maxwell Eqs. (\ref{eq:m1})-(\ref{eq:m4}) tambi\'en se pueden
escribir de forma integral. Para ver esto, recordemos  el teorema de 
 Gauss y el teorema de Stokes. Supongamos que tenemos una regi\'on de volumen $V$
cuya frontera es la superficie $S.$ Entonces, el {\bf Teorema de Gauss}
nos dice que, si $\vec F$ es un campo vectorial suave definido en $V,$ 
se cumple 
\begin{eqnarray}
\int_{V} \vec \nabla \cdot\vec F dV=\oint_{\partial V}\vec F\cdot d\vec a= \int_{S}\vec F\cdot \hat n da,
\label{eq:tgauss}
\end{eqnarray}
aqu\'{\i} $d\vec a$ es  el elemento de \'area de $S$ y $\hat n$ representa 
la normal exterior a  esta  superficie.\\

Ahora, supongamos que tenemos una superficie $S$ cuya frontera
est\'a dada por la curva $\Gamma.$ Entonces, el {\bf Teorema de Stokes} nos dice
que si $\vec F$ es un campo regular sobre $S,$ se cumple 
\begin{eqnarray}
\int_{S} (\vec \nabla \times \vec F) \cdot \hat nda=
\int_{\Gamma}\vec F\cdot d\vec l.
\label{eq:tstokes}
\end{eqnarray}
donde $\vec l$ es el vector tangente a $\Gamma$ y gira en el sentido
opuesto a las manecillas del reloj.\\

As\'{\i}, ocupando el teorema de Gauss Eq. (\ref{eq:tgauss}) y 
de Stokes Eq. (\ref{eq:tstokes}), las ecuaciones de Maxwell toman la forma
\begin{eqnarray}
\int_{\partial V} \vec E\cdot \hat n da&=&4\pi Q_{T},\\
\int_{\partial S}\vec E\cdot d\vec l&=&-\frac{1}{c}\frac{d\Phi_{m}}{dt},\\
\int_{\partial V} \vec B\cdot \hat n da&=&0,\\
\int_{\partial S}\vec B\cdot d\vec l&=&\frac{4\pi}{c}I+
\frac{1}{c}\frac{d\Phi_{e}}{dt}.
\end{eqnarray}
Donde 
\begin{eqnarray}
Q_{T}=\int_{V}\rho dv 
\end{eqnarray}
es la carga total contenida en el volumen $V.$  
Mientras que
\begin{eqnarray}
\Phi_{m}=\int_{S} \vec B\cdot \hat n da,\qquad 
\Phi_{e}=\int_{S} \vec E\cdot \hat n da
\end{eqnarray}
son, respectivamente, el flujo magn\'etico y
el\'ectrico que pasa por la superficie $S.$ 
Adicionalmente
\begin{eqnarray}
I=\int_{S} \vec J\cdot \hat n da
\end{eqnarray}
representa la corriente
total que pasa por la superficie $S.$\\

\section{Producto escalar y la delta de Kronecker}

Recordemos que un vector tridimensional se define como
\begin{eqnarray}
\vec A= (A_{x},A_{y},A_{z})=(A^{1}, A^{2}, A^{3}),
\end{eqnarray}
tambi\'en lo podemos representar como $A^{i}$ con $i=1,2,3,$ 
es decir, $A^{i}$ es la componente $i$-\'esima.\\

Una operaci\'on importante entre vectores es el producto escalar.
Si tenemos los vectores $\vec A$ y 
$\vec B=(B^{1}, B^{2}, B^{3}),$ el producto escalar se define como
\begin{eqnarray}
\vec A\cdot\vec B= A^{1} B^{1}+ A^{2}B^{2}+ A^{3}B^{3}=
\sum_{i=1}^{3} A^{i} B^{i}.
\end{eqnarray}
Por simplicidad es com\'un escribir 
\begin{eqnarray}
\vec A\cdot\vec B= A^{i} B^{i},
\end{eqnarray}
donde se entiende que los \'{\i}ndices repetidos se suman,
a esta regla se le llama convenci\'on de Einstein. 
Cuando dos \'{\i}ndices est\'an repetidos se dice que est\'an
{\it contraidos}. Por ejemplo, si
\begin{eqnarray}
\vec r =(x,y,z)=(x^{1}, x^{2}, x^{3})
\end{eqnarray}
es el vector posici\'on,  entonces el cuadrado de la distancia es
\begin{eqnarray}
r^{2}=\vec r\cdot\vec r= x^{i} x^{i}=x^{1} x^{1}+ x^{2}x^{2}+ x^{3}x^{3}= x^{2}+ y^{2}+ z^{2}.
\end{eqnarray}

Ahora, definamos la delta de Kronecker como el s\'{\i}mbolo tal que
\begin{equation}
\delta_{ij}=\left\{
\begin{array}{ccc}
1 & \,\, {\rm si} & i=j,\\
0 & \,\, {\rm si} & i\not =j.
\end{array}
\right. \label{eq:Kronecker}
\end{equation}
En realidad se est\'a defini\'endo una matriz, la matriz identidad $I,$ 
pues,
\begin{eqnarray}
\delta_{ij}=
\left( 
\begin{array}{rrrr}
 \delta_{11} &\delta_{12}& 
\delta_{13} \\
 \delta_{21} & \delta_{22}& 
\delta_{23} \\
 \delta_{31} & \delta_{32}& \delta_{33}
\end{array}
\right)=
\left( 
\begin{array}{rrrr}
 1 & 0& 0 \\
 0 & 1& 0 \\
0 & 0 & 1
\end{array}
\right)=I.
\end{eqnarray}
Con la delta de Kronecker y la convenci\'on de Einstein se tiene
\begin{eqnarray}
\delta_{1j}A_{j}=\sum_{j=1}^{3}\delta_{1j}A_{j}=A_{1},\quad   \delta_{2j}A_{j}=A_{2}\qquad 
 \delta_{3j}A_{j}=A_{3},
\end{eqnarray}
es decir 
\begin{eqnarray}
\delta_{ij}A_{j}=A_{i}.
\end{eqnarray}
Por lo que, el producto escalar se puede escribir como
\begin{eqnarray}
A^{i}\delta_{ij} B^{j}=A^{i}B^{i}=\vec A\cdot\vec B.
\end{eqnarray}
Adem\'as, con la convenci\'on de Einstein el s\'{\i}mbolo $\delta_{ii}$
significa
\begin{eqnarray}
\delta_{ii} =\sum_{i=1}^{3}\delta_{ii}=
\delta_{11}+\delta_{22}+\delta_{33}=3.
\end{eqnarray}
Un vector importante es el gradiente, el cual se define
como
\begin{eqnarray}
\vec \nabla=\left(\frac{\partial }{\partial x},\frac{\partial }{\partial y},
\frac{\partial }{\partial z} \right)=
\left(\frac{\partial }{\partial x^{1}},\frac{\partial }{\partial x^{2}},
\frac{\partial }{\partial x^{3}} \right).
\end{eqnarray}
Por simplicidad, en algunos casos solo escribiremos
\begin{eqnarray}
\left(\vec \nabla \right)_{i}=\frac{\partial }{\partial x^{i}}=\partial_{i}, \qquad i=1,2,3.
\end{eqnarray}
Veamos que significa 
\begin{eqnarray}
\frac{\partial x^{j}}{\partial x^{i}}=\partial_{i}x^{j},
\end{eqnarray}
para cada valor de $i$ y $j$ se tiene un valor de $\partial_{i}x^{j}$ por lo que
se tiene  la matriz
\begin{eqnarray}
\frac{\partial x^{j}}{\partial x^{i}}=
\left( 
\begin{array}{rrrr}
 \frac{\partial x^{1}}{\partial x^{1}} & \frac{\partial x^{2}}{\partial x^{1}}  & 
\frac{\partial x^{3}}{\partial x^{1}} \\
  \frac{\partial x^{1}}{\partial x^{2}}   & \frac{\partial x^{2}}{\partial x^{2}} & 
\frac{\partial x^{3}}{\partial x^{2}}  \\
\frac{\partial x^{1}}{\partial x^{3}}  & \frac{\partial x^{2}}{\partial x^{3}} & 
\frac{\partial x^{3}}{\partial x^{3}} 
\end{array}
\right)=
\left( 
\begin{array}{rrrr}
 1 & 0& 0 \\
 0 & 1& 0 \\
0 & 0 & 1
\end{array}
\right)=\delta_{ij}.
\end{eqnarray}
Ahora, considerando que $r=\sqrt{x^{i}x^{i}},$ se tiene
\begin{eqnarray}
\frac{\partial r}{\partial x^{j}}&=& \frac{\partial \sqrt{x^{i}x^{i}} } {\partial x^{j}}=
\frac{1}{2\sqrt{x^{i}x^{i}}} \frac{\partial \left(x^{i}x^{i}\right) }{\partial x^{j}}= \frac{1}{2r}
\left( \frac{\partial x^{i}}{\partial x^{j}}x^{i}+x^{i}\frac{\partial x^{i}}{\partial x^{j}}\right)\nonumber\\
&=&\frac{1}{2r}\left(\delta_{ij}x^{i}+x^{i}\delta_{ij}\right)
=\frac{x_{j}}{r}.
\end{eqnarray}
De donde 
\begin{eqnarray}
\frac{\partial r}{\partial x^{j}}=\frac{x_{j}}{r},\qquad 
\vec \nabla r= \frac{\vec r}{r}=\hat r.
\end{eqnarray}
Si $f(r)$ es una funci\'on que solo depende de $r$ se tiene
\begin{eqnarray}
\frac{\partial f(r)}{\partial x^{i}}=\frac{\partial f(r)}{\partial r}\frac{\partial r}{\partial x^{i}}
=\frac{\partial f(r)}{\partial r}\frac{x_{i}}{r},\qquad 
\vec \nabla f(r)= \frac{\partial f(r)}{\partial r} \frac{\vec r}{r}=\frac{\partial f(r)}{\partial r}\hat r.
\end{eqnarray}
Veamos otro ejemplo, consideremos la funci\'on 
\begin{eqnarray}
\phi(\vec r)=\frac{\vec P\cdot \vec r}{r^{3}},
\end{eqnarray}
con $\vec P$ un vector constante, entonces 
\begin{eqnarray}
\frac{\partial\phi(\vec r)}{\partial x^{i}}&=&\frac{\partial }{\partial x^{i}}\left(\frac{\vec P\cdot \vec r}{r^{3}}\right)
=\left(\frac{\partial \vec P\cdot \vec r  }{\partial x^{i}}\right) \frac{1}{r^{3}}+ \vec P\cdot \vec r 
\frac{\partial }{\partial x^{i}}\left(\frac{1}{r^{3}}\right) \nonumber\\
&=&
\left(\frac{\partial  P_{j} x_{j}  }{\partial x^{i}}\right) \frac{1}{r^{3}}-3 \vec P\cdot \vec r
\frac{1}{r^{4}} \frac{\partial r}{\partial x^{i}}\nonumber\\
&=&
\frac{P_{j}}{r^{3}}\frac{\partial x_{j}  }{\partial x^{i}}
-3 \vec P\cdot \vec r
\frac{1}{r^{4}} \frac{x_{i}}{r}\nonumber\\
&=& \frac{P_{j}\delta_{ij} }{r^{3}}-3 \vec P\cdot \vec r\frac{x_{i}}{r^{5}}
=\frac{P_{i} }{r^{3}}-3 \vec P\cdot \vec r\frac{x_{i}}{r^{5}}\nonumber\\
&=&\frac{P_{i}r^{2}-3\vec P\cdot \vec r x_{i} }{r^{5}},\nonumber
\end{eqnarray}
de donde
\begin{eqnarray}
\vec \nabla \left(\frac{\vec P\cdot \vec r}{r^{3}}\right)= \frac{\vec Pr^{2}-3 \left(\vec P\cdot \vec r \right)\vec r  }{r^{5}}.
\end{eqnarray}
Note que con esta notaci\'on  la divergencia de un vector $\vec E$ se puede escribir como
\begin{eqnarray}
\vec \nabla \cdot \vec E&=&\frac{\partial E_{x}}{\partial x}+
\frac{\partial E_{y} }{\partial y}+
\frac{\partial E_{z}}{\partial z} =
\frac{\partial E_{1} }{\partial x^{1}}+\frac{\partial E_{2} }{\partial x^{2}}
+\frac{\partial E_{3}}{\partial x^{3}}\nonumber\\
& =&\sum_{i=1}^{3}
\frac{\partial E_{i} }{\partial x^{i}}=\frac{\partial E_{i} }{\partial x^{i}}=\partial_{i} E_{i}.
\end{eqnarray}
Tambi\'en se puede probar la identidad
\begin{eqnarray}
\vec \nabla\cdot \left(fg\vec A\right)=g \left(\vec \nabla f\right)\cdot \vec A+
f \left(\vec \nabla g\right)\cdot \vec A +fg\left(\vec \nabla\cdot\vec A\right).
\end{eqnarray}
En efecto
\begin{eqnarray}
\vec \nabla\cdot (fg\vec A)&=&\partial_{i}\left(fg A_{i}\right)=\left(\partial_{i}f\right)gA_{i}+
f\left(\partial_{i}g\right)A_{i}+fg\left(\partial_{i}A_{i}\right)\nonumber\\
&=& g \left(\vec \nabla f\right)\cdot \vec A+
f \left(\vec \nabla g\right)\cdot \vec A +fg\left(\vec \nabla\cdot\vec A\right).
\end{eqnarray}

\section{Producto vectorial y el tensor de Levi-Civita}

Otra operaci\'on importante entre vectores es el producto vectorial:
\begin{eqnarray}
\vec A\times \vec B&=&\left| 
\begin{array}{rrrr}
 \hat i & \hat j& \hat k \\
 A_{1} & A_{2}& A_{3} \\
B_{1} & B_{2} & B_{3}
\end{array}
\right |\label{eq:provec} \\
&=& \left(A_{2} B_{3}-A_{3}B_{2}\right)\hat i+
\left(A_{3} B_{1}-A_{1}B_{3}\right)\hat j+
\left(A_{1} B_{2}-A_{2}B_{1}\right)\hat k \nonumber,
\end{eqnarray}
es decir
\begin{eqnarray}
\left(\vec A\times \vec B\right)_{1}&=&(A_{2} B_{3}-A_{3}B_{2}),
\label{eq:prodvec1}\\
\left(\vec A\times \vec B\right)_{2}&=&(A_{3} B_{1}-A_{1}B_{3}),
\label{eq:prodvec2} \\
\left(\vec A\times \vec B\right)_{3}&=&(A_{1} B_{2}-A_{2}B_{1}).
\label{eq:prodvec3}
\end{eqnarray}
Note que en la componente $\left(\vec A\times \vec B\right)_{1}$
no est\'a $A_{1}$ ni  $B_{1}.$  De hecho, esto tambi\'en ocurre para 
las dem\'as componentes, es decir, en la componente 
$\left(\vec A\times \vec B\right)_{i}$ no est\'a la componente  $A_{i}$ ni  
la componente $B_{i}.$\\

Anteriormente vimos que el producto escalar se puede escribir en t\'erminos
de una matriz, veamos si con el producto vectorial ocurre lo mismo. Consideremos 
la matriz antisim\'etrica  
\begin{eqnarray}
\epsilon_{ij}=\left( 
\begin{array}{rr}
 0 & 1 \\
-1 & 0
\end{array}
\right ).\label{eq:lev-ch}
\end{eqnarray}
Con esta matriz, las igualdades Eqs. (\ref{eq:prodvec1})-(\ref{eq:prodvec3}) 
se pueden escribir como
\begin{eqnarray}
\left(\vec A\times \vec B\right)_{1}&=&\left( 
\begin{array}{r}
A_{2}\quad A_{3}
\end{array}\right)
\left( 
\begin{array}{rr}
 0 & 1 \\
-1 & 0
\end{array}\right)
\left( 
\begin{array}{r}
B_{2}\\
B_{3}\\
\end{array}\right)\nonumber\\
&=&\epsilon_{ab}A_{a}B_{b}, \quad a,b=2,3,\\
\left(\vec A\times \vec B\right)_{2}&=&\left( 
\begin{array}{r}
A_{3}\quad
A_{1}\\
\end{array}\right)
\left( 
\begin{array}{rr}
 0 & 1 \\
-1 & 0
\end{array}\right)
\left( 
\begin{array}{r}
B_{3}\\
B_{1}\\
\end{array}\right)\nonumber\\
&=&\epsilon_{cd}A_{c}B_{d}, \quad c,d=3,1,\\
\left(\vec A\times \vec B\right)_{3}&=&\left( 
\begin{array}{r}
A_{1}\quad 
A_{2}\\
\end{array}\right)
\left( 
\begin{array}{rr}
 0 & 1 \\
-1 & 0
\end{array}\right)
\left( 
\begin{array}{r}
B_{1}\\
B_{2}\\
\end{array}\right)\nonumber\\
&=&\epsilon_{ef}A_{e}B_{f}, \quad e,f=1,2.
\end{eqnarray}
Note que la matriz Eq. (\ref{eq:lev-ch}) est\'a en 
dos dimensiones, mientras que el espacio es tridimensional.
As\'{\i} es m\'as conveniente ocupar una generalizaci\'on de 
Eq. (\ref{eq:lev-ch}) en tres dimensiones, la cual denotaremos con
\begin{eqnarray}
\epsilon_{ijk}, \qquad i,j,k=1,2,3.\label{eq:levi-civita}
\end{eqnarray}
En principio, para reproducir el productor vectorial basta que
$\epsilon_{ijk}$ sea antisim\'etrico en las dos \'ultimas
entradas. Sin embargo, como en
$\left(\vec A\times \vec B\right)_{i}$ no est\'a $A_{i}$ ni  $B_{i},$
pediremos que $\epsilon_{ijk}$ sea antisim\'etrico tambi\'en en las dos 
primeras entradas. Es decir, pediremos las propiedades
\begin{eqnarray}
\epsilon_{ijk}=-\epsilon_{ikj}, \quad \epsilon_{ijk}=-\epsilon_{jik},
\quad \epsilon_{123}=1.\label{eq:prop-levi-civita}
\end{eqnarray}
Note que esto implica que, para cualquier $i$ y  $k,$ se cumpla 
\begin{eqnarray}
\epsilon_{iik}=-\epsilon_{iik}=0,
\end{eqnarray}
de donde  
\begin{eqnarray}
\epsilon_{iki}=-\epsilon_{kii}=0.
\end{eqnarray}
Es decir, las componentes
de $\epsilon_{ijk}$ con dos \'{\i}ndices repetidos tienen valor cero.
Por lo tanto, las  componentes de $\epsilon_{ijk}$ no nulas  
tienen todos los \'{\i}ndices diferentes.  Ocupando las propiedades de 
$\epsilon_{ijk}$ Eq. (\ref{eq:prop-levi-civita}) se puede ver que 
\begin{eqnarray}
\epsilon_{123}&=&1,\\
\epsilon_{132}&=&-\epsilon_{123}=-1,\\
\epsilon_{213}&=&-\epsilon_{123}=-1,\\
\epsilon_{231}&=&-\epsilon_{213}=1,\\
\epsilon_{312}&=&-\epsilon_{132}=1,\\
\epsilon_{321}&=&-\epsilon_{312}=-1.
\end{eqnarray}
Claramente, todos estos valores se obtienen de permutar los 
\'{\i}ndices de
$\epsilon_{123}.$ Al s\'{\i}mbolo Eq. (\ref{eq:levi-civita}) se le llama 
{\bf tensor de Levi-Civita}. Ahora veremos que este tensor es \'util
para expresar el producto vectorial. Definamos
\begin{eqnarray}
\left(\vec A\times \vec B\right)_{i}=\sum_{j=1}^{3}\sum_{k=1}^{3}
\epsilon_{ijk}A_{j}B_{k}=\epsilon_{ijk}A_{j}B_{k}.\label{eq:prod-vec}
\end{eqnarray}
Veamos que esta igualdad es correcta, para la primera componente tenemos
\begin{eqnarray}
\left(\vec A\times \vec B\right)_{1}&=&\sum_{j=1}^{3}\sum_{k=1}^{3}
\epsilon_{1jk}A_{j}B_{k}=\sum_{k=1}^{3}\left(\epsilon_{11k}A_{1}B_{k}
+\epsilon_{12k}A_{2}B_{k}+\epsilon_{13k}A_{3}B_{k}\right)\nonumber\\
&=&\sum_{k=1}^{3}\left(\epsilon_{12k}A_{2}B_{k}+\epsilon_{13k}A_{3}
B_{k}\right)\\
&=&\epsilon_{121}A_{2}B_{1}+\epsilon_{122}A_{2}B_{2}+\epsilon_{123}A_{2}B_{3}\\
& &+\epsilon_{131}A_{3}B_{1}+\epsilon_{132}A_{3}B_{2}+\epsilon_{133}A_{3}
B_{3}\\
&=&\epsilon_{123}A_{2}B_{3}+\epsilon_{132}A_{3}B_{2}\\
&=&
A_{2}B_{3}-A_{3}B_{2}.
\end{eqnarray}
Este c\'alculo es m\'as f\'acil si se ocupan las propiedades
de $\epsilon_{ijk}.$ En efecto, si se tiene
$\epsilon_{1jk},$ los \'unicos valores que puede tomar $j$ son $2$ \'o
$3$ y $k$ solo puede tomar los valores $3$ \'o $2.$ Por lo que, 
\begin{eqnarray}
\left(\vec A\times \vec B\right)_{1}&=&\sum_{j=1}^{3}\sum_{k=1}^{3}
\epsilon_{1jk}A_{j}B_{k}=\epsilon_{123}A_{2}B_{3}+\epsilon_{132}A_{3}B_{2}
\nonumber\\
&=&
A_{2}B_{3}-A_{3}B_{2}.
\end{eqnarray} 
Para las dem\'as componentes tenemos
\begin{eqnarray}
\left(\vec A\times \vec B\right)_{2}&=&\sum_{j=1}^{3}\sum_{k=1}^{3}
\epsilon_{2jk}A_{j}B_{k}=\epsilon_{213}A_{1}B_{3}+\epsilon_{231}A_{3}B_{1}
\nonumber\\
&=&
A_{3}B_{1}-A_{1}B_{3},\\
\left(\vec A\times \vec B\right)_{3}&=&\sum_{j=1}^{3}\sum_{k=1}^{3}
\epsilon_{3jk}A_{j}B_{k}=\epsilon_{312}A_{1}B_{2}+\epsilon_{321}A_{2}B_{1}
\nonumber\\
&=&
A_{1}B_{2}-A_{2}B_{1}.
\end{eqnarray} 
Como podemos ver, la definici\'on de producto vectorial Eq. (\ref{eq:prod-vec}) coincide con
Eq. (\ref{eq:provec}).\\

Con el  s\'{\i}mbolo $\epsilon_{ijk}$ es m\'as econ\'omico escribir un producto vectorial. 
Por ejemplo el momento angular se puede escribir como
\begin{eqnarray}
L_{i}=(\vec r\times \vec p)_{i}=\epsilon_{ijk}r_{j}p_{k}.
\end{eqnarray}
Mientras que el rotacional se puede escribir como
\begin{eqnarray}
\left(\vec \nabla \times \vec A\right)_{i}=\epsilon_{ijk}\partial_{j}A_{k}.
\end{eqnarray}
Adem\'as el momento  angular cu\'antico, 
$\vec L=-i\hbar \left(\vec r\times \vec \nabla\right),$  se escribe
\begin{eqnarray}
 L_{i}=-i\hbar\epsilon_{ijk}r_{j}\partial_{k}.
\end{eqnarray}
El tensor de Levi-Civita no es solo otra forma de expresar el producto vectorial,
tambi\'en es \'util para simplificar los c\'alculos.

\section{El tensor de Levi-Civita y las matrices}

Vimos que con el tensor de Levi-Civita se puede 
expresar el producto vectorial de forma sencilla. 
Este tensor tambi\'en se puede relacionar con las
matrices. Se tienen resultados particularmente interesantes con
 matrices antisim\'etricas y sim\'etricas.

\subsection{El tensor de Levi-Civita y las matrices antisim\'etricas}

Cualquier matriz antim\'etrica de $3\times 3$ se puede escribir como
\begin{eqnarray}
M_{ij}=\left(
\begin{array}{rrrr}
0 & B_{3}& -B_{2} \\
-B_{3} & 0& B_{1} \\
B_{2} & -B_{1}& 0
\end{array}
\right).
\end{eqnarray}
Con las componentes no nulas de esta matriz se puede formar el vector $\vec B=(B_{1},B_{2},B_{3}).$
Note que si hacemos la contracci\'on de $\epsilon_{ijk}$ con $B_{i}$ se tiene $\epsilon_{ijk}B_{k}$
que es un objeto con dos \'{\i}ndices libres, es decir es una matriz. Considerando las propiedades
Eq. (\ref{eq:prop-levi-civita}) se tiene 
\begin{eqnarray}
\epsilon_{ijk}B_{k}=
\left(
\begin{array}{rrrr}
0 &\epsilon_{123}B_{3}& \epsilon_{132}B_{2} \\
\epsilon_{213}B_{3} & 0& \epsilon_{231}B_{1} \\
\epsilon_{312}B_{2} & \epsilon_{321}B_{1}& 0
\end{array}
\right)=
\left(
\begin{array}{rrrr}
0 & B_{3}& -B_{2} \\
-B_{3} & 0& B_{1} \\
B_{2} & -B_{1}& 0
\end{array}
\right)=M_{ij}.\nonumber
\end{eqnarray}
Por lo tanto, cualquier matriz antisim\'etrica $M$ de $3\times 3$ se puede poner en t\'ermino
del tensor de Levi-Civita y un vector $B$:
\begin{eqnarray}
M_{ij}=\epsilon_{ijk}B_{k}.\label{eq:levi-anti}
\end{eqnarray}

\subsection{El tensor de Levi-Civita y las matrices sim\'etricas}

Hasta aqu\'{\i} hemos ocupado $\epsilon_{ijk}$ con vectores. Pero tambi\'en 
lo podemos emplear con matrices de $3\times 3.$ En efecto, dada la matriz $M_{ij}$ 
podemos definir
\begin{eqnarray}
V_{i}=\sum_{j=1}^{3}\sum_{k=1}^{3}
\epsilon_{ijk}M_{jk}=\epsilon_{ijk}M_{jk}.
\end{eqnarray}
Para evitar confusiones notemos que la contracci\'on de los \'{\i}ndices
de $\epsilon_{ijk}$ se puede escribir de diferentes forma. Por ejemplo, 
\begin{eqnarray}
V_{i}=\sum_{j=1}^{3}\sum_{k=1}^{3}
\epsilon_{ijk}M_{jk}=
\sum_{k=1}^{3}\sum_{j=1}^{3} \epsilon_{ikj}M_{kj}.
\end{eqnarray}
Esta igualdad no se obtiene por un intercambio de \'{\i}ndices en
$\epsilon_{ijk}.$ Se obtiene por renombrar al mismo tiempo los dos 
\'ultimos \'{\i}ndices de $\epsilon_{rst}$ y los dos \'{\i}ndices 
de $M_{ab}.$ Con la conveci\'on de Einstein esta igualdad se escribe
como
\begin{eqnarray}
V_{i}=\epsilon_{ijk}M_{jk}=\epsilon_{ikj}M_{kj}.\label{eq:mudos}
\end{eqnarray}
Un resultado de este hecho trivial es que si $M_{ij}$ es una
matriz sim\'etrica entonces la contracci\'on con $\epsilon_{ijk}$ es cero, es decir,
\begin{eqnarray}
M_{ij}=M_{ji}\qquad \Longrightarrow \qquad
\epsilon_{ijk}M_{jk}=0.\label{eq:sim-ant}
\end{eqnarray}
Esto se debe a que
\begin{eqnarray}
\epsilon_{ijk}M_{jk}=-\epsilon_{ikj}M_{jk}=-\epsilon_{ikj}M_{kj}
=-\epsilon_{ijk}M_{jk}.
\end{eqnarray}
En la primera igualdad, se empleo que $\epsilon_{ijk}$ es antisim\'etrico,
en la segunda que $M_{jk}$ es sim\'etrica, en la tercera la igualdad 
Eq. (\ref{eq:mudos}). Por lo tanto, $\epsilon_{ijk}M_{jk}=-\epsilon_{ijk}M_{jk}$
y se cumple Eq. (\ref{eq:sim-ant}).\\

Por ejemplo,  con cualquier vector $A_{i}$ se puede forma
la matriz $M_{ij}=A_{i}A_{j}.$ Esta matriz es sim\'etrica, 
pues
\begin{eqnarray}
M_{ij}=
\left( 
\begin{array}{rrrr}
A_{1}A_{1} &A_{1}A_{2}& 
A_{1}A_{3} \\
A_{2}A_{1} & A_{2}A_{2}& 
A_{2}A_{3} \\
A_{3}A_{1} & A_{3}A_{2}& A_{3}A_{3}
\end{array}
\right)=
\left( 
\begin{array}{rrrr}
A_{1}A_{1} &A_{2}A_{1}& 
A_{3}A_{1} \\
A_{1}A_{2} & A_{2}A_{2}& 
A_{3}A_{2} \\
A_{1}A_{3} & A_{2}A_{3}& A_{3}A_{3}
\end{array}
\right).
\end{eqnarray}
Esto implica que 
\begin{eqnarray}
\vec A\times \vec A=0.
\end{eqnarray}
En efecto, ocupando la definici\'on de producto vectorial y que la matriz  
$$M_{ij}=A_{i}A_{j}$$
es sim\'etrica se cumple 
\begin{eqnarray}
0=\epsilon_{ijk}A_{j}A_{k}=(\vec A\times \vec A)_{i}. 
\end{eqnarray}
Otra matriz sim\'etrica est\'a dada por $M_{ij}=\partial_{i}\partial_{j}.$
Esto implica que
\begin{eqnarray}
\left(\vec \nabla\times \vec \nabla \phi\right)=0.
\end{eqnarray}
Pues,
\begin{eqnarray}
(\vec \nabla\times \vec\nabla  \phi)_{i}=
\epsilon_{ijk}\partial_{j}\left(\vec \nabla \phi\right)_{k}
=\epsilon_{ijk}\partial_{j}\partial_{k}\phi=0.
\end{eqnarray}
Tambi\'en se puede mostrar la identidad
\begin{eqnarray}
\vec \nabla\cdot (\vec \nabla\times \vec A)=0. \label{eq:carga}
\end{eqnarray}
En efecto, como $M_{ij}=\partial_{i}\partial_{j}$ es sim\'etrica 
se llega a
\begin{eqnarray}
\vec \nabla\cdot (\vec \nabla\times \vec A)&=&
\partial_{i}\left(\vec \nabla\times \vec A\right)_{i}
=\partial_{i}\left(\epsilon_{ijk}\partial_{j}A_{k}\right)
=\epsilon_{ijk}\partial_{i}\partial_{j}A_{k}\nonumber\\
&=&\epsilon_{kij}\partial_{i}\partial_{j}A_{k}=0.
\end{eqnarray}
Esta identidad tiene implicaciones en las ecuaciones de Maxwell.

\subsection{Conservaci\'on de carga}

Un hecho experimental bien conocido es que la carga el\'ectrica se conserva.
Veamos si las ecuaciones de Maxwell son compatibles con este
resultado. Para esto ocuparemos la ley de Gauss
Eq. (\ref{eq:m1}) y la ley de Amp\'ere Eq. (\ref{eq:m4}). De la ley de
Gauss obtenemos
\begin{eqnarray}
\frac{\partial \rho}{\partial t}=\frac{1}{4\pi}\vec \nabla \cdot
\frac{\partial \vec E}{\partial t}
\end{eqnarray}
y ocupando la ley de Amp\`ere se tiene
\begin{eqnarray}
\frac{\partial \rho}{\partial t}=
\vec \nabla \cdot\left(\vec \nabla\times \vec B-\vec J\right).
\end{eqnarray}
Ahora, considerando Eq. (\ref{eq:carga}) tenemos que
$\vec \nabla \cdot(\vec \nabla\times \vec B)=0,$
de donde
\begin{eqnarray}
\frac{\partial \rho}{\partial t}+\vec \nabla \cdot\vec J=0, 
\label{eq:vec-continui}
\end{eqnarray}
que es la llamada ecuaci\'on de continuidad. Integrando sobre un volumen $V$
la expresi\'on Eq.  (\ref{eq:vec-continui}) y ocupando el teorema de Gauss
Eq. (\ref{eq:tgauss}) se encuentra
\begin{eqnarray}
\frac{dQ_{T}}{dt}=
\frac{d}{dt}\left(\int dV \rho\right)=
\int dV \frac{\partial \rho}{\partial t}=-\int dV\vec \nabla \cdot\vec J=
-\int_{\partial V} da \vec J\cdot \hat n.\label{eq:car-ga}
\end{eqnarray}
Si el volumen de integraci\'on es suficientemente grande, de tal forma que en su frontera no haya corriente, el \'ultimo
t\'ermino de  Eq. (\ref{eq:car-ga}) es cero y se obtiene
\begin{eqnarray}
\frac{dQ_{T}}{dt}=0, 
\end{eqnarray}
es decir, la carga total se conserva en el tiempo.\\

\section{Triple producto escalar}

Algunas identidades vectoriales son f\'aciles de demostrar con la convenci\'on de 
Einstein y  el s\'{\i}mbolo de Levi-Civita. Por ejemplo el triple 
producto escalar 
\begin{eqnarray}
\vec A\cdot \left(\vec B\times \vec C\right)
=\vec C\cdot \left(\vec A\times \vec B\right)=\vec B\cdot 
\left(\vec C\times \vec A\right),
\label{eq:triple-escalar}
\end{eqnarray}
que se demuestra simplemente de la forma
\begin{eqnarray}
\vec A\cdot \left(\vec B\times \vec C\right)&=&
A_{i}\left(\vec B\times \vec C\right)_{i}
=A_{i}\epsilon_{ijk} B_{j} C_{k}=
C_{k}\left(\epsilon_{kij}A_{i} B_{j}\right)\nonumber\\
&=&C_{k}\left(\vec A\times \vec B\right)_{k}=
\vec C\cdot \left(\vec A\times \vec B\right),\nonumber\\
\vec A\cdot \left(\vec B\times \vec C\right)&=&
A_{i}\left(\vec B\times \vec C\right)_{i}
=A_{i}\epsilon_{ijk} B_{j} C_{k}=
-B_{j}\epsilon_{jik}A_{i}C_{k}\nonumber \\
&=& B_{j}\epsilon_{jki}C_{k}A_{i}= 
B_{j}\left(\vec C\times \vec A\right)_{j}=
\vec B\cdot \left(\vec C\times \vec A\right).
\end{eqnarray}
Por lo tanto, se cumple Eq. (\ref{eq:triple-escalar}).
Note que ocupando la regla del triple producto escalar y la igualdad $\vec A\times \vec A=0,$ 
se encuentra 
\begin{eqnarray}
\vec A\cdot \left(\vec A\times \vec C\right)=\vec C\cdot \left(\vec A\times \vec A\right)=0
.\label{eq:idenII}
\end{eqnarray}
Si en lugar del vector constante  $\vec C$ se tiene el operador $\vec \nabla,$
la identidad del triple producto escalar ya no es v\'alido. En este caso se cumple la 
identidad
\begin{eqnarray}
\left(\vec \nabla \times \vec A\right)\cdot \vec B=
 \vec \nabla \cdot\left(\vec A\times \vec B\right)+
\left(\vec \nabla \times \vec B\right)\cdot \vec A.
\label{eq:ener1}
\end{eqnarray}
En efecto,
\begin{eqnarray}
\left(\vec \nabla \times \vec A\right)\cdot \vec B&=&
\left(\vec \nabla \times \vec A\right)_{i} B_{i}=
\left(\epsilon_{ijk}\partial_{j}A_{k}\right) B_{i}
=\epsilon_{ijk}\left(\partial_{j}A_{k}\right)B_{i}\nonumber\\
&=&\epsilon_{ijk}\left[\partial_{j}(A_{k}B_{i})
-A_{k}\partial_{j}B_{i}\right]=\partial_{j}
\left(\epsilon_{jki}A_{k}B_{i}\right)+
\left(\epsilon_{kji}\partial_{j}B_{i}\right)A_{k}\nonumber\\
&=&
\vec \nabla \cdot\left(\vec A\times \vec B\right)+
\left(\vec \nabla \times \vec B\right)\cdot \vec A. \label{eq:ener}
\end{eqnarray}
Otra identidad que se puede mostrar ocupando s\'olo las propiedades
del tensor de Levi-Civita es
\begin{eqnarray}
\vec \nabla \times \left(\phi \vec A\right)=
\vec \nabla \phi \times \vec A+\phi \vec \nabla \times \vec A.
\label{eq.lemsto}
\end{eqnarray}
Pues 
\begin{eqnarray}
\left(\vec \nabla \times \left(\phi \vec A\right)\right)_{i}&=&
\epsilon_{ijk}\partial_{j}\left(\phi\vec A\right)_{k}=
\epsilon_{ijk}\partial_{j}\left(\phi A_{k}\right)=
\epsilon_{ijk}\left(\partial_{j}\phi\right)A_{k}+ 
\phi \epsilon_{ijk}\partial_{j}A_{k}\nonumber\\
&=&\left(\vec \nabla \phi \times \vec A\right)_{i}+
\left(\phi \vec \nabla \times \vec A\right)_{i}.
\end{eqnarray}
En la pr\'oxima secci\'on veremos implicaciones de estas propiedades vectoriales.

\section{Aplicaciones del triple producto escalar}

Las identidades de la secci\'on anterior tienen consecuencias
importantes para las ecuaciones de Maxwell, veamos cuales son.

\subsection{Energ\'{\i}a cin\'etica}

Ocupando Eq. (\ref{eq:idenII}) en la fuerza de Lorentz se encuentra
\begin{eqnarray}
\vec F\cdot \vec v= \left (q\vec E+\frac{q}{c}\vec v\times \vec B\right)\cdot
 \vec v=q\vec E\cdot \vec v,  
\end{eqnarray}
este resultado nos indica que el campo magn\'etico no hace trabajo.\\

Ahora, recordemos que la derivada temporal de la
energ\'{\i}a cin\'etica es
\begin{eqnarray}
\dot {\cal E}_{cin}=\frac{d}{dt}\left(\frac{m}{2}\vec v^{2}\right)=
m\frac{d \vec v}{dt}\cdot \vec v=m\vec a\cdot \vec v=\vec F\cdot \vec v.
\end{eqnarray}
Para el caso particular de la fuerza de Lorentz se tiene
\begin{eqnarray}
\dot {\cal E}_{cin}=\vec F\cdot \vec v=
q\left(\vec E+\frac{\vec v}{c}\times \vec B\right)\cdot \vec v=
q\vec E\cdot \vec v.
\end{eqnarray}
Adem\'as, suponiendo que no tenemos una carga si no una distribuci\'on de cargas
$\rho$ en un volumen $V$, como $dq=\rho d^{3}x,$
\begin{eqnarray}
dq\vec E\cdot \vec v=\vec E\cdot (\rho \vec v)dx^{3}=
\vec E\cdot \vec Jdx^{3}.
\end{eqnarray}
Por lo tanto
\begin{eqnarray}
\dot {\cal E}_{cin}=\int_{V}\vec E\cdot \vec Jdx^{3}.
\end{eqnarray}
Para este razonamiento solo hemos ocupado la fuerza de Lorentz.
Posteriormente veremos lo que dicen las ecuaciones de Maxwell respecto a la energ\'ia.

\subsection{Conservaci\'on de la energ\'{\i}a}

Veamos que implicaciones tiene  Eq. (\ref{eq:ener}) en las ecuaciones de 
Maxwell.\\

Para esto consideremos la ley de Faraday Eq. (\ref{eq:m2}) y
de Amp\`ere Eq. (\ref{eq:m4}). Haciendo el producto escalar
de $\vec B$ con la ley de Faraday encontramos que
\begin{eqnarray}
\left(\vec \nabla \times \vec E\right)\cdot \vec B=
-\frac{1}{c}\frac{\partial \vec B}{\partial t}\cdot \vec B=-\frac{1}{2c}
\frac{\partial \vec B^{2}}{\partial t}.
\end{eqnarray}                                                                
Si hacemos el producto escalar
de $\vec E$ con la ley de Amp\`ere se llega a
\begin{eqnarray}
\left(\vec \nabla \times \vec B\right)\cdot \vec E=
\frac{4\pi}{c}\vec J\cdot \vec E
+\frac{1}{c}\frac{\partial \vec E}{\partial t}\cdot \vec E=
\frac{4\pi}{c}\vec J\cdot \vec E+\frac{1}{2c}
\frac{\partial \vec E^{2}}{\partial t}.
\end{eqnarray}
Restando estas dos ecuaciones y ocupando la identidad  Eq. (\ref{eq:ener}) 
se obtiene
\begin{eqnarray}
\frac{1}{8\pi}\frac{\partial }{\partial t}\left(\vec E^{2}+\vec B^{2}\right)
&=&-\vec J\cdot \vec E+\frac{c}{4\pi}(\vec \nabla \times \vec B)\cdot \vec E-
\frac{c}{4\pi}(\vec \nabla \times \vec E)\cdot \vec B\nonumber\\
&=&-\vec J\cdot \vec E-
\frac{c}{4\pi}\vec \nabla \cdot \left(\vec E\times \vec B\right).
\label{eq:en}
\end{eqnarray}
Al vector
\begin{eqnarray}
 \vec S=\frac{c}{4\pi} \left(\vec E\times \vec B\right)
\end{eqnarray}
se le llama vector de Poynting.
Ahora, integrando  Eq. (\ref{eq:en}) sobre un volumen $V$ y ocupando la
teorema  de Gauss Eq. (\ref{eq:tgauss}) se encuentra
\begin{eqnarray}
\frac{d}{dt}\left({\cal E}_{em}+{\cal E}_{cin}\right)=
-\int_{V} d^{3}x \vec\nabla\cdot \vec S=
-\oint_{\partial V} \vec S\cdot \hat nda,
\label{eq:ener2}
\end{eqnarray}
donde
\begin{eqnarray}
{\cal E}_{em}=\frac{1}{8\pi}\int_{V} d^{3}x\left(\vec E^{2}+\vec B^{2}\right).
\end{eqnarray}
Como podemos ver, adem\'as de la energ\'{\i}a cin\'etica, las ecuaciones
de Maxwell nos dicen que hay otra energ\'{\i}a. Esta nueva energ\'ia se debe a los campos
el\'ectricos y magn\'eticos dada por $ {\cal E}_{em},$ a 
esta energ\'{\i}a se le llama energ\'{\i}a electromagn\'etica.  El t\'ermino
$$\oint_{\partial V} \vec S\cdot \hat nda $$
se interpreta como flujo de energ\'{\i}a. Si el volumen es suficientemente grande de tal forma que
no haya flujo de energ\'{\i}a en su frontera,  Eq. (\ref{eq:ener2})
implica
\begin{eqnarray}
{\cal E}_{T}={\cal E}_{em}+{\cal E}_{cin}={\rm constante},
\end{eqnarray}
es decir, la energ\'{\i}a total se conserva.

\section{Contracci\'on de dos tensores de Levi-Civita}

El tensor $\delta_{ij}$ es sim\'etrico
mientras que $\epsilon_{ijk}$ es totalmente antisim\'etrico.
Sin embargo estos dos tensores est\'an relacionados. Primero notemos
que en dos dimensiones se cumple 
\begin{eqnarray}
\epsilon_{ik}\epsilon_{kj}=\left(
\begin{array}{rr}
 0 & 1 \\
-1 & 0
\end{array}\right)\left(
\begin{array}{rr}
 0 & 1 \\
-1 & 0
\end{array}\right) =-\left(
\begin{array}{rr}
 1 & 0 \\
0 & 1
\end{array}\right) =-\delta_{ij}.
\end{eqnarray}
En tres dimensiones se cumple la identidad
\begin{eqnarray}
\sum_{k=1}^{3}\epsilon_{ijk}\epsilon_{klm}=\epsilon_{ijk}\epsilon_{klm}=
\delta_{il}\delta_{jm}-\delta_{im}\delta_{jl}.\label{eq:proeps}
\end{eqnarray}
Para mostrar esto definamos el s\'{\i}mbolo
\begin{eqnarray}
M_{ijlm}=
\delta_{il}\delta_{jm}-\delta_{im}\delta_{jl}.\label{eq:M}
\end{eqnarray}
Debido a que $\delta_{ab}$ es un tensor sim\'etrico, se cumple
\begin{eqnarray}
M_{ijlm}=-M_{jilm}=-M_{ijml},\label{eq:das1}
\end{eqnarray}
en efecto
\begin{eqnarray}
M_{jilm}&=&\delta_{jl}\delta_{im}-\delta_{jm}\delta_{il}=
- \left(\delta_{il}\delta_{jm}-\delta_{im}\delta_{jl}\right)
=-M_{ijlm},\\
M_{ijml}&=&\delta_{im}\delta_{jl}-\delta_{il}\delta_{jm}=
- \left(\delta_{il}\delta_{jm}-\delta_{im}\delta_{jl}\right)
=-M_{ijlm}.
\end{eqnarray}
En particular se tiene $M_{iilm}=M_{ijmm}=0.$
Tambi\'en se puede observar que   $\epsilon_{ijk}\epsilon_{klm}$
es antisim\'etrico si hacemos una permutaci\'on en $(ij)$
\'o $(lm),$ es decir
\begin{eqnarray}
\epsilon_{ijk}\epsilon_{klm}=- \epsilon_{jik}\epsilon_{klm}=-\epsilon_{ijk}\epsilon_{kml}.
\label{eq:das2}
\end{eqnarray}
Las igualdades Eqs. (\ref{eq:das1})-(\ref{eq:das2}) nos indican que si
Eq. (\ref{eq:proeps}) se cumple para una cuarteta ordenada $(ijml),$
tambi\'en se cumple para las cuartetas ordenadas 
$(jilm), (ijml).$\\

Note si $i=j,$ Eq. (\ref{eq:proeps}) toma la forma $0=0.$ As\'{\i},
los valores que falta por probar son $i\not = j.$ De la definici\'on Eq. (\ref{eq:M})
es claro que $M_{ijlm}$ es diferente de cero s\'olo si $i=l,j=m$ \'o $i=m,j=l$,
que son las cuartetas ordenadas $(ijij)$ y $(ijji).$ Esta propiedad 
tambi\'en la tiene la cantidad $\epsilon_{ijk}\epsilon_{klm}.$
En efecto, recordemos que los \'indices $i,j,k,l,m$ s\'olo
pueden tomar los valores $(1,2,3),$ adem\'as en la suma 
$\epsilon_{ijk}\epsilon_{klm}$ los \'unicos t\'erminos que contribuyen son tales
que $i\not =j,i\not =k,j\not=k$ y $k\not =l,k\not=m,l\not =m.$
Estas condiciones implican que $i=m,l=j$ \'o $i=l, j=m.$ Es decir las \'unica cuartetas
ordenadas que dan resultados no nulos en $\epsilon_{ijk}\epsilon_{klm}$ son 
$(ijij)$ y $(ijji).$\\  

Como probar Eq. (\ref{eq:proeps}) para el caso $(ijij)$ 
es equivalente a probarla para el caso $(ijji),$ s\'olo probaremos
los caso $(ijij)=(1212),(1313),(2323).$\\
 
Si $(ijij)=(1212),$ se tiene
\begin{eqnarray}
\epsilon_{12k}\epsilon_{k12}=\epsilon_{123}\epsilon_{312}=1,\quad  
M_{1212}=\delta_{11}\delta_{22}-\delta_{12}\delta_{21}=1.
\end{eqnarray}
Por lo tanto, Eq. (\ref{eq:proeps}) se cumple.\\

Si $(ijij)=(1313),$ se encuentra
\begin{eqnarray}
\epsilon_{13k}\epsilon_{k13}=\epsilon_{132}\epsilon_{213}=1,\quad  
M_{1313}=\delta_{11}\delta_{33}-\delta_{13}\delta_{23}=1.
\end{eqnarray}
Por lo tanto, Eq. (\ref{eq:proeps}) se cumple.\\

Si $(ijij)=(2323),$ se llega a
\begin{eqnarray}
\epsilon_{23k}\epsilon_{k23}=\epsilon_{231}\epsilon_{123}=1,\quad  
M_{2323}=\delta_{22}\delta_{33}-\delta_{23}\delta_{32}=1.
\end{eqnarray}
Por lo tanto, Eq. (\ref{eq:proeps}) se cumple.\\

En conclusi\'on la igualdad  Eq. (\ref{eq:proeps}) es v\'alida para cualquier 
$ijlm.$ Una implicaci\'on de Eq. (\ref{eq:proeps}) es
\begin{eqnarray}
\epsilon_{ilm}\epsilon_{mjs}+\epsilon_{sim}\epsilon_{mjl}=
\epsilon_{ijm}\epsilon_{mls}.
\end{eqnarray}
En efecto ocupando Eq. (\ref{eq:proeps}) se tiene
\begin{eqnarray}
\epsilon_{ilm}\epsilon_{mjs}+\epsilon_{sim}\epsilon_{mjl}&=&\left(
\delta_{ij}\delta_{ls}- \delta_{is}\delta_{lj}\right)+
\left(\delta_{sj}\delta_{il}- \delta_{sl}\delta_{ij}\right)\nonumber\\
&=&
\delta_{il}\delta_{sj}- \delta_{is}\delta_{lj}=
\epsilon_{ijm}\epsilon_{mls}.
\end{eqnarray}
En la pr\'oxima secci\'on veremos la importancia de la igualdad Eq. (\ref{eq:proeps}).\\

\section{Triple producto vectorial I}

Ocupando Eq. (\ref{eq:proeps}),  se puede probar el llamado triple producto
vectorial
\begin{eqnarray}
\vec A\times (\vec B\times \vec C)&=&
\vec B(\vec A\cdot \vec C)-\vec C(\vec A\cdot \vec B),
\label{eq:trivec}
\end{eqnarray}
pues
\begin{eqnarray}
\left[\vec A\times (\vec B\times \vec C)\right]_{i}&=&
\epsilon_{ijk}A_{j}\left(\vec B\times \vec C\right)_{k}
=\epsilon_{ijk}A_{j}(\epsilon_{klm}B_{l}C_{m})\nonumber\\
&=&\epsilon_{ijk}\epsilon_{klm}A_{j}B_{l}C_{m}=
\left(\delta_{il}\delta_{jm}-\delta_{im}\delta_{jl}\right)
A_{j}B_{l}C_{m}\nonumber\\
&=&(A_{m}C_{m})B_{i}-C_{i}(A_{l}B_{l})\nonumber\\
&=&(\vec A\cdot \vec C)B_{i}-(\vec A\cdot \vec B)C_{i}.
\end{eqnarray}

Si en lugar de un vector constante se tiene el operador $\vec \nabla,$ 
la identidad Eq. (\ref{eq:trivec}) ya no es v\'alidad. Por ejemplo,
si en lugar de $\vec B\times \vec C$ se considera $\vec \nabla\times \vec A,$
ahora se cumple
\begin{eqnarray}
\left[(\vec \nabla \times \vec A)\times \vec A\right]_{i}&=&
\partial_{j}\left(A_{i}A_{j}-\frac{1}{2}\delta_{ij}A^{2}\right)-
(\vec \nabla\cdot \vec A)A_{i}.\label{eq:momento}
\end{eqnarray}
En efecto,
\begin{eqnarray}
\left[(\vec \nabla \times \vec A)\times \vec A\right]_{i}&=&
\epsilon_{ijk}(\vec \nabla \times \vec A)_{j}A_{k}=
\epsilon_{ijk}(\epsilon_{jlm} \partial_{l}A_{m})A_{k}\nonumber\\
&=&-\epsilon_{ikj}\epsilon_{jlm}(\partial_{l}A_{m})A_{k}\nonumber\\
&=&-\left(\delta_{il} \delta_{km} -\delta_{im} \delta_{kl}\right)
(\partial_{l}A_{m})A_{k}\nonumber\\
&=&A_{l}\partial_{l}A_{i}-A_{m}\partial_{i}A_{m}\nonumber\\
&=&\partial_{l}(A_{i}A_{l})-A_{i}\partial_{l}A_{l} -\frac{1}{2}
\partial_{i}(A_{m}A_{m})\nonumber\\
&=&\partial_{l}\left( A_{i}A_{l}-\frac{1}{2}\delta_{il}A^{2}\right)
-A_{i}\vec \nabla \cdot \vec A. 
\end{eqnarray}
Con esta identidad posteriormente
veremos que se conserva el momento.\\

\section{Conservaci\'on del momento}

Para estudiar la conservaci\'on del momento veamos de nuevo
la fuerza de Lorentz, la cual se puede escribir como
\begin{eqnarray}
\frac{d \vec P_{cin}}{dt}=m\vec a=\vec F=
q\left(\vec E+\frac{\vec v}{c}\times \vec B\right).
\end{eqnarray}
Si tenemos una distribuci\'on $\rho$ de carga, un elemento de carga est\'a dado por  $dq=\rho d^{3}x$
y el elemento de fuerza es
\begin{eqnarray}
d\vec F=
\left(\rho\vec E +\frac{\rho \vec v}{c}\times \vec B\right)d^{3}x=
\vec {\cal F} d^{3}x,
\end{eqnarray}
con 
\begin{eqnarray}
\vec {\cal F}= \rho\vec E +\frac{\vec J}{c}\times \vec B
\end{eqnarray}
la densidad de fuerza mec\'anica. As\'{\i}, la fuerza total mec\'anica es  
\begin{eqnarray}
\frac{d \vec P_{cin}}{dt}=\vec F=
\int_{V}\left(\rho\vec E +\frac{\vec J}{c}\times \vec B\right)d^{3}x.
\label{eq:momen-cin}
\end{eqnarray}
En este resultado solo se ocupo la fuerza de Lorentz. Veamos que dicen
las ecuaciones de Maxwell.\\

Si hacemos el producto vectorial de $\vec E$ con la ley de Faraday  
Eq. (\ref{eq:m2}) se tiene
\begin{eqnarray}
\left(\vec \nabla \times \vec E\right)\times \vec E=
-\frac{1}{c}\left(\frac{\partial \vec B}{\partial t}\right)\times \vec E.
\end{eqnarray}
Si hacemos el producto vectorial de $\vec B$
con la ley de Amp\`ere Eq. (\ref{eq:m4}) encontramos
\begin{eqnarray}
\left(\vec \nabla \times \vec B\right)\times \vec B&=&
\left(\frac{4\pi}{c} \vec J+
\frac{1}{c}\frac{\partial \vec E}{\partial t}\right)\times \vec B\\
&=&\frac{4\pi}{c} \vec J\times \vec B+\frac{1}{c}\frac{\partial \vec E}
{\partial t}\times \vec B.
\end{eqnarray}
Sumando estas dos ecuaciones y considerando 
\begin{eqnarray}
\frac{\partial\left(\vec E\times \vec B\right)}{\partial t}
= \frac{\partial\vec E}{\partial t}\times \vec B+\vec E\times
\frac{\partial\vec B}{\partial t},
\end{eqnarray}
se llega a
\begin{eqnarray}
\left(\vec \nabla \times \vec E\right)\times \vec E+
\left(\vec \nabla \times \vec B\right)\times \vec B&=&
\frac{4\pi}{c} \vec J\times \vec B+\frac{1}{c}
\left(\frac{\partial \vec E}
{\partial t}\times \vec B-
\frac{\partial \vec B}{\partial t}\times \vec E \right)\nonumber\\
&=&\frac{4\pi}{c} \vec J\times \vec B+\frac{1}{c}
\left(\frac{\partial \vec E}
{\partial t}\times \vec B+\vec E\times
\frac{\partial \vec B}{\partial t} \right)\nonumber\\
&=&\frac{4\pi}{c} \vec J\times \vec B+ \frac{1}{c}
\frac{\partial\left(\vec E\times \vec B\right)}{\partial t}. 
\label{eq:preb-m}
\end{eqnarray}
Adem\'as, tomando en cuenta Eq. (\ref{eq:momento}), la ley de Gauss Eq. (\ref{eq:m1})
y la ley de inexistencia de monopolos magn\'eticos Eq. (\ref{eq:m3}), se tiene
\begin{eqnarray}
\left(\left(\vec \nabla \times \vec E\right)\times \vec E\right)_{i}&=&
\partial_{j}\left(E_{i}E_{j}-\frac{1}{2}\delta_{ij}E^{2}\right)-
(\vec \nabla\cdot \vec E)E_{i}\\
&=&\partial_{j}\left(E_{i}E_{j}-\frac{1}{2}\delta_{ij}E^{2}\right)-
4\pi \rho E_{i}, \\
\left(\left(\vec \nabla \times \vec B\right)\times \vec B\right)_{i}&=&
\partial_{j}\left(B_{i}B_{j}-\frac{1}{2}\delta_{ij}B^{2}\right)-
(\vec \nabla\cdot \vec B)B_{i}\\
&=&\partial_{j}\left(B_{i}B_{j}-\frac{1}{2}\delta_{ij}B^{2}\right).
\end{eqnarray}
Introduciendo estos resultados en Eq. (\ref{eq:preb-m}) se llega a
\begin{eqnarray}
\partial_{j}\left(E_{i}E_{j}+B_{i}B_{j}-
\frac{\delta_{ij}}{2}\left(E^{2}+B^{2}\right)\right)\nonumber\\
=4\pi\left(\rho\vec E +\frac{1}{c} \vec J\times \vec B +\frac{1}{4\pi c}
\frac{\partial}{\partial t} \left(\vec E\times \vec B\right)\right)_{i}
\label{eq:casi-mo}.
\end{eqnarray}
Antes de continuar definamos la densidad de momento electromagn\'etico como
\begin{eqnarray}
\vec {\cal P}= \frac{1}{4\pi c} \vec E\times \vec B=\frac{1}{c^{2}}\vec S
\end{eqnarray}
y el tensor de esfuerzos de Maxwell como 
\begin{eqnarray}
\tau_{ij}=\frac{1}{4\pi} \left(E_{i}E_{j}+B_{i}B_{j}-
\frac{\delta_{ij}}{2}\left(E^{2}+B^{2}\right)\right).
\end{eqnarray}
Entonces, la igualdad Eq. (\ref{eq:casi-mo}) toma la forma
\begin{eqnarray}
\frac{\partial {\cal P}_{i}}{\partial t}+ {\cal F}_{i}=
\partial_{j} \tau_{ji}.\label{eq:densidad-momento}
\end{eqnarray}
Integrando esta ecuaci\'on sobre un volumen $V$ se tiene
\begin{eqnarray}
\int_{V} dx^{3}\left(\frac{\partial {\cal P}_{i}}{\partial t}+ {\cal F}_{i}
\right)&=& \frac{d}{dt}\left( 
\int_{V} dx^{3} {\cal P}_{i}\right)+ \int_{V} dx^{3}{\cal F}_{i}
=\int_{V}dx^{3}
\partial_{j} \tau_{ji}\nonumber\\
&=& \oint_{\partial V}\tau_{ij}n_{j}da.
\label{eq:monf}
\end{eqnarray}
Definiremos el momento electromagn\'etico como
\begin{eqnarray}
\vec P_{em}=\frac{1}{4\pi c}\int_{V} dx^{3}\vec E\times \vec B,
\end{eqnarray}
entonces, considerando Eq. (\ref{eq:momen-cin}) se encuentra que
\begin{eqnarray}
\frac{d}{dt}\left(\vec P_{em}+\vec P_{cin}\right)_{i}=
\oint_{\partial V}\tau_{ij}n_{j}da \label{eq:monf-1}
\end{eqnarray}
Como podemos ver, adem\'as del momento cin\'etico,
las ecuaciones de Maxwell implican el momento electromagn\'etico $\vec P_{em}$
que solo depende de los campos. Tambi\'en podemos ver que 
$$\oint_{\partial V}\tau_{ij}n_{j}da$$ 
es un t\'ermino 
de fuerza y $\tau_{ij}n_{j}$ es una presi\'on. De hecho, si definimos
\begin{eqnarray}
\vec P_{T}=\vec P_{cin}+ \vec P_{em},\quad F_{i}=\oint_{\partial V}\tau_{ij}n_{j}da
\end{eqnarray}
se tiene la segunda ley de Newton
\begin{eqnarray}
\frac{d\vec P_{T} }{dt}=\vec F. 
\end{eqnarray}
Ahora, si el volumen de integraci\'on es suficientemente grande
de tal forma que el t\'ermino de la derecha sea nulo, la
ecuaci\'on Eq. (\ref{eq:monf-1}) implica
\begin{eqnarray}
\left( P_{cin}+ P_{em}\right)_{i}={\rm constante}
\end{eqnarray}
que significa que el momento total se conserva.

\subsection{Conservaci\'on del momento angular}

Tomando en cuenta la ecuaci\'on Eq. (\ref{eq:densidad-momento}) se
encuentra
\begin{eqnarray}
\epsilon_{ijk}x_{j}\frac{\partial {\cal P}_{k}}{\partial t}+ 
\epsilon_{ijk}x_{j}{\cal F}_{i}=
\epsilon_{ijk}x_{j}\partial_{l} \tau_{lk}.
\end{eqnarray}
Considerando que $\partial_{t} x_{j}=0$ y que $\tau_{lk}$ es sim\'etrico
se llega a 
\begin{eqnarray}
\frac{\partial }{\partial t}\left(\epsilon_{ijk}x_{j} {\cal P}_{k}\right)
+\left(\vec x\times \vec {\cal F}\right)_{i}&=&
\epsilon_{ijk}\left( \partial_{l}(x_{j}\tau_{lk})-
\tau_{lk}\partial_{l}x_{j} \right)\nonumber \\
&=&\epsilon_{ijk}\left( \partial_{l}(x_{j}\tau_{lk})-
\tau_{lk}\delta_{lj} \right)\nonumber \\
&=&\partial_{l}\left(\epsilon_{ijk}x_{j}\tau_{lk}\right)- 
\epsilon_{ijk}\tau_{jk}\nonumber \\
&=&\partial_{l}\left(\epsilon_{ijk}x_{j}\tau_{lk}\right).
\label{eq:momento-angular}
\end{eqnarray}
Definamos el momento angular electromagn\'etico como
\begin{eqnarray}
\vec L_{em}=\int_{V}dx^{3} \vec x\times\vec {\cal P}.
\end{eqnarray}
Adem\'as, note que, como $\vec {\cal F}$ es una densidad de fuerza, 
$\vec x\times \vec {\cal F}$ es una densidad de torca. Por lo que,
la torca, que es la deriva temporal del momento angular cin\'etico,
es
\begin{eqnarray}
\frac{d \vec L_{cin}}{dt}=\int_{V}dx^{3} \vec x\times\vec {\cal F}.
\end{eqnarray}
As\'{\i}, integrando Eq. (\ref{eq:momento-angular}) sobre un volumen $V$ y 
ocupando el teorema de Gauss, se tiene
\begin{eqnarray}
\frac{d }{dt}\left(\vec L_{em} +\vec L_{cin}\right)_{i}=
\oint_{\partial V} (\epsilon_{ijk}x_{j}\tau_{lk})n_{l} da.
\end{eqnarray}
Si $V$ es suficientemente grande de tal forma que no haya
campo electromagn\'etico en su frontera se tiene
\begin{eqnarray}
\frac{d }{dt}\left(\vec L_{em} +\vec L_{cin}\right)_{i}=0.
\end{eqnarray}
Es decir, el momento angular total se conserva.

\section{Triple producto vectorial II}

Hay otra versiones del triple producto escalar, una de ellas
es la identidad
\begin{eqnarray}
\vec \nabla \times \left(\vec \nabla \times \vec A\right)=
\vec \nabla \left(\vec \nabla \cdot \vec A\right)-\nabla^{2}\vec A.
\end{eqnarray}
Esta propiedad se demuestra de la siguiente forma
\begin{eqnarray}
\left[\vec \nabla \times \left(\vec \nabla \times \vec A\right)\right]_{i}&=&
\epsilon_{ijk}\partial_{j}\left(\vec \nabla \times \vec A\right)_{k}
=\epsilon_{ijk}\partial_{j}(\epsilon_{klm}\partial_{l}A_{m})
=\epsilon_{ijk}\epsilon_{klm}\partial_{j}\partial_{l}A_{m}\nonumber\\
&=&\left(\delta_{il}\delta_{jm}-\delta_{im}\delta_{jl}\right)
\partial_{j}\partial_{l}A_{m}= 
\partial_{i}\left( \partial_{j}A_{j}\right)-
\partial_{j}\partial_{j}A_{i}\nonumber\\
&=&\partial_{i}\left( \vec \nabla \cdot \vec A\right)-\nabla^{2}A_{i}.
\label{eq:onda}
\end{eqnarray}
Con esta identidas porteriomente
mostraremos que de las ecuaciones de Maxwell se puede obtener la
ecuaci\'on de onda.\\

Otro tipo de triple producto escalar es
\begin{eqnarray}
\vec \nabla \times\left(\vec A \times \vec B\right)=
\vec B\cdot \vec \nabla \vec A+\vec A \left(\vec \nabla \cdot \vec B\right)
- \vec A\cdot \nabla \vec B
-\vec B \left(\vec \nabla \cdot \vec A\right).
\label{eq:laquefalta}
\end{eqnarray}
Que se demuestra de la siguiente forma
\begin{eqnarray}
\left(\vec \nabla \times\left(\vec A \times \vec B\right)\right)_{i}&=&
\epsilon_{ijk}\partial_{j}\left(\vec A \times \vec B\right)_{k}
=\epsilon_{ijk}\partial_{j}\epsilon_{klm} (A_{l}B_{m})\nonumber\\
&=&\epsilon_{ijk}\epsilon_{klm} \partial_{j}(A_{l}B_{m})\nonumber\\
&=&\left(\delta_{il}\delta_{jm}-\delta_{im}\delta_{jl}\right)
\left(B_{m}\partial_{j}A_{l} +A_{l}\partial_{j}B_{m}\right)\nonumber\\
&=&\delta_{il}\delta_{jm} B_{m}\partial_{j}A_{l}+ 
\delta_{il}\delta_{jm}A_{l}\partial_{j}B_{m}\nonumber\\
& &-\left(
\delta_{im}\delta_{jl} B_{m}\partial_{j}A_{l}+
\delta_{im}\delta_{jl}A_{l}\partial_{j}B_{m}\right)\nonumber\\
&=&B_{j}\partial_{j}A_{i}+ A_{i}\partial_{j}B_{j}-
B_{i}\partial_{j}A_{j}-A_{j}\partial_{j}B_{i}\nonumber\\
&=&\Bigg[
\left(\vec B\cdot \vec \nabla\right) \vec A+\vec A \left(\vec \nabla \cdot \vec B\right)
- \left(\vec A\cdot \vec \nabla\right) \vec B-\vec B \left(\vec \nabla \cdot \vec A\right)
\Bigg]_{i},\nonumber
\end{eqnarray}
que es la prueba de Eq. (\ref{eq:laquefalta}).\\
 
\subsection{Ecuaci\'on de onda}

Con la identidad Eq. (\ref{eq:onda}) se puede probar que las ecuaciones de Maxwell implican
la ecuaci\'on de onda. Por simplicidad veremos solo el caso donde no hay fuentes, por lo que las
ecuaciones de Maxwell toman la forma
\begin{eqnarray}
\vec \nabla \cdot \vec E&=&0\label{eq:max-vac1},\\
\vec \nabla \times \vec E&=&-\frac{1}{c}\frac{\partial \vec B}{\partial t}, \label{eq:max-vac2}\\
\vec \nabla \cdot \vec B&=&0,\label{eq:max-vac3} \\
\vec \nabla \times \vec B&=&\frac{1}{c}\frac{\partial \vec E}{\partial t}.\label{eq:max-vac4}
\end{eqnarray}
De la ley de Faraday Eq. (\ref{eq:max-vac2}) se obtiene
\begin{eqnarray}
\vec\nabla \times(\vec\nabla\times\vec E)=
-\frac{1}{c}\frac{\partial(\vec\nabla\times\vec B)}{\partial t},
\end{eqnarray}
ocupando la ley de Amp\`ere sin fuentes Eq. (\ref{eq:max-vac4}) y la identidad 
Eq. (\ref{eq:onda}) se llega a
\begin{eqnarray}
\left(\nabla^{2} -\frac{1}{c^{2}}\frac{\partial^{2} }{\partial t^{2}}
\right)\vec E=0.\label{eq:ond-e}
\end{eqnarray}
Analogamente, aplicando el operador rotacional a la ley de Amp\`ere Eq. (\ref{eq:max-vac4}) sin fuentes y 
ocupando la ley de Faraday Eq. (\ref{eq:max-vac2}) se obtiene 
\begin{eqnarray}
\left(\nabla^{2} -\frac{1}{c^{2}}\frac{\partial^{2} }{\partial t^{2}}
\right)\vec B=0.\label{eq:ond-m}
\end{eqnarray}
Por lo tanto, en el vac\'{\i}o  los campos el\'ectrico y magn\'etico
satisfacen la ecuaci\'on de onda.\\

\section{Libertad de norma}

Recordemos dos teoremas de c\'alculo vectorial, el teorema de la divergencia y teorema del rotacional.
El teorema de la divergencia nos dice que  si $\vec a$ es un campo vectorial tal que 
\begin{eqnarray}
\vec \nabla \cdot \vec a=0\qquad \Longrightarrow \qquad \exists\quad
\vec b \qquad {\rm tal}\quad {\rm que} \qquad \vec a=\vec 
\nabla \times \vec b,\label{eq:tdivergencia}
\end{eqnarray}
donde $ \vec b$ es un campo vectorial. Mientras que el teorema del rotacional establece que si $\vec f$ es
un campo vectorial tal que
\begin{eqnarray}
\vec \nabla \times \vec c=0\qquad \Longrightarrow \qquad\exists\quad
h \qquad {\rm tal}\quad {\rm que} \qquad \vec c=-\vec \nabla h,
\label{eq:trotacional}
\end{eqnarray}
donde $h$ es un campo escalar.\\

De la ley de inexistencia de monopolos magn\'eticos  Eq. (\ref{eq:m3}) y 
del teorema de la divergencia Eq. (\ref{eq:tdivergencia})
se infiere que existe un campo vectorial 
$\vec A$ tal que 
\begin{eqnarray}
\vec B=\vec \nabla\times \vec A.
\end{eqnarray}
Sustituyendo este resultado en la ley de
Faraday  Eq. (\ref{eq:m2}) obtenemos que 
\begin{eqnarray}
\vec \nabla\times\left(\vec E +\frac{1}{c}\frac{\partial\vec A}{\partial t}\right)=0.
\end{eqnarray}
Ahora, por el teorema del rotacional Eq. (\ref{eq:trotacional})
se concluye que existe un campo escalar $\phi$ tal que
\begin{eqnarray}
\vec E +\frac{1}{c}\frac{\partial\vec A}{\partial t}=-\vec \nabla \phi.
\end{eqnarray}
Por lo tanto podemos decir que la inexistencia de monopolos magn\'eticos y
la ley de Faraday implican que existen los campos $\vec A$ y $\phi$ tal que
\begin{eqnarray}
\vec B=\vec \nabla\times \vec A,\qquad 
\vec E= -\left(\vec \nabla \phi+
\frac{1}{c}\frac{\partial\vec A}{\partial t}\right).\label{eq:n1}
\end{eqnarray}
Este es un resultado importante. Note que dados los
campos $\vec E, \vec B$ los campos $\vec A$ y $\phi$ no son \'unicos.
En efecto, sea $\chi=\chi(\vec x,t)$ un campo escalar arbitrario y 
definamos
\begin{eqnarray}
\vec A^{\prime}=\vec A +\vec \nabla \chi,\qquad 
\phi^{\prime}=\phi -\frac{1}{c}\frac{\partial\chi}{\partial t}.
\label{eq:tn1}
\end{eqnarray}
Entonces  Eq. (\ref{eq:n1}) implica
\begin{eqnarray}
\vec B^{\prime}=\vec B,\qquad 
\vec E^{\prime}=\vec E,
\end{eqnarray}
es decir los campos el\'ectrico y magn\'etico no cambian bajo las 
transformaciones Eq. (\ref{eq:tn1}). Debido a que $\chi$ depende del espacio
y del tiempo, se dice que las transformaciones Eq. (\ref{eq:tn1}) son locales
y se les suele llamar transformaciones de norma. \\

Ahora,  definamos $U=e^{-i\chi},$ entonces  Eq. (\ref{eq:tn1}) se puede
escribir como
\begin{eqnarray}
\vec A^{\prime}=U^{-1}\left(\vec A +i\vec \nabla \right)U,\qquad 
\phi^{\prime}=U^{-1}\left(\phi -i\frac{1}{c}\frac{\partial}{\partial t}
\right)U.
\end{eqnarray}
Esto es interesantes pues el conjunto de todas las funciones 
\begin{eqnarray}
U=e^{-i\chi},
\end{eqnarray}
forman un c\'{\i}rculo de tama\~no unitario y 
cumple las propiedades algebraica del grupo llamado 
$U(1)$ \cite{henneaux:gnus}. Por lo que, se dice que el grupo de norma de
la electrodin\'amica es $U(1).$ \\

El resto de las ecuaciones de Maxwell, 
la ley de Gauss Eq. (\ref{eq:m1}) y la ley de Amp\'ere Eq. (\ref{eq:m4}),
nos dan la din\'amica de los campos $\phi$ y $\vec A.$ En efecto,
si sustituimos
 Eq. (\ref{eq:n1}) en  Eqs. (\ref{eq:m1},\ref{eq:m4}) se obtiene
\begin{eqnarray}
-\nabla^{2}\phi-\frac{1}{c}\frac{\partial\vec \nabla\cdot \vec A }{\partial t}
&=&4\pi\rho,\\
\vec \nabla\times \left(\vec \nabla \times \vec A\right)&=&
\frac{4\pi}{c}\vec J-\frac{1}{c}\left(
\vec \nabla \frac{\partial \phi}{\partial t} +\frac{1}{c}
\frac{\partial^{2}}{\partial t^{2}}\vec A\right).
\end{eqnarray}
Ocupamos la identidad Eq. (\ref{eq:onda}) se encuentra
\begin{eqnarray}
\nabla^{2}\phi-\frac{1}{c}\frac{\partial\vec \nabla\cdot \vec A }{\partial t}
&=&-4\pi\rho,\label{eq:en1}\\
\left(\nabla^{2}-
\frac{1}{c^{2}}\frac{\partial^{2}}{\partial t^{2}}\right)\vec A
+\vec\nabla\left(\vec \nabla\cdot \vec A+
\frac{1}{c}\frac{\partial\phi}{\partial t}\right)&=&-\frac{4\pi}{c}\vec J.
\label{eq:en2}
\end{eqnarray}
Debido a que los campos de norma $\phi$ y $\vec A$ no son \'unicos,
para trabajar con estos debemos elegir un par de ellos de un n\'umero
infinito de posibilidades. Para hacer esto se suele imponer condiciones
sobre los campos de norma, una de las condiciones m\'as recurridas es la norma
de Lorentz que pide que se cumpla la condici\'on
\begin{eqnarray}
\vec \nabla\cdot \vec A+\frac{1}{c}\frac{\partial\phi}{\partial t}=0.
\end{eqnarray}
En este caso Eqs. (\ref{eq:en1}-\ref{eq:en2}) toman la forma
\begin{eqnarray}
\left(\nabla^{2}-\frac{1}{c^{2}}\frac{\partial^{2}}{\partial t^{2}}\right)\phi
&=&-4\pi\rho,\\
\left(\nabla^{2}-
\frac{1}{c^{2}}\frac{\partial^{2}}{\partial t^{2}}\right)\vec A
&=&-\frac{4\pi}{c}\vec J.
\end{eqnarray}
Que son ecuaciones de onda con fuentes.\\

Otra condici\'on de norma que se puede ocupar es la llamada condici\'on
de Coulomb 
\begin{eqnarray}
\vec \nabla\cdot \vec A=0.
\end{eqnarray}
En este caso las Eqs.(\ref{eq:en1})-(\ref{eq:en2}) toman la forma
\begin{eqnarray}
\nabla^{2}\phi
&=&-4\pi\rho,\\
\left(\nabla^{2}-
\frac{1}{c^{2}}\frac{\partial^{2}}{\partial t^{2}}\right)\vec A
&=&-\frac{4\pi}{c}\vec J+
\frac{1}{c}\vec \nabla\frac{\partial \phi}{\partial t}.
\end{eqnarray}
Un estudio m\'as detallado sobre las posibles condiciones
de norma de la electrodin\'amica se puede ver en \cite{henneaux:gnus}.\\

\section{Representaci\'on compleja de las
ecuaciones de Maxwell}

Cuando no hay fuentes, $\rho=0$ y $\vec J=0,$
las ecuaciones de Maxwell toman la forma 
\begin{eqnarray}
\vec\nabla\cdot \vec E&=&0,\\
\vec\nabla\times \vec E&=&-\frac{1}{c}\frac{\partial \vec B}{\partial t},\\
\vec \nabla\cdot \vec B&=&0,\\
\vec\nabla\times \vec B&=&\frac{1}{c}\frac{\partial \vec E}{\partial t}.
\end{eqnarray}
En este caso podemos definir el vector $\vec{\cal E}=\vec E+i\vec B,$
donde $i^{2}=-1.$ Por lo que las ecuaciones de Maxwell se pueden escribir
como
\begin{eqnarray}
\vec\nabla\cdot \vec {\cal E}=0,\qquad
\vec\nabla\times \vec {\cal E}=\frac{i}{c}
\frac{\partial \vec {\cal E}}{\partial t}.
\end{eqnarray}
Claramente estas ecuaciones son invariantes bajo la transformaci\'on
\begin{eqnarray}
\vec {\cal E}\quad \to \quad 
\vec {\cal E}^{\prime}=\left(\vec E^{\prime}+i \vec B^{\prime}\right)=e^{i\alpha}\vec {\cal E}, 
\qquad \alpha={\rm constante}.
\end{eqnarray}
Esta transformaci\'on expl\'{\i}citamente toma la forma  
\begin{eqnarray} 
\vec {\cal E}^{\prime}= \left(\vec E^{\prime}+i \vec B^{\prime}\right)
=e^{i\alpha}\vec {\cal E}=
(\vec E\cos\alpha-\vec B\sin\alpha )+i(\vec B\cos\alpha+\vec E\sin\alpha),
\nonumber
\end{eqnarray}
que se puede expresar como
\begin{eqnarray} 
\left(
\begin{array}{c}
\vec E^{\prime}\\
 \vec B^{\prime}
\end{array}
\right)
=\left(
\begin{array}{cc}
\cos\alpha& -\sin\alpha\\ 
\sin\alpha&   \cos\alpha\\
\end{array}
\right)
\left(
\begin{array}{c}
\vec E^{\prime}\\
 \vec B^{\prime}
\end{array}
\right).
\end{eqnarray}
Si hay fuentes, las ecuaciones de Maxwell no son invariantes bajo estas transformaciones. 
Para mantener esta invariancia, en 1931 P. A. M Dirac propuso la existencia de cargas y 
corrientes magn\'eticas. En este caso
se pueden proponer la ecuaciones de Maxwell generalizadas:
\begin{eqnarray}
\vec\nabla\cdot \vec E&=&4\pi \rho_{e},\\
\vec\nabla\times \vec E&=&-\left(\frac{4\pi}{c}\vec J_{m}+
\frac{1}{c}\frac{\partial \vec B}{\partial t}\right),\\
\vec \nabla\cdot \vec B&=&4\pi \rho_{m},\\
\vec\nabla\times \vec B&=&\frac{4\pi}{c}\vec J_{e}+
\frac{1}{c}\frac{\partial \vec E}{\partial t}.
\end{eqnarray}
Definamos  ${\bf \rho}= \rho_{e}+i\rho_{m}$ y 
${\bf \vec J}=\vec J_{e}+i\vec J_{m},$ entonces las ecuaciones de Maxwell con
monopolos magn\'eticos toman la forma
\begin{eqnarray}
\vec\nabla\cdot \vec {\cal E}=4\pi{\bf \rho} ,\qquad
\vec\nabla\times \vec {\cal E}=i\left(\frac{4\pi}{c}{\bf \vec J}+\frac{1}{c}
\frac{\partial \vec {\cal E}}{\partial t}\right).
\end{eqnarray}
Estas ecuaciones de Maxwell son invariantes bajo las transformaciones
\begin{eqnarray}
\vec {\cal E}\quad &\to& \quad e^{i\alpha}\vec {\cal E}=
(\vec E\cos\alpha-\vec B\sin\alpha )+i(\vec B\cos\alpha+\vec E\sin\alpha), \\
 {\bf \rho}\quad &\to& \quad e^{i\alpha} {\bf \rho}=
(\rho_{e}\cos\alpha-\rho_{m}\sin\alpha)+i(\rho_{m} \cos\alpha+
\rho_{e}\sin\alpha),\\
\vec {\bf J }\quad &\to& \quad e^{i\alpha}\vec {\bf J}=
(\vec J_{e}\cos\alpha-\vec J_{m}\sin\alpha )+i
(\vec J_{m} \cos\alpha+ \vec J_{e}\sin\alpha).
\end{eqnarray}
La fuerza de Lorentz con monopolos magn\'eticos toma la forma
\begin{eqnarray}
\vec F=q_{e}\vec E +q_{m}\vec B +
\frac{q_{e}}{c}\vec v\times \vec B
-\frac{q_{m}}{c}\vec v\times \vec E.
\end{eqnarray}

\section{Otros resultados de c\'alculo vectorial}

Antes de finalizar  este cap\'itulo veamos otros resultados de c\'alculo vectorial que ocuparemos 
posteriormente.\\

Ahora, demostraremos los siguientes lemas.\\
{\bf Lema de Gauss}: Si $\vec B$ es un campo vectorial suave sobre una regi\'on
de volumen $V$ y frontera $\partial V,$ se cumple
\begin{eqnarray}
\int_{V}\left(\vec \nabla \times \vec B\right)dv=
\oint_{\partial V} \left(d\vec a\times \vec B\right).
\label{eq:lemgauss2}
\end{eqnarray}
Para probar esta afirmaci\'on ocuparemos 
el teorema de Gauss Eq. (\ref{eq:tgauss}), el cual es es v\'alido para cualquier campo vectorial suave
$\vec F.$ En particular, si $\vec F=\vec B\times \vec C,$ el teorema de Gauss Eq. (\ref{eq:tgauss}) implica
\begin{eqnarray}
\int_{V} \vec \nabla \cdot\left(\vec B\times \vec C\right)dv=
\oint_{\partial V}\left(\vec B\times \vec C\right)\cdot \hat n da.
\label{eq:lemgauss}
\end{eqnarray}
Adem\'as ocupando la identidad del triple producto escalar Eq. (\ref{eq:triple-escalar})
se tiene 
\begin{eqnarray}
\oint_{\partial V}\left(\vec B\times \vec C\right)\cdot \hat n da
=\oint_{\partial V}\left(\hat n\times \vec B\right)\cdot\vec C da.
\end{eqnarray}
Para el caso en que  $\vec C$ es un vector constante, se encuentra 
\begin{eqnarray}
\oint_{\partial V}\left(\vec B\times \vec C\right)\cdot \hat n da
=\vec C\cdot \oint_{\partial V}\left(\hat n\times \vec B\right)da.
\label{eq:lemgauss1}
\end{eqnarray}
Adem\'as, si $\vec C$ es constante,  se cumple 
\begin{eqnarray}
\vec \nabla \cdot\left(\vec B\times \vec C\right)&=&
\partial_{i}\left(\vec B\times \vec C\right)_{i}=
\partial_{i}(\epsilon_{ijk}B_{j}C_{k})=
C_{k}\epsilon_{ijk}\partial_{i}B_{j}\nonumber\\
&=& C_{k}\epsilon_{kij}\partial_{i}B_{j}
=\vec C\cdot\left(\vec \nabla\times \vec B\right).
\nonumber
\end{eqnarray}
Por lo tanto, si $\vec C$ es constante
\begin{eqnarray}
\int_{V} \vec \nabla \cdot\left(\vec B\times \vec C\right)dv=
\vec C\cdot\int_{V}dv \left(\vec \nabla\times \vec B\right).
\label{eq:lemgauss1}
\end{eqnarray}
As\'{\i}, igualando Eq. (\ref{eq:lemgauss}) con Eq. (\ref{eq:lemgauss1}) para
el caso $\vec C$ constante se llega
\begin{eqnarray}
\vec C\cdot
\int_{V}\left(\vec \nabla \times \vec B\right)dv= 
\vec C\cdot \oint_{\partial V} \left(d\vec a\times \vec B\right),
\nonumber
\end{eqnarray}
es decir 
\begin{eqnarray}
\vec C\cdot\Bigg[
\int_{V}\left(\vec \nabla \times \vec B\right)dv-
\oint_{\partial V} \left(d\vec a\times \vec B\right)\Bigg]=0.\nonumber
\end{eqnarray}
Esta igualdad es v\'alida para cualquier vector $\vec C$ constante, por lo tanto
se debe cumplir Eq. (\ref{eq:lemgauss2}), que es el llamado lema de Gauss.\\

{\bf Lema de Stokes}: Si $\phi$ es un campo escalar que toma valores sobre una superficie
$S$ cuya frontera es  $\Gamma,$ entonces
\begin{eqnarray}
\int_{S}\left(\hat n\times\vec \nabla \phi\right) da
=\oint_{\Gamma} \phi   d\vec l.\label{eq:lema-gauss}
\end{eqnarray}
Probemos esta igualdad, supongamos que $\vec A$ es un vector constant, 
entonces de (\ref{eq.lemsto}) se llega a
\begin{eqnarray}
\vec \nabla \times (\phi\vec A) =  (\vec \nabla \phi \times \vec A).
\nonumber
\end{eqnarray}
Ocupando este resultado y la identidad del triple producto escalar
Eq. (\ref{eq:triple-escalar}), se encuentra 
\begin{eqnarray}
\int_{S} \left(\vec \nabla \times \left(\phi\vec A\right)\right) 
\cdot \hat nda&=&
\int_{S} \left(\vec \nabla \phi \times \vec A\right)\cdot \hat nda
=\int_{S}\left(\hat n\times\vec \nabla \phi\right)\cdot \vec A da
\nonumber\\
&=&\vec A \cdot\int_{S}\left(\hat n\times\vec \nabla \phi\right) da.
\label{eq:lemsto1}
\end{eqnarray}
Adem\'as, de acuerdo al teorema de Stokes Eq. (\ref{eq:tstokes}), se llega a
\begin{eqnarray}
\int_{S} \left(\vec \nabla \times \left(\phi\vec A\right)\right) 
\cdot \hat nda=
\oint_{\Gamma}  \left(\phi\vec A\right) \cdot d\vec l=
\vec A \cdot\oint_{\Gamma} \phi   d\vec l.
\label{eq:lemsto2}
\end{eqnarray}
Por lo tanto, igualando Eq. (\ref{eq:lemsto1}) con Eq. (\ref{eq:lemsto2}) se tiene
\begin{eqnarray}
\vec A \cdot 
\int_{S}\left(\hat n\times\vec \nabla \phi\right) da= 
\vec A \cdot\oint_{\Gamma} \phi   d\vec l.\nonumber
\end{eqnarray}
Como $\vec A$ es un vector arbitrario constante, se cumple  el lema de Stokes Eq. (\ref{eq:lema-gauss}).

\chapter{Operadores en Coordenadas Curvil\'{\i}neas}

En este cap\'{\i}tulo veremos la expresi\'on de  los operadores
gradiente, Laplaciano y rotacional en t\'erminos de coordenadas curvil\'{\i}neas.
Primero recordaremos algunos resultados  de c\'alculo
vectorial en coordenadas cartesianas y des\-pu\'es veremos el caso general 
en coordenadas curvil\'{\i}neas.

\section{Interpretaci\'on geom\'etrica de operaciones vectoriales}

Supongamos que tenemos los vectores $\vec A$ y $\vec B$  con magnitudes  $A$ y $B$. 
Recordemos que cualquiera dos vectores
los podemos poner en un plano, por lo que, sin perdida de generalidad,
podemos tomar 
$\vec A=A(\cos\theta_{1},\sin\theta_{1}),\vec B=B(\cos\theta_{2},\sin\theta_{2}).$
Entonces, 
\begin{eqnarray}
\vec A\cdot \vec B=AB(\cos\theta_{1}\cos\theta_{2}+\sin\theta_{1}\sin\theta_{2})=
AB\cos(\theta_{1}-\theta_{2}). \nonumber
\end{eqnarray}

Es decir, como $\theta=\theta_{1}-\theta_{2}$ es el \'angulo entre los vectores $\vec A$ y $\vec B,$
se tiene
\begin{eqnarray}
\vec A\cdot \vec B= AB\cos \theta.\label{eq:geoesca}
\end{eqnarray}
Tambi\'en recordemos que  dados los vectores $\vec A$ y $\vec B,$ siempre se puede
construir un paralelogramo. El \'area  de un paralelogramos es simplemente el producto de la
base por la altura $h,$ en este caso es \cite{courant-calculo:gnus}
\begin{eqnarray}
a=hA=AB\sin\theta.\nonumber
\end{eqnarray}
%
%
Esta cantidad se puede relacionar con  $\vec A\times\vec B.$
Primero notemos que ocupando el triple producto escalar Eq. (\ref{eq:triple-escalar}),
el triple producto vectorial Eq. (\ref{eq:trivec}) y el producto escalar Eq. (\ref{eq:geoesca}), se encuentra
\begin{eqnarray}
\left(\vec A\times \vec B\right)^{2}&=&\left(\vec A\times \vec B\right)\cdot \left(\vec A\times \vec B\right)
=\vec B\cdot\left(\left(\vec A\times\vec B\right)\times \vec A\right)\nonumber\\
& =& -\vec B\cdot \left(\vec A\times \left(\vec A\times \vec B\right)\right) \nonumber\\
&=&-\vec B\cdot \left(\vec A\left(\vec A\cdot B\right)- A^{2}\vec B\right)\nonumber\\
&=&A^{2}B^{2}-\left(\vec A\cdot  B\right)^{2}\nonumber\\
&=&A^{2}B^{2}-A^{2}B^{2}\cos^{2}\theta\nonumber\\
&=&A^{2}B^{2}\sin^{2}\theta=a^{2}.\nonumber
\end{eqnarray}
Entonces, el \'area del paralelogramo que forman $\vec A$ y $\vec B$ est\'a dada por
\begin{eqnarray}
a=AB\sin\theta=|\vec A\times \vec B|.\nonumber
\end{eqnarray}
Por esta raz\'on a $\vec A\times \vec B$ se le  suele llamar vector \'area.
Note que este vector  es normal al paralelogramo.\\

Otro resultado que es conveniente tener presente es que si tenemos tres vectores $\vec A,\vec B,\vec C$ con ellos podemos formar
un paralelep\'{\i}pedo. El volumen de un paralelep\'{\i}pedo
est\'a dado por el producto de la altura con el \'area de su base \cite{courant-calculo:gnus}
\begin{eqnarray}
V=ha.\nonumber
\end{eqnarray}
Donde $h=C\cos\gamma,$ con $\gamma$ el \'angulo que hace $\vec C$ con 
la normal de la base del paralelep\'{\i}pedo, es decir con $\vec A\times \vec B.$
Adem\'as, $a=|\vec A\times \vec B|,$ de donde
\begin{eqnarray}
V=C\cos\gamma |\vec A\times \vec B|=\bigg|\vec C\cdot\left( \vec A\times \vec B\right)\bigg|.
\nonumber
\end{eqnarray}

\section{Operadores en coordenadas cartesianas}
 
Un vector tridimensional $\vec r=(x,y,z)$ se puede escribir en diferentes
coordenadas. En la base Euclidiana tenemos 
\begin{eqnarray}
\vec r=x\hat i+y\hat j+z\hat k, \label{eq:pos-cart}
\end{eqnarray}
con
\begin{eqnarray}
\hat i=(1,0,0),\qquad \hat j=(0,1,0),\qquad \hat k=(0,0,1).
\end{eqnarray}
Para estos vectores el producto escalar es
\begin{eqnarray}
\hat i\cdot \hat i&=&  
\hat j\cdot \hat j=  
\hat k\cdot \hat k=1,\nonumber \\
\hat i\cdot \hat j&=&  \hat k\cdot \hat i=  
\hat j\cdot \hat k=0.\nonumber
\end{eqnarray}
Otro vector que se puede construir es
\begin{eqnarray}
d\vec r=dx\hat i+dy\hat j+dz\hat k,
\end{eqnarray}
que nos da el elemento de l\'{\i}nea
\begin{eqnarray}
ds^{2}=d\vec r\cdot d\vec r =dx^{2}+dy^{2} +dz^{2}.
\end{eqnarray}
Note que la energ\'{\i}a cin\'etica est\'a dada por
\begin{eqnarray}
T=\frac{m}{2}\left(\frac{ds}{dt}\right)^{2}=\frac{m}{2}\frac{d\vec r}{dt}\cdot \frac{d\vec r}{dt}
 =\frac{m}{2}\left(\dot x^{2}+\dot y^{2} +\dot z^{2}\right).\nonumber
\end{eqnarray}
En esta base se tiene los productos vectoriales
\begin{eqnarray}
\hat i\times \hat j= \hat k,\qquad 
\hat k\times \hat i= \hat j,\qquad 
\hat j\times \hat k= \hat i.\nonumber
\end{eqnarray}
As\'{\i}, se pueden plantear los elementos de \'area
\begin{eqnarray}
d\vec a_{xy}&=&d\vec r_{x}\times d\vec r_{y}=\left(dx\hat i\times dy\hat j\right)= dxdy \hat k,\nonumber \\ 
d\vec a_{yz}&=&d\vec r_{y}\times d\vec r_{z}=\left(dy\hat j\times dz\hat k\right)= dydz \hat i, \nonumber\\
d\vec a_{zx}&=&d\vec r_{z}\times d\vec r_{x}=\left(dz\hat k\times dx\hat i\right)= dzdx \hat j,\nonumber
\end{eqnarray}
que forman el vector
\begin{eqnarray}
d\vec a=dydz \hat i+dzdx \hat j+dxdy \hat k.\label{eq:area-cart}
\end{eqnarray}
Mientras que el elemento de volumen se plantea
como
\begin{eqnarray}
dV=d\vec r_{z}\cdot \left(d\vec r_{x}\times d\vec r_{y}\right)=dxdydz.
\label{eq:vol-cart}
\end{eqnarray}
En este caso el operador gradiente es
\begin{eqnarray}
\vec \nabla \phi=\frac{\partial \phi }{\partial x} \hat i+
\frac{\partial \phi }{\partial y} \hat j+
\frac{\partial \phi }{\partial z} \hat k. \label{eq:grad-cart}
\end{eqnarray}
Por lo que un  {\it elemento}  de $\phi$ se puede escribir como
\begin{eqnarray}
 d\phi=\frac{\partial \phi }{\partial x} dx+
\frac{\partial \phi }{\partial y} dy+
\frac{\partial \phi }{\partial z} dz= \vec \nabla \phi\cdot d\vec r.
\label{eq:elemento-funcion}
\end{eqnarray}
Adem\'as, la divergencia es 
\begin{eqnarray}
\vec \nabla \cdot \vec A=
\frac{\partial A_{x} }{\partial x} +
\frac{\partial A_{y} }{\partial y} +
\frac{\partial A_{z} }{\partial z},\label{eq:divergencia-carteciana}
\end{eqnarray}
tomando el caso $\vec A=\vec \nabla\phi$ se obtiene el Laplaciano: 
\begin{eqnarray}
\vec \nabla\cdot \vec \nabla \phi= \nabla^{2} \phi=
\frac{\partial^{2} \phi }{\partial x^{2}} +
\frac{\partial^{2} \phi }{\partial y^{2} } +
\frac{\partial^{2} \phi }{\partial z^{2}}.
\end{eqnarray}
Mientras que el rotacional es
\begin{eqnarray}
\vec \nabla \times \vec A=
\left(\frac{\partial A_{z} }{\partial y}- 
\frac{\partial A_{y} }{\partial z}\right) \hat i+
\left(\frac{\partial A_{x} }{\partial z}- 
\frac{\partial A_{z} }{\partial x}\right) \hat j+
\left(\frac{\partial A_{y} }{\partial x}- 
\frac{\partial A_{x} }{\partial y}\right) \hat k.
\label{eq:rotacional-carteciana} 
\end{eqnarray}
Estos resultados los ocuparemos para construir la versi\'on del gradiente, divergencia, Laplaciano y
rotacional en coordenadas curvil\'ineas.

\subsection{Coordenadas esf\'ericas}

Antes de estudiar las coordenadas curvil\'ineas en general veremos dos casos particulares.\\

Primero veamos las coordenadas esf\'ericas:
\begin{eqnarray}
x=r\cos\varphi \sin\theta,\quad y=r\sin\varphi\sin\theta,\quad z=r\cos\theta,
\end{eqnarray}
es decir
\begin{eqnarray}
\vec r&=&r(\cos\varphi \sin\theta,\sin\varphi\sin\theta,\cos\theta)\nonumber\\
&= & r(\cos\varphi \sin\theta\hat i+ \sin\varphi\sin\theta \hat j+ \cos\theta\hat k).
\label{eq:pos-esfe}
\end{eqnarray}
As\'{\i}, 
\begin{eqnarray}
d\vec r= \hat e_{r}dr+r \hat e_{\theta} d\theta +r\sin\theta \hat e_{\varphi}d\varphi,
\end{eqnarray}
con
\begin{eqnarray}
\hat e_{r}&=&(\cos\varphi \sin\theta,\sin\varphi\sin\theta,\cos\theta) \nonumber\\
&=& \cos\varphi \sin\theta\hat i+ \sin\varphi\sin\theta \hat j+\cos\theta\hat k,\label{eq:esfe-uni1} \\
\hat e_{\theta}&=&\frac{\partial \hat e_{r}}{\partial \theta}=(\cos\varphi \cos\theta, \sin\varphi \cos\theta, 
-\sin\theta),\nonumber\\
&=& \cos\varphi \cos\theta\hat i+ \sin\varphi \cos\theta \hat j+ 
-\sin\theta\hat k,\label{eq:esfe-uni2}\\
\hat e_{\varphi}&=&\frac{1}{\sin\theta} \frac{\partial \hat e_{r}}{\partial \varphi}=(-\sin\varphi, \cos\varphi)
\nonumber\\
&=& -\sin\varphi\hat i+ \cos\varphi\hat j.\label{eq:esfe-uni3}
\end{eqnarray}
Los vectores $\hat e_{r}, \hat e_{\theta}, \hat e_{\varphi}$  cumplen
\begin{eqnarray}
\hat e_{r}\cdot \hat e_{r}&=&
\hat e_{\theta}\cdot \hat e_{\theta}=\hat e_{\varphi}\cdot \hat e_{\varphi}=1,\\
\hat e_{r}\cdot \hat e_{\varphi}&=&
\hat e_{r}\cdot \hat e_{\theta}=\hat e_{\varphi}\cdot \hat e_{\theta}=0,
\end{eqnarray}
por lo que forman una base ortonormal.\\

Esto implica que el elemento de l\'{\i}nea toma la forma 
\begin{eqnarray}
ds^{2}&=&d\vec r\cdot d\vec r=dr^{2}+r^{2}d\theta^{2}+r^{2}\sin^{2}\theta d\varphi^{2}\nonumber\\
& =&dr^{2}+r^{2}\left(d\theta^{2}+\sin^{2}\theta d\varphi^{2}\right),
\end{eqnarray}
mientras que la energ\'{\i}a cin\'etica es 
\begin{eqnarray}
T=\frac{m}{2}\left(\frac{ds}{dt}\right)^{2}=\frac{m}{2}
\left(\dot r^{2}+r^{2}\left(\dot \theta^{2}+\sin^{2}\theta \dot \varphi^{2}\right)\right).
\end{eqnarray}
 Con el producto vectorial se tiene
\begin{eqnarray}
\hat e_{r}\times \hat e_{\theta}=\hat e_{\varphi},\qquad
\hat e_{\varphi}\times \hat e_{r}=\hat e_{\theta},\qquad
\hat e_{\theta}\times \hat e_{\varphi}=\hat e_{r}.
\end{eqnarray}
Por lo que los elementos de \'area son
\begin{eqnarray}
d\vec a_{r\theta}&=& d\vec r_{r}\times d\vec r_{\theta}= 
\left( \hat e_{r}dr \times r\hat e_{\theta} d\theta\right)=rdrd\theta\hat e_{\varphi},\\
d\vec a_{\varphi r}&=& d\vec r_{\varphi}\times d\vec r_{r}= 
\left(r \sin\theta\hat e_{\varphi} d\varphi\times \hat e_{r} dr\right)=r\sin\theta drd\varphi\hat e_{\theta},\\
d\vec a_{\theta\varphi}&=& d\vec r_{\theta}\times d\vec r_{\varphi}= 
\left( r\hat e_{\theta}d\theta \times r\sin\theta\hat e_{\varphi} d\varphi\right)=
r^{2}\sin\theta d\theta d\varphi\hat e_{r},
\end{eqnarray}
que forman el vector
\begin{eqnarray}
d\vec a=r^{2}\sin\theta d\theta d\varphi\hat e_{r}+r\sin\theta drd\varphi\hat e_{\theta}
+rdrd\theta\hat e_{\varphi}. \label{eq:area-esfe}
\end{eqnarray}
Para este caso el elemento de volumen es
\begin{eqnarray}
dV=d\vec r_{r}\cdot \left(d\vec r_{\theta}\times d\vec r_{\varphi}\right)=
r^{2}\sin\theta drd\theta d\varphi =r^{2}drd\Omega, 
\quad d\Omega=\sin\theta d\theta d\varphi,
\label{eq:vol-esfe}
\end{eqnarray}
a $d\Omega$ se le llama elemento de \'angulo s\'olido.\\

Adem\'as ocupando Eqs. (\ref{eq:esfe-uni1})-(\ref{eq:esfe-uni3}) se encuentra
\begin{eqnarray}
\hat i&=& \cos\varphi\sin\theta \hat e_{r}+
\cos\varphi\cos\theta \hat e_{\theta}-\sin\varphi\hat e_{\varphi},\\
\hat j&=& \sin\varphi\sin\theta \hat e_{r}+
\sin\varphi\cos\theta \hat e_{\theta}+\cos\varphi\hat e_{\varphi},\\
\hat k&=& \cos\theta \hat e_{r}-\sin\theta \hat e_{\theta}.
\end{eqnarray}

\subsection{Coordenadas cil\'{\i}ndricas}

Ahora veamos la transformaci\'on de coordenadas cil\'{\i}ndricas
\begin{eqnarray}
x=\rho \cos \varphi,\qquad
y=\rho\sin\varphi,\qquad
z=z,
\end{eqnarray}
es decir,
\begin{eqnarray}
\vec r &=&( \rho \cos\varphi, \rho \sin \varphi, z)\nonumber\\
 &=& \rho \cos\varphi\hat i +\rho \sin \varphi\hat j +z\hat k.
\end{eqnarray}
De donde, 
\begin{eqnarray}
d\vec r =\hat e_{\rho} d\rho+\rho \hat e_{\varphi}d\varphi+\hat e_{z}dz,
\end{eqnarray}
con
\begin{eqnarray}
\hat e_{\rho}&=&( \cos \varphi,\sin\varphi, 0)
= \cos \varphi \hat i+\sin\varphi \hat j,\\
\hat  e_{\varphi}&=&\frac{\partial \hat e_{\rho}}{\partial \varphi}=(- \sin \varphi,\cos\varphi, 0)
=- \sin \varphi\hat i+\cos\varphi \hat k,\\
\hat  e_{z}&=&(0,0,1)=\hat k.
\end{eqnarray}
Estos vectores son ortonormales, pues cumplen
\begin{eqnarray}
\hat e_{\rho}\cdot\hat  e_{\rho}&=&
\hat e_{\varphi}\cdot \hat e_{\varphi}=
\hat  e_{z}\cdot \hat e_{z}=1,\\
\hat e_{\rho}\cdot\hat  e_{\varphi}&=&
\hat e_{\rho}\cdot \hat e_{z}=
\hat  e_{\varphi}\cdot \hat e_{z}=0,
\end{eqnarray}
que  implica
\begin{eqnarray}
ds^{2}=d\vec r \cdot d \vec r = d\rho^{2}+\rho^{2} d\varphi^{2}+dz^{2}.
\end{eqnarray}
As\'{\i}, la energ\'{\i}a cin\'etica toma la forma
\begin{eqnarray}
T=\frac{m}{2}\left(\frac{ds}{dt}\right)^{2}=\frac{m}{2}
\left( \dot\rho^{2}+\rho^{2} \dot\varphi^{2}+\dot z^{2}\right).
\end{eqnarray}
Con el producto vectorial tenemos
\begin{eqnarray}
\hat e_{\varphi}\times \hat  e_{z}=\hat e_{\rho},\qquad
\hat e_{\rho}\times \hat  e_{\varphi}=\hat e_{z},\qquad
\hat e_{z} \times \hat e_{\rho}=\hat e_{\varphi}.
\end{eqnarray}
Por lo que los elementos de \'area son
\begin{eqnarray}
d\vec a_{\rho\varphi}&=& d\vec r_{\rho}\times d\vec r_{\varphi}=\rho d\rho d\varphi \hat e_{z},\nonumber\\
d\vec a_{z\rho}&=& d\vec r_{z}\times d\vec r_{\rho}=dz d\rho \hat e_{\varphi},\nonumber \\
d\vec a_{\varphi z}&=& d\vec r_{\varphi}\times d\vec r_{z}=\rho d\varphi dz\hat e_{\varphi},\nonumber
\end{eqnarray}
que definen el vector elemento \'area
\begin{eqnarray}
d\vec a= \rho d\varphi dz\hat e_{\rho}+dz d\rho \hat e_{\varphi}+\rho d\rho d\varphi \hat e_{z}.
\label{eq:area-cilin}
\end{eqnarray}
Mientras que el elemento de volumen est\'a dado por
\begin{eqnarray}
dV=d\vec r_{z} \cdot \left( d\vec r_{\rho}\times d\vec r_{\varphi} \right)=\rho d\rho d\varphi dz.
\label{eq:vol-cilin}
\end{eqnarray}

\section{Coordenadas curvil\'{\i}neas ortogonales}

Ya hemos practicado suficiente, ahora veamos el caso general. 
Supongamos que tenemos el cambio de coordenadas
\begin{eqnarray}
x=f_{1}\left(u_{1},u_{2},u_{3}\right),\quad y=f_{2}\left(u_{1},u_{2},u_{3}\right),
\quad z=f_{3}\left(u_{1},u_{2},u_{3}\right),
\end{eqnarray} 
es decir
\begin{eqnarray}
\vec r=\left( f_{1}\left(u_{1},u_{2},u_{3}\right),f_{2}\left(u_{1},u_{2},u_{3}\right),
f_{3}\left(u_{1},u_{2},u_{3}\right)\right).
\end{eqnarray}
De donde,
\begin{eqnarray}
d\vec r=h_{1}\hat e_{1}du^{1}+h_{2}\hat e_{2}du^{2}+h_{2}\hat e_{3}du^{3},\label{eq:dvec-curv}
\end{eqnarray}
con 
\begin{eqnarray}
\hat e_{1}&=&\frac{1}{h_{1}}\frac{\partial \vec r}{\partial u^{1}},\quad 
\hat e_{2}=\frac{1}{h_{2}}\frac{\partial \vec r}{\partial u^{2}},\quad 
\hat e_{3}=\frac{1}{h_{3}}\frac{\partial \vec r}{\partial u^{3}},\\
h_{1}&=& \bigg|\frac{\partial \vec r}{\partial u^{1}}\bigg|,\quad h_{2}=\bigg|\frac{\partial \vec r}{\partial u^{2}}\bigg|,\quad
h_{3}= \bigg|\frac{\partial \vec r}{\partial u^{3}}\bigg|.
\end{eqnarray}
Claramente los vectores $\hat e_{1},\hat e_{2},\hat e_{3}$ son unitarios. 
Supondremos que la base $\hat e_{1},\hat e_{2},\hat e_{3}$ forma una base ortonormal 
\begin{eqnarray}
\hat e_{1}\cdot \hat e_{1}&=&
\hat e_{2}\cdot \hat e_{2}=\hat e_{3}\cdot \hat e_{3}=1,\\
\hat e_{1}\cdot \hat e_{2}&=&\hat e_{1}\cdot \hat e_{3}= 
\hat e_{2}\cdot \hat e_{3}=0.
\end{eqnarray}
Esto implica que el elemento de l\'{\i}nea  
\begin{eqnarray}
ds^{2}&=&d\vec r\cdot d\vec r=\left(h_{1}\right)^{2}\left(du_{1}\right)^{2}+
\left(h_{2}\right)^{2}\left(du_{2}\right)^{2}+
\left(h_{3}\right)^{2}\left(du_{3}\right)^{2},
\end{eqnarray}
por lo que la energ\'ia cin\'etica toma forma 
\begin{eqnarray}
T&=&\frac{m}{2}\left(\frac{ds}{dt}\right)^{2}=\frac{m}{2}\left(\left(h_{1}\right)^{2}\dot u_{1}^{2}+
\left(h_{2}\right)^{2}\dot u_{2}^{2}+
\left(h_{3}\right)^{2}\dot u_{3}^{2}\right).
\end{eqnarray}
Tambi\'en supondremos que $\hat e_{1},\hat e_{2},\hat e_{3}$ forman una base derecha, es decir
\begin{eqnarray}
\hat e_{1}\times\hat e_{2}&=&\hat e_{3},\quad  \hat e_{2}\times \hat e_{3}=\hat e_{1} ,\quad 
\hat e_{3}\times \hat e_{1}= \hat e_{2},
\end{eqnarray}
que implica los elementos de \'area 
\begin{eqnarray}
d\vec a_{12}&=&d\vec r_{1}\times d\vec r_{2} =h_{1}h_{2}\left(\hat e_{1}\times\hat e_{2}\right) du^{1}du^{2}=
h_{1}h_{2}du^{1}du^{2}\hat e_{3},\\
d\vec a_{23}&=&d\vec r_{2}\times d\vec r_{3} = h_{2}h_{3}\left(\hat e_{2}\times\hat e_{3}\right) du^{2}du^{3}=
h_{2}h_{3}du^{2}du^{3}\hat e_{1},\\
d\vec a_{31}&=&d\vec r_{1}\times d\vec r_{2}= h_{3}h_{1}\left(\hat e_{3}\times\hat e_{1}\right) du^{3}du^{1}=
h_{3}h_{1}du^{3}du^{1}\hat e_{2},
\end{eqnarray}
con el cual se forma el vector
\begin{eqnarray}
d\vec a= h_{2}h_{3}du^{2}du^{3}\hat e_{1}+h_{3}h_{1}du^{3}du^{1}\hat e_{2}
+h_{1}h_{2}du^{1}du^{2}\hat e_{3}.\label{eq:area-curv}
\end{eqnarray}
Adem\'as, el elemento de volumen est\'a dado por
\begin{eqnarray}
d V&=&d\vec r_{1}\cdot\left( d\vec r_{2}\times\vec r_{3}\cdot\right)=
h_{1}du^{(1)} \hat e_{1}\cdot \left( h_{2}du^{(2)} \hat e_{2}\times h_{3}du^{(3)}\hat e_{3}\right)
\nonumber\\ 
&=&h_{1}h_{2} h_{3}du^{(1)}du^{(2)}du^{(3)}.\label{eq:vol-curv}
\end{eqnarray}
Estas cantidades son de gran utilidad para construir operadores diferenciales
en coordenadas curvil\'{\i}neas.

\subsection{Gradiente en coordenadas curvil\'{\i}neas}

Como $\hat e_{1},\hat e_{2},\hat e_{3}$ forma una base, cualquier vector $\vec A$  se 
puede escribir en t\'erminos de ella, es decir,
\begin{eqnarray}
\vec A =A_{1} \hat e_{1}+ A \hat e_{2}
+ A_{3}\hat e_{3}.\label{eq:vec-curv}
\end{eqnarray}
En particular, el  gradiente se debe escribir como 
\begin{eqnarray}
\vec \nabla \phi =\left(\nabla \phi\right)_{1} \hat e_{1}+
\left(\nabla \phi\right)_{2} \hat e_{2}
+\left(\nabla \phi\right)_{3}\hat e_{3}.
\end{eqnarray}
De donde, considerando Eq. (\ref{eq:elemento-funcion}), se tiene 
\begin{eqnarray}
& &d\phi=\vec \nabla \phi\cdot d\vec r\nonumber\\
&=& 
\Bigg[\left(\nabla \phi\right)_{1} \hat e_{1}+
\left(\nabla \phi\right)_{2} \hat e_{2}
+\left(\nabla \phi\right)_{3}\hat e_{3}\Bigg]\cdot 
\Bigg[h_{1}\hat e_{1}du^{1}+h_{2}\hat e_{2}du^{2}+h_{2}\hat e_{3}du^{3}\Bigg]\nonumber\\
&=&\left(\nabla \phi\right)_{1}h_{1}du^{1}+
\left(\nabla \phi\right)_{2}h_{2}du^{2}+
\left(\nabla \phi\right)_{3}h_{3}du^{3}.\label{eq:genelem2}
\end{eqnarray}
Esta cantidad tiene el mismo valor independientemente de las coordenadas en que se calcule.
Adem\'as en las variables $u_{1},u_{2},u_{3}$ se encuentra
\begin{eqnarray}
d\phi(u^{1},u^{2},u^{3})=\frac{\partial \phi}{\partial u^{1}}du^{1}+
\frac{\partial \phi}{\partial u^{2}}du^{2}+
\frac{\partial \phi}{\partial u^{3}}du^{3}.\label{eq:genelem1}
\end{eqnarray}
Por lo tanto, como las variables $u_{1},u_{2},u_{3}$ son idependientes,
igualando Eq. (\ref{eq:genelem1}) con Eq. (\ref{eq:genelem2}) se llega a
\begin{eqnarray}
\left(\nabla \phi\right)_{1}h_{1}=\frac{\partial \phi}{\partial u^{1}},
\quad 
\left(\nabla \phi\right)_{2}h_{2}=\frac{\partial \phi}{\partial u^{2}},
\quad 
\left(\nabla \phi\right)_{3}h_{3}=\frac{\partial \phi}{\partial u^{3}},
\end{eqnarray}
es decir, 
\begin{eqnarray}
\left(\nabla \phi\right)_{1}=
\frac{1}{h_{1}}\frac{\partial \phi}{\partial u^{1}},
\quad 
\left(\nabla \phi\right)_{2}=
\frac{1}{h_{2}}\frac{\partial \phi}{\partial u^{2}},
\quad 
\left(\nabla \phi\right)_{3}=
\frac{1}{h_{3}}\frac{\partial \phi}{\partial u^{3}}.
\end{eqnarray}
As\'{\i}, el gradiente en coordenadas curvil\'ineas es
\begin{eqnarray}
\vec\nabla \phi=\frac{1}{h_{1}}\frac{\partial \phi}{\partial u^{1}}\hat e_{1}
+\frac{1}{h_{2}}\frac{\partial \phi}{\partial u^{2}}\hat e_{1}+
\frac{1}{h_{3}}\frac{\partial \phi}{\partial u^{3}}\hat e_{3}.
\end{eqnarray}
En particular el gradiente en coordenadas esf\'ericas es
\begin{eqnarray}
\vec \nabla \phi=\frac{\partial \phi}{\partial r}\hat e_{r}+
\frac{1}{r}\frac{\partial \phi}{\partial \theta}\hat e_{\theta}+
\frac{1}{r\sin\theta}\frac{\partial \phi}{\partial \varphi}\hat e_{\varphi}.
\label{eq:grad-esfe}
\end{eqnarray}
Mientras que en coordenadas cil\'{\i}ndricas tenemos
\begin{eqnarray}
\vec \nabla \phi=\frac{\partial \phi}{\partial \rho}\hat e_{\rho}+
\frac{1}{\rho}\frac{\partial \phi}{\partial \varphi}\hat e_{\varphi}+
\frac{\partial \phi}{\partial z}\hat e_{z}.
\end{eqnarray}

\subsection{Divergencia en coordenadas curvil\'{\i}neas}

Para obtener la divergencia en coordenadas  curvil\'{\i}neas ocuparemos  
el teorema de Gauss Eq. (\ref{eq:tgauss}). Note que 
usando los elementos de \'area Eq. (\ref{eq:area-cart}) y volumen Eq. (\ref{eq:vol-cart}) en coordenadas cartesianas,
este teorema se puede escribir como
\begin{eqnarray}
\int_{V} 
\left(\frac{\partial A_{x}}{\partial x} 
+\frac{\partial A_{y}}{\partial y}
+\frac{\partial A_{z}}{\partial z}\right)
dxdydz=\oint_{\partial V} \left(A_{x}dydz+ 
A_{y}dzdx+A_{z}dxdy\right).\quad 
\label{eq:gaussexpl}
\end{eqnarray}
Con el elemento de volumen Eq. (\ref{eq:vol-curv})
en coordenadas curvil\'{\i}neas se tiene
\begin{eqnarray}
\int_{V} \vec \nabla \cdot \vec A dv=
\int_{V} \left(\vec \nabla \cdot \vec A\right)_{c} h_{1}h_{2}h_{3}du^{(1)}
du^{(2)}du^{(3)}.
\label{eq:gaussexpl3}
\end{eqnarray}
Mientras que ocupando $\vec A$ y el elemento de \'area en coordenadas  curvil\'{\i}neas, Eq. (\ref{eq:vec-curv}) y Eq.  (\ref{eq:area-curv}), se encuentra
\begin{eqnarray}
\oint_{\partial V}\vec A\cdot d\vec a&=&\oint_{\partial V} 
\Bigg[A_{1}\hat e_{1}+A_{2}\hat e_{2}+A_{3}\hat e_{3}\Bigg]
\cdot\nonumber\\
& &\Bigg[h_{2}h_{3}\hat e_{1}du^{(2)}du^{(3)} +
h_{3}h_{1}\hat e_{2}du^{(3)}du^{(1)}+
h_{1}h_{2}\hat e_{3}du^{(1)}du^{(2)}\Bigg]\nonumber\\
&=& \oint_{\partial V} \Bigg[A_{1}h_{3}h_{1}du^{(3)}du^{(2)}+
A_{2} h_{3}h_{1}du^{(3)}du^{(1)}+
A_{3}h_{1}h_{2}du^{(1)}du^{(2)}\Bigg].
\nonumber
\end{eqnarray}
Por lo tanto, definiendo
\begin{eqnarray}
\tilde A_{x}&=&A_{1}h_{2}h_{3}, \qquad \tilde A_{y}=A_{2}h_{3}h_{1},\qquad  
\tilde A_{z}=A_{3}h_{1}h_{2},\nonumber\\
d\tilde x&=& du^{(1)},\quad d\tilde y= du^{(2)},\quad
 d\tilde z= du^{(3)}.\nonumber
\end{eqnarray}
 y considerando la igualdad Eq. (\ref{eq:gaussexpl}) se llega a
\begin{eqnarray}
& &\oint_{\partial V}\vec A\cdot d\vec a  =\oint_{\partial V} 
\Bigg[\tilde A_{x}\tilde zd\tilde y+  \tilde A_{y} d\tilde zd\tilde x+\tilde A_{z}d\tilde xd\tilde y .\Bigg]\nonumber\\
&=& \int_{V}\Bigg(\frac{\partial \tilde A_{x}}{\partial \tilde x}+ 
\frac{\partial \tilde A_{y}}{\partial \tilde y}+
\frac{\partial \tilde A_{z}}{\partial \tilde z}\Bigg)d\tilde xd\tilde yd\tilde z,
\nonumber\\
& =&\int_{V} du^{(1)}du^{(2)}du^{(3)}
\Bigg(\frac{\partial (A_{1}h_{2}h_{3})}{\partial u^{(1)}}+ 
\frac{\partial (A_{2}h_{3}h_{1})}{\partial u^{(2)}}+
\frac{\partial (A_{3}h_{1}h_{2})}{\partial u^{(3)}}\Bigg).
\label{eq:gaussexpl4}
\end{eqnarray}
As\'{\i}, igualando Eq. (\ref{eq:gaussexpl3}) con 
Eq. (\ref{eq:gaussexpl4}) se llega a
\begin{eqnarray}
\left(\vec \nabla \cdot \vec A \right)_{c} h_{1}h_{2}h_{3}=
\Bigg(\frac{\partial (A_{1}h_{2}h_{3})}{\partial u^{(1)}}+ 
\frac{\partial (A_{2}h_{3}h_{1})}{\partial u^{(2)}}+
\frac{\partial (A_{3}h_{1}h_{2})}{\partial u^{(3)}}\Bigg),
\end{eqnarray}
es decir
\begin{eqnarray}
\left(\vec \nabla \cdot \vec A\right)_{c}=\frac{1}{h_{1}h_{2}h_{3}}
\Bigg(\frac{\partial (A_{1}h_{2}h_{3})}{\partial u^{(1)}}+ 
\frac{\partial (A_{2}h_{3}h_{1})}{\partial u^{(2)}}+
\frac{\partial (A_{3}h_{1}h_{2})}{\partial u^{(3)}}\Bigg).
\label{eq:divgen}
\end{eqnarray}
Esta es la expresi\'on de la divergencia en coordenadas  curvil\'{\i}neas.
Para coordenadas esf\'ericas obtenemos
\begin{eqnarray}
\vec \nabla\cdot \vec A=
\frac{1}{r^{2}}\frac{\partial}{\partial r}\left(r^{2} 
A_{r} \right)+
\frac{1}{r\sin\theta}\frac{\partial (\sin\theta A_{\theta})}{\partial \theta}+
\frac{1}{r\sin\theta}\frac{\partial  A_{\varphi}}{\partial \varphi}.
\end{eqnarray}
Mientras que en coordenadas cil\'{\i}ndricas se llega a
\begin{eqnarray}
\vec \nabla\cdot \vec A=
\frac{1}{\rho}\frac{\partial}{\partial \rho}\left(\rho A_{\rho} \right)+
\frac{1}{\rho}\frac{\partial  A_{\varphi}}{\partial \varphi}+
\frac{\partial  A_{z}}{\partial z}.
\end{eqnarray}

\subsection{Laplaciano en coordenadas curvil\'{\i}neas}

La expresi\'on Eq.  (\ref{eq:divgen}) es v\'alida para cualquier campo
vectorial $\vec A,$ en particular si es el gradiente de un campo
escalar $\phi:$ 
\begin{eqnarray}
\vec A=\vec \nabla \phi=
\frac{1}{h_{1}}\frac{\partial \phi}{\partial u^{1}}\hat e_{1}
+\frac{1}{h_{2}}\frac{\partial \phi}{\partial u^{2}}\hat e_{1}+
\frac{1}{h_{3}}\frac{\partial \phi}{\partial u^{3}}\hat e_{3}.
\end{eqnarray}
En este caso $\vec \nabla\cdot \vec A=\nabla^{2}\phi,$ de donde
\begin{eqnarray}
\nabla^{2}\phi&=&\frac{1}{h_{1}h_{2}h_{3}}
\Bigg(\frac{\partial }{\partial u^{(1)}} 
\left(\frac{h_{2}h_{3}}{h_{1}} \frac{\partial \phi}{\partial u^{(1)}}\right)
+ \frac{\partial}{\partial u^{(2)}} 
\left(\frac{h_{3}h_{1} }{h_{2}}\frac{\partial \phi}{\partial u^{(2)}} \right)\nonumber\\
& & 
+ \frac{\partial}{\partial u^{(3)}}
\left(\frac{h_{1}h_{2}}{h_{3}}\frac{\partial\phi }{\partial u^{(3)}}
\right)\Bigg).\nonumber
\end{eqnarray}
Para coordenadas esf\'ericas se tiene
\begin{eqnarray}
\nabla^{2}\phi=\frac{1}{r^{2}}\frac{\partial}{\partial r}
\left(r^{2}\frac{\partial}{\partial r}\phi \right)+
\frac{1}{r^{2}\sin\theta}\frac{\partial}{\partial\theta }
\left(\sin\theta \frac{\partial \phi}{\partial \theta}\right)+
\frac{1}{r^{2}\sin^{2}\theta}\frac{\partial^{2}\phi}{\partial \varphi^{2} },
\label{eq:lapla-esfe0}
\end{eqnarray}
tambi\'en se puede ocupar
\begin{eqnarray}
\frac{1}{r^{2}}\frac{\partial}{\partial r}
\left(r^{2}\frac{\partial}{\partial r}\phi \right)=
\frac{1}{r}\frac{\partial^{2}}{\partial r^{2}}(r\phi).
\label{eq:lapla-esfe}
\end{eqnarray}
Mientras que en coordenadas cil\'{\i}ndricas se encuentra
\begin{eqnarray}
\nabla^{2}\phi=\frac{1}{\rho}\frac{\partial }{\partial \rho}
\left(\rho \frac{\partial \phi}{\partial \rho} \right)+
\frac{1}{\rho^{2}}\frac{\partial^{2}\phi}{\partial \varphi^{2}}+
\frac{\partial^{2}\phi}{\partial z^{2}}.
\end{eqnarray}

\subsection{Rotacional en coordenadas curvil\'{\i}neas}
El rotacional es un vector, por lo que se debe poder escribir de la forma
\begin{eqnarray}
\left(\vec \nabla \times \vec A\right)
= \left(\vec \nabla \times \vec A\right)_{1}\hat e_{1}+ 
\left(\vec \nabla \times \vec A\right)_{2}\hat e_{2}+
\left(\vec \nabla \times \vec A\right)_{3}\hat e_{3}.\label{eq:rot-curv-gen}
\end{eqnarray}
Para encontrar los coeficientes $ \left(\vec \nabla \times \vec A\right)_{i},$
notemos que el teorema de Stokes Eq. (\ref{eq:tstokes}) en coordenadas 
cartesianas toma la forma
\begin{eqnarray}
\int_{S}\left(\vec \nabla \times \vec A\right)\cdot d\vec a&=&
\int_{S} \Bigg[\left(\vec \nabla \times \vec A\right)_{x}dydz
+\left(\vec \nabla \times \vec A\right)_{y}dzdx \nonumber\\
& &+ 
\left(\vec \nabla \times \vec A\right)_{z}dxdy\Bigg] \nonumber\\
&=&
\int_{S}\Bigg[
\left(\frac{\partial A_{z} }{\partial y}- 
\frac{\partial A_{y} }{\partial z}\right) dydz+
\left(\frac{\partial A_{x} }{\partial z}- 
\frac{\partial A_{z} }{\partial x}\right)dxdy +\nonumber\\
& &
\left(\frac{\partial A_{y} }{\partial x}-
 \frac{\partial A_{x} }{\partial y}\right) dxdy\Bigg]\nonumber\\
&=& \int_{\partial S} \vec A\cdot d\vec r=
\int_{\partial S} (A_{x},A_{y},A_{z})\cdot 
(dx,dy,dz)\nonumber\\
&=&\int_{\partial S} (A_{x}dx+A_{y}dy+A_{z}dz)
.\label{eq:compa-curv}
\end{eqnarray}
Ahora, considerando $\vec A$ y $d\vec r $ en coordenadas generalizadas,
 Eq. (\ref{eq:vec-curv}) y  Eq. (\ref{eq:dvec-curv}), tenemos
\begin{eqnarray}
\int_{\partial S} \vec A\cdot d\vec r&=&
\int_{\partial S} \left(A_{1}\hat e_{1}+
A_{2}\hat e_{2}+A_{3}\hat e_{3}\right)
\cdot \left(h_{1}\hat e_{1}du^{(1)}+h_{2}\hat e_{2}du^{(2)}
+h_{3}\hat e_{3}du^{(3)}\right)\nonumber\\
&=&\int_{\partial S} \left(A_{1}h_{1}du^{(1)}+A_{2}h_{2}du^{(2)}+
A_{3}h_{3}du^{(3)}\right).\nonumber
\end{eqnarray}
Por lo que definiendo 
\begin{eqnarray}
(A_{x}^{\prime}, A_{y}^{\prime},A_{z}^{\prime})=(A_{1}h_{1},A_{2}h_{2}, 
A_{3}h_{3}),\qquad (dx^{\prime},dy^{\prime},dz^{\prime})=(du^{(1)}, du^{(2)}, du^{(3)})\nonumber
\end{eqnarray}
y ocupando el teorema de Stokes Eq. (\ref{eq:compa-curv}), se llega a
\begin{eqnarray}
\int_{\partial S} \vec A\cdot d\vec r&=&
\int_{\partial S} \left( A^{\prime}_{x}dx^{\prime}+ A^{\prime}_{y}dy^{\prime}+ A^{\prime}_{z}d z^{\prime}\right)\nonumber\\
&=&
\int_{S}\Bigg[
\left(\frac{\partial A^{\prime}_{z} }{\partial y^{\prime} }- 
\frac{\partial A^{\prime}_{y} }{\partial z^{\prime}}\right) dy^{\prime}dz^{\prime}+
\left(\frac{\partial A_{x}^{\prime} }{\partial z^{\prime}}- 
\frac{\partial A_{z}^{\prime} }{\partial x^{\prime}}\right)dx^{\prime}dy ^{\prime}+\nonumber\\
& &
\left(\frac{\partial A_{y}^{\prime} }{\partial x^{\prime}}-
 \frac{\partial A_{x}^{\prime} }{\partial y^{\prime}}\right) dx^{\prime}dy^{\prime}\Bigg]\nonumber\\
&=&\int_{S}\Bigg[
\left(\frac{\partial (A_{3}h_{3})}{\partial u^{(2)}}-  
\frac{\partial (A_{2}h_{2})}{\partial u^{(3)}}\right)du^{(2)}
du^{(3)}\nonumber\\
& &+ \left(\frac{\partial (A_{1}h_{1})}{\partial u^{(3)}}-  
\frac{\partial (A_{3}h_{3})}{\partial u^{(1)}}\right)du^{(1)}
du^{(3)}\nonumber\\
& &
+ \left(\frac{\partial (A_{2}h_{2})}{\partial u^{(1)}}-  
\frac{\partial (A_{1}h_{1})}{\partial u^{(2)}}\right)du^{(1)}
du^{(2)}\Bigg].\label{eq:rot-fin1}
\end{eqnarray}
Adem\'as, usando el elemento de \'area y $\left(\vec \nabla\times \vec A\right)$ en coordenadas
generalizadas, Eq. (\ref{eq:area-curv}) y Eq. (\ref{eq:rot-curv-gen}), tenemos 
\begin{eqnarray}
\int_{S}\left(\vec \nabla \times \vec A\right)\cdot d\vec a&=&
\int_{S} 
\Bigg[\left(\vec \nabla \times \vec A\right)_{1}\hat e_{1}+
\left(\vec \nabla \times \vec A\right)_{2}\hat e_{2}+
\left(\vec \nabla \times \vec A\right)_{3}\hat e_{3}\Bigg]
\cdot \nonumber\\
 & &\Bigg[ h_{2}h_{3}du^{2}du^{3}\hat e_{1}+h_{3}h_{1}du^{3}du^{1}\hat e_{2}
+h_{1}h_{2}du^{1}du^{2}\hat e_{3}\Bigg]\nonumber\\
&=&\int_{S}\Bigg[
\left(\vec \nabla \times \vec A\right)_{1}h_{2}h_{3}du^{2}du^{3}
+\left(\vec \nabla \times \vec A\right)_{2}h_{3}h_{1}du^{3}du^{1}\nonumber\\
& &+
\left(\vec \nabla \times \vec A\right)_{3}h_{1}h_{2}du^{1}du^{2}\Bigg],
\label{eq:rot-fin2}
\end{eqnarray}
igualando Eq. (\ref{eq:rot-fin1}) con Eq. (\ref{eq:rot-fin2}) se obtiene
\begin{eqnarray}
\left(\vec \nabla \times \vec A\right)_{1}h_{2}h_{3}&=&
\left(\frac{\partial (A_{3}h_{3})}{\partial u^{(2)}}-  
\frac{\partial (A_{2}h_{2})}{\partial u^{(3)}}\right),\nonumber\\
\left(\vec \nabla \times \vec A\right)_{2}h_{3}h_{1}&=&
\left(\frac{\partial (A_{1}h_{1})}{\partial u^{(3)}}-  
\frac{\partial (A_{3}h_{3})}{\partial u^{(1)}}\right),\nonumber\\
\left(\vec \nabla \times \vec A\right)_{3}h_{1}h_{2}&=&
\left(\frac{\partial (A_{2}h_{2})}{\partial u^{(1)}}-  
\frac{\partial (A_{1}h_{1})}{\partial u^{(2)}}\right).
\end{eqnarray}
Es decir,
\begin{eqnarray}
\left(\vec \nabla \times \vec A\right)_{1}&=&
\frac{1}{h_{2}h_{3}} 
\left(\frac{\partial (A_{3}h_{3})}{\partial u^{(2)}}-  
\frac{\partial (A_{2}h_{2})}{\partial u^{(3)}}\right),\nonumber\\
\left(\vec \nabla \times \vec A\right)_{2}&=&
\frac{1}{h_{3}h_{1}}
\left(\frac{\partial (A_{1}h_{1})}{\partial u^{(3)}}-  
\frac{\partial (A_{3}h_{3})}{\partial u^{(1)}}\right),\nonumber\\
\left(\vec \nabla \times \vec A\right)_{3}&=&
\frac{1}{h_{1}h_{2}}
\left(\frac{\partial (A_{2}h_{2})}{\partial u^{(1)}}-  
\frac{\partial (A_{1}h_{1})}{\partial u^{(2)}}\right),
\end{eqnarray}
que son las componentes del rotacional en coordenadas generalizadas.\\

En particular para  coordenas esf\'ericas tenemos
\begin{eqnarray}
\vec \nabla \times \vec A&=&
\frac{1}{r\sin\theta}
\left[
\frac{\partial (\sin \theta A_{\varphi}) }{\partial \theta}
- \frac{\partial \left(r A_{\theta}\right)}{\partial \varphi} \right]\hat e_{r}+
\left[\frac{1}{r\sin\theta} \frac{\partial  A_{r}}{\partial \varphi}-
\frac{1}{r}\frac{\partial (r A_{\varphi})}{\partial \varphi}\right]
\hat e_{\theta}+\nonumber\\
& &\frac{1}{r}\left[\frac{\partial (rA_{\theta} )}{\partial r}-
\frac{\partial  A_{r}}{\partial \theta}\right]\hat e_{\varphi}.
\label{eq:rota-esfe}
\end{eqnarray}
Mientras que en  coordenadas cil\'{\i}ndricas se llega a
\begin{eqnarray}
\vec \nabla \times \vec A&=&
\left[\frac{1}{\rho} 
\frac{\partial A_{z}}{\partial \varphi}
- \frac{\partial A_{\varphi}}{\partial z} \right]\hat e_{\rho}+
\left[\frac{\partial  A_{r}}{\partial z}-
\frac{\partial  A_{z}}{\partial \rho}\right] \hat e_{\varphi}+\nonumber\\
& &\frac{1}{\rho}\left[\frac{\partial (\rho A_{\varphi})}{\partial \rho}-
\frac{\partial  A_{r}}{\partial \varphi}\right]\hat e_{z}.
\end{eqnarray}
%

\section{Operador momento angular}

Para obtener mayor pr\'actica en el manejo de vectores veamos el operador 
\begin{eqnarray}
\vec L=-i\left(\vec r\times \vec \nabla\right).
\end{eqnarray}
Este operador surge de manera natural en mec\'anica cu\'antica, pero tambi\'en es importante
en la teor\'{\i}a del potencial, en el grupo de rotaciones y para el estudio de las ondas 
electromagn\'eticas.\\

En coordenadas cartesianas, usando Eq. (\ref{eq:pos-cart}) y Eq. (\ref{eq:grad-cart}), se encuentra
\begin{eqnarray}
\vec L&=& \hat L_{x}\hat i+ \hat L_{y}\hat j+\hat L_{z}\hat k\nonumber\\
&=&-i \left(\vec r\times \vec \nabla\right)=-i
\left(x\hat i+y\hat j+z\hat k\right)\times\left(\frac{\partial }{\partial x}\hat i+  \frac{\partial }{\partial y}\hat j
+\frac{\partial }{\partial z}\hat k\right)\nonumber\\
&=&-i\Bigg[x \frac{\partial }{\partial y}\left(\hat i \times\hat j\right) + 
x \frac{\partial }{\partial z}\left(\hat i \times\hat k\right)
+  y \frac{\partial }{\partial x}\left(\hat j \times\hat i\right) + 
y \frac{\partial }{\partial z}\left(\hat j \times\hat k\right)\nonumber\\
& &+ z \frac{\partial }{\partial x}\left(\hat k \times\hat i\right)+
z \frac{\partial }{\partial y}\left(\hat k \times\hat j\right)\Bigg]\nonumber\\
&=&-i\Bigg[\left( y \frac{\partial }{\partial z}-z \frac{\partial }{\partial y}\right)\hat i+
\left( z \frac{\partial }{\partial x}-x \frac{\partial }{\partial z}\right)\hat j+
\left( x \frac{\partial }{\partial y}-y \frac{\partial }{\partial x}\right)\hat k\Bigg].\nonumber
\end{eqnarray}
Por lo que,
\begin{eqnarray}
\hat L_{x}=-i\left( y \frac{\partial }{\partial z}-z \frac{\partial }{\partial y}\right),
\hat L_{y}=-i \left( z \frac{\partial }{\partial x}-x \frac{\partial }{\partial z}\right),
\hat L_{z}=-i\left( x \frac{\partial }{\partial y}-y \frac{\partial }{\partial x}\right).\quad 
\label{eq:vec-momento-angular}
\end{eqnarray}
En coordenadas esf\'ericas, considerando Eq. (\ref{eq:pos-esfe}) y Eq. (\ref{eq:grad-esfe}), se llega a
\begin{eqnarray}
\vec L&=&-i \left(\vec r\times \vec \nabla\right)=-i
\left(r\hat e_{r}\right)\times\left(\hat e_{r}\frac{\partial }{\partial r}+ 
\hat e_{\theta}  \frac{1}{r}\frac{\partial }{\partial \theta}+ \hat e_{\varphi}
\frac{1}{r\sin\theta}\frac{\partial }{\partial \varphi}\right)\nonumber\\
&=& -i\left(\left(\hat e_{r}\times\hat e_{\theta}\right) \frac{\partial }{\partial \theta}+
\left(\hat e_{r}\times\hat e_{\varphi}\right)\frac{1}{\sin\theta}\frac{\partial }{\partial \varphi}\right)
\nonumber\\
&=& -i\left( \hat e_{\varphi}\frac{\partial }{\partial \theta}-
\hat e_{\theta}\frac{1}{\sin\theta}\frac{\partial }{\partial \varphi}\right),\nonumber
\end{eqnarray}
es decir
\begin{eqnarray}
\vec L=-i\left( \hat e_{\varphi}\frac{\partial }{\partial \theta}-
\hat e_{\theta}\frac{1}{\sin\theta}\frac{\partial }{\partial \varphi}\right).
\label{eq:esfe-moment}
\end{eqnarray}
De esta expresi\'on se puede obtener $\hat L_{x},\hat L_{y},\hat L_{z}$ en coordenadas esf\'ericas. 
En efecto,  utilizando Eqs. (\ref{eq:esfe-uni1})-(\ref{eq:esfe-uni3}), en (\ref{eq:esfe-moment}) se obtiene
\begin{eqnarray}
\vec L&=& -i\Bigg[\left(-\sin\varphi \hat i+\cos\varphi \hat j\right) \frac{\partial }{\partial \theta}
\nonumber\\
& & - \frac{1}{\sin\theta}\left(\cos\theta\cos\varphi\hat i+\cos\theta \sin\varphi\hat j-\sin\theta \hat k\right)
 \frac{\partial }{\partial \varphi}\Bigg]\nonumber\\
&=&-i \Bigg[\left(-\sin\varphi \frac{\partial }{\partial \theta}-\frac{\cos\theta}{\sin\theta}\cos\varphi 
\frac{\partial }{\partial \varphi} \right)\hat i\nonumber\\
& &+\left(\cos\varphi \frac{\partial }{\partial \theta}-\frac{\cos\theta}{\sin\theta}\sin\varphi 
\frac{\partial }{\partial \varphi} \right)\hat j+ \frac{\partial }{\partial \varphi}\hat k\Bigg],\nonumber
\end{eqnarray}
por lo que, 
\begin{eqnarray}
\hat L_{x}&=&i\left(\sin\varphi \frac{\partial }{\partial \theta}+\cot\theta\cos\varphi 
\frac{\partial }{\partial \varphi} \right), \label{eq:x-momen-esfe}\\
\hat L_{y}&=&-i \left(\cos\varphi \frac{\partial }{\partial \theta}-\cot\theta\sin\varphi 
\frac{\partial }{\partial \varphi} \right),\label{eq:y-momen-esfe} \\
\hat L_{z}&=& -i\frac{\partial }{\partial \varphi}.\label{eq:z-momen-esfe}
\end{eqnarray}
Un operador importante es 
\begin{eqnarray}
L^{2}=\vec L\cdot \vec L=\hat L_{x}^{2}+ \hat L_{y}^{2}+\hat L_{z}^{2}.
\end{eqnarray}
Este operador es m\'as sugerente en coordenadas esf\'ericas, usando Eq. (\ref{eq:esfe-moment}) se encuentra 
\begin{eqnarray}
\hat L^{2}&=&\vec L\cdot \vec L=-i\left( \hat e_{\varphi}\frac{\partial }{\partial \theta}-
\hat e_{\theta}\frac{1}{\sin\theta}\frac{\partial }{\partial \varphi}\right)\cdot \vec L\nonumber\\
&=&-i\left( \hat e_{\varphi}\cdot \frac{\partial\vec L}{\partial \theta}-
 \frac{\hat e_{\theta}}{\sin\theta}\cdot\frac{\partial \vec L}{\partial \varphi}\right).\label{eq:cua-esfe}
\end{eqnarray}
Antes de continuar notemos que si $\vec A$ y $\vec B$ dependen de la variable $u,$ entonces
\begin{eqnarray}
\frac{\partial \left(\vec A\cdot \vec B\right)}{\partial u}= \frac{\partial \vec A}{\partial u}\cdot \vec B+ 
\vec A\cdot \frac{\partial \vec B}{\partial u},
\end{eqnarray}
adem\'as de Eqs. (\ref{eq:esfe-uni1})-(\ref{eq:esfe-uni2}) se llega a 
\begin{eqnarray}
\frac{\partial \hat e_{\theta}}{\partial \varphi}&=&(-\cos\theta\sin\varphi, \cos\theta\cos\varphi,0)=
\cos\theta(-\sin\varphi,\cos\varphi,0)=\cos\theta \hat e_{\varphi},\nonumber
\end{eqnarray}
Tomando en cuenta esta dos ultimas igualdades, las expresiones Eqs. (\ref{eq:esfe-moment}),   (\ref{eq:esfe-uni1})-(\ref{eq:esfe-uni3}),  se tiene 
\begin{eqnarray}
\hat e_{\varphi}\cdot \frac{\partial\vec L}{\partial \theta}&=&\frac{\partial\left( \hat e_{\varphi}\cdot\vec L\right)}{\partial \theta}=-i\frac{\partial^{2} }{\partial \theta^{2}}\\
\hat e_{\theta}\cdot\frac{\partial \vec L}{\partial \varphi}&=& 
\frac{\partial\left( \hat e_{\theta}\cdot\vec L\right)}{\partial \varphi} - \frac{\partial\left( \hat e_{\theta}\right)}{\partial \varphi}\cdot \vec L
=\frac{\partial}{\partial \varphi}\left( i\frac{1}{\sin\theta} \frac{\partial}{\partial \varphi}\right)-\cos\theta \hat e_{\varphi}\cdot \vec L\nonumber\\
&=&i\left( \frac{1}{\sin\theta}  \frac{\partial^{2}}{\partial \varphi^{2}}+ \cos\theta \frac{\partial}{\partial \theta}\right)
\nonumber
\end{eqnarray}
Sustituyendo estos resultado en Eq. (\ref{eq:cua-esfe}) se consigue
\begin{eqnarray}
\hat L^{2}
&=&-\left( \frac{\partial^{2} }{\partial \theta^{2}}+
\frac{\cos\theta }{\sin \theta}\frac{\partial }{\partial \theta}+
\frac{1}{\sin^{2}\theta}\frac{\partial^{2} }{\partial \varphi^{2}}\right),\nonumber
\end{eqnarray}
adem\'as utilizando la igualdad
\begin{eqnarray}
\frac{1}{\sin\theta} \frac{\partial }{\partial \theta}
\left(\sin\theta\frac{\partial }{\partial \theta}\right)= \frac{\partial^{2} }{\partial \theta^{2}}+
\frac{\cos\theta }{\sin \theta}\frac{\partial }{\partial \theta}
\end{eqnarray}
se llega a
\begin{eqnarray}
\hat L^{2}=-\left[\frac{1}{\sin\theta} \frac{\partial }{\partial \theta}
\left(\sin\theta\frac{\partial }{\partial \theta}\right)+
\frac{1}{\sin^{2}\theta}\frac{\partial^{2} }{\partial \varphi^{2}}\right].\label{eq:op-moment-esfe}
\end{eqnarray}
Note que con este resultado el Laplaciano en coordenadas esf\'ericas Eq. (\ref{eq:lapla-esfe}) se puede escribir como
\begin{eqnarray}
\nabla^{2}\phi=\frac{1}{r^{2}}\frac{\partial }{\partial r}
\left(r^{2}\frac{\partial \phi}{\partial r}\right)-\frac{\hat L^{2}\phi}{r^{2}}.
\end{eqnarray}
Esta versi\'on del Laplaciano es de gran utilidad para resolver la ecuaci\'on de Laplace,$\nabla^{2}\phi=0,$ en coordenadas esf\'ericas. Este problema lo estudiaremos
en otro cap\'itulo.

\chapter{El Factorial y la Funci\'on Gamma}

En este cap\'itulo veremos una funci\'on que generaliza
el factorial de un n\'umero natural, que es la funci\'on Gamma.
El estudio de este tema no es exhaustivo pero es suficiente 
para resolver diferentes problemas interesantes.

\section{Funci\'on Gamma}

Dado un n\'umero natural, $n,$ se define el factorial como
\begin{eqnarray}
n!=n\cdot (n-1)\cdot (n-2)\cdots 2\cdot 1.
\end{eqnarray}
Tambi\'en se puede definir un producto similar para los primeros $n$ n\'umeros pares
mediante
\begin{eqnarray}
(2n)!!=(2n)\cdot 2(n-1)\cdot 2(n-2)\cdots \cdot 6\cdot 4\cdot 2,
\end{eqnarray}
factorizando un $2$ de cada t\'ermino se encuentra 
\begin{eqnarray}
(2n)!!=2^{n}\left( n\cdot (n-1)\cdot (n-2)\cdots 2\cdot 1\right)=2^{n}n!,
\end{eqnarray}
es decir 
\begin{eqnarray}
(2n)!!=2^{n}n!.
\end{eqnarray}
Adem\'as, el producto de los primeros $(n+1)$ n\'umeros impares es
\begin{eqnarray}
(2n+1)!!=(2n+1)\cdot (2n-1)\cdot (2n-3)\cdots \cdot 5\cdot 3\cdot 1.
\end{eqnarray}
Ahora, multiplicando y dividiendo este n\'umero por $(2n)!!$ se encuentra
\begin{eqnarray}
(2n+1)!!&=&\frac{ (2n+1)\cdot {\bf(2n)}\cdot (2n-1)\cdot {\bf 2(n-1)}\cdots (5)\cdot
{\bf (4)}\cdot(3)\cdot{\bf (2)}\cdot(1)}
{(2n)\cdot 2(n-1)\cdots (4)\cdot (2)}\nonumber\\
&=&\frac{ (2n+1)!}{(2n)!!}= \frac{ (2n+1)!}{2^{n}n!}\nonumber,
\end{eqnarray}
es decir
\begin{eqnarray}
(2n+1)!!= \frac{ (2n+1)!}{2^{n}n!}.\label{eq:doble-factorial-impar}
\end{eqnarray}

Adicionalmente se puede definir el factorial para cualquier n\'umero real o complejo.
Para hacer esa definici\'on  ocuparemos la funci\'on Gamma
\begin{eqnarray}
\Gamma(z)=\int_{0}^{\infty}e^{-t}t^{z-1}dt,\qquad {\rm Re}(z)>0.
\label{eq:gamma}
\end{eqnarray}
Primero veamos dos valores de esta funci\'on. Notemos que si $z=1,$ se tiene
\begin{eqnarray}
\Gamma(1)=\int_{0}^{\infty}e^{-t}dt=-e^{-t}\bigg|_{0}^{\infty}=1,\nonumber
\end{eqnarray}
mientras que si $z=\frac{1}{2},$ con el cambio de variable $u=t^{1/2},$ se encuentra 
\begin{eqnarray}
\Gamma\left(\frac{1}{2}\right)&=&\int_{0}^{\infty}e^{-t}t^{-1/2}dt=
2\int_{0}^{\infty}e^{-u^{2}}du=\int_{-\infty}^{\infty}e^{-u^{2}}du \nonumber\\
&=&\left(\int_{-\infty}^{\infty}e^{-u^{2}}du \int_{-\infty}^{\infty}e^{-v^{2}}dv\right)^{\frac{1}{2}}\nonumber\\
&=&\left(\int_{-\infty}^{\infty}\int_{-\infty}^{\infty} dudve^{-(u^{2}+v^{2})} \right)^{\frac{1}{2}}=
\left(\int_{0}^{2\pi }d\varphi \int_{0}^{\infty}dr r e^{-r^{2}} \right )^{1/2}
\nonumber\\
&=&
\left(2\pi \frac{(-)}{2}\int_{0}^{\infty}dr\frac{d e^{-r^{2}}}{dr} \right )^{1/2}=\sqrt{\pi},\nonumber
\end{eqnarray}
es decir,
\begin{eqnarray}
\Gamma\left(\frac{1}{2}\right)=\sqrt{\pi}.
\end{eqnarray}
La funci\'on $\Gamma(z)$ tiene las mismas propiedades que el factorial, en efecto observemos 
que se cumple
\begin{eqnarray}
e^{-t}t^{z}=zt^{z-1}e^{-t}-\frac{d}{dt}\left(e^{-t}t^{z}\right),
\end{eqnarray}
que implica
\begin{eqnarray}
\Gamma(z+1)= \int_{0}^{\infty}e^{-t}t^{z}dt=z
\int_{0}^{\infty}t^{z-1}e^{-t}=z\Gamma(z),
\end{eqnarray}
es decir
\begin{eqnarray}
\Gamma(z+1)=z\Gamma(z).
\label{eq:recursiva}
\end{eqnarray}
Usando de forma reiterada  (\ref{eq:recursiva}) se encuentra
\begin{eqnarray}
\Gamma(z+1)&=&z\Gamma(z)=z(z-1)\Gamma(z-1)= z(z-1)(z-2)\Gamma(z-2) \nonumber\\
&=& z(z-1)(z-2)\cdots (z-k)\Gamma(z-k),\quad {\rm Re}(z-k)>0.\label{eq:gamma-recurrencia}
\end{eqnarray}
En particular si $z$ es un natural, $n,$ el m\'aximo valor que puede tomar $k$ es $n-1,$ por lo que
\begin{eqnarray}
\Gamma(n+1)&=& n(n-1)(n-2)\cdots (n-(n-1))\Gamma(1)= n(n-1)(n-2)\cdots 2\cdot 1\nonumber\\
&=&n!,\nonumber
\end{eqnarray}
entonces
\begin{eqnarray}
\Gamma(n+1)=n!.
\end{eqnarray}
As\'i, para cualquier n\'umero complejo con  ${\rm Re}(z-k)>0,$ definiremos el factorial como
\begin{eqnarray}
z!=\Gamma(z+1)=z(z-1)(z-2)\cdots (z-k)\Gamma(z-k),\quad {\rm Re}(z-k)>0.
\label{eq:factorial}
\end{eqnarray}
Por ejemplo,
\begin{eqnarray}
\left(\frac{1}{2}\right)!=\Gamma\left(\frac{1}{2}+1\right)=\frac{1}{2} \Gamma\left(\frac{1}{2}\right)= \frac{\sqrt{\pi}}{2}.
\end{eqnarray}
Si queremos saber cuanto vale $\left(n+\frac{1}{2}\right)!$ debemos ocupar la definici\'on (\ref{eq:factorial}).
Para este caso es claro que el m\'aximo valor que puede tomar $k$ es $n,$  de donde
\begin{eqnarray}
\left(n+\frac{1}{2}\right)!&=&\left(n+\frac{1}{2}\right)\left(n-\frac{1}{2}\right)\left(n-\frac{3}{2}\right)\cdots \cdot\left(\frac{1}{2}\right)\Gamma\left(\frac{1}{2}\right) \nonumber\\
&=& \frac{(2n+1)}{2}\frac{(2n-1)}{2} \frac{(2n-3)}{2}\cdots \frac{1}{2} \sqrt{\pi}\nonumber\\
&=&\frac{ \sqrt{\pi}(2n+1)!!}{2^{n+1}}= \frac{ \sqrt{\pi}(2n+1)!}{2^{2n+1}n!}\nonumber,
\end{eqnarray}
por lo que
\begin{eqnarray}
\left(n+\frac{1}{2}\right)!= \frac{ \sqrt{\pi}(2n+1)!}{2^{2n+1}n!}\nonumber.
\end{eqnarray}
De este resultado se tiene
\begin{eqnarray}
\left(n-\frac{1}{2}\right)!&=&\left(n-1+\frac{1}{2}\right)!= \frac{\sqrt{\pi} (2(n-1)+1)!}{2^{2(n-1)+1}(n-1)!}= \frac{\sqrt{\pi} (2n-1)!}{2^{2n-1}(n-1)!}\nonumber\\
&=&\frac{ \sqrt{\pi}(2n-1)!(2n)}{2^{2n-1}(n-1)!(2n)}= \frac{ \sqrt{\pi}(2n)!}{2^{2n}n!}\nonumber,
\end{eqnarray}
de donde
\begin{eqnarray}
\left(n-\frac{1}{2}\right)!= \frac{ \sqrt{\pi}(2n)!}{2^{2n}n!}\nonumber.
\end{eqnarray}
Tambi\'en se puede calcular  el factorial para n\'umeros negativos, por ejemplo
\begin{eqnarray}
\left(\frac{-1}{2}\right)!&=&\Gamma\left(1-\frac{1}{2}\right)=\Gamma\left(\frac{1}{2}\right)=\sqrt{\pi}.
\end{eqnarray}
De hecho, ocupando que la ecuaci\'on (\ref{eq:recursiva}) implica 
\begin{eqnarray}
\Gamma(z)=\frac{\Gamma(z+1)}{z}, \label{eq:reversa}
\end{eqnarray}
se puede definir $\Gamma(z)$ para n\'umeros negativos. 
Por ejemplo, si $z=-\frac{1}{2},$ se  encuentra
\begin{eqnarray}
\Gamma\left(-\frac{1}{2}\right)=\frac{\Gamma\left(1-\frac{1}{2}\right)}{\left(\frac{-1}{2} \right)}=-2\Gamma\left(\frac{1}{2}\right)=
-2\sqrt{\pi}.
\end{eqnarray}
Similarmente, si $0<\epsilon<1,$ podemos definir 
\begin{eqnarray}
\Gamma(-\epsilon)=\frac{\Gamma(1-\epsilon)}{-\epsilon}. \label{eq:cero-izquierda}
\end{eqnarray}
La parte  derecha de esta igualdad tiene sentido pues $(1-\epsilon)>0,$ as\'i la parte izquierda tiene sentido.\\

Ocupando de forma reiterada  (\ref{eq:reversa}) se llega a 
\begin{eqnarray}
\Gamma(z)&=&\frac{\Gamma(z+1)}{z}=\frac{\Gamma(z+2)}{z(z+1)} =\frac{\Gamma(z+3)}{z(z+1)(z+2)}=\cdots=\nonumber\\
&=& \frac{\Gamma(z+k)}{z(z+1)(z+2)(z+3)\cdots (z+(k-1))}.\label{eq:gamma-negativo}
\end{eqnarray}
Si  $z+k>0$ la parte derecha de esta igualdad tiene sentido, por lo tanto la parte izquierda est\'a bien definida, a\'un si $z<0.$
Por ejemplo, si $z$ es de la forma $z=-n+\epsilon,$  con $n$ un natural y $\epsilon\in (0,1)$, tomando $k=n$  
se cumple $z+k>0$  y  la cantidad
\begin{eqnarray}
\Gamma(z)&=&\Gamma(-n+\epsilon)\nonumber\\
&=& \frac{\Gamma(\epsilon)}{(-n+\epsilon)(-(n-1)+\epsilon)(-(n-2)+\epsilon)\cdots (-1+\epsilon)}
\label{eq:gamma-negativa}
\end{eqnarray}
est\'a bien definida.\\

Ahora, es claro que 
\begin{eqnarray}
\lim_{\epsilon \to 0}  (-n+\epsilon)(-(n-1)+\epsilon)(-(n-2)+\epsilon)\cdots (-1+\epsilon)=(-)^{n}n!.
\end{eqnarray}
Mientras que de la integral (\ref{eq:gamma}) se tiene  $\Gamma(0^{+})=\infty$ y de (\ref{eq:cero-izquierda}) se encuentra  $\Gamma(0^{-})=-\infty.$
Por lo tanto, si $n$ es un natural se cumple
\begin{eqnarray}
\Gamma(-n^{\pm})&=&(-)^{n}(\pm)\infty ,
\end{eqnarray}
en ambos caso 
\begin{eqnarray}
\frac{1}{\Gamma(-n)}&=&0.
\end{eqnarray}
Existen m\'as propiedades de la funci\'on $\Gamma(z),$ pero las que hemos visto nos bastan para estudiar las funciones de Bessel.
 

\chapter{Repaso de Ecuaciones Diferenciales Ordinarias}

En este cap\'itulo veremos una serie de resultado sobre ecuaciones diferenciales que aplicaremos posteriormente.
 
\section{Teorema de existencia y unicidad}

El primer resultado es sobre la existencia y unicidad de las soluciones de ecuaciones diferenciales 
de la forma
\begin{eqnarray}
\frac{d^{2} Y(x)}{dx^{2}}+ P(x)\frac{d Y(x)}{dx}+Q(x)Y(x)=R(x).
\label{eq:general-SL}
\end{eqnarray}
Si $P(x),Q(x)$ y $R(x)$ son funciones continuas en el intervalo $[a,b]$ y  $x_{0}\in [a,b],$
entonces existe una \'unica soluci\'on de Eq. (\ref{eq:general-SL}) que cumple las condiciones iniciales
\begin{eqnarray}
Y(x_{0})=a_{1},\qquad \frac{d Y(x)}{dx}\bigg|_{x_{0}}=a_{2},
\end{eqnarray}
donde $a_{1}$ y $a_{2}$ son constantes. Este resultado lo usaremos sin demostrar, la demostraci\'on se puede ver en 
\cite{simmos:gnus}.

\section{El Wronskiano}

Un concepto de mucha utilidad en el estudio de la independencia de las soluciones de las ecuaciones diferenciales es el Wronskiano. 
Supongamos que tenemos dos funciones $f$ y $g,$ el Wronskiano se define como   
\begin{eqnarray}
W(f,g)(x)=\left| 
\begin{array}{rr}
 f& g\\
 \frac{df}{dx}& \frac{dg}{dx}  
\end{array}
\right|(x)=f(x)\frac{dg(x)}{dx}-\frac{df(x)}{dx}g(x),
\end{eqnarray}
note que 
\begin{eqnarray}
\frac{d W(f,g)(x)}{dx}=f(x)\frac{d^{2}g(x)}{dx^{2}}-\frac{d^{2}f(x)}{dx^{2}}g(x).
\end{eqnarray}

\section{Independencia lineal}

Ahora, si  $Y_{1}(x)$ y $Y_{2}(x)$ son soluciones de la ecuaci\'on diferencial 
\begin{eqnarray}
\frac{d^{2} Y(x)}{dx^{2}}+ P(x)\frac{d Y(x)}{dx}+Q(x)Y(x)=0,
\label{eq:sturm-homogenea}
\end{eqnarray}
es decir, si se cumple 
\begin{eqnarray}
\frac{d^{2} Y_{1}(x)}{dx^{2}}+ P(x)\frac{d Y_{1}(x)}{dx}+Q(x)Y_{1}(x)=0,\\
\frac{d^{2} Y_{2}(x)}{dx^{2}}+ P(x)\frac{d Y_{2}(x)}{dx}+Q(x)Y_{2}(x)=0,
\end{eqnarray}
se obtiene 
\begin{eqnarray}
Y_{2}(x)\frac{d^{2} Y_{1}(x)}{dx^{2}}+ P(x)Y_{2}(x)\frac{d Y_{1}(x)}{dx}+Q(x)Y_{2}(x)Y_{1}(x)=0,\\
Y_{1}(x)\frac{d^{2} Y_{2}(x)}{dx^{2}}+ P(x)Y_{1}(x)\frac{d Y_{2}(x)}{dx}+Q(x)Y_{1}(x)Y_{2}(x)=0.
\end{eqnarray}
Al restar estas ecuaciones se llega a
\begin{eqnarray}
Y_{2}(x)\frac{d^{2} Y_{1}(x)}{dx^{2}} -Y_{2}(x)\frac{d Y_{1}(x)}{dx}+ 
P(x)\left( Y_{1}(x)\frac{d Y_{1}(x)}{dx}-Y_{2}(x) \frac{dY_{1}(x)}{dx} \right)=0,\nonumber
\end{eqnarray}
esta \'ultima ecuaci\'on  se puede escribir como
\begin{eqnarray}
\frac{dW(Y_{1},Y_{2})(x)}{dx}+P(x)W(Y_{1},Y_{2})(x)=0,
\end{eqnarray}
cuya   soluci\'on es 
\begin{eqnarray}
W(Y_{1},Y_{2})(x)=Ce^{-\int P(x)dx},\qquad C={\rm constante}.
\end{eqnarray}
Como la funci\'on exponencial nunca se anula, si el Wronskiano es cero en un punto, implica que $C=0.$ Por lo tanto, 
si el Wronskiano es cero en un punto, es cero en cualquier otro punto. 
Claramente tambi\'en es cierto que si el Wronskiano es diferente de cero en un punto, es diferentes de cero en cualquier otro punto. 
Adem\'as, si el Wronskiano es diferente de cero  no puede cambiar de signo, pues de lo contrario tendr\'ia que pasar por cero.\\

Con el  Wronskiano se puede obtener informaci\'on sobre $Y_{1}(x)$ y $Y_{2}(x).$ En efecto, 
si $W(Y_{1},Y_{2})(x)=0,$ entonces el sistema de ecuaciones lineales
\begin{eqnarray}
 \left( 
\begin{array}{rr}
Y_{1}(x)& Y_{2}(x)\\
 \frac{dY_{1}(x)}{dx} & \frac{dY_{2}(x)}{dx}
\end{array}
\right)\left(
\begin{array}{r}
a_{1}\\
a_{2} 
\end{array}
\right)=
\left(
\begin{array}{r}
0\\
0 
\end{array}
\right)\label{eq:sistema-wronkiano}
\end{eqnarray}
tiene soluci\'on no trivial y las funciones $Y_{1}(x),Y_{2}(x)$ son linealmente dependientes.
Ahora, si $W(Y_{1},Y_{2})(x)\not=0,$ la \'unica soluci\'on a (\ref{eq:sistema-wronkiano}) es la trivial y por lo tanto,  
$Y_{1}(x)$ y $Y_{2}(x)$ son linealmente independientes.\\

Supongamos que  $Y_{1}(x)$ y $Y_{2}(x)$ son soluciones linealmente independientes de Eq. (\ref{eq:sturm-homogenea}), entonces podemos afirmar que estas funciones no se pueden anular en un mismo punto. Esto es verdad, pues si existe $x_{0}$ tal que $Y_{1}(x_{0})=Y_{2}(x_{0})=0,$ entonces $W(Y_{1},Y_{2})(x_{0})=0,$ que no puede ser posible pues $Y_{1}(x)$ y $Y_{2}(x)$ son linealmente independientes.\\

\section{Los ceros de las soluciones}

Diremos que $a_{1}$ y $a_{2}$ son ceros sucesivos de $Y_{1}(x),$ si para toda $x$ en el intervalo $(a_{1},a_{2})$ se cumple $Y_{1}(x)\not =0$ y $Y_{1}(a_{1})=Y_{1}(a_{2})=0.$\\

El Wronskiano nos da informaci\'on sobre los puntos donde se anulan las soluciones linealmente independientes de Eq. (\ref{eq:sturm-homogenea}). En efecto, supongamos que $Y_{1}(x)$ y $Y_{2}(x)$ son soluciones linealmente independientes de Eq. (\ref{eq:sturm-homogenea}) y  que $a_{1}$ y $a_{2}$ son ceros sucesivos de $Y_{1}(x),$ entonces  podemos afirmar que $Y_{2}(x)$ tiene un cero en el intervalo $(a_{1},a_{2}).$ Para probar esta afirmaci\'on, notemos que la derivada de $Y_{1}(x)$ no puede tener el mismo signo en $a_{1}$ y $a_{2},$ adem\'as 
\begin{eqnarray}
W(Y_{1},Y_{2})(a_{1})&=&-\frac{d Y_{1}(x)}{dx}\bigg|_{a_{1}} Y_{1}(a_{1}), \\
W(Y_{1},Y_{2})(a_{2})&=&-\frac{d Y_{1}(x)}{dx}\bigg|_{a_{2}} Y_{1}(a_{2}).
\end{eqnarray}
Ahora como $\frac{d Y_{1}(x)}{dx}\bigg|_{a_{2}}$ tiene signo diferente a $\frac{d Y_{1}(x)}{dx}\bigg|_{a_{1}}$ y el Wronskiano no cambia de signo, entonces $Y_{2}(a_{1})$ y $Y_{2}(a_{2})$ tienen signos diferentes. Como $Y_{2}(x)$ es continua, existe un punto 
$a_{3}\in (a_{1},a_{2})$ tal que  $Y_{2}(a_{3})=0,$   que es lo que quer\'iamos demostrar. \\

De hecho, podemos afirmar que si $Y_{1}(x)$ y $Y_{2}(x)$ son soluciones linealmente independientes de Eq. (\ref{eq:sturm-homogenea}), $Y_{2}(x)$ tiene un \'unico cero entre dos ceros sucesivos de $Y_{1}(x).$

\subsection{Forma normal}

Para poder estudiar la ecuaci\'on diferencial Eq. (\ref{eq:sturm-homogenea}) en muchos caso 
es mejor expresarla en una forma m\'as conveniente. Por ejemplo, supongamos que $Y(x)=u(x)v(x),$ de donde 
\begin{eqnarray}
\frac{dY(x)}{dx}&=& \frac{du(x)}{dx}v(x)+ u(x)\frac{d v(x)}{dx},\\
\frac{d^{2}Y(x)}{dx^{2}}&=& \frac{d^{2}u(x)}{dx^{2}}v(x)+ 2 \frac{du(x)}{dx}\frac{d v(x)}{dx}+ u(x)\frac{d^{2} v(x)}{dx^{2}},
\end{eqnarray}
por lo que  Eq. (\ref{eq:sturm-homogenea}) se puede escribir como
\begin{eqnarray}
  & & \frac{d^{2}u(x)}{dx^{2}}v(x)+ \left(2\frac{d v(x)}{dx}+P(x)v(x)\right)\frac{d u(x)}{dx}\nonumber\\
  & &+\left( \frac{d^{2}v(x)}{dx^{2}}+P(x)\frac{d v(x)}{dx}+ Q(x)v(x)\right)u(x)=0.\label{eq:sturm-ordinarias-4}
\end{eqnarray}
Si pedimos que
\begin{eqnarray}
2\frac{d v(x)}{dx}+P(x)v(x)=0,\label{eq:sturm-ordinarias-3}
\end{eqnarray}
se obtiene,
\begin{eqnarray}
\frac{d v(x)}{dx}=-\frac{P(x)v(x)}{2},\qquad \frac{d^{2} v(x)}{dx^{2}}=\left( -\frac{1}{2}\frac{dP(x)}{dx}+\frac{P^{2}(x)}{4}\right)v(x).
\end{eqnarray}
Sustituyendo estos resultados en Eq. (\ref{eq:sturm-ordinarias-4}) se llega a 
\begin{eqnarray}
\frac{d^{2} u(x)}{d x^{2}}+\left(Q(x)-\frac{P^{2}(x)}{4}-\frac{1}{2}\frac{dP(x)}{dx}\right)u(x)=0.
\label{eq:sturm-ordinarias-7}
\end{eqnarray}
A esta ecuaci\'on se le llama forma norma de Eq. (\ref{eq:sturm-homogenea}). Note que la soluci\'on de 
Eq. (\ref{eq:sturm-ordinarias-3}) es 
\begin{eqnarray}
v(x)=ce^{-\frac{1}{2}\int dx P(x)}, \qquad c={\rm constante}
\end{eqnarray}
y esta funci\'on nunca se anula. As\'i la informaci\'on de los ceros de $Y(x)$ est\'a contenida en
$u(x).$ Por lo tanto, para estudiar los ceros de las soluciones de Eq. (\ref{eq:sturm-homogenea}) es m\'as conveniente estudiar 
su forma normal  (\ref{eq:sturm-ordinarias-7}) .\\

Notablemente la ecuaci\'on normal Eq. (\ref{eq:sturm-ordinarias-7}) es un caso particular de 
\begin{eqnarray}
\frac{d^{2} Y(x)}{d x^{2}}+q(x) Y(x)=0. \label{eq:sturm-ordinarias-5}
\end{eqnarray}
Por lo tanto, para estudiar los ceros de las soluciones de la ecuaci\'on diferencial 
Eq. (\ref{eq:sturm-homogenea}) basta
estudiar los ceros de la ecuaci\'on diferencial Eq. (\ref{eq:sturm-ordinarias-5}). Antes de entrar en detalles formales observemos que Eq. (\ref{eq:sturm-ordinarias-5}) se puede escribir como
\begin{eqnarray}
\frac{d^{2} Y(x)}{d x^{2}}=-q(x) Y(x) \label{eq:sturm-ordinarias-9}
\end{eqnarray}
que se puede ver como la segunda ley de Newton donde $Y(x)$ representa la posici\'on de una part\'icula y $q(x)$ una fuerza que cambia punto a punto.\\

Primero veamos un caso sencillo. Supongamos que $q(x)=\beta$, con $\beta$ una constante, en este caso Eq. (\ref{eq:sturm-ordinarias-9}) toma la forma 
\begin{eqnarray}
\frac{d^{2} Y(x)}{d x^{2}}=-\beta Y(x), \label{eq:sturm-ordinarias-8}
\end{eqnarray}
que es la segunda ley de Newton con una fuerza constante centrada en el origen. Si $\beta >0,$ la fuerza es atractiva y una part\'icula bajo su influencia pasa una cantidad infinita de veces por el cero. Es decir, si $\beta >0$ las soluciones de Eq. (\ref{eq:sturm-ordinarias-8}) tienen un n\'umero infinito de ceros. Ahora, si $\beta <0$ tenemos una fuerza repulsiva y una part\'icula bajo su influencia a lo m\'as puede pasar una vez por el cero. Por lo tanto podemos, afirmar que, si $\beta <0$ las soluciones de Eq. (\ref{eq:sturm-ordinarias-8}) tienen a lo m\'as un cero.\\

Ahora veamos un caso m\'as general donde $q(x)$  tiene signo definido. Primero supongamos que $q(x)<0$ para cualquier $x$ positiva. Entonces afirmamos que la soluciones de Eq. (\ref{eq:sturm-ordinarias-5}) a lo m\'as tienen un cero. Para demostrar esta afirmaci\'on, primero notemos que, desde el punto de vista f\'isico  Eq. (\ref{eq:sturm-ordinarias-9}) representa una part\'icula bajo una fuerza repulsiva. Por lo que, si existe $x_{0}$ tal que $Y(x_{0})=0,$  la part\'icula no puede regresar a la posici\'on $Y(x_{0})=0.$ Por lo tanto, si $q(x)<0$ las soluciones de  Eq. (\ref{eq:sturm-ordinarias-5}) a lo m\'as tienen un cero.\\

Si $q(x)>0,$ podemos afirmar que la soluciones de Eq. (\ref{eq:sturm-ordinarias-5}) tiene un n\'umero infinito de ceros. 
Primero notemos que, desde el punto de vista f\'isico,  la ecuaci\'on Eq. (\ref{eq:sturm-ordinarias-9}) representa una part\'icula bajo una fuerza atractiva. Supongamos que $Y(x)$ es una soluci\'on con un n\'umero finito de cero. 
Si $\alpha$ es el m\'aximo de los ceros, entonces si $x>\alpha$ la posici\'on $Y(x)$ debe tener signo defindio. Ahora,  como la fuerza es tractiva, 
la part\'icula debe regresar de nuevo a la posici\'on que ten\'ia en $\alpha,$ es decir debe regresar a cero. Esto implica que  debe existir $x_{1}>\alpha$ donde $Y(x_{1})=0.$ Por lo tanto, $\alpha$ no es el  m\'aximo de los ceros de $Y(x)$ y  esta funci\'on no puede tener un n\'umero finito de ceros. \\

Otra forma de mostrar esta afirmaci\'on es la siguiente, como $\alpha$ es el m\'aximo de los ceros de $Y(x),$ entonces si $x>\alpha,$ la funci\'on $Y(x)$ tiene signo definido. De   Eq. (\ref{eq:sturm-ordinarias-9}) se puede observar que si $Y(x)>0,$ entonces la segunda  derivada es negativa, lo que quiere decir que la taza de crecimiento disminuye, es decir $Y(x)$ decrece y eventualmente llega a cero.  Que contradice el hecho de que $\alpha$ sea el m\'aximo de los ceros de $Y(x).$ Ahora si  $Y(x)<0,$ entonces de 
Eq. (\ref{eq:sturm-ordinarias-9}) se puede observar que la segunda  derivada es positiva, lo que quiere decir que la taza de crecimiento aumenta. Por lo tanto, $Y(x)$ crece y eventualmente llega a cero. Esto contradice el hecho de que $\alpha$ sea el m\'aximo de los cero de $Y(x).$\\
  
Tambi\'en podemos afirmar que en un intervalo cerrado y acotado las soluciones de 
Eq. (\ref{eq:sturm-ordinarias-9}) s\'olo pueden tener un n\'umero finito de ceros. Para probar esta afirmaci\'on recordemos el principio de Weierstrass, el cual no dice que una sucesi\'on acotada de n\'umeros reales tiene una subsucesi\'on convergente. Ahora, supongamos que $Y(x)$ es soluci\'on no trivial de 
Eq. (\ref{eq:sturm-ordinarias-9}) y que tiene un n\'umero infinito de ceros en el intervalo $[a,b].$ Con ese conjunto infinito de ceros se puede formar una sucesi\'on acotada. Por lo que existe 
una subsucesi\'on, $\{ x_{i} \}_{i=0}^{\infty},$ de ceros de $Y(x)$ que converge en un punto $x_{0}$ de $[a,b].$ Como $Y(x)$ es continua, se debe cumplir $Y(x_{0})=0,$ adem\'as
\begin{eqnarray}
\frac{dY(x)}{dx}\bigg|_{x_{0}}=\lim_{i\to \infty}\frac{Y(x_{i})-Y(x_{0})}{x_{i}-x_{0}}=\lim_{i\to \infty}\frac{0-0}{x_{i}-x_{0}}=0.
\end{eqnarray}
As\'i, tenemos una soluci\'on de Eq. (\ref{eq:sturm-ordinarias-9}) que cumple $Y(x_{0})=0, \frac{dY(x)}{dx}\bigg|_{x_{0}}=0,$
por el teorema de unicidad, este resultado implica que $Y(x)=0.$ Esto es absurdo, pues supusimos que $Y(x)$ es una soluci\'on no trivial de
 Eq. (\ref{eq:sturm-ordinarias-9}). As\'i, en un intervalo cerrado y acotado las soluciones de  Eq. (\ref{eq:sturm-ordinarias-9}) solo puede tener un n\'umero finito de ceros, que implica que los ceros de $Y(x)$ deben formar un conjunto numerable.\\

\section{Teorema de comparaci\'on de Sturm}

Ahora, supongamos que $\tilde q(x)<q(x)$ y que $Y(x)$ y $\tilde Y(x)$ son soluciones de las ecuaciones
\begin{eqnarray}
\frac{d^{2} Y(x)}{d x^{2}}+q(x) Y(x) =0,\label{eq:comparacion-sturm-1}\\
\frac{d^{2} \tilde Y(x)}{d x^{2}}+\tilde q(x) \tilde Y(x) =0 \label{eq:comparacion-sturm-2} .
\end{eqnarray}
Entonces se puede afirmar que $Y(x)$ tiene un cero entre dos cero consecutivos de $\tilde Y(x).$ A esta afirmaci\'on se le llama el {\bf Teorema de Comparaci\'on de Sturm}, para su demostraci\'on ocuparemos el Wronskiano 
\begin{eqnarray}
W(Y,\tilde Y)(x)=Y(x)\frac{d\tilde Y(x)}{dx}-\frac{dY(x)}{dx}\tilde Y(x).
\end{eqnarray}
Se puede probar que considerando  Eq. (\ref{eq:comparacion-sturm-1}) y Eq. (\ref{eq:comparacion-sturm-2}) se encuentra
\begin{eqnarray}
\frac{d W(Y,\tilde Y)(x)}{dx}=Y(x)\frac{d^{2}\tilde Y(x)}{dx^{2}}-\frac{d^{2}Y(x)}{dx^{2}}\tilde Y(x)=\left(q(x)-\tilde q(x)\right)\tilde Y(x) Y(x).\label{eq:comparacion-sturm-3}
\end{eqnarray}
Ahora, supongamos que $a_{1},a_{2}$ son ceros consecutivos de $\tilde Y(x).$ Sin perdida de generalidad,
podemos suponer que $\tilde Y(x)>0$ en $(a_{1},a_{2}),$ esto implica
\begin{eqnarray}
 \frac{d\tilde Y(x)}{dx}\bigg|_{a_{1}}>0 \qquad 
\frac{d\tilde Y(x)}{dx}\bigg|_{a_{2}}<0. 
\end{eqnarray}
Tambi\'en se cumple
\begin{eqnarray}
W(Y,\tilde Y)(a_{1})=Y(a_{1})\frac{d\tilde Y(x)}{dx}\bigg|_{a_{1}},\qquad W(Y,\tilde Y)(a_{2})=Y(a_{2})\frac{d\tilde Y(x)}{dx}\bigg|_{a_{2}}.\label{eq:comparacion-sturm-4}
\end{eqnarray}
De Eq. (\ref{eq:comparacion-sturm-3}) es claro que el signo de $\frac{d W(Y,\tilde Y)(x)}{dx}$ en  $[a_{1},a_{2}]$ solo depende del signo de $Y(x).$
Supongamos que $Y(x)$ no tiene ceros en ese intervalo, si $Y(x)>0$  entonces  $\left(q(x)-\tilde q(x)\right)Y(x)\tilde Y(x)>0.$ Note que al integrar Eq. (\ref{eq:comparacion-sturm-3}) se encuentra que $W(a_{2})>W(a_{1}).$ Mientras que de 
Eq. (\ref{eq:comparacion-sturm-4}) se tiene  $W(a_{1})>0$ y $W(a_{2})<0,$ lo cual es absurdo. As\'i, $Y(x)$ no puede tener signo positivo en el intervalo  $[a_{1},a_{2}].$ Ahora, si   $Y(x)<0$ en $[a_{1},a_{2}],$ entonces $\left(q(x)-\tilde q(x)\right)Y(x)\tilde Y(x)<0$ y al integrar Eq. (\ref{eq:comparacion-sturm-3}) se encuentra que $W(a_{2})<W(a_{1}).$ Pero de Eq. (\ref{eq:comparacion-sturm-4}) se tiene $W(a_{1})<W(a_{2}),$ lo cual es absurdo. As\'i, $Y(x)$ no puede tener solo signo negativo en el intervalo  $[a_{1},a_{2}].$ Esto implica que debe cambiar de signo en el intervalo $[a_{1},a_{2}]$ y por lo tanto debe tener un cero en ese intervalo.\\

Por ejemplo, supongamos que  tenemos las ecuaciones
\begin{eqnarray}
\frac{d^{2} Y(x)}{d x^{2}}+q(x) Y(x)& =&0,\label{eq:comparacion-sturm-5} \\
\frac{d^{2} \tilde Y(x)}{d x^{2}}+k^{2} \tilde Y(x) &=&0,\qquad k={\rm constante}, \label{eq:comparacion-sturm-6}
\end{eqnarray}
y se cumple $q(x)>k^{2}.$ Como las soluciones de  Eq. (\ref{eq:comparacion-sturm-6}) tienen ceros en los intervalos $\left[\frac{n\pi}{k}, \frac{(n+1)\pi}{k}\right],$ podemos
afirmar que las soluciones de Eq. (\ref{eq:comparacion-sturm-5}) tambi\'en tienen ceros en esos intervalos.\\

Los resultados que hemos visto nos sirven para estudiar los ceros de las soluciones de la ecuaci\'on de Bessel
\begin{eqnarray}
\frac{d^{2} Y(x)}{dx^{2}}+\frac{1}{x}\frac{dY(x)}{dx}+
\left(1-\frac{\nu^{2}}{x^{2}}\right)Y(x)=0,
\end{eqnarray}
a las soluciones de esta ecuaci\'on se les llaman funciones de Bessel.
En este caso la forma normal es 
\begin{eqnarray}
\frac{d^{2} u(x)}{dx^{2}}+q(x) u(x)=0,
\label{eq:normal-bessel}
\end{eqnarray}
con  
\begin{eqnarray}
q(x)=1+\frac{1-4\nu^{2}}{4x^{2}}.
\end{eqnarray}
Note que si $x>(\sqrt{4\nu^{2}-1})/2,$ se tiene $q(x)>0.$ Por lo tanto, las funciones de Bessel tienen un
n\'umero infinito de ceros.\\ 

Las funciones de Bessel las podemos comparar con las soluciones
de la ecuaci\'on
\begin{eqnarray}
\frac{d^{2} u(x)}{dx^{2}}+u(x)=0,\label{eq:normal-bessel-1}
\end{eqnarray}
cuyas soluciones son $\{\sin x, \cos x\}.$ La distancia entre dos ceros consecutivos para estas funciones es $\pi.$\\

Ahora, si $-\frac{1}{2}\leq \nu <\frac{1}{2},$ se cumple
\begin{eqnarray}
1<1+\frac{1-4\nu^{2}}{4x^{2}}.
\end{eqnarray}
Entonces cada intervalo de longitud $\pi$ tiene al menos un cero de las soluciones de la ecuaci\'on de Bessel. 
Para el caso $\nu  =\frac{1}{2},$ la ecuaci\'on normal de Bessel Eq. (\ref{eq:normal-bessel}) se reduce a Eq. (\ref{eq:normal-bessel-1}) 
y la distancia entre los ceros es exactamente $\pi.$ Ahora, si $\frac{1}{2}<\nu,$ se cumple
\begin{eqnarray}
1+\frac{1-4\nu^{2}}{4x^{2}}<1,\label{eq:normal-bessel-2}.
\end{eqnarray}
Esto implica que entre dos ceros sucesivos de las soluciones de  Eq. (\ref{eq:normal-bessel-1}) hay a lo m\'as un cero de las funciones
de Bessel. En efecto, supongamos que $\alpha_{1}$ y $\alpha_{2}$ son ceros sucesivos de Eq. (\ref{eq:normal-bessel-1}) y que en 
$(\alpha_{1}, \alpha_{2})$ hay dos ceros de las funciones de Bessel. Como se cumple Eq. (\ref{eq:normal-bessel-2}), debe haber un cero
de las soluciones de Eq. (\ref{eq:normal-bessel-1}), lo cual es absurdo, pues supusimos que $\alpha_{1}$ y $\alpha_{2}$ son ceros sucesivos
de las soluciones de Eq. (\ref{eq:normal-bessel-1}). Por lo tanto, si  $\frac{1}{2}<\nu,$ en cada intervalo de longitud $\pi$ hay a lo m\'as un cero de 
las funciones de Bessel.\\

\section{Problema de Sturm-Liuoville}

Una ecuaci\'on diferencial que surge en diferentes problemas de f\'isica y matem\'aticas es la
ecuaci\'on de Sturm-Liuoville:
\begin{eqnarray}
\frac{d}{dx}\left(p(x)\frac{d\psi(x)}{dx}\right)+\left(\lambda q(x)+r(x)\right)\psi(x)=0.
\label{eq:sturm}
\end{eqnarray}
Donde $q(x),p(x),r(x)$ son funciones reales, $q(x)$ es una funci\'on positiva en el intervalo $(a,b)$ y $\lambda$ es una
constante real. El problema consiste en encontrar las constantes $\lambda$ y funciones $\psi(x)$ que resuelven 
Eq. (\ref{eq:sturm}).\\ 

La ecuaci\'on Eq. (\ref{eq:sturm}) se suele resolver con  las condiciones de Dirichlet  
\begin{eqnarray}
\psi(a)=\psi(b)=0,\label{eq:vec-dirichlet}
\end{eqnarray}
o las de Neumann   
\begin{eqnarray}
\frac{d\psi(x)}{dx}\bigg|_{x=a}=\frac{d\psi(x)}{dx}\bigg|_{x=b}=0.\label{eq:vec-neumann}
\end{eqnarray}
Si no se cumple ninguna de estas condiciones se puede pedir que $p(x)$  cumpla 
\begin{eqnarray}
p(a)=p(b)=0.\label{eq:vec-libre}
\end{eqnarray}
La afirmaci\'on importante aqu\'i es que si $\psi_{\lambda_{1}}(x)$ es soluci\'on de Eq. (\ref{eq:sturm}) con $\lambda_{1}$ y
$\psi_{\lambda_{2}}(x)$  es soluci\'on de Eq. (\ref{eq:sturm}) con $\lambda_{2}$ y adem\'as se satisfacen una de las condiciones Eqs. 
(\ref{eq:vec-dirichlet})-(\ref{eq:vec-libre}),
entonces se cumple
\begin{eqnarray}
\left(\lambda_{1}-\lambda_{2}\right)\int_{a}^{b}dx q(x) \psi_{\lambda_{2}}^{*}(x)\psi_{\lambda_{1}}(x)=0.\nonumber
\end{eqnarray}
Para probar esta afirmaci\'on ocuparemos que se satisface
\begin{eqnarray}
\frac{d}{dx}\left(p(x)\frac{d\psi_{\lambda_{1}}(x)}{dx}\right)+\left(\lambda_{1} q(x)+r(x)\right)\psi_{\lambda_{1}}(x)&=&0,
\label{eq:sturm1} \\
\frac{d}{dx}\left(p(x)\frac{d\psi_{\lambda_{2}}(x)}{dx}\right)+\left(\lambda_{2} q(x)+r(x)\right)\psi_{\lambda_{2}}(x)&=&0.
\nonumber
\end{eqnarray}
El complejo conjugado de la segunda  ecuaci\'on es
\begin{eqnarray}
\frac{d}{dx}\left(p(x)\frac{d\psi_{\lambda_{2}}^{*}(x)}{dx}\right)+\left(\lambda_{2} q(x)+r(x)\right)\psi_{\lambda_{2}}^{*}(x)=0.
\label{eq:sturm2}
\end{eqnarray}
Adem\'as, multiplicando $\psi_{\lambda_{2}}^{*}(x)$ por Eq. (\ref{eq:sturm1}) y $\psi_{\lambda_{1}}(x)$ por Eq. (\ref{eq:sturm2}) se llega a 
\begin{eqnarray}
\psi_{\lambda_{2}}^{*}(x)\frac{d}{dx}\left(p(x)\frac{d\psi_{\lambda_{1}}(x)}{dx}\right)+\left(\lambda_{1} q(x)+r(x)\right)\psi_{\lambda_{2}}^{*}(x)\psi_{\lambda_{1}}(x)=0,
\nonumber\\
\psi_{\lambda_{1}}(x)\frac{d}{dx}\left(p(x)\frac{d\psi_{\lambda_{2}}^{*}(x)}{dx}\right)+\left(\lambda_{2} q(x)+r(x)\right)\psi_{\lambda_{1}}(x)\psi_{\lambda_{2}}^{*}(x)=0.
\nonumber
\end{eqnarray}
Adicionalmente, considerando
\begin{eqnarray}
f(x)\frac{dg(x)}{dx}=\frac{d f(x)g(x)}{dx}-\frac{df(x)}{dx}g(x)\nonumber
\end{eqnarray}
se encuentra
\begin{eqnarray}
\frac{d}{dx}\left(p(x)\psi_{\lambda_{2}}^{*}(x)\frac{d\psi_{\lambda_{1}}(x)}{dx}\right)- p(x)\frac{d\psi_{\lambda_{2}}^{*}(x)}{dx} \frac{d\psi_{\lambda_{1}}(x)}{dx}\nonumber\\
+\Bigg(\lambda_{1} q(x) +r(x)\Bigg)\psi_{\lambda_{2}}^{*}(x)\psi_{\lambda_{1}}(x)=0,\nonumber\\
 \frac{d}{dx}\left(p(x)\psi_{\lambda_{1}}(x)\frac{d\psi_{\lambda_{2}}^{*}(x)}{dx}\right)- p(x)\frac{d\psi_{\lambda_{1}}(x)}{dx} \frac{d\psi_{\lambda_{2}}^{*}(x)}{dx}\nonumber\\
+\Bigg(\lambda_{2} q(x)
+r(x)\Bigg)\psi_{\lambda_{2}}^{*}(x)\psi_{\lambda_{1}}(x)=0.\nonumber
\end{eqnarray}
Restando estas dos \'ultimas ecuaciones se llega a 
\begin{eqnarray}
& &\frac{d}{dx}
\left( p(x) \left( \psi_{\lambda_{2}}^{*}(x) \frac{d\psi_{\lambda_{1}}(x)}{dx}- \psi_{\lambda_{1}}(x)\frac{d\psi_{\lambda_{2}}^{*}(x)}{dx}\right)\right) \nonumber\\
& &+\left(\lambda_{1}-\lambda_{2}\right) q(x)\psi_{\lambda_{2}}^{*}(x)\psi_{\lambda_{1}}(x)=0.\nonumber
\end{eqnarray}
Integrando esta ecuaci\'on en el intervalo $[a,b],$ se obtiene
\begin{eqnarray}
 & & \left(p(x)\left( \psi_{\lambda_{2}}^{*}(x)\frac{d\psi_{\lambda_{1}}(x)}{dx}- \psi_{\lambda_{1}}(x) \frac{d\psi_{\lambda_{2}}^{*}(x)}{dx}\right)\right) \Bigg|_{a}^{b} \nonumber\\
& &+ \left(\lambda_{1}-\lambda_{2}\right)\int_{a}^{b}dx q(x) \psi_{\lambda_{2}}^{*}(x)\psi_{\lambda_{1}}(x)=0.\nonumber
\end{eqnarray}
Suponiendo que las soluciones satisfacen las condiciones de Dirichlet, de Neumann o bien que $p(x)$ se anule en la frontera, se consigue
\begin{eqnarray}
\left(\lambda_{1}-\lambda_{2}\right)\int_{a}^{b}dx q(x) \psi_{\lambda_{2}}^{*}(x)\psi_{\lambda_{1}}(x)=0,\nonumber
\end{eqnarray}
que es lo que queriamos probar.\\

En particular note que si $\lambda_{1}\not =\lambda_{2}$ se infiere que  
\begin{eqnarray}
\int_{a}^{b}dx q(x) \psi_{\lambda_{2}}^{*}(x)\psi_{\lambda_{1}}(x)=0,
\end{eqnarray}
Adem\'as, se puede ver que la integral 
\begin{eqnarray}
\int_{a}^{b}dx q(x) \psi_{\lambda_{1}}^{*}(x)\psi_{\lambda_{1}}(x)=\alpha_{\lambda},
\end{eqnarray}
es positiva, es decir $\alpha_{\lambda}>0.$ Por lo que, si $\alpha_{\lambda}<\infty, $ el conjunto de funciones 
\begin{eqnarray}
\frac{\tilde \psi_{\lambda}(x)}{\sqrt{\alpha_{\lambda}}}
\end{eqnarray}
cumplen
\begin{eqnarray}
 \int_{a}^{b}dx q(x) \psi^{*}_{\lambda_{1}}(x)\tilde \psi_{\lambda_{2}}(x)=\delta_{\lambda_{1}\lambda_{2} }.
\end{eqnarray}
Se dice que las soluciones de Eq. (\ref{eq:sturm}) que satisfacen alguna de las condiciones Eqs. (\ref{eq:vec-dirichlet})-(\ref{eq:vec-libre})
son un conjunto de funciones ortonormales con funci\'on de peso $q(x)$. En los pr\'oximos cap\'itulos  veremos varias aplicaciones de este resultado.

\chapter{Funciones de Bessel}

En este cap\'itulo estudiaremos la ecuaci\'on de Bessel y sus soluciones, las cuales se llaman funciones de Bessel. Las funciones de Bessel tienen aplicaciones en diversos problemas de mec\'anica cu\'antica, electrodin\'amica
y otras disciplinas. 

\section{Ecuaci\'on de Bessel}

La ecuaci\'on de Bessel es
\begin{eqnarray}
\frac{d^{2} R(z)}{dz^{2}}+\frac{1}{z}\frac{dR(z)}{dz}+
\left(1-\frac{\nu^{2}}{z^{2}}\right)R(z)=0,\label{eq:bessel-0}
\end{eqnarray}
que se puede escribir de la forma
\begin{eqnarray}
z^{2}\frac{d^{2} R(z)}{dz^{2}}+z \frac{dR(z)}{dz}+
\left(z^{2}-\nu^{2}\right)R(z)=0.\label{eq:bessel-1}
\end{eqnarray}
Para resolver esta ecuaci\'on ocuparemos 
el {\it M\'etodo de Frobenius} \cite{simmos:gnus}, es decir propondremos soluciones de la forma
\begin{eqnarray}
R(z)=z^{m}\sum_{n\geq 0} a_{n}z^{n}=\sum_{n\geq 0} a_{n}z^{n+m},\qquad a_{0}\not =0.
\label{eq:solbess}
\end{eqnarray}
De donde
\begin{eqnarray}
-\nu^{2}R(z)&=& \sum_{n\geq 0} -\nu^{2}a_{n}z^{n+m}= -\nu^{2}a_{0}z^{m}- \nu^{2}a_{1}z^{m+1}-\sum_{n\geq 2} \nu^{2}a_{n}z^{n+m} ,\nonumber\\
z^{2}R(z)&=& z^{2}\sum_{n\geq 0} a_{n}z^{n+m}=\sum_{n\geq 0} a_{n}z^{n+m+2}= \sum_{n\geq 2} a_{n-2}z^{n+m},\nonumber\\
z\frac{dR(z)}{dz}&=&z\sum_{n\geq 0}(n+m) a_{n}z^{n+m-1}= \sum_{n\geq 0}(n+m) a_{n}z^{n+m}\nonumber\\
&=& ma_{0}z^{m}+(m+1)a_{1}z^{m+1} + \sum_{n\geq 2}(n+m) a_{n}z^{n+m} ,  \nonumber\\
z^{2}\frac{d^{2}R(z)}{dz^{2}}&=&z^{2}\sum_{n\geq 0}(n+m)(n+m-1) a_{n}z^{n+m-2}\nonumber\\
&=& \sum_{n\geq 0}(n+m)(n+m-1) a_{n}z^{n+m}\nonumber\\
&=& m(m-1)a_{0}z^{m}+(m+1)m a_{1}z^{m+1}\nonumber\\
& &+  \sum_{n\geq 2}(n+m)(n+m-1) a_{n}z^{n+m}. \nonumber
\end{eqnarray}
Considerando estas cuatro igualdades en Eq. (\ref{eq:bessel-1}) y tomando en cuenta que
\begin{eqnarray}
 (n+m)(n+m-1)+(n+m)=(n+m)^{2},\nonumber
\end{eqnarray}
se tiene
\begin{eqnarray}
& & z^{2}\frac{d^{2} R(z)}{dz^{2}}+z \frac{dR(z)}{dz}+\left(z^{2}-\nu^{2}\right)R(z)=\nonumber\\
 & &= a_{0}\left( -\nu^{2}+m+m(m-1)\right)z^{m}+a_{1}
\left( -\nu^{2} +(m+1)+(m+1)m  \right)z^{m+1}\nonumber\\
& &+\sum_{n\geq 2}\bigg[\left[(n+m)(n+m-1)+(n+m) -\nu^{2}\right]a_{n}+ a_{n-2} \bigg]z^{n+m}\nonumber\\
& &= a_{0}\left(m^{2}-\nu^{2}\right)z^{m}+a_{1}\left((m+1)^{2}-\nu^{2}\right)z^{m+1}\nonumber\\
& & +\sum_{n\geq 2}\left(a_{n-2}+\left((n+m)^{2}-\nu^{2}\right)a_{n}\right)z^{n+m}=0,\label{eq:bessel-serie1}
\end{eqnarray}
que se debe cumplir para cualquier $z.$ Esto implica 
\begin{eqnarray}
a_{0}(m^{2}-\nu^{2})&=&0,\label{eq:bessel-primera}\\
a_{1}\left( (1+m)^{2}-\nu^{2}\right)&=&0,\label{eq:bessel-segunda}\\
a_{n-2}+a_{n} \left((n+m)^{2}-\nu^{2}\right)&=&0.\label{eq:bessel-tercera}
\end{eqnarray}
Como $a_{0}\not =0,$ Eq. (\ref{eq:bessel-primera}) induce
\begin{eqnarray}
m^{2}=\nu^{2},\qquad m=\pm \nu  ,\label{eq:relacion}
\end{eqnarray}
introduciendo este resultado en Eq. (\ref{eq:bessel-segunda}) se llega a
\begin{eqnarray}
a_{1}=0.\label{eq:primer-cero}
\end{eqnarray}
Adem\'as, considerando
\begin{eqnarray}
(n+m)^{2}-\nu^{2}=(n\pm \nu)^{2}-\nu^{2}=n^{2}\pm 2n\nu +\nu^{2}-\nu^{2}=n(n\pm 2\nu)
\end{eqnarray}
en Eq. (\ref{eq:bessel-tercera}) obtiene
\begin{eqnarray}
a_{n}=-\frac{a_{n-2}}{n(n\pm 2\nu)}.\label{eq:recurrencia}
\end{eqnarray}
Apartir de esta igualdad y ocupando  Eq. (\ref{eq:primer-cero}) se infiere que $a_{3}=0,$ que a su vez implica $a_{5}=0.$ Es claro que en general  $a_{2n+1}=0.$ As\'i, los \'unicos $a_{n}$ diferentes de cero son de la forma  
\begin{eqnarray}
a_{2n}=-\frac{a_{2(n-1)}}{2n(2n\pm 2 \nu)}= \frac{(-)}{2^{2}n(n\pm  \nu)}a_{2(n-1)}.
\label{eq:recubess}
\end{eqnarray}
Note que hay un problema si  $\nu$ es un natural y se considera el signo negativo en (\ref{eq:recubess}),
despu\'es trataremos esta cuesti\'on. Observe que Eq. (\ref{eq:recubess}) se puede escribir como 
\begin{eqnarray}
a_{2n}&=&\frac{(-)}{2^{2}n(n\pm  \nu)} a_{2(n-1)}=\frac{(-)(n-1)!(n-1\pm \nu)! }{2^{2}n!(n\pm  \nu)!} a_{2(n-1)}\nonumber\\
&=& \left(\frac{(-)(n-1)!(n-1\pm \nu)! }{2^{2}n!(n\pm  \nu)!} \right)
\left(\frac{(-)(n-2)!(n-2\pm \nu)! }{2^{2}(n-1)!(n-1\pm  \nu)!}\right) a_{2(n-2)} \nonumber\\
&=&\left(\frac{(-)^{2}(n-2)!(n-2\pm \nu)! }{2^{2\cdot 2}n!(n\pm  \nu)!}\right)  a_{2(n-2)}
\nonumber\\
&=&\left(\frac{(-)^{2}(n-2)!(n-2\pm \nu)! }{2^{2\cdot 2}n!(n\pm  \nu)!}\right) \left(\frac{(-)(n-3)!(n-3\pm \nu)! }{2 (n-2)!(n-2\pm  \nu)!}\right)  a_{2(n-3)}\nonumber\\
&=& \left(\frac{(-)^{3}(n-3)!(n-3\pm \nu)! }{2^{2\cdot 3}n!(n\pm  \nu)!}\right) a_{2(n-3)}\nonumber\\
&\vdots&\nonumber\\
&=&\left(\frac{(-)^{k}(n-k)!(n-k\pm \nu)! }{2^{2\cdot k}n!(n\pm  \nu)!}\right) a_{2(n-k)}.
\end{eqnarray}
El m\'aximo valor que puede tomar $k$  en la expresi\'on anterior es $n,$ entonces
\begin{eqnarray}
a_{2n}=\frac{(-)^{n}(\pm \nu)! }{2^{2n}n!(n\pm  \nu)!} a_{0},
\end{eqnarray}
tomando 
\begin{eqnarray}
a_{0}=\frac{1}{2^{\pm\nu}(\pm \nu)!}, 
\end{eqnarray}
%

%
%
se tiene
\begin{eqnarray}
a_{2n}=\frac{(-)^{n}}{2^{2n\pm \nu}n!(n\pm \nu)!}.
\end{eqnarray}
Sustituyendo este resultado en Eq. (\ref{eq:solbess})
se encuentra
\begin{eqnarray}
R(z)=z^{\pm \nu}\sum_{n\geq 0} \frac{(-)^{n}}{2^{2n\pm\nu}n!(n\pm \nu)!} z^{2n}
= \sum_{n\geq 0} \frac{(-)^{n}}{n!(n\pm \nu)!} \left(\frac{z}{2}\right)^{2n\pm \nu},
\end{eqnarray}
que son las  llamadas funciones  de Bessel. Se puede observar que ocupando la funci\'on Gamma, $\Gamma(z),$ las funciones de Bessel se pueden escribir como
\begin{eqnarray}
J_{\nu}(z)=\left(\frac{z}{2}\right)^{\nu}\sum_{n \geq 0}
\frac{ (-1)^{n}}{\Gamma(n+1)\Gamma(n+\nu+1)}\left(\frac{z}{2}\right)^{2n}, \label{eq:def-bessel}\\
J_{-\nu}(z)=\left(\frac{z}{2}\right)^{-\nu}\sum_{n \geq 0}
\frac{ (-1)^{n} }{\Gamma(n+1) \Gamma(n-\nu+1)} \left(\frac{z}{2}\right)^{2n}.
\end{eqnarray}
Note que si $\nu>0,$ se cumple $J_{\nu}(0)=0$ y $J_{-\nu}(0)=\infty.$\\ 

Ahora, para el caso en que $\nu$ es un natural, $\nu=m,$ probaremos que se cumple 
\begin{eqnarray}
J_{-m}(z)=(-)^{m} J_{m}(z).
\end{eqnarray}
Primero notemos que $J_{-m}(z)$ est\'a bien definida y recordemos que 
si $l$ es un natural $1/ \Gamma(-l)=0.$ Por lo que, el t\'ermino $1/\Gamma(n-m+1)$ es nulo si $n-m+1<0,$ entonces
\begin{eqnarray}
J_{-m}(z)&=& \sum_{n \geq 0}\frac{ (-1)^{n} }{\Gamma(n+1)\Gamma(n-m+1)} \left(\frac{z}{2}\right)^{2n-m}\nonumber\\
&=&\sum_{n \geq m}\frac{ (-1)^{n} }{\Gamma(n+1)\Gamma(n-m+1)} \left(\frac{z}{2}\right)^{2n-m}\nonumber\\
&=&\sum_{n \geq 0}\frac{ (-1)^{n+m} }{\Gamma(n+m+1)\Gamma(n+m-m+1)} \left(\frac{z}{2}\right)^{2(n+m)-m}\nonumber\\
&=&(-)^{m}\sum_{n \geq 0}\frac{ (-1)^{n} }{\Gamma(n+m+1)\Gamma(n+1)} \left(\frac{z}{2}\right)^{2n+m}\nonumber\\
&=&(-)^{m} J_{m}(z),
\end{eqnarray}
que es lo que queriamos probar. Es decir, si $m$ es natural, $J_{-m}(z)$ es soluci\'on de la ecuaci\'on de Bessel, pero no es  linealmente independientes  de $J_{m}(z).$ Por esta raz\'on en lugar de usar la funciones de Bessel del tipo $J_{-\nu}(z), \nu >0,$ se suelen ocupar las  funciones de Neumman 
\begin{eqnarray}
N_{\nu}(z)=\frac{ J_{\nu}(z) \cos\nu\pi - J_{-\nu}(z)}{\sin\nu\pi},
\end{eqnarray}
o las funciones de Hankel 
\begin{eqnarray}
H_{\nu}^{(1,2)}(z)&=&J_{\nu} (z)\pm iN_{\nu}(z).
\end{eqnarray}

\section{Funci\'on generatriz}

Existe una funci\'on de la cual se pueden extraer todas las funciones de Bessel de orden $n.$
A esta funci\'on se le llama  funci\'on generatriz y es:
\begin{eqnarray}
e^{\frac{z}{2}\left(t-\frac{1}{t}\right)}=\sum_{n\in Z} J_{n}(z) t^{n}.
\label{eq:generadora-bessel}
\end{eqnarray}
Para probar esta igualdad primero note que
\begin{eqnarray}
e^{\frac{zt}{2}}&=&\sum_{k\geq 0} \frac{1}{k!}\left(\frac{zt}{2}\right)^{k}= \sum_{k\geq 0} \frac{t^{k}}{k!}\left(\frac{z}{2}\right)^{k},\nonumber\\
e^{\frac{-z}{2t}}&=&\sum_{j\geq 0} \frac{1}{j!}\left(\frac{-z}{2t}\right)^{j}= \sum_{j\geq 0} \frac{(-)^{j}t^{-j}}{j!}\left(\frac{z}{2}\right)^{j},\nonumber
\end{eqnarray}
estas series implican
\begin{eqnarray}
e^{\frac{z}{2}\left(t-\frac{1}{t}\right)}&=&e^{\frac{zt}{2}}e^{\frac{-z}{2t}}=\left( \sum_{k\geq 0} \frac{t^{k}}{k!}\left(\frac{z}{2}\right)^{k}\right)\left(\sum_{j\geq 0} \frac{(-)^{j}t^{-j}}{j!}\left(\frac{z}{2}\right)^{j}\right)\nonumber\\
&=& \sum_{k\geq 0}\sum_{j\geq 0}\frac{t^{k-j}(-)^{j}}{k!j!}\left(\frac{z}{2}\right)^{k+j}.\nonumber
\end{eqnarray}
Ahora, definamos $n=k-j,$ por lo que $k=n+j$ y $k+j=2j+n,$ con este cambio de variable se llega a 
\begin{eqnarray}
e^{\frac{z}{2}\left(t-\frac{1}{t}\right)}&=& \sum_{n\in Z}\sum_{j\geq 0}\frac{t^{n}(-)^{j}}{j!(j+n)!}\left(\frac{z}{2}\right)^{2j+n}=
\sum_{n\in Z}t^{n} \sum_{j\geq 0}\frac{(-)^{j}}{j!(j+n)!}\left(\frac{z}{2}\right)^{2j+n}\nonumber\\
&=& \sum_{n\in Z} t^{n} J_{n}(z). \nonumber
\end{eqnarray}
Por lo tanto, se cumple Eq. (\ref{eq:generadora-bessel}). En particular si $t=e^{i\theta}$ se encuentra
\begin{eqnarray}
\frac{z}{2}\left(t-\frac{1}{t}\right)=iz\sin\theta,
\end{eqnarray}
de donde
\begin{eqnarray}
e^{iz\sin\theta}=\sum_{n\in Z} J_{n}(z) e^{in\theta}.\label{eq:generatriz-angulo}
\end{eqnarray}
Adicionalmente, como $\sin(\theta+\frac{\pi}{2})=\cos\theta$ y $ e^{i\frac{\pi}{2}}=i,$ se llega a
\begin{eqnarray}
e^{iz\cos\theta}=\sum_{n\in Z} (i)^{n}J_{n}(z) e^{in\theta},
\end{eqnarray}
esta es la lamada propiedad de Jacobi-Anger.\\

Adem\'as, considerando que si $m$ y $n$ son enteros se tiene 
\begin{eqnarray}
\int_{-\pi}^{\pi}d\theta  e^{-im\theta} e^{in\theta}=2\pi \delta_{nm}
\end{eqnarray}
y recurriendo  a Eq. (\ref{eq:generatriz-angulo}) se consigue
\begin{eqnarray}
\int_{-\pi}^{\pi}d\theta   e^{i(z\sin\theta- m\theta) }&=&
\int_{-\pi}^{\pi}d\theta   e^{-im\theta}e^{iz\sin\theta}=
 \int_{-\pi}^{\pi}d\theta   e^{-im\theta} \sum_{n\in Z} J_{n}(z) e^{in\theta}\nonumber\\
 &=&  \sum_{n\in Z} J_{n}(z) \int_{-\pi}^{\pi}d\theta  
 e^{-im\theta}e^{in\theta}= \sum_{n\in Z} J_{n}(z) 2\pi \delta_{nm}\nonumber\\
 &=& 2\pi J_{m}(z),\nonumber
\end{eqnarray}
entonces
\begin{eqnarray}
J_{n}(z)=\frac{1}{2\pi} \int_{-\pi}^{\pi}d\theta   e^{i(z\sin\theta- n\theta) }.
\end{eqnarray}
Tomando en cuenta la paridad de las funciones $\{\sin u,\cos u\}$ y la f\'ormula de Euler, 
esta integral toma la forma 
\begin{eqnarray}
J_{n}(z)&=&\frac{1}{2\pi} \int_{-\pi}^{\pi}d\theta   e^{i(z\sin\theta- n\theta) }\nonumber\\
&=& 
\frac{1}{2\pi} \int_{-\pi}^{\pi}d\theta  \left( \cos(z\sin\theta- n\theta) +i \sin (z\sin\theta- n\theta)\right) \nonumber\\
& =& \frac{2}{2\pi} \int_{0}^{\pi}d\theta  \cos(z\sin\theta- n\theta),
\end{eqnarray}
es decir 
\begin{eqnarray}
J_{n}(z)=\frac{1}{\pi} \int_{0}^{\pi}d\theta  \cos(z\sin\theta- n\theta).
\end{eqnarray}
Esta expresi\'on de las funciones de Bessel fue la  que originalmente encontr\'o 
F. W. Bessel. 

\section{Relaciones de recurrencia}

Ahora veremos que las funciones de Bessel satisfacen las relaciones de recu\-rrencia
\begin{eqnarray}
\frac{d}{dz}\left(z^{\nu}J_{\nu}(z)\right)&=& z^{\nu}J_{\nu-1}(z),\label{eq:iden-1-bessel}\\
\frac{d}{dz}\left(z^{-\nu}J_{\nu}(z)\right)&=& -z^{-\nu}J_{\nu+1}(z), \label{eq:iden-2-bessel}\\
\left(\frac{1}{z} \frac{d }{dz}\right)^{n}
\left(z^{\nu}J_{\nu}(z)\right)&=&z^{\nu-n}J_{\nu-n}(z), \label{eq:iden-3-bessel}\\
\left(\frac{1}{z} \frac{d }{dz}\right)^{n}
\left(z^{-\nu}J_{\nu}(z)\right)&=&(-)^{n}z^{-(\nu+n)}J_{\nu+n}(z).\label{eq:iden-4-bessel}
\end{eqnarray}
Para probar la primera identidad notemos que 
\begin{eqnarray}
z^{\nu}J_{\nu}(z)=z^{\nu}\sum_{n\geq 0}\frac{(-)^{n} }{n!(n+\nu)!}\frac{z^{2n+\nu}}{2^{2n+\nu}}=  
\sum_{n\geq 0}\frac{(-)^{n} }{n!(n+\nu)!}\frac{z^{2(n+\nu)}}{2^{2n+\nu}},\nonumber
\end{eqnarray}
entonces 
\begin{eqnarray}
\frac{d\left( z^{\nu}J_{\nu}(z)\right)}{d z}&=&  
\sum_{n\geq 0}\frac{(-)^{n}2(n+\nu) }{n!(n+\nu)!}\frac{z^{2(n+\nu)-1}}{2^{2n+\nu}}=
\sum_{n\geq 0}\frac{(-)^{n}}{n!(n+\nu -1)!}\frac{z^{2n+\nu-1}}{2^{2n+\nu-1}}z^{\nu}\nonumber\\
&=&
z^{\nu}\sum_{n\geq 0}\frac{(-)^{n}}{n!(n+\nu -1)!}\left(\frac{z}{2}\right)^{2n+\nu-1}
=z^{\nu} J_{\nu-1}(z),\nonumber
\end{eqnarray}
por lo tanto se cumple la identidad Eq. (\ref{eq:iden-1-bessel}).\\

Ahora,
\begin{eqnarray}
z^{-\nu}J_{\nu}(z)=z^{-\nu}\sum_{n\geq 0}\frac{(-)^{n} }{n!(n+\nu)!}\frac{z^{2n+\nu}}{2^{2n+\nu}}=  
\sum_{n\geq 0}\frac{(-)^{n} }{n!(n+\nu)!}\frac{z^{2n}}{2^{2n+\nu}},\nonumber
\end{eqnarray}
de donde
\begin{eqnarray}
\frac{d \left(z^{-\nu}J_{\nu}(z)\right)}{d z}&=&  
\sum_{n\geq 0}\frac{(-)^{n}2n }{n!(n+\nu)!}\frac{z^{2n-1}}{2^{2n+\nu}}=\sum_{n\geq 1}\frac{(-)^{n}2n }{n!(n+\nu)!}\frac{z^{2n-1}}{2^{2n+\nu}}\nonumber\\
&=& \sum_{n\geq 0}\frac{(-)^{n+1}2(n+1) }{(n+1)!(n+\nu+1)!}\frac{z^{2n+1}}{2^{2n+\nu+2}}\nonumber\\
&=&(-)\sum_{n\geq 0}\frac{(-)^{n}}{n!(n+\nu+1)!}\frac{z^{2n+\nu+1}}{2^{2n+\nu+1}}z^{-\nu }\nonumber\\
&=&(-)z^{-\nu}\sum_{n\geq 0}\frac{(-)^{n}}{n!(n+\nu +1)!}\left(\frac{z}{2}\right)^{2n+\nu+1}\nonumber\\
&=&(-)z^{-\nu} J_{\nu+1}(z),\nonumber
\end{eqnarray}
as\'i, se cumple la identidad Eq. (\ref{eq:iden-2-bessel}).\\

Para probar las identidades Eqs. (\ref{eq:iden-3-bessel})-(\ref{eq:iden-4-bessel})
ocuparemos inducci\'on.  Primero haremos la prueba de Eq. (\ref{eq:iden-3-bessel}). Para $n=0$
esta igualdad es correcta, por lo que la base inductiva est\'a demostrada.  Para el paso 
inductivo debemos suponer Eq. (\ref{eq:iden-3-bessel}) y probar
\begin{eqnarray}
\left(\frac{1}{z} \frac{d }{dz}\right)^{n+1}\left(z^{\nu}J_{\nu}(z)\right)&=&z^{\nu-(n+1)}J_{\nu-(n+1)}(z).
\nonumber
\end{eqnarray}
Note que ocupando la hip\'otesis inductiva y Eq. (\ref{eq:iden-1-bessel}) se tiene 
\begin{eqnarray}
\left(\frac{1}{z} \frac{d }{dz}\right)^{n+1}\left(z^{\nu}J_{\nu}(z)\right)&=&\frac{1}{z} \frac{d }{dz}
\left(\left(\frac{1}{z} \frac{d }{dz}\right)^{n}\left(z^{\nu}J_{\nu}(z)\right)\right)\nonumber\\
&=& \frac{1}{z} \frac{d }{dz}\left( z^{\nu-n}J_{\nu-n}(z)    \right)\nonumber\\
&=& \frac{1}{z}\left( z^{\nu-n}J_{\nu-n-1}(z)\right)=  z^{\nu-(n+1)}J_{\nu-(n+1)}(z)\nonumber
\end{eqnarray}
que es lo que queriamos demostrar. As\'i, la igualdad (\ref{eq:iden-3-bessel}) es correcta
para cualquier $n.$\\

Ahora probaremos Eq. (\ref{eq:iden-4-bessel}). Para $n=0$
esta igualdad es correcta, por lo que la base inductiva est\'a demostrada.  Para el paso 
inductivo debemos suponer Eq. (\ref{eq:iden-4-bessel}) y demostrar la igualdad 
\begin{eqnarray}
\left(\frac{1}{z} \frac{d }{dz}\right)^{n+1}\left(z^{-\nu}J_{\nu}(z)\right)&=&(-)^{n+1} z^{-(\nu+n+1)}J_{\nu+n+1}(z).
\nonumber
\end{eqnarray}
Usando la hip\'otesis inductiva y Eq. (\ref{eq:iden-2-bessel}) se encuentra 
\begin{eqnarray}
\left(\frac{1}{z} \frac{d }{dz}\right)^{n+1}\left(z^{-\nu}J_{\nu}(z)\right)&=&\frac{1}{z} \frac{d }{dz}
\left(\left(\frac{1}{z} \frac{d }{dz}\right)^{n}\left(z^{-\nu}J_{\nu}(z)\right)\right)\nonumber\\
&=& \frac{1}{z} \frac{d }{dz}\left((-)^{n} z^{-\nu-n}J_{\nu+n}(z) \right)\nonumber\\
& =& (-)^{n} \frac{1}{z} \frac{d }{dz}\left(z^{-(\nu+n)}J_{\nu+n}(z) \right)\nonumber\\
&=& (-)^{n}(-)\frac{1}{z}\left( z^{-(\nu+n)}J_{\nu+n+1}(z)\right)\nonumber\\
&=&(-)^{n+1}  z^{-(\nu+n+1)}J_{\nu+n+1}(z)\nonumber
\end{eqnarray}
esto es lo que queriamos demostrar. Por lo tanto la igualdad Eq. (\ref{eq:iden-4-bessel}) es v\'alida para cualquier $n.$\\

Las identidades Eqs. (\ref{eq:iden-1-bessel})-(\ref{eq:iden-4-bessel}) 
tambi\'en se pueden escribir como
\begin{eqnarray}
J_{\nu-1}(z)&=&\frac{dJ_{\nu}(z)}{dz}+ \frac{\nu}{z}J_{\nu}(z), \label{eq:iden-iden-1}\\
J_{\nu+1}(z)&=&\frac{dJ_{\nu}(z)}{dz}- \frac{\nu}{z}J_{\nu}(z), \label{eq:iden-iden-2} \\
J_{\nu-n}(z)&=&z^{n-\nu}\left(\frac{1}{z} \frac{d }{dz}\right)^{n} \left(z^{\nu}J_{\nu}(z)\right), \label{eq:iden-iden-3}\\
J_{\nu+n}(z)&=& (-)^{n}z^{\nu+n}\left(\frac{1}{z} \frac{d }{dz}\right)^{n}
\left(z^{-\nu}J_{\nu}(z)\right).\label{eq:iden-iden-4}
\end{eqnarray}
Estas identidades son importantes para las aplicaciones.

\section{Funciones de Bessel de orden $\left(n+\frac{1}{2}\right)$}

Las funciones de Bessel de orden $(n+\frac{1}{2})$ son particularmente importantes para las
aplicaciones, por lo que vale la pena estudiar sus propiedades.  Primero observemos que ocupando 
Eq. (\ref{eq:def-bessel}) y la serie de Taylor de la funci\'on $\sin z$ se llega a 
\begin{eqnarray}
J_{\frac{1}{2}}(z)&=&\left(\frac{z}{2}\right)^{\frac{1}{2}}\sum_{n \geq 0}
\frac{ (-1)^{n}}{n!\left(n+\frac{1}{2}\right)!}\left(\frac{z}{2}\right)^{2n}
= \left(\frac{z}{2}\right)^{\frac{1}{2}}\sum_{n \geq 0}
\frac{(-1)^{n} } {n! \left(\frac{(2n+1)!\sqrt{\pi}}{2^{2n+1} n!} \right) } 
\left(\frac{z}{2}\right)^{2n}\nonumber\\
&=& \left(\frac{z}{2}\right)^{\frac{1}{2}}\sum_{n \geq 0}
\frac{2(-1)^{n} } {( 2n+1)!\sqrt{\pi} } z^{2n}
= \left(\frac{2z}{\pi}\right)^{\frac{1}{2}} \frac{1}{z} 
\sum_{n \geq 0}
\frac{(-1)^{n} } {( 2n+1)!} z^{2n+1}\nonumber\\
&=& \left(\frac{2}{\pi z}\right)^{\frac{1}{2}}  \sum_{n \geq 0}
\frac{(-1)^{n} } {( 2n+1)!} z^{2n+1},\nonumber
\end{eqnarray}
es decir 
\begin{eqnarray}
J_{\frac{1}{2}}(z)=\left(\frac{2}{\pi z}\right)^{1/2} \sin z.
\label{eq:bessel-semi-entero}
\end{eqnarray}
Adem\'as, considerando 
Eq. (\ref{eq:def-bessel}) y la serie de Taylor de la funci\'on $\cos z,$ se obtiene
\begin{eqnarray}
J_{-\frac{1}{2}}(z)&=&\left(\frac{z}{2}\right)^{-\frac{1}{2}}\sum_{n \geq 0}
\frac{ (-1)^{n}}{n!\left(n-\frac{1}{2}\right)!}\left(\frac{z}{2}\right)^{2n}
= \left(\frac{z}{2}\right)^{-\frac{1}{2}}\sum_{n \geq 0}
\frac{(-1)^{n} } {n! \left(\frac{(2n)!\sqrt{\pi}}{2^{2n} n!} \right) } 
\left(\frac{z}{2}\right)^{2n}\nonumber\\
&=& \left(\frac{2}{\pi z}\right)^{\frac{1}{2}}\sum_{n \geq 0}
\frac{(-1)^{n} } {(2n)!} z^{2n},
\end{eqnarray}
por lo que 
\begin{eqnarray}
J_{-\frac{1}{2}}(z)=\left(\frac{2}{\pi z}\right)^{\frac{1}{2}} \cos z.
\label{eq:bessel-semi-entero-negativo}
\end{eqnarray}
Usando Eqs.  (\ref{eq:iden-iden-4})-(\ref{eq:bessel-semi-entero})  se encuentra
\begin{eqnarray}
J_{n+\frac{1}{2}}(z)=(-)^{n}z^{\left(n+\frac{1}{2}\right)} \left(\frac{2}{\pi}\right)^{\frac{1}{2} }
\left(\frac{1}{z}\frac{d}{dz} \right)^{n}\left(\frac{\sin z}{z}\right).
\end{eqnarray}
De forma analoga, apelando a (\ref{eq:iden-iden-4})-(\ref{eq:bessel-semi-entero}) se llega a 
\begin{eqnarray}
J_{-\left(n+\frac{1}{2}\right)}(z)=z^{\left(n+\frac{1}{2}\right)} \left(\frac{2}{\pi}\right)^{\frac{1}{2}}
\left(\frac{1}{z}\frac{d}{dz} \right)^{n}\left(\frac{\cos z}{z}\right).
\end{eqnarray}
Adicionalmente, ocupando el resultado
\begin{eqnarray}
\cos\left(n+\frac{1}{2}\right)\pi=0,\qquad \sin\left(n+\frac{1}{2}\right)\pi=(-)^{n},
\end{eqnarray}
se encuentra
\begin{eqnarray}
N_{\left(n+\frac{1}{2}\right)}(z)&=&\frac{ J_{n+\frac{1}{2}}(z) \cos\left(n+\frac{1}{2}\right)\pi - J_{-\left(n+\frac{1}{2}\right)}(z) }
{\sin\left(n+\frac{1}{2}\right)\pi}\nonumber\\
&=&(-)^{n+1} J_{-\left(n+\frac{1}{2}\right)}(z),\nonumber
\end{eqnarray}
es decir 
\begin{eqnarray}
N_{\left(n+\frac{1}{2}\right)}(z)&=&(-)^{n+1}
z^{\left(n+\frac{1}{2}\right)} \left(\frac{2}{\pi}\right)^{\frac{1}{2} }
\left(\frac{1}{z}\frac{d}{dz} \right)^{n}\left(\frac{\cos z}{z}\right). \nonumber
\end{eqnarray}
Las funciones de Hankel de orden $\left(n+\frac{1}{2}\right)$ tienen la forma
\begin{eqnarray}
H^{(1,2)}_{\left(n+\frac{1}{2}\right)}(z)&=&  J_{\left(n+\frac{1}{2}\right)}(z)\pm i N_{\left(n+\frac{1}{2}\right)}(z)\nonumber\\
& =& (-)^{n}z^{\left(n+\frac{1}{2}\right)} \left(\frac{2}{\pi}\right)^{\frac{1}{2} }
\left(\frac{1}{z}\frac{d}{dz} \right)^{n}\left(\frac{\sin z}{z}\right) \nonumber\\
 & & \pm i(-)^{n+1}
z^{\left(n+\frac{1}{2}\right)} \left(\frac{2}{\pi}\right)^{\frac{1}{2} }
\left(\frac{1}{z}\frac{d}{dz} \right)^{n}\left(\frac{\cos z}{z}\right) \nonumber \\
&=  & (-)^{n}z^{\left(n+\frac{1}{2}\right)} \left(\frac{2}{\pi}\right)^{\frac{1}{2} }
\left(\frac{1}{z}\frac{d}{dz} \right)^{n} \left( \frac{\sin z}{z}\mp i \frac{\cos z}{z}\right)  \nonumber \\
&=& (-)^{n}(\mp i) z^{\left(n+\frac{1}{2}\right)} \left(\frac{2}{\pi}\right)^{\frac{1}{2} }
\left(\frac{1}{z}\frac{d}{dz} \right)^{n} \left(\frac{\cos z \pm i \sin z}{z}\right)\nonumber\\
&=&(\mp i)(-)^{n}\left(\frac{2}{\pi}\right)^{\frac{1}{2}}
z^{\left(n+\frac{1}{2}\right)} \left(\frac{1}{z}\frac{d}{dz} \right)^{n}
\left(\frac{e^{\pm iz}}{z}\right),\nonumber
\end{eqnarray}
entonces 
\begin{eqnarray}
H^{(1,2)}_{\left(n+\frac{1}{2}\right)}(z)=(\mp i)(-)^{n}\left(\frac{2}{\pi}\right)^{\frac{1}{2}}
z^{\left(n+\frac{1}{2}\right)} \left(\frac{1}{z}\frac{d}{dz} \right)^{n}
\left(\frac{e^{\pm iz}}{z}\right).\nonumber
\end{eqnarray}
Definiremos las funciones esf\'ericas de Bessel como
\begin{eqnarray}
j_{l}(z)&=&\left(\frac{\pi}{2z}\right)^{\frac{1}{2}} J_{\left(l+\frac{1}{2}\right)}(z),\label{eq:esfericas-bessel-1} \nonumber\\    n_{l}(z)&=&\left(\frac{\pi}{2z}\right)^{\frac{1}{2}} N_{\left(l+\frac{1}{2}\right)}(z),\label{eq:esfericas-bessel-2}\\
h^{(1,2)}_{l}(z)&=& \left(\frac{\pi}{2z}\right)^{\frac{1}{2}}H^{(1,2)}_{\left(l+\frac{1}{2}\right)}(z) ,\label{eq:esfericas-bessel-3}\nonumber
\end{eqnarray}
de donde
\begin{eqnarray}
j_{l}(z)&=&(-z)^{l} \left(\frac{1}{z}\frac{d}{dz} \right)^{l}\left(\frac{\sin z}{z}\right), \\
n_{l}(z)&=& -(-z)^{l}\left(\frac{1}{z}\frac{d}{dz} \right)^{l}\left(\frac{\cos z}{z}\right),\nonumber\\
 h^{(1,2)}_{l}(z)&=& 
(\mp i)(-z)^{l} \left(\frac{1}{z}\frac{d}{dz} \right)^{l}
\left(\frac{e^{\pm iz}}{z}\right).
\end{eqnarray}
Estas funciones se usan en  mec\'anica cu\'antica y  electrodin\'amica.

\section{Ortonormalidad}

En el cap\'itulo anterior vimos que cada  funci\'on de Bessel $J_{\nu}(z)$ tienen un n\'umero numerable de ra\'ices, $\lambda_{ n},$ que satisfacen $ J_{\nu}(\lambda_{n})=0.$
Ocuparemos este resultado para probar que la integral del producto de dos funciones de Bessel satisfacen
una propiedad que llamaremos de ortonormalidad.\\

La ecuaci\'on 
\begin{eqnarray}
\frac{d^{2} R_{\alpha}(x)}{dx^{2}}+\frac{1}{x} \frac{d R_{\alpha}(x)}{dx}+\left(\alpha^{2}-\frac{\nu^{2}}{x^{2}}\right)R_{\alpha}(x)=0,
\label{eq:bessel-orto-1}
\end{eqnarray}
con el cambio de variable $z=\alpha x$ se convierte en la ecuaci\'on de Bessel Eq. (\ref{eq:bessel-0}) que tiene la soluciones $J_{\nu}(z),$
por lo que  $R_{\alpha}(x)= J_{\nu}(\alpha x).$ Adem\'as, Eq. (\ref{eq:bessel-orto-1}) se puede escribir como
\begin{eqnarray}
\frac{ 1}{x}\frac{d}{dx} \left(x \frac{d R_{\alpha}(x)}{dx}\right)+\left(\alpha^{2}-\frac{\nu^{2}}{x^{2}}\right)R_{\alpha}(x)=0,
\end{eqnarray}
es decir 
\begin{eqnarray}
\frac{d}{dx} \left(x \frac{d R_{\alpha}(x)}{dx}\right)+\left(x\alpha^{2}-\frac{\nu^{2}}{x}\right)R_{\alpha}(x)=0.
\label{eq:bessel-orto-1-alpha}
\end{eqnarray}
En particular si $\alpha=\lambda_{n}$ se tiene la ecuaci\'on 
\begin{eqnarray}
\frac{d}{dx} \left(x \frac{d R_{\nu}(x)}{dx}\right)+\left(x\lambda_{n} ^{2}-\frac{\nu^{2}}{x}\right)R_{\nu}(x)=0,
\end{eqnarray}
que tiene las soluciones $R_{\nu}(x)= J_{\nu}(\lambda_{ n} x).$ Note que 
esta ecuaci\'on es del tipo Sturm-Liuoville Eq. (\ref{eq:sturm}) y  si $\nu\geq 0,$
se cumplen las condiciones de borde de Dirichlet 
\begin{eqnarray}
R_{\nu}(0)=R_{\nu}(1)=0.
\end{eqnarray}
Por lo tanto, usando el teorema de Sturm-Liuoville, mostrado en el cap\'itulo anterior, 
se llega a 
\begin{eqnarray}
(\lambda_{n}^{2}-\lambda_{m}^{2}) \int_{0}^{1}dx x J_{\nu}(\lambda_{n} x)J_{\nu}(\lambda_{m} x)=0.
\end{eqnarray}
En particular si $\lambda_{n}\not =\lambda_{m},$ se debe cumplir 
\begin{eqnarray}
\int_{0}^{1}dx x J_{\nu}(\lambda_{n} x)J_{\nu}(\lambda_{m} x)=0,
\end{eqnarray}
de donde 
\begin{eqnarray}
\int_{0}^{1}dx x J_{\nu}(\lambda_{n} x)J_{\nu}(\lambda_{m} x)=\delta_{nm}a^{2},\qquad  a= {\rm constante}. 
\end{eqnarray}
A esta propiedad se le llamada de ortogonalidad, se dice que las funciones de Bessel son ortogonales con peso $x.$\\ 

Para calcular la constante $a$ multilplicaremos Eq. (\ref{eq:bessel-orto-1}) por $2x^{2}\frac{d R_{\alpha}(x)}{dx},$
de donde
\begin{eqnarray}
0&=&2x^{2}\frac{dR_{\alpha}(x) }{dx} \frac{d^{2} R_{\alpha}(x)}{dx^{2}}+2x \left(\frac{dR_{\alpha}(x) }{dx}\right)^{2}
+\left(x^{2}\alpha^{2}-\nu^{2} \right)2 \frac{dR_{\alpha}(x) }{dx}   R_{\alpha}(x) \nonumber\\
 &=& x^{2} \frac{d}{dx}\left(\frac{dR_{\alpha}(x) }{dx}\right)^{2} +2x  \left(\frac{dR_{\alpha}(x)}{dx}\right)^{2} + 
 \left(x^{2}\alpha^{2}-\nu^{2} \right)  \frac{d }{dx}  \left( R_{\alpha}(x)\right)^{2}\nonumber\\
 & =& \frac{d}{dx}\left(x^{2} \left(\frac{dR_{\alpha}(x) }{dx}\right)^{2}\right)-  \nu^{2}  \frac{d }{dx}  \left( R_{\alpha}(x)\right)^{2}
 + x^{2}\alpha^{2} \frac{d }{dx}  \left( R_{\alpha}(x)\right)^{2}\nonumber,
\end{eqnarray}
ocupando que 
\begin{eqnarray}
x^{2} \frac{d}{dx} \left(R_{\alpha}(x)\right)^{2}= \frac{ d}{dx} \left(  x^{2}\left(R_{\alpha}(x)\right)^{2}\right)- 2x \left(R_{\alpha}(x)\right)^{2}, \nonumber
\end{eqnarray}
se tiene 
\begin{eqnarray}
\frac{d}{dx}\left(x^{2} \left( \frac{dR_{\alpha}(x) }{dx}\right)^{2}+(x^{2}\alpha^{2}-  \nu^{2} )  \left( R_{\alpha}(x)\right)^{2}  \right)-  
2\alpha^{2}x  \left( R_{\alpha}(x)\right)^{2}=0.
\end{eqnarray}
Por lo tanto, 
\begin{eqnarray}
2\alpha^{2}\int_{0}^{1} dx x  \left( R_{\alpha}(x)\right)^{2}=
\left(x^{2} \left( \frac{dR_{\alpha}(x) }{dx}\right)^{2}+(x^{2}\alpha^{2}-  \nu^{2} )  \left( R_{\alpha}(x)\right)^{2}  \right)\Bigg|_{0}^{1}.
\end{eqnarray}
En particular, como $R_{\alpha}(x)=J_{\nu}(\alpha x),$ si $\nu >0$ y $\alpha=\lambda_{n}$ con $J_{\nu}(\lambda_{n})=0,$  se tiene
\begin{eqnarray}
2\lambda_{n}^{2} \int_{0}^{1} dx x  \left( J_{\nu}(\lambda_{n} x)\right)^{2}= \left( \frac{dJ_{\nu}(\lambda_{n} x) }{dx}\right)^{2}\Bigg|_{x=1}
= \lambda_{n}^{2}\left( \frac{dJ_{\nu}(\lambda_{n} x) }{d (\lambda_{n} x) }\right)^{2}\Bigg|_{x=1},
\end{eqnarray}
considerando la identidad Eq. (\ref{eq:iden-iden-2}) se llega a 
\begin{eqnarray}
\int_{0}^{1} dx x  \left( J_{\nu}(\lambda_{n} x)\right)^{2}= \frac{1}{2} \left( J_{\nu+1}(\lambda_{n} )\right)^{2}.
\end{eqnarray}
Por lo que, si $\nu>0$ y $\lambda_{n},\lambda_{m}$ son raices de la funci\'on de Bessel $J_{\nu}( z)$ se cumple
\begin{eqnarray}
\int_{0}^{1} dz z  J_{\nu}(\lambda_{n} z) J_{\nu}(\lambda_{m} z)= \frac{\delta_{nm} }{2} \left( J_{\nu+1}(\lambda_{n} )\right)^{2}.
\label{eq:ortonormalidad-bessel}
\end{eqnarray}
Entonces, para cualquier funci\'on $f(z)$ definida en el intervalo $(0,1)$ se puede expresar en t\'erminos
de la funci\'on de Bessel $J_{\nu}(\lambda_{n}z).$ En efecto, supongamos que 
\begin{eqnarray}
f(z)=\sum_{m\geq 0}a_{m}J_{\nu}(\lambda_{m} z),
\end{eqnarray}
entonces
\begin{eqnarray}
\int_{0}^{1}dz z J_{\nu}(\lambda_{n} z)f(z)&=&\int_{0}^{1}dz zJ_{\nu}(\lambda_{n} z) \sum_{m\geq 0}a_{m}J_{\nu}(\lambda_{m} z)\nonumber\\
&=&
\sum_{m\geq 0}a_{m}\int_{0}^{1}dzz J_{\nu}(\lambda_{n} z) J_{\nu}(\lambda_{m} z)\nonumber\\
&=&  \sum_{m\geq 0}a_{m} \frac{\delta_{nm} }{2} \left( J_{\nu+1}(\lambda_{n} )\right)^{2}= \frac{a_{n} }{2} \left( J_{\nu+1}(\lambda_{n} )\right)^{2},
\end{eqnarray}
por lo que
\begin{eqnarray}
a_{n} = \frac{2}{ \left( J_{\nu+1}(\lambda_{n} )\right)^{2} } \int_{0}^{1}dz z J_{\nu}(\lambda_{n} z)f(z).
\end{eqnarray}
En la pr\'oxima secci\'on veremos una aplicaci\'on de este resultado.

\section{La ecuaci\'on de Laplace en coordenadas cil\'indricas }

Ahora veremos algunas aplicaciones de las funciones de Bessel.\\
 
Primero estudiaremos las soluciones de la ecuaci\'on de Laplace en coordenadas cil\'indricas
\begin{eqnarray}
\nabla^{2}\phi(\rho,\varphi,z)=\frac{1}{\rho}\frac{\partial }{\partial \rho}
\left(\rho \frac{\partial \phi(\rho,\varphi, z)}{\partial \rho}\right) +\frac{1}{\rho^{2}} \frac{\partial^{2} \phi(\rho,\varphi, z)}{\partial \varphi^{2}}+ \frac{\partial^{2} \phi(\rho,\varphi,z)} {\partial z^{2}}=0.\nonumber
\end{eqnarray}
Para resolver esta ecuaci\'on propondremos $\phi(\rho,\varphi, z)=R(\rho)\Phi(\varphi)Z(z),$ de donde
\begin{eqnarray}
\nabla^{2}\phi(\rho,\varphi,z)&=&
\frac{1}{\rho}\frac{\partial }{\partial \rho}
\left(\rho \frac{\partial R(\rho)\Phi(\varphi)Z(z)}{\partial \rho}\right) +\frac{1}{\rho^{2}} \frac{\partial^{2} 
 R(\rho)\Phi(\varphi)Z(z)}{\partial \varphi^{2}}\nonumber\\
 & & + \frac{\partial^{2} R(\rho)\Phi(\varphi)Z(z)} {\partial z^{2}}\nonumber\\
 &=&  \frac{ \Phi(\varphi)Z(z) }{\rho}\frac{\partial }{\partial \rho}
\left(\rho \frac{\partial R(\rho)}{\partial \rho}\right)+ \frac{R(\rho)Z(z) }{\rho^{2}} \frac{\partial^{2} \Phi(\varphi)}{\partial \varphi^{2}}\nonumber\\
& &+ R(\rho)\Phi(\varphi) \frac{\partial^{2} Z(z)} {\partial z^{2}}=0,\nonumber
\end{eqnarray}
por lo que 
\begin{eqnarray}
\frac{\nabla^{2}\phi(\rho,\varphi, z)}{ \phi(\rho,\varphi,z)} = 
\frac{1}{ \rho R(\rho) }\frac{\partial }{\partial \rho}\left(\rho \frac{\partial R(\rho)}{\partial \rho}\right)+ 
 \frac{1}{\rho^{2}\Phi(\varphi) } \frac{\partial^{2} \Phi(\varphi)}{\partial \varphi^{2}}
+ \frac{1}{ Z(z)}\frac{\partial^{2} Z(z)} {\partial z^{2}}=0.\quad 
 \label{eq:pre-bessel-1}
\end{eqnarray}
Entonces 
\begin{eqnarray}
\frac{\partial }{\partial z}\left( \frac{\nabla^{2}\phi(\rho,\varphi,z)}{ \phi(\rho,\varphi,z)}\right) &=& 
 \frac{\partial }{\partial z}\left( 
 \frac{1}{ Z(z)}\frac{\partial^{2} Z(z)} {\partial z^{2}}\right)=0,\nonumber
\end{eqnarray}
que induce
\begin{eqnarray}
\frac{1}{ Z(z)}\frac{\partial^{2} Z(z)} {\partial z^{2}}=\alpha^{2},\qquad 
\frac{\partial^{2} Z(z)} {\partial z^{2}}= \alpha^{2}Z(z),
\label{eq:bessel-z}
\end{eqnarray}
la soluci\'on general a esta ecuaci\'on es 
\begin{eqnarray}
Z(z)=a_{\alpha}e^{\alpha z}+ b_{\alpha}e^{-\alpha z}.
\end{eqnarray}
Sustituyendo (\ref{eq:bessel-z}) en (\ref{eq:pre-bessel-1}) se  tiene
\begin{eqnarray}
\frac{\nabla^{2}\phi(\rho,\varphi,z)}{ \phi(\rho,\varphi,z)} &=& 
\frac{1}{ \rho R(\rho) }\frac{\partial }{\partial \rho}\left(\rho \frac{\partial R(\rho)}{\partial \rho}\right)+ 
 \frac{1}{\rho^{2}\Phi(\varphi) } \frac{\partial^{2} \Phi(\varphi)}{\partial \varphi^{2}}+ \alpha^{2}=0,\nonumber
\end{eqnarray}
as\'{\i}
\begin{eqnarray}
\frac{\partial }{\partial \varphi}
\left( \rho^{2}\frac{\nabla^{2}\phi(\rho,\varphi,z)}{ \phi(\rho,\varphi,z)}\right) &=& 
 \frac{\partial }{\partial \varphi}
\left(\frac{1}{\Phi(\varphi) } \frac{\partial^{2} \Phi(\varphi)}{\partial \varphi^{2}}\right)=0,
\end{eqnarray}
que implica
\begin{eqnarray}
\frac{1}{ \Phi(\varphi)} \frac{\partial^{2} \Phi(\varphi)} {\partial \varphi^{2}}=-\nu^{2},\qquad 
\frac{\partial^{2} \Phi(\varphi)} {\partial \varphi^{2}}= -\nu^{2}\Phi(\varphi),\label{eq:bessel-angulo}
\end{eqnarray}
cuya  soluci\'on general  es
\begin{eqnarray}
\Phi(\varphi)=A_{\nu}e^{i\nu \varphi }+ B_{\nu}e^{-i\nu \varphi }.
\end{eqnarray}
Introduciendo  (\ref{eq:bessel-angulo}) en (\ref{eq:pre-bessel-1}) se llega a
\begin{eqnarray} 
\frac{1}{ \rho R(\rho) }\frac{\partial }{\partial \rho}\left(\rho \frac{\partial R(\rho)}{\partial \rho}\right)- 
 \frac{\nu^{2}}{\rho^{2}} + \alpha^{2}=0,\nonumber
\end{eqnarray}
es decir,
\begin{eqnarray} 
\frac{1}{ \rho }\frac{\partial }{\partial \rho}\left(\rho \frac{\partial R(\rho)}{\partial \rho}\right)+
\left(\alpha^{2}- \frac{\nu^{2}}{\rho^{2}}\right) R(\rho) =0.\nonumber
\end{eqnarray}
Con el cambio de variable $\zeta=\alpha\rho$ se encuentra
\begin{eqnarray} 
\frac{1}{ \zeta }\frac{d }{d \zeta}\left(\zeta  \frac{d R(\zeta )}{d \zeta }\right)+
\left(1- \frac{\nu^{2}}{\zeta ^{2}}\right) R(\zeta ) =0,
\label{eq:bessel}
\end{eqnarray}
que es la  ecuaci\'on de Bessel. De donde
\begin{eqnarray}
R(\rho )=C_{\nu} J_{\nu}(\alpha \rho) + D_{\nu} J_{-\nu}(\alpha \rho) \nonumber.
\end{eqnarray}
Por lo que las soluciones de la ecuaci\'on de Laplace en coordenadas cil\'indricas son de la forma
\begin{eqnarray}
\phi(\rho,\varphi,z)_{\alpha,\nu}= \left(a_{\alpha}e^{\alpha  z}+ b_{\alpha}e^{-\alpha z}\right) 
\left(A_{\nu}e^{i\nu \varphi }+ B_{\nu}e^{-i\nu \varphi }\right)\left(C_{\nu} J_{\nu}(\alpha \rho) + D_{\nu} J_{-\nu}(\alpha \rho)\right)
\nonumber.
\end{eqnarray}
Las constantes $a_{\alpha}, b_{\alpha}, A_{\nu}, B_{\nu}, C_{\nu}, D_{\nu}$ de determina seg\'un las condiciones de borde del problema.\\

Por ejemplo, si en $\rho=0$ el potencial debe ser finito, como la funci\'on de Bessel $ J_{-\nu}(\alpha\rho)$ diverge en $\rho=0,$   debe ocurrir que $D_{\nu}=0.$ En ese caso la
soluci\'on es de la forma
\begin{eqnarray}
\phi(\rho,\varphi,z)_{\alpha,\nu}= \left(a_{\alpha}e^{\alpha z}+ b_{\alpha}e^{-\alpha z}\right) 
\left(A_{\nu}e^{i\nu \varphi }+ B_{\nu}e^{-i\nu \varphi }\right)J_{\nu}(\alpha \rho) 
\nonumber.
\end{eqnarray}
Adem\'as, para muchos problemas es importante que $\phi(\rho,\varphi, z)$ sea una funci\'on univaluada. 
As\'i, como $(\rho,\varphi , z )$ y $(\rho,\varphi+2\pi , z )$ representan el mismo
punto, se debe cumplir
\begin{eqnarray}
\phi(\rho,\varphi+2\pi , z )= \phi(\rho,\varphi , z ),
\end{eqnarray}
en consecuencia 
\begin{eqnarray}
\Phi(\varphi+2\pi)=A_{\nu}e^{i\nu (\varphi +2\pi)}+ B_{\nu}e^{-i\nu (\varphi+2\pi) }
=\Phi(\varphi)=A_{\nu}e^{i\nu \varphi }+ B_{\nu}e^{-i\nu \varphi },
\end{eqnarray}
que induce 
\begin{eqnarray}
e^{i2\pi\nu}=1.
\end{eqnarray}
Por lo tanto, $\nu$ debe ser un n\'umero natural $n.$ Esto implica que $R(\rho)$ debe  ser de la forma
\begin{eqnarray}
R(\rho)=C_{n} J_{n}(\alpha \rho) + D_{n} J_{-n}(\alpha \rho).
\end{eqnarray}
As\'i, para este caso se tiene las soluciones
\begin{eqnarray}
\phi_{\alpha,n} (\rho,\varphi,z)= \left(a_{\alpha}e^{\alpha z}+ b_{\alpha}e^{-\alpha z}\right) 
\left(A_{n}e^{in\varphi }+ B_{n}e^{-in \varphi }\right)\left(C_{n} J_{n}(\alpha \rho) + D_{n} J_{-n}(\alpha \rho)\right)
\nonumber.
\end{eqnarray}
Por lo tanto, si el potencial es univaluado y adem\'as  finito en el origen, debe ser una combinaci\'on lineal de potenciales
de la forma
\begin{eqnarray}
\phi _{\alpha,n}(\rho,\varphi,z)= \left(a_{\alpha}e^{\alpha z}+ b_{\alpha}e^{-\alpha z}\right) 
\left(A_{n}e^{in\varphi }+ B_{n}e^{-in \varphi }\right) J_{n}(\alpha \rho) 
\nonumber.
\end{eqnarray}

\subsection{Ejemplo}

Veamos un problema de electrost\'atica.\\
%
%
Supongamos que tenemos un cilindro de radio $\tilde R$ y altura $h.$ La tapa inferior del cilindro y la superficie lateral tiene pontencial cero, mientras que la tapa superior tiene potencial $V(\rho,\varphi).$ Calcularemos el potencial el\'ectrico en el interior del cilindro suponiendo que no hay cargas en esa regi\'on.\\

Por simplicidad, pondremos el eje del cilindro en el eje $z$ y la tapa inferior la pondremos sobre el plano $x-y.$ En este sistema 
las condiciones de borde son 
\begin{eqnarray}
\phi(\rho,\varphi, 0 )=0, \qquad \phi(\tilde R,\varphi, z )=0, \qquad \phi(\rho ,\varphi, h )=  V(\rho,\varphi).
\end{eqnarray}
Como no hay fuentes dentro del cilindro, el potencial deber ser finito en el interior. Adem\'as como el potencial debe ser univaluado, \'este debe ser de la forma 
\begin{eqnarray}
\phi_{\alpha,n} (\rho,\varphi,z)= \left(a_{\alpha}e^{\alpha  z}+ b_{\alpha}e^{-\alpha z}\right) 
\left(A_{n}e^{in\varphi }+ B_{n}e^{-in \varphi }\right) J_{n}(\alpha \rho) 
\nonumber.
\end{eqnarray}
\begin{eqnarray}
R(\rho)=C_{n} J_{n}(\alpha \rho).
\end{eqnarray}
Adicionalmente, como se debe cumplir la condici\'on de borde $\phi(\tilde R,\varphi, z )=0,$
se tiene que 
\begin{eqnarray}
R(\tilde R)=C_{n} J_{n}(\alpha \tilde R)=0,
\end{eqnarray}
que implica 
\begin{eqnarray}
\alpha \tilde R=\lambda_{nm},\quad \alpha= \frac{\lambda_{nm}}{\tilde R}.
\label{eq:raiz-cilindro}
\end{eqnarray}
Donde $\lambda_{nm}$ es la ra\'iz $m$-\'esima la funci\'on de Bessel de orden $n.$ As\'i, la funciones $R(\rho)$
son de la forma
\begin{eqnarray}
R(\rho)=C_{n} J_{n}\left( \frac{\lambda_{nm}\rho }{\tilde R}\right).
\end{eqnarray}
Note que Eq. (\ref{eq:raiz-cilindro}) implica que $Z(z)$ tome la forma
\begin{eqnarray}
Z(z)=a_{nm}e^{\frac{\lambda_{nm} z}{\tilde R} }+ b_{nm}e^{-\frac{\lambda_{nm} z}{\tilde R} } 
\end{eqnarray}
Mientras que  la condici\'on de borde $\phi(\rho,\varphi, 0 )=0$ implica que 
\begin{eqnarray}
Z(0)=\left(a_{nm}+ b_{nm}\right)=0,
\end{eqnarray}
por lo tanto,
\begin{eqnarray}
Z(z)=a_{nm}\left( e^{\frac{\lambda_{nm}  z}{\tilde R} }- e^{-\frac{\lambda_{nm} z}{\tilde R}}\right)=A_{nm}\sinh\left(\frac{\lambda_{nm} z}{\tilde R} \right).
\end{eqnarray}
As\'i, la soluci\'on m\'as general de la ecuaci\'on de Laplace que satisface las condiciones de borde 
$\phi(\rho,\varphi, 0 )=\phi(\tilde R,\varphi, z )=0,$ es 
\begin{eqnarray}
\phi(\rho,\varphi, z )=\sum_{n\geq 0}\sum_{m \geq 0} \sinh\left(\frac{\lambda_{nm} z}{\tilde R}\right)J_{n}\left( \frac{\lambda_{nm}\rho }{\tilde R}\right)\left(A_{nm} \cos n\phi +B_{nm}\sin n\phi\right)\nonumber.
\end{eqnarray}
Para determinar los coeficientes $A_{nm}, B_{nm}$ debemos imponer la condici\'on de borde faltante:
\begin{eqnarray}
& &\phi(\rho,\varphi, h )=V(\rho,\varphi)\nonumber\\
& & = \sum_{n\geq 0}
\sum_{m \geq 0} \sinh\left(\frac{\lambda_{nm} h }{\tilde R} \right)  J_{n}\left( \frac{\lambda_{nm}\rho }{\tilde R}\right)
\left(A_{nm} \cos n\phi +B_{nm}\sin n\phi\right)  \nonumber.
\end{eqnarray}
Ahora, se puede probar que si $k$ y $l$ son naturales se cumplen las integrales 
\begin{eqnarray}
\int_{0}^{2\pi}d\varphi \cos k\varphi  \cos l\varphi =\int_{0}^{2\pi}d\varphi \sin k\varphi  \sin l\varphi=\pi \delta_{kl},\quad \int_{0}^{2\pi}d\varphi \cos k\varphi  \sin l\varphi=0,\nonumber 
\end{eqnarray}
por lo que, usando Eq. (\ref{eq:ortonormalidad-bessel}), se encuentra
\begin{eqnarray}
& &\int_{0}^{2\pi} d\varphi \int_{0}^{1} d\left(\frac{\rho}{\tilde R}\right)\sin k\varphi J_{k}\left(\frac{\lambda_{kl} \rho}{\tilde R} \right) V(\rho, \varphi)= \sum_{n\geq 0}
\sum_{m \geq 0} \sinh\left(\frac{\lambda_{nm} h }{\tilde R} \right)  B_{nm} \nonumber\\
& &\times\int_{0}^{2\pi} d\varphi \sin k\varphi \sin n\varphi
\int_{0}^{1} d\left(\frac{\rho}{\tilde R}\right) J_{k}\left(\frac{\lambda_{kl} \rho}{\tilde R} \right) J_{n}\left( \frac{\lambda_{nm}\rho }{\tilde R}\right)\nonumber\\
&=& \sum_{n\geq 0}
\sum_{m \geq 0} \sinh\left(\frac{\lambda_{nm} h }{\tilde R} \right)  B_{nm}\pi \delta_{kn}
\int_{0}^{1} d\left(\frac{\rho}{\tilde R}\right) J_{k}\left(\frac{\lambda_{kl} \rho}{\tilde R} \right) J_{n}\left( \frac{\lambda_{nm}\rho }{\tilde R}\right)\nonumber\\
&=&\sum_{m \geq 0} \sinh\left(\frac{\lambda_{km} h }{\tilde R} \right)  B_{km}\pi 
\int_{0}^{1} d\left(\frac{\rho}{\tilde R}\right) J_{k}\left(\frac{\lambda_{kl} \rho}{\tilde R} \right) J_{k}\left( \frac{\lambda_{km}\rho }{\tilde R}\right)\nonumber\\
&=&\sum_{m \geq 0} \sinh\left(\frac{\lambda_{km} h }{\tilde R} \right)  B_{km}\pi \delta_{lm} \frac{1}{2} \left(J_{k+1}(\lambda_{kl}\right)^{2}
= \sinh\left(\frac{\lambda_{kl} h }{\tilde R} \right)  B_{kl}\pi \left(J_{k+1}(\lambda_{kl})\right)^{2},\nonumber
\end{eqnarray}
entonces
\begin{eqnarray}
B_{kl}=\frac{ 2}{\pi \sinh\left(\frac{\lambda_{kl} h }{\tilde R} \right)  \left(J_{k+1}(\lambda_{kl}\right)^{2}}
\int_{0}^{2\pi} d\varphi \int_{0}^{1} d\left(\frac{\rho}{\tilde R}\right)\sin k\varphi J_{k}\left(\frac{\lambda_{kl} \rho}{\tilde R} \right) V(\rho, \varphi)  .\nonumber
\end{eqnarray}
De la misma forma se obtiene
\begin{eqnarray}
A_{kl}=\frac{ 2}
{\pi \sinh\left(\frac{\lambda_{kl} h }{\tilde R} \right)  \left(J_{k+1}(\lambda_{kl})\right)^{2}}\int_{0}^{2\pi} d\varphi \int_{0}^{1} d\left(\frac{\rho}{\tilde R}\right)\cos k\varphi J_{k}\left(\frac{\lambda_{kl} \rho}{\tilde R} \right) V(\rho, \varphi) .\nonumber
\end{eqnarray}

\section{Ecuaciones tipo Bessel}

Existen varias ecuaciones que se pueden reducir a la ecuaci\'on de Bessel. 
Por ejemplo, si $R(z)$ es soluci\'on de la ecuaci\'on de Bessel, 
 la funci\'on
\begin{eqnarray}
u(z)=z^{-c}R(az^{b}) \nonumber
\end{eqnarray}
es una soluci\'on de la ecuaci\'on
\begin{eqnarray}
z^{2}\frac{d^{2} u(z)}{dz^{2}}+(2c +1)z\frac{du(z)}{dz}+\left(a^{2}b^{2} z^{2b}+\left(c^{2}-\nu^{2}b^{2}\right)\right)u(z)=0.
\label{eq:tipo-bessel}
\end{eqnarray}
Para probar esta afirmaci\'on tomaremos el cambio de variable $w=az^{b},$ 
de donde
\begin{eqnarray}
z&=&\left(\frac{w}{a}\right)^{\frac{1}{b}},\\
\frac{dz}{dw}&=&\frac{z}{bw}=\frac{z^{1-b}}{ba},\\
R(w)&=&z^{c}u(z).\nonumber
\end{eqnarray}
Por lo que
\begin{eqnarray}
\left(w^{2}-\nu^{2}\right)R(w)&=&\left(a^{2}z^{2b} -\nu^{2}\right)z^{c}u(z)=\frac{z^{c}}{b^{2}}
\left(a^{2}b^{2}z^{2b} -\nu^{2}b^{2}\right)u(z),\nonumber\\
\frac{d R(w)}{dw} &=&\frac{d}{dw}\left(z^{c}u(z)\right)=\frac{d z}{dw}\frac{d}{dz}\left(z^{c}u(z)\right)\nonumber\\
&=&\left(\frac{z^{1-b}}{ab}\right) \left(cz^{c-1}u(z)+z^{c}\frac{d u(z)}{dz} \right)\nonumber\\
&=&\frac{1}{ab}\left( cz^{c-b}u(z)+z^{c-b+1}\frac{d u(z)}{dz} \right)
\nonumber\\
w\frac{d R(w)}{dw}&=&\frac{z^{c}}{b}\left( cu(z)+z\frac{du(z)}{dz}\right)= 
\frac{z^{c}}{b^{2}}\left( cbu(z)+zb\frac{du(z)}{dz}\right)\nonumber\\
\frac{d^{2} R(w)}{dw^{2}}&=& \frac{dz}{dw}\frac{d}{dz}\left(\frac{1}{ab}\left( cz^{c-b}u(z)+z^{c-b+1}\frac{d u(z)}{dz} \right)\right)\nonumber\\
&=& \frac{1}{ab} \Bigg(c(c-b)z^{c-b-1}u(z)+(2c-b+1)z^{c-b} \frac{d u(z)}{dz}\nonumber\\
 & & + z^{c-b+1} \frac{d^{2} u(z)}{dz^{2}}\Bigg) \nonumber\\
w^{2} \frac{d^{2} R(w)}{dw^{2}}&=& \frac{z^{c}}{b^{2}}
 \left( c(c-b)u(z) +(2c+1-b) z\frac{d u(z)}{dz}+ z^{2}\frac{d^{2} u(z)}{dz^{2}}\right).\nonumber
\end{eqnarray}
De donde, como $R(w)$ satisface la ecuaci\'on de Bessel, se encuentra 
\begin{eqnarray}
0&=&w^{2} \frac{d^{2} R(w)}{dw^{2}}+w \frac{d R(w)}{dw}+ \left(w^{2}-\nu^{2}\right)R(w)\nonumber\\
&=& \frac{z^{c}}{b^{2}}
 \left( c(c-b)u(z) +(2c+1-b) z\frac{d u(z)}{dz}+ z^{2}\frac{d^{2} u(z)}{dz^{2}}\right)\nonumber\\
 &+ & \frac{z^{c}}{b^{2}}\left( cbu(z)+zb\frac{du(z)}{dz}\right)+ \frac{z^{c}}{b^{2}}
\left(a^{2}b^{2}z^{2b} -\nu^{2}b^{2}\right)u(z)\nonumber\\
&=&\frac{z^{c}}{b^{2}}\left( z^{2}\frac{d^{2} u(z)}{dz^{2}}+(2c +1)z\frac{du(z)}{dz}+\left(a^{2}b^{2} z^{2b}+\left(c^{2}-\nu^{2}b^{2}\right)\right)u(z)\right)=0.\nonumber
\end{eqnarray}
Lo que implica que la funci\'on $u(z)$ es soluci\'on de Eq. (\ref{eq:tipo-bessel}). 
Este resultado tiene varias aplicaciones. Por ejemplo, consideremos la ecuaci\'on de Airy 
\begin{eqnarray}
\frac{d^{2} u(z)}{dz^{2}}+zu(z)=0.
\label{eq:airy}
\end{eqnarray}
Note que si 
\begin{eqnarray}
c=-\frac{1}{2},\quad b=\frac{3}{2},\qquad a=\frac{2}{3},\qquad \nu=\pm \frac{1}{3} 
\end{eqnarray}
Eq. (\ref{eq:tipo-bessel}) se convierte en Eq. (\ref{eq:airy}). Por lo tanto,
la soluci\'on general de la ecuaci\'on de Airy es
\begin{eqnarray}
u(z)=|z|^{\frac{1}{2}} \left( A J_{\frac{1}{3}}\left( \frac{2 |z|^{\frac{3}{2}}}{3}\right) +B  J_{-\frac{1}{3}}\left( \frac{2 |z|^{\frac{3}{2}}}{3}\right)\right),
\label{eq:Airy}
\end{eqnarray}
con $A$ y $B$ constantes.

\subsection{ Part\'icula cu\'antica en una fuerza constante }

La ecuaci\'on de Schr$\ddot {\rm o}$dinger para una part\'icula  en una fuerza constante, $F,$ es
\begin{eqnarray}
\left(-\frac{\hbar^{2}}{2m} \frac{\partial^{2}}{\partial x^{2}}  -Fx \right)\psi(x)=E\psi(x).
\label{eq:bessel-campo-constante}
\end{eqnarray}
Con el cambio de variable 
\begin{eqnarray}
z=\left(\frac{2mF}{\hbar^{2}}\right)^{\frac{1}{3}}\left(x+\frac{E}{F}\right)
\end{eqnarray}
se obtiene 
\begin{eqnarray}
x=z\left(\frac{\hbar^{2}}{2mF}\right)^{\frac{1}{3}}-\frac{E}{F},\qquad    \frac{\partial }{\partial x}=\left(\frac{2mF}{\hbar^{2}}\right)^{\frac{1}{3}}\frac{\partial}{\partial x},
\end{eqnarray}
por lo que Eq. (\ref{eq:bessel-campo-constante}) toma la forma
\begin{eqnarray}
\left( \frac{\partial^{2}}{\partial z^{2}}  +z \right)\psi(z)=0,
\end{eqnarray}
que es la ecuaci\'on de Airy. Entonces, considerando (\ref{eq:Airy}), se tiene 
\begin{eqnarray}
\psi(x)&=&\left(\frac{2mF}{\hbar^{2}}\right)^{\frac{1}{3}} \left|x+\frac{E}{F}\right|^{\frac{1}{2}}
\Bigg[ A J_{\frac{1}{3}} \left(\left|\frac{8mF}{9\hbar^{2}}\right|^{\frac{1}{2}} \left|x+\frac{E}{F}\right|^{\frac{3}{2}} \right)      \nonumber\\
& & +    B J_{-\frac{1}{3}} \left(\left|\frac{8mF}{9\hbar^{2}}\right|^{\frac{1}{2}} \left|x+\frac{E}{F}\right|^{\frac{3}{2}} \right)        \Bigg]
\end{eqnarray}
que es la funci\'on de onda del sistema.

\section{Mec\'anica cu\'antica conforme}

La ecuaci\'on de Schr\"odinger para la llamada cu\'antica conforme es \cite{jackiw:gnus,jackiw1:gnus} 
\begin{eqnarray}
i\hbar \frac{\partial \psi(x,t)}{\partial t}& =&\left(-\frac{\hbar^{2}}{2m} \frac{\partial^{2} }{\partial x^{2}} +\frac{g}{x^{2}}\right)\psi(x,t).
\label{eq:cqmh}
\end{eqnarray}
Este sistema tiene aplicaciones en diferentes \'areas de la f\'isica. En este caso  tomaremos la propuesta
\begin{eqnarray}
 \psi(x,t)=e^{-i\frac{Et}{\hbar} }  \phi(x),
 \end{eqnarray}
de donde (\ref{eq:cqmh}) toma la forma 
\begin{eqnarray}
E \phi(x)& =&\left(-\frac{\hbar^{2}}{2m} \frac{\partial^{2} }{\partial x^{2}} +\frac{g}{x^{2}}\right)\phi(x),\label{eq:dus2}
\end{eqnarray}
la cual se puede escribir como
\begin{eqnarray}
 \left(\frac{\partial^{2} }{\partial x^{2}} +\left( \frac{2mE}{\hbar^{2}}-\frac{2mg}{\hbar^{2} x^{2}}\right)\right) \phi(x)=0. \label{eq:alfaro}
\end{eqnarray}
Se puede observar que tomando las constantes 
\begin{eqnarray}
c=-\frac{1}{2},\quad b=1,\qquad a=\sqrt{ \frac{2mE}{\hbar^{2}}},\qquad \nu=\pm\sqrt{ \frac{1}{4}+\frac{2mg}{\hbar^{2}}} 
\end{eqnarray}
en  la ecuaci\'on  (\ref{eq:tipo-bessel}) se obtiene   (\ref{eq:airy}). Por lo tanto,
\begin{eqnarray}
\phi(x)=A|x|^{\frac{1}{2}}  J_{\pm \sqrt{\frac{1}{4} +\frac{2mg}{\hbar^{2}} } } \left( \sqrt{\frac{2mE }{\hbar^{2}} } x\right).
\end{eqnarray}
con $A$ una  constante. Entonces,  la funci\'on de onda de la mec\'anica cu\'antica conforme es
\begin{eqnarray}
 \psi(x,t)=Ae^{-i\frac{Et}{\hbar} } |x|^{\frac{1}{2}}  J_{\pm \sqrt{\frac{1}{4} +\frac{2mg}{\hbar^{2}} } } \left( \sqrt{\frac{2mE }{\hbar^{2}} } x\right).
  \end{eqnarray}
\section{Ecuaci\'on de  Fick-Jacobs }

En diferentes sistemas es importante el estudio de difusi\'on de part\'iculas en un medio. El modelo de disfusi\'on m\'as simple est\'a dado 
por la ecuaci\'on de Fick
\begin{eqnarray}
\frac{ \partial C(x,t)}{\partial t}=D \frac{\partial^{2} C(x,t) }{\partial x^{2}}.\label{eq:dus}
\end{eqnarray}
Aqu\'i $C(x,t)$ es la concentraci\'on de part\'iculas y  $D$ es el coeficiente de difusi\'on.  Si $D$ es una constante,  la ecuaci\'on de Fick 
es equivalente a la ecuaci\'on de calor (\ref{eq:TFcalor}),  as\'i la ecuaci\'on de Fick se puede resolver con las t\'ecnicas usadas para resolver 
esta \'ultima ecuaci\'on. M\'as adelante se estudiar\'a la ecuaci\'on de calor.\\

Cuando la difusi\'on se da en canales la geometr\'ia de \'este es importante y la ecuaci\'on de Fick ya no es v\'alida. Para el caso en que el canal tiene la
forma de una superficie de revoluci\'on, donde el \'area de la secci\'on transversal es $A(x),$  la ecuaci\'on de Fick se debe cambiar por la llamada ecuaci\'on de 
Fick-Jacobs \cite{jacobs:gnus}
\begin{eqnarray}
\frac{ \partial C(x,t)}{\partial t}&=&\frac{\partial }{\partial x}
\left[D(x) A(x) \frac{\partial }{\partial x} \left(
\frac{C(x,t)}{A(x)} \right)\right].\label{eq:fick0} 
\end{eqnarray}
Claramente esta ecuaci\'on es m\'as sofisticada que la ecuaci\'on de Fick (\ref{eq:dus}). \\

Existen caso en los cuales la ecuaci\'on de Fick-Jacobs se puede escribir como una ecuaci\'on de Schr\"odinger. Por ejemplo,  
si 
$$D(x)=D_{0}={\rm constante},$$ 
usando la propuesta 
$$C(x,t)=\sqrt{A(x)}\psi(x,t)$$ 
se obtiene  
\begin{eqnarray}
\frac{ \partial \psi(x,t)}{\partial t}=\left[D_{0} \frac{\partial^{2} }{\partial x^{2}} 
-\frac{D_{0}}{2\sqrt{A(x)}}
\frac{\partial}{\partial x}\left(\frac{1}{\sqrt{A(x)}} \frac{\partial A(x)}{\partial x}\right)\right]\psi(x,t).
\end{eqnarray}
Por  lo tanto proponiendo la soluci\'on $\psi(x,t)=e^{-Et}\phi(x),$ se llega a la ecuaci\'on de Schr\"odinger
\begin{eqnarray}
E\phi(x)=H\phi(x),\label{eq:static}
\end{eqnarray}
con
\begin{eqnarray}
H=-D_{0} \frac{\partial^{2} }{\partial x^{2}} +
\frac{D_{0}}{2\sqrt{A(x)}} \frac{\partial}{\partial x}\left(\frac{1}{\sqrt{A(x)}} \frac{\partial A(x)}{\partial x}\right).
\label{eq:fjhc}
\end{eqnarray}
Para el caso particular de  canales con  secci\'on transversal  de la forma
\begin{eqnarray}
A(x)= (a+\lambda x)^{2\nu}
\end{eqnarray}
el Hamiltoniano es 
\begin{eqnarray}
H=-D_{0} \frac{\partial^{2} }{\partial x^{2}} +\frac{g}{\left(a+\lambda x\right)^{2}}, \qquad g=\lambda^{2}D_{0}\nu\left(\nu-1\right).  
\end{eqnarray}
Por lo que, usando el cambio de variable 
\begin{eqnarray}
z= a+\lambda x,  
\end{eqnarray}
se llega a 
\begin{eqnarray}
H=-D_{0}\lambda^{2} \frac{\partial^{2} }{\partial z^{2}} +\frac{g}{z^{2}}, \qquad g=\lambda^{2}D_{0}\nu\left(\nu-1\right). 
\end{eqnarray}
Por lo tanto, la ecuaci\'on a resolver es 
\begin{eqnarray}
E\phi(x)=\left(-D_{0}\lambda^{2} \frac{\partial^{2} }{\partial z^{2}} +\frac{g}{z^{2}}\right) \phi(x), \qquad g=\lambda^{2}D_{0}\nu\left(\nu-1\right),
\end{eqnarray}
que  tiene la misma forma que la ecuaci\'on (\ref{eq:dus2}).  Entonces, usando los resultados de la secci\'on previa, se encuentra que concentraci\'on de part\'iculas est\'a dada por 
\begin{eqnarray}
C_{\nu}(x,t)=Be^{-Et}\left(a+\lambda x\right)^{\frac{2\nu+1}{2}} J_{\pm\left(\frac{2\nu-1}{2}\right)} \left( \pm \sqrt{\frac{E}{\lambda^{2} D_{0}}} \left(a+\lambda x\right)\right),
\end{eqnarray}
donde $B$ es una constante.

\chapter{Elementos de \'Algebra Lineal }

En este cap\'itulo veremos algunas herramientas del \'algebra lineal.
Este cap\'itulo es importante para
entender los cap\'itulos posteriores. Tambi\'en es importante para entender los principios 
de la mec\'anica cu\'antica, as\'i como para resolver ecuaciones diferenciales. 

\section{Espacios vectoriales}

Un espacio vectorial se define con un conjunto ${\bf V},$ un campo $\bf K$ y dos operaciones
\begin{eqnarray}
&+&:{\bf V}\times {\bf V}\to {\bf V},\\
&\mu& :{\bf K}\times {\bf V}\to {\bf V}.
\end{eqnarray}
Esta operaciones deben cumplir que si $u,v$ pertenecen a ${\bf V},$ entonces $u+v$ pertenece a ${\bf V}$
y si $\alpha$ pertenece a $\bf K,$ entonces $\mu(\alpha,v)=\alpha v$ pertenece a $\bf V.$
Adem\'as, se debe  cumplir 
\begin{eqnarray}
&1)& \forall u,v \in {\bf V},\qquad u+v=v+u,\\
&2)& \forall u,v,w \in {\bf V},\qquad (u+v)+w=u+(v+w),\\
&3)& \forall u,v \in {\bf V}, \forall \alpha,\in {\bf K},\quad  \alpha(u+v)=\alpha u+\alpha v,\\
&4)& \forall v \in {\bf V}, \forall \alpha, \beta \in {\bf K}, \qquad (\alpha +\beta)v=\alpha v+\beta v,\\
&5)& \forall v \in {\bf V}, \forall \alpha, \beta \in {\bf K},\qquad (\alpha \beta)v=\alpha (\beta v),\\
&6)& \exists \quad  0 \in {\bf V} \quad {\rm tal~\ que}\quad \forall v\in {\bf V},\qquad  0+v=v,\\
&7)& \forall v\in {\bf V}, \quad \exists -v\in {\bf V},\quad {\rm tal ~\ que}\quad  v+(-v)=0,\\
&8)& \forall v\in {\bf V},\quad ev=v,
\end{eqnarray}
aqu\'{\i} $e$ representa el neutro multiplicativo de $\bf K.$\\

\section{Ejemplos}

Ahora, veremos algunos ejemplos de espacios vectoriales. El lector puede verificar que 
los siguientes espacios cumplen las reglas de espacios vectoriales.

\subsection{${\bf C}^{n}$ }

Supongamos que ${\bf C}$ es el conjunto de los n\'umeros complejos. 
Un ejemplo de espacio vectorial son los arreglos de
la forma
\begin{eqnarray}
{\bf C}^{n}=\{ (c_{1},c_{2}, \cdots, c_{n}), c_{i}\in {\bf C} \}.
\end{eqnarray}
Si se tienen dos vectores de ${\bf C}^{n},$
\begin{eqnarray}
(c_{1},c_{2},\cdots, c_{n}),\quad (d_{1},d_{2},\cdots, d_{n})
\end{eqnarray}
la suma se define como 
\begin{eqnarray}
(c_{1}+d_{1},c_{2}+d_{2},\cdots, c_{n}+d_{n}).
\end{eqnarray}
Mientras que si $\lambda$ es un n\'umero complejo, el producto por un escalar se define como
\begin{eqnarray}
\lambda(c_{1},c_{2},\cdots, c_{n})= (\lambda c_{1},\lambda c_{2},\cdots, \lambda c_{n}).
\end{eqnarray}

\subsection{Sucesiones}

Una generalizaci\'on de ${\bf C}^{n}$ es tomar el l\'imite $n\to \infty,$ que nos da  
el espacio de sucesiones 
\begin{eqnarray}
\{ a_{n}\}_{n= 0}^{\infty},\qquad  a_{n} \in {\bf C}.
\end{eqnarray}
Para este caso debemos pedir que 
\begin{eqnarray}
\sum_{n\geq 0} |a_{n}|^{2} <\infty.
\end{eqnarray}
As\'i, si se tienen dos sucesiones
\begin{eqnarray}
\{ a_{n}\}_{n= 0}^{\infty}, \quad \{ b_{n}\}_{n=0}^{\infty},  \qquad a_{n},b_{n} \in {\bf C}
\end{eqnarray}
la suma se define como 
\begin{eqnarray}
\{ a_{n}+b_{n}\}_{n=0}^{\infty}.
\end{eqnarray}
Mientras que si $\lambda$ es un n\'umero complejo, el producto por un escalar se define como
\begin{eqnarray}
\lambda \{ a_{n}\}_{n=0}^{\infty} = \{ \lambda a_{n}\}_{n=0}^{\infty}.
\end{eqnarray}
\subsection{ Matrices}

Otra generalizaci\'on de ${\bf C}^{n}$ es el espacio de matrices $M_{(nm)}$ de entradas complejas
\begin{eqnarray}
M=\left(
\begin{array}{rrrr}
M_{11}& M_{12}& \cdots & M_{1m}\\
M_{21}&  M_{22}& \cdots& M_{2m}\\
\vdots & \vdots &  \ddots & \vdots\\
M_{n1}& M_{n2} & \cdots& M_{nm}
\end{array}\right),\quad M_{ij}\in {\bf C},\quad i=1,\cdots n, j=1,\cdots, m.\nonumber
\end{eqnarray}
Si se tienen dos matrices

\begin{eqnarray}
M=\left(
\begin{array}{rrrr}
M_{11}& M_{12}& \cdots & M_{1m}\\
M_{21}&  M_{22}& \cdots& M_{2m}\\
\vdots & \vdots &  \ddots & \vdots\\
M_{n1}& M_{n2} & \cdots& M_{nm}
\end{array}\right),\quad 
N=\left(
\begin{array}{rrrr}
N_{11}& M_{12}& \cdots & N_{1m}\\
N_{21}&  M_{22}& \cdots& N_{2m}\\
\vdots & \vdots &  \ddots & \vdots\\
N_{n1}& N_{n2} & \cdots& N_{nm}
\end{array}\right),
\end{eqnarray}
La suma se define como
\begin{eqnarray}
M+N=\left(
\begin{array}{rrrr}
M_{11}+N_{11}& M_{12}+N_{12}& \cdots & M_{1m}+N_{1m}\\
M_{21}+N_{21}&  M_{22}+N_{22}& \cdots& M_{2m}+N_{2m}\\
\vdots & \vdots &  \ddots & \vdots\\
M_{n1}+N_{n1}& M_{n2}+N_{n2} & \cdots& M_{nm}+N_{nm}
\end{array}\right).
\end{eqnarray}
Mientras que el producto por un escalar, $\lambda \in {\bf C},$ se define como
\begin{eqnarray}
\lambda M=\left(
\begin{array}{rrrr}
\lambda M_{11}& \lambda M_{12}& \cdots & \lambda M_{1m}\\
\lambda M_{21}&  \lambda M_{22}& \cdots& \lambda M_{2m}\\
\vdots & \vdots &  \ddots & \vdots\\
\lambda M_{n1}& \lambda M_{n2} & \cdots& \lambda M_{nm}
\end{array}\right).\nonumber
\end{eqnarray}

\subsection{Funciones}

Otro ejemplo de espacio vectorial es el espacio de  funciones $f:[a,b]\to {\bf C}$.
Supongamos que tenemos dos funciones
\begin{eqnarray}
f:[a,b]\to {\bf C},\qquad g:[a,b]\to {\bf C},
\end{eqnarray}
la suma se define como
\begin{eqnarray}
f+g:[a,b]\to {\bf C},
\end{eqnarray}
con la regla de correspondencia
\begin{eqnarray}
\left(f+g\right)(x)=f(x)+g(x),\qquad x\in [a,b].
\end{eqnarray}
Mientras que el producto por un escalar, $\lambda \in {\bf C},$ se define como
\begin{eqnarray}
\lambda f:[a,b]\to {\bf C},
\end{eqnarray}
con la regla de correspondencia
\begin{eqnarray}
\left(\lambda f\right)(x)=\lambda f(x),\qquad x\in [a,b].
\end{eqnarray}

\section{Producto escalar}

Una operaci\'on importante entre vectores es el producto escalar. Este producto 
manda dos vectores a un n\'umero complejo
\begin{eqnarray}
\left<|\right>:{\bf V}\times {\bf V}\to {\bf C},
\end{eqnarray}
y debe satisfacer los axiomas:
\begin{eqnarray}
&\cdot) &\forall v\in {\bf V} \qquad \left<v|v\right>\geq 0,\qquad \left<v|v\right>=0 \Longleftrightarrow v=0, \label{eq:producto-escalar1}\\
& \cdot\cdot)& \forall v,u, w\in {\bf V} \qquad \left<v+u|w\right>=\left<v|w\right>+\left<u|w\right>, \label{eq:producto-escalar2}\\
& \cdots)& \forall v,u\in {\bf V}, \lambda \in C \qquad \left<v|\lambda u\right>=\lambda \left<v|u\right>, \label{eq:producto-escalar3}\\
& \cdots\cdot)& \forall v,u\in {\bf V}, \qquad \left<v| u\right>=\left(\left<u|v\right>\right)^{*} \label{eq:producto-escalar4} .
\end{eqnarray}
Existen diferentes implicaciones de estos axiomas. Por ejemplo, 
para cualquier vector $v$ se cumple $\left<v|0\right>=0.$ En efecto sabemos que $v-v=0,$ entonces 
\begin{eqnarray}
\left<v|0\right>=\left<v|v-v \right>=\left <v|v\right>-\left<v|v\right>=0.
\end{eqnarray}
Otra implicaci\'on es que si $\lambda$ es un n\'umero complejo y  $v_{1},v_{2}$ dos vectores,
entonces se cumple 
\begin{eqnarray}
\left <\lambda v_{1}|v_{2}\right>=\lambda^{*} \left < v_{1}|v_{2}\right>.
\end{eqnarray}
Esta igualdad es correcta pues considerando Eq. (\ref{eq:producto-escalar3}) y  Eq. (\ref{eq:producto-escalar4})
se encuentra
\begin{eqnarray}
\left <\lambda v_{1}|v_{2}\right> &=& \left( \left< v_{2}|\lambda v_{1}\right>\right)^{*}= \left(\lambda \left< v_{2}|v_{1}\right>\right)^{*}=
 \lambda^{*} \left( \left< v_{2}|v_{1}\right>\right)^{*} \nonumber\\
 &=&\lambda^{*} \left< v_{1}|v_{2}\right> .\nonumber
\end{eqnarray}
Adem\'as, si $v$ y $w$ son dos vectores, entonces
\begin{eqnarray}
& & \left<v+w|v+w\right> +\left<v-w|v-w\right> =\left<v+w|v+w\right>\nonumber\\
& &  +\left<v-w|v-w\right>\nonumber\\
&& =\left<v|v+w\right>+\left<w|v+w\right>+\left<v|v-w\right>-\left<w|v-w\right>,\nonumber\\
&& = \left<v|v\right>+\left<v|w\right>+\left<w|v\right>+\left<w|w\right>+ \left<v|v\right>-\left<v|w\right>\nonumber\\
& &-\left<w|v\right>+\left<w|w\right>\nonumber\\
&& =2\left(\left<v|v\right>+\left<w|w\right>\right),\nonumber
\end{eqnarray}
es decir 
\begin{eqnarray}
 \left<v+w|v+w\right> +\left<v-w|v-w\right> =2\left(\left<v|v\right>+\left<w|w\right>\right),
\end{eqnarray}
que es la llamada igualdad del paralelogramo.\\

Antes de ver otras propiedades del producto escalar veremos algunos ejemplos de ellos.

\section{Ejemplos de producto escalar}

\subsection{Producto escalar en  ${\bf C}^{n}$}

Si tenemos dos vectores en ${\bf C}^{n},$ $v=(v_{1},v_{2},\dots , v_{n})$ y $w=(w_{1},w_{2},\dots , w_{n}),$ el producto escalar
se define como
\begin{eqnarray}
\left<v|w\right>=\sum_{i=1}^{n}v_{i}^{*}w_{i}.
\end{eqnarray}
Note que a los vectores $v$ y $w$ se les puede asignar las matrices columna
\begin{eqnarray}
v=
\left( \begin{array}{r}
v_{1}\\
v_{2}\\
\vdots\\
v_{n} 
\end{array}\right),\qquad 
w=\left(\begin{array}{r}
w_{1}\\
w_{2}\\
\vdots\\
w_{n} 
\end{array}\right),
\end{eqnarray}
por lo que 
\begin{eqnarray}
\left<v|w\right>=v^{*T}w=\sum_{i=1}^{n}v_{i}^{*}w_{i}=
\left( \begin{array}{r}
v_{1}^{*}\quad
v_{2}^{*}\quad
\cdots\quad 
v_{n} ^{*}
\end{array}\right)
\left(\begin{array}{r}
w_{1}\\
w_{2}\\
\vdots\\
w_{n} 
\end{array}\right).
\end{eqnarray}

\subsection{Sucesiones}

Si tenemos dos sucesiones $s_{1}=\{ a_{n}\}_{n=0}^{\infty}$ y $ s_{2}=\{b_{n}\}_{n=0}^{\infty}$ donde $a_{n},b_{n}\in {\bf C}$ y $ \sum_{n\geq 0}|a_{n}|^{2}<\infty, \sum_{n\geq 0}|b_{n}|^{2}<\infty,$  se puede definir el producto escalar como
\begin{eqnarray}
\left<s_{1}|s_{2}\right>=\sum_{n\geq 0}a_{n}^{*}b_{n},
\end{eqnarray}
note que \'este es una generalizaci\'on del producto escalar entre vectores.\\

\subsection{Matrices}

En el espacio vectorial de las matrices de entradas complejas de $n\times n$ tambi\'en es posible definir un producto escalar.
Antes de definir este producto recordemos que si $M$ es una matriz de entradas $M_{ij},$ la traza se define como 
$Tr(M)=\sum_{i=1}^{n}M_{ii}.$ Tambi\'en recordemos que las entradas de la matriz transpuesta $M^{T}$ se definen como $(M^{T})_{ij}=M_{ji}.$ 
Adem\'as si $N$ es otra matriz de $n\times n$ las entradas de la matriz producto, $MN,$ son $(MN)_{ij}=\sum_{k=1}^{n}M_{ik}N_{kj}.$
De estas definiciones es claro que
\begin{eqnarray}
Tr\left(M^{T}\right)&=&Tr\left(M\right),\quad 
Tr(M+N)=Tr(M)+Tr(N),\nonumber\\ 
 Tr(NM)&=&Tr(MN),\quad (MN)^{T} =N^{T}M^{T},\nonumber
\end{eqnarray}
en efecto
\begin{eqnarray}
Tr\left(M^{T}\right)&=&\sum_{i=1}^{n}\left(M^{T}\right)_{ii}= \sum_{i=1}^{n}M_{ii}=Tr\left(M\right),\nonumber\\
Tr(M+N)&=&\sum_{i=1}^{n}\left(M+N\right)_{ii}=\sum_{i=1}^{n}\left(M_{ii}+N_{ii}\right)= 
\sum_{i=1}^{n}M_{ii}+\sum_{i=1}^{n}N_{ii}\nonumber\\
&=&Tr(M)+Tr(N),\nonumber\\
Tr\left(MN\right)&=&\sum_{a=1}^{n}(MN)_{aa}= \sum_{a=1}^{n}\sum_{b=1}^{n} M_{ab}N_{ba}= \sum_{a=1}^{n}\sum_{b=1}^{n} N_{ba}M_{ab}\nonumber\\
&=& \sum_{b=1}^{n}\left(\sum_{a=1}^{n} N_{ba}M_{ab}\right)=\sum_{b=1}^{n}\left( NM\right)_{bb}=Tr(NM), \nonumber\\
\left((MN)^{T}\right)_{ij}&=&\left(MN\right)_{ji}=\sum_{k=1}^{n} M_{jk}N_{ki}=\sum_{k=1}^{n} N_{ki} M_{jk}\nonumber\\
&=&\sum_{k=1}^{n} \left(N^{T}\right)_{ik} \left(M^{T}\right)_{kj}= \left(N^{T}M^{T}\right)_{ij}.\nonumber
\end{eqnarray}

Definiremos el producto escalar entre matrices como
\begin{eqnarray}
\left<M|N\right>=Tr\left( M^{*T} N\right).\label{eq:escalar-matrices}
\end{eqnarray}
El primer axioma se cumple, pues
\begin{eqnarray}
\left<M|M\right>&=&Tr\left( M^{*T} M\right)=\sum_{i=i}^{n}\left( M^{*T} M\right)_{ii}= \sum_{i=1}^{n}\sum_{k=1}^{n} \left(M^{*T}\right)_{ik} M_{ki}\nonumber\\
&=& \sum_{i=1}^{n}\sum_{k=1}^{n}  M^{*}_{ki} M_{ki}=\sum_{i=1}^{n}\sum_{k=1}^{n}  |M_{ik}|^{2}\geq 0.
\end{eqnarray}
De donde, si $\left<M|M\right>=0,$ entonces $M_{ik}=0,$ es decir $M=0.$\\

Adem\'as, si $N_{1}$ y $N_{2}$ son matrices de $n\times n,$
\begin{eqnarray}
\left<M|N_{1}+N_{2}\right>&=&Tr\left( M^{*T} \left(N_{1}+N_{2}\right)\right)=Tr\left( M^{*T}N_{1}+M^{*T}N_{2}\right)\nonumber\\
&=&Tr \left( M^{*T}N_{1} \right)+Tr\left(M^{*T}N_{2}\right)\nonumber\\
&= &\left<M|N_{1}\right>+\left<M|N_{2}\right>.\nonumber
\end{eqnarray}
Por lo que se cumple el segundo axioma de producto escalar.\\

Tambi\'en se puede observar que  si $\lambda$ es un n\'umero complejo, entonces 
\begin{eqnarray}
\left<M|N\right>&=&Tr\left( M^{*T} \lambda N\right)=\sum_{i=i}^{n}\left( M^{*T} \lambda N\right)_{ii}= \lambda \sum_{i=i}^{n}\left( M^{*T}N\right)_{ii}
\nonumber\\
&=& \lambda Tr\left( M^{*T} N\right)=\lambda \left<M| N\right>.
\end{eqnarray}
Adicionalmente, se encuentra
\begin{eqnarray}
\left<M|N\right>&=&Tr\left( M^{*T} N\right)=Tr\left(\left(M^{*T}N\right)^{T}\right)=Tr\left(N^{T}M^{*}\right)\nonumber\\
&=& \left(\left[Tr\left( N^{T}  M^{*}\right)\right]^{*}\right)^{*}=\left(Tr\left( N^{*T}  M\right)\right)^{*}= \left(\left<N|M\right>\right)^{*}
\nonumber.
\end{eqnarray}
Por lo tanto, (\ref{eq:escalar-matrices}) es un producto escalar para el espacio vectorial de las matrices de $n\times n.$\\


\subsection{Funciones}

Si $q(x)$ es una funci\'on  real, continua y positiva en el intervalo $(a,b),$ para el espacio vectorial de las funciones continuas, $\{f\},$
que va del intervalo $[a,b]$ a los complejos, tales que 
\begin{eqnarray}
\int_{a}^{b}dx q(x)f^{*}(x)f(x)  <\infty,
\end{eqnarray}
se puede definir el producto escalar como
\begin{eqnarray}
\left<f|g\right>=\int_{a}^{b}dx q(x)f^{*}(x)g(x).\label{eq:escalar-funcione}
\end{eqnarray}
Considerando las propiedades del $q(x)$ y $f(x),$ el primer axioma de producto escalar se cumple pues 
\begin{eqnarray}
\left<f|f\right>&=&\int_{a}^{b}dx q(x)f^{*}(x)f(x)= \int_{a}^{b}dx q(x)|f(x)|^{2}\geq 0,\nonumber\\
\left<f|f\right>&=&\int_{a}^{b}dx q(x)f^{*}(x)f(x)= \int_{a}^{b}dx q(x)|f(x)|^{2}= 0 \Longleftrightarrow f(x)=0.\nonumber
\end{eqnarray}
El segundo axioma de producto escalar se cumple, en efecto
\begin{eqnarray}
\left<f|g_{1}+g_{2}\right>&=&\int_{a}^{b}dx q(x)f^{*}(x)\left(g_{1}(x)+ g_{2}(x)\right)\nonumber\\
&=& \int_{a}^{b}dx q(x)\left(f(x)^{*}g_{1}(x)+ f(x)^{*}g_{2}(x)\right)\nonumber\\
&=&\int_{a}^{b}dx q(x)f(x)^{*}g_{1}(x)+\int_{a}^{b}dx q(x)f(x)^{*}g_{2}(x)\nonumber\\
&=&\left<f|g_{1}\right>+\left<f|g_{2}\right>.\nonumber
\end{eqnarray}
Los axiomas restantes tambi\'en se cumplen, para probar esta afirmaci\'on supongamos que $\lambda$ es un n\'umero complejo,
entonces
\begin{eqnarray}
\left<f|\lambda g\right>&=&\int_{a}^{b}dx q(x)f^{*}(x)\lambda g(x)= \lambda \int_{a}^{b}dx q(x)f^{*}(x)g(x)= \lambda \left<f| g\right>,\nonumber\\
\left<f|g\right>&=&\int_{a}^{b}dx q(x)f^{*}(x)g(x)=
\left(\int_{a}^{b}dx\left( q(x)f^{*}(x) g(x)\right)^{*}\right)^{*}\nonumber\\
&=&  \left(\int_{a}^{b}dx q(x)g^{*}(x) f(x) \right)^{*}=\left( \left<g|f\right> \right)^{*}.
\nonumber
\end{eqnarray}
Por lo tanto, Eq. (\ref{eq:escalar-funcione}) es un producto escalar para el espacio vectorial de las funciones.\\

\section{Ortonormalidad e independencia lineal}

Se dice que un conjunto de vectores $\{v_{i}\}_{i=1}^{n}=\{v_{1}, v_{2},\cdots, v_{n}\}$ es linealmente independiente
si cualquier combinaci\'on lineal de la forma
\begin{eqnarray}
0=\sum_{i=1}^{n}a_{i}v_{i},\label{eq:independencia}
\end{eqnarray}
implica $a_{i}=0.$ Adem\'as, ee dice que un conjunto de vectores, $\{v_{i}\}_{i=1}^{n},$ es ortonormal si 
\begin{eqnarray}
\left<v_{i}|v_{j}\right>=\delta_{ij},\qquad i,j=1,2,\dots, n.
\end{eqnarray}
Si un conjunto de vectores es ortonomal, entonces es linealmente independiente. En efecto, 
supongamos que $\{a_{1}, a_{2},\cdots, a_{n}\}$ son un conjunto de escalares
tales que se cumple Eq. (\ref{eq:independencia}), entonces como los vectores son
ortonormales,
\begin{eqnarray}
0&=&\left<v_{j}|0\right>=\left<v_{j}\Bigg|\sum_{i=1}^{n}a_{i}v_{i}\right>=\sum_{i=1}^{n}\left<v_{j}|a_{i}v_{i}\right>=\sum_{i=1}^{n}a_{i}\left<v_{j}|v_{i}\right>\nonumber\\
&=&
\sum_{i=1}^{n}a_{i}\delta_{ij}=a_{j}, \nonumber
\end{eqnarray}
es decir $a_{i}=0,$ lo que completa la prueba. \\

Un conjunto de vectores ornormal,  $\{v_{i}\}_{i=1}^{n},$ tiene varias propiedades interesantes. Por ejemplo,
si $v$ es una combinaci\'on lineal de estos vectores, es decir si  $v=\sum_{i=1}^{n}a_{i}v_{i},$ se cumple
\begin{eqnarray}
\left<v|v\right>=\sum_{i=1}^{n} |a_{i}|^{2}.
\end{eqnarray}
En efecto
\begin{eqnarray}
\left<v|v\right>&=&\left< \sum_{i=1}^{n}a_{i}v_{i}\Bigg |\sum_{j=1}^{n}a_{j}v_{j} \right>=\sum_{i=1}^{n} \sum_{j=1}^{n}\left<a_{i}v_{i}|a_{j}v_{j}\right>\nonumber\\
&=& \sum_{i=1}^{n} \sum_{j=1}^{n}a_{i}^{*}a_{j}\left<v_{i}|v_{j}\right>
= \sum_{i=1}^{n} \sum_{j=1}^{n}a_{i}^{*}a_{j}\delta_{ij}=
\sum_{i=1}^{n} |a_{i}|^{2}.
\end{eqnarray}

\section{Teorema de Pit\'agoras} 

Supongamos que $v$ es un vector y $\{v_{i}\}_{i=1}^{n}$ es un conjunto de vectores ortonormales, entonces
el vector 
$$w_{1}=\sum_{i=1}^{n}\left<v_{i}|v\right>v_{i}$$ 
es ortonormal a   
$$w_{2}=v-\sum_{j=1}^{n}\left<v_{j}|v\right>v_{j}.$$
Esta afirmaci\'on se satisface, pues
\begin{eqnarray}
\left<w_{1}|w_{2}\right>&=&\left<\left(\sum_{i=1}^{n}\left<v_{i}|v\right>v_{i}\right)|\left(v-\sum_{j=1}^{n}\left<v_{j}|v\right>v_{j}\right)\right>\nonumber\\
&=&\left<\left(\sum_{i=1}^{n}\left<v_{i}|v\right> v_{i}\right)|v\right>\nonumber\\
& &-\left<\left(\sum_{i=1}^{n}\left<v_{i}|v\right>v_{i}\right) |\left(\sum_{j=1}^{n}\left<v_{j}|v\right>v_{j}\right)\right>\nonumber\\
&=&\sum_{i=1}^{n} \left<v_{i}\left(\left<v_{i}|v\right>\right)|v\right>\nonumber\\
& &- \sum_{i=1}^{n}\sum_{j=1}^{n} \left<\left(\left<v_{i}|v\right>v_{i}\right)|\left(\left<v_{j}|v\right>v_{j}\right)\right>\nonumber\\
&=& \sum_{i=1}^{n}\left<v_{i}|v\right>^{*} \left<v_{i}|v\right>- \sum_{i=1}^{n}\sum_{j=1}^{n} \left<v_{i}|v\right>^{*} \left<v_{j}|v\right>\left<v_{i}|v_{j}\right>\nonumber\\
&=& \sum_{i=1}^{n}\left<v_{i}|v\right>^{*} \left<v_{i}|v\right>- \sum_{i=1}^{n}\sum_{j=1}^{n} \left<v_{i}|v\right>^{*} \left<v_{j}|v\right>\delta_{ij}\nonumber\\
&=&  \sum_{i=1}^{n}\left<v_{i}|v\right>^{*} \left<v_{i}|v\right>- \sum_{i=1}^{n} \left<v_{i}|v\right>^{*} \left<v_{i}|v\right>=0.\nonumber
\end{eqnarray}
Adem\'as, ocupando que $\{v_{i}\}_{i=1}^{n}$ es un conjunto de vectores ortonormales se cumple 
\begin{eqnarray}
\left<w_{1}|w_{1}\right>=\sum_{i=1}^{n}|\left<v_{i}|v\right>|^{2}.
\end{eqnarray}
Adicionalmente, se puede notar que 
\begin{eqnarray}
v= w_{1}+w_{2},
\end{eqnarray}
considerando que  $w_{1}$ y $w_{2}$ son vectores ortonormales, se puede probar que 
\begin{eqnarray}
\left<v|v\right>=\left<w_{1}|w_{1}\right>+\left<w_{2}|w_{2}\right>.
\end{eqnarray}
Tomando en cuenta los resultados anteriores, claramente se cumple que  si $\{v_{n}\}$ 
es un conjunto de vectores ortonormales, para cualquier vector $v$ se tiene que
\begin{eqnarray}
\left<v|v\right> =\sum_{i=1}^{n}|\left<v_{i}|v\right>|^{2} 
 +
\left<\left(v-\sum_{i=1}^{n}\left<v_{i}|v\right>v_{i}\right)\Bigg | \left(v-\sum_{i=1}^{n}\left<v_{i}|v\right>v_{i}\right) \right>.\quad 
\end{eqnarray}

\subsection{Desigualdad de Bessel}

Note que el teorema de Pit\'agoras implica la desigualdad 
\begin{eqnarray}
 \sum_{i=1}^{n}|\left<v_{i}|v\right>|^{2} \leq \left<v|v\right>,
\label{eq:desigualdad-bessel}
\end{eqnarray}
que es la llamada desigualdad de Bessel. 

\subsection{Desigualdad de Schwarz}

Sea $w$ un vector diferente de cero, claramente el conjunto formado por $\{\frac{w}{\sqrt{\left<w|w\right>}} \}$ es ortonormal. 
Entonces, de acuerdo a la desigualdad de Bessel  Eq. (\ref{eq:desigualdad-bessel}), para cualquier vector $v$ se cumple
\begin{eqnarray}
\left|\left<\frac{w}{\sqrt{\left<w|w\right>}  }\Bigg|v\right>\right|^{2} \leq \left<v|v\right>,\quad \Longrightarrow \quad 
\left|\left<w|v\right>\right|^{2} \leq  \left<w|w\right> \left<v|v\right>
\end{eqnarray}
que implica 
\begin{eqnarray}
\left|\left<w|v\right>\right| \leq \sqrt{\left<w|w\right>}\sqrt{ \left<v|v\right>}.
\label{eq:desigualdad-schwarz}
\end{eqnarray}
Esta es la llamada desigualdad de Schwarz.

\subsection{Desigualdad del tri\'angulo}

Sean $v$ y $w$ dos vectores, entonces 
\begin{eqnarray}
\left<v+w|v+w\right>&=&\left<v+w|v+w\right>=\left<v|v+w\right>+\left<w|v+w\right>\nonumber\\
& =& \left<v|v\right>+\left<v|w\right>+\left<w|v\right>+\left<w|w\right>\nonumber\\
&=&\left<v|v\right>+\left<w|w\right>+\left(\left<w|v\right>^{*}+\left<w|v\right>\right)\nonumber\\
&=& \left<v|v\right>+\left<w|w\right>   +2{\rm Re}\left(\left<w|v\right>\right)\nonumber\\
&\leq &  \left<v|v\right>+\left<w|w\right> +2|\left<w|v\right>|.
\end{eqnarray}
De donde, ocupando desigualdad de Schwarz Eq. (\ref{eq:desigualdad-schwarz})
se tiene
\begin{eqnarray}
 \left<v+w|v+w\right>  &\leq& \left<v|v\right>+\left<w|w\right> +2\sqrt{\left<w|w\right>}\sqrt{\left<v|v\right>}\nonumber\\
 &=&\left( \sqrt{\left<w|w\right>}+ \sqrt{\left<v|v\right>} \right)^{2},
\end{eqnarray}
es decir 
\begin{eqnarray}
\left<v+w|v+w\right>  &\leq&\left( \sqrt{\left<w|w\right>}+ \sqrt{\left<v|v\right>} \right)^{2}
\end{eqnarray}
esta es la llamada desigualdad del tri\'angulo.

\section{Espacios normados }

Sea  $V$ un espacio vectorial y $||\quad ||$ una funci\'on de $V$ en los reales. Se dice que $(V,|| \quad ||)$ es un espacio normado si
\begin{eqnarray}
&I)&||v|| \geq 0,\nonumber \\
&II)& ||v||=0, \qquad  \Longleftrightarrow \qquad  v=0, \nonumber\\
&III)& ||\alpha v||=|\alpha| ||v||,\nonumber \\
&IV)&|| v+w||\leq ||v||+||w||. \nonumber\\
\end{eqnarray}

Note que si $V$ es un espacio vectorial con producto escalar, entonces se puede definir un espacio normado con la norma dada por
\begin{eqnarray}
||v||=\sqrt{\left<v|v\right>}.
\end{eqnarray}
En efecto, las propiedades $I)$ y $II)$ se cumplen, pues por los axiomas de producto escalar se tiene que $\left<v|v\right>\geq 0$ y 
$\left<v|v\right>= 0 \quad \Longleftrightarrow \quad  v=0.$ La propiedad $III)$ se cumple pues, si $\alpha$ es un escalar, se tiene
\begin{eqnarray}
||\alpha v||&=&\sqrt{\left<\alpha v|\alpha v\right>}=\sqrt{\alpha^{*}\alpha \left<v|v\right>}=\sqrt{|\alpha|^{2} \left<v|v\right>}\nonumber\\
&=&|\alpha|\sqrt{ \left<v|v\right>}= |\alpha|\quad  ||v||.
\end{eqnarray}
La propiedad $IV)$ tambi\'en se cumple pues, ocupando la desigualdad del tri\'angulo se encuentra 
\begin{eqnarray}
|| v+w||^{2}=\left<v+w|v+w\right> \leq&\left( \sqrt{\left<w|w\right>}+ \sqrt{\left<v|v\right>} \right)^{2}=\left( ||w||+ ||v|| \right)^{2},\nonumber
\end{eqnarray}
es decir
\begin{eqnarray}
|| v+w|| \leq ||w||+ ||v|| \nonumber.
\end{eqnarray}
Por lo tanto, un espacio vectorial con producto escalar es un espacio normado con la norma dada por $||v||=\sqrt{\left<v|v\right>}.$\\

Note que ocupando la notaci\'on de espacios normados la igualdad del pa\-ralelogramos toma la forma 
\begin{eqnarray}
||v+w||^{2}+||v-w||^{2} =2\left(||v||^{2}+||w||^{2}\right).
\end{eqnarray}
Mientras que, si $\{v_{i}\}_{i=1}^{n}$ es un conjunto de vectores ortonormales,  el teorema de Pit\'agoras
se puede escribir como 
\begin{eqnarray}
||v||^{2}&=&\sum_{i=1}^{n}|\left<v_{i}|v\right>|^{2} +  \left|\left|v-\sum_{i=1}^{n}\left<v_{i}|v\right>v_{i}\right|\right|^{2}.
\end{eqnarray}
Adem\'as,  la desigualdad de Bessel toma la forma
\begin{eqnarray}
 \sum_{i=1}^{n}|\left<v_{i}|v\right>|^{2} &\leq& ||v||^{2}, \label{eq:desigualdad-bessel}
\end{eqnarray}
y la desigualdad de Schwarz es
\begin{eqnarray}
\left|\left<w|v\right>\right| &\leq & ||w||\quad ||v||. \label{eq:desigualdad-schwarz}
\end{eqnarray}

\subsection{Espacios m\'etricos}

Sea un conjunto $M$ y $d$ una funci\'on  de $M\times M$ en los reales. Se dice que $(M,d)$ es un espacio m\'etrico si satisface
que $\forall x, y \in M$ se cumple 
\begin{eqnarray}
&A)&d(x,y)\geq 0,\nonumber \\
&B)& d(x,y)=0, \qquad  \Longleftrightarrow \qquad  x=y, \nonumber\\
&C)&d(x,y)=d(y,x),\nonumber \\
&D)& d(x,z)\leq d(x,y)+d(y,z). \nonumber\\
\end{eqnarray}
A la funci\'on $d$ se le llama distancia.\\

Si  $V$ es un espacio normado, entonces se tiene un espacio m\'etrico con la distancia definida
por 
\begin{eqnarray}
d(v_{1},v_{2})=||v_{1}-v_{2}||.
\end{eqnarray}
Los dos primeros axiomas de distancia se cumplen, pues
\begin{eqnarray}
d(v_{1},v_{2})=||v_{1}-v_{2}||\geq 0 \quad {\rm y }\quad 
d(v_{1},v_{2})=||v_{1}-v_{2}||= 0 \quad \Longleftrightarrow \quad v_{1}-v_{2}=0,\nonumber
\end{eqnarray}
es decir 
$v_{1}=v_{2}.$ Tambi\'en se cumple,
\begin{eqnarray}
 d(v_{1},v_{2})&=&||v_{1}-v_{2}||= ||(-)(v_{2}-v_{1})||=|(-)|\quad ||v_{2}-v_{1}||=||v_{2}-v_{1}||\nonumber\\
 &=&d(v_{2},v_{1}),\nonumber
\end{eqnarray}
por lo tanto, se cumple el axioma $C)$.
Adem\'as, por la desigualdad del tri\'angulo, se tiene
\begin{eqnarray}
 d(v_{1},v_{3})&=&||v_{1}-v_{3}||= ||(v_{1}-v_{2})+(v_{2}-v_{3})||\leq ||v_{1}-v_{2}||+||v_{2}-v_{3}||\nonumber\\
 &=&d(v_{1},v_{2})+d(v_{2},v_{3}),\nonumber
\end{eqnarray}
de donde se cumple el axioma $D).$\\

As\'i, un espacio normado es m\'etrico. Lo que quiere decir que cualquier espacio con producto escalar es normado y por lo tanto m\'etrico.

\section{Ejemplos de bases ortonormales}

En esta secci\'on veremos diferentes conjuntos de funciones que forman una base ortonormal.

\subsection{Exponencial compleja}

Sea el conjunto de funciones 
\begin{eqnarray}
\Phi_{n}(\varphi)=\frac{e^{in\varphi}}{\sqrt{2\pi}}
\label{eq:exponenciales}
\end{eqnarray}
definidas en el intervalo $[0,2\pi],$ aqu\'i $n$ es un n\'umero entero. Este conjunto de funciones es
ortonormal. En efecto
\begin{eqnarray}
\left<\Phi_{n}(\varphi)|\Phi_{m}(\varphi)\right>=\int_{0}^{2\pi} d\varphi \left( \Phi_{n}(\varphi)\right)^{*}\Phi_{m}(\varphi)=\frac{1}{2\pi}
 \int_{0}^{2\pi}d\varphi  e^{i(m-n)\varphi}.
\end{eqnarray}
Si $m=n,$ es claro que 
\begin{eqnarray}
\frac{1}{2\pi}\int_{0}^{2\pi}d\varphi  e^{i(n-n)\varphi}=\frac{1}{2\pi} \int_{0}^{2\pi}d\varphi =1.
\end{eqnarray}
Adicionalmente, si  $n\not =m$ se tiene
\begin{eqnarray}
\frac{1}{2\pi}\int_{0}^{2\pi}d\varphi  e^{i(m-n)\varphi}&=&\frac{1}{2\pi}\frac{1}{i(n-m)}e^{i(m-n)\varphi}\bigg|_{0}^{2\pi}\nonumber\\
& =&\frac{1}{2\pi}\frac{1}{i(m-n)}\left( (-1)^{2(m-n)}-1\right)=0.\nonumber
\end{eqnarray}
Por lo tanto,
\begin{eqnarray}
\left<\Phi_{n}(\varphi)|\Phi_{m}(\varphi)\right>=\delta_{mn},
\label{eq:lineal-exponencial-ortonormal}
\end{eqnarray}
es decir, el conjunto Eq. (\ref{eq:exponenciales}) es ortonormal y entonces linealmente independiente.  

\subsection{Ecuaciones tipo Sturm-Liouville}

Anteriormente  vimos que si $\left(\psi_{\lambda_{1}}(x),\lambda_{1}\right), \left(\psi_{\lambda_{2}}(x),\lambda_{2}\right)$ son soluciones de la ecuaci\'on de Sturm-Liouville 
\begin{eqnarray}
\frac{d}{dx}\left(p(x)\frac{d\psi(x)}{dx}\right)+\left(\lambda q(x)+r(x)\right)\psi(x)=0
\label{eq:sturm-II}
\end{eqnarray}
que satisfacen la condiciones  de Dirichlet  
\begin{eqnarray}
\psi(a)=\psi(b)=0 \label{eq:vec-dirichlet}
\end{eqnarray}
\'o las de Neumann   
\begin{eqnarray}
\frac{d\psi(x)}{dx}\bigg|_{x=a}=\frac{d\psi(x)}{dx}\bigg|_{x=b}=0 \label{eq:vec-neumann-II}
\end{eqnarray}
\'o bien que $p(x)$  cumpla 
\begin{eqnarray}
p(a)=p(b)=0,\label{eq:vec-libre-II}
\end{eqnarray}
se encuentra que 
\begin{eqnarray}
\left(\lambda_{1}-\lambda_{2}\right)\int_{a}^{b}dx q(x) \psi_{\lambda_{2}}^{*}(x)\psi_{\lambda_{1}}(x)=0.\nonumber
\end{eqnarray}
En particular note que si $\lambda_{1}\not =\lambda_{2}$ se tiene 
\begin{eqnarray}
\int_{a}^{b}dx q(x) \psi_{\lambda_{2}}^{*}(x)\psi_{\lambda_{1}}(x)=0,
\end{eqnarray}
Adem\'as, se puede ver que la integral 
\begin{eqnarray}
\int_{a}^{b}dx q(x) \psi_{\lambda_{1}}^{*}(x)\psi_{\lambda_{1}}(x)=\alpha_{\lambda},
\end{eqnarray}
es positiva, es decir $\alpha_{\lambda}>0.$ Por lo que, si $\alpha_{\lambda}< \infty, $ el conjunto de funciones 
\begin{eqnarray}
\frac{\tilde \psi_{\lambda}(x)}{\sqrt{\alpha_{\lambda}}}
\end{eqnarray}
cumplen
\begin{eqnarray}
 \int_{a}^{b}dx q(x) \psi^{*}_{\lambda_{1}}(x)\tilde \psi_{\lambda_{2}}(x)=\delta_{\lambda_{1}\lambda_{2} }.
\end{eqnarray}
Se dice que las soluciones de Eq. (\ref{eq:sturm-II}) que satisfacen alguna de las condiciones Eqs. (\ref{eq:vec-dirichlet})-(\ref{eq:vec-libre-II})
son un conjunto de funciones ortonormales con funci\'on de peso $q(x)$. 

\subsection{Ecuaci\'on de Schr$\ddot {\rm o}$dinger en una dimension}

Supongamos que $V(x)$ es una funci\'on real, la ecuaci\'on de Schr$\ddot {\rm o}$dinger en el intervalo $[a,b]$ es 
\begin{eqnarray}
H\psi(x)=\left(-\frac{\hbar^{2}}{2m} \frac{\partial ^{2}}{\partial x^{2}} +V(x)\right)\psi(x) =E\psi(x),
\end{eqnarray}
donde $E$ es una constante real a determinar y se deben satisfacer las condiciones de borde $\psi(a)=\psi(b)=0.$ 
Claramente este es un problema tipo Sturm-Lioville con condiciones de Dirichlet.
Por lo que si $\psi_{E}(x)$ es soluci\'on con la constante $E$ y  $\psi_{E^{\prime}}(x)$ es soluci\'on con la contante $E^{\prime},$
entonces 
\begin{eqnarray}
\left<\psi_{E}(x)|\psi_{E^{\prime}}(x)\right>=\int_{a}^{b}dx \psi_{E}^{*}(x)\psi_{E^{\prime}}(x)=\delta_{EE^{\prime}}.
\end{eqnarray}
Por lo tanto, las soluciones de la ecuaci\'on de Schr$\ddot {\rm o}$dinger en una dimensión  forman un conjunto de funciones ortonormales. \\

\subsection{Ecuaci\'on de Schr$\ddot {\rm o}$dinger en tres dimensiones}

Para la ecuaci\'on de Schr$\ddot {\rm o}$dinger en  tres dimensiones se tiene el mismo resultado. Veamos este caso, si $U(x,y,z)$ es una funci\'on real, la ecuaci\'on de Schr$\ddot {\rm o}$dinger es 
\begin{eqnarray}
H\psi(x,y,z)=\left(-\frac{\hbar^{2}}{2m} \nabla^{2} +U(x,y,z)\right)\psi(x,y,z) =E\psi(x,y,z).
\end{eqnarray}
Supondremos que esta ecuaci\'on est\'a definida en una regi\'on de volumen $V$ cuya frontera es $\Sigma,$ por lo que la condici\'on de 
Dirichlet es  $\psi(x,y,z)|_{\Sigma}=0.$ \\  

Si $\psi_{E}(x,y,z)$ es soluci\'on con la constante $E$ y  $\psi_{E^{\prime}}(x,y,z)$ es soluci\'on con la contante $E^{\prime},$
entonces 
\begin{eqnarray}
\left(-\frac{\hbar^{2}}{2m} \nabla^{2} +U(x,y,z)\right)\psi_{E}(x,y,z) &=&E\psi_{E}(x,y,z),\label{eq:tres-Schrodinger1}\\
\left(-\frac{\hbar^{2}}{2m} \nabla^{2} +U(x,y,z)\right)\psi_{E^{\prime}}(x,y,z) &=&E^{\prime}\psi_{E^{\prime}}(x,y,z)
\label{eq:tres-Schrodinger2}.
\end{eqnarray}
El complejo conjugado de la seguda ecuaci\'on es
\begin{eqnarray}
\left(-\frac{\hbar^{2}}{2m} \nabla^{2} +U(x,y,z)\right)\psi_{E^{\prime}}^{*}(x,y,z) &=&E^{\prime}\psi_{E^{\prime}}^{*} (x,y,z),\label{eq:tres-Schrodinger22}
\end{eqnarray}
Por lo tanto, multiplicando $\psi_{E^{\prime}}^{*}$ por Eq. (\ref{eq:tres-Schrodinger1}) y $\psi_{E}$ por Eq. (\ref{eq:tres-Schrodinger22})
se encuentra
\begin{eqnarray}
-\frac{\hbar^{2}}{2m} \psi_{E^{\prime}}^{*} \nabla^{2} \psi_{E}+U(x,y,z)\psi_{E^{\prime}}^{*} \psi_{E} &=&E\psi_{E^{\prime}}^{*}\psi_{E},\nonumber\\
-\frac{\hbar^{2}}{2m} \psi_{E}\nabla^{2}\psi_{E^{\prime}}^{*} +U(x,y,z) \psi_{E}\psi_{E^{\prime}}^{*}  &=&
E^{\prime}\psi_{E}\psi_{E^{\prime}}^{*} \nonumber,
\end{eqnarray}
Ahora, considerando la igualdad 
\begin{eqnarray}
\vec \nabla\cdot \left(f\vec \nabla g\right)=\left( \vec \nabla f\right)\cdot\left( \vec \nabla g\right)+ f\nabla^{2}g,
\end{eqnarray}
se llega a 
\begin{eqnarray}
f\nabla^{2}g= \vec \nabla\cdot \left(f\vec \nabla g\right)-\left( \vec \nabla f\right)\cdot\left( \vec \nabla g\right),
\end{eqnarray}
por lo tanto 
\begin{eqnarray}
-\frac{\hbar^{2}}{2m}\left(\vec \nabla\cdot \left(  \psi_{E^{\prime}}^{*} \vec \nabla \psi_{E}\right) - 
\vec \nabla \psi_{E^{\prime}}^{*} \cdot \vec \nabla \psi_{E}\right)+U\psi_{E^{\prime}}^{*} \psi_{E} &=&E\psi_{E^{\prime}}^{*}\psi_{E},\nonumber\\
-\frac{\hbar^{2}}{2m} \left( \vec \nabla\cdot \left(\psi_{E} \vec \nabla \psi_{E^{\prime}}^{*}\right) - \vec \nabla \psi_{E} \cdot \vec \nabla \psi_{E^{\prime}}^{*} \right)+U\psi_{E}\psi_{E^{\prime}}^{*}  &=&
E^{\prime}\psi_{E}\psi_{E^{\prime}}^{*} \nonumber,
\end{eqnarray}
restando estas ecuaciones se encuentra
\begin{eqnarray}
-\frac{\hbar^{2}}{2m} \vec \nabla\cdot \left( \psi_{E^{\prime}}^{*} \vec \nabla \psi_{E} - \psi_{E} \vec\nabla \psi_{E^{\prime}}^{*} \right)=\left(E-E^{\prime}\right)\psi_{E^{\prime}}^{*}\psi_{E}.\nonumber
\end{eqnarray}
Integrando esta  \'ultima ecuaci\'on sobre el volumen $V$ y usando el teorema de Gauss Eq. (\ref{eq:tgauss}), se tiene 
\begin{eqnarray}
\left(E-E^{\prime}\right)\int_{V} dv \psi_{E^{\prime}}^{*}\psi_{E}=
\frac{\hbar^{2}}{2m} \oint_{\Sigma} da 
\left( \psi_{E^{\prime}}^{*} \vec \nabla \psi_{E} - \psi_{E} \vec \nabla \psi_{E^{\prime}}^{*} \right)\cdot\hat n=0,
\end{eqnarray}
es decir
\begin{eqnarray}
\left(E-E^{\prime}\right)\int_{V} dv \psi_{E^{\prime}}^{*}\psi_{E}=0.
\end{eqnarray}
Por lo tanto, si $E\not =E^{\prime},$ se llega a
\begin{eqnarray}
\int_{V} dv \psi_{E^{\prime}}^{*}(x,y,z)\psi_{E}(x,y,z)=0.
\end{eqnarray}
Si la funciones de onda son tales que $\int_{V} dv \psi_{E}^{*}\psi_{E}=\alpha<\infty,$ siempre se puede tener un conjunto de funciones tales que
\begin{eqnarray}
\int_{V} dv \psi_{E^{\prime}}^{*}(x,y,z)\psi_{E}(x,y,z)=\delta_{EE^{\prime}}.
\end{eqnarray}
Por lo tanto con las soluciones de la ecuaci\'on de Schr$\ddot {\rm o}$dinger se puede formar un conjunto ortonormal de funciones. Este resultado es fundamental para la mec\'anica cu\'antica.

\subsection{Arm\'onicos esf\'ericos}

Otro ejemplo est\'a en las funciones propias del operado $\hat L^{2},$ 
\begin{eqnarray}
\hat L^{2}Y_{\lambda}(\theta,\varphi)=\lambda Y_{\lambda}(\theta,\varphi), \nonumber
\end{eqnarray}
que deben satisfacer
\begin{eqnarray}
\hat L^{2}Y_{\lambda}(\theta,\varphi)=-\left[\frac{1}{\sin\theta} \frac{\partial }{\partial \theta}
\left(\sin\theta\frac{\partial Y_{\lambda}(\theta,\varphi) }{\partial \theta}\right)+
\frac{1}{\sin\theta^{2}}\frac{\partial^{2}Y_{\lambda}(\theta,\varphi) }{\partial \varphi^{2}}\right]=
\lambda Y_{\lambda}(\theta,\varphi). \nonumber
\end{eqnarray}
Por ahora no resolveremos esta ecuaci\'on, pero veremos algunas de sus propiedades respecto a la ortonormalidad. \\

Propondremos $Y_{\lambda m}(\theta,\varphi)=\Theta(\theta)\Phi(\varphi),$ de donde
\begin{eqnarray}
\hat L^{2}Y_{\lambda}(\theta,\varphi)=-\left[\frac{\Phi(\varphi)}{\sin\theta} \frac{\partial }{\partial \theta}
\left(\sin\theta\frac{\partial \Theta(\theta) }{\partial \theta}\right)+
\frac{\Theta(\theta)}{\sin\theta^{2}}\frac{\partial^{2}\Phi(\varphi) }{\partial \varphi^{2}}\right]=
\lambda \Theta(\theta) \Phi(\varphi), \nonumber
\end{eqnarray}
por lo que 
\begin{eqnarray}
\left(\frac{\sin^{2}\theta \hat L^{2}Y_{\lambda}(\theta,\varphi)}{Y_{\lambda m}(\theta,\varphi)}\right)
&=&
-\left[\frac{\sin\theta}{\Theta(\theta)} \frac{\partial }{\partial \theta}
\left(\sin\theta\frac{\partial \Theta(\theta) }{\partial \theta}\right)+
\frac{1}{\Phi(\varphi)}\frac{\partial^{2}\Phi(\varphi) }{\partial \varphi^{2}}\right]\nonumber\\
&=&
\lambda \sin^{2}\theta,  \quad \label{eq:theta}
\end{eqnarray}
que implica
\begin{eqnarray}
\frac{\partial }{\partial \varphi} \left(\frac{\sin^{2}\theta \hat L^{2}Y_{\lambda}(\theta,\varphi)}{Y_{\lambda m}(\theta,\varphi)}\right)
=-\frac{\partial }{\partial \varphi}\left(
\frac{1}{\Phi(\varphi)}\frac{\partial^{2}\Phi(\varphi) }{\partial \varphi^{2}}\right)=0,
\end{eqnarray}
entonces 
\begin{eqnarray}
\frac{1}{\Phi(\varphi)}\frac{\partial^{2}\Phi(\varphi) }{\partial \varphi^{2}}=-m^{2}={\rm constante},
\label{eq:varphi}
\end{eqnarray}
es decir 
\begin{eqnarray}
\frac{\partial^{2}\Phi(\varphi) }{\partial \varphi^{2}}=-m^{2}\Phi(\varphi).
\label{eq:lineal-azimutal-esfericos}
\end{eqnarray}
Sustituyendo Eq. (\ref{eq:varphi}) en Eq. (\ref{eq:theta}), se llega a 
\begin{eqnarray}
-\left[\frac{\sin\theta}{\Theta(\theta)} \frac{\partial }{\partial \theta}
\left(\sin\theta\frac{\partial \Theta(\theta) }{\partial \theta}\right)- m^{2}\right]=\lambda \sin^{2}\theta,
\end{eqnarray}
que se puede escribir como 
\begin{eqnarray}
\frac{\partial }{\partial \theta}\left(\sin\theta\frac{\partial \Theta(\theta) }{\partial \theta}\right)+ \left(\lambda \sin\theta -\frac{m^{2}}{\sin\theta} \right)\Theta(\theta)=0.
\label{eq:lineal-sturm-legendre}
\end{eqnarray}
Esta ecuaci\'on depende de los p\'arametros $\lambda$ y $m$ por lo que redefiniremos $\Theta(\theta)$ como $\Theta(\theta)=P_{\lambda}^{m}(\cos\theta).$
Note que Eq.  (\ref{eq:lineal-sturm-legendre}) es tipo Sturm-Liouville con 
\begin{eqnarray}
p(\theta)=\sin\theta,\quad q(\theta)=\sin\theta,\quad r(\theta) =-\frac{m^{2}}{\sin\theta}.
\end{eqnarray}
Adem\'as como $p(0))=\sin 0=p(\pi)=\sin\pi=0,$ las soluciones de  Eq. (\ref{eq:lineal-sturm-legendre}) son ortormales en el intervalo $[0,\pi]$ con funci\'on de peso $\sin\theta,$ es decir   
\begin{eqnarray}
\int_{0}^{\pi}d\theta \sin\theta P_{\lambda^{\prime} }^{m}(\cos \theta) P_{\lambda}^{m}(\cos \theta) =\alpha_{\lambda m}\delta_{\lambda^{\prime}\lambda},\qquad \alpha_{\lambda m}={\rm constante}> 0. 
\end{eqnarray}
Note que las funciones 
\begin{eqnarray}
\Phi_{m}(\varphi)=A_{0}e^{im\varphi} 
\end{eqnarray}
son soluciones de Eq. (\ref{eq:lineal-azimutal-esfericos}). En particular si todas las $m$ son enteros el conjunto de funciones
\begin{eqnarray}
\Phi_{m}(\varphi)=\frac{e^{im\varphi}}{\sqrt{2\pi}} 
\end{eqnarray}
son ortonormales en intervalo $[0,2\pi],$ como fue mostrado en Eq. (\ref{eq:lineal-exponencial-ortonormal}). 
As\'i, si el conjunto de las $m$ est\'a en los enteros,
las funciones 
\begin{eqnarray}
Y_{\lambda m}(\theta,\varphi)=\frac{1}{\sqrt{\alpha_{\lambda m}}\sqrt{2\pi}} e^{im\varphi}P_{\lambda}^{m}(\cos\theta)
\end{eqnarray}
son ortonormales. En efecto, considerando la ortonormalidad de las funciones $\frac{e^{im\varphi}}{\sqrt{2\pi}}$ y 
$P_{\lambda}^{m}(\cos\theta),$ se tiene
\begin{eqnarray}
& &\left<Y_{\lambda^{\prime} m^{\prime}}(\theta,\varphi)|Y_{\lambda m}(\theta,\varphi)\right>
=\int d\Omega Y_{\lambda^{\prime} m^{\prime}}^{*}(\theta,\varphi)Y_{\lambda m}(\theta,\varphi)\quad \nonumber\\
& =&\int_{0}^{2\pi}d\varphi  \int_{0}^{\pi}d\theta \sin\theta  
Y_{\lambda^{\prime} m^{\prime}}^{*}(\theta,\varphi)Y_{\lambda m}(\theta,\varphi)\nonumber\\
&=& 
\int_{0}^{2\pi}d\varphi  \int_{0}^{\pi}d\theta \sin\theta 
\left(\frac{1}{\sqrt{\alpha_{\lambda^{\prime} m^{\prime}}}\sqrt{2\pi}} e^{im^{\prime}\varphi}P_{\lambda^{\prime}}^{m^{\prime}}(\cos\theta)\right)^{*}
\frac{1}{\sqrt{\alpha_{\lambda m}}\sqrt{2\pi}} e^{im\varphi}P_{\lambda}^{m}(\cos\theta)\quad \nonumber\\
&=&
\int_{0}^{2\pi}d\varphi  \int_{0}^{\pi}d\theta \sin\theta 
\frac{1}{\sqrt{\alpha_{\lambda^{\prime} m^{\prime} }}\sqrt{2\pi}} e^{-im^{\prime} \varphi}P_{\lambda^{\prime}}^{m^{\prime}}(\cos\theta)
\frac{1}{\sqrt{\alpha_{\lambda m}}\sqrt{2\pi}} e^{im\varphi}P_{\lambda}^{m}(\cos\theta)\nonumber\\
&=  &\frac{1}{\sqrt{\alpha_{\lambda^{\prime} m^{\prime} }} \sqrt{\alpha_{\lambda m}} }\left(\frac{1}{2\pi} \int_{0}^{2\pi}d\varphi e^{-im^{\prime} \varphi}e^{im\varphi}\right) \int_{0}^{\pi}d\theta \sin\theta P_{\lambda^{\prime}}^{m^{\prime}}(\cos\theta)P_{\lambda}^{m}(\cos\theta)\nonumber\\
&=& \frac{1}{\sqrt{\alpha_{\lambda^{\prime} m^{\prime} }} \sqrt{\alpha_{\lambda m}} }\delta_{mm^{\prime}} 
\int_{0}^{\pi}d\theta \sin\theta P_{\lambda^{\prime}}^{m^{\prime}}(\cos\theta)P_{\lambda}^{m}(\cos\theta)\nonumber\\
&=& \frac{1}{\sqrt{\alpha_{\lambda^{\prime} m }} \sqrt{\alpha_{\lambda m}} }\delta_{mm^{\prime}} 
\int_{0}^{\pi}d\theta \sin\theta P_{\lambda^{\prime}}^{m}(\cos\theta)P_{\lambda}^{m}(\cos\theta)\nonumber\\
&=&\frac{1}{\sqrt{\alpha_{\lambda^{\prime} m } \alpha_{\lambda m}}}\delta_{mm^{\prime}} \alpha_{\lambda m}\delta_{\lambda^{\prime} \lambda}
\nonumber\\
&=&\frac{1}{\sqrt{\alpha_{\lambda m}} \sqrt{\alpha_{\lambda m }}} \alpha_{\lambda m}\delta_{\lambda^{\prime} \lambda}
= \delta_{mm^{\prime}}\delta_{\lambda^{\prime} \lambda},\nonumber
\end{eqnarray}
es decir 
\begin{eqnarray}
\left<Y_{\lambda^{\prime} m^{\prime}}(\theta,\varphi)|Y_{\lambda m}(\theta,\varphi)\right>=
 \delta_{mm^{\prime}}\delta_{\lambda^{\prime} \lambda}.
\end{eqnarray}
A\'un no sabemos cual es la  forma expl\'icita las funciones $Y_{\lambda m}(\theta,\varphi),$ 
pero podemos decir que son un conjunto de funciones ortornales. Posteriormente ocuparemos este hecho
para encontrar  la forma expl\'icita de estas funciones. Cabe señalar que las funciones   $Y_{\lambda m}(\theta,\varphi)$ son importantes para diferentes \'areas de la f\'isica, como la mec\'anica cu\'antica y la electrodin\'amica.

\section{Polinomios trigonom\'etricos}

Supongamos que $\{\psi_{n}(x)\}_{n=0}^{\infty}$ es un conjunto de funciones ortonormales
en el intervalo $[a,b],$ es decir 
\begin{eqnarray}
\left<\psi_{n}|\psi_{m}\right>=\int_{a}^{b} dx q(x) \psi^{*}_{n}(x)\psi_{m}(x)=\delta_{nm},
\end{eqnarray}
aqu\'i $q(x)$ es una funci\'on de peso positiva en el intervalo $(a,b).$
Con este conjunto de funciones podemos hacer las combinaciones lineales
\begin{eqnarray}
T_{n}(x)=\sum_{i=1}^{n}b_{i}\psi_{i}(x),
\end{eqnarray}
las cuales llamaremos polinomios trigonom\'etricos.
Note que debido a que $\{\psi_{n}(x)\}_{n=0}^{\infty}$ es un conjunto de funciones ortonormales,  la norma de $T_{n}(x)$ es
\begin{eqnarray}
||T_{n}||^{2}=\sum_{i=1}^{n}|b_{i}|^{2}.
\end{eqnarray}
Sea $F(x)$ una funci\'on tal que $\left<F|F\right>=||F||^{2}=\int_{a}^{b}dx q(x)|F(x)|^{2}<\infty,$ entonces definiremos 
los coeficientes de Fourier de $F$ como
\begin{eqnarray}
a_{n}=\left<\psi_{n}|F\right>=\int_{a}^{b}dx q(x)\psi_{n}^{*}(x)F(x).
\end{eqnarray}
Ahora veremos que tanto se puede aproximar la funci\'on $F(x)$ con polinomios
de la forma $T_{n}(x).$ El sentido de la distancia en este espacio est\'a
dada por la norma de la funciones. As\'{\i}, el problema es encontrar los polinomios tales que 
\begin{eqnarray}
d^{2}(F,T_{n})=||F-T_{n}||^{2}=\int_{a}^{b}dxq(x)|F(x)-T_{n}(x)|^{2}\label{eq:distancia-funciones}
\end{eqnarray}
es m\'{\i}nimo. Basicamente se trata de encontrar los coeficientes $b_{i}$ que hacen m\'inimo 
(\ref{eq:distancia-funciones}). Podemos iniciar notando que
\begin{eqnarray}
d^{2}(F,T_{n})&=&||F-T_{n}||^{2}=\left<F-T_{n}|F-T_{n}\right>\nonumber\\
& =&\left<F|F\right>-\left<F|T_{n}\right>-\left<T_{n}|F\right>+\left<T_{n}|T_{n}\right>\nonumber\\
& =&||F||^{2}+||T_{n}||^{2}-\left<F|T_{n}\right>-\left<T_{n}|F\right>\nonumber\\
& =&||F||^{2}+||T_{n}||^{2}-\left<F|\sum_{i=1}^{n}b_{i}\psi_{i}\right>-
\left<\sum_{i=1}^{n}b_{i}\psi_{i}|F\right>\nonumber\\
& =&||F||^{2}+||T_{n}||^{2}-\sum_{i=1}^{n}b_{i}^{*}\left<F|\psi_{i}\right>-\sum_{i=1}^{n}b_{i}
\left<\psi_{i}|F\right>,\nonumber
\end{eqnarray}
considerando la norma de $T_{n}(x)$ y la definici\'on de los coeficientes de Fourier se llega a
\begin{eqnarray}
||F-T_{n}||^{2}
&=&||F||^{2}+\sum_{i=1}^{n}
\left(|b_{i}|^{2}- b_{i}a_{i}^{*}-b^{*}_{i}a_{i}\right).
\end{eqnarray}
Adem\'as, como
\begin{eqnarray}
|b_{i}-a_{i}|^{2}= (b_{i}-a_{i})(b_{i}-a_{i})^{*}=|b_{i}|^{2}+|a_{i}|^{2}-(b_{i}a_{i}^{*}+b^{*}_{i}a_{i}),
\end{eqnarray}
se tiene
\begin{eqnarray}
|b_{i}-a_{i}|^{2}-|a_{i}|^{2} = |b_{i}|^{2}-(b_{i}a_{i}^{*}+b^{*}_{i}a_{i}).
\end{eqnarray}
De donde, 
\begin{eqnarray}
||F-T_{n}||^{2}&=&||F||^{2}+ \sum_{i=1}^{n}|b_{i}-a_{i}|^{2}-
\sum_{i=1}^{n}|a_{i}|^{2}.
\end{eqnarray}
Claramente la distancia entre esta dos funciones es m\'{\i}nima
cuando $b_{i}=a_{i},$ es decir, cuando el polinomio tienen los coeficientes de Fourier.\\

Si $b_{i}=a_{i}$ se encuentra
\begin{eqnarray}
||F-T_{n}||^{2}=||F||^{2}-\sum_{i=1}^{n}|a_{i}|^{2} \geq 0.
\end{eqnarray}
Por lo que, para cualquier  $n$
\begin{eqnarray}
||T_{n}||^{2}=\sum_{i=1}^{n}|a_{i}|^{2} \leq ||F||^{2}<\infty.
\end{eqnarray}
Esta desigualdad se llama la desigualdad de Bessel, la cual implica que la sucesi\'on 
$||T_{n}||^{2}=\sum_{i=1}^{n}|a_{i}|^{2}$ est\'a acotada. \\

Un resultado de c\'alculo diferencial nos dice que si una sucesi\'on es monotona creciente y est\'a 
acotada, converge \cite{courant-calculo:gnus}. Note que la
sucesi\'on $||T_{n}||^{2}$ es monotona creciente y est\'a acotada, entonces
converge. La pregunta es hacia donde converge, supondremos sin demostrar que converge a
$||F||^{2},$ es decir 
\begin{eqnarray}
\lim_{n\to \infty}||T_{n}||^{2}=\lim_{n\to \infty}\sum_{i=1}^{n}|a_{i}|^{2}=
\sum_{n\geq 0}|a_{n}|^{2}=||F||^{2}. \label{eq:persival}
\end{eqnarray}
A esta igualdad se llama igualdad de Parseval. Demostrar esta igualdad es un problema importante \cite{reed:gnus,kolmogorov:gnus}, 
pero altamente no trivial y rebasa el prop\'osito de este escrito por lo que solo  tocaremos
este tema en casos particulares. \\

\section{Espacios completos}

Cuando se cumple la igualdad de Parseval se dice que el conjunto de funciones $\{\psi_{n}(x)\}_{n\geq 0}$
es completo. En este caso cualquier funci\'on, $F(x),$ con $||F||<\infty$ se puede escribir como combinaci\'on
lineal de $\{\psi_{n}(x)\}_{n\geq 0},$ es decir
\begin{eqnarray}
F(x)=\sum_{n\geq 0} a_{n} \psi_{n}(x).
\end{eqnarray}
Un resultado de c\'alculo diferencial es que si $\sum_{n\geq 0}|a_{n}|^{2}$ converge, entonces  $a_{n}\to 0.$\\

Esto tiene diferentes implicaciones f\'isicas. Por ejemplo, en mec\'anica cu\'antica significa que es m\'as probable que el sistema est\'e en estado base. Mientras que en electrost\'atica, significa que en un sistema de cargas los t\'erminos m\'as importantes son el monopolo, dipolo, cu\'adrupolo. Posteriormente veremos ejemplos concretos de esta afirmaci\'on.

\section{Operadores lineales}

Sea $V$ un espacio vectorial, una funci\'on 
$O:V\to V$ es un operador lineal,  o transformaci\'on lineal, si 
\begin{eqnarray}
\forall v_{1},v_{2}\in V, \forall \alpha,\beta \in K \qquad 
O\left(\alpha v_{1}+\beta v_{2}\right)= \alpha O\left(v_{1} \right)+\beta O\left( v_{2}\right).
\end{eqnarray}
Por ejemplo, el operador derivada es lineal, pues
\begin{eqnarray}
\frac{\partial }{\partial x}\left(\alpha f_{1}(x)+\beta f_{2}(x)\right)= 
\alpha \frac{\partial }{\partial x}f_{1}(x)+\beta \frac{\partial }{\partial x}f_{2}(x).
\end{eqnarray}
Usando el producto por un escalar, con una funci\'on $f(x)$ podemos definir un operador lineal, $O,$ de la forma:
\begin{eqnarray}
O(v_{1})=f(x)v_{1}, \nonumber
\end{eqnarray}
claramente este operador es lineal, pues 
\begin{eqnarray}
O(\alpha v_{1}+\beta v_{2})=f(x)(\alpha v_{1}+\beta v_{2})=\alpha f(x)v_{1}+f(x) \beta v_{2})= 
\alpha O\left(v_{1} \right)+\beta O\left( v_{2}\right). \nonumber
\end{eqnarray}
Dada una funci\'on $g(k,x)$ se puede definir una transformaci\'on  con la
integral 
\begin{eqnarray}
\tilde f(k)=\int_{a}^{b} dx g(k,x) f(x). \label{eq:trans-int}
\end{eqnarray}
Esta transformaci\'on es lineal, pues si definimos $h(x)= \alpha f_{1}(x)+\beta f_{2}(x)$ 
se encuentra
\begin{eqnarray}
\tilde h(k)&=&\int_{a}^{b} dx g(k,x) h(x)=
\int_{a}^{b} dx g(k,x) \left(\alpha f_{1}(x)+\beta f_{2}(x)\right)\nonumber\\
&=&\alpha \int_{a}^{b} dx g(k,x) f_{1}(x)
+\beta \int_{a}^{b} dx g(k,x) f_{2}(x)= \alpha \tilde f_{1}(k)+\beta \tilde f_{2}(k).
\nonumber
\end{eqnarray}
A Eq. (\ref{eq:trans-int}) se le llama transformada integral en base $g(k,x)$.\\

Por ejemplo, con $g(k,x)=e^{-ikx}$ se define la transformada de Fourier
\begin{eqnarray}
\tilde f(k)=\int_{-\infty}^{\infty} dx  e^{-ikx}f(x). \label{eq:trans-int}
\end{eqnarray}
Para cada funci\'on, $g(k,x),$ bien portada se puede definir una transformada integral.\\

Si tenemos dos transformaciones lineales, $O_{1}$ y $O_{2},$ cualquier  combinaci\'on lineal 
de ellas tambi\'en es lineal. En efecto, si $a$ y $b$ son dos escalares podemos construir 
la combinaci\'on lineal 
\begin{eqnarray}
O=aO_{1}+bO_{2},
\end{eqnarray}
entonces 
\begin{eqnarray}
O \left(\alpha v_{1}+\beta v_{2}\right)& =&\left(aO_{1}+bO_{2}\right)\left(\alpha v_{1}+\beta v_{2}\right)\nonumber\\
&=& aO_{1}\left(\alpha v_{1}+\beta v_{2}\right) +bO_{2}\left(\alpha v_{1}+\beta v_{2}\right)\nonumber\\
&=&a\left( \alpha O_{1}\left( v_{1}\right)+ \beta O_{1}\left( v_{2}\right)\right)+
b\left( \alpha O_{2}\left( v_{1}\right)+ \beta O_{2}\left( v_{2}\right)\right)\nonumber\\
&=&\alpha\left(aO_{1}\left(v_{1}\right) +bO_{2}\left(v_{1}\right) \right)+ \beta\left(aO_{1}\left(v_{2}\right) +bO_{2}\left(v_{2}\right) \right)\nonumber\\
&=& \alpha\left(aO_{1} +bO_{2} \right)\left(v_{1}\right)+
 \beta\left(aO_{1} +bO_{2}\right)\left(v_{2}\right)\nonumber\\
&=& \alpha O\left(v_{1}\right)+\beta O \left(v_{2}\right).\nonumber
\end{eqnarray}
As\'i, cualquier combinaci\'on lineal de dos operadores lineales nos da otro operador lineal.\\

Ahora veamos el productor de dos operadores lineales, definamos 
\begin{eqnarray}
O=O_{1}O_{2},
\end{eqnarray}
entonces 
\begin{eqnarray}
O \left(\alpha v_{1}+\beta v_{2}\right)& =&\left(O_{1}O_{2}\right)\left(\alpha v_{1}+\beta v_{2}\right)\nonumber\\
&=& O_{1}\left(O_{2}\left(\alpha v_{1}+\beta v_{2}\right)\right)= O_{1}
\left( \alpha O_{2}\left( v_{1}\right)+ \beta O_{2}\left( v_{2}\right)\right)\nonumber\\
&=& \alpha O_{1}\left(O_{2}\left( v_{1}\right)\right) + \beta O_{1}\left(O_{2}\left( v_{2}\right)\right)\nonumber\\
&=&\alpha \left( O_{1} O_{2}\right) \left( v_{1}\right) + \beta\left( O_{1}O_{2}\right)\left( v_{2}\right)\nonumber\\
&=&\alpha O \left( v_{1}\right) + \beta O\left( v_{2}\right).\nonumber
\end{eqnarray}
Por lo tanto, el producto de dos operadores lineales tambi\'en nos da otro operador lineal.
Por ejemplo, sabemos que los operadores 
\begin{eqnarray}
\frac{\partial }{\partial x},\qquad  \frac{\partial }{\partial y},\qquad  \frac{\partial }{\partial z}
\end{eqnarray}
son lineales, entonces tambi\'en lo son 
\begin{eqnarray}
\frac{\partial^{2} }{\partial x^{2}},\qquad  \frac{\partial^{2} }{\partial y^{2}},\qquad  \frac{\partial^{2} }{\partial z^{2}}.
\end{eqnarray}
Esto implica que el operador Laplaciano
\begin{eqnarray}
\nabla^{2}=\frac{\partial^{2} }{\partial x^{2}}+\frac{\partial^{2} }{\partial y^{2}}+ \frac{\partial^{2} }{\partial z^{2}}
\end{eqnarray}
sea lineal. Cualquier funci\'on $V(x,y,z)$ como operador es lineal, entonces el operador Hamiltoniano 
\begin{eqnarray}
H=-\frac{\hbar^{2}}{2m} \nabla^{2}+V(x,y,z)
\end{eqnarray}
es lineal, esto se debe a que es combinaci\'on lineal de opeadores lineales. Adem\'as las variables 
\begin{eqnarray}
x ,\qquad y ,\qquad  z
\end{eqnarray}
como operadores son lineales. Entonces los operadores 
\begin{eqnarray}
L_{x}=-i\left( y \frac{\partial }{\partial z}-z \frac{\partial }{\partial y}\right),\quad
L_{y}=-i \left( z \frac{\partial }{\partial x}-x \frac{\partial }{\partial z}\right),\quad
L_{z}=-i\left( x \frac{\partial }{\partial y}-y \frac{\partial }{\partial x}\right),\nonumber
\end{eqnarray}
son lineales, puesto que son combinaciones lineales de productos de operadores lineales. Por la misma raz\'on
el operador 
\begin{eqnarray}
L^{2}=L_{x}^{2}+L_{y}^{2}+L_{z}^{2}\nonumber
\end{eqnarray}
es lineal.

\section{Operador adjunto}

Dado un operador $A$ definiremos el operador adjunto, $A^{\dagger},$ como el operador que satisface
\begin{eqnarray}
\left <Av|u\right>=\left<v|A^{\dagger}u\right>,
\end{eqnarray}
con $u$ y $v$ dos vectores arbitrarios.\\

\subsection{Matrices}
Por ejemplo, para ${\bf C}^{n}$ se tiene 
\begin{eqnarray}
\left<Av|u\right>=(Av)^{*T}u=v^{*T}A^{*T}u=\left<v|A^{\dagger}u\right>=v^{*T}A^{\dagger}u,
\end{eqnarray}
de donde, para una matriz cuadrada con entradas complejas la matriz adjunta es
\begin{eqnarray}
A^{\dagger}=A^{*T}.\label{eq:adj-mat}
\end{eqnarray}

\subsection{Derivada}

Para el espacio vectorial de las funciones, el adjunto de un operador depende 
fuertemente del dominio, las condiciones de borde que se satisfacen 
y del producto escalar. En el espacio de las funciones suaves e integrables 
${\psi(x)}$ definidas en el intervalo $\left[a,b\right]$ y que cumplen $\psi(a)=\psi(b)=0.$
se puede definir el operador 
\begin{eqnarray}
A=\alpha\frac{\partial }{\partial x}, 
\end{eqnarray}
con $\alpha$ un n\'umero complejo. Veamos cual es el operador adjunto de este ope\-rador, 
para ello tomaremos el producto escalar  Eq. (\ref{eq:escalar-funcione}) con funci\'on de peso $q(x)=1.$
En este caso se puede notar que
\begin{eqnarray}
\left<A\psi_{1}|\psi_{2}\right>&=&\int_{a}^{b}dx (A\psi_{1}(x))^{*}\psi_{2}(x)=\int_{a}^{b}dx 
\left(\alpha\frac{\partial }{\partial x} \psi_{1}(x)\right)^{*}\psi_{2}(x)\nonumber\\ 
& =& \alpha^{*}\int_{a}^{b}dx \frac{\partial\psi_{1}^{*}(x) }{\partial x} \psi_{2}(x)\nonumber\\
&=&\alpha^{*}\int_{a}^{b}dx \left(
\frac{\partial\left(\psi_{1}^{*}(x) \psi_{2}(x)\right)}{\partial x}- \psi_{1}^{*}(x) 
\frac{\partial\psi_{2}(x)}{\partial x}\right)\nonumber \\
&=&\alpha^{*} \left(\psi_{1}^{*}(x) \psi_{2}(x)\right)\Bigg|_{a}^{b}+\int_{a}^{b}dx \psi_{1}^{*}(x)\left(-\alpha^{*} 
\frac{\partial\psi_{2}(x)}{\partial x}\right)\nonumber\\
&=&\int_{a}^{b}dx  \psi_{1}^{*}(x) 
\left(-\alpha^{*}\frac{\partial}{\partial x}\right)\psi_{2}(x),\nonumber
\end{eqnarray}
ahora
\begin{eqnarray}
\left<\psi_{1}|A^{\dagger}\psi_{2}\right>=\int_{a}^{b}dx \psi^{*}_{1}(x) A^{\dagger}\psi_{2}(x),\nonumber
\end{eqnarray}
entonces
\begin{eqnarray}
\left(\alpha\frac{\partial }{\partial x}\right)^{\dagger}= -\alpha^{*}\frac{\partial}{\partial x}.
\label{eq:adj-der}
\end{eqnarray}
Note que este resultado  depende fuertemente de que se cumpla $\psi(a)=\psi(b)=0.$

\subsection{Derivada con peso}

Se puede observar que el adjunto del operador $A=\alpha\frac{\partial }{\partial x}$ no est\'a bien definido con un producto escalar general
\begin{eqnarray}
\left<f|g\right>=\int_{a}^{b}dx q(x)f^{*}(x)g(x), \qquad q(x) >0.
\end{eqnarray}
Para este caso es m\'as conveniente ocupar el operador
\begin{eqnarray}
\tilde A=\frac{\alpha}{q(x)}\frac{\partial }{\partial x}, 
\end{eqnarray}
quien s\'i tiene bien definido su adjunto. En efecto,
\begin{eqnarray}
\left<\tilde A\psi_{1}|\psi_{2}\right>&=&\int_{a}^{b}dx q(x)(\tilde A\psi_{1}(x))^{*}\psi_{2}(x)\nonumber\\
&=&\int_{a}^{b}dx q(x)
\left(\frac{\alpha}{q(x)}\frac{\partial }{\partial x} \psi_{1}(x)\right)^{*}\psi_{2}(x)\nonumber\\ 
& =& \alpha^{*}\int_{a}^{b}dx \frac{\partial\psi_{1}^{*}(x) }{\partial x} \psi_{2}(x)
=\int_{a}^{b}dx  \psi_{1}^{*}(x) 
\left(-\alpha^{*}\frac{\partial}{\partial x}\right)\psi_{2}(x),\nonumber\\
&=&\int_{a}^{b}dx q(x) \psi_{1}^{*}(x) 
\left(\frac{-\alpha^{*}}{q(x)}\frac{\partial}{\partial x}\right)\psi_{2}(x)
\end{eqnarray}
ahora
\begin{eqnarray}
\left<\psi_{1}|\tilde A^{\dagger}\psi_{2}\right>=\int_{a}^{b}dx q(x)\psi^{*}_{1}(x) \tilde A^{\dagger}\psi_{2}(x),\nonumber
\end{eqnarray}
de donde
\begin{eqnarray}
\tilde A^{\dagger}=-\frac{\alpha^{*}}{q(x)}\frac{\partial }{\partial x}.
\end{eqnarray}

\subsection{Propiedades del operador adjunto}

Hay dos propiedades importantes de los operadores adjuntos. La primera propiedad est\'a relacionada con las suma.
Si $A,B$ son dos operadores lineales, se tiene 
\begin{eqnarray}
\left<\left(A+B\right)v|u\right>=\left<v|\left(A+B\right)^{\dagger}u\right>,
\end{eqnarray}
pero 
\begin{eqnarray}
\left<\left(A+B\right)v|u\right>&=&\left<\left(Av+Bv\right)|u\right>=\left<Av|u\right>+\left<Bv|u\right>\nonumber \\
&=&\left<v|A^{\dagger}u\right>+\left<v|B^{\dagger}u\right>= 
\left<v|\left(A^{\dagger}+B^{\dagger}\right)u\right>.\nonumber
\end{eqnarray}
Por lo tanto, 
\begin{eqnarray}
\left<v|\left(A+B\right)^{\dagger}|u\right>=\left<v|\left(A^{\dagger}+B^{\dagger}\right)u\right>,\nonumber
\end{eqnarray}
este resultado es v\'alido para cualquier par de vectores $v$ y $u,$ entonces
\begin{eqnarray}
\left(A+B\right)^{\dagger}= A^{\dagger}+B^{\dagger}.\label{eq:sum-hermit}
\end{eqnarray}
La otra propiedad est\'a relacionada con el producto de dos operadores:
\begin{eqnarray}
\left<\left(AB\right)v|u\right>&=&\left<A\left(Bv\right)|u\right>=\left<Bv|A^{\dagger}u\right>=
\left<v|B^{\dagger}A^{\dagger}u\right>= 
\left<v|\left(AB\right)^{\dagger}u\right>,\nonumber
\end{eqnarray}
que implica 
\begin{eqnarray}
\left(AB\right)^{\dagger}= B^{\dagger}A^{\dagger}.\label{eq:prod-hermit}
\end{eqnarray}

\section{Operadores Herm\'{\i}ticos}

Una clase importante de operadores son los autoadjuntos, que satisfacen
\begin{eqnarray}
A^{\dagger}= A,
\end{eqnarray}
a estos operadores tambi\'en se les llama Herm\'{\i}ticos. \\

De (\ref{eq:sum-hermit}) se puede ver que la suma de dos operadores Herm\'{\i}ticos 
nos da otro operador Herm\'{\i}tico. Adem\'as de  Eq. (\ref{eq:prod-hermit})
es claro que si $A$ y $B$  son operadores Herm\'{\i}ticos y {\bf conmutan}, es decir
$AB=BA,$ entonces
\begin{eqnarray}
\left(AB\right)^{\dagger}= B^{\dagger}A^{\dagger}=BA=AB.
\end{eqnarray}
Por lo tanto, el producto de dos operadores Herm\'{\i}ticos que {\bf conmutan}
es Herm\'{\i}tico.\\

\subsection{ Ejemplos de matrices Hem\'iticas}

De (\ref{eq:adj-mat}) se puede ver que una matriz de ${\bf C}^{n}$ Herm\'{\i}tica  cumple
\begin{eqnarray}
\left(\Lambda\right)^{*T}=\Lambda.
\end{eqnarray}
En ${\bf C}^{2}$ un ejemplo trivial una matriz Herm\'{\i}tica es la matriz identidad
\begin{eqnarray}
I=
\left(
\begin{array}{rr}
1& 0 \\
0& 1
\end{array}\right).
\end{eqnarray}
En ese mismo espacio, se puede ver que las matrices de Pauli
\begin{eqnarray}
\sigma_{1}=
\left(
\begin{array}{rr}
0& 1 \\
1& 0
\end{array}\right),\quad 
\sigma_{2}=
\left(
\begin{array}{rr}
0& -i\\
i& 0
\end{array}\right),\quad 
\sigma_{3}=
\left(
\begin{array}{rr}
1& 0\\
0& -1
\end{array}\right)
\end{eqnarray}
son Herm\'{\i}ticas.\\

\subsection{Ejemplos de operadores Herm\'{\i}ticos}

En el espacio de funciones, claramente cualquier funci\'on real $f(\vec r)$ es un operador
Herm\'{\i}tico. De Eq. (\ref{eq:adj-der}) se puede ver que si $\alpha$ es imaginario, por ejemplo
si $\alpha=-i\hbar$ el operador 
\begin{eqnarray}
 P_{x}=-i\hbar \frac{\partial }{\partial x}
\end{eqnarray}
es Herm\'{\i}tico. Adem\'as, como $P_{x}$ conmuta con $P_{x},$ entonces 
$ P_{x} P_{x}=P_{x}^{2}$ es un operador Herm\'{\i}tico. Si $V(x)$ es un
potencial real, el operador Hamiltoniano de la mec\'anica cu\'antica
\begin{eqnarray}
 H=\frac{1}{2m} P_{x}^{2}+V(x)
\end{eqnarray}
es Herm\'{\i}tico, pues es la suma de dos operadores Herm\'{\i}ticos.\\

Claramente los operadores
\begin{eqnarray}
P_{x}=-i\hbar \frac{\partial }{\partial x},\quad  P_{y}=-i\hbar \frac{\partial }{\partial y},\quad 
 P_{z}=-i\hbar \frac{\partial }{\partial z}
\end{eqnarray}
son Herm\'{\i}ticos. Entonces, si $V(\vec r)$ es una funci\'on real, el operador
\begin{eqnarray}
 H=\frac{1}{2m} P^{2}+V(\vec r)=-\frac{\hbar^{2}}{2m}\nabla^{2} +V(\vec r) \nonumber
\end{eqnarray}
es Herm\'{\i}tico. \\

Si $\eta$ y $\xi$ son variables independientes, entonces  definiendo 
$$O_{1}=\eta,\qquad O_{2}=-i\frac{\partial }{\partial \xi}$$
se tiene 
$$O_{1}O_{2}f=\eta(-i)\frac{\partial f}{\partial \xi}= (-i)\frac{\partial }{\partial \xi}\eta f= O_{2}O_{1}f,$$
es decir $\eta$ y $\left(-i\frac{\partial }{\partial \xi}\right)$ conmutan. As\'i los operadores 
\begin{eqnarray}
xp_{y}=x(-i)\frac{\partial }{\partial y},\quad xp_{z}= x(-i)\frac{\partial }{\partial z},\nonumber\\
yp_{x}=y(-i)\frac{\partial }{\partial x},\quad  yp_{z}=y(-i)\frac{\partial }{\partial z},\nonumber\\
zp_{x}=z(-i)\frac{\partial }{\partial x},\quad  zp_{y}=z(-i)\frac{\partial }{\partial y},\nonumber
\end{eqnarray}
son Herm\'{\i}ticos. Por lo tanto, tambi\'en los operadores 
\begin{eqnarray}
 L_{x}=-i\left( y \frac{\partial }{\partial z}-z \frac{\partial }{\partial y}\right),\quad
 L_{y}=-i \left( z \frac{\partial }{\partial x}-x \frac{\partial }{\partial z}\right),\quad
 L_{z}=-i\left( x \frac{\partial }{\partial y}-y \frac{\partial }{\partial x}\right).\nonumber
\end{eqnarray}
son Herm\'{\i}ticos. Esto implica que el operador 
$$L^{2}=L_{x}^{2}+L_{y}^{2}+L_{z}^{2}$$
es Herm\'{\i}tico. 

\section{Conmutador}

Supongamos que tenemos los operadores lineales
$A$ y $B,$ entonces definiremos el conmutador como
\begin{eqnarray}
\left[A,B\right]=AB-BA.
\end{eqnarray}
Por ejemplo, si $f$ es una funci\'on de prueba y tenemos los operadores $x,y$ se tiene
\begin{eqnarray}
(xy)f=xyf=yxf=(yx)f, 
\end{eqnarray}
de donde 
\begin{eqnarray}
\left(xy-yx\right)f=0.
\end{eqnarray}
Este resultado es v\'alido para cualquier funci\'on, por lo que se suele escribir 
\begin{eqnarray}
[x,y]=0.
\end{eqnarray}
Por lo misma raz\'on se encuentra
\begin{eqnarray}
[x,z]=[z,y]=0.
\end{eqnarray}

Adem\'as si $f$ es una funci\'on bien portada, se cumple 
\begin{eqnarray}
\left(\frac{\partial^{2} }{\partial x \partial y}\right)f=\left( \frac{\partial^{2} }{\partial x\partial y}\right)f,
\nonumber
\end{eqnarray}
entonces si 
\begin{eqnarray}
p_{x}=-i\frac{\partial }{\partial x}, \qquad p_{y}=-i\frac{\partial }{\partial y},
\end{eqnarray}
se llega a
\begin{eqnarray}
(p_{y}p_{x})f&=&\left( (-i) \frac{\partial }{\partial y}(-i) \frac{\partial }{\partial x}\right)f
=(-i)(-i) \left(\frac{\partial^{2} }{\partial x \partial y}\right)f\nonumber\\
&=&\left( (-i) \frac{\partial }{\partial y}(-i) \frac{\partial }{\partial x}\right)f=p_{x}p_{y}f,\nonumber
\end{eqnarray}
que implica
\begin{eqnarray}
[p_{x},p_{y}]=0.
\end{eqnarray}
De la misma forma se tiene
\begin{eqnarray}
[p_{x},p_{z}]=[p_{y},p_{z}]=0.
\end{eqnarray}
Con los operadores $y,p_{x}$ se tiene
\begin{eqnarray}
(yp_{x})f&=&\left(y(-i) \frac{\partial }{\partial x}\right)f= -iy\frac{\partial  }{\partial x}f
=-i\frac{\partial\left( y f \right) }{\partial x}= (p_{x}y)f,\nonumber
\end{eqnarray}
por lo tanto,
\begin{eqnarray}
[y,p_{x}]=0.
\end{eqnarray}
Con los mismos argumentos se encuentra
\begin{eqnarray}
[y,p_{z}]=[x,p_{y}]=[x,p_{z}]=[z,p_{x}]=[z,p_{y}]=0.
\end{eqnarray}
Ahora, si $x,p_{x}=-i\frac{\partial }{\partial x},$ entonces 
\begin{eqnarray}
(xp_{x})f&=&\left(x(-i) \frac{\partial }{\partial x}\right)f= -ix\frac{\partial f }{\partial x}, \nonumber\\ 
(p_{x} x)f&=&\left((-i) \frac{\partial }{\partial x} x\right)f= -i\frac{\partial  }{\partial x}\left(x f\right)=-if
-i x\frac{\partial f }{\partial x},\nonumber
\end{eqnarray}
es decir
\begin{eqnarray}
\left(xp_{x}-p_{x}x\right)f=if.
\end{eqnarray}
Como este resultado es v\'alido para cualquier funci\'on, se escribe 
\begin{eqnarray}
[x,p_{x}]=i.
\end{eqnarray}
Tambi\'en se tiene
\begin{eqnarray}
[y,p_{y}]=i,\quad [z,p_{z}]=i.
\end{eqnarray}
Si definimos 
\begin{eqnarray}
&x_{1}=x,&\qquad x_{2}=y,\qquad x_{3}=z,\nonumber\\
p_{1}&=&-i \frac{\partial }{\partial x_{1}},\qquad p_{2}=-i\frac{\partial }{\partial x_{2}},\qquad 
p_{3}=-i \frac{\partial }{\partial _{3}},
\end{eqnarray}
las anteriores reglas de conmutaci\'on se pueden escribir como
\begin{eqnarray}
\left[x_{i},x_{j}\right]&=&\left[p_{i},p_{j}\right]=0\qquad  \left[x_{i},p_{j}\right]=i\delta_{ij}. \label{eq:reglas-basicas}
\end{eqnarray}
Salvo un factor de $\hbar,$ estas son las reglas de conmutaci\'on que propuso Heisenberg como base de la
mec\'anica cu\'antica.

\subsection{Propiedades de los conmutadores}

Los conmutadores tiene varias propiedades algebraicas que los hacen importantes
para la F\'{\i}sica y las Matem\'aticas, a continuaci\'on veremos algunas de ellas.
Sea $c$ un n\'umero, entonces si $A,B,C$ son operadores lineales se cumple
\begin{eqnarray}
\left[A,c\right]&=&0, \label{eq:conmu-1}\\
\left[A,B\right]&=&-\left[B,A\right],\label{eq:conmu-2}\\
\left[A,B+C\right]&=&\left[A,B\right]+ \left[A,C\right],\label{eq:conmu-3}\\
\left[A,BC\right]&=&\left[A,B\right]C+ B\left[A,C\right],\label{eq:conmu-4}\\
\left[A,\left[B,C\right]\right]+ \left[C,\left[A,B\right]\right]+\left[B,\left[A,C\right]\right]&=&0.
\label{eq:conmu-5}
\end{eqnarray}
La identidad Eq. (\ref{eq:conmu-1}) es v\'alida pues  $c$ es un n\'umero. La prueba de Eq. (\ref{eq:conmu-2}) es
\begin{eqnarray}
\left[B,A\right]=BA-AB=-\left(AB-BA\right)=-\left[A,B\right].\nonumber
\end{eqnarray}
Mientras que la prueba de Eq. (\ref{eq:conmu-3}) es
\begin{eqnarray}
\left[A,B+C\right]&=&A\left(B+C\right)- \left(B+C\right)A
=\left(AB+AC\right)-\left(BA+CA\right)\nonumber\\
&=&\left(AB-BA\right)+ \left(AC-CA\right)= \left[A,B\right]+\left[A,C\right].\nonumber
\end{eqnarray}
Adem\'as,
\begin{eqnarray}
\left[A,BC\right]&=&A\left(BC\right)- \left(BC\right)A
=\left(ABC\right)-\left(BAC\right)+\left(BAC\right)-  \left(BCA\right) \nonumber\\
&=&\left(AB-BA\right)C+ B\left(AC-CA\right)= \left[A,B\right]C+B\left[A,C\right],
\nonumber
\end{eqnarray}
que prueba Eq. (\ref{eq:conmu-4}).\\

Ahora tenemos que
\begin{eqnarray}
\left[A,\left[B,C\right]\right]&=& \left[A,\left(BC-CB\right)\right]= \left[A,BC\right]-\left[A,CB\right]\nonumber\\
&=&B \left[A,C\right]+ \left[A,B\right]C-\left( C \left[A,B\right]+ \left[A,C\right]B\right)\nonumber\\
&=& B\left(AC-CA\right) + \left(AB-BA\right)C\nonumber\\
& &-\left( C \left(AB-BA\right)+ \left(AC-CA\right)B\right)\nonumber \\
&=& BAC+ABC+CBA+CAB\nonumber\\
& & -\left( BCA+BAC+CAB+ACB\right)\nonumber\\
&=& ABC+CBA-\left(BCA+ACB\right),\nonumber
\end{eqnarray}
de la misma forma
\begin{eqnarray}
\left[C,\left[A,B\right]\right]&=& CAB+BAC-\left(ABC+CBA\right),\nonumber\\
\left[B,\left[C,A\right]\right]&=& BCA+ACB-\left(CAB+BAC\right),\nonumber
\end{eqnarray}
sumando estas tres igualdades se llega a Eq. (\ref{eq:conmu-5}).\\

\subsection{Ejercicio}

Para tener un poco de pr\'actica con los conmutadoremos calcularemos las reglas de conmutaci\'on
del momento angular Eq. (\ref{eq:vec-momento-angular}), cuyas componentes se pueden escribir como
\begin{eqnarray}
L_{x}&=&yp_{z}-zp_{y},\nonumber \\
L_{y}&=&zp_{x}-xp_{z},\nonumber\\
L_{z}&=&xp_{y}-yp_{x}.\nonumber
\end{eqnarray}
De donde aplicando las propiedades de los conmutadores y considerando Eq. (\ref{eq:reglas-basicas})
se llega 
\begin{eqnarray}
\left[L_{x},L_{y}\right]&=&\left[yp_{z}-zp_{y},zp_{x}-xp_{z}\right]=\left[ yp_{z},zp_{x}-xp_{z}\right]+
\left[-zp_{y},zp_{x}-xp_{z}\right]\nonumber\\
&=&\left[ yp_{z},zp_{x}-xp_{z}\right]-\left[zp_{y},zp_{x}-xp_{z}\right]\nonumber\\
&=&\left[ yp_{z},zp_{x}\right]-\left[yp_{z},xp_{z}\right]-
\left(\left[zp_{y},zp_{x}\right]-\left[zp_{y},xp_{z}\right]\right)\nonumber\\
&=&y\left[ p_{z},zp_{x}\right]+\left[y,zp_{x}\right]p_{z}-y\left[p_{z},xp_{z}\right]-\left[y,xp_{z}\right]p_{z}\nonumber\\
& &-\left(z \left[p_{y},zp_{x}\right]+\left[z,zp_{x}\right]p_{y}- z\left[p_{y},xp_{z}\right]
- \left[z,xp_{z}\right]p_{y}\right)\nonumber\\
&= &yz\left[ p_{z},p_{x}\right]+y\left[ p_{z},z\right]p_{x}+z\left[y,p_{x}\right]p_{z}+\left[y,z\right]p_{x}p_{z}
- yx\left[p_{z},p_{z}\right]\nonumber\\
& &-y\left[p_{z},x\right]p_{z}- x\left[y,p_{z}\right]p_{z}-\left[y,x\right]p_{z}p_{z}- 
z \left[p_{y},p_{x}\right]-z \left[p_{y},z\right]p_{x}\nonumber\\
& &
-z\left[z,p_{x}\right]p_{y}-z\left[z,p_{y}\right]p_{x}+zx\left[p_{y},p_{z}\right]+z\left[p_{y},x\right]p_{z}\nonumber\\
& &+ x\left[z,p_{z}\right]p_{y}+\left[z,x\right]p_{z}p_{y}\nonumber\\
& =&i\left(xp_{y}-yp_{x}\right)=iL_{z}
\end{eqnarray}
Adem\'as
\begin{eqnarray}
\left[L_{y},L_{z}\right]&=&
\left[zp_{x}-xp_{z},xp_{y}-yp_{x}\right]=
\left[ zp_{x},xp_{y}-yp_{x}\right]-
\left[xp_{z},xp_{y}-yp_{x}\right]\nonumber\\
&=& \left[ zp_{x},xp_{y}\right]+\left[xp_{z},yp_{x}\right]
=z\left[ p_{x},x\right]p_{y}+y\left[x,p_{x}\right]p_{z}\nonumber\\
&=&i\left(yp_{z}-zp_{y}\right)=iL_{x},\nonumber\\
\left[L_{z},L_{x}\right]&=&\left[xp_{y}-yp_{x},yp_{z}-zp_{y}\right]=
\left[xp_{y},yp_{z}-zp_{y}\right]-\left[yp_{x},yp_{z}-zp_{y}\right]\nonumber\\
&=&\left[xp_{y},yp_{z}\right]+\left[yp_{x},zp_{y}\right]=
x\left[p_{y},y\right]p_{z}+z\left[y,p_{y}\right]p_{x}\nonumber\\
&=& i\left(zp_{x}-xp_{z}\right)=iL_{y}.
\end{eqnarray}
Es decir,
\begin{eqnarray}
\left[L_{x},L_{y}\right]=L_{z},\qquad \left[L_{z},L_{x}\right]=iL_{y},\qquad 
\left[L_{y},L_{z}\right]=iL_{x}.\label{eq:algebra-lie}
\end{eqnarray}
Ahora, considerando $L^{2}=L_{x}^{2}+L_{y}^{2}+L_{z}^{2}$ y Eq. (\ref{eq:algebra-lie})
se tiene 
\begin{eqnarray}
\left[L^{2},L_{x},\right]&=&\left[ L_{x}^{2}+L_{y}^{2}+L_{z}^{2},  L_{x}\right]=
\left[ L_{x}^{2},L_{x}\right]+\left[L_{y}^{2},L_{x}\right]+\left[L_{z}^{2},L_{x}\right]\nonumber\\
&=&L_{y}\left[L_{y},L_{x}\right]+\left[L_{y},L_{x}\right]L_{y}+L_{z}\left[L_{z},L_{x}\right]+
\left[L_{z},L_{x}\right]L_{z}\nonumber\\
&=&-iL_{y}L_{z}-iL_{z}L_{y}+iL_{z}L_{y}+iL_{y}L_{z}=0,\nonumber\\
\left[L^{2},L_{y},\right]&=&\left[ L_{x}^{2}+L_{y}^{2}+L_{z}^{2},  L_{y}\right]=
\left[ L_{x}^{2},L_{y}\right]+\left[L_{y}^{2},L_{y}\right]+\left[L_{z}^{2},  L_{y}\right]\nonumber\\
&=&L_{x}\left[ L_{x},L_{y}\right]+ \left[ L_{x},L_{y}\right]L_{x}+ L_{z}\left[L_{z},  L_{y}\right]
+\left[L_{z},  L_{y}\right]L_{z}\nonumber\\
&=&iL_{x}L_{z}+iL_{z}L_{x}-i L_{z}L_{x}-iL_{x}L_{z}=0,\nonumber\\
\left[L^{2},L_{z},\right]&=&\left[ L_{x}^{2}+L_{y}^{2}+L_{z}^{2},  L_{z}\right]=
\left[ L_{x}^{2},L_{z}\right]+\left[L_{y}^{2},L_{z}\right]+\left[L_{z}^{2},L_{z}\right]\nonumber\\
&=& L_{x}\left[ L_{x},L_{z}\right]+\left[ L_{x},L_{z}\right]L_{x}+
L_{y}\left[L_{y},L_{z}\right]+\left[L_{y},L_{z}\right]L_{y}\nonumber\\
&=&-iL_{x}L_{y}-iL_{y}L_{x}+iL_{y}L_{x}+iL_{x}L_{y}=0. \nonumber
\end{eqnarray}
Por lo tanto, $L^{2}$ conmuta con cualquier componente del momento angular
\begin{eqnarray}
\left[L^{2},L_{x}\right]=\left[L^{2},L_{y}\right]=\left[L^{2},L_{z}\right]=0.
\end{eqnarray}

\section{Conmutadores y la derivada}

Una de las propiedades del conmutador entre dos operadores
es que en algunos casos act\'ua como derivada. En efecto
supongamos que 
\begin{eqnarray}
\left[A,B\right]=\alpha,\qquad \alpha={\rm constante},
\label{eq:deri-mat}
\end{eqnarray}
entonces
\begin{eqnarray}
\left[A,B^{2}\right]=\left[A,BB\right]=B\left[A,B\right]+\left[A,B\right]B=
B\alpha +B\alpha=2\alpha B.
\end{eqnarray}
En general se cumple que 
\begin{eqnarray}
\left[A,B\right]=\alpha \qquad \Longrightarrow \qquad \left[A,B^{n}\right]=\alpha nB^{n-1}.
\label{eq:derivada-mat}
\end{eqnarray}
Probaremos esta afirmaci\'on por inducci\'on. La hip\'otesis de inducci\'on 
ya la hemos probado. Para probar el paso inductivo debemos mostrar que
\begin{eqnarray}
\left[A,B^{k}\right]=\alpha kB^{k-1} \qquad \Longrightarrow \qquad
\left[A,B^{k+1}\right]=\alpha (k+1)B^{k}.
\end{eqnarray}
Esta \'ultima igualdad es correcta, pues ocupando la hip\'otesis de inducci\'on y  Eq. (\ref{eq:conmu-3}) 
se encuentra
\begin{eqnarray}
\left[A,B^{k+1}\right]&=& \left[A,BB^{k}\right]=B\left[A,B^{k}\right]+\left[A,B\right]B^{k}=
\alpha k BB^{k-1}+\alpha B^{k}\nonumber\\
&=& \alpha (k+1)B^{k},\nonumber
\end{eqnarray}
que es lo que se queria demostrar. Por lo tanto, la propiedad (\ref{eq:derivada-mat}) es v\'alida
para cualquier natural $n.$ As\'{\i}, en este caso el conmutador act\'ua como derivada.\\

Para ver de forma m\'as expl\'{\i}cita esta afirmaci\'on, supongamos que tenemos la funci\'on
\begin{eqnarray}
f(x)=\sum_{n\geq 0} b_{n} x^{n},\nonumber
\end{eqnarray}
con la cual podemos formar el operador 
\begin{eqnarray}
f(B)=\sum_{n\geq 0} b_{n} B^{n}.\nonumber
\end{eqnarray}
De donde, 
\begin{eqnarray}
\left[A,f(B)\right]&=&\left[A,\sum_{n\geq 0} b_{n}B^{n}\right]=
\sum_{n\geq 0} b_{n}\left[A, B^{n}\right]= \sum_{n\geq 0} b_{n}\alpha nB^{n-1}\nonumber\\
& =& \alpha \sum_{n\geq 0} b_{n}nB^{n-1}=\alpha \frac{df(B)}{dB}.\nonumber
\end{eqnarray}
Por lo tanto, para cualquier funci\'on
\begin{eqnarray}
\left[A,B\right]=\alpha,\qquad \Longrightarrow \qquad \left[A,f(B)\right]=\alpha \frac{df(B)}{dB}.
\label{eq:derivada-mat-1}
\end{eqnarray}
Por ejemplo, sabemos que con $p_{x}=-i\frac{\partial }{\partial x}$ se cumple $[x,p_{x}]=i,$ entonces si
$f$ y $g$ son dos funciones se encuentra
\begin{eqnarray}
\left[f(x),p_{x}\right]&=&i \frac{\partial f(x)}{\partial x},\nonumber\\
\left[x,g(p_{x})\right]&=&i \frac{\partial g(p_{x})}{\partial p_{x}}.\nonumber 
\end{eqnarray}
Estas igualdades son de gran utilidad para obtener ecuaciones de movimiento en mec\'anica cu\'antica.
En efecto, dado un  Hamiltoniano $ H$ las ecuaciones de movimiento se definen como
\begin{eqnarray}
i\dot  x&=&\left[ x, H\right],\nonumber\\
i\dot  p&=&\left[ p, H\right],
\end{eqnarray}
que son las ecuaciones de Heisenger. En particular con el Hamiltoniano
$$H=\frac{1}{2m}p^{2}+V(x),$$ 
se tiene
\begin{eqnarray}
i\dot  x&=&\left[ x, H\right]=\left[ x, \frac{1}{2m}p^{2}\right]= \frac{i}{m}p,\nonumber\\
i\dot  p&=&\left[ p, H\right]= \left[ p,V(x)\right]=-i\frac{\partial V(x)}{\partial x},
\end{eqnarray}
es decir 
\begin{eqnarray}
\dot  x=\frac{1}{m}p,\qquad 
\dot  p=-\frac{\partial V(x)}{\partial x}.\nonumber
\end{eqnarray}

\section{Vectores propios}

Sea $A$ un operador lineal, si $v$ es un vector tal que  $Av=\lambda v,$ se dice que $v$ es un vector propio de $A$ con valor propio $\lambda.$ Al conjunto de valores propios de $A$ se le llama el espectro de $A.$\\

Por ejemplo, los vectores propios del operador derivada deben cumplir
\begin{eqnarray}
\frac{\partial }{\partial x} f(x)=\lambda f(x), 
\label{eq:vec-prop-deri}
\end{eqnarray}
que son las funciones de la forma $f(x)=A_{0} e^{\lambda x}, A_{0}={\rm constante}.$
En la f\'isica matem\'atica es importante obtener las funciones propias y  valores propios
de diferentes o\-peradores, como el momento angular Eq. (\ref{eq:op-moment-esfe})
\begin{eqnarray}
L^{2}Y_{\lambda m}=\lambda Y_{\lambda m},\qquad 
L_{z}Y_{\lambda m}=m Y_{\lambda m},
\end{eqnarray}
y el operador Hamiltoniano
\begin{eqnarray}
H \psi\left(\vec r\right)=\left(-\frac{\hbar^{2}}{2m}\nabla^{2}+V(r)\right) 
\psi\left(\vec r\right)=
E \psi\left(\vec r\right).
\end{eqnarray}

\subsection{Espectro de operadores Herm\'iticos}
Los operadores Herm\'{\i}ticos tienen valores propios reales. En efecto, si $v$ es un vector propio de un operador Herm\'itico $A$ con valor propio  $\lambda$, es decir $Av=\lambda v,$ se tiene 
\begin{eqnarray}
\left<Av|v\right>&=&\left<\lambda v|v\right>=\lambda^{*}\left<v|v\right>\nonumber\\
&=&\left<v|A^{\dagger}v\right>=\left<v|Av\right>=\left<v|\lambda v\right>=
\lambda \left<v|v\right>, \nonumber
\end{eqnarray}
que nos conduce al resultado
\begin{eqnarray}
\lambda^{*}=\lambda.
\end{eqnarray}
As\'{\i}, los valores propios de los operadores Herm\'{\i}ticos son reales.\\

Como el operador $ H$ es Herm\'itico, la ecuaci\'on diferencial 
\begin{eqnarray}
 H\psi(\vec r)=
\left(\frac{1}{2m} P^{2}+V(\vec r)\right)\psi(\vec r) =
\left(-\frac{\hbar^{2}}{2m}\nabla^{2} +V(\vec r) \right)\psi(\vec r)=E\psi(\vec r)\nonumber
\end{eqnarray}
s\'olo tiene soluci\'on si  $E$  es un valor real. Posteriormente veremos como resolver
 esta ecuaci\'on para algunos casos particulares
de $V(\vec r).$ \\

De la misma forma, como $ L^{2}$ es Herm\'itico, la ecuaci\'on diferencial
\begin{eqnarray}
 L^{2}Y_{\lambda}(\theta,\varphi)=-\left[\frac{1}{\sin\theta} \frac{\partial }{\partial \theta}
\left(\sin\theta\frac{\partial }{\partial \theta}\right)+
\frac{1}{\sin\theta^{2}}\frac{\partial^{2} }{\partial \varphi^{2}}\right]Y_{\lambda}(\theta,\varphi)=\lambda Y_{\lambda}(\theta,\varphi)
\nonumber
\end{eqnarray}
solo tiene soluci\'on si $\lambda$ es un n\'umero real. Las soluciones de esta ecuaci\'on se llaman arm\'onicos esf\'ericos 
y son de suma importancia para
la electrost\'atica y la mec\'anica cu\'antica.\\

\subsection{Operadores que conmutan}

Se dice que un operador $A$ tiene espectro no degenerado si 
\begin{eqnarray}
Av=\lambda v\quad {\rm y} \quad Au=\lambda u\Longrightarrow \exists \alpha, \quad v=\alpha u. 
\end{eqnarray}
Por ejemplo, el operador derivada tiene espectro no degenerado,
pues todas las soluciones de Eq. (\ref{eq:vec-prop-deri}) son de la forma
$e^{\lambda x},$ en particular 
$$L_{z}=-i\frac{\partial }{\partial \varphi}$$
tiene espectro no degenerado.\\

Supongamos que  $A$ y $B$ son dos operadores lineales que conmutan, es decir 
$$[A,B]=0$$ 
y que $A$ tiene espectro no degenerado.  Entonces podemos afirmar que si $v$ es vector propio de $A,$ es decir $Av=\lambda v,$ tambi\'en lo es de $B.$ Esta afirmaci\'on es correcta pues si $AB=BA,$ entonces 
\begin{eqnarray}
ABv=BAv=B(Av)=\lambda Bv\quad \Longrightarrow A(Bv)=\lambda (Bv). 
\end{eqnarray}
As\'{\i},  $(Bv)$ es vector propio de $A$ con valor propio $\lambda$. De donde, como $A$ 
tiene espectro no degenerado, existe $\beta$ tal que 
$$Bv=\beta v.$$
Esto quiere decir que $v$ es vector propio de $B,$ lo que completa la prueba.\\

Por ejemplo, $L_{z}$ tiene espectro no degenerado y conmuta con $L^{2}.$ Por lo tanto,
estos operadores comparten funciones propias. En el lenguaje de la
mec\'anica cu\'antica significa que estas dos cantidades se pueden medir al mismo tiempo. \\

\chapter{Prueba de Feynman de las Ecuaciones de Maxwell}

Como un ejercicio para reforzar el usos de conmutadores y vectores, en este cap\'itulo  veremos la prueba
de Feynman de dos ecuaciones de Maxwell.\\

Apesar de que las ecuaciones de Maxwell se obtienen de mediciones de
la naturaleza, Feynman encontr\'o que la ley de inexistencia de monopolos
magn\'eticos y la ley de Faraday se pueden deducir de la din\'amica de una part\'{\i}cula.
La demostraci\'on parte de suponer la segunda ley de Newton
\begin{eqnarray}
m\ddot x_{i}=F_{i}(x,\dot x,t)
\label{eq:newton-fey}
\end{eqnarray}
y las reglas de conmutaci\'on
\begin{eqnarray}
[x_{i},x_{j}]=0,\qquad m[x_{i},\dot x_{j}]=i\hbar\delta_{ij}.
\label{eq:feynman-conmuta}
\end{eqnarray}
Primero veremos como se obtiene la fuerza de Lorentz de
estas hip\'otesis, para despu\'es hacer la deducci\'on 
de la ley de inexistencia de monopolos magn\'eticos y
la ley de Faraday.

\section{Fuerza de Lorentz}

Para iniciar, notemos que 
\begin{eqnarray}
x_{i}\ddot x_{j}=\frac{d}{dt} \left(x_{i}\dot x_{j}\right)-
\dot x_{i}\dot x_{j},
\end{eqnarray}
entonces,
\begin{eqnarray}
[x_{i},\ddot x_{j}]&=&x_{i}\ddot x_{j}-\ddot x_{j}x_{i}
=\left(\frac{d}{dt} \left(x_{i}\dot x_{j}\right)-
\dot x_{i}\dot x_{j}\right)-
\left(\frac{d}{dt} \left(\dot x_{j} x_{i}\right)-
\dot x_{j}\dot x_{i}\right)\nonumber\\
&=&\frac{d}{dt}\left(x_{i}\dot x_{j}-\dot x_{j}\dot x_{i}\right)+
\left(\dot x_{j}\dot x_{i}- \dot x_{i}\dot x_{j}\right)\nonumber\\
&=&\frac{d}{dt}\left([x_{i},\dot x_{j}]\right)+[\dot x_{j},\dot x_{i}]
=[\dot x_{j},\dot x_{i}],
\end{eqnarray}
es decir,
\begin{eqnarray}
[x_{i},\ddot x_{j}]=[\dot x_{j},\dot x_{i}].
\end{eqnarray}
Por lo tanto, de la segunda ley de Newton Eq. (\ref{eq:newton-fey}) se obtiene
\begin{eqnarray}
[x_{i},F_{j}]=[x_{i},m\ddot x_{j}]=m[\dot x_{j},\dot x_{i}].
\label{eq:fey1}
\end{eqnarray}
Usando esta ecuaci\'on y la propiedad antisim\'etrica de los conmutadores,
se llega a
\begin{eqnarray}
[x_{i},F_{j}]=m[\dot x_{j},\dot x_{i}]=- m[\dot x_{i},\dot x_{j}]
=-[x_{j},F_{i}].
\end{eqnarray}
De donde, $[x_{i},F_{j}]$ es una matriz antisim\'etrica. Ahora, sabemos que cualquier
matriz antisim\'etrica de $3\times 3$ se puede escribir en t\'erminos 
del tensor de Levi-Civita  y un vector, ver Eq. (\ref{eq:levi-anti}). As\'{\i}, sin perdida de generalidad podemos
proponer 
\begin{eqnarray}
[x_{i},F_{j}]=-\frac{i\hbar}{m} \epsilon_{ijk}B_{k}.
\label{eq:fey2}
\end{eqnarray}
Note que de Eq. (\ref{eq:fey1}) y Eq. (\ref{eq:fey2}) se obtiene
\begin{eqnarray}
[\dot x_{i},\dot x_{j}]=\frac{i\hbar}{m^{2}} \epsilon_{ijk}B_{k}.
\label{eq:fey-penultima}
\end{eqnarray}
En principio $B_{k}$ puede depender de la velocidad, sin embargo,
por la identidad de Jacobi
\begin{eqnarray}
[x_{i},[\dot x_{j}, \dot x_{k} ]]
+[\dot x_{k},[x_{i},\dot x_{j}]]
+[\dot x_{j},[\dot x_{k}, x_{i}]]
=0,
\end{eqnarray}
las reglas de conmutaci\'on Eqs. (\ref{eq:feynman-conmuta}), (\ref{eq:fey1}) y (\ref{eq:fey2}) se obtiene
\begin{eqnarray}
[x_{i},[ x_{j}, F_{k} ]]= -\frac{i\hbar}{m}[x_{i}, \epsilon_{jkl}B_{l}]=0,
\end{eqnarray}
por lo tanto, $B_{k}$ solo depende de las coordenadas.
As\'{\i}, la igualdad Eq. (\ref{eq:fey2}) implica que $F_{i}$
debe ser de la forma $F_{i}=\epsilon_{ijk}\dot x_{j}B_{k}.$
Note que la igualdad Eq. (\ref{eq:fey2}) se sigue cumpliendo  si sumamos a $F_{i}$
una funci\'on, $E_{i},$ que solo depende de las coordenadas. As\'{\i},
la forma m\'as general de la fuerza es  
\begin{eqnarray}
F_{i}=E_{i}+ \epsilon_{ijk}\dot x_{j}B_{k},
\end{eqnarray}
que es exactamente la fuerza de Lorentz.\\

\section{Inexistencia de monopolos magn\'eticos}

Ahora, veremos la ley de inexistencia de monopolos magn\'eticos.
De Eq. (\ref{eq:fey1}) y  Eq. (\ref{eq:fey2})
se obtiene
\begin{eqnarray}
\epsilon_{lij}[\dot x_{i},\dot x_{j}]&=&-\frac{\epsilon_{lij}}{m}[x_{i},F_{j}]
=-\frac{\epsilon_{lij}}{m}\left(\frac{-i\hbar}{m}\epsilon_{ijk}B_{k}\right)=
\frac{i\hbar}{m^{2}} \epsilon_{lij}\epsilon_{ijk}B_{k}\nonumber\\
&=&-\frac{i\hbar}{m^{2}}  \epsilon_{lij}\epsilon_{jik}B_{k}= 
-\frac{i\hbar}{m^{2}}\left(\delta_{li}\delta_{ik}-\delta_{lk}\delta_{ii}\right)B_{k}
=-\frac{i\hbar}{m^{2}} \left(\delta_{lk}-3\delta_{lk}\right)B_{k}\nonumber\\
&=&\frac{2i\hbar}{m^{2}}B_{l},
\end{eqnarray}
es decir,
\begin{eqnarray}
B_{l}=\frac{-im^{2}}{2\hbar}\epsilon_{lij}[\dot x_{i},\dot x_{j}].
\label{eq:fey3}
\end{eqnarray}
Adem\'as, notemos que con  los operadores $A_{1},A_{2},A_{3}$
la identidad de Jacobi se puede escribir como 
\begin{eqnarray}
\epsilon_{ijk}[A_{i},[ A_{j}, A_{k} ]]=0.
\end{eqnarray}
Por lo tanto, tomando $A_{i}=\dot x_{i}$ y considerando Eq. (\ref{eq:fey3}),  se tiene
\begin{eqnarray}
0=\epsilon_{ijk}[\dot x_{i},[ \dot x_{j}, \dot x_{k} ]]=
[\dot x_{i},\epsilon_{ijk}[ \dot x_{j}, \dot x_{k} ]]
=-\frac{2\hbar}{im^{2}}[\dot x_{i}, B_{i}]=\frac{i2\hbar}{im^{2}}\partial_{i}B_{i}
= \frac{2\hbar}{m^{2}}\vec \nabla \cdot \vec B,\nonumber
\end{eqnarray}
es decir, se cumple
\begin{eqnarray}
\vec \nabla \cdot \vec B=0.
\end{eqnarray}
Que es la ley de inexistencia de monopolos magn\'eticos.

\section{Ley de Faraday}

Veamos la deducci\'on de la ley de Faraday.
Primero notemos que ocupando las propiedades antisim\'etricas
del tensor de Levi-Civita y del conmutador, adem\'as de renombrar \'{\i}ndices  
se encuentra,
\begin{eqnarray}
\epsilon_{lij}[\dot x_{i},\ddot x_{j}]=\epsilon_{lji}[\ddot x_{j},\dot x_{i}]=
\epsilon_{lij}[\ddot x_{i},\dot x_{j}], 
\end{eqnarray}
es decir,
\begin{eqnarray}
\epsilon_{lij}[\dot x_{i},\ddot x_{j}]=\epsilon_{lij}[\ddot x_{i},\dot x_{j}].
\label{eq:fey-ultima}
\end{eqnarray}
Tambi\'en se puede ver que se cumple
\begin{eqnarray}
\dot B_{l}&=&\frac{\partial B_{l}}{\partial t}
+\dot x_{m}\frac{\partial B_{l}}{\partial x_{m}}.
\label{eq:fey-ultima1}
\end{eqnarray}
As\'i, considerando  Eqs. (\ref{eq:fey3}), (\ref{eq:fey-ultima}) y 
(\ref{eq:fey-penultima}) se encuentra
\begin{eqnarray}
\dot B_{l}&=&\frac{\partial B_{l}}{\partial t}
+\dot x_{m}\frac{\partial B_{l}}{\partial x_{m}} =\frac{d}{dt}
\left( \frac{-im^{2}}{2\hbar}\epsilon_{lij}[\dot x_{i},\dot x_{j}]\right)
=-\frac{im^{2}}{2\hbar}\epsilon_{lij}\frac{d}{dt}
\left( \dot x_{i}\dot x_{j}-\dot x_{j}\dot x_{i}\right)\nonumber\\
&=&
\frac{-im^{2}}{2\hbar}\epsilon_{lij}\left(\ddot x_{i}\dot x_{j}+\dot x_{i}\ddot x_{j}-
\ddot x_{j}\dot x_{i}- \dot x_{j}\ddot x_{i}\right)\nonumber\\
&=&-\frac{im^{2}}{2\hbar}\epsilon_{lij}\left(
[\ddot x_{i},\dot x_{j}]+[\dot x_{i},\ddot x_{j}]\right)
=-\frac{im}{\hbar}\epsilon_{lij}[m\ddot x_{i},\dot x_{j}]\nonumber\\
&=&-\frac{im}{\hbar}\epsilon_{lij}\left[ F_{i}, \dot x_{j}\right]
=-\frac{im}{\hbar}\epsilon_{lij}\left[ E_{i}+\epsilon_{irs}\dot x_{r}B_{s}, 
\dot x_{j}\right] \nonumber\\
&=&-\frac{im}{\hbar}\epsilon_{lij}\left(\left[ E_{i},\dot x_{j}\right]+
\epsilon_{irs}\left[\dot x_{r}B_{s}, 
\dot x_{j}\right]\right)\nonumber\\
&=&\epsilon_{lij}\partial_{j}E_{i}
+\frac{im}{\hbar}\epsilon_{lji}\epsilon_{irs}
\left( \dot x_{r}[B_{s},\dot x_{j}]+[\dot x_{r},\dot x_{j}]B_{s}\right)
\nonumber\\
&=&-\left(\vec \nabla \times \vec E\right)_{l}
+\frac{im}{\hbar}(\delta_{lr}\delta_{js} -\delta_{ls} \delta_{jr})
\left( \frac{i\hbar}{m}\dot x_{r}\frac{\partial B_{s}}{\partial x_{j}}
+[\dot x_{r},\dot x_{j}]B_{s}\right)\nonumber\\
&=&-\left(\vec \nabla \times \vec E\right)_{l}
-\left(\dot x_{l} \frac{\partial B_{s}}{\partial x_{s}}-\dot x_{j}
\frac{\partial B_{l}}{\partial x_{j}}\right)
+\frac{im}{\hbar}[\dot x_{l},\dot x_{j}]B_{j}-\frac{im}{\hbar}
[\dot x_{r},\dot x_{r}]B_{l}\nonumber\\
&=&-\left(\vec \nabla \times \vec E\right)_{l}-
\dot x_{l}\left(\vec \nabla \cdot \vec B\right)+
 \dot x_{m}\frac{\partial B_{l}}{\partial x_{m}}
 -\frac{1}{m} \epsilon_{ljr}B_{r}B_{j}\nonumber\\
&=&-\left(\vec \nabla \times \vec E\right)_{l}+
 \dot x_{m}\frac{\partial B_{l}}{\partial x_{m}}.
\label{eq:fey-ultima2} 
\end{eqnarray}
Por lo tanto, igualando Eq. (\ref{eq:fey-ultima1}) con Eq. (\ref{eq:fey-ultima2}) 
se tiene
\begin{eqnarray}
\vec \nabla \times \vec E=-\frac{\partial \vec B}{\partial t},
\end{eqnarray}
que es la ley de Faraday. \\

Es notable que la fuerza de Lorentz y dos ecuaciones de Maxwell se puedan 
obtener partiendo solo de las reglas de conmutaci\'on y la segunda ley de Newton.
Hay diversas opiniones sobre este hecho. Algunos piensan que hay algo profundo
en este resultado, el cual a\'un no hemos comprendido. Pero tambi\'en hay quienes
piensan que es un resultado sin importancia y otros que creen tener una explicaci\'on
de esta prueba. Feynman encontr\'o este prueba pero no la public\'o, solo se la cont\'o a 
algunos de sus amigos. Fue Dyson quien  mand\'o a publicar el resultado despu\'es de la muerte de Feynman
\cite{dyson:gnus}.


\chapter{Series de Fourier}

En este cap\'itulo mostraremos que, eligiendo de manera adecuada los par\'ametros y el intervalo, las funciones trigonom\'etricas $\sin \alpha$ y $\cos \alpha$ forman una base ortonormal en el espacio de funciones.
Este cap\'itulo es de especial importancia, pues se aplican varias de las herramientas desarrolladas en el cap\'itulo 7. Al final del cap\'itulo se muestran varios ejercicios.

\section{Funciones trigonom\'etricas}

Recordemos que las funciones trigonom\'etrica  $\sin\alpha$ y $ \cos\alpha$ satisfacen 
\begin{eqnarray}
\sin(-\alpha)&=&-\sin\alpha ,\label{eq:fourier-paridads} \\ 
\cos(-\alpha)&=&\cos\alpha, \label{eq:fourier-paridadc}\\
\sin(\alpha+2\pi)&=&\sin\alpha , \\
\cos(\alpha+2\pi)=\cos\alpha, \label{eq:periodo} \\
\sin n\pi&=&0,\qquad n=0,\pm 1,\pm 2, \pm 3, \cdots,  \\
\sin\left(\frac{2n+1}{2}\pi\right)&=&(-)^{n},\\
\cos n\pi&=&(-)^{n},\\
\cos\left(\frac{2n+1}{2}\pi\right)&=&0,\\
\int_{-\pi}^{\pi} dx \cos lx&=&\int_{0}^{2\pi} dx \cos lx= 0,\quad l=\pm 1,\pm 2,\pm 3,\cdots ,\label{eq:fourier-i1pi}\\ 
 \int_{-\pi}^{\pi} dx \sin lx&=& \int_{0}^{2\pi}dx  \sin lx= 0,\label{eq:fourier-i2pi}\\
\int_{0}^{\pi} dx \cos l x&=&0
\end{eqnarray}
Tambi\'en es \'util recordar la f\'ormula de Euler
\begin{eqnarray}
e^{i\alpha}=\cos\alpha+i\sin\alpha,
\end{eqnarray}
que induce 
\begin{eqnarray}
e^{-i\alpha}=\cos\alpha-i\sin\alpha,
\end{eqnarray}
de estas dos igualdades se obtiene
\begin{eqnarray}
\cos\alpha&=&\frac{1}{2}\left(e^{i\alpha}+e^{-i\alpha}\right),\\
\sin\alpha&=&\frac{1}{2i}\left(e^{i\alpha}-e^{-i\alpha}\right).
\end{eqnarray}
La representacion de Euler es muy \'util, pues nos permite 
obtener propiedades de las funciones trigonom\'etricas de forma
sencilla. Por ejemplo, como
\begin{eqnarray}
e^{i\alpha}e^{i\beta}=e^{i(\alpha+\beta)},
\end{eqnarray}
se encuentra que 
\begin{eqnarray}
e^{i\alpha}e^{i\beta}&=&\left(\cos\alpha+i\sin\alpha\right)
\left(\cos\beta+i\sin\beta\right) \nonumber\\
&=& 
\left(\cos\alpha\cos\beta-\sin\alpha\sin\beta\right)+i
\left(\sin\alpha\cos\beta+ \cos\alpha\sin\beta\right) \nonumber \\
&=&e^{i(\alpha+\beta)}=\cos(\alpha+\beta)+i\sin(\alpha+\beta),
\end{eqnarray}
es decir,
\begin{eqnarray}
\cos(\alpha+\beta)&=&\cos\alpha\cos\beta-\sin\beta\sin\alpha,\label{eq:sum1}\\
\sin(\alpha+\beta)&=&\sin\alpha\cos\beta +\sin\beta\cos\alpha.\label{eq:sum2}
\end{eqnarray}
De estas identidades y ocupando (\ref{eq:fourier-paridads})-(\ref{eq:fourier-paridadc}) se tiene
\begin{eqnarray}
\cos(\alpha-\beta)&=&\cos\alpha\cos\beta+\sin\beta\sin\alpha,\label{eq:sum3}\\
\sin(\alpha-\beta)&=&\sin\alpha\cos\beta -\sin\beta\cos\alpha.\label{eq:sum4}
\end{eqnarray}
Adem\'as, de   Eqs. (\ref{eq:sum1})-(\ref{eq:sum4}) se encuentra
\begin{eqnarray}
\cos(\alpha)\cos(\beta)&=&\frac{1}{2}
\left[\cos(\alpha-\beta)+\cos(\alpha+\beta)\right],\label{eq:sum5}\\
\sin(\alpha)\sin(\beta)&=&\frac{1}{2}
\left[\cos(\alpha-\beta)-\cos(\alpha+\beta)\right],\label{eq:sum6}\\
\sin(\alpha)\cos(\beta)&=&\frac{1}{2}
\left[\sin(\alpha+\beta)+\sin(\alpha-\beta)\right].\label{eq:sum7}
\end{eqnarray}

\section{Relaciones de ortonormalidad}

Con las funciones trigonom\'etricas 
\begin{eqnarray}
\left\{ \psi(x)\right\} =\left\{ \frac{1}{\sqrt{2\pi}}, \frac{\cos nx}{\sqrt{\pi}},
\frac{\sin nx}{\sqrt{\pi}} \right\}, \quad n=1,2,3 \cdots , \label{eq:base-fourier}
\end{eqnarray}
se puede construir una base de funciones ortonormal en el intervalo $[-\pi,\pi].$ Para mostrar esta afirmaci\'on, primero notemos que 
\begin{eqnarray}
\int_{-\pi}^{\pi}dx \left(\frac{1}{\sqrt{2\pi}}\right)^{2}=\
\frac{1}{2\pi}\int_{-\pi}^{\pi}dx=\frac{2\pi}{2\pi}=1.\nonumber
\end{eqnarray}
Adem\'as, considerado Eqs. (\ref{eq:fourier-i1pi})-(\ref{eq:fourier-i2pi}) se encuentra 
\begin{eqnarray}
\int_{-\pi}^{\pi}dx\frac{1}{\sqrt{2\pi}}\frac{\sin nx}{\sqrt{\pi}}
=\int_{-\pi}^{\pi}dx\frac{1}{\sqrt{2\pi}}\frac{\cos nx}{\sqrt{\pi}}=0.
\end{eqnarray}
Esto nos indica que la funci\'on constante $1/\sqrt{2\pi}$ es ortogonal
a las funciones  $\{\cos nx, \sin nx\}.$\\

Ahora, definamos $l=n-m$ y $k=n+m,$ entonces considerando Eqs. (\ref{eq:sum5})-(\ref{eq:sum7}) y 
ocupando Eqs. (\ref{eq:fourier-i1pi})-(\ref{eq:fourier-i2pi})
se llega a 
\begin{eqnarray}
\int_{-\pi}^{\pi}dx\frac{\sin mx}{\sqrt{\pi}}\frac{\cos nx}{\sqrt{\pi}}
&=&\frac{1}{\pi}\int_{-\pi}^{\pi}dx\sin mx\cos nx\nonumber\\
&=&
\frac{1}{2\pi}\int_{-\pi}^{\pi}dx\bigg(\sin kx+\sin lx\bigg)=0.
\end{eqnarray}
Para el caso en que $n\not =m,$ usando de nuevo Eqs. (\ref{eq:sum5})-(\ref{eq:sum7}) y 
ocupando Eqs. (\ref{eq:fourier-i1pi})-(\ref{eq:fourier-i2pi}),  se llega a 
\begin{eqnarray}
\int_{-\pi}^{\pi}dx\frac{\cos nx}{\sqrt{\pi}}\frac{\cos mx}{\sqrt{\pi}}
&=&\frac{1}{\pi}\int_{-\pi}^{\pi}dx\cos mx\cos nx\nonumber\\
&=&
\frac{1}{2\pi}\int_{-\pi}^{\pi}dx\bigg(\cos(n-m)x+\cos(n+m)x\bigg)\nonumber\\
&=&\frac{1}{2\pi}\int_{-\pi}^{\pi}dx\bigg(\cos lx+\cos kx\bigg)=0,\nonumber\\
\int_{-\pi}^{\pi}dx\frac{\sin nx}{\sqrt{\pi}}\frac{\sin mx}{\sqrt{\pi}}
&=&\frac{1}{\pi}\int_{-\pi}^{\pi}dx\sin mx\sin nx \nonumber\\
&=&
\frac{1}{2\pi}\int_{-\pi}^{\pi}dx\bigg(\cos(n-m)x-\cos(n+m)x\bigg)\nonumber\\
&=&\frac{1}{2\pi}\int_{-\pi}^{\pi}dx\bigg(\cos lx-\cos kx\bigg)=0.\nonumber
\end{eqnarray}
Adem\'as, se cumplen
\begin{eqnarray}
\int_{-\pi}^{\pi}dx\frac{\cos nx}{\sqrt{\pi}}\frac{\cos nx}{\sqrt{\pi}}
&=&\frac{1}{\pi}\int_{-\pi}^{\pi}dx\cos nx\cos nx \nonumber\\
&=&
\frac{1}{2\pi}\int_{-\pi}^{\pi}dx\bigg(\cos(n-n)x+\cos(2n)x\bigg) \nonumber\\
&=& \frac{1}{2\pi}\int_{-\pi}^{\pi}dx+  
\frac{1}{2\pi}\int_{-\pi}^{\pi}dx\cos(2n)x=1,\nonumber\\
\int_{-\pi}^{\pi}dx\frac{\sin nx}{\sqrt{\pi}}\frac{\sin nx}{\sqrt{\pi}}
&=&\frac{1}{\pi}\int_{-\pi}^{\pi}dx\sin nx\sin nx \nonumber\\
&=&
\frac{1}{2\pi}\int_{-\pi}^{\pi}dx\bigg(\cos(n-n)x-\cos(2n)x\bigg) \nonumber\\
&=& \frac{1}{2\pi}\int_{-\pi}^{\pi}dx-
\frac{1}{2\pi}\int_{-\pi}^{\pi}dx\cos(2n)x=1. \nonumber
\end{eqnarray}
En resumen tenemos
\begin{eqnarray}
\int_{-\pi}^{\pi}dx\frac{\sin mx}{\sqrt{\pi}}\frac{\sin nx}{\sqrt{\pi}}
&=&\delta_{nm}, \label{eq:fourier1pi}\\
\int_{-\pi}^{\pi}dx\frac{\cos mx}{\sqrt{\pi}}\frac{\cos nx}{\sqrt{\pi}}
&=&\delta_{nm},\\
\int_{-\pi}^{\pi}dx\left(\frac{1}{\sqrt{2\pi}}\right)^{2}&=&1,\\
\int_{-\pi}^{\pi}dx\frac{\sin mx}{\sqrt{\pi}}\frac{\cos nx}{\sqrt{\pi}}
&=&0,\\
\int_{-\pi}^{\pi}dx\frac{1}{\sqrt{2\pi}}\frac{\cos nx}{\sqrt{\pi}}&=&0,\\
\int_{-\pi}^{\pi}dx\frac{1}{\sqrt{2\pi}}\frac{\sin nx}{\sqrt{\pi}}&=&0.\label{eq:fourier2pi}
\end{eqnarray}
Por lo tanto, el conjunto de funciones Eq. (\ref{eq:base-fourier}) son  un conjunto de funciones ortornales 
en el intervalo $[-\pi,\pi].$

\subsection{Series de Fourier}

Como el conjunto de funciones Eq. (\ref{eq:base-fourier}) es ortornal 
en el intervalo $[-\pi,\pi],$ entonces dada una funci\'on $f$ en dicho intervalo se puede hacer la expansi\'on 
\begin{eqnarray}
f(x)=\frac{a_{0}}{\sqrt{2\pi}}+\sum_{n\geq 1}\left(a_{n}\frac{\cos nx}{\sqrt{\pi}}+b_{n}\frac{\sin nx}{\sqrt{\pi}} \right), 
\end{eqnarray}
con los coeficientes de Fourier dados por 
\begin{eqnarray}
a_{0}&=&\left<\frac{1}{\sqrt{2\pi}}\Bigg|f(x)\right>= \int_{-\pi}^{\pi}dx \frac{1}{\sqrt{2\pi}}f(x),\nonumber\\
 a_{n}&=&\left<\frac{\cos nx}{\sqrt{\pi}}\Bigg|f(x)\right>= \int_{-\pi}^{\pi}dx \frac{\cos nx}{\sqrt{\pi}}f(x),\nonumber\\
 b_{n}&=&\left<\frac{\sin nx}{\sqrt{\pi}}\Bigg|f(x)\right>= \int_{-\pi}^{\pi}dx \frac{\sin nx}{\sqrt{\pi}}f(x). 
\end{eqnarray}

\section{Ejemplos}

En esta secci\'on veremos  varios ejemplos de series de Fourier

\subsection{Caso $f(x)=x$}

Consideremos la funci\'on $f(x)=x$ en el intervalo $[-\pi, \pi].$  Como esta funci\'on es impar, para este caso se tiene
\begin{eqnarray}
a_{0}&=&\left< \frac{1}{\sqrt{2\pi}}\bigg|x\right>=\frac{1}{\sqrt{2\pi}}\int_{-\pi}^{\pi}dx x=0, \nonumber\\
a_{n}&=&\left< \frac{\cos nx}{\sqrt{\pi}}\bigg|x\right>=\frac{1}{\sqrt{\pi}}\int_{-\pi}^{\pi}dx \left(\cos nx \right)x=0.\nonumber
\end{eqnarray}
Adem\'as
\begin{eqnarray}
b_{n}&=&\left< \frac{\sin nx}{\sqrt{\pi}}|x\right>=\frac{1}{\sqrt{\pi}}\int_{-\pi}^{\pi}dx \left(\sin nx \right)x,\nonumber
\end{eqnarray}
por lo que empleando el resultado
\begin{eqnarray}
x\sin nx=-\frac{1}{n}\left(\frac{d}{dx}(x\cos nx)- \cos nx\right),\nonumber
\end{eqnarray}
se llega a
\begin{eqnarray}
b_{n}= -\frac{1}{ n\sqrt{\pi} }\left(x\cos nx- \frac{1}{n} \sin nx\right) \Bigg|_{-\pi}^{\pi}=-\frac{2\pi}{n\sqrt{\pi}}\cos n\pi=
\frac{2\sqrt{\pi}}{n} (-)^{n}.
\nonumber
\end{eqnarray}
As\'i,
\begin{eqnarray}
x=\sum_{n\geq 1} \frac{2\sqrt{\pi}}{n}(-)^{n} \frac{\sin nx}{\sqrt{\pi}}= 2\sum_{n\geq 1} \frac{(-)^{n}}{n} \sin nx.
\nonumber
\end{eqnarray}
La igualdad de Parseval Eq. (\ref{eq:persival}) para este caso es
\begin{eqnarray}
||x||^{2}=\int_{-\pi}^{\pi} x^{2}= \sum_{n\geq 1} b_{n}^{2},
\nonumber
\end{eqnarray}
es decir, 
\begin{eqnarray}
\int_{-\pi}^{\pi} x^{2}= \frac{x^{3}}{3}\Bigg |_{-\pi}^{\pi}= \frac{2\pi^{3}}{3}= \sum_{n\geq 1} \frac{4 \pi}{n^{2}},
\nonumber
\end{eqnarray}
de donde 
\begin{eqnarray}
\sum_{n\geq 1} \frac{1}{n^{2}}=\frac{\pi^{2}}{6}.
\nonumber
\end{eqnarray}
Por otros m\'etodos, este resultado fue obtenido primero por Euler.\\

Tambi\'en note que 
\begin{eqnarray}
\sum_{n\geq 1} \frac{1}{n^{2}}&=& \sum_{n\geq 1} \frac{1}{(2n)^{2}}+  \sum_{n\geq 1} \frac{1}{(2n-1)^{2}}=
\frac{1}{4}\sum_{n\geq 1} \frac{1}{n^{2}}+  \sum_{n\geq 1} \frac{1}{(2n-1)^{2}}\nonumber\\
&=&\frac{\pi^{2}}{24}+  \sum_{n\geq 1} \frac{1}{(2n-1)^{2}}= \frac{\pi^{2}}{6},
\nonumber
\end{eqnarray}
as\'i
\begin{eqnarray}
\sum_{n\geq 1} \frac{1}{(2n-1)^{2}}= \frac{\pi^{2}}{8}.
\nonumber
\end{eqnarray}

\subsection{Caso $f(x)=x^{2}$}

Ahora consideremos la funci\'on $f(x)=x^{2}$ en el intervalo $[-\pi, \pi].$  Como esta funci\'on es par, se encuentra
\begin{eqnarray}
b_{n}&=&\left< \frac{\sin nx}{\sqrt{\pi}}|x ^{2}\right>=\frac{1}{\sqrt{\pi}}\int_{-\pi}^{\pi}dx \left(\sin nx \right)x^{2}=0,\nonumber\\
a_{0}&=&\left< \frac{1}{\sqrt{2\pi}}|x^{2}\right>=\frac{1}{\sqrt{2\pi}}\int_{-\pi}^{\pi}dx x^{2}=\frac{2 \pi^{3}}{3 \sqrt{2\pi}}.
\end{eqnarray}
Adicionalmente, usando la igualdad 
\begin{eqnarray}
x^{2}\cos nx=x^{2}\frac{d}{dx}\left(\frac{\sin nx}{n}\right)=\frac{d}{dx}\left(\frac{x^{2}\sin x}{n}\right)-\frac{2}{n} x\sin nx, 
\end{eqnarray}
se encuentra
\begin{eqnarray}
a_{n}&=&\left< \frac{\cos nx}{\sqrt{\pi}}|x^{2}\right>=\frac{1}{\sqrt{\pi}}\int_{-\pi}^{\pi}dx \left(\cos nx \right)x^{2},\nonumber\\
 &=&-\frac{2}{n\sqrt{\pi}}\int_{-\pi}^{\pi} dx \left(\sin nx\right) x= -\frac{2}{n\sqrt{\pi}}  \frac{2\pi (-)^{n+1}}{n}=\frac{4 \sqrt{\pi} (-)^{n}}{n^{2}} . \nonumber
\end{eqnarray}
Por lo tanto,
\begin{eqnarray}
x^{2}&=& \frac{2 \pi^{3}}{3 \sqrt{2\pi}} \frac{1}{\sqrt{2\pi}} +\sum_{n\geq 1} \frac{4 \sqrt{\pi} (-)^{n}}{n^{2}}\frac{\cos nx}{\sqrt{\pi}}
\nonumber\\
&=& \frac{\pi^{2}}{3} + \sum_{n\geq 1} \frac{4(-)^{n}} {n^{2}} \cos nx .\nonumber
\end{eqnarray}
Note que al evaluar esta funci\'on en $x=0$ se tiene 
\begin{eqnarray}
0= \frac{\pi^{2}}{3} + \sum_{n\geq 1} \frac{4(-)^{n}} {n^{2}},\nonumber
\end{eqnarray}
de donde 
\begin{eqnarray}
\sum_{n\geq 1} \frac{(-)^{n+1}} {n^{2}}=\frac{\pi^{2}}{12}.\nonumber
\end{eqnarray}

\subsection{Caso $f(x)= \cos(\mu x)$}

Ahora estudiaremos la funci\'on $f(x)=\cos(\mu x)$ en el intervalo $[-\pi, \pi].$  Como esta funci\'on es par,  se tienen que
\begin{eqnarray}
b_{n}&=&\left< \frac{\sin nx}{\sqrt{\pi}}\bigg|\cos(\mu x) \right>=\frac{1}{\sqrt{\pi}}\int_{-\pi}^{\pi}dx \left(\sin nx \right)\cos(\mu x)=0,\nonumber\\
a_{0}&=&\left< \frac{1}{\sqrt{2\pi}}\bigg| \cos(\mu x)\right>=\frac{1}{\sqrt{2\pi}}\int_{-\pi}^{\pi}dx \cos(\mu x) =\frac{2 \sin (\mu \pi)}{\mu \sqrt{2\pi}}.
\nonumber
\end{eqnarray}
Tambi\'en se encuentra
\begin{eqnarray}
a_{n}&=&\left< \frac{\cos nx}{\sqrt{\pi}}\bigg|\cos(\mu x)\right>=\frac{1}{\sqrt{\pi}}\int_{-\pi}^{\pi}dx \left(\cos nx \right)\cos(\mu x)\nonumber\\
&=&\frac{1}{2\sqrt{\pi}}\int_{-\pi}^{\pi}dx\left(\cos(n+\mu)x+ \cos(n-\mu)x\right)\nonumber\\
&=& \frac{1}{2\sqrt{\pi}} \left(\frac{ \sin(n+\mu)x}{n+\mu}+ \frac{\sin(n-\mu)x}{n-\mu} \right)\Bigg|_{-\pi}^{\pi}\nonumber\\
&=& \frac{1}{\sqrt{\pi}} \left(\frac{ \sin(n+\mu)\pi}{n+\mu}+ \frac{\sin(n-\mu)\pi}{n-\mu} \right)\nonumber\\
&=&\frac{1}{\sqrt{\pi}} \left(\frac{ \sin \mu \pi \cos n\pi+\sin n\pi \cos \mu \pi }{n+\mu}+ \frac{\sin n \pi \cos \mu \pi-\sin \mu \pi \cos n\pi}{n-\mu} \right)\nonumber\\
&=&  \frac{(-)^{n}\sin\mu \pi}{\sqrt{\pi}} \left(\frac{1}{n+\mu}- \frac{1}{n-\mu} \right)= \frac{2(-)^{n+1}\sin\mu \pi}{\sqrt{\pi}} \left(\frac{\mu}{n^{2}-\mu^{2}}\right)\nonumber\\
& =&\frac{2(-)^{n}\sin\mu \pi}{\sqrt{\pi}} \left(\frac{\mu}{\mu^{2}-n^{2}}\right).\nonumber
\end{eqnarray}
Por lo tanto,
\begin{eqnarray}
\cos(\mu x)&=& \frac{2 \sin \mu \pi}{\mu \sqrt{2\pi}} \frac{1}{\sqrt{2\pi}}+\sum_{n\geq 1} \frac{2(-)^{n}\sin\mu \pi}{\sqrt{\pi}} \left(\frac{\mu}{\mu^{2}-n^{2}}\right) \frac{\cos nx}{\sqrt{\pi}}\nonumber\\
&= &\frac{2 \mu \sin \mu \pi}{\pi} \left( \frac{1}{2\mu^{2}}+\sum_{n\geq 1} \frac{(-)^{n}\cos n x}{\mu^{2}-n^{2}}\right),
\end{eqnarray}
de este resultado se obtiene
\begin{eqnarray}
\frac{\cos\mu x}{\sin\mu \pi} &= &\frac{1}{\pi} \left( \frac{1}{\mu}+\sum_{n\geq 1} \frac{\mu 2(-)^{n}\cos n x}{\mu^{2}-n^{2}}\right). 
\end{eqnarray}
Para el caso particular $x=\pi,$ se llega a 
\begin{eqnarray}
\frac{\cos\mu \pi}{\sin\mu \pi} = \frac{1}{\pi} \left( \frac{1}{\mu}+\sum_{n\geq 1} \frac{2\mu }{\mu^{2}-n^{2}}\right), 
\end{eqnarray}
de donde 
\begin{eqnarray}
\left(\frac{\cos\mu \pi}{\sin\mu \pi}-\frac{1}{\pi \mu} \right)\pi  =\sum_{n\geq 1} \frac{2\mu }{\mu^{2}-n^{2}}. \label{eq:foucosigua1}
\end{eqnarray}
Ahora, note que 
\begin{eqnarray}
\int_{0}^{x} \left(\frac{\cos\mu \pi}{\sin\mu \pi}-\frac{1}{\pi \mu}\right) \pi d\mu &=& 
\int_{0}^{\pi x} \left(\frac{\cos u}{\sin u }-\frac{1}{u}\right)  du \nonumber\\
&=&\int_{0}^{\pi x} \frac{d}{du} \left(\ln \sin u-\ln u\right) \nonumber\\
&=&\int_{0}^{\pi x} \frac{d}{du} \ln \left(\frac{\sin u}{u}\right)\nonumber\\
&=&\ln \left(\frac{\sin \pi x}{\pi x}\right). \label{eq:foucosigua2}
\end{eqnarray}
Tambi\'en se obtiene
\begin{eqnarray}
\int_{0}^{x} \sum_{n\geq 1} \frac{2\mu }{\mu^{2}-n^{2}} d\mu &=&\sum_{n\geq 1}
\int_{0}^{x} \left(\frac{-2}{n^{2}}\right)\frac{d\mu } {1- \frac{\mu^{2}} {n^{2}}  } =
\sum_{n\geq 1} \int_{0}^{x}\frac{d}{d\mu}\ln  \left(1- \frac{\mu^{2}} {n^{2}}\right)\nonumber\\
&=&
\sum_{n\geq 1} \ln  \left(1- \frac{\mu^{2}} {n^{2}}\right)\Bigg|_{0}^{x}=\sum_{n\geq 1} \ln  \left(1- \frac{x^{2}} {n^{2}}\right)\nonumber\\
&=& \ln \Pi_{n=1}^{\infty}\left(1-\frac{x^{2}}{n^{2}}\right). \label{eq:foucosigua3}
\end{eqnarray}
La expresi\'on Eq. (\ref{eq:foucosigua1}) nos indica que Eq.   (\ref{eq:foucosigua2}) debe ser igual a Eq. (\ref{eq:foucosigua3}). De donde  se obtiene 
\begin{eqnarray}
\ln \left(\frac{\sin \pi x}{\pi x}\right)= \ln \Pi_{n=1}^{\infty}\left(1-\frac{x^{2}}{n^{2}}\right),
\end{eqnarray}
es decir
\begin{eqnarray}
\frac{\sin \pi x}{\pi x}= \Pi_{n=1}^{\infty}\left(1-\frac{x^{2}}{n^{2}}\right).
\end{eqnarray}
Definiendo $z=\pi x,$ se llega a 
\begin{eqnarray}
\frac{\sin z}{z}= \Pi_{n=1}^{\infty}\left[1-\left(\frac{z}{n\pi}\right)^{2}\right].
\end{eqnarray}
Esta igualdad la dedujo por primera vez Euler. Sorprendentemente Feynman encontr\'o 
que esta expresi\'on tiene aplicaciones en mec\'anica cu\'antica \cite{in-trayectoria:gnus} 

\subsection{Funci\'on $f(x)=e^{x}$}

Para la funci\'on $f(x)=e^{x}$ se tiene
\begin{eqnarray}
a_{0}=\frac{1}{\sqrt{2\pi}}\int_{-\pi}^{\pi} dx e^{x}=\frac{e^{\pi}-e^{-\pi}}{\sqrt{\pi}}=\frac{2 \sinh \pi}{\sqrt{2\pi}}.
\end{eqnarray}
Adem\'as 
\begin{eqnarray}
a_{n}=\frac{1}{\sqrt{\pi}}\int_{-\pi}^{\pi} dx e^{x}\cos n x, \qquad b_{n}=\frac{1}{\sqrt{\pi}}\int_{\pi}^{\pi} dx e^{x}\sin nx,
\end{eqnarray}
de donde 
\begin{eqnarray}
a_{n}+ib_{n}&=&\frac{1}{\sqrt{\pi}}\int_{-\pi}^{\pi} dx e^{x}\left(\cos n x+i \sin nx\right)=\frac{1}{\sqrt{\pi}}\int_{-\pi}^{\pi} dx e^{x(1+in)}\nonumber\\
&=&
\frac{1}{\sqrt{\pi}} \frac{e^{x(1+in)}}{1+in} \Bigg|_{-\pi}^{\pi}\nonumber\\
&=& \frac{1}{\sqrt{\pi}} \frac{e^{\pi}e^{in\pi } - e^{-\pi}e^{-in\pi }}{1+in}= \frac{1-in}{\sqrt{\pi} (1+n^{2})} \left(e^{\pi}(-)^{n} -(-)^{n} e^{-\pi}\right)\nonumber\\
& =& \frac{(1-in)(-)^{n} }{\sqrt{\pi} (1+n^{2})} \left(e^{\pi} -e^{-\pi}\right)= \frac{(1-in)(-)^{n} }{\sqrt{\pi} (1+n^{2})} 2\sinh\pi. 
\end{eqnarray}
Por lo tanto, 
\begin{eqnarray}
a_{n}=\frac{2(-)^{n}\sinh\pi  }{\sqrt{\pi} (1+n^{2})} ,\qquad 
b_{n}=\frac{2n(-)^{n+1} \sinh\pi}{\sqrt{\pi} (1+n^{2})},\nonumber
\end{eqnarray}
que implican 
\begin{eqnarray}
e^{x}= \frac{ \sinh \pi}{\pi}+ \frac{2\sinh \pi}{\pi} \sum_{n \geq 1} \left( \frac{(-)^{n} \cos n x}{1+n^{2}}+  \frac{n(-)^{n+1} \sin n x}{1+n^{2}}\right).
\end{eqnarray}
De esta serie se encuentran las series 
\begin{eqnarray}
e^{-x}&=& \frac{ \sinh \pi}{\pi}\left[1+2 \sum_{n \geq 1} \left( \frac{(-)^{n} \cos n x}{1+n^{2}}+  \frac{n(-)^{n} \sin n x}{1+n^{2}}\right)\right],\nonumber\\
\cosh x &=&\frac{e^{x}+e^{-x}}{2}= \frac{ \sinh \pi}{\pi}\left[1+2 \sum_{n \geq 1}  \frac{(-)^{n} \cos n x}{1+n^{2}}\right],\nonumber\\
\sinh x &=& \frac{e^{x}-e^{-x}}{2}= \frac{ 2\sinh \pi}{\pi} \sum_{n \geq 1}  \frac{(-)^{n+1} \sin n x}{1+n^{2}}.\nonumber
\end{eqnarray}
\section{Serie tipo coseno}

Para  el espacio de  funciones el intervalo donde \'estas son definidas es muy importante. Por ejemplo, en el intervalo $[0,\pi]$ el conjunto 
(\ref{eq:base-fourier}) ya no es ortonormal. En este intervalo consideraremos el conjunto de funciones
\begin{eqnarray}
\{\psi(x)\}= \left\{\frac{1}{\sqrt{\pi}}, \sqrt{\frac{2}{\pi}}\cos nx\right \}.
\end{eqnarray}
Ocupando las identidades Eqs. (\ref{eq:sum5})-(\ref{eq:sum7}), se encuentra 
\begin{eqnarray}
\int_{0}^{\pi}dx\left(\frac{1}{\sqrt{\pi}}\right)^{2}&=&1,\\
\int_{0}^{\pi}dx\frac{1}{\sqrt{\pi}}\sqrt{\frac{2}{\pi}}\cos nx
&=&0,\\
\int_{0}^{\pi}dx \sqrt{\frac{2}{\pi}}\cos nx \sqrt{\frac{2}{\pi}}\cos mx 
&=&\delta_{nm}.
\end{eqnarray}
Entonces, si $f$ es una funci\'on en el intervalo $[0,\pi],$ se puede expresar como
\begin{eqnarray}
f(x)=\frac{a_{0}}{\sqrt{\pi}}+\sum_{n\geq 1} a_{n}  \sqrt{\frac{2}{\pi}}\cos nx,
\end{eqnarray}
con los coeficientes de Fourier dados por
\begin{eqnarray}
a_{0}&=& \int_{0}^{\pi} \frac{dx}{\sqrt{\pi}} f(x),\nonumber\\
 a_{n}&=&\int_{0}^{\pi}dx   \sqrt{\frac{2}{\pi}}\cos nx f(x). \nonumber
\end{eqnarray}
Por ejemplo, para la funci\'on $f(x)=x$ se tienen 
\begin{eqnarray}
a_{0}&=& \int_{0}^{\pi} \frac{dx}{\sqrt{\pi}} x=\frac{1}{\sqrt{\pi}}\frac{x^{2}}{2}\Bigg|_{0}^{\pi}= \frac{\pi^{2}}{2\sqrt{\pi}} ,\nonumber
\end{eqnarray}
adem\'as
\begin{eqnarray}
a_{n}&=&\int_{0}^{\pi}dx   \sqrt{\frac{2}{\pi}} x\cos nx =\sqrt{\frac{2}{\pi}}\frac{1}{n} \int_{0}^{\pi}dx   x\frac{d}{dx}\sin nx\nonumber\\
&=&\sqrt{\frac{2}{\pi}}\frac{1}{n} \int_{0}^{\pi}dx   \left(\frac{d}{dx} \left(x\sin nx\right)- \sin nx\right)=
\sqrt{\frac{2}{\pi}}\frac{1}{n^{2}}\int_{0}^{\pi}dx   \frac{d}{dx}\cos nx\nonumber\\
&=& \sqrt{\frac{2}{\pi}}\frac{1}{n^{2}}\left[(-)^{n}-1\right],
  \nonumber
\end{eqnarray}
de donde $a_{2n}=0$ y 
\begin{eqnarray}
a_{2n-1}&=&- \sqrt{\frac{2}{\pi}}\frac{2}{(2n-1)^{2}}.
  \nonumber
\end{eqnarray}
Por lo tanto,
\begin{eqnarray}
x=\frac{\pi}{2}- \frac{4}{\pi} \sum_{n\geq 1} \frac{\cos (2n-1)x}{(2n-1)^{2}}.
\end{eqnarray}

\section{Serie tipo seno}

En un mismo intervalo puede haber diferentes base para el espacio de funciones. Por ejemplo, usando las identidades Eqs. (\ref{eq:sum5})-(\ref{eq:sum7}), se puede mostrar que se cumple 
\begin{eqnarray}
\int_{0}^{\pi}dx \sqrt{\frac{2}{\pi}}\sin nx \sqrt{\frac{2}{\pi}}\sin mx 
&=&\delta_{nm}.
\end{eqnarray}
Por lo tanto, el conjunto de funciones
\begin{eqnarray}
\{\psi(x)\}= \left\{ \sqrt{\frac{2}{\pi}}\sin nx\right \}
\label{eq:fou-sin-base}
\end{eqnarray}
es ortonormal en el intervalo $[0,\pi].$ 
\\

Entonces, si $f$ es una funci\'on en el intervalo $[0,\pi],$ se encuentra 
\begin{eqnarray}
f(x)=\sum_{n\geq 1} b_{n}  \sqrt{\frac{2}{\pi}}\sin nx,
\end{eqnarray}
en este caso los coeficientes de Fourier son 
\begin{eqnarray}
 b_{n}=\int_{0}^{\pi}dx   \sqrt{\frac{2}{\pi}}\sin nx f(x).
\end{eqnarray}
Por ejemplo, para la funci\'on constante $f(x)=C$ se tienen los coeficientes de Fourier
\begin{eqnarray}
b_{n}= C\sqrt{\frac{2}{\pi}} \int_{0}^{\pi}dx  \sin nx=  -\frac{C}{n}\sqrt{\frac{2}{\pi}} \cos n x \Bigg|_{0}^{\pi}= 
-\frac{C}{n}\sqrt{\frac{2}{\pi}}\left[(-)^{n}-1\right],\nonumber
\end{eqnarray}
de donde $b_{2n}=0$ y 
\begin{eqnarray}
b_{2n-1}=\frac{2C}{2n-1}\sqrt{\frac{2}{\pi}}.\nonumber
\end{eqnarray}
As\'i,
\begin{eqnarray}
C= \frac{4C}{\pi} \sum_{n\geq 1} \frac{\sin (2n-1)x}{2n-1}.
\end{eqnarray}

\section{Intervalo arbitrario}

Ahora estudiaremos el espacio de funciones en el intervalo $[-L,L].$ En este intervalo 
mostraremos que  el conjunto de funciones
\begin{eqnarray}
\left\{ \psi(x)\right\} =\left\{ \frac{1}{\sqrt{2L}}, \frac{1}{\sqrt{L}} \cos \frac{n\pi x}{L},
\frac{1}{\sqrt{L}} \sin \frac{n\pi x}{L} \right\}, \quad n=1,2,3 \cdots , \label{eq:base-fourier-L}
\end{eqnarray}
es ortonormal. Primero notemos que 
\begin{eqnarray}
\int_{-L}^{L}dx \left(\frac{1}{\sqrt{2L}}\right)^{2}&=&1, \label{eq:base-fourier-L-1}\\ 
\int_{-L}^{L}dx\frac{1}{\sqrt{2L}}  \frac{1}{\sqrt{L}} \sin \frac{n\pi x}{L} &=&0,\\
\int_{-L}^{L}dx\frac{1}{\sqrt{2L}}  \frac{1}{\sqrt{L}} \cos \frac{n\pi x}{L} &=&0.\label{eq:base-fourier-L-2} 
\end{eqnarray}
Adem\'as, con el cambio de variable
\begin{eqnarray}
u=\frac{\pi x}{L} \label{eq:fourier-cambiodevariable}
\end{eqnarray}
se encuentra 
\begin{eqnarray}
\int_{-L}^{L}dx f\left( \frac{\pi x}{L}\right)= \frac{L}{\pi}\int_{-\pi}^{\pi}du f(u) 
\label{eq:cambiodevariable-fourier}.
\end{eqnarray}
Para este caso es conveniente tomar las siguientes definiciones
\begin{eqnarray}
\left<f\left( \frac{\pi x}{L}\right)\Bigg|g\left( \frac{\pi x}{L}\right)\right> _{L}&=& 
\int_{-L}^{L}dx  f^{*}\left( \frac{\pi x}{L}\right)g\left( \frac{\pi x}{L}\right), \nonumber\\
\left<f(x)|g(x)\right> _{\pi}&=& \int_{-\pi}^{\pi}dx  f^{*}(x)g(x).
\end{eqnarray}
Ocupando el resultado (\ref{eq:cambiodevariable-fourier}) se encuentra 
\begin{eqnarray}
\left<\frac{1}{\sqrt{L}} f\left( \frac{\pi x}{L}\right)\Bigg|\frac{1}{\sqrt{L}} g\left( \frac{\pi x}{L}\right)\right> _{L}= 
\left<\frac{1}{\sqrt{\pi}} f(u)\Bigg |\frac{1}{\sqrt{\pi}}g(u)\right> _{\pi}.
\end{eqnarray}
Entonces, utilizando las ecuaciones   (\ref{eq:fourier1pi})-(\ref{eq:fourier2pi}) se llega a
\begin{eqnarray}
\left<\frac{1}{\sqrt{L}} \sin\left( \frac{n\pi x}{L}\right)\Bigg|\frac{1}{\sqrt{L}} \sin\left( \frac{m\pi x}{L}\right)\right> _{L}&=&\delta_{nm} ,\nonumber\\
\left<\frac{1}{\sqrt{L}} \cos\left( \frac{n\pi x}{L}\right)\Bigg|\frac{1}{\sqrt{L}} \cos\left( \frac{m\pi x}{L}\right)\right> _{L}&=&\delta_{nm} 
,\nonumber\\
\left<\frac{1}{\sqrt{L}} \sin\left( \frac{n\pi x}{L}\right)\Bigg|\frac{1}{\sqrt{L}} \cos\left( \frac{m\pi x}{L}\right)\right> _{L}&=&0.
\end{eqnarray}
Estas ecuaciones junto con Eqs. (\ref{eq:base-fourier-L-1})-(\ref{eq:base-fourier-L-1}) 
nos indican que el conjunto de funciones Eq. (\ref{eq:base-fourier-L}) forman un conjunto de funciones ortornales 
en el intervalo $[-L,L].$\\

Este resultado implica que  si $f$ es una funci\'on continua en el intervalo $[-L,L],$ se puede escribir como
\begin{eqnarray}
f(x)= \frac{ a_{0} }{\sqrt{2L}}+\sum_{n\geq 1} \left( \frac{a_{n}}{\sqrt{L}} \cos \frac{n\pi x}{L}+
\frac{b_{n} }{\sqrt{L}} \sin \frac{n\pi x}{L} \right) , \label{eq:serie-fourier-L}
\end{eqnarray}
aqu\'i  los coeficientes de Fourier son
\begin{eqnarray}
a_{0}&=&\left<\frac{1}{\sqrt{2L}} \Bigg|f(x)\right> _{L}=\int_{-L}^{L}dx\frac{1}{\sqrt{2L}}f(x),\label{eq:coeficiente0-fourier-L} \\
a_{n}&=&\left<\frac{1}{\sqrt{L}} \cos\left( \frac{n\pi x}{L}\right)\Bigg|f(x)\right> _{L}= \int_{-L}^{L} dx \frac{1}{\sqrt{L}} \cos\left( \frac{n\pi x}{L}\right)  f(x)
,  \label{eq:coeficientea-fourier-L} \\
b_{n} &=&\left<\frac{1}{\sqrt{L}} \sin\left( \frac{n\pi x}{L}\right)\Bigg|f(x)\right> _{L}=\int_{-L}^{L} dx \frac{1}{\sqrt{L}} \sin\left( \frac{n\pi x}{L}\right)  f(x) .  \label{eq:coeficienteb-fourier-L} 
\end{eqnarray}

\subsection{ Delta de Dirac}

Ahora veremos que las series de Fourier permiten generalizar la delta de Kronecker, esta generalizaci\'on se llama delta de Dirac y es de utilidad para resolver problemas de mec\'anica cu\'antica y electromagnetismos.\\

Sustituyendo los coeficientes de Fourier Eqs. (\ref{eq:coeficiente0-fourier-L})-(\ref{eq:coeficienteb-fourier-L}) en Eq. (\ref{eq:serie-fourier-L}) se encuentra
\begin{eqnarray}
f(x)&=& \int_{-L}^{L}dx^{\prime} \frac{f(x^{\prime})}{2L}+ 
\sum_{n\geq 1} \frac{1}{L} \left[ \int_{-L}^{L}dx^{\prime} f(x^{\prime})\cos\left(\frac{n\pi x^{\prime}}{L}\right)  \cos\left(\frac{n\pi x}{L}\right)\right] \nonumber\\
& & + \sum_{n\geq 1} \frac{1}{L} \left[ \int_{-L}^{L}dx^{\prime} f(x^{\prime})\sin\left(\frac{n\pi x^{\prime}}{L}\right)  \sin\left(\frac{n\pi x}{L}\right)\right],\nonumber\\
& =& \int_{-L}^{L}dx^{\prime} \frac{f(x^{\prime})}{L}\nonumber\\
& &\left[ \frac{1}{2}+  
\sum_{n\geq 1} \left[\cos\left(\frac{n\pi x^{\prime}}{L}\right)  \cos\left(\frac{n\pi x}{L}\right) + \sin\left(\frac{n\pi x^{\prime}}{L}\right)
\sin\left(\frac{n\pi x}{L}\right)\right] \right], \nonumber\\
& = & \int_{-L}^{L}dx^{\prime} \frac{f(x^{\prime})}{L}\left[ \frac{1}{2}+  
\sum_{n\geq 1} \cos\left(\frac{n\pi }{L}(x-x^{\prime}) \right)   \right].
\end{eqnarray}
En el intervalo $[-L,L]$ definiremos la delta de Dirac como
\begin{eqnarray}
\delta\left(x-x^{\prime}\right)= \frac{1}{2L}+ \frac{1}{L} \sum_{n\geq 1} \cos\left(\frac{n\pi }{L}(x-x^{\prime}) \right),   
\end{eqnarray}
de donde
\begin{eqnarray}
f(x)= \int_{-L}^{L}dx^{\prime} f(x^{\prime})\delta\left(x-x^{\prime}\right).
\end{eqnarray}
Note que, bajo una integral,  la delta de Dirac cambia la variable $x^{\prime}$ por $x.$
En ese sentido la delta de Dirac es la generalizaci\'on continua de la delta de Kronecker 
Eq. (\ref{eq:Kronecker}).\\

Ahora, recordemos la identidad trigonom\'etrica 
\begin{eqnarray}
\frac{1}{2}+  \cos\alpha +    \cos2\alpha+\cdots +\cos n\alpha= \frac{\sin\left(n+\frac{1}{2}\right)\alpha}{2\sin\frac{\alpha}{2}}, 
\end{eqnarray}
por lo tanto la delta de Dirac se puede expresar como
\begin{eqnarray}
\delta\left(x-x^{\prime}\right)= \lim_{n\to \infty}   \frac{\sin\left[\left(n+\frac{1}{2}\right)\frac{\left(x-x^{\prime}\right)\pi}{L}\right] }
{2L \sin \frac{\left(x-x^{\prime}\right)\pi}{2L}}.  
\end{eqnarray}

\section{Serie coseno en el intervalo $[0,L]$}

Ahora, mostraremos que  el conjunto de funciones
\begin{eqnarray}
\left\{ \psi(x)\right\} =\left\{ \frac{1}{\sqrt{L}}, \sqrt{\frac{2}{L}} \cos \frac{n\pi x}{L} 
\right\}, \quad n=1,2,3 \cdots , \label{eq:base-fourier-0L}
\end{eqnarray}
es ortonormal en el intervalo $[0,L].$ Primero notemos que con el cambio de variable 
Eq. (\ref{eq:cambiodevariable-fourier}) se encuentra 
\begin{eqnarray}
\int_{0}^{L}dx f\left( \frac{\pi x}{L}\right)= \frac{L}{\pi}\int_{0}^{\pi}du f(u) 
\label{eq:cambiodevariable-fourier0L}.
\end{eqnarray}
Por lo que 

\begin{eqnarray}
\int_{0}^{L}dx \left(\frac{1}{\sqrt{L}}\right)^{2}&=&1, \nonumber \\ 
\int_{0}^{L}dx  \frac{1}{\sqrt{L}}\sqrt{\frac{2}{L}} \cos \frac{n\pi x}{L} &=&0, \nonumber \\
\int_{0}^{L}dx  \sqrt{\frac{2}{L}} \cos \frac{n\pi x}{L} \sqrt{\frac{2}{L}}\cos \frac{m\pi x}{L} &=&  \int_{0}^{\pi}dx  \sqrt{\frac{2}{\pi}} \cos nx \sqrt{\frac{2}{\pi}}\cos m x= \delta_{nm}.\nonumber
\end{eqnarray}
As\'i,  el conjunto de funciones Eq. (\ref{eq:base-fourier-0L}) forman un conjunto de funciones ortornales 
en el intervalo $[0,L].$\\

Entonces, si $f$ es una funci\'on continua en el intervalo $[0,L],$ se puede escribir como
\begin{eqnarray}
f(x)= \frac{ a_{0} }{\sqrt{L}}+\sum_{n\geq 1} a_{n}  \sqrt{\frac{2}{L}} \cos \frac{n\pi x}{L}
\label{eq:serie-fourier-0L}
\end{eqnarray}
con los coeficientes de Fourier dados por 
\begin{eqnarray}
a_{0}&=&\int_{0}^{L}\frac{1}{\sqrt{L}}f(x),\label{eq:coeficiente0-fourier-0L}  \\
a_{n}&=& \int_{0}^{L}dx \sqrt{\frac{2}{L}}  \cos\left( \frac{n\pi x}{L}\right)  f(x). \label{eq:coeficientes-fourier-0L}
\end{eqnarray}

\subsection{Delta de Dirac}

Sustituyendo los coeficientes de Fourier Eqs. (\ref{eq:coeficiente0-fourier-0L})-(\ref{eq:coeficientes-fourier-0L})  en 
Eq. (\ref{eq:serie-fourier-0L})
se encuentra
\begin{eqnarray}
f(x)&=& \int_{0}^{L}dx^{\prime} \frac{f(x^{\prime})}{L}+ 
\sum_{n\geq 1} \frac{2}{L} \left( \int_{0}^{L}dx^{\prime} f(x^{\prime})\cos\left(\frac{n\pi x^{\prime}}{L}\right)  \cos\left(\frac{n\pi x}{L}\right)\right) \nonumber\\
& =& \int_{0}^{L}dx^{\prime} \frac{f(x^{\prime})}{L}
\left[ 1+ 2 \sum_{n\geq 1} \cos\left(\frac{n\pi x^{\prime}}{L}\right)  \cos\left(\frac{n\pi x}{L}\right) \right].
\end{eqnarray}
En el intervalo $[0,L],$  con la base (\ref{eq:base-fourier-0L}) ,  definiremos la delta de Dirac como
\begin{eqnarray}
\delta\left(x-x^{\prime}\right)= \frac{1}{L}+ \frac{2}{L} \sum_{n\geq 1} \cos\left(\frac{n\pi x^{\prime}}{L}\right)  \cos\left(\frac{n\pi x}{L}\right), 
\end{eqnarray}
por lo tanto
\begin{eqnarray}
f(x)= \int_{0}^{L}dx^{\prime} f(x^{\prime})\delta\left(x-x^{\prime}\right).
\end{eqnarray}

\section{Serie seno en el intervalo $[0,L]$}

Ahora estudiaremos el conjunto de funciones
\begin{eqnarray}
\left\{ \psi(x)\right\} =\left\{ \sqrt{\frac{2}{L}} \sin \frac{n\pi x}{L} 
\right\}, \quad n=1,2,3 \cdots , \label{eq:base-fourier-seno0L}
\end{eqnarray}
mostraremos que este conjunto de funciones es ortonormal en el intervalo $[0,L].$ Ocupando el cambio de variable 
Eq. (\ref{eq:cambiodevariable-fourier}) y la igualdad Eq. (\ref{eq:cambiodevariable-fourier0L}), se encuentra
\begin{eqnarray}
\int_{0}^{L}dx  \sqrt{\frac{2}{L}} \sin \frac{n\pi x}{L} \sqrt{\frac{2}{L}}\sin \frac{m\pi x}{L} &=&  \int_{0}^{\pi}dx  \sqrt{\frac{2}{\pi}} \sin nx \sqrt{\frac{2}{\pi}}\sin m x= \delta_{nm}.\nonumber
\end{eqnarray}
As\'i,  el conjunto de funciones Eq. (\ref{eq:base-fourier-seno0L}) forman un conjunto de funciones ortornales 
en el intervalo $[0,L].$\\

Entonces, si $f$ es una funci\'on continua en el intervalo $[0,L],$ se puede escribir como
\begin{eqnarray}
f(x)= \sum_{n\geq 1} b_{n}  \sqrt{\frac{2}{L}} \sin \frac{n\pi x}{L} \label{eq:serie-fourier-seno0L}
\end{eqnarray}
con los coeficientes de Fourier dados por 
\begin{eqnarray}
b_{n}&=& \int_{0}^{L} dx \sqrt{\frac{2}{L}}  \sin\left( \frac{n\pi x}{L}\right)  f(x).\label{eq:coeficientes-fourier-seno0L}
\end{eqnarray}

\subsection{Delta de Dirac}

Sustituyendo los coeficientes de Fourier Eq. (\ref{eq:coeficientes-fourier-seno0L}) en Eq. (\ref{eq:serie-fourier-seno0L}) se encuentra
\begin{eqnarray}
f(x)&=& \int_{0}^{L}dx^{\prime} f\left(x^{\prime}\right)
\sum_{n\geq 1} \frac{2}{L} \left( \int_{0}^{L}dx^{\prime} f(x^{\prime})\sin\left(\frac{n\pi x^{\prime}}{L}\right)  \sin\left(\frac{n\pi x}{L}\right)\right) \nonumber\\
& =& \int_{0}^{L}dx^{\prime} f(x^{\prime})
\left[\frac{ 2}{L} \sum_{n\geq 1} \sin\left(\frac{n\pi x^{\prime}}{L}\right)  \sin\left(\frac{n\pi x}{L}\right) \right].
\end{eqnarray}
En el intervalo $[0,L],$  con la base  (\ref{eq:base-fourier-seno0L}),  definiremos la delta de Dirac como
\begin{eqnarray}
\delta\left(x-x^{\prime}\right)=\frac{2}{L} \sum_{n\geq 1} \sin\left(\frac{n\pi x^{\prime}}{L}\right)  \sin\left(\frac{n\pi x}{L}\right), 
\end{eqnarray}
por lo tanto
\begin{eqnarray}
f(x)= \int_{0}^{L}dx^{\prime} f(x^{\prime})\delta\left(x-x^{\prime}\right).
\end{eqnarray}

\section{Representaci\'on compleja}

En el intervalo $[-L,L]$ tambi\'en podemos usar  el conjunto de funciones
\begin{eqnarray}
\left\{ \psi_{n}(x)\right\} =\left\{  \frac{e^{i \frac{n\pi x}{L}}}{\sqrt{2L}}\right\}, \qquad n=0, \pm 1,\pm 2, \pm 3, \cdots . 
\label{eq:representacion-compleja-fourier}
\end{eqnarray}
Note que si $n\not =m$
\begin{eqnarray}
\int_{-L}^{L}  dx \psi_{n}^{*}(x)\psi_{m}(x)&=&\int_{-L}^{L} \frac{e^{i \frac{(m-n)\pi x}{L}} }{2L}= \frac{L}{i(m-n)\pi} \frac{e^{i(m-n)\pi } - e^{-i(m-n)\pi } }{2L}\nonumber\\
&=&\frac{L}{i(n-m)\pi} \frac{(-)^{n-m} - (-)^{m-n} }{2L}=0,
\end{eqnarray}
si $m=n$ se cumple 
\begin{eqnarray}
\int_{-L}^{L} dx\psi_{n}^{*}(x)\psi_{n}(x)=1.
\end{eqnarray}
As\'i, el conjunto de funciones  Eq. (\ref{eq:representacion-compleja-fourier}) es  ortonormal.
Entonces, si $f$ es una funci\'on continua en el intervalo $[0,L],$ se puede escribir como
\begin{eqnarray}
f(x)= \sum_{-\infty}^{\infty}  \alpha_{n}  \frac{e^{i \frac{n\pi x}{L}}}{\sqrt{2L}}
\label{eq:serie-representacion-compleja-fourier}
\end{eqnarray}
con los coeficientes de Fourier dados por 
\begin{eqnarray}
\alpha_{n}&=& \int_{-L}^{L} dx  \frac{e^{-i \frac{n\pi x}{L}}}{\sqrt{2L}} f(x).
\label{eq:coeficietes-representacion-compleja-fourier}
\end{eqnarray}
Note que si la funci\'on $f$ es real, se cumple 
\begin{eqnarray}
\alpha_{n}^{*}=\alpha_{-n} .
\end{eqnarray}

\subsection{ Delta de Dirac}

Sustituyendo los coeficientes de Fourier Eq. (\ref{eq:coeficietes-representacion-compleja-fourier}) en 
Eq. (\ref{eq:serie-representacion-compleja-fourier}) se encuentra
\begin{eqnarray}
f(x)&=& \sum_{-\infty}^{\infty}  \frac{1}{2L}  \int_{-L}^{L}dx^{\prime} f(x^{\prime}) e^{i\frac{n\pi(x- x^{\prime})}{L}} =\int_{-L}^{L}dx^{\prime} f(x^{\prime})\sum_{-\infty}^{\infty}  \frac{ e^{i\frac{n\pi(x- x^{\prime})}{L}}  }{2L}  .\nonumber
\end{eqnarray}
As\'i, la delta de Dirac en el intervalo $[-L,L]$ toma la forma
\begin{eqnarray}
\delta\left(x-x^{\prime}\right)=\sum_{-\infty}^{\infty}  \frac{ e^{i\frac{n\pi(x- x^{\prime})}{L}}  }{2L}  ,   
\end{eqnarray}
de donde
\begin{eqnarray}
f(x)= \int_{-L}^{L}dx^{\prime} f(x^{\prime})\delta\left(x-x^{\prime}\right).
\end{eqnarray}

\section{ Ecuaci\'on de Laplace en dos dimensiones}

Ahora veremos algunas aplicaciones de las series de Fourier. Primero estudiaremos problemas en dos dimensiones.\\

Usando coordenadas cartesianas,  la ecuaci\'on de Laplace en dos dimensiones es 
\begin{eqnarray}
\nabla^{2}_{2D}\phi(x,y) =\left(\frac{\partial^{2} }{\partial x^{2}}+  \frac{\partial^{2} }{\partial y^{2}}\right)\phi(x,y)=0.
\label{eq:Flaplace2d}
\end{eqnarray}
Para resolver esta ecuaci\'on propondremos 
\begin{eqnarray}
\phi(x,y)=X(x)Y(y),
\end{eqnarray}
sustituyendo esta propuesta en Eq. (\ref{eq:Flaplace2d}) se encuentra
\begin{eqnarray}
\nabla^{2}_{2D}\phi(x,y) =Y(y)\frac{\partial^{2} X(x)}{\partial x^{2}}+  X(x) \frac{\partial^{2}Y(y) }{\partial y^{2}}=0,
\end{eqnarray}
de donde 
\begin{eqnarray}
\frac{ \nabla^{2}_{2D}\phi(x,y)}{\phi(x,y)} =\frac{1}{X(x)} \frac{\partial^{2} X(x)}{\partial x^{2}}+   \frac{1}{Y(y)}\frac{\partial^{2}Y(y) }{\partial y^{2}}=0.\label{eq:Flaplace2d-1}
\end{eqnarray}
Por lo tanto, 
\begin{eqnarray}
\frac{ \partial }{\partial x} \left(\frac{ \nabla^{2}_{2D}\phi(x,y)}{\phi(x,y)}\right) =\frac{ \partial }{\partial x} \left( \frac{1}{X(x)} \frac{\partial^{2} X(x)}{\partial x^{2}}\right) =0,
\end{eqnarray}
que implica
\begin{eqnarray}
 \frac{1}{X(x)} \frac{\partial^{2} X(x)}{\partial x^{2}} =-\alpha^{2},\quad  \frac{\partial^{2} X(x)}{\partial x^{2}} =-\alpha^{2}X(x),\quad \alpha={\rm constante}. \label{eq:Flaplace2d-2}
\end{eqnarray}
Usando este resultado en Eq. (\ref{eq:Flaplace2d-1}) se llega a 
\begin{eqnarray}
-\alpha^{2}+   \frac{1}{Y(y)}\frac{\partial^{2}Y(y) }{\partial y^{2}}=0,
\end{eqnarray}
es decir
\begin{eqnarray}
\frac{\partial^{2}Y(y) }{\partial y^{2}}=\alpha^{2}Y(y).\label{eq:Flaplace2d-3}
\end{eqnarray}
Si $\alpha=0,$ entonces  Eq. (\ref{eq:Flaplace2d-2}) y  Eq. (\ref{eq:Flaplace2d-3})  toman 
la forma
\begin{eqnarray}
X_{0}(x)=\left( A+Bx\right),\quad Y_{0}(y)=\left( C+Dy\right), \quad A,B,C,D={\rm constante}.\nonumber
\end{eqnarray}
Si $\alpha\not =0$  las soluciones a las ecuaciones (\ref{eq:Flaplace2d-2})  y (\ref{eq:Flaplace2d-3}) son 
\begin{eqnarray}
X_{\alpha}(x)&=&\left( a_{\alpha}\cos\alpha x+b_{\alpha} \sin \alpha x \right),\nonumber\\
 Y_{\alpha}(y)&=&
\left( c_{\alpha} e^{\alpha y} +d_{\alpha} e^{-\alpha y} \right), \quad a_{\alpha},b_{\alpha}, c_{\alpha},d_{\alpha}={\rm constante}.\nonumber
\end{eqnarray}
As\'i, en coordenadas cartesianas, las soluciones generales de la ecuaci\'on de Laplace en dos dimensiones  son
\begin{eqnarray}
\phi_{0}(x,y)&=& \left( A+Bx\right)\left( C+Dy\right),\\
\phi_{\alpha}(x,y)&=&\left( a_{\alpha}\cos\alpha x+b_{\alpha} \sin \alpha x \right)
\left( c_{\alpha} e^{\alpha y} +d_{\alpha} e^{-\alpha y} \right).
\end{eqnarray}
\subsection{Ejemplo}

Suponga que tiene un sistema en dos dimensiones cuyo potencial el\'ectrico satisface las condiciones de borde
\begin{eqnarray}
\phi(x,0)&=&V, \qquad V={\rm constante}, \label{eq:12pisson}\\
\phi(x,\infty )&=&0, \label{eq:22pisson} \\
\phi(0,y)&=&\phi(L,y)=0. \label{eq:32pisson}
\end{eqnarray}
Encontrar el potencial el\'ectrico en todo el espacio y  mostrar que se puede escribir como
\begin{eqnarray}
\phi (x,y)= \frac{2V}{\pi} \tan^{-1}\left( \frac{ \sin \frac{2\pi x}{L}  }{\sinh \frac{\pi y}{L} }\right) .
\end{eqnarray}

Primero notemos que la condici\'on de borde Eq. (\ref{eq:32pisson})
\begin{eqnarray}
\phi_{0}(0,y)&=&A\left( C+Dy\right)=0
\end{eqnarray}
implica $A=0.$ As\'i,  $\phi_{0}(x,y)=Bx\left( C+Dy\right).$ Adem\'as, si pedimos que
\begin{eqnarray}
\phi_{0}(L,y)&=&BL\left( C+Dy\right)=0,
\end{eqnarray}
se encuentra que $B=0.$ Por lo tanto $\phi_{0}(x,y)=0.$ De las misma condici\'on de borde  Eq. (\ref{eq:32pisson})
se puede ver que 
\begin{eqnarray}
\phi_{\alpha}(0,y)&=&a_{\alpha}\left( c_{\alpha} e^{\alpha y} +d_{\alpha} e^{-\alpha y} \right) =0
\end{eqnarray}
entonces $a_{\alpha}=0.$ As\'i,  $\phi_{\alpha}(x,y)=b_{\alpha}\sin\alpha x\left( c_{\alpha} e^{\alpha y} +d_{\alpha} e^{-\alpha y} \right).$ 
Adem\'as, si pedimos que
\begin{eqnarray}
\phi(L,y)&=& b_{\alpha}\sin\alpha L\left( c_{\alpha} e^{\alpha y} +d_{\alpha} e^{-\alpha y} \right)=0
\end{eqnarray}
se debe cumplir
\begin{eqnarray}
\sin\alpha L=0,
\end{eqnarray}
que se satisface  si 
\begin{eqnarray}
\alpha=\frac{n\pi}{ L}, \quad n=1,2,3,\cdots .
\end{eqnarray}
De donde, las soluciones deben ser de la forma
\begin{eqnarray}
\phi_{n}(x,y)=\left( c_{n} e^{\frac{n\pi }{ L}  y} +d_{n} e^{- \frac{n\pi }{ L} y}\right) \sqrt{\frac{2}{L}}\sin\frac{n\pi }{ L}x.
\end{eqnarray}
Ahora, se puede observar que la  condici\'on de borde Eq. (\ref{eq:22pisson}) implica $ c_{n}=0.$ As\'i, las soluciones deben ser de la forma
\begin{eqnarray}
\phi_{n}(x,y)= a_{n}  e^{- \frac{n\pi }{ L} y} \sqrt{\frac{2}{L}}\sin\frac{n\pi }{ L}x.
\end{eqnarray}
Por lo tanto, las soluci\'on m\'as general que cumple las condiciones de borde Eqs. (\ref{eq:22pisson})-(\ref{eq:32pisson}) 
es
\begin{eqnarray}
\phi (x,y)= \sum_{n\geq 1} a_{n}  e^{- \frac{n\pi }{ L} y} \sqrt{\frac{2}{L}}\sin\frac{n\pi }{ L}x.
\label{eq:e1pfourier}
\end{eqnarray}
Para cumplir la condici\'on de borde Eq. (\ref{eq:12pisson}) se debe pedir que 
\begin{eqnarray}
\phi (x,0)=V =\sum_{n\geq 1} a_{n}  \sqrt{\frac{2}{L}}\sin\frac{n\pi }{ L}x,
\end{eqnarray}
que se cumple siempre y cuando 
\begin{eqnarray}
a_{n}&=&\sqrt{\frac{2}{L}}\int_{0}^{L} dx  V \sin\frac{n\pi }{ L}x= \sqrt{\frac{2}{L}}V\int_{0}^{L} dx  \sin\frac{n\pi }{ L}x\nonumber\\
&=& \sqrt{\frac{2}{L}}V \frac{L}{n\pi}(-) \int_{0}^{L} dx  \frac{d}{dx}\cos\frac{n\pi }{ L}x\nonumber \\
&=& \sqrt{\frac{2}{L}}V \frac{L}{n\pi}(-) \cos\frac{n\pi }{ L}x\Bigg|_{0}^{L}= \sqrt{\frac{2}{L}}V \frac{L}{n\pi}(-)\left[(-)^{n}-1\right],
\end{eqnarray}
de donde
\begin{eqnarray}
a_{2n-1}=\sqrt{\frac{2}{L}}\frac{2VL}{(2n-1)\pi},\qquad a_{2n}=0.
\end{eqnarray}
Sustituyendo estos resultados en Eq. (\ref{eq:e1pfourier}), se obtiene 
\begin{eqnarray}
\phi (x,y)= \frac{4V}{\pi} \sum_{n\geq 1}  \frac{1}{2n-1} e^{- \frac{(2n-1)\pi }{ L} y } \sin\frac{(2n-1)\pi }{ L}x.
\end{eqnarray}
Ahora, note que 
\begin{eqnarray}
e^{- \frac{(2n-1)\pi }{ L} y } \sin\frac{(2n-1)\pi }{ L}x&=&{\rm Im}\left( e^{- \frac{(2n-1)\pi }{ L} y } e^{i\frac{(2n-1)\pi }{ L}x}\right)=
{\rm Im}\left( e^{\frac{(2n-1)\pi }{ L} (ix-y) } \right)\nonumber\\
& =&{\rm Im}\left( e^{i\frac{(2n-1)\pi }{ L} (x+iy) } \right)= {\rm Im}\left( e^{i\frac{\pi }{ L} (x+iy) } \right)^{(2n-1)}, \nonumber
\end{eqnarray}
por lo tanto, definiendo 
\begin{eqnarray}
\omega=  e^{i\frac{\pi }{ L} (x+iy) } , \nonumber
\end{eqnarray}
se encuentra 
\begin{eqnarray}
\phi (x,y)= \frac{4V}{\pi} {\rm Im}\left(\sum_{n\geq 1}  \frac{\omega^{2n-1}}{2n-1}\right).
\end{eqnarray}
Ahora, recordemos las series 
\begin{eqnarray}
\frac{1}{1-\omega}=\sum_{n\geq 0} \omega^{n},\qquad \frac{1}{1+\omega}=\sum_{n\geq 0} (-)^{n} \omega^{n},
\end{eqnarray}
de las cuales se deduce 
\begin{eqnarray}
\int  \frac{d\omega }{1-\omega}=-\ln(1-\omega)=\sum_{n\geq 0} \frac{\omega^{n+1}}{n+1}=\sum_{n\geq 1} \frac{\omega^{n}}{n},\nonumber\\
\int \frac{d\omega}{1+\omega}=\ln(1+\omega)=\sum_{n\geq 0} \frac{(-)^{n} \omega^{n+1}}{n+1}=\sum_{n\geq 1} (-)^{n+1}\frac{\omega^{n}}{n},
\end{eqnarray}
que a su vez implican 
\begin{eqnarray}
\ln\left(\frac{1+\omega}{1-\omega}\right)&=& \ln(1+\omega)- \ln(1-\omega)=\sum_{n\geq 1} \frac{1+(-)^{n+1} \omega^{n}}{n}\nonumber\\
&=&2\sum_{n\geq 1} \frac{ \omega^{2n-1}}{2n-1}.\label{eq:Fouinde1}
\end{eqnarray}
Entonces
\begin{eqnarray}
\phi (x,y)= \frac{2V}{\pi} {\rm Im}\left[\ln  \left(\frac{1+\omega}{1-\omega}\right) \right].
\end{eqnarray}
Adem\'as, si definimos
\begin{eqnarray}
z=\frac{1+\omega}{1-\omega}=|z|e^{i\theta},\quad \tan \theta= \frac{{\rm Im}(z)}{{\rm Re}(z)},
\end{eqnarray}
se tiene que 
\begin{eqnarray}
\ln z=\ln|z|+ i\theta. 
\end{eqnarray}
As\'i, 
\begin{eqnarray}
\phi (x,y)= \frac{2V}{\pi} \theta.
\end{eqnarray}
Considerando la definici\'on de $\omega,$  tenemos 
\begin{eqnarray}
z&=&\frac{1+\omega}{1-\omega}=\frac{ (1+\omega)(1-\omega^{*}) }{|1-\omega|^{2}}= 
\frac{ 1+\omega-\omega^{*}-\omega\omega^{*}}{|1-\omega|^{2}}\nonumber\\
&=& 
\frac{ 1+e^{i\frac{\pi}{L}(x+iy)} -e^{-i\frac{\pi}{L}(x-iy)}-e^{i\frac{\pi}{L}(x+iy)}e^{-i\frac{\pi}{L}(x-iy)} }{|1-\omega|^{2}}\nonumber\\
&=& \frac{ 1+e^{-\frac{\pi y}{L}} \left(e^{i\frac{\pi}{L}x} -e^{-i\frac{\pi}{L}x}\right)-e^{-\frac{2\pi}{L}} }{|1-\omega|^{2}}\nonumber\\
&=&\frac{ 1- e^{-\frac{2\pi y}{L}} +2i e^{-\frac{\pi y}{L}}\sin \frac{2\pi x}{L} }{|1-\omega|^{2}},
\end{eqnarray}
entonces 
\begin{eqnarray}
\tan \theta= \frac{2 e^{-\frac{\pi y}{L}}\sin \frac{2\pi x}{L}  }{1- e^{-\frac{2\pi y}{L} }}= 
\frac{2 \sin \frac{2\pi x}{L}  }{ e^{\frac{\pi y}{L}} - e^{-\frac{\pi y}{L}} }= \frac{ \sin \frac{2\pi x}{L}  }{\sinh \frac{\pi y}{L} }.
\end{eqnarray}
Por lo tanto,
\begin{eqnarray}
\phi (x,y)= \frac{2V}{\pi} \tan^{-1}\left( \frac{ \sin \frac{2\pi x}{L}  }{\sinh \frac{\pi y}{L} }\right) .
\end{eqnarray}

\section{Ecuaci\'on de Poisson en dos dimensiones con coordenas polares}

En coordenas polares, la ecuaci\'on de Laplace en dos dimensiones es
\begin{eqnarray}
\nabla^{2}_{2D}\phi= \frac{1}{\rho}\frac{\partial  }{\partial \rho}\left(\rho \frac{\partial \phi }{\partial \rho}\right) +
\frac{1}{\rho^{2}} \frac{\partial^{2} \phi }{\partial \varphi^{2}}=0.\label{eq:Flapacepo0}
\end{eqnarray}
Para resolverla propondremos $\phi(\rho,\varphi)=R(\rho)\Psi(\varphi),$ de donde   
\begin{eqnarray}
\nabla^{2}_{2D}\phi= \frac{ \Psi(\varphi)}{\rho}\frac{\partial  }{\partial \rho}\left(\rho \frac{\partial R(\rho) }{\partial \rho}\right) +
\frac{R(\rho) }{\rho^{2}} \frac{\partial^{2} \Psi(\varphi) }{\partial \varphi^{2}}=0,
\end{eqnarray}
que implica
\begin{eqnarray}
\frac{\rho^{2} \nabla^{2}_{2D}\phi}{\phi} = \frac{ \rho}{R(\rho)}\frac{\partial  }{\partial \rho}\left(\rho \frac{\partial R(\rho) }{\partial \rho}\right) +
\frac{1 }{\Psi(\varphi) } \frac{\partial^{2} \Psi(\varphi) }{\partial \varphi^{2}}=0.\label{eq:Flapacepo}
\end{eqnarray}
Por lo tanto, 
\begin{eqnarray}
\frac{\partial }{\partial \varphi}\left(\frac{\rho^{2} \nabla^{2}_{2D}\phi}{\phi}\right) = 
\frac{\partial }{\partial \varphi}\left(
\frac{1 }{\Psi(\varphi) } \frac{\partial^{2} \Psi(\varphi) }{\partial \varphi^{2}}\right)=0,
\end{eqnarray}
as\'i
\begin{eqnarray}
\frac{1 }{\Psi(\varphi) } \frac{\partial^{2} \Psi(\varphi) }{\partial \varphi^{2}}=-\alpha^{2},\quad  \frac{\partial^{2} \Psi(\varphi) }{\partial \varphi^{2}}=-\alpha^{2}\Psi(\varphi) ,\quad \alpha={\rm constate}.
\end{eqnarray}
Sustituyendo este resultado en Eq. (\ref{eq:Flapacepo}) se llega a 
\begin{eqnarray}
\frac{ \rho}{R(\rho)}\frac{\partial  }{\partial \rho}\left(\rho \frac{\partial R(\rho) }{\partial \rho}\right) -\alpha^{2}
=0,\qquad  \rho\frac{\partial  }{\partial \rho}\left(\rho \frac{\partial R(\rho) }{\partial \rho}\right)=\alpha^{2} R(\rho).
\end{eqnarray}
En consecuencia las ecuaciones a resolver son
\begin{eqnarray}
  \frac{\partial^{2} \Psi(\varphi) }{\partial \varphi^{2}}=-\alpha^{2}\Psi(\varphi) , \qquad  \rho\frac{\partial  }{\partial \rho}\left(\rho \frac{\partial R(\rho) }{\partial \rho}\right)=\alpha^{2} R(\rho).\label{eq:Flapacepo1}
\end{eqnarray}
Si $\alpha=0,$ para el sector angular se tiene
\begin{eqnarray}
\Psi(\varphi)=\left( A +B \varphi \right), \quad A,B={\rm constante}.
\end{eqnarray}
Para la parte radial se tiene la ecuaci\'on 
\begin{eqnarray}
\frac{\partial  }{\partial \rho}\left(\rho \frac{\partial R(\rho) }{\partial \rho}\right)=0, 
\end{eqnarray}
es decir 
\begin{eqnarray}
\rho \frac{\partial R(\rho) }{\partial \rho}=C, \qquad C={\rm constante},  
\end{eqnarray}
entonces
\begin{eqnarray}
R(\rho) =C\ln \rho+D, \qquad D={\rm constante}.  
\end{eqnarray}
Por lo tanto,
\begin{eqnarray}
\phi_{0}(\rho,\varphi)= \left( A +B \varphi \right)\left(C\ln \rho+D\right), \qquad A,B,C,D={\rm constante}.  
\end{eqnarray}
Si $\alpha\not =0,$ la parte angular tiene como soluci\'on
\begin{eqnarray}
\Psi(\varphi)=a_{\alpha} \cos\alpha \varphi  +b_{\alpha}  \sin\alpha \varphi, \quad a_{\alpha},b_{\alpha}={\rm constante}.
\end{eqnarray}
Para la parte radial propondremos  $R(\rho)=E\rho^{\lambda},$ con $E$ una contante. Al sustituir esta propuesta en Eq. (\ref{eq:Flapacepo1})
se encuentra $\alpha^{2}=\lambda ^{2},$ es decir $\lambda=\pm \alpha.$ As\'i, la soluci\'on radial es 
\begin{eqnarray}
R(\rho) =c_{\alpha} \rho^{\alpha}+d_{\alpha}\rho^{-\alpha} , \qquad c_{\alpha},d_{\alpha}={\rm constante}.  
\end{eqnarray}
De donde, si $\alpha\not =0,$ las soluciones son
\begin{eqnarray}
\phi_{\alpha}(\rho, \varphi)=\left(a_{\alpha} \cos\alpha \varphi  +b_{\alpha}  \sin\alpha \varphi\right)\left(c_{\alpha} \rho^{\alpha}+d_{\alpha}\rho^{-\alpha}\right).  
\end{eqnarray}
como $(\rho,\varphi)$ y $(\rho,\varphi+2\pi)$ representan el mismo punto en el espacio, se debe cumplir que
\begin{eqnarray}
\phi(\rho, \varphi)=\phi(\rho, \varphi+2\pi).\label{eq:Flapacepo2}
\end{eqnarray}
Para el caso $\alpha=0$ esta condici\'on implica $B=0,$ de donde 
\begin{eqnarray}
\phi_{0}(\rho,\varphi)= C\ln \rho+D, \qquad C,D={\rm constante}.  
\end{eqnarray}
Si $\alpha \not =0,$ la condici\'on Eq. (\ref{eq:Flapacepo2}) impone que  
\begin{eqnarray}
\cos\alpha \left( \varphi+2\pi\right)=  \cos\alpha \varphi,\qquad  \sin\alpha \left( \varphi+2\pi\right)=\sin\alpha \varphi,
\end{eqnarray}
que se  satisfacen si $n=\alpha,$ con $n=1,2,3,\cdots.$ Por lo tanto, las soluciones son 
\begin{eqnarray}
\phi_{n}(\rho, \varphi)=\left(a_{n} \cos n \varphi  +b_{n}  \sin n\varphi\right)\left(c_{n} \rho^{n}+d_{n}\rho^{-n}\right).  
\end{eqnarray}
Esto nos indica que, usando coordenas polares, la soluci\'on general a la ecuaci\'on de Poisson en dos dimensiones 
es
\begin{eqnarray}
\phi(\rho, \varphi)= C\ln \rho+\frac{a_{0}}{\sqrt{2\pi}} +\sum_{n\geq 1} \left(\frac{a_{n}}{\sqrt{\pi}}  \cos n \varphi  +\frac{b_{n}}{\sqrt{\pi}}  \sin n\varphi\right)\left(c_{n} \rho^{n}+d_{n}\rho^{-n}\right). \qquad  \label{eq:Flapacepo3}
\end{eqnarray}

\subsection{F\'ormula de Poisson en dos dimensiones}

Un cilindro infinito de radio $R$ est\'a a potencial $V(\varphi)$ en su superficie. Suponiendo que el potencial es finito en cualquier punto 
del espacio, muestre  que el potencial se puede escribir como
\begin{eqnarray}
\phi(\rho,\varphi)=\frac{1}{2\pi} \left[1-\left(\frac{\rho_{<}}{\rho_{>}}\right)^{2}\right]
\int_{0}^{2\pi}  \frac{ d\varphi^{\prime}V(\varphi^{\prime})}{1-2\left(\frac{\rho_{<}}{\rho_{>}}\right) \cos(\varphi-\varphi^{\prime})+ \left(\frac{\rho_{<}}{\rho_{>}}\right) ^{2}},
\end{eqnarray}
con $\rho_{<}={\rm menor}\{\rho, R\}$ y $\rho_{>}={\rm mayor}\{\rho, R\}.$\\

Para este caso se debe ocupar la ecuaci\'on de Poisson en tres dimensiones con coordenadas cil\'indricas. Pero en  el cilindro es infinito
el potencial no depende de $z,$ por lo tanto la ecuaci\'on que se debe satisfacer es la ecuaci\'on de Poisson en dos dimensiones en coordenadas polares
Eq. (\ref{eq:Flapacepo0}). As\'i, el potencial  debe ser de forma Eq. (\ref{eq:Flapacepo3}). Como el potencial debe ser finito en el interior del cilindro, en esta regi\'on debe tomar la forma
\begin{eqnarray}
\phi_{int}(\rho, \varphi)= \frac{a_{0}}{\sqrt{2\pi}} +\sum_{n\geq 1} \left(\frac{a_{n}}{\sqrt{\pi}}  \cos n \varphi  
+\frac{b_{n}}{\sqrt{\pi}}  \sin n\varphi\right)\left(\frac{\rho}{R}\right)^{n}.
\end{eqnarray}
Ahora, como el potencial debe ser finito en el exterior del cilindro, en esta regi\'on debe tomar la forma
\begin{eqnarray}
\phi_{ext}(\rho, \varphi)= \frac{A_{0}}{\sqrt{2\pi}} +\sum_{n\geq 1} \left(\frac{A_{n}}{\sqrt{\pi}}  \cos n \varphi  
+\frac{B_{n}}{\sqrt{\pi}}  \sin n\varphi\right)\left(\frac{R}{\rho}\right)^{n}.
\end{eqnarray}
Adem\'as, si $\rho=R$ se debe cumplir  
\begin{eqnarray}
V(\phi)&=&\phi_{ext}(R, \varphi)= \frac{A_{0}}{\sqrt{2\pi}} +\sum_{n\geq 1} \left(\frac{A_{n}}{\sqrt{\pi}}  \cos n \varphi  
+\frac{B_{n}}{\sqrt{\pi}}  \sin n\varphi\right)\nonumber\\
&=&\phi_{int}(R, \varphi)= \frac{a_{0}}{\sqrt{2\pi}} +\sum_{n\geq 1} \left(\frac{a_{n}}{\sqrt{\pi}}  \cos n \varphi  
+\frac{b_{n}}{\sqrt{\pi}}  \sin n\varphi\right),\nonumber
\end{eqnarray}
que implica
\begin{eqnarray}
A_{0}&=&a_{0}=\int_{-\pi}^{\pi} \frac{V(\varphi^{\prime})}{\sqrt{2\pi}} d\varphi^{\prime},\label{eq:FouPisson1}\\
 a_{n}&=&A_{n}=\int_{-\pi}^{\pi} V(\varphi^{\prime}) \frac{\cos n\varphi^{\prime} }{\sqrt{\pi}} d\varphi^{\prime} ,\\
 b_{n}&=&B_{n}= \int_{-\pi}^{\pi} V(\varphi^{\prime}) \frac{ \sin n\varphi^{\prime}}{\sqrt{\pi}} d\varphi^{\prime}.\label{eq:FouPisson3}
\end{eqnarray}
As\'i, los potenciales son
\begin{eqnarray}
\phi_{int}(\rho, \varphi)&=& \frac{a_{0}}{\sqrt{2\pi}} +\sum_{n\geq 1} \left(\frac{a_{n}}{\sqrt{\pi}}  \cos n \varphi  
+\frac{b_{n}}{\sqrt{\pi}}  \sin n\varphi\right)\left(\frac{\rho}{R}\right)^{n},\nonumber\\
\phi_{ext}(\rho, \varphi)&=& \frac{a_{0}}{\sqrt{2\pi}} +\sum_{n\geq 1} \left(\frac{a_{n}}{\sqrt{\pi}}  \cos n \varphi  
+\frac{b_{n}}{\sqrt{\pi}}  \sin n\varphi\right)\left(\frac{R}{\rho}\right)^{n},\nonumber
\end{eqnarray}
estos dos potenciales se puenden escribir como
\begin{eqnarray}
\phi(\rho, \varphi)&=& \frac{a_{0}}{\sqrt{2\pi}} +\sum_{n\geq 1} \left(\frac{a_{n}}{\sqrt{\pi}}  \cos n \varphi  
+\frac{b_{n}}{\sqrt{\pi}}  \sin n\varphi\right)\left(\frac{\rho_{<}}{\rho_{>}}\right)^{n}.\nonumber
\end{eqnarray}
Ahora, sustituyendo los coeficientes de Fourier Eq. (\ref{eq:FouPisson1}) en Eq. (\ref{eq:FouPisson3}) se encuentra 
\begin{eqnarray}
& &\phi(\rho, \varphi)= \frac{1}{\sqrt{2\pi}} \int_{-\pi}^{\pi} \frac{V(\varphi^{\prime})}{\sqrt{2\pi}} d\varphi^{\prime}+ \nonumber\\
& &+\sum_{n\geq 1} \Bigg(\frac{1}{\sqrt{\pi}}  \cos n \varphi  \int_{-\pi}^{\pi} d \varphi^{\prime} V(\varphi^{\prime}) \frac{\cos n\varphi^{\prime} }{\sqrt{\pi}} d\varphi^{\prime}  \nonumber\\
& & +\frac{1}{\sqrt{\pi}}  \sin n\varphi \int_{-\pi}^{\pi} d \varphi^{\prime}  V(\varphi^{\prime}) \frac{ \sin n\varphi^{\prime}}{\sqrt{\pi}} d\varphi^{\prime} \Bigg)\left(\frac{\rho_{<}}{\rho_{>}}\right)^{n}\nonumber\\
&=& \int_{-\pi}^{\pi} d \varphi^{\prime}  V(\varphi^{\prime})\left( \frac{1}{2\pi}+\frac{1}{\pi} \sum_{n\geq 1} \left( \cos n \varphi^{\prime} \cos n \varphi+ 
\sin n \varphi^{\prime} \sin n \varphi\right) \left(\frac{\rho_{<}}{\rho_{>}}\right)^{n}\right)\nonumber\\
&=& \int_{-\pi}^{\pi} d \varphi^{\prime}  \frac{V(\varphi^{\prime})}{\pi}\left( \frac{1}{2}+\sum_{n\geq 1} \left(\frac{\rho_{<}}{\rho_{>}}\right)^{n} \cos n (\varphi-\varphi^{\prime})  \right).\nonumber
\end{eqnarray}
Definamos 
\begin{eqnarray}
z=\left(\frac{\rho_{<}}{\rho_{>}}\right) e^{i (\varphi-\varphi^{\prime})}<1,\label{eq:Foudef1}
\end{eqnarray}
de donde
\begin{eqnarray}
{\rm Re} z=\left(\frac{\rho_{<}}{\rho_{>}}\right) \cos (\varphi-\varphi^{\prime}).
\end{eqnarray}
As\'i, 
\begin{eqnarray}
\phi(\rho, \varphi)&=& \int_{-\pi}^{\pi} d \varphi^{\prime}  \frac{V(\varphi^{\prime})}{\pi} {\rm Re}\left( \frac{1}{2}+\sum_{n\geq 1} z^{n}  \right) \nonumber\\
&=&  \int_{-\pi}^{\pi} d \varphi^{\prime}  \frac{V(\varphi^{\prime})}{\pi} {\rm Re} \left( \frac{1}{2}-1+1+\sum_{n\geq 1} z^{n}  \right)\nonumber\\
 &=& \int_{-\pi}^{\pi}  d \varphi^{\prime} \frac{V(\varphi^{\prime})}{\pi} {\rm Re}\left( -\frac{1}{2}+\sum_{n\geq 0} z^{n}  \right)\nonumber\\
 &=& \int_{-\pi}^{\pi} d \varphi^{\prime} \frac{V(\varphi^{\prime})}{\pi} {\rm Re}\left( -\frac{1}{2}+\frac{1}{1+z}  \right).\label{eq:FouFin}
\end{eqnarray}
Ahora, notemos que
\begin{eqnarray}
 -\frac{1}{2}+\frac{1}{1+z}= \frac{1}{2}\frac{1-z}{1+z}=  \frac{1}{2}\frac{(1-z)(1+z^{*})}{|1+z|^{2}}=\frac{1}{2} \frac{1+z^{*}-z-z^{*}}{|1+z|^{2}}.
\end{eqnarray}
Adem\'as, considerando la definici\'on Eq. (\ref{eq:Foudef1}) se tiene
\begin{eqnarray}
1+z^{*}-z-z^{*}z&=&1-2i\left(\frac{\rho_{<}}{\rho_{>}}\right) \sin(\varphi- \varphi^{\prime})- \left(\frac{\rho_{<}}{\rho_{>}}\right)^{2},\nonumber\\
1+z&=&1+\left(\frac{\rho_{<}}{\rho_{>}}\right) \cos(\varphi- \varphi^{\prime})+i \left(\frac{\rho_{<}}{\rho_{>}}\right) \sin(\varphi- \varphi^{\prime})\nonumber\\
 |1+z|^{2}&=&\left(1+\left(\frac{\rho_{<}}{\rho_{>}}\right) \cos(\varphi- \varphi^{\prime})\right)^{2} + \left(\frac{\rho_{<}}{\rho_{>}}\right)^{2} \sin^{2}(\varphi- \varphi^{\prime})\nonumber\\
 &=&1+2 \left(\frac{\rho_{<}}{\rho_{>}}\right) \cos(\varphi- \varphi^{\prime})+ \left(\frac{\rho_{<}}{\rho_{>}}\right)^{2}.
\end{eqnarray}
Por lo tanto

\begin{eqnarray}
 {\rm Re}\left(-\frac{1}{2}+\frac{1}{1+z}\right)= \left(\frac{1}{2}\right)\frac{1- \left(\frac{\rho_{<}}{\rho_{>}}\right)^{2} }{ 1+2 \left(\frac{\rho_{<}}{\rho_{>}}\right) \cos(\varphi- \varphi^{\prime})+ \left(\frac{\rho_{<}}{\rho_{>}}\right)^{2}}.
\end{eqnarray}
Sustituyendo este resultado en Eq. (\ref{eq:FouFin}) se llega a 
\begin{eqnarray}
\phi(\rho,\varphi)=\frac{1}{2\pi} \left[1-\left(\frac{\rho_{<}}{\rho_{>}}\right)^{2}\right]
\int_{0}^{2\pi}  \frac{ d\varphi^{\prime}F(\varphi^{\prime})}{1-2\left(\frac{\rho_{<}}{\rho_{>}}\right) \cos(\varphi-\varphi^{\prime})+ \left(\frac{\rho_{<}}{\rho_{>}}\right) ^{2}},\nonumber
\end{eqnarray}
que es lo que se queria demostrar.

\subsection{ Cilindro infinito}

Suponga que un cilindro infinito de radio $R$ tiene en su superficie el potencial
\begin{equation}
V(\varphi)=\left\{
\begin{array}{cccc}
-V & \,\, \varphi \in \left[-\pi,-\frac{\pi}{2}\right] ,\\
V & \,\, \varphi \in  \left[-\frac{\pi}{2}, 0\right] ,\\
-V & \,\, \varphi \in  \left[0, \frac{\pi}{2}\right] ,\\
V & \,\, \varphi \in  \left[\frac{\pi}{2}, \pi \right].
\end{array}
\right. 
\end{equation}
Suponiendo que el potencial es finito en cual quier punto del espacio, mostrar que el potencial en todo el espacio es 
\begin{eqnarray}
\phi (x,y)= -\frac{2V}{\pi} \tan^{-1}\left( \frac{2 \left(\frac{\rho_{<}}{\rho_{>}}\right)^{2}\sin 2\varphi  }
{ 1- \left(\frac{\rho_{<}}{\rho_{>}}\right)^{4}} \right),
\end{eqnarray}
 con $\rho_{<}={\rm menor}\{\rho, R\}$ y $\rho_{>}={\rm mayor}\{\rho, R\}.$\\ 

Por las caracter\'isticas del sistema, el potencial es de la forma Eq. (\ref{eq:Flapacepo3}) con los coeficientes de Fourier dados por Eqs. (\ref{eq:FouPisson1})-(\ref{eq:FouPisson3}). Adem\'as, como $V(\varphi)$ es una funci\'on impar se encuentra que $a_{0}=a_{n}=0$ y 
\begin{eqnarray}
b_{n}&=&\frac{1}{\sqrt{\pi}}\int_{-\pi}^{\pi} d\varphi V(\varphi) \sin n\varphi= \frac{2}{\sqrt{\pi}}\int_{0}^{\pi} d\varphi V(\varphi) \sin n\varphi\nonumber\\
& =&\frac{2V}{n\sqrt{\pi}}\left( \cos n\varphi\Bigg|_{0}^{\frac{\pi}{2}} -  \cos n\varphi\Bigg|_{ \frac{\pi}{2}}^{\pi}\right)= \frac{2V}{n\sqrt{\pi}}\left( (-)^{n+1} -1+ 2\cos\frac{n \pi}{2} \right). \nonumber
\end{eqnarray}
Como $\cos\frac{(2n-1) \pi}{2}=0,$ se encuentra que $b_{2n-1}=0,$ mientras que 
\begin{eqnarray}
b_{2n}= \frac{2V}{n\sqrt{\pi}}\left( -1+ \cos n \pi \right)= \frac{2V}{n\sqrt{\pi}}\left( -1+ (-)^{n} \right), \nonumber
\end{eqnarray}
de donde $b_{2(2n)}=0$ y 
\begin{eqnarray}
b_{2(2n-1)}= \frac{-4V}{(2n-1)\sqrt{\pi}}.
\end{eqnarray}
Por lo tanto el potencial el\'ectrico es
\begin{eqnarray}
\phi(\rho, \varphi)= \frac{-4V}{\pi} \sum_{n\geq 1} \left(\frac{\rho_{<}}{\rho_{>}}\right)^{2(2n-1)} \frac{1}{2n-1}  \sin 2(2n-1)\varphi .\nonumber
\end{eqnarray}
Con la definici\'on
\begin{eqnarray}
\omega= \left(\frac{\rho_{<}}{\rho_{>}}\right)^{2} e^{i 2\varphi},
\end{eqnarray}
el potencial se puede escribir de la forma
\begin{eqnarray}
\phi(\rho, \varphi)= \frac{-4V}{\pi} \sum_{n\geq 1}  \frac{{\rm Im} \omega ^{2n-1}}{2n-1} =   \frac{-4V}{\pi} {\rm Im}
\left( \sum_{n\geq 1}  \frac{\omega ^{2n-1}}{2n-1}\right).\nonumber
\end{eqnarray}
Adem\'as, dem\'as usando la serie 
$$
\ln  \left(\frac{1+\omega}{1-\omega}\right)=2 \sum_{n\geq 1}  \frac{\omega ^{2n-1}}{2n-1}
$$
se obtiene
\begin{eqnarray}
\phi (x,y)= -\frac{2V}{\pi} {\rm Im}\left[\ln  \left(\frac{1+\omega}{1-\omega}\right) \right].
\end{eqnarray}
Ahora, si definimos
\begin{eqnarray}
z=\frac{1+\omega}{1-\omega}=|z|e^{i\theta},\quad \tan \theta= \frac{{\rm Im}(z)}{{\rm Re}(z)},
\end{eqnarray}
se encuentra
\begin{eqnarray}
\ln z=\ln|z|+ i\theta, 
\end{eqnarray}
de donde
\begin{eqnarray}
\phi (x,y)= \frac{2V}{\pi} \theta.
\end{eqnarray}
Ocupando  la definici\'on de $\omega$  tenemos que
\begin{eqnarray}
z&=&\frac{1+\omega}{1-\omega}=\frac{ (1+\omega)(1-\omega^{*}) }{|1-\omega|^{2}}= 
\frac{ 1+\omega-\omega^{*}-\omega\omega^{*}}{|1-\omega|^{2}}\nonumber\\
&=& 
\frac{ 1+2i\left(\frac{\rho_{<}}{\rho_{>}}\right)^{2}\sin 2\varphi- \left(\frac{\rho_{<}}{\rho_{>}}\right)^{4}   }{|1-\omega|^{2}}= 
\frac{ 1- \left(\frac{\rho_{<}}{\rho_{>}}\right)^{4}+2i\left(\frac{\rho_{<}}{\rho_{>}}\right)^{2}\sin 2\varphi}{|1-\omega|^{2}}\nonumber
\end{eqnarray}
entonces 
\begin{eqnarray}
\tan \theta= \frac{2 \left(\frac{\rho_{<}}{\rho_{>}}\right)^{2}\sin 2\varphi  }
{ 1- \left(\frac{\rho_{<}}{\rho_{>}}\right)^{4}},
\end{eqnarray}
de donde 
\begin{eqnarray}
\phi (x,y)= -\frac{2V}{\pi} \tan^{-1}\left( \frac{2 \left(\frac{\rho_{<}}{\rho_{>}}\right)^{2}\sin 2\varphi  }
{ 1- \left(\frac{\rho_{<}}{\rho_{>}}\right)^{4}} \right),
\end{eqnarray}
que es lo que se queria mostrar.

\section{ Ecuaci\'on de Schr$\ddot {\rm o}$dinger en una dimensi\'on}

La ecuaci\'on de Schr${\rm \ddot o}$dinger en una dimensi\'on es
\begin{eqnarray}
i\hbar \frac{ \partial \psi (x,t)}{\partial t}= H\psi(x,t),\qquad H=-\frac{\hbar^{2}}{2m}\frac{\partial^{2}}{\partial x^{2}}+ V(x).
\label{eq:FouSho}
\end{eqnarray}
Para resolver esta ecuaci\'on, propondremos que $\psi(x,t)=T(t)\phi(x),$ sustituyendo esta propuesta en Eq. (\ref{eq:FouSho}) se encuentra
\begin{eqnarray}
i\hbar \phi(x)\frac{ \partial T(t)}{\partial t}= T(t)H\phi(x),
\end{eqnarray}
de donde
\begin{eqnarray}
i\hbar \frac{1}{T(t)} \frac{ \partial T(t)}{\partial t}= \frac{1}{\phi(x)} H\phi(x).\label{eq:FouSho1}
\end{eqnarray}
Por lo tanto, 
\begin{eqnarray}
i\hbar \frac{\partial }{\partial t}\left(\frac{1}{T(t)} \frac{ \partial T(t)}{\partial t}\right)= 0,
\end{eqnarray}
es decir 
\begin{eqnarray}
i\hbar \left(\frac{1}{T(t)} \frac{ \partial T(t)}{\partial t}\right)= E,\qquad 
i\hbar \frac{ \partial T(t)}{\partial t}=E T(t),  \qquad E={\rm constante},\nonumber
\end{eqnarray}
de donde 
\begin{eqnarray}
T(t)=e^{-i\frac{Et}{\hbar}}.
\end{eqnarray}
Sustituyendo este resultado en Eq. (\ref{eq:FouSho1}) se llega a 
\begin{eqnarray}
H\phi(x)=E\phi(x),\label{eq:FouSho2}
\end{eqnarray}
como $H$ es un operador Herm\'itico, $E$ debe ser una constante real.\\

\subsection{Pozo infinito}

Suponga que el potencial es dado por 
\begin{eqnarray}
V(x)= \left\{
\begin{array}{cc}
0 & \,\,  0\leq x\leq L ,\\
\infty  & \,\,   x<0, \quad x>L. \\

\end{array}
\right. 
\end{eqnarray}
Este potencial representa una part\'icula encerrada en una l\'inea de longitud $L.$ Como la part\'icula est\'a encerrada,
se deben cumplir las condiciones de frontera $\phi(0)=\psi(L)=0,$ que son condiciones de Dirichlet. Por lo tanto, Eq. (\ref{eq:FouSho2}) es tipo Sturm-Llioville y sus soluciones deben formar una base ortonormal.\\

Dentro de la l\'inea $[0,L],$ se tiene la ecuaci\'on 
\begin{eqnarray}
H\phi(x)=-\frac{\hbar^{2}}{2m}\frac{\partial^{2} \phi(x)}{\partial x^{2}}=E\phi(x),
\end{eqnarray}
es decir
\begin{eqnarray}
\frac{\partial^{2} \phi(x)}{\partial x^{2}}=-\frac{2m E}{\hbar^{2}} \phi(x),
\end{eqnarray}
cuya soluci\'on general es 
\begin{eqnarray}
\phi(x)=A_{E}\sin\sqrt{\frac{2m E}{\hbar^{2}}}x + B_{E}\cos\sqrt{\frac{2m E}{\hbar^{2}}}x, \quad A_{E},B_{E}={\rm constante}.
\end{eqnarray}
La condici\'on de borde $\phi(0)=0$ implica $B_{E}=0.$ Mientras que la condici\'on $\phi(L)=0$ implica
\begin{eqnarray}
\sin\sqrt{\frac{2m E}{\hbar^{2}}}L=0,
\end{eqnarray}
es decir 
\begin{eqnarray}
\sqrt{\frac{2m E}{\hbar^{2}}}L=n\pi.
\end{eqnarray}
Por lo tanto las \'unicas energ\'ias permitidas son
\begin{eqnarray}
E_{n}=\frac{\hbar^{2} }{2m} \left(\frac{n\pi}{L}\right)^{2},
\end{eqnarray}
adem\'as las funciones de onda normalizadas son 
\begin{eqnarray}
\phi_{n}(x)=\sqrt{\frac{2}{L}} \sin\frac{n\pi}{L}x.
\end{eqnarray}
Entonces, las funciones de onda del sistema son 
\begin{eqnarray}
\psi(x,t)=\sum_{n\geq 1} a_{n} e^{-i\frac{E_{n} t}{\hbar}} \sqrt{\frac{2}{L}} \sin\frac{n\pi}{L}x.
\end{eqnarray}

\section{Ecuaci\'on de Onda}

La ecuaci\'on de onda en una dimensi\'on es 
\begin{eqnarray}
\left( \frac{\partial ^{2}}{\partial x^{2}}-\frac{1}{c^{2}} \frac{\partial ^{2}}{\partial t^{2}}\right)\phi(x,t)=0.\label{eq:fourier-onda}
\end{eqnarray}
Para resolver esta ecuaci\'on propondremos la funci\'on $\phi(x,t)=T(t)X(x),$
al sustituirla en Eq. (\ref{eq:fourier-onda}) se obtiene
\begin{eqnarray}
T(t)\frac{\partial ^{2} X(x)}{\partial x^{2}}-\frac{X(x)}{c^{2}} \frac{\partial ^{2} T(t) }{\partial t^{2}}=0,
\end{eqnarray}
que implica
\begin{eqnarray}
\frac{1}{X(x)} \frac{\partial ^{2} X(x)}{\partial x^{2}}=\frac{1}{T(t) c^{2}} \frac{\partial ^{2} T(t) }{\partial t^{2}}, 
\label{eq:fourier-onda-1} 
\end{eqnarray}
de donde 
\begin{eqnarray}
\frac{\partial }{\partial x}\left(\frac{1}{X(x)} \frac{\partial ^{2} X(x)}{\partial x^{2}}\right)=0, 
\end{eqnarray}
es decir 
\begin{eqnarray}
\frac{1}{X(x)} \frac{\partial ^{2} X(x)}{\partial x^{2}}=-\alpha^{2},\qquad \alpha={\rm constante},
\end{eqnarray}
al considerar este resultado en Eq. (\ref{eq:fourier-onda-1}) se llega a 
\begin{eqnarray}
\frac{1}{T(t) c^{2}} \frac{\partial ^{2} T(t) }{\partial t^{2}}=-\alpha^{2}. 
\end{eqnarray}
Por lo tanto, las ecuaciones a resolver son 
\begin{eqnarray}
\frac{\partial ^{2} T(t) }{\partial t^{2}}=-\alpha^{2}c^{2}T(t),\qquad    \frac{\partial ^{2} X(x)}{\partial x^{2}}=-\alpha^{2} X(x).
\end{eqnarray}
Si $\alpha=0,$ se tienen las soluciones 
\begin{eqnarray}
T(t)=\left(A+Bt\right),\qquad X(x)=\left(C+Dx\right),\qquad A,B,C,D={\rm constante},\nonumber
\end{eqnarray}
es decir
\begin{eqnarray}
\phi_{0}(x,t)= \left(A+Bt\right)\left(C+Dx\right).
\end{eqnarray}
Si $\alpha\not =0$ se encuentra que
\begin{eqnarray}
X(x)= \left(a_{\alpha}\cos \alpha x + b_{\alpha} \sin \alpha x \right),\qquad 
T(t)=\left(c_{\alpha}\cos \alpha ct + d_{\alpha} \sin \alpha ct\right), \nonumber
\end{eqnarray}
con $a_{\alpha},b_{\alpha},c_{\alpha},d_{\alpha}$ son constantes. De estas soluciones se tiene 
\begin{eqnarray}
\phi_{\alpha}(x,t)=\left(a_{\alpha}\cos \alpha x + b_{\alpha} \sin \alpha x\right)\left(c_{\alpha}\cos \alpha ct + d_{\alpha} \sin \alpha ct \right).
\end{eqnarray}

\subsection{ Cuerda con extremos fijos}

Supongamos que tenemos una cuerda de longitud $L$ con los extremos fijos.
Para este caso, las soluciones de la ecuaci\'on de onda  deben satisfacer las condiciones de borde
\begin{eqnarray}
\phi(0,t)=\phi(L,t)=0.
\end{eqnarray}
Estas condiciones implican
\begin{eqnarray}
\phi_{0}(0,t)=A(C+Dt)=0,
\end{eqnarray}
es decir $A=0,$ adem\'as
\begin{eqnarray}
\phi_{0}(L,t)=BL(C+Dt)=0,
\end{eqnarray}
que induce $B=0.$ Por lo tanto $\phi_{0}(x,t)=0.$ Para el caso $\alpha \not =0,$ se tiene 
\begin{eqnarray}
\phi_{\alpha}(0,t)=a_{\alpha} \left(c_{\alpha}\cos \alpha ct + d_{\alpha} \sin \alpha ct \right)=0,
\end{eqnarray}
es decir $a_{\alpha}=0,$ adem\'as
\begin{eqnarray}
\phi_{\alpha}(L,t)=b_{\alpha} \sin \alpha L \left(c_{\alpha}\cos \alpha ct + d_{\alpha} \sin \alpha ct  \right)=0,
\end{eqnarray}
que implica $\sin \alpha L=0,$ por lo que $\alpha L=n\pi.$ De donde, en este caso  las soluciones deben ser de la forma
\begin{eqnarray}
\phi_{n}(x,t)=\sqrt{\frac{2}{L}} \sin \left(\frac{n\pi x}{L}\right) \left[c_{n}\cos  \left(\frac{n\pi ct}{L}\right) + 
d_{n} \sin \left(\frac{n\pi ct}{L}\right)  \right].
\end{eqnarray}
As\'i, la soluci\'on general es 
\begin{eqnarray}
\phi(x,t)=\sum_{n\geq 1} \sqrt{\frac{2}{L}} \sin \left(\frac{n\pi x}{L}\right) \left[c_{n}\cos\left(\frac{n\pi ct}{L}\right) + 
d_{n} \sin  \left(\frac{n\pi ct}{L}\right) \right].
\end{eqnarray}

\subsection{ Condiciones iniciales}

Supongamos que se cumplen las condiciones iniciales  
\begin{eqnarray}
\phi(x,0)=f(x),\quad \frac{\partial \phi(x,t)}{\partial t}\Bigg |_{t=0}=g(x),
\end{eqnarray}
entonces
\begin{eqnarray}
\phi(x,0)=f(x)=\sum_{n\geq 1} c_{n} \sqrt{\frac{2}{L}} \sin \left(\frac{n\pi x}{L}\right),
\end{eqnarray}
por lo tanto
\begin{eqnarray}
c_{n}= \int_{0}^{L}dx  \sqrt{\frac{2}{L}} \sin \left(\frac{n\pi x}{L}\right) f(x)=\int_{0}^{L}dx  \sqrt{\frac{2}{L}} \sin \left(\frac{n\pi x}{L}\right) \phi(x,0).
\end{eqnarray}
Adem\'as,
\begin{eqnarray}
\frac{\partial \phi(x,t)}{\partial t}\Bigg |_{t=0}=g(x)=\sum_{n\geq 1}  d_{n}\left(\frac{n\pi c}{L}\right)\sqrt{\frac{2}{L}} \sin \left(\frac{n\pi x}{L}\right),
\end{eqnarray}
por lo tanto
\begin{eqnarray}
d_{n}&=&\left(\frac{L}{n\pi c}\right) \int_{0}^{L}dx  \sqrt{\frac{2}{L}} \sin \left(\frac{n\pi x}{L}\right) g(x)\nonumber\\
&=&\left(\frac{L}{n\pi c}\right)\int_{0}^{L}dx  \sqrt{\frac{2}{L}} \sin \left(\frac{n\pi x}{L}\right) \frac{\partial \phi(x,t)}{\partial t}\Bigg |_{t=0}.\nonumber
\end{eqnarray}
\subsection{ Energ\'ia}

La energ\'ia para una cuerda est\'a dada por 
\begin{eqnarray}
H=\frac{1}{2}\int_{0}^{L}dx\left[ \frac{1}{c^{2}}\left(\frac{\partial \phi(x,t)}{\partial t}\right)^{2}+ \left(\frac{\partial \phi(x,t)}{\partial x}\right)^{2}\right]
\end{eqnarray}
Para calcular esta cantidad note que
\begin{eqnarray}
\frac{\partial \phi(x,t)}{\partial t}&=&\sum_{n\geq 1} \sqrt{\frac{2}{L}} \left(\frac{n\pi c}{L}\right)\sin \left(\frac{n\pi x}{L}\right) \left[-c_{n}\sin\left(\frac{n\pi ct}{L}\right) + 
d_{n} \cos \left(\frac{n\pi ct}{L}\right) \right],\nonumber\\
\frac{\partial \phi(x,t)}{\partial x}&=&\sum_{n\geq 1} \sqrt{\frac{2}{L}} \left(\frac{n\pi }{L}\right)\cos \left(\frac{n\pi x}{L}\right) \left[c_{n}\cos\left(\frac{n\pi ct}{L}\right) + 
d_{n} \sin  \left(\frac{n\pi ct}{L}\right) \right],\nonumber
\end{eqnarray}
de donde
\begin{eqnarray}
& & \frac{1}{c^{2}} \left(\frac{\partial \phi(x,t)}{\partial t}\right)^{2}=\sum_{n\geq 1} \sum_{l\geq 1}\left(\frac{n\pi }{L}\right)\left(\frac{l\pi }{L}\right) \sqrt{\frac{2}{L}} \sin \left(\frac{n\pi x}{L}\right) \sqrt{\frac{2}{L}}\sin \left(\frac{l\pi x}{L}\right)\nonumber\\
& & \left[-c_{n}\sin\left(\frac{n\pi ct}{L}\right) + 
d_{n} \cos \left(\frac{n\pi ct}{L}\right) \right]\left[-c_{l}\sin\left(\frac{l\pi ct}{L}\right) + 
d_{l} \cos \left(\frac{l\pi ct}{L}\right) \right],\nonumber\\
& & \left(\frac{\partial \phi(x,t)}{\partial x}\right)^{2}=\sum_{n\geq 1} \sum_{l\geq 1}\left(\frac{n\pi }{L}\right)\left(\frac{l\pi }{L}\right) \sqrt{\frac{2}{L}} \cos \left(\frac{n\pi x}{L}\right) \sqrt{\frac{2}{L}}\cos \left(\frac{l\pi x}{L}\right)\nonumber\\
& & \left[c_{n}\cos \left(\frac{n\pi ct}{L}\right) + 
d_{n} \sin \left(\frac{n\pi ct}{L}\right) \right]\left[c_{l}\cos\left(\frac{l\pi ct}{L}\right) + 
d_{l} \sin \left(\frac{l\pi ct}{L}\right) \right],\nonumber
\end{eqnarray}
ocupando las relaciones de ortormalidad se encuentra 

\begin{eqnarray}
\int_{0}^{L} dx\frac{1}{c^{2}} \left(\frac{\partial \phi(x,t)}{\partial t}\right)^{2}&=&\sum_{n\geq 1} \left(\frac{n\pi }{L}\right)^{2} 
\left[-c_{n}\sin\left(\frac{n\pi ct}{L}\right) + d_{n} \cos \left(\frac{n\pi ct}{L}\right) \right]^{2} ,\nonumber\\
\int_{0}^{L} dx \left(\frac{\partial \phi(x,t)}{\partial x}\right)^{2}&=&\sum_{n\geq 1} \left(\frac{n\pi }{L}\right)^{2}
\left[c_{n}\cos \left(\frac{n\pi ct}{L}\right) + d_{n} \sin \left(\frac{n\pi ct}{L}\right) \right]^{2}\nonumber.
\end{eqnarray}
Por lo tanto, la energ\'ia es 
\begin{eqnarray}
H=\frac{1}{2} \sum_{n\geq 1} \left(\frac{n\pi }{L}\right)^{2}
\left[(c_{n})^{2}+(d_{n})^{2}  \right].
\end{eqnarray}

\chapter{El Oscilador Arm\'onico y los Polinomios de Hermite}

En este cap\'itulo obtendremos las funciones de onda propias del oscilador arm\'onico cu\'antico.
En principio para obtener estas funciones se debe resolver una ecuaci\'on diferencial ordinaria
de segundo orden. Sin embargo, J. G. Darboux desaroll\'o el llamado m\'etodo de factorizaci\'on, el cual permite resolver diferentes ecuaciones diferenciales de forma algebraica. Al surgir la mec\'anica cu\'antica
este m\'etodo mostr\'o su gran potencial, pues con el se pueden obtener soluciones exactas de sistemas sofisticados.\\

El material de este cap\'itulo es importante desde el punto de vista f\'isico, pues se presentan los operadores de ascenso  y descenso, los cuales nos  abre una ventana que nos permite ver m\'as all\'a del oscilador arm\'onico. Tambi\'en es importante  como herramienta matem\'atica,
pues muestra como herramientas algebraicas permiten resolver ecuaciones diferenciales.

\section{Hamiltoniano}

El operador Hamiltoniano para el oscilador arm\'onico 
en una dimensi\'on es 
\begin{eqnarray}
\hat H=\frac{1}{2m}\hat p^{2}+\frac{m\omega^{2}}{2}\hat x^{2},\qquad 
\hat x=x ,\quad  \hat p=-i\hbar \frac{\partial }{\partial x},
\end{eqnarray}
de donde la ecuaci\'on de onda estacionaria es
\begin{eqnarray}
\hat H\psi(x)=\left(-\frac{\hbar^{2}}{2m}\frac{\partial ^{2}}{\partial x^{2}}+
\frac{m\omega^{2}}{2}x^{2}\right)\psi(x)=E\psi(x).
\label{eq:os-scho}
\end{eqnarray}
Esta ecuaci\'on está definida en el intervalo $(-\infty,\infty)$ y se debe cumplir las
condiciones de Dirichlet $\psi(-\infty)=\psi(\infty)=0.$ Resolver este problema equivale a
encontrar las funciones propias $\psi(x)$ y los valores propios $E.$\\

A\'un sin resolver el problema sabemos que $E$ debe ser real, pues $\hat p$ y $\hat x$ son operadores
Herm\'iticos. Adem\'as para cualquier funci\'on $f$ se cumple
\begin{eqnarray}
0&\leq& \left<\hat pf|\hat pf\right>=\left<f|\hat p^{\dagger}\hat pf\right>=\left<f|\hat p^{2}f\right>, \nonumber\\
 0&\leq &\left<\hat xf|\hat xf\right>=\left<f|\hat x^{\dagger}\hat xf\right>=\left<f|\hat x^{2}f\right>,\nonumber
\end{eqnarray}
por lo que
\begin{eqnarray}
0\leq \left<f|Hf\right>=\left<f\Bigg|\left(\frac{\hat p^{2}}{2m}+\frac{m\omega^{2}}{2} \hat x^{2}\right)f\right>.
\end{eqnarray}
En  particular si $f=\psi$ y $\hat H\psi=E\psi$ se tiene 
\begin{eqnarray}
0\leq E\left<\psi|\psi\right>.  
\end{eqnarray}
De este resultado es claro que si $E<0,$ entonces $\left<\psi|\psi\right>=0,$ que implica que $\psi(x)=0.$ Adem\'as, si $\left<\psi|\psi\right>\not = 0,$ 
entonces $E\geq 0.$ Es decir, la \'unica soluci\'on con valor propio negativo es $\psi(x)=0,$ en cualquier otro caso se debe cumplir que $E \geq 0.$

\subsection{Ortonormalidad}

Tambi\'en podemos ver que Eq. (\ref{eq:os-scho}) es tipo Sturm-Liouville, 
por lo que sus soluciones que satisfancen la condiciones de Dirichlet $\psi(-\infty)=\psi(\infty)=0$
con valores propios diferentes, son ortogonales. Es decir, si $\psi_{a}(x)$ y $\psi_{b}(x)$ son soluciones 
de Eq.  (\ref{eq:os-scho}) con los valores propios $E_{a}$ y $E_{b}$  y adem\'as $\psi_{a}(\pm\infty)=\psi_{b}(\pm\infty)=0,$ entonces se cumple
\begin{eqnarray}
\left(E_{a}-E_{b}\right) \int_{-\infty}^{\infty} dx \psi_{a}(x)\psi_{b}^{*}(x)=0.
\end{eqnarray}
En particular, si $E_{a}\not =E_{b},$ tenemos
\begin{eqnarray}
\int_{-\infty}^{\infty} dx \psi_{a}(x)\psi_{b}^{*}(x)=0.
\end{eqnarray}
Por lo que las funciones de onda con diferente valor propio son ortonormales.
Si $E_{a} =E_{b},$ entonces la integral 
$$\int_{-\infty}^{\infty} dx \psi_{a}(x)\psi_{b}^{*}(x)=
\int_{-\infty}^{\infty} dx \psi_{a}(x)\psi_{a}^{*}(x)$$
es la norma de la funci\'on de onda, la cual supondremos que est\'a normalizada, es decir,
\begin{eqnarray}
\int_{-\infty}^{\infty} dx |\psi_{a}(x)|^{2}=1.
\end{eqnarray}
Como este resultado es v\'alido para cualquier valor propio, se encuentra
\begin{eqnarray}
\left<\psi_{a}|\psi_{a}\right>=\int_{-\infty}^{\infty} dx\psi_{a}^{*}(x) \psi_{b}(x)=\delta_{ab}.
\end{eqnarray}
Por lo tanto, las funciones de onda del oscilador arm\'onico
forman una base ortonormal.

\section{Operadores de acenso y decenso}

Recordemos que se cumplen la reglas de conmutaci\'on
\begin{eqnarray}
\left[\hat x, \hat p\right]=\hat x\hat p-\hat p \hat x=i\hbar,\quad
 \left[\hat x, \hat x\right]=\left[\hat p, \hat p\right]=0.\label{eq:conmu-os}
\end{eqnarray}
Ahora, definamos el operador 
\begin{eqnarray}
\hat a= \frac{1}{\sqrt{2m\hbar \omega}}\left(\hat p-im\omega \hat x\right). 
\end{eqnarray}
Como $\hat p$ y $\hat x$ son operadores herm\'{\i}ticos se tiene
\begin{eqnarray}
\hat a^{\dagger}= \frac{1}{\sqrt{2m\hbar \omega}}\left(\hat p+im\omega \hat x\right).
\end{eqnarray}
De donde
\begin{eqnarray}
\hat a^{\dagger} \hat a&=&\frac{1}{2m\hbar \omega}\left(\hat p+im\omega \hat x\right)
\left(\hat p-im\omega \hat x\right)\nonumber\\
&=&
\frac{1}{2m\hbar \omega}
\left[\hat p\left(\hat p-im\omega \hat x\right)+
im\omega \hat x\left(\hat p-im\omega \hat x\right)\right]\nonumber\\
&=& \frac{1}{2m\hbar \omega}\left( \hat p\hat p -im \omega \hat p\hat x+im\omega \hat x \hat p
+m^{2} \omega^{2}\hat x\hat x\right)\nonumber\\
& =& \frac{1}{2m\hbar \omega}\left( \hat p^{2} +m^{2}\omega^{2}\hat x^{2}+
im \omega \left(\hat x\hat p-\hat p \hat x\right) \right)\nonumber\\
&=& \frac{1}{2m\hbar \omega}\left( \hat p^{2} +m^{2}\omega^{2}\hat x^{2}
+im\omega [\hat x,\hat p]\right)\nonumber\\
&=& \frac{2m}{2m\hbar \omega}\left( \frac{1}{2m}\hat p^{2} +\frac{m\omega^{2}}{2}\hat x^{2}
-\frac{1}{2} \hbar\omega \right)\nonumber\\
&=& \frac{1}{\hbar \omega}
 \left( \hat H
-\frac{\hbar \omega}{2}\right),
\end{eqnarray}
por lo tanto,
\begin{eqnarray}
\hat H=\hbar \omega\left( \hat a^{\dagger}\hat a +\frac{1}{2}\right).
\end{eqnarray}
Adem\'as, se cumple
\begin{eqnarray}
\left[\hat a, \hat a^{\dagger}\right]=1,
\end{eqnarray}
en efecto, tomando en cuenta (\ref{eq:conmu-os}) se encuentra
\begin{eqnarray}
\left[\hat a, \hat a^{\dagger}\right]&=&\Bigg[
\frac{\left(\hat p-im\omega \hat x\right)}{\sqrt{2m\hbar \omega}},
\frac{\left(\hat p+im\omega \hat x\right) }{\sqrt{2m\hbar \omega}}\Bigg]
=\frac{1}{2m\hbar \omega}\left[\hat p-im\omega \hat x,\hat p+im\omega \hat x\right]\nonumber\\
&=&
\frac{1}{2m\hbar \omega}\left(\left[\hat p,\hat p+im\omega \hat x\right]- im\omega 
\left[\hat x,\hat p+im\omega \hat x\right]\right)
\nonumber\\
&=&
\frac{1}{2m\hbar \omega}\left(\left[\hat p,\hat p\right]+im \omega\left[\hat p,\hat x\right]-
im\omega\left(\left[\hat x,\hat p\right]+im\omega \left[\hat x, \hat x\right]\right)\right)\nonumber\\
&=&
\frac{1}{2m\hbar \omega}\left(im(-i\hbar\omega)-im\omega(i\hbar)\right)=
\frac{2m\hbar \omega}{2m\hbar \omega}
=1.\nonumber
\end{eqnarray}
Tambi\'en se cumplen las reglas de conmutaci\'on
\begin{eqnarray}
\left[\hat H, \hat a\right]=-\hbar \omega \hat a,\qquad\left[\hat H, \hat a^{\dagger}\right]=
\hbar \omega \hat a^{\dagger},
\label{eq:escalera-os}
\end{eqnarray}
pues
\begin{eqnarray}
\left[\hat H, \hat a\right]&=&\left[\hbar \omega\left(\hat a^{\dagger}\hat a +\frac{1}{2}\right), \hat a\right]=
\hbar \omega \left[\hat a^{\dagger}\hat a, \hat a\right]=
\hbar \omega\left( \hat a^{\dagger}\left[\hat a,\hat a\right]+ 
\left[\hat a^{\dagger}, \hat a\right]\hat a \right)\nonumber\\
&=&\hbar \omega \left[\hat a^{\dagger},\hat a\right] \hat a
=-\hbar \omega \hat a,\nonumber\\
\left[\hat H, \hat a^{\dagger} \right]&=& 
\left[\hbar \omega\left(\hat a^{\dagger} \hat a +\frac{1}{2}\right), \hat a^{\dagger}\right]=
\hbar \omega \left[\hat a^{\dagger}\hat a, \hat a^{\dagger}\right]
=\hbar \omega\left(\hat a^{\dagger}\left[\hat a,\hat a^{\dagger}\right]+ 
\left[\hat a^{\dagger}, \hat a^{\dagger}\right]\hat a\right)\nonumber\\
&=&\hbar \omega \hat a^{\dagger}\left[\hat a,\hat a^{\dagger}\right]
=\hbar \omega \hat a^{\dagger}.\nonumber
\end{eqnarray}
Estas dos igualdades  se pueden escribir como 
\begin{eqnarray}
\hat H \hat a&=&\hat a\hat H-\hbar \omega \hat a,\label{eq:des-esca-os}\\ 
\hat H\hat a^{\dagger}&=& \hat a^{\dagger}\hat H +\hbar \omega \hat a^{\dagger}.
\label{eq:as-esca-os}
\end{eqnarray}
Con la ayuda de estas identidades encontraremos los valores propios $E$ y las funciones
propias $\psi(x)$ del operador Hamiltoniano $\hat H$. Primero supongamos que $\psi(x)$ satisface 
la ecuaci\'on $\hat H\psi(x)=E\psi(x),$ entonces las funciones
\begin{eqnarray}
 \psi_{-(1)}(x)=\hat a \psi(x),\qquad \psi_{+(1)}(x)=\hat a^{\dagger} \psi(x),
\end{eqnarray}
satisfacen las ecuaciones 
\begin{eqnarray}
\hat H\psi_{-(1)}(x)&=&\left( E-\hbar\omega\right)\psi_{-(1)}(x),\\
\hat H\psi_{+(1)}(x)&=&\left( E+\hbar\omega\right)\psi_{+(1)}(x).
\end{eqnarray}
Para probar estas afirmaciones  primero notemos que de  Eq. (\ref{eq:des-esca-os})
se obtiene  
\begin{eqnarray}
\hat H \psi_{-(1)}(x)&=& \hat H \hat a\psi(x)=
\left(\hat a\hat H-\hbar \omega \hat a\right)\psi(x)
=\hat a\hat H\psi(x)- \hbar \omega \hat a\psi(x)\nonumber\\
&=& E\hat a\psi(x)-\hbar \omega \hat a\psi(x)
= \left(E-\hbar \omega\right) \hat a\psi(x)=\left(E-\hbar \omega\right)\psi_{-(1)}(x),
\nonumber
\end{eqnarray}
mientras que de Eq. (\ref{eq:as-esca-os}) se encuentra
\begin{eqnarray}
\hat H \psi_{+(1)}(x)&=& \hat H \hat a^{\dagger}\psi(x)=
\left(\hat a^{\dagger}\hat H+\hbar \omega \hat a^{\dagger}\right)\psi(x)
=\hat a^{\dagger}\hat H\psi(x)+ \hbar \omega \hat a^{\dagger}\psi(x)\nonumber\\
&=& E\hat a^{\dagger}\psi(x)+\hbar \omega \hat a^{\dagger}\psi(x)
= \left(E+\hbar \omega\right) \hat a^{\dagger}\psi(x)=\left(E+\hbar \omega\right)\psi_{+(1)}(x).
\nonumber
\end{eqnarray}
Por lo tanto, la funci\'on $\psi_{-(1)}(x)$ es funci\'on propia de $\hat H$ con valor propio
$E-\hbar\omega,$ mientras que 
$\psi_{+(1)}(x)$ es funci\'on propia de $\hat H$
con valor propio $E+\hbar\omega.$ Es decir, si $\psi$ es soluci\'on de la ecuaci\'on Schr$\ddot {\rm o}$dinger del oscilador arm\'onico, la funciones 
$\psi_{-(1)}=a\psi$ y $\psi_{+(1)}=a^{\dagger} \psi$ tambi\'en son soluciones  de esta ecuaci\'on. \\

Por inducci\'on se puede probar que las funciones
\begin{eqnarray}
\psi_{-(n)}(x)=\left(\hat a\right)^{n} \psi(x),\qquad \psi_{+(n)}(x)=\left(\hat a^{\dagger}\right)^{n} \psi(x),
\end{eqnarray}
satisfacen 
\begin{eqnarray}
\hat H\psi_{-(n)}(x) &=&\left( E-n\hbar\omega\right)\psi_{-(n)}(x),\label{eq:espec-os}\\
\hat H\psi_{+(n)}(x)&=&\left( E+n\hbar\omega\right)\psi_{+(n)}(x). \label{eq:espec-0-os}
\end{eqnarray}
Para ambos casos la base inductiva ya est\'a probada, falta probar el paso inductivo. Primero probaremos el paso inductivo 
para las funciones $\psi_{-(n)}(x),$ en este
caso debemos suponer que se cumple Eq. (\ref{eq:espec-os}) y probar 
\begin{eqnarray}
\hat H\psi_{-(n+1)}(x) &=&\left[ E-(n+1)\hbar\omega\right]\psi_{-(n+1)}(x),\qquad {\rm con} \nonumber\\
\qquad \psi_{-(n+1)}(x)&=&a\psi_{-(n)}(x)= \left(\hat a\right)^{n+1}\psi(x).\nonumber
\end{eqnarray}
Esta igualdad es cierta, pues de  Eq. (\ref{eq:des-esca-os}) y  Eq. (\ref{eq:espec-os}) se llega a
\begin{eqnarray}
\hat H\psi_{-(n+1)}(x)& =&\hat H\hat a \psi_{-(n)}(x)=\left(\hat a\hat H-\hbar \omega \hat a\right)\psi_{-(n)}(x)\nonumber\\
& =& \hat a\hat H\psi_{-(n)}(x)- \hbar \omega \hat a\psi_{-(n)}(x)\nonumber\\
&=& \left( E-n\hbar\omega\right) \hat a\psi_{-(n)}(x)- \hbar \omega \hat a\psi_{-(n)}(x)\nonumber\\
&=& \left( E-n\hbar\omega-\hbar \omega  \right)\hat a\psi_{-(n)}(x)\nonumber\\
&=&\left[ E-(n+1)\hbar\omega\right]\psi_{-(n+1)}(x),\nonumber
\end{eqnarray}
as\'{\i} la igualdad Eq. (\ref{eq:espec-os}) se satisface para cualquier $n.$\\

Para el  caso de $\psi_{+n}(x)$ debemos suponer Eq. (\ref{eq:espec-0-os}) y probar
que se cumple 
\begin{eqnarray}
\hat H\psi_{+(n+1)}(x) &=&\left[ E+(n+1)\hbar\omega\right]\psi_{+(n+1)}(x),\quad {\rm con}\\
 \psi_{+(n+1)}(x)&=&a^{\dagger}\psi_{+(n)}(x)= \left(\hat a^{\dagger}\right)^{n+1}\psi(x).\nonumber
\end{eqnarray}
Esta igualdad tambi\'en es cierta, pues de  Eq. (\ref{eq:as-esca-os}) y Eq. (\ref{eq:espec-0-os}) se
encuentra 
\begin{eqnarray}
\hat H\psi_{+(n+1)}(x)& =&\hat H\hat a^{\dagger} \psi_{+(n)}(x)=
\left(\hat a^{\dagger}\hat H+\hbar \omega \hat a^{\dagger}\right)\psi_{+(n)}(x)\nonumber\\
&=& \hat a^{\dagger}\hat H\psi_{+(n)}(x)+\hbar \omega \hat a^{\dagger}\psi_{+(n)}(x)\nonumber\\
&=& \left( E+n\hbar\omega\right) \hat a^{\dagger}\psi_{+(n)}(x)+ 
\hbar \omega \hat a^{\dagger}\psi_{+(n)}(x)\nonumber\\
&=& \left( E+n\hbar\omega+\hbar \omega  \right)
\hat a^{\dagger}\psi_{+(n)}(x)\nonumber\\
&=&\left[ E+(n+1)\hbar\omega\right]\psi_{+(n+1)}(x).\nonumber
\end{eqnarray}
Por lo tanto la igualdad Eq. (\ref{eq:espec-0-os}) es v\'alidas para cualquier natural $n.$\\ 

Lo que hemos demostrado es que si $\psi(x)$ es funci\'on propia de $\hat H,$ con valor  propio $E,$ 
entonces, dado cualquier natural $n,$ las funciones 
$ \left(\hat a\right)^{n} \psi(x)$ y $\left(\hat a^{\dagger}\right)^{n} \psi(x) $ tambi\'en son
funciones propias del mismo operador, con valores propios  $E- n\hbar\omega$ y  $E+ n\hbar\omega,$ 
respectivamente.\\

Ahora, note que para cualquier valor $E$ existen un natural $\tilde n$ tal que el valor propio 
$\tilde E_{\tilde n}= E-\tilde n\hbar \omega,$
con funci\'on propia $ \psi_{-(\tilde n)}(x)=\left(\hat a\right)^{\tilde n} \psi(x),$ 
satisfce $\tilde E_{-(\tilde n)}= E-\tilde n\hbar \omega<0.$ Esto implica que $\psi_{\tilde n}(x)=0.$
Es decir, si $\psi(x)$ es una funci\'on propia de $\hat H$ existe un natural  $\tilde n$ tal que 
$(\hat a)^{\tilde n}\psi(x)=0.$ Note que en realidad existe un n\'umero infinito de posibles valores
de $\tilde n,$ pues si $(\hat a)^{\tilde n}\psi(x)=0,$ entonces tambi\'en se cumple
$(\hat a)^{\tilde n+1}\psi(x)=0.$\\

Supongamos $n^{\prime}$ es el m\'{\i}nimo de los valores posibles de $\tilde n$ tal que $(a)^{\tilde n}\psi(x)=0$ 
y definamos $N=n^{\prime}-1.$ Entonces,
la funci\'on $\psi_{0}(x)=\left(\hat a\right)^{n^{\prime}-1}\psi(x)= \left(\hat a\right)^{N}\psi(x)$ 
satisface $ \hat a \psi_{0}(x)=0.$ Note que $\psi_{0}(x)\not =0,$ pues de lo contrario $n^{\prime}$  no ser\'ia el m\'inimo
de los valores de $\tilde n,$ note tambi\'en que el valor de la energ\'ia de $\psi_{0}(x)$ es 
$$E_{0}=E-N\hbar \omega.$$
As\'i, podemos afirmar que  existe una funci\'on, $\psi_{0}(x)$, tal que 
\begin{eqnarray}
\hat a \psi_{0}(x)=0,\qquad \psi_{0}(x)\not =0\label{eq:aniqui-os}
\end{eqnarray}
a esta funci\'on le llamaremos estado base.  Con el estado base se encuentra
\begin{eqnarray}
\hat H \psi_{0}(x)=\hbar \omega\left(\hat a^{\dagger}\hat a +\frac{1}{2}\right)\psi_{0}(x)=
\frac{\hbar \omega}{2}\psi_{0}(x)= E_{0}\psi_{0}(x).
\end{eqnarray}
Por lo que $E_{0}=\hbar \omega/2$ es el valor propio de $\psi_{0}(x),$ pero este valor propio debe ser igual a  $E_{0}=E-N\hbar \omega,$
de donde 
\begin{eqnarray}
 E=E_{N}=\hbar\omega\left( N+\frac{1}{2}\right).
\end{eqnarray}
Como $\psi(x)$ es cualquier estado propio de $\hat H,$ los valores propio  $E$ son discretos.\\

Adem\'as, si definimos 
$\psi_{N}(x)=\left(\hat a^{\dagger}\right)^{N}\psi_{0}(x)$ y consideramos 
(\ref{eq:espec-0-os}), se llega a
\begin{eqnarray}
\hat H\psi_{N}(x)= (E_{0}+N\hbar \omega) \psi_{N}(x)= \hbar\omega\left( N+\frac{1}{2}\right)\psi_{N}(x).
\end{eqnarray}
As\'{\i}, $\psi_{N}(x)$ tiene el mismo valor propio que $ \psi(x).$ Por lo tanto, $\psi(x)$ y $\psi_{N}(x)$ satisfacen la misma ecuaci\'on diferencial de segundo orden con las mismas condiciones de borde, esto implica que estas funciones deben ser iguales.\\

\section{Estado base y ortonormalidad}

En el caso que estamos estudiado, cualquier soluci\'on de la ecuaci\'on de Schr$\rm \ddot o$dinger se puede expresar en t\'erminos del estado base. Entonces, basta conocer est\'a funci\'on para obtener todas las soluciones. El estado base lo hemos definido como la funci\'on que satisface
\begin{eqnarray}
\hat a\psi_{0}(x)&=&\frac{1}{\sqrt{2m\omega \hbar}}\left(\hat p-im\omega \hat x\right)\psi_{0}(x)=
\frac{1}{\sqrt{2m\omega \hbar}}\left(-i\hbar\frac{\partial }{\partial x}-im\omega x\right)\psi_{0}(x)\nonumber\\
&=&\frac{-i\hbar}{\sqrt{2m\omega \hbar}}\left(\frac{\partial }{\partial x}+\frac{m\omega}{\hbar} x\right)\psi_{0}(x)=0,
\nonumber
\end{eqnarray}
es decir, el estado base debe ser soluci\'on de la ecuaci\'on diferencial

\begin{eqnarray}
\frac{\partial \psi_{0}(x)}{\partial x}=-\frac{m\omega}{\hbar} x\psi_{0}(x),\nonumber
\end{eqnarray}
cuya soluci\'on es
\begin{eqnarray}
\psi_{0}(x)=Ae^{-\frac{m\omega}{\hbar}\frac{x^{2}}{2}}, \qquad A={\rm constante.}\nonumber
\end{eqnarray}
Para determinar la constante $A$ pediremos  
\begin{eqnarray}
\int_{-\infty}^{\infty}dx \psi_{0}(x)\psi_{0}(x)=1,\nonumber
\end{eqnarray}
que quiere decir que el estado base tiene norma unitaria.
De esta condici\'on tenemos
\begin{eqnarray}
\int_{-\infty}^{\infty}dx \psi_{0}(x)\psi_{0}(x)=\int_{-\infty}^{\infty}dx 
\left(Ae^{-\frac{m\omega}{\hbar}\frac{x^{2}}{2}}\right)^{2}
=A^{2}\int_{-\infty}^{\infty}dxe^{-\frac{m\omega}{\hbar}x^{2}}=1,\nonumber
\end{eqnarray}
entonces
\begin{eqnarray}
A=\frac{1}{\sqrt{\int_{-\infty}^{\infty}dxe^{-\frac{m\omega}{\hbar}x^{2}}}}.\nonumber
\end{eqnarray}
Ahora recordemos como calcular una integral de la forma
\begin{eqnarray}
I=\int_{-\infty}^{\infty}dxe^{-\alpha x^{2}},\qquad \alpha>0,\quad  \alpha={\rm constante}.\nonumber
\end{eqnarray}
Notablemente, es m\'as f\'acil calcular $I^{2},$ que es
\begin{eqnarray}
I^{2}&=&\left(\int_{-\infty}^{\infty}dxe^{-\alpha x^{2}}\right)\left(\int_{-\infty}^{\infty}dye^{-\alpha y^{2}}\right)
=\int_{-\infty}^{\infty}dx \int_{-\infty}^{\infty} dye^{-\alpha x^{2}}e^{-\alpha y^{2}}\nonumber\\
&=&
\int_{-\infty}^{\infty}dx \int_{-\infty}^{\infty} dye^{-\alpha\left(x^{2}+ y^{2}\right)},\nonumber
\end{eqnarray}
ocupando coordenadas polares tenemos
\begin{eqnarray}
I^{2}&=&\int_{0}^{\infty}drr \int_{0}^{2\pi}d\varphi e^{-\alpha r^{2}}=2\pi\int_{0}^{\infty}drre^{-\alpha r^{2}}
=2\pi \int_{0}^{\infty}dr\frac{(-1)}{2\alpha}\frac{d \left(e^{-\alpha r^{2}} \right)}{dr}\nonumber\\
&=&-\frac{\pi}{\alpha} e^{-\alpha r^{2}}\Bigg|_{0}^{\infty}=\frac{\pi}{\alpha}.\nonumber
\end{eqnarray}
De donde 
\begin{eqnarray}
I=\int_{-\infty}^{\infty}dxe^{-\alpha x^{2}}=\sqrt{\frac{\pi}{\alpha}},\nonumber
\end{eqnarray}
considerando este resultado encontramos 
\begin{eqnarray}
A=\left(\frac{m\omega}{\pi\hbar}\right)^{1/4}.\nonumber
\end{eqnarray}
As\'{\i}, el estado base normalizado es
\begin{eqnarray}
\psi_{0}(x)=\left(\frac{m\omega}{\pi\hbar}\right)^{1/4} e^{-\frac{m\omega}{\hbar}\frac{x^{2}}{2}}.
\end{eqnarray}
Con esta funci\'on y $a^{\dagger}$ podemos construir el resto de las funciones propias de $\hat H,$ que tienen  la forma
\begin{eqnarray}
\psi_{n}(x)=A_{n}\left(\hat a^{\dagger}\right)^{n}\psi_{0}(x),
\end{eqnarray}
donde, $A_{n}$ es una constante de normalizaci\'on. Antes de calcular expl\'icitamente
las funciones de onda $\psi_{n}(x),$ calcularemos la constante de normalizaci\'on $A_{n}.$
 Recordemos que las funciones $\psi_{n}(x)$ deben tener  norma unitaria, es decir,
\begin{eqnarray}
\left<\psi_{n}(x)|\psi_{n}(x)\right>=1.
\end{eqnarray}
Ahora,
\begin{eqnarray}
 \left<\psi_{n}|\psi_{n}\right>&=&\left<A_{n}\left(\hat a^{\dagger}\right)^{n}\psi_{0}(x)|A_{n}\left(\hat a^{\dagger}\right)^{n}\psi_{0}(x)\psi_{n}\right>\nonumber\\
&=&A_{n}^{*} A_{n} \left<\left(\hat a^{\dagger}\right)^{n}\psi_{0}(x)|\left(\hat a^{\dagger}\right)^{n}\psi_{0}(x)\psi_{n}\right>
\nonumber\\
&=& A_{n}^{*}A_{n} \left<\psi_{0}(x)|a^{n}\left(\hat a^{\dagger}\right)^{n}\psi_{0}(x)\right>\nonumber\\
&=&A_{n}^{*}A_{n} \int_{-\infty}^{\infty}dx \psi_{0}(x) (\hat a)^{n} 
\left(\hat a^{\dagger}\right)^{n}\psi_{0}(x). \label{eq:pp-oscilador}
\end{eqnarray}
Como $\left[\hat a,\hat a^{\dagger}\right]=1,$ se tiene 
$\left[\hat a,\left(\hat a^{\dagger}\right)^{n}\right]=n\left(\hat a^{\dagger}\right)^{n-1}$ y entonces
\begin{eqnarray}
(\hat a)^{n} \left(\hat a^{\dagger}\right)^{n}\psi_{0}(x)&=&(\hat a)^{n-1} \hat a\left(\hat a^{\dagger}\right)^{n}\psi_{0}\nonumber\\
&=&(\hat a)^{n-1} \left(\hat a\left(\hat a^{\dagger}\right)^{n}- \left(\hat a^{\dagger}\right)^{n} \hat a
+  \left(\hat a^{\dagger}\right)^{n} \hat a \right)\psi_{0}(x)\nonumber\\
&=& (\hat a)^{n-1} \left(\left[\hat a,\left(\hat a^{\dagger}\right)^{n}\right]+ \left(\hat a^{\dagger}\right)^{n} \hat a \right)\psi_{0}(x)\nonumber\\
&=&(\hat a)^{n-1} \left(n \left(\hat a^{\dagger}\right)^{n-1}+ \left(\hat a^{\dagger}\right)^{n} \hat a \right)\psi_{0}(x)
\nonumber\\
&=&
(\hat a)^{n-1} \left(n \left(\hat a^{\dagger}\right)^{n-1}\psi_{0}(x)+ 
\left(\hat a^{\dagger}\right)^{n} \hat a \psi_{0}(x)\right)\nonumber\\
&=& n (\hat a)^{n-1}  \left(\hat a^{\dagger}\right)^{n-1}\psi_{0}(x).
\end{eqnarray}
Este resultado lo podemos aplicar de forma reiterada $k$ veces, donde $k\leq n,$ por lo que
\begin{eqnarray}
(\hat a)^{n} 
\left(\hat a^{\dagger}\right)^{n}\psi_{0}(x)&=& n(n-1)(n-2)\cdots (n-(k-1)) (\hat a)^{n-k}  
\left(\hat a^{\dagger}\right)^{n-k}\psi_{0}(x)\nonumber \\
&=&\frac{n!}{(n-k)!}a^{n-k}\left(\hat a^{\dagger}\right)^{n-k}\psi_{0}(x).
\end{eqnarray}
El m\'aximo valor que puede tomar $k$ es $n,$ de donde  
\begin{eqnarray}
(\hat a)^{n} 
\left(\hat a^{\dagger}\right)^{n}\psi_{0}(x)=n!\psi_{0}(x).
\end{eqnarray}
Introduciendo esta igualdad en (\ref{eq:pp-oscilador}), se obtiene
\begin{eqnarray}
\left< \psi_{n}(x)|\psi_{n}(x)\right>=|A_{n}|^{2}n!\int_{-\infty}^{\infty} \psi_{0}(x)\psi_{0}(x)=1,
\end{eqnarray}
as\'{\i}
\begin{eqnarray}
A_{n}=\frac{1}{\sqrt{ n!}}.
\end{eqnarray}
Por lo tanto, las funciones de onda normalizadas son 
\begin{eqnarray}
\psi_{n}(x)= \frac{1}{\sqrt{ n!}} \left(\hat a^{\dagger}\right)^{n}\psi_{0}(x).
\label{eq:funcion-onda-oscilador}
\end{eqnarray}

\section{Polinomios de Hermite}

Veamos la forma expl\'icita de las funciones $\psi_{n}(x),$ primero consideraremos  que  
\begin{eqnarray}
\psi_{0}(x)&=&\left(\frac{m\omega}{\pi \hbar}\right)^{1/4}e^{-\frac{m\omega x^{2}}{2\hbar}},\nonumber\\
a^{\dagger}&=&\frac{1}{\sqrt{2m\omega\hbar}} \left(\hat p+im\omega \hat x\right)=
\frac{1}{\sqrt{2m\omega\hbar}} \left(-i\hbar\frac{\partial }{\partial x}+im\omega \hat x\right)\nonumber\\
&=&\frac{-i\hbar}{\sqrt{2m\omega\hbar}} \left(\frac{\partial }{\partial x}-\frac{m\omega}{\hbar} \hat x\right),
\nonumber
\end{eqnarray}
por lo que, ocupando el cambio de variable
\begin{eqnarray}
\zeta^{2}= \frac{m\omega x^{2}}{\hbar},\qquad \zeta= \sqrt{\frac{m\omega }{\hbar}} x, \qquad
\frac{d}{d\zeta}= \sqrt{\frac{\hbar}{m\omega }}\frac{d}{dx},\label{eq:cambio-hermite}
\end{eqnarray}
se tiene
\begin{eqnarray}
\psi_{0}(\zeta)=\left(\frac{m\omega}{\pi \hbar}\right)^{1/4}e^{-\frac{\zeta^{2}}{2}},\qquad 
a^{\dagger}=\frac{-i}{\sqrt{2}} \left(\frac{\partial }{\partial \zeta }- \zeta \right).
\nonumber
\end{eqnarray}
Por lo tanto, introduciendo estos resultados en Eq. (\ref{eq:funcion-onda-oscilador}) se llega a 
\begin{eqnarray}
\psi_{n}(\zeta)=\left(\frac{m\omega}{\pi \hbar}\right)^{1/4} 
\frac{(-i)^{n}}{\sqrt{n!2^{n}}}\left( \frac{\partial }{\partial \zeta }- \zeta\right)^{n} e^{-\frac{\zeta^{2}}{2}}.
\label{eq:fun-onda-os-1}
\end{eqnarray}
Note que
\begin{eqnarray}
e^{\frac{\zeta^{2}}{2}}\frac{d}{d\zeta }\left(fe^{-\frac{\zeta^{2}}{2}}\right)&=&e^{\frac{\zeta^{2}}{2}}
\left( \frac{d f}{d\zeta }e^{-\frac{\zeta^{2}}{2}}+f \frac{d\left( e^{-\frac{\zeta^{2}}{2}}\right)}{d\zeta }\right)\nonumber\\
 &=&e^{\frac{\zeta^{2}}{2}}\left( \frac{d f}{d\zeta }e^{-\frac{\zeta^{2}}{2}}-f\zeta e^{-\frac{\zeta^{2}}{2}}\right)=
\left( \frac{d f}{d\zeta }-f\zeta\right),\nonumber
\end{eqnarray}
es decir, 
\begin{eqnarray}
e^{\frac{\zeta^{2}}{2}}\frac{d}{d\zeta }\left(fe^{-\frac{\zeta^{2}}{2}}\right)=
\left( \frac{d }{d\zeta }-\zeta\right)f.\label{eq:hermi-1}
\end{eqnarray}
En general se cumple 
\begin{eqnarray}
e^{\frac{\zeta^{2}}{2}}\frac{d^{n}}{d\zeta^{n} }\left(fe^{-\frac{\zeta^{2}}{2}}\right)=
\left( \frac{d }{d\zeta }-\zeta\right)^{n}f.\label{eq:hermi-2}
\end{eqnarray}
Probaremos esta afirmaci\'on por inducci\'on. La base inductiva, $n=1$ ya ha sido demostrada, falta demostrar el paso inductivo.
Aqu\'i debemos suponer que se cumple Eq. (\ref{eq:hermi-2}) y demostrar
\begin{eqnarray}
e^{\frac{\zeta^{2}}{2}}\frac{d^{n+1}}{d\zeta^{n+1} }\left(fe^{-\frac{\zeta^{2}}{2}}\right)=
\left( \frac{d }{d\zeta }-\zeta\right)^{n+1}f.\label{eq:hermi-3}
\end{eqnarray}
Ocupando la hip\'otesis de inducci\'on,
se encuentra
\begin{eqnarray}
\left( \frac{d }{d\zeta }-\zeta\right)^{n+1}f=\left( \frac{d }{d\zeta }-\zeta\right)\left( \frac{d }{d\zeta }-\zeta\right)^{n}f
= \left( \frac{d }{d\zeta }-\zeta\right)e^{\frac{\zeta^{2}}{2}}\frac{d^{n}}{d\zeta^{n} }\left(fe^{-\frac{\zeta^{2}}{2}}\right),
\nonumber
\end{eqnarray}
as\'{\i} definiendo  $\tilde f= e^{\frac{\zeta^{2}}{2}}\frac{d^{n}}{d\zeta^{n} }\left(fe^{-\frac{\zeta^{2}}{2}}\right)$
y usando Eq. (\ref{eq:hermi-1}) se obtiene
\begin{eqnarray}
\left( \frac{d }{d\zeta }-\zeta\right)^{n+1}f&=&
 \left( \frac{d }{d\zeta }-\zeta\right)\tilde f= e^{\frac{\zeta^{2}}{2}}\frac{d}{d\zeta }
\left(e^{-\frac{\zeta^{2}}{2}}\tilde f\right)\nonumber\\
&=& e^{\frac{\zeta^{2}}{2}}\frac{d}{d\zeta }
\left(e^{-\frac{\zeta^{2}}{2}}e^{\frac{\zeta^{2}}{2}}\frac{d^{n}}{d\zeta^{n} }\left(fe^{-\frac{\zeta^{2}}{2}}\right) \right)
\nonumber\\
&=& e^{\frac{\zeta^{2}}{2}}\frac{d}{d\zeta }
\left(\frac{d^{n}}{d\zeta^{n} }\left(fe^{-\frac{\zeta^{2}}{2}}\right) \right)= e^{\frac{\zeta^{2}}{2}}
\frac{d^{n+1}}{d\zeta^{n+1} }\left(fe^{-\frac{\zeta^{2}}{2}}\right),
\nonumber
\end{eqnarray}
que es lo que se queria demostrar, esto implica que Eq. (\ref{eq:hermi-2}) es v\'alida
para cualquier n\'umero natural $n.$ Introduciendo  el resultado  Eq. (\ref{eq:hermi-2})  en Eq. (\ref{eq:fun-onda-os-1}) se llega a 
\begin{eqnarray}
\psi_{n}(\zeta)&=&\left(\frac{m\omega}{\pi \hbar}\right)^{1/4} 
\frac{(-i)^{n}}{\sqrt{n!2^{n}}}e^{\frac{\zeta^{2}}{2}}\frac{\partial^{n} }{\partial \zeta^{n}}\left( e^{-\frac{\zeta^{2}}{2}}
e^{-\frac{\zeta^{2}}{2}}\right)\nonumber\\
&=&\left(\frac{m\omega}{\pi \hbar}\right)^{1/4} 
\frac{(i)^{n}}{\sqrt{n!2^{n}}}e^{-\frac{\zeta^{2}}{2}} (-)^{n} e^{\zeta^{2}}\frac{\partial^{n}  }{\partial \zeta^{n}}
e^{-\zeta^{2}}.
\label{eq:fun-onda-os-2}
\end{eqnarray}
Definiremos el  polinomio de Hermite de grado $n$ como
\begin{eqnarray}
H_{n}(\zeta)= (-)^{n} e^{\zeta^{2}}\frac{\partial^{n}  }{\partial \zeta^{n}}
e^{-\zeta^{2}}.\label{eq:rodriguez-hermite}
\end{eqnarray}
A esta expresi\'on tambi\'en se le llama f\'ormula de Rodrigues para los polinomios
de Hermite. En particular se tiene 
\begin{eqnarray}
H_{0}(\zeta)=1,\quad  H_{1}(\zeta)=2\zeta,\quad  H_{2}(\zeta)=-2+4\zeta^{2}, \cdots.
\label{eq:primeros-hermite}
\end{eqnarray}
Por lo tanto, la funci\'on de onda Eq. (\ref{eq:fun-onda-os-2}) se escribe como
\begin{eqnarray}
\psi_{n}(\zeta)
=\left(\frac{m\omega}{\pi \hbar}\right)^{1/4} 
\frac{(i)^{n}}{\sqrt{n!2^{n}}}e^{-\frac{\zeta^{2}}{2}} H_{n}(\zeta)
\end{eqnarray}
Debido a que las funciones de onda forman un conjunto ortonormal de funciones, 
ocupando el cambio de variable Eq. (\ref{eq:cambio-hermite}), tenemos
\begin{eqnarray}
\delta_{nl}&=&<\psi_{n}(x)|\psi_{l}(x)>=\int_{-\infty}^{\infty}dx\psi_{n}^{*}(x)\psi_{l}(x)=\int_{-\infty}^{\infty}
d\zeta \left(\sqrt{\frac{\hbar}{m\omega}}\right)\psi_{n}^{*}(\zeta)\psi_{l}(\zeta)\nonumber\\
&=& \sqrt{\frac{\hbar}{m\omega}} \left(\frac{m\omega}{\pi \hbar}\right)^{1/2} 
\frac{(-i)^{n}}{\sqrt{n!2^{n}}} 
\frac{(i)^{l}}{\sqrt{l!2^{l}}}\int_{-\infty}^{\infty}d\zeta e^{-\zeta^{2}}H_{n}(\zeta)H_{l}(\zeta)\nonumber, 
\end{eqnarray}
es decir 
\begin{eqnarray}
\int_{-\infty}^{\infty}d\zeta e^{-\zeta^{2}}H_{n}(\zeta)H_{l}(\zeta)= \sqrt{\pi}2^{n}n!\delta_{nl}.
\end{eqnarray}
De donde, los polinomios de Hermite, forman un conjunto de funciones ortonormales con 
funci\'on de peso $e^{-\zeta^{2}}.$

\newpage

\section{Funci\'on generadora}

Ahora veremos una funci\'on que est\'a intimamente relacionada
con los polinomios de Hermite, la llamada funci\'on generadora.
Primero recordemos que cualquier funci\'on, $f(z),$ bien comportada se puede expresar en
su serie de Taylor 
\begin{eqnarray}
f(z)=\sum_{n\geq 0}\frac{z^{n}}{n!}\left(\frac{d^{n}f(z)}{dz^{n}}\Bigg|_{z=0}\right),
\end{eqnarray}
en particular
\begin{eqnarray}
W(\zeta,t)=e^{2\zeta t -t^{2}}=\sum_{n\geq 0}\frac{t^{n}}{n!}
\left(\frac{\partial ^{n}W(\zeta,t)}{\partial t^{n}}\Bigg|_{t=0}\right).
\label{eq:generadora-hermite}
\end{eqnarray}
Note que, como 
\begin{eqnarray}
2\zeta t -t^{2}=-\left(t^{2}-2\zeta t+\zeta^{2}-\zeta^{2}\right)=-\left(\left(t-\zeta\right)^{2}-\zeta^{2}\right)=
-\left(t-\zeta\right)^{2}+\zeta^{2}, \nonumber
\end{eqnarray}
entonces 
\begin{eqnarray}
\frac{\partial ^{n}W(\zeta,t)}{\partial t^{n}}=\frac{\partial ^{n} e^{2\zeta t -t^{2}} }{\partial t^{n}}=
 \frac{\partial ^{n} e^{\zeta ^{2}-(t-\zeta)^{2} } }{\partial t^{n}}=
e^{\zeta^{2}}\frac{\partial ^{n} e^{-(t-\zeta)^{2}}}{\partial t^{n}}.
\end{eqnarray}
Adem\'as, con el cambio de variable $u=\zeta-t$ se tiene
\begin{eqnarray}
\frac{\partial }{\partial t}=\frac{\partial u}{\partial t}\frac{\partial }{\partial u}=
(-)\frac{\partial }{\partial u}, \qquad u|_{t=0}=\zeta.
\end{eqnarray}
Por lo tanto, considerando Eq. (\ref{eq:rodriguez-hermite}) se tiene
\begin{eqnarray}
\frac{\partial ^{n}W(\zeta,t)}{\partial t^{n}}\Bigg|_{t=0}&=&e^{\zeta^{2}}\frac{\partial ^{n} e^{-(t-\zeta)^{2}}}{\partial t^{n}}\Bigg|_{t=0}
=e^{\zeta^{2}}(-)^{n}\frac{\partial ^{n} e^{-u^{2}}}{\partial u^{n}}\Bigg|_{u=\zeta}\nonumber\\
&=&e^{\zeta^{2}}(-)^{n}\frac{\partial  ^{n} e^{-\zeta^{2}}}{\partial \zeta^{n}}
= (-)^{n}e^{\zeta^{2}}\frac{\partial ^{n} e^{-\zeta^{2}}}{\partial \zeta^{n}}=H_{n}(\zeta),
\end{eqnarray}
sustituyendo este resultado en Eq. (\ref{eq:generadora-hermite}) se llega a
\begin{eqnarray}
W(\zeta,t)=e^{2\zeta t -t^{2}}=\sum_{n\geq 0}\frac{t^{n}}{n!} H_{n}(\zeta),
\label{eq:genera-herm}
\end{eqnarray}
que es la llamada funci\'on generadora de los polinomios de Hermite. Con  esta funci\'on 
podemos obtener la forma expl\'{\i}cita de las funciones $ H_{n}(\zeta).$ 
Para esto recordemos los resultados 
\begin{eqnarray}
(a+b)^{N}&=&\sum_{k=0}^{N}C_{k}^{N}a^{N-k}b^{k},\qquad C_{k}^{N}=\frac{N!}{k!(N-k)!},\nonumber\\
e^{z}&=&\sum_{N\geq 0}\frac{z^{N}}{N!}.\nonumber
\end{eqnarray}
Por lo tanto, 
\begin{eqnarray}
W(\zeta,t)=e^{2\zeta t -t^{2}}&=&\sum_{N\geq 0}\frac{\left(2\zeta t -t^{2}\right)^{N}}{N!}=
\sum_{N\geq 0}\frac{1}{N!}\sum_{k=0}^{N}C_{k}^{N}(2\zeta t)^{N-k}\left(-t^{2}\right)^{k}\nonumber\\ 
&=&\sum_{N\geq 0}\frac{1}{N!}\sum_{k=0}^{N}C_{k}^{N}(2\zeta)^{N-k}(-)^{k}t^{2k+N-k}\nonumber\\
&=&\sum_{N\geq 0}\sum_{k=0}^{N}\frac{(-)^{k}}{N!}C_{k}^{N}(2\zeta)^{N-k}t^{N+k}.\nonumber
\end{eqnarray}
Para hacer m\'ar f\'acil el c\'alculo, definamos $n=N+k,$ entonces $N=n-k$ y $N-k=n-2k.$ Note que el m\'aximo valor que puede tener
$k$ es $N,$ es decir $N-k=n-2k\geq 0,$ que implica $k\leq n/2.$ As\'{\i}, si $n$ es par, 
el m\'aximo valor que puede tomar $k$ es $n/2.$ Pero si $n$ es impar, $k$ no puede tomar 
el valor $n/2$ porque \'este no es un n\'umero natural. En este caso el m\'aximo valor que puede
tomar $k$ es $(n-1)/2.$ Definiremos como $\left[\frac{n}{2}\right]$ como el m\'aximo entero menor o igual
a $n/2.$ Entonces, el m\'aximo valor que puede tomar $k$ es $\left[\frac{n}{2}\right].$ Por lo tanto,
si tomamos como variable de suma a $n$ en lugar de $N,$ se tiene
\begin{eqnarray}
W(\zeta,t)&=&e^{2\zeta t -t^{2}}
=\sum_{n\geq 0}\sum_{k=0}^{\left[\frac{n}{2}\right]}\frac{(-)^{k}}{(n-k)!}C_{k}^{n-k}(2\zeta)^{n-k-k}t^{n}\nonumber\\
&=&\sum_{n\geq 0}\frac{t^{n}}{n!}\sum_{k=0}^{\left[\frac{n}{2}\right]}\frac{(-)^{k}n!}{(n-k)!}C_{k}^{n-k}(2\zeta)^{n-2k}
\nonumber\\
&=&\sum_{n\geq 0}\frac{t^{n}}{n!}\sum_{k=0}^{\left[\frac{n}{2}\right]}
\frac{(-)^{k}n!}{(n-k)!}\frac{(n-k)!}{k!(n-2k)!}(2\zeta)^{n-2k}
\nonumber\\
&=&\sum_{n\geq 0}\frac{t^{n}}{n!}\sum_{k=0}^{\left[\frac{n}{2}\right]}
\frac{(-)^{k}n!}{k!(n-2k)!}(2\zeta)^{n-2k}
\nonumber\\
&=&\sum_{n\geq 0}\frac{t^{n}}{n!}H_{n}(\zeta).
\end{eqnarray}
Entonces, 
\begin{eqnarray}
H_{n}(\zeta)=\sum_{k=0}^{\left[\frac{n}{2}\right]}\frac{(-)^{k}n!}{k!(n-2k)!}(2\zeta)^{n-2k}.
\end{eqnarray}
Con la funci\'on generatriz se pueden probar varias propiedades de los polinomios de Hermite,
por ejemplo, note que
\begin{eqnarray}
W(-\zeta,t)=e^{2(-\zeta)t -t^{2}}=e^{2\zeta(-t) -(-t)^{2}}= W(\zeta,-t).
\end{eqnarray}
Utilizando Eq. (\ref{eq:genera-herm}) se tiene 
\begin{eqnarray}
W(-\zeta,t)=\sum_{n\geq 0}\frac{t^{n}}{n!} H_{n}(-\zeta)= W(\zeta,-t)=
\sum_{n\geq 0}\frac{(-t)^{n}}{n!} H_{n}(\zeta)=\sum_{n\geq 0}\frac{t^{n}}{n!} (-)^{n}H_{n}(\zeta),
\nonumber
\end{eqnarray}
igualdando t\'ermino a t\'ermino se encuentra
\begin{eqnarray}
H_{n}(-\zeta)=(-)^{n} H_{n}(\zeta).
\end{eqnarray}
Por lo tanto, los polinomios de Hermite, son pares o impares dependiendo de su grado $n.$ 
En particular $H_{2n+1}(-0)=(-)^{2n+1} H_{2n+1}(0),$ es decir 
$$H_{2n+1}(0)=0.$$ 
Adem\'as, 
\begin{eqnarray}
W(\zeta=0,t)&=&e^{-t^{2}}=\sum_{n\geq 0}\frac{\left(-t^{2}\right)^{n}}{n!}=\sum_{n\geq 0}\frac{\left(-\right)^{n}t^{2n}}{n!}\nonumber\\
W(\zeta=0,t)&=&\sum_{n\geq 0}\frac{t^{n}}{n!} H_{n}(0)=\sum_{n\geq 0}\frac{H_{2n}(0)}{(2n)!} t^{2n}.\nonumber
\end{eqnarray}
Igualando t\'ermino a t\'ermino estas dos expresiones se llega a 
$$ H_{2n}(0)=(-)^{n}\frac{(2n)!}{n!}.$$
Tambi\'en se tiene el resultado
\begin{eqnarray}
\frac{\partial W(\zeta,t)}{\partial \zeta}=\frac{\partial e^{2\zeta t-t^{2}}}{\partial \zeta}=2t e^{2\zeta t-t^{2}}=2tW(\zeta,t).
\end{eqnarray}
Adicionalmente, considerando Eq. (\ref{eq:genera-herm}) y que $H_{0}(\zeta)=1$ se encuentra
\begin{eqnarray}
\frac{\partial W(\zeta,t)}{\partial \zeta}&=&\sum_{n\geq 0}\frac{t^{n}}{n!} \frac{dH_{n}(\zeta)}{d\zeta}=
\sum_{n\geq 1}\frac{t^{n}}{n!} \frac{dH_{n}(\zeta)}{d\zeta},\nonumber\\
2tW(\zeta,t)&=&2t\sum_{n\geq 0}\frac{t^{n}}{n!} H_{n}(\zeta)= \sum_{n\geq 0}\frac{2t^{n+1}}{n!} H_{n}(\zeta)\nonumber \\
&=&
\sum_{n\geq 1}\frac{2t^{n}}{(n-1)!} H_{n-1}(\zeta)
= \sum_{n\geq 1}\frac{t^{n}}{n!} 2nH_{n-1}(\zeta).
\end{eqnarray}
De donde, 
\begin{eqnarray}
\frac{dH_{n}(\zeta)}{d\zeta}= 2nH_{n-1}(\zeta).\label{eq:recurr-hermite}
\end{eqnarray}

\subsection{Ecuaci\'on de Hermite}

De la f\'ormula de Rodr\'{\i}guez para los polinomios de Hermite Eq. (\ref{eq:rodriguez-hermite})
se tiene
\begin{eqnarray}
H_{n}(\zeta)&=&(-)^{n}e^{\zeta^{2}}\frac{d^{n} e^{-\zeta^{2}}}{d\zeta^{n}}=(-)^{n} e^{\zeta^{2}}
\frac{d}{d\zeta}\left(\frac{d^{n-1} e^{-\zeta^{2}}}{d\zeta^{n-1}}\right)\nonumber\\
&=&(-) e^{\zeta^{2}}\frac{d}{d\zeta}\left[e^{-\zeta^{2}}\left((-)^{n-1}e^{\zeta^{2}}\frac{d^{n-1} e^{-\zeta^{2}}}{d\zeta^{n-1}}\right)\right]\nonumber\\
&=&(-) e^{\zeta^{2}} \frac{d}{d\zeta}\left(e^{-\zeta^{2}}H_{n-1}(\zeta)\right)\nonumber\\
&=&(-) e^{\zeta^{2}}\left(-2\zeta e^{-\zeta^{2}}H_{n-1}(\zeta)+e^{-\zeta^{2}}\frac{dH_{n-1}(\zeta)}{d\zeta}\right)\nonumber\\
&=&
2\zeta H_{n-1}(\zeta)-\frac{dH_{n-1}(\zeta)}{d\zeta}.\nonumber
\end{eqnarray}
es decir, 
\begin{eqnarray}
H_{n}(\zeta)=2\zeta H_{n-1}(\zeta)-\frac{dH_{n-1}(\zeta)}{d\zeta}.
\end{eqnarray}
Derivando esta igualdad con respecto a $\zeta,$ se llega a
\begin{eqnarray}
\frac{dH_{n}(\zeta)}{d\zeta}=2 H_{n-1}(\zeta)+2\zeta\frac{d H_{n-1}(\zeta)}{d\zeta}-\frac{d^{2}H_{n-1}(\zeta)}{d\zeta^{2}}.
\end{eqnarray}
Adem\'as, usando Eq. (\ref{eq:recurr-hermite}) se encuentra
\begin{eqnarray}
2nH_{n-1}(\zeta)=2 H_{n-1}(\zeta)+2\zeta\frac{d H_{n-1}(\zeta)}{d\zeta}-\frac{d^{2}H_{n-1}(\zeta)}{d\zeta^{2}}
\end{eqnarray}
que se puede escribir como 
\begin{eqnarray}
\frac{d^{2}H_{n}(\zeta)}{d\zeta^{2}}-2\zeta\frac{d H_{n}(\zeta)}{d\zeta}+2n H_{n}(\zeta)=0, 
\label{eq:eq-de-hermite}
\end{eqnarray}
esta es la llamada ecuaci\'on  de Hermite.

\section{M\'etodo tradicional}

Ahora ocuparemos el m\'etodo tradicional para resolver la ecuaci\'on de onda
del oscilador arm\'onico
\begin{eqnarray}
\left(-\frac{\hbar^{2}}{2m} \frac{\partial^{2}}{\partial x^{2}} +
\frac{m\omega^{2}}{2} x^{2}\right)\psi(x)=E\psi(x).
\label{eq:ecua-oscilador1}
\end{eqnarray}
Con el cambio de variable Eq. (\ref{eq:cambio-hermite}) la ecuaci\'on (\ref{eq:ecua-oscilador1}) toma la forma
\begin{eqnarray}
\left(- \frac{\partial^{2}}{\partial \zeta^{2}} +
\zeta^{2}\right)\psi(\zeta )=\frac{2E}{\hbar \omega}\psi(\zeta).
\label{eq:ecua-oscilador}
\end{eqnarray}
Note que en el l\'{\i}mite $\zeta\to \infty$ se tiene la ecuaci\'on asint\'otica
\begin{eqnarray}
\left(- \frac{\partial^{2}}{\partial \zeta^{2}} +
\zeta^{2}\right)\psi(\zeta )\approx 0.
\end{eqnarray}
Proponemos como soluci\'on  asint\'otica a la funci\'on $\psi(\zeta)=e^{-\frac{\zeta^{2}}{2}},$
que satisface
\begin{eqnarray}
\frac{d\psi}{d \zeta}&=& -\zeta\psi(\zeta), \nonumber\\
 \frac{d^{2}\psi}{d \zeta^{2}}&=&-\frac{d}{d\zeta}\left(\zeta\psi(\zeta)\right)
=-\left(\psi(\zeta)-\zeta^{2}\psi(\zeta)\right)\approx -\zeta^{2}\psi(\zeta)\nonumber
\end{eqnarray}
Por lo tanto, si $\zeta >>1$ se tiene 
\begin{eqnarray}
\left( \frac{d^{2}}{d \zeta ^{2}}-\zeta^{2}\right)\psi(\zeta)\approx 0.
\end{eqnarray}
As\'{\i}, cuando $\zeta \to \infty$ las soluciones de Eq. (\ref{eq:ecua-oscilador}) deben ser de la forma
$$\psi(\zeta)\approx e^{-\frac{\zeta^{2}}{2}},$$ 
para el caso general, propondremos como
soluci\'on 
\begin{eqnarray}
\psi(\zeta)= e^{-\frac{\zeta^{2}}{2}}\phi(\zeta),
\end{eqnarray}
con $\phi(\zeta)$ una funci\'on que crece menos r\'apido que $e^{-\frac{\zeta^{2}}{2}}$ cuando $\zeta\to \infty.$
Para esta propuesta se tiene 
\begin{eqnarray}
\frac{d\psi(\zeta)}{d\zeta}&=& e^{-\frac{\zeta^{2}}{2}}\left(-\zeta \phi(\zeta)+
\frac{d\phi(\zeta)}{d\zeta}\right),\nonumber\\
\frac{d^{2}\psi(\zeta)}{d\zeta^{2}}&=& e^{-\frac{\zeta^{2}}{2}}
\left(\frac{d^{2} \phi(\zeta)}{d\zeta^{2}}
 -2\zeta \frac{d\phi(\zeta)}{d\zeta}+\left(\zeta^{2}-1\right)\phi(\zeta)\right).
\label{eq:talacha-hermite}
\end{eqnarray}
Sustituyendo Eq. (\ref{eq:talacha-hermite}) en Eq. (\ref{eq:ecua-oscilador}) se encuentra
\begin{eqnarray}
& & e^{-\frac{\zeta^{2}}{2}}(-)
\left(\frac{d^{2} \phi(\zeta)}{d\zeta^{2}}
 -2\zeta \frac{d\phi(\zeta)}{d\zeta}+\left(\zeta^{2}-1\right)\phi(\zeta)\right)
+e^{-\frac{\zeta^{2}}{2}}\zeta^{2}\psi(\zeta)\nonumber\\
&=&
\frac{2E}{\hbar \omega}e^{-\frac{\zeta^{2}}{2}}\zeta^{2}\phi(\zeta).
\nonumber
\end{eqnarray}
De donde 
\begin{eqnarray}
-\frac{d^{2} \phi(\zeta)}{d\zeta^{2}}
+2\zeta \frac{d\phi(\zeta)}{d\zeta} +\phi(\zeta)=\frac{2E}{\hbar \omega}\phi(\zeta),
\end{eqnarray}
es decir 
\begin{eqnarray}
\frac{d^{2} \phi(\zeta)}{d\zeta^{2}}
-2\zeta \frac{d\phi(\zeta)}{d\zeta} +
\left(\frac{2E}{\hbar \omega}-1\right) \phi(\zeta)=0.
\label{eq:ecua-oscilador3}
\end{eqnarray}
Para resover esta ecuaci\'on propondremos la soluci\'on en serie de potencia
\begin{eqnarray}
\phi(\zeta)=\sum_{n\geq 0} a_{n}\zeta^{n},
\end{eqnarray}
de la cual se obtiene
\begin{eqnarray}
 \left(\frac{2E}{\hbar \omega}-1\right) \phi(\zeta)&=&\sum_{n\geq 0} \left(\frac{2E}{\hbar \omega}-1\right) a_{n}\zeta^{n},
\nonumber\\
-2\zeta \frac{d\phi(\zeta)}{d\zeta}&=&-2\zeta \sum_{n\geq 0} n a_{n}\zeta^{n-1}=\sum_{n\geq 0} (-2n) a_{n}\zeta^{n},
\nonumber\\
\frac{d^{2}\phi(\zeta)}{d\zeta^{2}}&=&\sum_{n\geq 0} n(n-1) a_{n}\zeta^{n-2}= \sum_{n\geq 2} n(n-1) a_{n}\zeta^{n-2}
\nonumber\\
&=& \sum_{n\geq 0} (n+2)(n+1) a_{n+2}\zeta^{n}. \nonumber
\end{eqnarray}
Por lo tanto, 
\begin{eqnarray}
& &\frac{d^{2} \phi(\zeta)}{d\zeta^{2}}
-2\zeta \frac{d\phi(\zeta)}{d\zeta} +
\left(\frac{2E}{\hbar \omega}-1\right) \phi(\zeta)\nonumber\\
&=&\sum_{n\geq 1} \left( \left[ \left(\frac{2E}{\hbar \omega}-1\right) -2n \right]a_{n}+(n+2)(n+1) a_{n+2}\right)\zeta^{n}=0,
\nonumber\\
\nonumber
\end{eqnarray}
de donde 
\begin{eqnarray}
a_{n+2}&=&\frac{\left( 2n -\left(\frac{2E}{\hbar \omega}-1\right) \right)a_{n}}{ (n+2)(n+1)},
\end{eqnarray}
es decir
\begin{eqnarray}
\frac{a_{n+2}}{a_{n}}&=&\frac{\left( 2n -\left(\frac{2E}{\hbar \omega}-1\right) \right)}{ (n+2)(n+1)}.
\end{eqnarray}
Note que esta relaci\'on de recurrencia separa los t\'erminos pares e impares, por lo que se tendr\'an series de
la forma $\sum_{n\geq 0}a_{2n}\zeta^{2n}$ y $\sum_{n\geq 0}a_{2n+1}\zeta^{2n+1}.$ 
Para el caso par, si $n\to \infty,$  se tiene
\begin{eqnarray}
\frac{a_{2(n+1)}}{a_{2n}}\to \frac{4n}{ (2n+2)(2n+1)}\approx\frac{1}{n},
\end{eqnarray}
que es el mismo comportamiento que tiene los coeficientes de Taylor 
de la serie
\begin{eqnarray}
e^{\zeta^{2}}=\sum_{n\geq 0}\frac{\zeta^{2n}}{n!}.
\end{eqnarray}
Por lo tanto, en este caso la soluci\'on $\phi(\zeta)$ tiene el mismo comportamiento asint\'otico que $e^{\frac{\zeta^{2}}{2}}.$
Para el caso impar, la serie tiene el mismo comportamiento asint\'otico que $\zeta e^{\frac{\zeta^{2}}{2}}.$
Ninguno de estos dos caso debe ocurrir, pues $\phi(\zeta)$ debe estar dominada por $e^{-\frac{\zeta^{2}}{2}}.$
El problema se resuelve si  $\phi(\zeta)$ no es una serie, sino un polinomio.
Esto se cumple si despu\'es de cierto n\'umero todos los t\'erminos de la serie son cero, es decir $a_{n+2}/a_{n}=0.$ 
Lo que implica
\begin{eqnarray}
2n -\left(\frac{2E}{\hbar \omega}-1\right)=0,
\end{eqnarray}
de donde,
\begin{eqnarray}
E=E_{n}=\hbar\omega\left(n+\frac{1}{2}\right).
\end{eqnarray}
Entonces la energ\'{\i}a debe ser  discreta y las funciones
$\phi(\zeta)$ son polinomios. Sustituyendo $E_{n}$ en
Eq. (\ref{eq:ecua-oscilador3}) se obtiene
\begin{eqnarray}
\frac{d^{2} \phi(\zeta)}{d\zeta^{2}}
-2\zeta \frac{d\phi(\zeta)}{d\zeta} +2n \phi(\zeta)=0,
\end{eqnarray}
que es la ecuaci\'on de Hermite (\ref{eq:eq-de-hermite}).
Por lo tanto, las funciones $\phi(\zeta)$ son los
polinomios de Hermite.\\

As\'{\i}, las soluciones de la ecuaci\'on de onda para el
oscilador arm\'onico son  de la forma
\begin{eqnarray}
\psi_{n}(\zeta)=A_{n} e^{-\frac{\zeta^{2}}{2}}H_{n}(\zeta),
\end{eqnarray}
con $H_{n}(\zeta)$ el polinomio de Hermite de grado $n.$

\section{Oscilador en campo el\'ectrico constante}

Ahora veamos como se resuelve el problema de un oscilador arm\'onico
en un campo el\'ectrico constante.\\

El operador Hamiltoniano para un oscilador en un campo magn\'etico constante, $\cal E,$ es 
\begin{eqnarray}
\hat H= \frac{1}{2m}\hat p^{2}+\frac{m\omega^{2}}{2}\hat x^{2}-q{\cal E}\hat x,
\end{eqnarray}
note que este operador se puede escribir como
\begin{eqnarray}
\hat H&=& \frac{1}{2m}\hat p^{2}+\frac{m\omega^{2}}{2}\left(\hat x^{2}-\frac{2q{\cal E}}{m\omega^{2}}\hat x\right),\nonumber\\
&=& \frac{1}{2m}\hat p^{2}+
\frac{m\omega^{2}}{2}\left( \hat x^{2}-2\frac{q {\cal E}}{m\omega^{2}}\hat x+\left(\frac{q{\cal E}}{m\omega^{2}}\right)^{2}- \left(\frac{q{\cal E}}{m\omega^{2}}\right)^{2} \right)\nonumber\\
&=& \frac{1}{2m}\hat p^{2}+
\frac{m\omega^{2}}{2}\left( \hat x-\frac{q {\cal E}}{m\omega^{2}}\right)^{2}-\frac{q^{2}{\cal E}^{2}}{2m\omega^{2}} 
\end{eqnarray}
Por lo que, con el cambio de variable $\eta= x-\frac{q {\cal E}}{m\omega^{2}},$ se tiene 
\begin{eqnarray}
\hat H= \frac{1}{2m}\hat p^{2}+
\frac{m\omega^{2}}{2}\hat \eta^{2}-\frac{q^{2}{\cal E}^{2}}{2m\omega^{2}} .
\end{eqnarray}
Entones, la ecuaci\'on de valores propios 
\begin{eqnarray}
\hat H\psi(x)=\left( \frac{1}{2m}\hat p^{2}+\frac{m\omega^{2}}{2}\hat x^{2}-q{\cal E}x\right)\psi(x)=E\psi(x)
\end{eqnarray}
se puede escribir como 
\begin{eqnarray}
\hat H\psi(\eta)=\left( \frac{1}{2m}\hat p^{2}+\frac{m\omega^{2}}{2}\hat \eta^{2}-\frac{q^{2}{\cal E}^{2}}{2m\omega^{2}} \right)\psi(\eta)=E\psi(\eta),
\end{eqnarray}
de donde 
\begin{eqnarray}
\left( \frac{1}{2m}\hat p^{2}+\frac{m\omega^{2}}{2} \hat \eta^{2} \right)\psi(\eta)=
\left(E+\frac{q^{2}{\cal E}^{2}}{2m\omega^{2}}\right)\psi(\eta).
\end{eqnarray}
Esta \'ultima ecuaci\'on es la ecuaci\'on del oscilador arm\'onico, por lo que 
\begin{eqnarray}
E+\frac{q^{2}{\cal E}^{2}}{2m\omega^{2}}=\hbar \omega\left(n+\frac{1}{2}\right),
\end{eqnarray}
es decir los \'unicos valores de la energ\'ia permitidos son 
\begin{eqnarray}
E_{n}=\hbar \omega\left(n+\frac{1}{2}\right)-\frac{q^{2}{\cal E}^{2}}{2m\omega^{2}}
\end{eqnarray}
mientras que las funciones de onda son 
\begin{eqnarray}
\psi_{n}(\zeta)
=\left(\frac{m\omega}{\pi \hbar}\right)^{1/4} 
\frac{(i)^{n}}{\sqrt{n!2^{n}}}e^{-\frac{\zeta^{2}}{2}} H_{n}(\zeta),\qquad  
\zeta=\sqrt{\frac{m\omega}{\hbar}}\left(x-\frac{q {\cal E}}{m\omega^{2}}\right).
\end{eqnarray}
\section{ Suma de osciladores y el oscilador en $D$ dimensiones}

Supongamos que tenemos un sistema que consiste en dos osciladores des\-acoplados.
En este caso el Hamiltoniano est\'a dado por
\begin{eqnarray}
H=\frac{\hat p_{1}^{2}}{2m_{1}}+ \frac{\hat p_{2}^{2}}{2m_{1}}
+\frac{m_{1}\omega_{1}^{2}}{2} \hat x_{1}^{2}+\frac{m_{2}\omega_{2}^{2}}{2} \hat x_{2}^{2}, 
\label{eq:Hamiltoniano2}
\end{eqnarray}
con
\begin{eqnarray}
\hat p_{1}=-i\hbar \frac{\partial }{\partial x_{1}},\qquad 
\hat p_{2}=-i\hbar \frac{\partial }{\partial x_{2}}.\nonumber
\end{eqnarray}
Como los osciladores son independientes, estos operadores satisfance las reglas de conmutaci\'on
\begin{eqnarray}
\left[\hat x_{k},\hat x_{l}\right]= \left[\hat p_{k},\hat p_{l}\right]=0, \quad  \left[\hat x_{k},\hat p_{l}\right]=i\hbar\delta_{kl}, \quad k,l=1,2.
\end{eqnarray}

Para obtener los valores y  funciones propias de $H$ definamos los Hamiltonianos
\begin{eqnarray}
\hat H_{1}=\frac{\hat p_{1}^{2}}{2m_{1}}+
\frac{m_{1}\omega_{1}^{2}}{2} \hat x_{1}^{2},\quad \hat H_{2}= \frac{\hat p_{2}^{2}}{2m_{1}}+\frac{m_{2}\omega_{2}^{2}}{2} \hat x_{2}^{2}.\nonumber
\end{eqnarray}
Tambi\'en definamos los operadores 
\begin{eqnarray}
\hat a_{1}&=&\frac{1}{\sqrt{2m_{1}\omega_{1}\hbar}}\left(\hat p_{1}-im_{1}\omega_{1}\hat x_{1}\right),\quad  \hat a_{2}=\frac{1}{\sqrt{2m_{2}\omega_{2}\hbar}}\left(\hat p_{2}-im_{2}\omega_{1}\hat x_{2}\right)\nonumber\\
\hat a^{\dagger}_{1}&=&\frac{1}{\sqrt{2m_{1}\omega_{1}\hbar}}\left(\hat p_{1}+im_{1}\omega_{1}\hat x_{1}\right),\quad  \hat a_{2}^{\dagger}=\frac{1}{\sqrt{2m_{2}\omega_{2}\hbar}}\left(\hat p_{2}+im_{2}\omega_{1}\hat x_{2}\right),\nonumber
\end{eqnarray}
que satisfacen las reglas de conmutaci\'on
\begin{eqnarray}
\left[\hat a_{1},\hat a_{1}^{\dagger}\right]&=&1,\qquad \left[\hat a_{2},\hat a_{2}^{\dagger}\right]=1\nonumber
\end{eqnarray}
y cero en cualquier otro caso. Tambi\'en tenemos que 
\begin{eqnarray}
H_{1}=\hbar \omega_{1}\left(a^{\dagger}_{1}a_{1}+\frac{1}{2}\right),\qquad  H_{2}=\hbar \omega_{2}\left(a^{\dagger}_{2}a_{2}+\frac{1}{2} \right),
\nonumber
\end{eqnarray}
as\'i, el Hamiltoniano Eq. (\ref{eq:Hamiltoniano2}) es 
\begin{eqnarray}
\hat H=\hat H_{1}+\hat H_{2}= \hbar \omega_{1}\left(\hat a^{\dagger}_{1}\hat a_{1}+\frac{1}{2}\right)+\hbar \omega_{2}\left(\hat a^{\dagger}_{2}\hat a_{2}+\frac{1}{2}\right).
\end{eqnarray}
Ahora, recordemos que para los Hamiltonianos $\hat H_{1}$ y $\hat H_{2}$ se tiene
\begin{eqnarray}
\hat H_{k}\psi_{n_{k}}(x_{k})&=&E_{n_{k}}\psi_{n_{k}}(x_{k}),\quad E_{n_{k}}=\hbar\omega_{k}\left(n_{k}+\frac{1}{2}\right),\quad k=1, 2, \nonumber\\
 \psi_{n_{k}}(x_{k})&=&
\psi_{n_{k}}(\zeta_{k})
=\left(\frac{m_{k}\omega_{1}}{\pi \hbar}\right)^{1/4} 
\frac{(i)^{n_{k}}}{\sqrt{n_{k}!2^{n_{k}}}}e^{-\frac{\zeta_{k}^{2}}{2}} H_{n_{k}}(\zeta_{k}),\nonumber\\
& & \zeta_{k}=\sqrt{\frac{m_{k}\omega_{k}}{\hbar}}x_{k},\nonumber\\
\left<\psi_{n_{k}}|\psi_{l_{k}}\right>&=&\int_{-\infty}^{\infty}dx_{k}\psi_{n_{k}}^{*}(x_{k})\psi_{l_{k}}(x_{k})= \delta_{n_{k}l_{k}}.
\nonumber
\end{eqnarray}
Entonces, definiremos  
\begin{eqnarray}
\psi_{n_{1}n_{2}}(x_{1},x_{2})=\psi_{n_{1}}(x_{1})\psi_{n_{2}}(x_{2}),\label{eq:oscialdor-2D}
\end{eqnarray}
que  satisface
\begin{eqnarray}
\hat H\psi_{n_{1}n_{2}}(x_{1},x_{2})&=&\left(\hat H_{1}+\hat H_{2}\right)\psi_{n_{1}}(x_{1})\psi_{n_{2}}(x_{2})\nonumber\\
&=&
\hat H_{1}\psi_{n_{1}}(x_{1})\psi_{n_{2}}(x_{2})+\hat H_{2}\psi_{n_{1}}(x_{1})\psi_{n_{2}}(x_{2})\nonumber\\
&=&\psi_{n_{2}}(x_{2})\hat H_{1}\psi_{n_{1}}(x_{1})+ \psi_{n_{1}}(x_{1})\hat H_{2}\psi_{n_{2}}(x_{2})\nonumber\\
&=& E_{n_{1}}\psi_{n_{2}}(x_{2})\psi_{n_{1}}(x_{1})+ E_{n_{2}}\psi_{n_{1}}(x_{1})\psi_{n_{2}}(x_{2})\nonumber\\
& =&  \left( E_{n_{1}}+ E_{n_{2}}\right)\psi_{n_{1}}(x_{1})\psi_{n_{2}}(x_{2})\nonumber\\
&=&\left(\hbar\omega_{1}\left(n_{1}+\frac{1}{2}\right) +\hbar\omega_{2}\left(n_{2}+\frac{1}{2}\right)\right) \psi_{n_{1}n_{2}}(x_{1},x_{2})\nonumber.
\end{eqnarray}
Por lo que, las funciones propias de $\hat H$ son Eq. (\ref{eq:oscialdor-2D}) y sus valores propios son
\begin{eqnarray}
E_{n_{1}n_{2}}=\hbar\omega_{1}\left(n_{1}+\frac{1}{2}\right) +\hbar\omega_{2}\left(n_{2}+\frac{1}{2}\right) 
\end{eqnarray}
Las funciones propias Eq. (\ref{eq:oscialdor-2D}) son otonormales, pues
\begin{eqnarray}
\left<\psi_{n_{1}n_{2}}| \psi_{l_{1}l_{2}}\right>&=&\int_{-\infty}^{\infty}dx_{1}\int_{-\infty}^{\infty}dx_{2}
\left(\psi_{n_{1}}(x_{1})\psi_{n_{2}}(x_{2})\right)^{*} \psi_{l_{1}}(x_{1})\psi_{l_{2}}(x_{2})\nonumber\\
&=& \int_{-\infty}^{\infty}dx_{1} \psi_{n_{1}}(x_{1})^{*} \psi_{l_{1}}(x_{1}) \int_{-\infty}^{\infty}dx_{2} \psi_{n_{2}}(x_{2})^{*} \psi_{l_{2}}(x_{2})\nonumber\\
&=&\delta_{n_{1}l_{1}}\delta_{n_{2}l_{2}}.
\end{eqnarray}

\subsection{Cadena de osciladores}

El resultado  anterior  se puede generalizar para un n\'umero $N$ de  osciladores desacoplados. En efecto, consideremos el  Hamiltoniano 

\begin{eqnarray}
\hat H=\sum_{k=1}^{N}\left(\frac{\hat p_{k}^{2}}{2m_{k}}+\frac{m_{k}\omega_{k}^{2}}{2} \hat x_{k}^{2}\right), \quad \hat p_{k}=-i\hbar \frac{\partial }{\partial x_{k}}.\label{eq:osciladorN}
\end{eqnarray}
Como los osciladores son independientes, estos operadores satisfance las reglas de conmutaci\'on
\begin{eqnarray}
\left[\hat x_{k},\hat x_{l}\right]= \left[\hat p_{k},\hat p_{l}\right]=0, \quad  \left[\hat x_{k},\hat p_{l}\right]=i\hbar\delta_{kl}, \quad k,l=1,2\cdots N.
\end{eqnarray}
Para obtener los valores y  funciones propias de $H$ definamos los Hamiltonianos
\begin{eqnarray}
\hat H_{k}=\frac{\hat p_{k}^{2}}{2m_{k}}+
\frac{m_{k}\omega_{k}^{2}}{2} \hat x_{k}^{2}\nonumber
\end{eqnarray}
y los operadores 
\begin{eqnarray}
\hat a_{k}&=&\frac{1}{\sqrt{2m_{k}\omega_{k}\hbar}}\left(\hat p_{k}-im_{k}\omega_{k}\hat x_{k}\right),\quad  \hat a_{k}^{\dagger}=\frac{1}{\sqrt{2m_{k}\omega_{k}\hbar}}\left(\hat p_{k}+im_{k}\omega_{k}\hat x_{k}\right),\nonumber
\end{eqnarray}
que satisfacen las reglas de conmutaci\'on
\begin{eqnarray}
\left[\hat a_{k},\hat a_{l}^{\dagger}\right]&=&\delta_{kl}
\end{eqnarray}
y cero en cualquier otro caso. Tambi\'en tenemos que 
\begin{eqnarray}
\hat H_{k}=\hbar \omega_{k}\left(\hat a^{\dagger}_{k}\hat a_{k}+\frac{1}{2}\right),
\nonumber
\end{eqnarray}
as\'i el Hamiltoniano Eq. (\ref{eq:osciladorN}) es 
\begin{eqnarray}
\hat H=\sum_{k=1}^{N}\hbar \omega_{k}\left(\hat a^{\dagger}_{k}\hat a_{k}+\frac{1}{2}\right).
\end{eqnarray}
Ahora, recordando de nuevo que para cada $\hat H_{k}$ se tiene
\begin{eqnarray}
\hat H_{k}\psi_{n_{k}}(x_{k})&=&E_{n_{k}}\psi_{n_{k}}(x_{k}),\quad E_{n_{k}}=\hbar\omega_{k}\left(n_{k}+\frac{1}{2}\right),\quad k=1, 2, \cdots, N, \nonumber\\
 \psi_{n_{k}}(x_{k})&=&
\psi_{n_{k}}(\zeta_{k})
=\left(\frac{m_{k}\omega_{1}}{\pi \hbar}\right)^{1/4} 
\frac{(i)^{n_{k}}}{\sqrt{n_{k}!2^{n_{k}}}}e^{-\frac{\zeta_{k}^{2}}{2}} H_{n_{k}}(\zeta_{k}),\nonumber\\
& & \zeta_{k}=\sqrt{\frac{m_{k}\omega_{k}}{\hbar}}x_{k},\nonumber\\
\left<\psi_{n_{k}}|\psi_{l_{k}}\right>&=&\int_{-\infty}^{\infty}dx_{k}\psi_{n_{k}}^{*}(x_{k})\psi_{l_{k}}(x_{k})= \delta_{n_{k}l_{k}},
\nonumber
\end{eqnarray}
definiremos  
\begin{eqnarray}
\psi_{n_{1}n_{2}\cdots n_{N}}(x_{1},x_{2},\cdots ,x_{N})=\psi_{n_{1}}(x_{1})\psi_{n_{2}}(x_{2})\cdots\psi_{n_{N}}(x_{N}) .\label{eq:osciladorN}
\end{eqnarray}
Estas funciones satisfacen las relaciones de ortonormalidad 
\begin{eqnarray}
\left<\psi_{n_{1}n_{2}\cdots n_{N}}|\psi_{l_{1}l_{2}\cdots l_{N}}\right>=\delta_{n_{1}l_{1}}\delta_{n_{2}l_{2}} \cdots \delta_{n_{N}l_{N}}. 
\end{eqnarray}
Tambi\'en cumplen que 
\begin{eqnarray}
H \psi_{n_{1}n_{2}\cdots n_{N}}(x_{1},x_{2},\cdots ,x_{N})  &=&E_{n_{1}n_{2}\cdots n_{N}}\psi_{n_{1}n_{2}\cdots n_{N}}(x_{1},x_{2},\cdots ,x_{N}), \nonumber 
\end{eqnarray}
con 
\begin{eqnarray}
E_{n_{1}n_{2}\cdots n_{N}}= \sum_{k=1}^{N} \hbar\omega_{k}\left(n_{k}+\frac{1}{2}\right) .
\end{eqnarray}

\subsection{Oscilador en $D$ dimensiones}

El  Hamiltoniano de un oscilador en $D$ dimensiones es 
\begin{eqnarray}
H=\sum_{k=1}^{D}\left(\frac{\hat p_{k}^{2}}{2m}+\frac{m\omega^{2}}{2}\hat  x_{k}^{2}\right), \quad p_{k}=-i\hbar \frac{\partial }{\partial x_{k}}.\label{eq:osciladorD}
\end{eqnarray}
Matematicamente este es un caso del problema de la subsecci\'on anterior, por lo que los valores propios son
\begin{eqnarray}
E_{n_{1}n_{2}\cdots n_{D}}= \sum_{k=1}^{D} \hbar\omega\left(n_{k}+\frac{1}{2}\right). 
\end{eqnarray}
Para este sistema las funciones propias son un caso particular de Eq. (\ref{eq:osciladorN}).\\

En particular para tres dimensiones se tiene 
\begin{eqnarray}
E_{n_{1}n_{2}n_{3}}= \hbar\omega\left(n_{1}+n_{2}+n_{3}+\frac{3}{2}\right),
\end{eqnarray}
con las funciones de onda 
\begin{eqnarray}
& &\psi_{n_{1}n_{2}n_{3}}(x,y, z)=\left(\frac{m\omega}{\pi \hbar}\right)^{3/4} \frac{(i)^{n_{1}+n_{2}n_{3}}} {\sqrt{n_{1}!n_{2}!n_{3}! 2^{n_{1}+n_{2}+n_{3}}}}\nonumber\\
& & e^{-\frac{m\omega \left(x^{2}+y^{2}+z^{2}\right)}{2\hbar}}
H_{n_{1}}\left(\sqrt{\frac{m\omega }{\hbar}}x\right) H_{n_{2}}\left(\sqrt{\frac{m\omega }{\hbar}}y\right)H_{n_{3}}\left(\sqrt{\frac{m\omega }{\hbar}}z\right).\nonumber
\end{eqnarray}

\section{Niveles de Landau, part\'icula en un campo magn\'etico constante}

Ahora veamos como obtener los estados propios de una part\'icula cargada en un campo 
magn\'etico constante. Este problema primero fue resuelto por Landau, por lo que algunos
autores los llaman el problema de los niveles de Landau.\\

El Hamiltoniano para una part\'icula de carga $e$ en un campo mag\'etico $\vec B$ es 
\begin{eqnarray}
\hat H=\frac{1}{2m}\left(\vec P-e\vec A\right)^{2},
\end{eqnarray}
donde 
\begin{eqnarray}
\vec P=-i\hbar \vec \nabla,\quad \vec B=\vec \nabla \times \vec A.  
\end{eqnarray}
Antes de obtener el espectro de este sistema es conveniente obtener algunos resultados
previos. Perimero note que ocupando el tensor de Levi-Civita Eq. (\ref{eq:prop-levi-civita}), se tiene
\begin{eqnarray}
B_{i}=\epsilon_{ijk}\frac{\partial A_{j}}{\partial x^{k}},  
\end{eqnarray}
adem\'as con la propiedad Eq. (\ref{eq:proeps}) se llega a
\begin{eqnarray}
\epsilon_{ijk}B_{k}=\frac{\partial A_{j}}{\partial x^{i}}- \frac{\partial A_{i}}{\partial x^{j}}.
\label{eq:contrac-os}
\end{eqnarray}
Tambi\'en es conveniente definir la derivada covariante
\begin{eqnarray}
D_{i}=P_{i}-eA_{i}, 
\end{eqnarray}
este operador satisface 
\begin{eqnarray}
[ D_{i},D_{j}]=ie\hbar \epsilon_{ijk}B_{k}. \label{eq:curv-osc}
\end{eqnarray}
En efecto, 
\begin{eqnarray}
[ D_{i},D_{j}]&=& [ P_{i}-e A_{i},P_{j}-e A_{j} ]=[  P_{i},P_{j}]-e[  P_{i},A_{j}]-e[ A_{i},P_{j}]+e^{2}
[ A_{i},A_{j}],\nonumber\\
&=& e \left([ A_{j},P_{i}]- [ A_{i},P_{j}]\right)=ie\hbar\left(\frac{\partial A_{j}}{\partial x^{i}}- 
\frac{\partial A_{i}}{\partial x^{j}}\right),
\end{eqnarray}
por lo tanto, ocupando  Eq. (\ref{eq:contrac-os}) se obtiene Eq. (\ref{eq:curv-osc}).\\

Con la derivada covariante el Hamiltoniano toma la forma
\begin{eqnarray}
\hat H=\frac{1}{2m} \vec D^{2}= \frac{1}{2m} \left(D_{1}^{2}+ D_{2}^{2}+D_{3}^{2}\right)
\end{eqnarray}
Note que, sin perdida de generalidad, si el campo magn\'etico es constante se puede escribir como
\begin{eqnarray}
\vec B=(0,0,B_{3}),\qquad B_{3}={\rm constante}.
\end{eqnarray}
En este caso las reglas de conmutaci\'on Eq. (\ref{eq:curv-osc}) implican
\begin{eqnarray}
[ D_{i},D_{3}]=0,
\end{eqnarray}
de donde 
\begin{eqnarray}
[ \hat H,D_{3}]=0.
\end{eqnarray}
Por lo tanto,  $D_{3}$ es un operador constante y comparte funciones propias con el Hamiltoniano. Es decir, existe $\psi_{E_{\gamma}}$ tal que 
\begin{eqnarray}
\hat H\psi_{E_\gamma}= E \psi_{E_\gamma},\qquad D_{3}\psi_{E_\gamma}=\gamma \psi_{E_\gamma}.
\end{eqnarray}
Esto implica que
\begin{eqnarray}
\hat H\psi_{E_\gamma}&=&\left(    \frac{1}{2m} \left(D_{1}^{2}+ D_{2}^{2}+D_{3}^{2}\right)       \right)
 \psi_{E_\gamma} = \frac{1}{2m} \left(D_{1}^{2}+ D_{2}^{2}\right) \psi_{E_\gamma}+D_{3}^{2}\psi_{E_\gamma}\nonumber \\
 &=&
 \frac{1}{2m} \left(D_{1}^{2}+ D_{2}^{2}\right) \psi_{E_\gamma}+\gamma^{2} \psi_{E_\gamma}   = E \psi_{E_\gamma},
\end{eqnarray}
entonces 
\begin{eqnarray}
\frac{1}{2m} \left(D_{1}^{2}+ D_{2}^{2}\right) \psi_{E_\gamma} = \left(E -\gamma^{2}\right)\psi_{E_\gamma}.
\label{eq:landau0-os}
\end{eqnarray}
Adicionalmente, definamos los operadores 
\begin{eqnarray}
D_{\pm}=\frac{1}{\sqrt{|e|2\hbar B}}\left( D_{1}\mp i\hat e D_{2}\right),
\end{eqnarray}
con $\hat e$ el signo de $e,$ es decir $\hat e |e|=e.$ Estos operadores act\'uan como 
operadores de ascenso y descenso, en efecto usando las relaciones de conmutaci\'on Eq. (\ref{eq:curv-osc}) se encuentra que
\begin{eqnarray}
D_{+}D_{-}&=&\frac{1}{|e|\hbar 2B}\left(D_{1}-i\hat e D_{2}\right)\left(D_{1}+i\hat e D_{2}\right) \nonumber\\
&= &\frac{1}{|e|\hbar 2B}\left(D_{1}^{2} +D_{2}^{2}+i\hat e \left(D_{1}D_{2}-D_{2}D_{1}\right)\right) 
\nonumber \\
&=& \frac{1}{|e|\hbar 2B}\left(D_{1}^{2} +D_{2}^{2}+i\hat e [D_{1},D_{2}]\right) \nonumber\\
&=&  \frac{1}{|e|\hbar 2B}\left(D_{1}^{2} +D_{2}^{2}+i\hat e ei\hbar B   \right)= 
\frac{1}{|e|\hbar 2B}\left(D_{1}^{2} +D_{2}^{2}\right)-\frac{\hat e e}{2|e|} \nonumber\\
&=&  \frac{1}{|e|\hbar 2B} \left(D_{1}^{2} +D_{2}^{2}\right)-\frac{1}{2}, \label{eq:landau-osc} 
\end{eqnarray}
de la misma forma se obtiene 
\begin{eqnarray}
D_{-}D_{+}&=& \frac{1}{|e|\hbar 2B} \left(D_{1}^{2} +D_{2}^{2}\right)+\frac{1}{2},
\end{eqnarray}
de donde
\begin{eqnarray}
[D_{-},D_{+}]&=& 1.
\end{eqnarray}
Adem\'as, de Eq. (\ref{eq:landau-osc}) se obtiene
\begin{eqnarray}
\frac{1}{2m}\left(D_{1}^{2} +D_{2}^{2}\right)&=&\frac{|e| \hbar B}{m}\left(D_{+}D_{-}+\frac{1}{2}\right).  
\label{eq:landau1-os}
\end{eqnarray}
As\'i, los operadores $D_{1},D_{2}$ tienen la misma \'algebra que los operadores de ascenso 
y descenso del oscilador arm\'onico. Esto implica que definiendo 
\begin{eqnarray}
\omega=\frac{|e|  B}{m}
\end{eqnarray}
se tiene que
\begin{eqnarray}
\hbar \omega\left(D_{+}D_{-}+\frac{1}{2}\right)\psi_{E_\gamma}=\hbar \omega\left(n+\frac{1}{2}\right)\psi_{E_\gamma} .  
\end{eqnarray}
Entonces, usando este resultado y Eq. (\ref{eq:landau0-os}),  Eq. (\ref{eq:landau1-os}) se encuentra
\begin{eqnarray}
E=E_{n\gamma}=\frac{\gamma^{2}}{2m}+\hbar\omega \left(n+\frac{1}{2}\right).  
\end{eqnarray}
estos son los niveles de energ\'ia, los cuales se llaman niveles de Landau.

\section{Ecuaci\'on de Fokker-Planck,  caso libre y homog\'eneo}

La ecuaci\'on de Fokker-Planck es
\begin{eqnarray}
\frac{\partial P(x,t,\nu)}{\partial t}&=& -\nu \frac{\partial P(x,t,\nu)}{\partial x}+
\frac{\partial }{\partial \nu}\left( \left(\frac{\gamma \nu}{m}-\frac{F(x)}{m}\right) P(x,t,\nu)\right) \nonumber\\
& &+\frac{g}{2m^{2}}\frac{\partial^{2} P(x,t,\nu)}{\partial \nu^{2} } \label{eq:fokker-planck}
\end{eqnarray}
donde $\gamma$ es una constante de fricci\'on, $m$ es la masa y $F(x)=-\frac{\partial V(x)}{\partial x}$ es la fuerza. \\

Para el caso libre $V(x)=0$ y homog\'eneo, $\frac{\partial P(x,\nu,\tau)}{\partial x}=0,$  la ecuaci\'on 
de Fokker-Planck  (\ref{eq:fokker-planck}) toma la forma 
\begin{eqnarray}
\frac{\partial P(t,\nu)}{\partial t}&=&\frac{\gamma }{m}
\frac{\partial }{\partial \nu}\left( \nu P(t,\nu)\right) +\frac{g}{2m^{2}}\frac{\partial^{2} P(t,\nu)}{\partial \nu^{2} }\nonumber\\
&=&\frac{\partial }{\partial \nu}\left( \frac{\gamma }{m}\nu P(t,\nu)+ \frac{g}{2m^{2}}\frac{\partial P(t,\nu)}{\partial \nu }\right).
\label{eq:fokker-planck3}
\end{eqnarray}
Esta ecuaci\'on no est\'a en t\'erminos de operadores herm\'iticos, para expresarla con operadores herm\'iticos  usaremos la transformaci\'on 
\begin{eqnarray}
P(\nu,t)=e^{-\frac{m\gamma \nu^{2}}{2g}} \psi(\nu,t).
\end{eqnarray}
De donde
\begin{eqnarray}
\frac{\partial P(\nu,t)}{\partial \nu} =e^{-\frac{m\gamma \nu^{2}}{2g}}\left(-\frac{m\gamma \nu}{g} \psi(\nu,t)+
\frac{\partial \psi(\nu,t)}{\partial \nu}\right),
\end{eqnarray}
por lo que
\begin{eqnarray}
& & \frac{\gamma }{m}\nu P(t,\nu)+ \frac{g}{2m^{2}}\frac{\partial P(t,\nu)}{\partial \nu } = \nonumber \\
& & =
\frac{\gamma \nu}{m}e^{-\frac{m\gamma \nu^{2}}{2g}} \psi(\nu,t) +\frac{g}{2m^{2}}e^{-\frac{m\gamma \nu^{2}}{2g}}\left(-\frac{m\gamma \nu}{g} \psi(\nu,t)+\frac{\partial \psi(\nu,t)}{\partial \nu}\right)\nonumber\\
& & =e^{-\frac{m\gamma \nu^{2}}{2g}}\left(\frac{\gamma \nu }{2m}\psi(\nu,t)+\frac{g}{2m^{2}}
\frac{\partial \psi(\nu,t)}{\partial \nu}\right).\nonumber
\end{eqnarray}
Entonces,
\begin{eqnarray}
& &\frac{\partial }{\partial \nu}\left(\frac{\gamma }{m}\nu P(t,\nu)+ \frac{g}{2m^{2}}\frac{\partial P(t,\nu)}{\partial \nu }\right)\nonumber\\
& &= \frac{\partial }{\partial \nu}\left(e^{-\frac{m\gamma \nu^{2}}{2g}}\left(\frac{\gamma \nu }{2m}\psi(\nu,t)+\frac{g}{2m^{2}}
\frac{\partial \psi(\nu,t)}{\partial \nu}\right) \right)\nonumber\\
& &=e^{-\frac{m\gamma \nu^{2}}{2g}}\Bigg[ -\frac{m\gamma\nu}{g} \left(\frac{\gamma \nu }{2m}\psi(\nu,t)+\frac{g}{2m^{2}}
\frac{\partial \psi(\nu,t)}{\partial \nu}\right)\nonumber\\
& &  +\frac{\partial }{\partial \nu} \left(\frac{\gamma \nu }{2m}\psi(\nu,t)+\frac{g}{2m^{2}}
\frac{\partial \psi(\nu,t)}{\partial \nu}\right) \Bigg]\nonumber\\
& & =e^{-\frac{m\gamma \nu^{2}}{2g}}\Bigg[ -\frac{\gamma^{2}\nu^{2}}{2g} \psi(\nu,t)- 
\frac{\gamma \nu}{2m} \frac{\partial \psi(\nu,t)}{\partial \nu}+ \frac{\gamma}{2m} \psi(\nu,t)\nonumber\\
& &+\frac{\gamma \nu}{2m} \frac{\partial \psi(\nu,t)}{\partial \nu}+\frac{g}{2m^{2}} \frac{\partial^{2} \psi(\nu,t)}{\partial \nu^{2}}\Bigg]\nonumber\\
& &=e^{-\frac{m\gamma \nu^{2}}{2g}}
\left[ \left(\frac{\gamma}{2m}-\frac{\gamma^{2} \nu^{2}}{2g}\right) \psi(\nu,t)+\frac{g}{2m^{2}}
\frac{\partial^{2} \psi(\nu,t)}{\partial \nu^{2}}\right]\nonumber\\
& &=-e^{-\frac{m\gamma \nu^{2}}{2g}}\hat H \psi(\nu,t)
\end{eqnarray}
con
\begin{eqnarray}
\hat H= \frac{g}{2m^{2}} \hat p^{2}+ \frac{1}{2}\left( \frac{\gamma^{2}\nu^{2}}{g} -\frac{\gamma}{m}\right),\qquad 
\hat p=-i \frac{\partial }{\partial \nu},
\end{eqnarray}
claramente $\hat H$ es un operador herm\'itico. Considerando este resultado en (\ref{eq:fokker-planck3}) se llega a
\begin{eqnarray}
-\frac{\partial \psi(\nu,t )}{\partial t}=\hat H\psi(\nu,t).
\end{eqnarray}
Si se propone como soluci\'on a $\psi(\nu,\tau)=e^{-\lambda \tau} \phi(\nu),$ se encuentra
\begin{eqnarray}
\hat H\phi(\nu)=\lambda \phi(\nu),
\end{eqnarray}
es decir se tiene la ecuaci\'on de valores propios 
\begin{eqnarray}
\left( \frac{g}{2m^{2}} \hat p^{2} +\frac{\gamma^{2}}{2g}\nu^{2} -\frac{\gamma}{m}\right) \phi(\nu)=\lambda \phi(\nu).
\label{eq:fokker-planck4}
\end{eqnarray}
Note que renombrando 
\begin{eqnarray}
\frac{\hbar^{2}}{2m}=\frac{g}{2m^{2}},\qquad m\omega^{2}=\frac{\gamma^{2}}{g},\qquad E=\lambda +\frac{\gamma}{m}
\end{eqnarray}
la  ecuaci\'on (\ref{eq:fokker-planck4}) toma la forma 
\begin{eqnarray}
\left( -\frac{\hbar^{2}}{2m} \frac{\partial^{2}}{\partial \nu^{2}} +\frac{m\omega^{2}}{2}\nu^{2}\right) \phi(\nu)=E \phi(\nu).
\end{eqnarray}
que es la ecuaci\'on del oscilador arm\'onico. Por lo  tanto, las soluciones que satisfacen las condiciones de Dirichlet, $\phi(\pm \infty)=0,$
implican 
\begin{eqnarray}
E=\hbar\omega \left(n+\frac{1}{2}\right)=\lambda +\frac{\gamma}{m},
\end{eqnarray}
es decir 
\begin{eqnarray}
\lambda_{n}=\frac{\gamma}{m}\left( n-\frac{1}{2}\right).
\end{eqnarray}
Adem\'as, definiendo 
\begin{eqnarray}
\zeta= \sqrt{\frac{m\omega}{\hbar}}\nu= \sqrt{\frac{m\gamma}{g}}\nu
\end{eqnarray}
se tienen las soluciones 
\begin{eqnarray}
\phi_{n}(\zeta)=\frac{(i)^{n}}{\sqrt{2^{n}n!}}\left(\frac{m\gamma}{\pi g}\right)^{\frac{1}{4}} e^{-\frac{\zeta^{2}}{2}}H_{n}(\zeta).
\end{eqnarray}
Por lo tanto,
\begin{eqnarray}
P_{n}(\nu,t)=e^{-\frac{m\gamma \nu^{2}}{2g} } e^{-\lambda_{n} t} \phi_{n}(\nu)=\frac{(i)^{n}}{\sqrt{2^{n}n!}}\left(\frac{m\gamma}{\pi g}\right)^{\frac{1}{4}} e^{-\zeta^{2}}H_{n}(\zeta).
\end{eqnarray}

\chapter{El Grupo de Rotaciones y los Arm\'onicos Esf\'ericos}

En este cap\'itulo estudiaremos los polinomios de Legendre, los polinomios asociados de Legendre y los Arm\'onicos esf\'ericos, quienes son importantes para resolver la ecuaci\'on de Laplace y diversos problemas de electrodin\'amica y mec\'anica cu\'antica. Normalmente estas funciones se obtienen resolviendo ecuaciones diferenciales. Sin embargo, tambi\'en es posible obtenerlas usando el grupo de rotaciones. Usaremos este \'ultimo m\'etodo debido a que nos introduce a las aplicaci\'on de la teor\'ia de grupos en la f\'isica. Este cap\'itulo se puede ver como una invitaci\'on al estudio de las aplicaciones de la teor\'ia de grupos.

\section{Transformaciones de coordenadas lineales}

Sea  $f$ una funci\'on de la variable $x,$ con el cambio de variable 
\begin{eqnarray}
x^{\prime}=\alpha x,\quad  {\rm con}\quad  \alpha={\rm constante},
\end{eqnarray}
se encuentra
\begin{eqnarray}
\frac{df}{dx^{\prime}}=\frac{dx}{dx^{\prime}}\frac{df}{dx}=
\frac{1}{\alpha}\frac{df}{dx}.
\end{eqnarray}
Entonces, si la variable $x$ transforma con $\alpha,$ el operador
derivada transforma con $1/\alpha,$ es decir,
\begin{eqnarray}
&x& \to x^{\prime}=\alpha x, \nonumber\\
&\frac{d}{dx}& \to \frac{d}{dx^{\prime}}=\frac{1}{\alpha}\frac{d}{dx}.
\end{eqnarray}
Veamos ahora que pasa en dos dimensiones. Consideremos la
matriz de $2\times 2$ con entradas constantes
\begin{eqnarray}
\Lambda= 
\left(
\begin{array}{cc}
 a_{1}   & a_{2} \\
 a_{3} &  a_{4}
\end{array} \right),
\end{eqnarray}
cuya inversa es 
\begin{eqnarray}
\Lambda^{-1}=\frac{1}{|\Lambda|}
\left(
\begin{array}{cc}
 a_{4}   & -a_{2} \\
 -a_{3} &  a_{1} 
\end{array} \right), \qquad |\Lambda|=a_{1}a_{4}-a_{2}a_{3}.
\end{eqnarray}
También definamos los vectores columna
\begin{eqnarray}
X=  \left(
\begin{array}{c}
 x^{1}\\
 x^{2} 
\end{array}
 \right), 
\qquad
\nabla=  \left(
\begin{array}{c}
 \frac{\partial }{\partial x^{1}} \\
 \frac{\partial }{\partial x^{2}} 
\end{array}
 \right).
\end{eqnarray}
Entonces, podemos hacer una transformaci\'on lineal de coordenadas de la forma
$X^{\prime}=\Lambda X,$ es decir
\begin{eqnarray}
\left(
\begin{array}{c}
 x^{\prime 1}\\
 x^{\prime 2} 
\end{array}
 \right)= 
\left(
\begin{array}{cc}
 a_{1}   & a_{2} \\
 a_{3} &  a_{4} 
\end{array} \right)
\left(\begin{array}{c}
 x^{1}\\
 x^{2} 
\end{array}
\right).\label{eq:trans}
\end{eqnarray}
La cual tiene la transformaci\'on inversa, $X=\Lambda^{-1} X^{\prime},$ 
\begin{eqnarray}
\left(
\begin{array}{c}
x^{1}\\
 x^{2} 
\end{array}
\right)&=&
\frac{1}{|\Lambda|}
\left(
\begin{array}{cc}
 a_{4}   & -a_{2} \\
 -a_{3} &  a_{1} 
\end{array} \right)\left(
\begin{array}{c}
 x^{\prime 1}\\
 x^{\prime 2} 
\end{array}
 \right),
\end{eqnarray}
es decir,
\begin{eqnarray}
 x^{1}&=&\frac{a_{4}x^{\prime 1} -a_{2}x^{\prime 2} }{|\Lambda|},  \\
 x^{2}&=& \frac{-a_{3}x^{\prime 1} +a_{1}x^{\prime 2} }{|\Lambda|}.
\label{eq:transinv}
\end{eqnarray}
Adem\'as, por la regla de la cadena se tiene 
\begin{eqnarray}
\frac{\partial }{\partial x^{\prime 1}}
&=&\frac{\partial  x^{1}}{\partial x^{\prime 1}}
\frac{\partial }{\partial x^{1}}
+\frac{\partial  x^{2}}{\partial x^{\prime 1}}
\frac{\partial }{\partial x^{2}},\nonumber\\
\frac{\partial }{\partial x^{\prime 2}}
&=&\frac{\partial  x^{1}}{\partial x^{\prime 2}}
\frac{\partial }{\partial x^{1}}
+\frac{\partial  x^{2}}{\partial x^{\prime 2}}
\frac{\partial }{\partial x^{2}}.
\end{eqnarray}
De la regla de transformaci\'on  Eq. (\ref{eq:transinv}) se encuentra
\begin{eqnarray}
\frac{\partial }{\partial x^{\prime 1}}
&=&\frac{1}{|\Lambda|}\left(a_{4}\frac{\partial }{\partial x^{1}}-
a_{3}\frac{\partial }{\partial x^{\prime 2}}\right),\nonumber\\
\frac{\partial }{\partial x^{ 2}}
&=&\frac{1}{|\Lambda|}\left(-a_{2}\frac{\partial }{\partial x^{1}}
+a_{1}\frac{\partial }{\partial x^{ 2}}\right),
\end{eqnarray}
que se puede expresar como
\begin{eqnarray}
\left(
\begin{array}{c}
\frac{\partial }{\partial x^{\prime 1}}\\
\frac{\partial }{\partial x^{\prime 2}}
\end{array}
\right)=
\frac{1}{|\Lambda|}
\left(
\begin{array}{cc}
 a_{4}   & -a_{3} \\
 -a_{2} &  a_{1} 
\end{array} \right)
\left(
\begin{array}{c}
\frac{\partial }{\partial x^{1}}  \\
\frac{\partial }{\partial x^{2}}
\end{array}
 \right).
\end{eqnarray}
De esta ecuaci\'on podemos ver que la matriz involucrada en la transformaci\'on de las
derivadas parciales es la transpuesta de la matriz inverza de $\Lambda,$
es decir $\left(\Lambda^{-1}\right)^{T}.$
Otra forma de escribir esta ecuaci\'on es
\begin{eqnarray}
 \nabla^{\prime}=\left(\Lambda^{-1}\right)^{T} \nabla,
\end{eqnarray}
por lo que   
\begin{eqnarray}
 \nabla=\left(\Lambda \right)^{T}\nabla^{\prime}.
\end{eqnarray}
Este resultado se puede generalizar a m\'as dimensiones. En efecto, en general una transformaci\'on de coordenadas se escribe como 
\begin{eqnarray}
x^{\prime i}=\Lambda_{ij}x^{j},\qquad x^{i}=\left(\Lambda^{-1}\right)_{ij}x^{\prime j}.
\end{eqnarray}
De donde, por la regla de la cadena, se tiene 
\begin{eqnarray}
\frac{\partial }{\partial x^{\prime i}}&=&\frac{\partial x^{j}}{\partial x^{\prime i}}
\frac{\partial }{\partial x^{j}}=\frac{\partial \left(\Lambda^{-1}\right)_{jk}x^{\prime k}} 
{\partial x^{\prime i}}
\frac{\partial }{\partial x^{j}}
= \left(\Lambda^{-1}\right)_{jk}\frac{\partial x^{\prime k}} 
{\partial x^{\prime i}}
\frac{\partial }{\partial x^{j}}\nonumber\\
&=& \left(\Lambda^{-1}\right)_{jk}\delta_{ki}
\frac{\partial }{\partial x^{j}}
=  \left(\Lambda^{-1}\right)_{ji}
\frac{\partial }{\partial x^{j}}= \left(\left(\Lambda^{-1}\right)^{T}\right)_{ij}\frac{\partial }{\partial x^{j}}.
\end{eqnarray}
Por lo tanto, para cualquier dimensi\'on  se cumple
\begin{eqnarray}
 \nabla^{\prime}=\left(\Lambda^{-1}\right)^{T} \nabla.
\label{eq:rot-grad-trans}
\end{eqnarray}
Esta ley de transformaci\'on ser\'a de gran utilidad para
obtener las simetr\'{\i}as de la ecuaci\'on de Laplace.

\section{ Laplaciano y elemento de l\'{\i}nea}

Supongamos que la  matriz $\tilde \eta,$ de $n\times n,$  satisface 
\begin{eqnarray}
\tilde \eta\tilde \eta=I,  
\end{eqnarray}
con $I$ la matriz identidad de $n\times n$. Entonces,  podemos definir  un  elemento de l\'{\i}nea como
\begin{eqnarray}
ds^{2}=dX^{T}\tilde \eta dX, \quad dX=\left(
\begin{array}{c}
 dx^{1}\\
 dx^{2}\\
 \vdots\\
 dx^{n} 
\end{array}
 \right).
\label{eq:rot-lin}
\end{eqnarray}
A la matriz $\tilde \eta$ se le llama m\'etrica, en dos dimensiones un ejemplo de estas matrices 
son
\begin{eqnarray}
\left(
\begin{array}{rr}
1& 0\\
0& 1
\end{array}\right),
\qquad 
\left(
\begin{array}{rr}
-1& 0\\
0& 1
\end{array}\right).
\end{eqnarray}
Con la matriz $\tilde \eta$ el "Laplaciano" se define como 
\begin{eqnarray}
\nabla ^{2}=\nabla^{T}\tilde \eta \nabla.\label{eq:rot-lapla}
\end{eqnarray}
Notablemente, el elemento de l\'{\i}nea est\'a intimamente
relacionado con el Laplaciano, en particular tienen las mismas simetr\'{\i}as.
Esta relaci\'on es importante y se da tambi\'en  para espacios no euclidianos. 
Veamos como se da esta relaci\'on.\\

Bajo una transformaci\'on lineal de coordenadas  se tiene 
\begin{eqnarray}
ds^{\prime 2}&=&dX^{\prime T}\tilde \eta dX^{\prime}= \left(\Lambda dX\right)^{T} \tilde \eta \Lambda dX
=dX^{T}\left(\Lambda ^{T}\tilde \eta\Lambda \right)  dX. \label{eq:rot-lin2}
\end{eqnarray}
Las transformaciones, $\Lambda,$ que dejan invariante al elemento de l\'{\i}nea deben cumplir
$ds^{2}=ds^{\prime 2}.$ Igualando Eq. (\ref{eq:rot-lin}) con Eq. (\ref{eq:rot-lin2}) se llega a la condici\'on
\begin{eqnarray}
\Lambda ^{T}\tilde \eta\Lambda =\tilde \eta. \label{eq:rot-con-trans1} 
\end{eqnarray}
Adem\'as, considerando que bajo una transformaci\'on lineal de coordenadas el gradiente
transforma como Eq. (\ref{eq:rot-grad-trans}), se obtiene 
\begin{eqnarray}
\nabla ^{\prime 2}&=&\nabla^{\prime T}\tilde \eta \nabla^{\prime}=
\left(\left(\Lambda^{-1}\right)^{T} \nabla\right)^{T} \tilde \eta \left(\Lambda^{-1}\right)^{T} \nabla \nonumber\\
&=& \nabla^{T}\left(\left(\left(\Lambda^{-1}\right)^{T}\right)^{T} \tilde \eta \left(\Lambda^{-1}\right)^{T}\right) \nabla =
 \nabla^{T}\left(\Lambda^{-1}\tilde \eta \left(\Lambda^{-1}\right)^{T}\right) \nabla. \qquad \label{eq:rot-lapla-trans}
\end{eqnarray}
Las transformaciones que dejan invariante al Laplaciano deben cumplir
$$\nabla^{2}=\nabla^{\prime 2},$$
entonces igualando Eq. (\ref{eq:rot-lapla}) con Eq. (\ref{eq:rot-lapla-trans}) se tiene la condici\'on
\begin{eqnarray}
\Lambda^{-1}\tilde \eta \left(\Lambda^{-1}\right)^{T}=\tilde \eta. \label{eq:rot-con-trans}
\end{eqnarray}
Como $\left(\tilde \eta\right)^{-1}= \tilde \eta,$ la condici\'on Eq. (\ref{eq:rot-con-trans}) tiene la forma
\begin{eqnarray}
\tilde \eta=\left(\tilde \eta\right)^{-1}=\left(\Lambda^{-1}\tilde \eta \left(\Lambda^{-1}\right)^{T}\right)^{-1}
=\left(\left(\Lambda^{-1}\right)^{T}\right)^{-1}\tilde \eta \left(\Lambda^{-1}\right)^{-1}
= \Lambda ^{T}\tilde \eta \Lambda,  
\end{eqnarray}
que coincide con Eq. (\ref{eq:rot-con-trans1}).
Por lo tanto, las transformaciones lineales, $\Lambda,$ que dejan invariante
al elemento de l\'{\i}nea Eq. (\ref{eq:rot-lin}) tambi\'en dejan invariante al Laplaciano 
Eq. (\ref{eq:rot-lapla}), claramente la afirmaci\'on inversa tambi\'en es correcta.

\section{Grupo de transformaciones}

Antes de continuar recordemos lo que es un grupo. 
Sea $G$ un conjunto con una operaci\'on $\cdot: G\times G\to G$. El par $(G,\cdot)$ es un grupo si 
 cumple \\
 
1) Axioma de cerradura:
\begin{eqnarray}
g_{1}\in G,g_{2}\in G \qquad \Longrightarrow \qquad g_{1}\cdot g_{2}\in G
\end{eqnarray}

2) Axioma de asociatividad:  
\begin{eqnarray}
g_{1}\in G, g_{2}\in G,g_{3}\in G,\qquad 
\Longrightarrow \qquad g_{1}\cdot\left( g_{2}\cdot g_{3}\right)=\left(g_{1}\cdot g_{2}\right)\cdot g_{3}.
\end{eqnarray}

3) Axioma del neutro:
\begin{eqnarray}
\exists e\in  G, \qquad  g_{1}\in G\Longrightarrow \qquad g_{1}\cdot e=e\cdot g_{1}=g_{1}.
\end{eqnarray}

4) Axioma del inverso:
\begin{eqnarray}
\forall g_{1} \in G,   \exists g_{1}^{-1}\in G, \quad 
g_{1}\cdot g_{1}^{-1}=g_{1}^{-1}\cdot g_{1}=e.
\end{eqnarray}

Definamos a $T$ como el conjunto de transformaciones $\Lambda$ que dejan invariante al 
Laplaciano. Estas transformaciones cumplen
\begin{eqnarray}
\Lambda ^{T}\tilde \eta\Lambda =\tilde \eta,\qquad \tilde \eta^{2}=I. \label{eq:rot-con-trans10} 
\end{eqnarray}
Probaremos que  $T$ es  un grupo. \\

Supongamos que $\Lambda_{1}\in T$ y  $\Lambda_{2}\in T,$ entonces cumplen 
\begin{eqnarray}
\Lambda_{1}^{T}\tilde \eta \Lambda_{1}=\tilde \eta,\quad 
\Lambda_{2}^{T}\tilde \eta \Lambda_{2}=\tilde \eta,
\end{eqnarray}
de donde 
\begin{eqnarray}
\left(\Lambda_{1}\Lambda_{2}\right)^{T}\tilde \eta \left(\Lambda_{1}\Lambda_{2}\right)=
\Lambda_{2}^{T}\Lambda_{1}^{T}\tilde \eta \Lambda_{1}\Lambda_{2}=\Lambda_{2}^{T}\tilde \eta \Lambda_{2}
=\tilde \eta,
\end{eqnarray}
esto implica $\Lambda_{1}\Lambda_{2}\in T$, es decir
\begin{eqnarray}
\Lambda_{1}\in T,\Lambda_{2}\in T\quad  \Longrightarrow \quad \Lambda_{1}\Lambda_{2}\in T.
\end{eqnarray}
Por lo tanto,  se cumple el axioma la cerradura.\\

El producto de matrices es asociativo, en particular el producto de las matrices que satisfacen Eq. (\ref{eq:rot-con-trans10}) .
Adem\'as, la identidad $I$ satisface  Eq. (\ref{eq:rot-con-trans10}), es decir, $I\in T.$ As\'{\i} se cumplen
el axioma de la asociatividad y el del elemento neutro.\\

Ahora, como $\tilde \eta^{2}=I,$ si 
$\Lambda$ est\'a en $ T$ entonces  se cumple $ \Lambda^{T}\tilde \eta\Lambda \tilde \eta=I.$
De donde, 
\begin{eqnarray}
\tilde \eta\Lambda \tilde \eta=\left(\Lambda^{T}\right)^{-1}= \left(\Lambda^{-1}\right)^{T}.
\end{eqnarray}
Por lo tanto,
\begin{eqnarray}
\left(\Lambda^{-1}\right)^{T}\tilde \eta\Lambda^{-1}=\left(
\tilde \eta\Lambda \tilde \eta\right)\tilde \eta \Lambda^{-1}= 
\tilde \eta \Lambda\Lambda^{-1}=\tilde\eta.
\end{eqnarray}
As\'{\i},  cuando $\Lambda$ est\'a en $T$, tambi\'en $\Lambda^{-1}$ est\'a en $T.$ Esto nos indica que se cumple el axioma del inverso.\\

En consecuencia el conjunto de matrices que satisface  Eq. (\ref{eq:rot-con-trans10})  es un grupo.
Es decir, el conjunto de tranformaciones que dejan invariante al elemento de l\'{\i}nea
Eq. (\ref{eq:rot-lin}) forma un grupo, que es el mismo grupo que deja invariante al Laplaciano Eq. (\ref{eq:rot-lapla}).

\section{El grupo de rotaciones}

Sea $\vec x=(x_{1},\cdots, x_{n})$ un vector en ${\bf R}^{n}$
y definamos la forma cuadr\'atica $l^{2}=x_{1}^{2}+x_{2}^{2}+\cdots+x_{n}^{2},$ la cual 
representa la distancia de $\vec x$ al origen. 
Note que si definimos la matriz columna
\begin{eqnarray}
X=\left(
\begin{array}{r}
x_{1} \\
x_{2} \\
\vdots \\
x_{n}
\end{array}\right)
\end{eqnarray}
y  la matriz rengl\'on
\begin{eqnarray}
X^{T}=\left(
\begin{array}{r}
x_{1}\quad 
x_{2}\quad 
\cdots \quad 
x_{n}
\end{array}\right),
\end{eqnarray}
la distancia se puede escribir como
\begin{eqnarray}
l^{2}=X^{T}X=X^{T}IX.
\end{eqnarray}
Ahora, si $\Lambda$ es una matriz de $n\times n$
y se hace la transformaci\'on de coordenadas
\begin{eqnarray}
X^{\prime}=\Lambda X,
\end{eqnarray}
se tiene la distancia
\begin{eqnarray}
l^{\prime 2}=X^{\prime T}IX^{\prime}=X^{T}\left(\Lambda^{T}I\Lambda\right) X.
\end{eqnarray}
Por lo tanto, si $\Lambda$ es tal que  deja la distancia invariante,
es decir que $l^{2}=l^{\prime 2},$  debe cumplir 
\begin{eqnarray}
\Lambda^{T}I\Lambda=I. \label{eq:fina}
\end{eqnarray}
Otra forma de expresar esta igualdad es $\Lambda^{T}=\Lambda^{-1}.$ 
Claramente las matrices que cumplen (\ref{eq:fina}) forman un grupo,
a este grupo de matrices se le llama $O(n).$\\

Recordemos que para cualquier matriz $A$ se cumple  ${\rm det}A={\rm det}A^{T}.$
Entonces, las matrices que satisfacen Eq. (\ref{eq:fina}) deben cumplir
$\left({\rm det}\Lambda\right)^{2}=1,$ es decir
${\rm det}\Lambda=\pm 1.$ El subconjunto de matrices $\Lambda$ que
cumplen ${\rm det}\Lambda=-1$ no forman un grupo, por ejemplo,
la identidad no est\'a en ese subconjunto. Sin embargo, las matrices
$\Lambda$ que cumplen ${\rm det}\Lambda=1$ s\'{\i} forman un grupo,
este es el grupo $SO(n).$\\

Note que la matriz de $n\times n$ 
\begin{eqnarray}
\Lambda=\left(
\begin{array}{rrrr}
\Lambda_{11}& \Lambda_{12}& \cdots & \Lambda_{1n}\\
\Lambda_{21}&  \Lambda_{22}& \cdots& \Lambda_{2n}\\
\vdots & \vdots &  \ddots & \vdots\\
\Lambda_{n1}& \Lambda_{n2} & \cdots& \Lambda_{nn}
\end{array}\right)
\end{eqnarray}
se pueden formar con los vectores columna
\begin{eqnarray}
\vec C_{1}=
\left( \begin{array}{r}
\Lambda_{11}\\
\Lambda_{21}\\
\vdots\\
\Lambda_{n1} 
\end{array}\right),
\vec C_{2}=\left(\begin{array}{r}
\Lambda_{12}\\
\Lambda_{22}\\
\vdots\\
\Lambda_{n2} 
\end{array}\right),
\cdots,
\vec C_{n}=
\left(\begin{array}{r}
\Lambda_{1n}\\
\Lambda_{2n}\\
\vdots\\
\Lambda_{nn} 
\end{array}\right).
\end{eqnarray}
Claramente, para la matriz traspuesta, $\Lambda^{T},$
estos vectores representan los renglones. Por lo tanto,
la condici\'on Eq. (\ref{eq:fina}) se puede escribir como 
\begin{eqnarray}
\Lambda ^{T}\Lambda&=&
\left(\begin{array}{rrrr}
\Lambda_{11}  &  \Lambda_{21}& \cdots& \Lambda_{n1}\\
\Lambda_{12}&  \Lambda_{22}& \cdots&  \Lambda_{n2}\\
\vdots& \vdots& \ddots & \vdots\\
\Lambda_{1n}&  \Lambda_{2n}& \cdots &\Lambda_{nn}
\end{array}\right)
\left(
\begin{array}{rrrr}
\Lambda_{11}& \Lambda_{12}& \cdots & \Lambda_{1n}\\
\Lambda_{21}&  \Lambda_{22}& \cdots& \Lambda_{2n}\\
\vdots & \vdots &  \ddots & \vdots\\
\Lambda_{n1}& \Lambda_{n2} & \cdots& \Lambda_{nn}
\end{array}\right)\nonumber\\
&=&\left(
\begin{array}{rrrr}
\vec C_{1}\cdot \vec C_{1}& \vec C_{1}\cdot \vec C_{2}&\cdots& 
\vec C_{1}\cdot C_{n} \\
\vec C_{2}\cdot \vec C_{1}& \vec C_{2}\cdot \vec C_{2}&\cdots& 
\vec C_{n}\cdot \vec C_{2} \\
\vdots & \vdots & \ddots & \vdots \\
\vec C_{n}\cdot \vec C_{1} & \vec C_{n}\cdot \vec C_{2}& \cdots&  
\vec C_{n}\cdot \vec C_{n}
\end{array}\right)=I.\label{eq:yesi}
\end{eqnarray}
Otra forma de expresar esta igualdad es 
\begin{eqnarray}
\vec C_{i}\cdot \vec C_{j}=\delta_{ij},
\end{eqnarray}
es decir si una matriz  
satisface Eq. (\ref{eq:fina}), tiene sus columnas  ortonormales. 
Ahora, si $\Lambda$ satisface la condici\'on (\ref{eq:fina}), entonces $\Lambda^{-1}$ tambi\'en
la satisface. Por lo tanto, $\Lambda^{-1}$ tiene sus columnas ortonormales entre si. 
Pero se debe cumplir $\Lambda^{-1}=\Lambda^{T},$ entonces las columnas
de $\Lambda^{T}$ son ortonormales entre si. Considerando que  las columnas de $\Lambda^{T}$
son los renglones de  $\Lambda,$ podemos ver que los renglones de $\Lambda$
son ortonormales entre si. En conclusi\'on, si $\Lambda$ satisface Eq. (\ref{eq:fina})
sus columnas y renglones son ortonormales entre si.\\

Una matriz de $n\times n$ tiene $n^{2}$ par\'ametros libres,
pero si satisface (\ref{eq:yesi}) no todos
sus par\'ametros son libres. De (\ref{eq:yesi}) se puede 
ver que $\Lambda ^{T}\Lambda$ es una matriz sim\'etrica, por lo que
(\ref{eq:yesi}) s\'olo tiene $\frac{n(n+1)}{2}$ ecuaciones
independientes. As\'{\i}, los par\'ametros libres de una matriz
que satisface (\ref{eq:yesi})  son
$$n^{2}-\frac{n(n+1)}{2}=\frac{n(n-1)}{2}.$$
\\

Para el caso $n=3$ hay tres par\'ametros libres, para esta dimensi\'on 
cualquier matriz se puede escribir como
\begin{eqnarray}
\Lambda&=&
\left(
\begin{array}{rr}
a_{1}\quad  b_{1}\quad c_{1}\\
a_{2}\quad  b_{2}\quad c_{2}\\
a_{3}\quad  b_{3}\quad c_{3}
\end{array}\right).
\end{eqnarray}
Si esta matriz satisface (\ref{eq:yesi}), debe cumplir 
\begin{eqnarray}
\vec a\cdot \vec a=\vec b\cdot \vec b=\vec c\cdot \vec c=1,
\quad \vec a\cdot \vec b=\vec a\cdot \vec c=\vec b\cdot \vec c=0,
\label{eq:bety-fina}
\end{eqnarray}
esto nos dice que la punta de los vectores 
$\vec a, \vec b,\vec c$ est\'an en una esfera y que son ortonormales
entre si. Un ejemplo de estas matrices son
\begin{eqnarray}
\Lambda_{x}(\theta)&=&
\left(
\begin{array}{rrrr}
1 & 0& 0 \\
 0& \cos \theta & \sin \theta  \\
0& -\sin \theta & \cos \theta  
\end{array}\right),\\
\Lambda_{y}(\psi)&=&
\left(
\begin{array}{rrrr}
\cos \psi & 0& \sin \psi \\
0  & 1   &0\\
-\sin \psi & 0&\cos \psi  
\end{array}\right),\\
\Lambda_{z}(\phi)&=&
\left(
\begin{array}{rrrr}
\cos \phi & \sin \phi & 0 \\
-\sin \phi & \cos \phi  & 0  \\
0& 0 & 1
\end{array}\right). 
\end{eqnarray}
La matriz $\Lambda_{x}(\theta)$ representa una rotaci\'on sobre el eje $x,$
$\Lambda_{y}(\psi)$ representa una rotaci\'on sobre el eje $y,$
mientras que $\Lambda_{z}(\phi)$ representa una rotaci\'on sobre el eje $z.$
Por lo tanto, las rotaciones dejan invariante la distancia y al Laplaciano. 
Note que estas tres matrices son linealmente independientes. \\

Existen otras formas de reparametrizar una matriz de rotaci\'on, por ejemplo 
la base unitaria en coordenadas esf\'ericas Eqs. (\ref{eq:esfe-uni1})-(\ref{eq:esfe-uni2})
satisfacen la condici\'on  Eq. (\ref{eq:bety-fina}), 
pero no son la soluci\'on m\'as general,
pues solo dependen de dos par\'ametros mientra que  la soluci\'on general de 
Eq. (\ref{eq:bety-fina}) depende de tres. 
Pero estos vectores nos sirven para obtener la
soluci\'on general. Propondremos a $\vec c$ como
\begin{eqnarray}
\vec c=\left( \sin\psi \sin\theta, \cos\psi\sin \theta, \cos\theta\right).
\end{eqnarray}
Los vectores $\vec b$ y $\vec a$ deben ser tales que si 
$\phi=0$ se cumple 
\begin{eqnarray}
\vec b|_{\phi=0}&=&\hat e_{\theta}=
\left(\sin\psi\cos\theta, \cos\psi\cos \theta, -\sin\theta\right),\\
\vec c|_{\phi=0}&=&\hat e_{\psi}=
\left(\cos\psi, -\sin\psi, 0\right).
 \end{eqnarray}
Un par de vectores que satisfacen estas condiciones son:
\begin{eqnarray}
\vec a&=&\left(\sin\phi \sin\psi\cos\theta +\cos\phi\cos\psi,-\sin\phi\cos\psi\cos\theta -\cos\phi\sin\psi,
\sin\phi \sin\theta\right),\nonumber\\
\vec b&=&\left(\sin\phi \cos\psi+\cos\phi\sin\psi\cos\theta,-\sin\phi \sin\psi+\cos\phi\cos\psi\cos \theta,
-\cos\phi\sin\theta\right).\nonumber 
\end{eqnarray}
Se puede probar que los vectores 
$\vec a, \vec b$ y $\vec c$  cumplen Eq. (\ref{eq:bety-fina}).
As\'{\i} $\Lambda$ se puede escribir como
\begin{eqnarray}
\Lambda\left(\phi,\theta,\psi\right)&=&
\left(
\begin{array}{rrrr}
c\phi c\psi- s\phi s\psi c\theta&  
s\phi c\psi+c\phi s\psi c\theta  & s\psi s\theta \\
-c\phi s\psi-s\phi c\psi c\theta&  
- s\phi s\psi+c\phi c\psi c \theta & c\psi s \theta  \\
s\phi s\theta&-c\phi s\theta & c\theta
\end{array}\right)\label{eq:fina-fina}. 
\end{eqnarray}
Claramente a\'un hay cierta arbitrariedad,
pues podemos cambiar el lugar de los vectores $\vec a, \vec b, \vec c,$
tambi\'en podemos cambiar renglones por columnas y se seguir\'a cumpliendo
Eq. (\ref{eq:fina}). La ventaja de escribir $\Lambda$ de la forma (\ref{eq:fina-fina}) es que se 
puede expresar como el producto de tres matrices
\begin{eqnarray}
\Lambda\left(\phi,\theta,\psi\right)=\Lambda_{1}(\phi)\Lambda_{2}(\theta)\Lambda_{3}(\psi),
\end{eqnarray}
con 
\begin{eqnarray}
\Lambda_{1}(\phi)&=&
\left(
\begin{array}{rrrr}
\cos \phi & \sin \phi & 0 \\
-\sin \phi & \cos \phi  & 0  \\
0& 0 & 1
\end{array}\right),\\
\Lambda_{2}(\theta)&=&
\left(
\begin{array}{rrrr}
1 & 0& 0 \\
 0& \cos \theta & \sin \theta  \\
0& -\sin \theta & \cos \theta  
\end{array}\right),\\
\Lambda_{3}(\psi)&=&
\left(
\begin{array}{rrrr}
\cos \psi & \sin \psi & 0 \\
-\sin \psi & \cos \psi  & 0  \\
0& 0 & 1
\end{array}\right). 
\end{eqnarray}
Esta descomposici\'on es muy \'util para el estudio
del movimiento del cuerpo r\'igido. A los  par\'ametros $\theta,\psi,\phi$
se les llama \'angulos de Euler.\\

Las rotaciones no son la \'unicas tranformaciones de $O(3).$
Por ejemplo, las matrices  
\begin{eqnarray}
\left(
\begin{array}{rrrr}
-1 & 0& 0 \\
 0& 1 & 0  \\
0& 0 &1   
\end{array}\right), 
\qquad 
\left(
\begin{array}{rrrr}
1 & 0& 0 \\
 0& -1 & 0  \\
0& 0 &-1   
\end{array}\right),
\qquad 
\left(
\begin{array}{rrrr}
-1 & 0& 0 \\
 0& -1 & 0  \\
0& 0 &-1   
\end{array}\right)
\end{eqnarray}
cumplen Eq. (\ref{eq:bety-fina}) pero no son de rotaci\'on, tampoco son de $SO(3).$ Las rotaciones
representan las transformaciones de $O(3)$ que se pueden conectar con la matriz unidad $I.$
Pues al hacer $\phi=0,\theta=0,\psi=0$ se obtiene la unidad $I.$

\subsection{Transformaciones infinitesimales}

Las transformaciones  de $SO(n)$ que est\'an infinitesimalmente cercanas
a la unidad, es decir que cumplen 
\begin{eqnarray}
\Lambda\approx I+\epsilon M    \qquad \epsilon<<1, \label{eq:jose}
\end{eqnarray}
son particularmente importantes. Para este tipo de transformaciones la condici\'on 
Eq. (\ref{eq:fina}) implica
\begin{eqnarray}
I=\Lambda^{T}\Lambda &\approx& \left(I+\epsilon M\right)^{T}\left(I+\epsilon M\right)\\
&=&\left(I+\epsilon M^{T}\right)\left(I+\epsilon M\right)\approx I +\epsilon \left(M^{T}+M\right). 
\end{eqnarray}
Por lo tanto, 
\begin{eqnarray}
 M=-M^{T}, 
\end{eqnarray}
es decir, $M$ debe ser antisim\'etrica. 
Note que una matriz antisim\'etrica de $n\times n$ s\'olo puede tener
$$\frac{n(n-1)}{2}$$ 
par\'ametros libres, este n\'umero de grados de libertad coincide con los 
par\'ametros libres del grupo $SO(n)$. Para el caso 
particular  $n=3,$ cualquier matriz antisim\'etrica se puede escribir como
\begin{eqnarray}
M=
\left(
\begin{array}{rrrr}
0 & -\alpha_{3} & \alpha_{2}\\
\alpha_{3} & 0 & -\alpha_{1} \\
-\alpha_{2} & \alpha_{1} & 0
\end{array}\right).
\end{eqnarray}
Definamos los vectores $\vec r=\left(x,y,z\right), 
\delta \vec \alpha=\epsilon\left(\alpha_{1}, \alpha_{2},\alpha_{3}\right),$ entonces
\begin{eqnarray}
\delta X&=& \epsilon MX=
\epsilon\left(
\begin{array}{rrrr}
0 & -\alpha_{3} & \alpha_{2}\\
\alpha_{3} & 0 & -\alpha_{1} \\
-\alpha_{2} & \alpha_{1} & 0
\end{array}\right)\left(
\begin{array}{r}
x\\
y\\
z
\end{array}\right)=\epsilon \left(
\begin{array}{r}
 \alpha_{2}z -\alpha_{3}y\\
\alpha_{3}x-\alpha_{1}z\\
\alpha_{1}x-\alpha_{2}y
\end{array}\right)\nonumber\\
&=&  \delta\vec \alpha\times \vec r.
\end{eqnarray}
Por lo tanto, una rotaci\'on infinitesimal est\'a dada por
\begin{eqnarray}
X^{\prime}=\Lambda X\approx \left(I+\epsilon M\right)X= \left(IX+\epsilon MX\right),
\label{eq:josefina1}
\end{eqnarray}
es decir
\begin{eqnarray}
\vec r^{\prime}=\vec r+ \delta\vec \alpha\times \vec r.
\label{eq:josefina2}
\end{eqnarray}
Estas rotaciones son importantes porque definen al resto de las transformaciones de $SO(3),$ para
ver esto primero notemos que
\begin{eqnarray}
M=
\left(
\begin{array}{rrrr}
0 & -\alpha_{3} & \alpha_{2}\\
\alpha_{3} & 0 & -\alpha_{1} \\
-\alpha_{2} & \alpha_{1} & 0
\end{array}\right)=\alpha_{1}m_{1}+\alpha_{2}m_{2}+\alpha_{3}m_{3}=\vec\alpha \cdot \vec m,
\label{eq:armonico-antisimetrica}
\end{eqnarray}
con
\begin{eqnarray}
m_{1}=
\left(
\begin{array}{rrrr}
0 & 0 & 0\\
0 & 0 & -1 \\
0 & 1 & 0
\end{array}\right),\quad
m_{2}=
\left(
\begin{array}{rrrr}
 0& 0 & 1 \\
 0& 0 & 0 \\
-1& 0 & 0  
\end{array}\right),\quad
m_{3}=
\left(
\begin{array}{rrrr}
0 & -1 & 0 \\
1 &  0 & 0  \\
0 &  0 & 0
\end{array}\right).  \nonumber
\end{eqnarray}
Se puede observar  que la matriz (\ref{eq:armonico-antisimetrica}) no se modifica  si hacemos el cambio 
$\vec \alpha\to -i\vec \alpha$ y $\vec m\to i\vec m=\vec M,$
es decir 
\begin{eqnarray}
M_{1}=i
\left(
\begin{array}{rrrr}
0 & 0 & 0\\
0 & 0 & -1 \\
0 & 1 & 0
\end{array}\right),\quad
M_{2}=i
\left(
\begin{array}{rrrr}
 0& 0 & 1 \\
 0& 0 & 0 \\
-1& 0 & 0  
\end{array}\right),\quad
M_{3}=i
\left(
\begin{array}{rrrr}
0 & -1 & 0 \\
1 &  0 & 0  \\
0 &  0 & 0
\end{array}\right).  \nonumber
\end{eqnarray}
La ventaja de ocupar la matrices $M_{i}$ es que son Herm\'{\i}ticas, es decir
$M^{\dagger}_{i}=M_{i}.$ Esto implica que sus valores propios 
son reales, por lo tanto las matrices $\vec M=(M_{1},M_{2},M_{3})$ 
pueden representar cantidades f\'{\i}sicas. \\

Una transformaci\'on finita se debe hacer como producto infinito
de transformaciones infinitesimales. Por ejemplo, ocupando
\begin{eqnarray}
X^{\prime}\approx\left(\Lambda\left(\vec \alpha/N\right)\right)^{N} X
=(I-i\delta \vec \alpha \cdot\vec  M)^{N}x=
\left(I-i\frac{1}{N} \vec \alpha \cdot \vec M\right)^{N}X
\end{eqnarray}
 y considerando  el resultado
\begin{eqnarray}
\lim_{N\to \infty}\left(
I+\frac{\beta x}{N} \right)^{N}=e^{\beta x},
\label{eq:carmen}
\end{eqnarray}
se tiene 
\begin{eqnarray}
\lim_{N\to \infty}\left(I-i\frac{1}{N} \vec \alpha \cdot M\right)^{N}=
e^{-i\vec \alpha \cdot \vec M}.
\end{eqnarray}
Por lo tanto, una transformaci\'on finita est\'a dada por 
\begin{eqnarray}
X^{\prime}=e^{-i\vec \alpha \cdot \vec M}X.
\end{eqnarray}
As\'{\i}, cualquier transformaci\'on infinitesimal conectada
con la identidad tiene la forma
\begin{eqnarray}
\Lambda\left(\vec \alpha\right)=e^{-i\vec \alpha \cdot \vec M}. 
\label{eq:beli}
\end{eqnarray}
En este sentido se dice que $M_{1},M_{2},M_{3}$ son los generadores del grupo $SO(3).$\\

Ahora veamos de forma expl\'{\i}cita  la expresi\'on  (\ref{eq:beli}) para  algunos caso
particulares. Primero notemos que si $n\geq 1,$ se tiene
\begin{eqnarray}
M_{1}^{2n}&=&\left(
\begin{array}{rrrr}
0 & 0 & 0\\
0 & 1 & 0 \\
0 & 0 & 1
\end{array}\right)=T_{1},\quad M_{1}^{2n+1}=M_{1},\\
 M_{2}^{2n}&=&\left(
\begin{array}{rrrr}
1 & 0 & 0\\
0 & 0 & 0 \\
0 & 0 & 1
\end{array}\right)=T_{2},\quad M_{2}^{2n+1}=M_{2},\\
 M_{3}^{2n}&=&\left(
\begin{array}{rrrr}
1 & 0 & 0\\
0 & 1 & 0 \\
0 & 0 & 0
\end{array}\right)=T_{3},\quad M_{3}^{2n+1}=M_{2}.
\end{eqnarray}
Entonces, considerando estos resultados, juntos con   las series 
de $\cos\beta$ y $\sin\beta,$ se tiene
\begin{eqnarray}
e^{-i\beta M_{i}}&=&\sum_{n\geq 0} \frac{1}{n!}\left(-i\beta M_{i}\right)^{n}
\nonumber\\
&=&\sum_{n\geq0} \frac{1}{(2n)!}\left(-i\beta\right)^{2n}\left(M_{i}\right)^{2n}+ 
\sum_{n\geq 0} \frac{1}{(2n+1)!}\left(-i\beta\right)^{2n+1} 
\left(M_{i}\right)^{2n+1}\nonumber\\
&=&I +T_{i} \sum_{n\geq 1} \frac{1}{(2n)!}\left(-i\beta\right)^{2n}+
M_{i}\sum_{n\geq 0} \frac{1}{(2n+1)!}\left(-i\beta\right)^{2n+1}\nonumber\\
&=&I-T_{i}+T_{i}+\sum_{n\geq 1} \frac{(-1)^{n}}{(2n)!}\beta^{2n}
-iM_{i}   \sum_{n\geq 0} \frac{(-1)^{n}}{(2n+1)!}\beta^{2n+1}\nonumber\\
&=&I-T_{i}+T_{i}\sum_{n\geq 0} \frac{(-)^{n}\beta^{2n}}{(2n)!}
-iM_{i}\sin \beta  \nonumber\\
&=&I-T_{i}+T_{i}\cos\beta-i M_{i}\sin\beta.
\end{eqnarray}
De donde,   
\begin{eqnarray}
e^{-i\alpha_{1} M_{1}}&=&
\left(
\begin{array}{rrrr}
1 & 0 & 0\\
0 & \cos\alpha_{1} & -\sin\alpha_{1}\\
0 & \sin\alpha_{1} & \cos\alpha_{1}
\end{array}\right),\quad 
e^{-i\alpha_{2} M_{2}}=
\left(
\begin{array}{rrrr}
\cos\alpha_{2} & 0 & \sin\alpha_{2}\\
0 & 1 &  0\\
-\sin\alpha_{2} & 0 & \cos\alpha_{2}
\end{array}\right), \nonumber \\
e^{-i\alpha_{3} M_{3}}&=&
\left(
\begin{array}{rrrr}
\cos\alpha_{3} &- \sin\alpha_{3} & 0\\
\sin\alpha_{3} & \cos\alpha_{3} & 0\\
0 & 0 & 1
\end{array}\right).
\end{eqnarray}
As\'{\i}, $e^{-i\alpha_{1} M_{1}}$ representa una rotaci\'on sobre el eje $x,$
$e^{-i\alpha_{2} M_{2}}$ representa una rotaci\'on sobre el eje $y,$
mientras que $e^{-i\alpha_{3} M_{3}}$ representa una rotaci\'on sobre 
el eje $z.$\\

Veamos que reglas de conmutaci\'on cumplen
los generadores de las rotaciones $M_{1},M_{2},M_{3}.$ Primero notemos que
\begin{eqnarray}
 M_{1}M_{2}&=&
-\left(
\begin{array}{rrrr}
0 & 0 & 0\\
1 & 0 & 0\\
0 & 0 & 0
\end{array}\right),\quad 
 M_{2}M_{1}=
-\left(
\begin{array}{rrrr}
0 & 1 & 0\\
0 & 0 &  0\\
0 & 0 & 0
\end{array}\right), \nonumber \\
M_{1}M_{3}&=&
-\left(
\begin{array}{rrrr}
 0&0 & 0\\
0 & 0 & 0\\
1 & 0 & 0
\end{array}\right),\quad 
 M_{3}M_{1}=
-\left(
\begin{array}{rrrr}
0 & 0 & 1\\
0 & 0 &  0\\
0 & 0 & 0
\end{array}\right), \nonumber \\
M_{2}M_{3}&=&
-\left(
\begin{array}{rrrr}
 0&0 & 0\\
0 & 0 & 0\\
0 & 1 & 0
\end{array}\right),\quad 
 M_{3}M_{2}=
-\left(
\begin{array}{rrrr}
0 & 0 & 0\\
0 & 0 &  1\\
0 & 0 & 0
\end{array}\right).
\end{eqnarray}
Por lo que
\begin{eqnarray}
[M_{1},M_{2}]=iM_{3},\quad [M_{3},M_{1}]=iM_{2},\quad [M_{2},M_{3}]=iM_{1}.
\end{eqnarray}
Estas reglas de conmutaci\'on se pueden escribir como
\begin{eqnarray}
[M_{i},M_{j}]=i\epsilon_{ijk} M_{k}.\label{eq:marta}
\end{eqnarray}
Por el hecho de que el conmutador de dos generadores de rotaci\'on nos de otro generador de rotaci\'on
se dice que estos generadores forman un \'algebra de Lie, el \'algebra de Lie de $SO(3).$ Note que, como el 
conmutador entre dos generadores no es cero, dos generadores no se pueden diagonalizar simultaneamente. 
Ahora, definamos 
\begin{eqnarray}
M^{2}=M_{1}^{2}+M_{2}^{2}+M_{3}^{2},
\end{eqnarray}
de donde
\begin{eqnarray}
M^{2}=2I,
\end{eqnarray}
por lo tanto
\begin{eqnarray}
[M^{2},M_{i}]=0.
\end{eqnarray}
As\'i, los valores propios de $M^{2}$ se pueden obtener al mismo tiempo que cualquiera de los generadores $M_{i}.$\\
 
Los valores propios de los generadores $M_{i}$ se pueden calcular
directamente y est\'an dados por 
\begin{eqnarray}
M_{1}V=\lambda V \quad \Rightarrow \quad \lambda=\pm 1\quad 
V=a\frac{1}{\sqrt{2}}\left(
\begin{array}{r} 
0\\
1\\
\pm i
\end{array}
\right), \\
M_{2}V=\lambda V \quad \Rightarrow \quad \lambda=\pm 1\quad 
V=a\frac{1}{\sqrt{2}}\left(
\begin{array}{r} 
1\\
0\\
\pm i
\end{array}
\right), \\
M_{3}V=\lambda V \quad  \Rightarrow \quad \lambda=\pm 1\quad 
V=a\frac{1}{\sqrt{2}}\left(
\begin{array}{r} 
1\\
\pm i\\
0
\end{array}
\right).
\end{eqnarray}
Si $a=1,$ en cada caso se tienen vectores propios ortonormales.
Estos resultados no son dif\'{\i}ciles de obtener, posteriormente
veremos que los generadores de las rotaciones en el espacio
de las funciones de $\vec x$ est\'an relacionados con operadores diferenciales.\\

Con el s\'{\i}mbolo $\epsilon_{ijk}$ las  matrices $\vec M=(M_{1},M_{2},M_{3})$ se pueden
escribir de forma m\'as econ\'omica. En efecto, en componentes tenemos 

\begin{eqnarray}
(M_{1})_{ij}=i\epsilon_{i1j},\quad (M_{2})_{ij}=i\epsilon_{i2j},\quad 
(M_{3})_{ij}=i\epsilon_{i3j}.
\end{eqnarray}
Por ejemplo,
\begin{eqnarray}
(M_{1})_{ij}=i\epsilon_{i1j}
=i\left(
\begin{array}{rrrr}
\epsilon_{111} & \epsilon_{112} & \epsilon_{113}\\
\epsilon_{211} & \epsilon_{212} & \epsilon_{213} \\
\epsilon_{311}&\epsilon_{312}  & \epsilon_{313}
\end{array}\right)
=i\left(
\begin{array}{rrrr}
0 & 0 & 0\\
0 & 0 & -1 \\
0 & 1 & 0
\end{array}\right),
\end{eqnarray}
se puede probar que las dem\'as igualdades se cumplen.\\

As\'{\i}, las componentes de cualquier matriz antisim\'etrica se
pueden escribir como
\begin{eqnarray}
M_{ik}=-i\alpha_{j}(M_{j})_{ik}=\alpha_{j}\epsilon_{ijk}.
\end{eqnarray}
Ahora, si $\alpha_{i}$ es infintesimal, por ejemplo 
$\delta\alpha_{i}=\alpha_{i}/N$ con $N$ grande, una transformaci\'on
infinitesimal en el espacio de $\vec x$ est\'a dada por
\begin{eqnarray}
X^{\prime}=\Lambda X= (I-i\delta \vec \alpha \cdot \vec M)X.
\end{eqnarray}
En componentes se tiene
\begin{eqnarray}
x^{\prime}_{i}&\approx
&\Lambda_{ik} x_{k}= (I-i \delta\alpha_{j} M_{j})_{ik}x_{k}
= (\delta_{ik}-i\delta\alpha_{j} (M_{j})_{ik})x_{k}\nonumber\\
&=&(\delta_{ik}+\delta\alpha_{j} \epsilon_{ijk})x_{k}
= x_{i}+ \epsilon_{ijk}\delta\alpha_{j}x_{k}=x_{i}+
\left( \delta\vec \alpha\times \vec x\right)_{i},
\end{eqnarray}
es decir 
\begin{eqnarray}
x^{\prime}_{i}\approx x_{i}+
\left( \delta\vec \alpha\times \vec x\right)_{i}.
\label{eq:voli}
\end{eqnarray}

\section{Arm\'onicos esf\'ericos}
Hasta el momento nos hemos enfocado en las trasformaciones que pasan
del espacio $\vec r$ al $\vec r^{\prime}.$ Ahora veamos que pasa con las
funciones que act\'uan en estos espacios. 
Supongamos que tenemos una funci\'on, $F,$ de ${\bf R}^{3}$ a  ${\bf R}.$
Entonces, al evaluar esta funci\'on en un punto  $\vec r^{\prime}$ y 
ocupando la transformaci\'on infinitesimal Eq. (\ref{eq:josefina2}), se encuentra
\begin{eqnarray}
F\left(\vec r^{\prime}\right)&\approx&
F\left(\vec r+ \delta\vec \alpha\times \vec r\right)
= F\left(\vec r\right)+ \left(\delta \vec \alpha\times \vec 
r\right)\cdot \vec \nabla F(\vec r).
\end{eqnarray}
Adem\'as usando la propiedad c\'{\i}clica del triple producto escalar, se tiene
\begin{eqnarray}
\left(\delta \vec \alpha\times \vec r\right)\cdot \vec \nabla F(\vec r)= 
\left(\vec \nabla F(\vec r)\times \delta \vec \alpha\right)\cdot \vec r = 
\left(\vec r\times \vec \nabla F(\vec r)\right)\cdot \delta \vec \alpha=
i\left(\vec \delta \alpha\cdot \vec L\right)  F(\vec r) \nonumber
\end{eqnarray}
con 
\begin{eqnarray}
\vec L= -i\vec r\times \vec \nabla.
\end{eqnarray}
As\'{\i},
\begin{eqnarray}
F\left(\vec r^{\prime}\right)&\approx&\left(1+i\delta \vec \alpha\cdot \vec L \right)F(\vec r). 
\end{eqnarray}
Por lo tanto, el operador $\vec L$ es el generador de las rotaciones en el espacio de las funciones.
Anteriormente vimos que este operador es Herm\'{\i}tico, por lo que sus valores propios son reales.\\

Veamos que forma tiene una rotaci\'on finita. Al igual que
en caso del espacio $\vec r,$ tomaremos
$\delta \vec \alpha =\vec \alpha /N$  y
consideraremos que para tener una transformaci\'on
finita debemos hacer el producto de un n\'umero infinito
de transformaciones infinitesimales. Entonces,
\begin{eqnarray}
F\left(\vec r^{\prime}\right)&\approx&
\left( 1+ i \frac{\vec \alpha}{N} \cdot \vec L \right)^{N}F(\vec r).
\end{eqnarray}
As\'{\i}, ocupando el resultado  (\ref{eq:carmen}), se encuentra
\begin{eqnarray}
\lim_{N\to \infty}\left(I+i\frac{1}{N} \vec \alpha 
\cdot\vec L\right)^{N}=U\left(\vec \alpha\right)=
e^{i\vec \alpha \cdot \vec L}.
\end{eqnarray}
Por lo tanto, una transformaci\'on finita est\'a dada por
\begin{eqnarray}
F\left(\vec r^{\prime}\right)=
e^{i\vec \alpha \cdot \vec L}F\left(\vec r \right).
\end{eqnarray}
Note que el operador $U\left(\vec \alpha\right)$ 
satisface 
\begin{eqnarray}
U\left(\vec \alpha\right)^{\dagger}=
U\left(-\vec \alpha\right)= U^{-1}\left(\vec \alpha\right).
\end{eqnarray}
Cuando un operador, $A,$ cumple  
$$A^{\dagger}=A^{-1}$$
se dice que es un operador unitario. As\'i,  $U\left(\vec \alpha\right)$ es unitario.

\section{Reglas de conmutaci\'on del momento angular}

Anteriormente vimos que los generadores $M_{i}$ satisfacen 
las reglas de conmutaci\'on Eq. (\ref{eq:marta}). Veamos que
reglas de conmutaci\'on cumplen los operadores $L_{i}.$ Primero notemos que
definiendo $\vec p=-i\vec \nabla$ se tiene $\vec L=\vec r \times \vec p.$
En componentes se encuentra $L_{i}=\epsilon_{ijk}x_{j}p_{k}.$  De donde
\begin{eqnarray}
[L_{i},L_{j}]&=&[\epsilon_{ilm}x_{l}p_{m},\epsilon_{jrs}x_{r}p_{s}]
 =\epsilon_{ilm}\epsilon_{jrs}[x_{l}p_{m},x_{r}p_{s}]\nonumber \\
&= &\epsilon_{ilm}\epsilon_{jrs}\left(x_{l}[p_{m},x_{r}]p_{s}+
x_{r}[x_{l},p_{s}]p_{m}]\right) \nonumber\\
&=&\epsilon_{ilm}\epsilon_{jrs}\left(-ix_{l}p_{s}\delta_{mr}+ix_{r}p_{m}\delta_{ls}\right)\nonumber\\
&=&i\left(- \epsilon_{ilr}\epsilon_{jrs}x_{l}p_{s} +\epsilon_{ism}\epsilon_{jrs}x_{r}p_{m} \right)\nonumber\\
&=&i\left(\epsilon_{ilr}\epsilon_{rjs}x_{l}p_{s}-\epsilon_{ims}\epsilon_{sjr}x_{r}p_{m}\right)\nonumber\\
&=&i\left(\epsilon_{iar}\epsilon_{rjb}x_{a}p_{b}-\epsilon_{ibs}\epsilon_{sja}x_{r}p_{b}\right)\nonumber\\
&=&i\left(\epsilon_{iar}\epsilon_{rjb}-\epsilon_{ibs}\epsilon_{sja}\right) x_{a}p_{b},
\end{eqnarray}
ahora, note que 
\begin{eqnarray}
\epsilon_{iar}\epsilon_{rjb}-\epsilon_{ibs}\epsilon_{sja}&=&\left(\delta_{ij}\delta_{ab}-\delta_{ib}\delta_{aj}\right)-
\left(\delta_{ij}\delta_{ba}-\delta_{ia}\delta_{bj}\right)\nonumber\\
&=&\delta_{ia}\delta_{bj}-\delta_{ib}\delta_{aj}=\epsilon_{ijk}\epsilon_{kab}.
\end{eqnarray}
Ocupando este resultado se encuentra 
\begin{eqnarray}
[L_{i},L_{j}]&=& i \epsilon_{ijk}\epsilon_{kab} x_{a}p_{b},\nonumber
\end{eqnarray}
es decir
\begin{eqnarray}
[L_{i},L_{j}]=i\epsilon_{ijk}L_{k}.
\end{eqnarray}
Se puede observar que son las mismas reglas de conmutaci\'on que cumple $M_{i}$
Eq. (\ref{eq:marta}). A estas reglas de comutaci\'on se les llama el \'algebra de Lie de $SO(3).$\\

Anteriormente vimos que la matriz $M^{2}=M_{1}^{2}+M_{2}^{2}+M_{3}^{2}$
conmuta con todas las matrices $M_{i}.$ Para los operadores $L_{i}$
el operador equivalente a $M^{2}$ es 
\begin{eqnarray}
L^{2}=L_{x}^{2}+L_{y}^{2}+L_{z}^{2}=L_{j}L_{j}.
\end{eqnarray}
El cual cumple
\begin{eqnarray}
[L^{2},L_{i}]&=&[L_{j}L_{j},L_{i}]=L_{j}[L_{j},L_{i}]+[L_{j},L_{i}]L_{j}
\nonumber\\
& =&i\epsilon_{jil}L_{j}L_{l}+i\epsilon_{jil}L_{l}L_{j}
=-i\left(\epsilon_{ijl}L_{j}L_{l}+\epsilon_{ijl}L_{l}L_{j}\right),
\label{eq:arm-conm1}
\end{eqnarray}
ahora, renombrando  \'{\i}ndices se encuentra
\begin{eqnarray}
\epsilon_{ijl}L_{j}L_{l}=\sum_{j=1}^{3}\sum_{l=1}^{3}\epsilon_{ijl}L_{j}L_{l}
=\sum_{l=1}^{3} \sum_{j=1}^{3}\epsilon_{ilj}L_{l}L_{j}=
\epsilon_{ilj}L_{l}L_{j},
\end{eqnarray}
introduciendo esta igualdad en Eq. (\ref{eq:arm-conm1}) se tiene
\begin{eqnarray}
[L^{2},L_{i}]=-i\left(\epsilon_{ilj}L_{l}L_{j}+\epsilon_{ijl}L_{l}L_{j}\right)=
-i\left(-\epsilon_{ijl}L_{l}L_{l}+\epsilon_{ijl}L_{l}L_{j}\right)=0.
\end{eqnarray}
Por lo tanto, $L^{2}$ conmuta con cualquier $L_{i}.$ A $L^{2}$ se le llama el Casimir del \'algebra de Lie
de $SO(3).$ Como $L^{2}$ conmuta con cualquier $L_{i},$ este operador
comparte vectores propios con estos tres operadores.\\

\section{Ecuaci\'on de valores propios de $L^{2}$}

Para obtener los vectores y valores propios de $M_{i}$ y $M^{2}$
resolvimos un problema de \'algebra lineal, mas para
obtener los vectores propios de $L^{2}$ y los de, por ejemplo, $L_{z}$ se deben plantear las
ecuaciones
\begin{eqnarray}
L^{2}Y_{lm}=\lambda Y_{\lambda m},\quad 
L_{z}Y_{\lambda m}=m Y_{\lambda m}.\label{eq:alo}
\end{eqnarray}
Estas son dos ecuaciones diferenciales. En efecto, considerando las expresiones de $L^{2}$ y $L_{z}$ en coordenadas esf\'ericas, Eq. (\ref{eq:op-moment-esfe}) y Eq. (\ref{eq:z-momen-esfe}), se encuentra
\begin{eqnarray}
L^{2}Y_{\lambda m}(\theta, \varphi)&=&-\left[\frac{1}{\sin\theta}
\frac{\partial }{\partial \theta}
\left(\sin\theta \frac{\partial Y_{\lambda m}(\theta,\varphi) }{\partial \theta}\right)
+\frac{1}{\sin^{2}\theta}\frac{\partial^{2} Y_{\lambda m}(\theta,\varphi)}{\partial \varphi^{2}}\right]
\nonumber\\
&=&\lambda Y_{\lambda m}(\theta,\varphi), \label{eq:armo-esfe-eq}\\
L_{z}Y_{\lambda m}(\theta,\varphi)&=&-i\frac{\partial Y_{\lambda m}(\theta,\varphi)
(\theta,\varphi)}{\partial \varphi}=m Y_{\lambda m}(\theta,\varphi).
\end{eqnarray}
De la segunda ecuaci\'on es claro que $Y_{\lambda m}(\theta,\varphi)$ es de la forma
\begin{eqnarray}
Y_{\lambda m}(\theta,\varphi)=\alpha_{\lambda m}
e^{im\varphi}P_{\lambda}^{m}(\theta).\label{eq:armo-esfe-sol}
\end{eqnarray}
Si queremos que la funci\'on $Y_{\lambda m}(\theta,\varphi)$  no sea multivaluada debemos pedir  
$Y_{\lambda m}(\theta,\varphi+2\pi )=Y_{\lambda m}(\theta,\varphi).$ Esto implica 
$e^{im\phi}=e^{im(\phi+2\pi)},$ lo cual se cumple s\'olo si 
\begin{eqnarray}
m=0, \pm 1,\pm 2,\pm 3,\cdots.
\end{eqnarray}
Por lo tanto, $m$ debe ser un entero. Posteriormente veremos los posibles valores de $\lambda.$\\

Sustituyendo Eq. (\ref{eq:armo-esfe-sol}) en Eq. (\ref{eq:armo-esfe-eq}) se encuentra
\begin{eqnarray}
\frac{1}{\sin\theta}
\frac{\partial }{\partial \theta}
\left(\sin\theta \frac{\partial P_{\lambda}^{ m}(\theta) }
{\partial \theta}\right)- \frac{m^{2}}{\sin^{2}\theta}  P_{\lambda}^{ m}(\theta)
=-\lambda P_{\lambda}^{ m}(\theta).
\label{eq:asociada}
\end{eqnarray}
Con el cambio de variable $u=\cos \theta,$ tenemos 
\begin{eqnarray}
\sin\theta=\sqrt{1-u^{2}},\qquad 
\partial_{\theta}=-\sqrt{1-u^{2}}\partial_{u}.
\end{eqnarray}
De donde, Eq. (\ref{eq:asociada}) toma la forma 
\begin{eqnarray}
\frac{d}{du}\left((1-u^{2})\frac{d P_{\lambda}^{m}(u)}{du}\right)+
\left(\lambda-\frac{m^{2}} {1-u^{2}}\right) P_{\lambda}^{m}(u)=0.
\label{eq:giovana}
\end{eqnarray}
Esta es la llamada ecuaci\'on asociada de Legendre.
Para el caso $m=0$ se define $P_{\lambda}^{0}(u)=P_{\lambda}(u),$ que debe
satisfacer 
\begin{eqnarray}
\frac{d}{du}\left((1-u^{2})\frac{d P_{\lambda}(u)}{du}\right)+
\lambda P_{\lambda}(u)=0,
\label{eq:giovana-1}
\end{eqnarray}
que es la llamada ecuaci\'on de Legendre. En lo que sigue, estudiando la  estructura del grupo de rotaciones, obtendremos las soluciones de estas ecuaciones. Primero veremos la 
ortormalidad de las soluciones de la ecuaci\'on (\ref{eq:asociada}).

\section{Relaciones de ortonormalidad}

Supongamos que $m$ y $m^{\prime}$ son enteros y  que $m\not = m^{\prime},$
entonces
\begin{eqnarray}
\int_{0}^{2\pi} d\varphi \left(e^{im^{\prime}\varphi}\right)^{*}e^{im\varphi}
&=&\int_{0}^{2\pi} d\varphi e^{-im^{\prime}\varphi}e^{im\varphi}= 
\int_{0}^{2\pi} d\varphi e^{i\left(m-m^{\prime}\right)\varphi}\nonumber\\
&=&\frac{ e^{i\left(m-m^{\prime}\right)\varphi} }{i\left(m-m^{\prime}\right)}\Bigg|_{0}^{2\pi}
=\frac{ e^{i\left(m-m^{\prime}\right)2\pi} -1}{i\left(m-m^{\prime}\right)}=0.
\end{eqnarray}
Si $m= m^{\prime},$ se encuentra 
\begin{eqnarray}
\int_{0}^{2\pi} d\varphi \left(e^{im^{\prime}\varphi}\right)^{*}e^{im\varphi}
&=&\int_{0}^{2\pi} d\varphi e^{-im\varphi}e^{im\varphi}= \int_{0}^{2\pi} d\varphi=2\pi.
\end{eqnarray}
De donde, si  $m$ y $m^{\prime}$ son enteros se tiene
\begin{eqnarray}
\int_{0}^{2\pi} d\varphi \left(e^{im^{\prime}\varphi}\right)^{*}e^{im\varphi}=
2\pi\delta_{mm^{\prime}}.
\end{eqnarray}
Por lo tanto las funciones $e^{im\varphi}$ son ortogonales en el intervalo $[0,2\pi].$\\

Note que la ecuaci\'on (\ref{eq:asociada}) se puede escribir como
\begin{eqnarray}
\frac{\partial }{\partial \theta}
\left(\sin\theta \frac{\partial P_{\lambda}^{m}(\theta) }
{\partial \theta}\right)+\left( \lambda \sen\theta- \frac{m^{2}}
{\sin\theta}\right)P_{\lambda}^{m}(\theta)=0.
\end{eqnarray}
Como se puede observar, esta ecuaci\'on es tipo Sturm-Liouville. En este caso $p(\theta)=\sin\theta, r(\theta)=\frac{-m^{2}}{\sin\theta}.$ Considerando los resultados para las ecuaciones tipo Sturm-Louville, como $p(0)=p(\pi)=0,$ se llega a 
\begin{eqnarray}
\left(\lambda^{\prime}-\lambda\right)\int_{0}^{\pi}d\theta \sin\theta
 P_{\lambda^{\prime}}^{m}(\theta)P_{\lambda}^{m}(\theta)=0.
\nonumber
\end{eqnarray}
En particular si $\lambda^{\prime}\not =\lambda$ 
\begin{eqnarray}
\int_{0}^{\pi}d\theta \sin\theta P_{\lambda^{\prime}}^{m}(\theta)P_{\lambda}^{m}(\theta)=0.
\end{eqnarray}
En general se tiene 
\begin{eqnarray}
\int_{0}^{\pi}d\theta \sin\theta P_{\lambda^{\prime}}^{m}(\theta)P_{\lambda}^{m}(\theta)=\delta_{\lambda\lambda^{\prime}}
\beta_{\lambda m},\qquad \beta_{\lambda m}={\rm constante}>0.
\end{eqnarray}
Por lo tanto, las funciones $P_{\lambda}^{m}(\theta)$ son ortogonales.\\

Empleando la ortogonalidad de las funciones $e^{im\varphi}$ y 
$P_{\lambda}^{m}(\theta),$ se puede escojer $\alpha_{lm}$ de tal forma que los
arm\'onicos esf\'ericos sean ortonormales. En efecto, como los arm\'onicos
esf\'ericos tienen la forma Eq. (\ref{eq:armo-esfe-sol}), se encuentra
\begin{eqnarray}
& &\int d\Omega Y^{*}_{\lambda^{\prime}m^{\prime}}(\theta,\varphi) 
Y_{\lambda m}(\theta,\varphi)=\nonumber\\
& &=\int_{0}^{2\pi} d\varphi\int_{0}^{\pi}d\theta \Bigg(\sin \theta 
\alpha^{*}_{\lambda^{\prime}m^{\prime}}e^{-im^{\prime}\varphi}
P_{\lambda^{\prime}}^{m^{\prime}}(\cos\theta)
\alpha_{\lambda m}e^{im\varphi}P_{\lambda}^{m}(\cos\theta)\Bigg)\nonumber \\
&=&\alpha^{*}_{\lambda^{\prime}m^{\prime}}\alpha_{\lambda m}
\int_{0}^{2\pi}d\varphi e^{i(m^{\prime}-m)\varphi}d\varphi 
\int_{0}^{\pi}d\theta \sin\theta P_{\lambda^{\prime}}^{m^{\prime}}(\cos\theta)P_{\lambda}^{m}(\cos \theta)
\nonumber \\
&=&\alpha^{*}_{\lambda^{\prime}m^{\prime}}\alpha_{\lambda m}2\pi 
\delta_{mm^{\prime}}
\int_{0}^{\pi}d\theta \sin\theta P_{\lambda^{\prime}}^{m^{\prime}}(\cos\theta)P_{\lambda}^{m}(\cos\theta)
\nonumber \\
&=&\alpha^{*}_{\lambda^{\prime}m}\alpha_{\lambda m}2\pi 
\delta_{mm^{\prime}}
\int_{0}^{\pi}d\theta \sin\theta P_{\lambda^{\prime}}^{m}(\cos\theta)P_{\lambda}^{m}(\cos\theta)\nonumber\\
&=&\alpha^{*}_{\lambda^{\prime}m}\alpha_{\lambda m}2\pi 
\delta_{mm^{\prime}}\beta_{\lambda m}\delta_{\lambda^{\prime}\lambda}
\nonumber \\
&=&|\alpha_{\lambda^{\prime}m}|^{2}2\pi \beta_{\lambda m}
\delta_{mm^{\prime}}\delta_{\lambda^{\prime}\lambda}.
\label{eq:p8}
\end{eqnarray}
Entonces, si 
\begin{eqnarray}
|\alpha_{\lambda^{\prime}m}|^{2}=\frac{1}{2\pi \beta_{\lambda m}},
\end{eqnarray}
se cumple 
\begin{eqnarray}
<Y_{\lambda^{\prime}m^{\prime}}(\theta,\varphi)|Y_{\lambda m}(\theta,\varphi)>
=\int d\Omega Y^{*}_{\lambda^{\prime}m^{\prime}}(\theta,\varphi) 
Y_{\lambda m}(\theta,\varphi)=\delta_{mm^{\prime}}\delta_{\lambda^{\prime}\lambda}.\label{eq:p80}
\end{eqnarray}
As\'i los arm\'onicos esf\'ericos son ortonormales.

\section{Operadores escalera y espectro de $L^{2}$}

Antes de resolver la ecuaci\'on diferencial (\ref{eq:armo-esfe-eq})  
veamos algunas propiedades de las funciones propias $Y_{\lambda m}$
y sus valores propios $\lambda ,m.$ Primero notemos que para cualquier funci\'on $f$ se tiene
\begin{eqnarray}
0&\leq& \left<L_{x}f| L_{x}f\right>= \left<f|L^{\dagger}_{x}L_{x}f\right>=\left<f|L_{x}^{2}f\right>\nonumber\\
0&\leq& \left<L_{y}f| L_{y}f\right>= \left<f|L^{\dagger}_{y}L_{y}f\right>=\left<f|L_{y}^{2}f\right>,\nonumber
\end{eqnarray} 
por lo que 
\begin{eqnarray}
0&\leq& \left<f|\left(L_{x}^{2}+L_{y}^{2}\right)f\right>= \left<f|\left(L^{2}-L_{z}^{2}\right)f\right>.
\nonumber
\end{eqnarray} 
En particular si $f=Y_{\lambda m},$ se llega a
\begin{eqnarray}
\left(L^{2}-L_{z}^{2}\right)Y_{\lambda m}=
\left(\lambda-m^{2}\right)Y_{\lambda m},
\end{eqnarray} 
entonces 
\begin{eqnarray}
0\leq \left(\lambda-m^{2}\right) \left<Y_{lm}|Y_{lm}\right>.
\end{eqnarray} 
Note que la \'unica forma de que se cumpla la desigualdad 
$\left(\lambda-m^{2}\right)<0$
es que 
$\left<Y_{lm}|Y_{lm}\right>=0,$ 
es decir $Y_{lm}=0.$ Si $Y_{lm}\not= 0$ se tiene $\left(\lambda-m^{2}\right)\geq 0,$ entonces 
\begin{eqnarray}
-\sqrt{\lambda} \le m \le \sqrt{\lambda}.
\label{eq:co}
\end{eqnarray} 
Esta restricci\'on nos permitir\'a encontrar los valores de $\lambda.$\\

En el estudio del oscilador arm\'onico fueron de gran utilidad los operadores de ascenso y descenso.
En el caso que ahora estudiamos hay dos operadores equivalentes, los cuales son 
\begin{eqnarray}
L_{\pm}=L_{x}\pm iL_{y}.
\end{eqnarray}
Note que 
\begin{eqnarray}
L_{\mp}L_{\pm}&=& \left(L_{x}\mp iL_{y}\right)\left(L_{x}\pm iL_{y}\right)=
L_{x}^{2}+L^{2}_{y}\pm i\left(L_{x}L_{y} - L_{y}L_{x}\right)\nonumber\\
&=& L_{x}^{2}+L^{2}_{y}+L_{z}^{2}-L_{z}^{2}\mp L_{z}
=L^{2}-L_{z}^{2}\mp L_{z}=L^{2}-\left(L_{z}^{2}\pm L_{z}\right),\nonumber
\end{eqnarray}
es decir,
\begin{eqnarray}
L_{\mp}L_{\pm}=L^{2}-\left(L_{z}^{2}\pm L_{z}\right).
\label{eq:mia}
\end{eqnarray}
Claramente $L_{\pm}$ conmuta con $L^{2}:$ 
\begin{eqnarray}
[L^{2},L_{\pm}]&=&0.
\end{eqnarray}
Esto indica que $L_{\pm}$ tiene funciones propias comunes con $L^{2}.$
Adem\'as
\begin{eqnarray}
[L_{z},L_{x}\pm iL_{y}]=[L_{z},L_{x}]\pm[L_{z}, L_{y}]=iL_{y}\pm L_{x}=\pm\left(L_{x}\pm iL_{y}\right),
\end{eqnarray}
es decir
\begin{eqnarray}
[L_{z},L_{\pm}]=\pm L_{\pm}.\label{eq:viky}
\end{eqnarray}
Por lo que, $L_{\pm}$ no tiene funciones propias comunes con 
$L_{z}.$\\

Veamos que efecto
tiene el operador $L_{\pm}$ sobre las funciones propias de  $L_{z}$.
Definamos la funci\'on
\begin{eqnarray}
\tilde Y=L_{\pm}Y_{\lambda m},
\end{eqnarray}
entonces tomando en cuenta que $L^{2}$ y $L_{\pm}$ conmutan, 
se tiene 
\begin{eqnarray}
L^{2}\tilde Y=L^{2}L_{\pm}Y_{\lambda m}=L_{\pm}L^{2}Y_{\lambda m}=
\lambda L_{\pm} Y_{\lambda m}=\lambda \tilde Y.
\end{eqnarray}
Por lo tanto, $L_{\pm}Y_{\lambda m}$ tambi\'en es funci\'on propia de 
$L^{2}$ con el mismo valor propio, $\lambda,$ que $Y_{\lambda m}.$ 
Adem\'as, considerando el conmutador Eq. (\ref{eq:viky}) 
tenemos 
\begin{eqnarray}
L_{z}\tilde Y&=&L_{z}L_{\pm}Y_{\lambda m}=
\left(L_{\pm}L_{z} \pm L_{\pm}\right)Y_{\lambda m}=
\left(L_{\pm}L_{z}Y_{\lambda m} \pm L_{\pm}Y_{\lambda m}\right) \nonumber\\
&=& \left(mL_{\pm}Y_{\lambda m} \pm L_{\pm}Y_{\lambda m}\right)=
(m\pm 1) L_{\pm}Y_{\lambda m}=(m\pm 1)\tilde Y ,
\end{eqnarray} 
es decir,
\begin{eqnarray}
L_{z}L_{\pm}Y_{\lambda m}=(m\pm 1) L_{\pm}Y_{\lambda m}.
\end{eqnarray} 
Por lo tanto, $\tilde Y=L_{\pm}Y_{\lambda m}$ es vector
propio de $L_{z},$ pero no con el valor propio de $Y_{\lambda m}$ si no
con el de  $Y_{\lambda m\pm 1}.$ De donde, como $L_{z}$ 
tiene un espectro  no degenerado, se debe cumplir
\begin{eqnarray}
L_{\pm}Y_{\lambda m}=\alpha(\lambda,m)Y_{\lambda m\pm 1},\qquad 
\alpha(\lambda,m)={\rm constante}. \label{eq:katy}
\end{eqnarray}
El valor de $\alpha(\lambda,m)$ lo obtendremos posteriormente.\\

Como podemos ver, el operador $L_{\pm}$ es muy \'util, pues
nos permite pasar de una funci\'on propia $Y_{\lambda m}$ a otra
$Y_{\lambda m\pm 1}.$ Ahora si aplicamos $n$ veces este operador
obtenemos
\begin{eqnarray}
(L_{\pm})^{n}Y_{\lambda m}=\tilde \alpha
(\lambda,m)Y_{\lambda m\pm n},
\end{eqnarray}
con 
\begin{eqnarray}
L_{z}Y_{\lambda m\pm n}= (m\pm n) Y_{\lambda m\pm n}
\label{eq:escalera}.
\end{eqnarray}
As\'{\i}, $Y_{\lambda m\pm n}$ es vector propio de $L_{z}$  con valor propio $(m\pm n),$ sin importar el valor de $n.$ 
Ahora, es claro que dado cualquier dos n\'umeros $\lambda$ y  $m$ existen otros dos n\'umeros, $n_{a}$ y $n_{b},$ 
tales que 
\begin{eqnarray}
 \sqrt{\lambda} < (m+n_{a}) ,\qquad   (m-n_{b})<   -\sqrt{\lambda},\nonumber
\end{eqnarray}
En estos casos no se cumple la restricci\'on Eq. (\ref{eq:co}), por lo que $Y_{\lambda m+n_{a}}=0$
y $Y_{\lambda m-n_{b}}=0.$ Esto quiere decir que 
\begin{eqnarray}
\left( L_{+} \right)^{n_{a}} Y_{\lambda m}=0,\qquad \left( L_{-}\right)^{n_{b}}Y_{\lambda m}=0.
\end{eqnarray}
Definamos el conjunto
\begin{eqnarray}
A_{+}=\{n|  \left( L_{+} \right)^{n} Y_{\lambda m}=0\}
\end{eqnarray}
y sea $N_{+}$ el m\'inimo de $A_{+}.$ Note que 
\begin{eqnarray}
\left( L_{+} \right)^{N_{+}-1} Y_{\lambda m}\not =0,
\end{eqnarray}
esto es cierto, pues de lo contrario $N_{+}$ no ser\'ia el m\'inimo de $A_{+}.$ Tambi\'en  definamos $l=N_{+}-1+m,$
es decir $l-m=N_{+}-1,$ se puede observar que 
\begin{eqnarray}
\left( L_{+} \right)^{N_{+}-1} Y_{\lambda m}= \left( L_{+} \right)^{l-m} Y_{\lambda m} =\gamma Y_{\lambda l}\not =0, \quad \gamma={\rm constante}.
\nonumber
\end{eqnarray}
Por lo tanto, existe $l$ tal que $Y_{\lambda l}\not =0$ y 
\begin{eqnarray}
L_{+} Y_{\lambda l}=0,
\nonumber
\end{eqnarray}
note que $l$ es el m\'aximo valor que puede tomar $m$ de tal forma que $Y_{\lambda m}\not= 0.$\\

Ahora, definamos 
\begin{eqnarray}
A_{-}=\{n|  \left( L_{-} \right)^{n} Y_{\lambda m}=0\}
\end{eqnarray}
y sea $N_{-}$ el m\'inimo de $A_{-}.$ Se puede observar que  
\begin{eqnarray}
\left( L_{-} \right)^{N_{-}-1} Y_{\lambda m}\not =0,
\end{eqnarray}
pues de lo contrario $N_{-}$ no ser\'ia el m\'inimo de $A_{-}.$ Tambi\'en  definamos 
$$l^{\prime}=m-\left(N_{-}-1\right)$$
es decir 
$$N_{1}-1=l^{\prime}-m,$$
 se puede observar que 
\begin{eqnarray}
\left( L_{+} \right)^{N_{-}-1} Y_{\lambda m}= \left( L_{+} \right)^{l^{\prime}-m} Y_{\lambda m} =\gamma^{\prime} Y_{\lambda l^{\prime}}\not =0, \quad \gamma^{\prime}={\rm constante}.
\nonumber
\end{eqnarray}
Por lo tanto, existe $l^{\prime}$ tal que $Y_{\lambda l^{\prime}}\not =0$ y 
\begin{eqnarray}
L_{-} Y_{\lambda l^{\prime}}=0,
\nonumber
\end{eqnarray}
note que $l^{\prime}$ es el m\'inimo  valor que puede tomar $m$ de tal forma que $Y_{\lambda m}\not= 0.$
De estos resultados tenemos que si $Y_{\lambda m}\not= 0,$ entonces se cumplen  $l^{\prime} \le m \le l.$\\

Ahora, como $l$ es el valor propio m\'aximo que
puede tener $L_{z},$ entonces considerando 
Eq. (\ref{eq:mia}),  se tiene 
\begin{eqnarray}
L_{-}L_{+}Y_{\lambda l}=\left(L^{2}-(L^{2}_{z}+L_{z})\right)Y_{\lambda l}
=[\lambda-l(l+1)]Y_{\lambda l}=0,\nonumber
\end{eqnarray}
esto implica que 
$$\lambda=l(l+1).$$
Adem\'as, como $l^{\prime}$
es el valor propio m\'{\i}nimo que puede tener $L_{z},$ se tiene 
\begin{eqnarray}
L_{+}(L_{-}Y_{\lambda l^{\prime}})=
\left(L^{2}-(L^{2}_{z}-L_{z})\right)Y_{\lambda l^{\prime}}
=[l(l+1)-l^{\prime}(l^{\prime}-1)]Y_{\lambda l^{\prime}}=0,
\nonumber
\end{eqnarray}
de donde 
\begin{eqnarray}
l(l+1)-l^{\prime}(l^{\prime}-1)&=&l^{2}-l^{\prime 2}+l+l^{\prime}=(l+l^{\prime})(l-l^{\prime})+l+l^{\prime}\nonumber\\
&=&(l+l^{\prime})(l+1-l^{\prime})=0.
\end{eqnarray}
Entonces,  
\begin{eqnarray}
l^{\prime}_{+}=(l+1) \qquad {\rm o}\qquad  l^{\prime}_{-}=-l,
\end{eqnarray}
claramente el  \'unico valor permitido es $l^{\prime}_{-}=-l$ y el  m\'{\i}nimo valor 
que puede tomar $m$ es $-l$. As\'i los valores propios de $L_{z}$ deben cumplir 
\begin{eqnarray}
-l \le m \le l\qquad l=0,\pm 1,\pm 2,\cdots, \pm l.
\label{eq:co2}
\end{eqnarray}
Note que hay $(2l+1)$ funciones propias con el mismo valor propio
$l(l+1).$ Es decir,  para cada  propio $l(l+1)$ hay $2l+1$ funciones que
satisfacen 
\begin{eqnarray}
L^{2}Y_{lm}(\theta,\varphi)=l(l+1)Y_{lm}(\theta,\varphi)
\end{eqnarray}
con valores propios de $L_{z}$ que cumplen Eq. (\ref{eq:co2}).\\ 

Un resultado importante de todo este desarrollo es que, dado un valor $l,$ basta obtener un 
arm\'onico esf\'erico, $Y_{lm}(\theta ,\varphi),$ para que, mediante los operadores $L_{\pm},$ obtener los $2l$ restantes. 
Por ejemplo podemos obtener
el arm\'onico esf\'erico $Y_{ll}(\theta,\varphi)$ y con el operador $L_{-}$ obtener los restantes. De la misma forma si conocemos
funci\'on $Y_{l0}$ cualquier otro arm\'onico esf\'erico 
est\'a dado por 
\begin{eqnarray}
(L_{-})^{m}Y_{l0}=\alpha Y_{l, -m}
\end{eqnarray}
o por
\begin{eqnarray}
(L_{+})^{m}Y_{l0}=\alpha Y_{l, m}.
\end{eqnarray}
Posteriormente ocuparemos estos resultados para obtener de forma
expl\'{\i}cita los arm\'onicos esf\'ericos. 

\section{Resultados preliminares}

Antes de continuar veremos dos resultados que nos permitir\'an obtener los arm\'onicos
esf\'ericos.

\subsection{Constante $\alpha$ y reglas de recurrencia}

Determinemos $\alpha$ definida en Eq. (\ref{eq:katy}). Ahora, sabemos que 
 los arm\'onicos esf\'ericos son ortonormales  y que se cumple $L_{\pm}Y_{lm}=\alpha_{\pm}Y_{lm\pm 1},$
entonces usando las propiedades del producto escalar se tiene 
\begin{eqnarray}
<L_{\pm} Y_{lm}|L_{\pm}Y_{lm}>&=&<\alpha_{\pm} Y_{lm\pm 1}|\alpha_{\pm}Y_{lm\pm 1}>=\alpha^{*}_{\pm}\alpha_{\pm}
< Y_{lm\pm 1}|Y_{lm\pm 1}>\nonumber\\
&=& |\alpha_{\pm}|^{2},\label{eq:arm-cons}
\end{eqnarray}
adem\'as
\begin{eqnarray}
<L_{\pm} Y_{lm}|L_{\pm}Y_{lm}>&=&<Y_{lm}|L^{\dagger}_{\pm}L_{\pm}Y_{lm}>
=<Y_{lm}|L_{\mp}L_{\pm}Y_{lm}>\nonumber\\
&=&
<Y_{lm}|\left[L^{2}-(L_{z}^{2}\pm L_{z})\right]Y_{lm}>\nonumber\\
&=&<Y_{lm}|\left[l(l+1)-(m^{2}\pm m)\right]Y_{lm}>\nonumber\\
&=&\left(l^{2}+l-m^{2}\mp m\right) <Y_{lm}| Y_{lm}>\nonumber\\
&=& \left( l^{2}-m^{2}+l\mp m\right)=
(l\pm m)(l\mp m )+(l\mp m)\nonumber\\
&=&(l\mp m)(l\pm m +1) .\label{eq:arm-cons1}
\end{eqnarray}
Igualando Eq. (\ref{eq:arm-cons}) con Eq. (\ref{eq:arm-cons1}) se encuentra
\begin{eqnarray}
|\alpha_{\pm}|^{2}=(l\mp m)(l\pm m +1).
\end{eqnarray}
Por lo tanto,
\begin{eqnarray}
\alpha_{\pm}=\sqrt{(l\mp m)(l\pm m +1)}.
\end{eqnarray}
As\'{\i}, Eq. (\ref{eq:katy}) tiene la forma
\begin{eqnarray}
L_{\pm}Y_{lm}(\theta \varphi)=\sqrt{(l\mp m)(l\pm m +1)}Y_{lm\pm 1}(\theta, \varphi).
\label{eq:escalera-norm}
\end{eqnarray}

\subsection{Relaciones de recurrencia de $L_{\pm}$}

Como aplicaremos varias veces el operador $L_{\pm},$ veremos algunas reglas 
de recurrencia de estos operadores. \\

Primero observemos que en coordenadas esf\'ericas, Eqs. (\ref{eq:x-momen-esfe})-(\ref{eq:z-momen-esfe}), se encuentra
\begin{eqnarray}
L_{\pm}&=&L_{x}\pm iL_{y}\nonumber\\
&=&
i\left(\sin\varphi \frac{\partial }{\partial \theta}+\cot\theta \cos\varphi
 \frac{\partial }{\partial \varphi}\right)\pm i(-i)
\left(\cos\varphi \frac{\partial }{\partial \theta}-\cot\theta \sin\varphi
 \frac{\partial }{\partial \varphi}\right)\nonumber\\
&=&\left((\pm \cos\varphi+i \sin\varphi) 
\frac{\partial }{\partial \theta}+\cot\theta (i\cos\varphi\mp\sin\varphi) 
 \frac{\partial }{\partial \varphi}\right)\nonumber\\
&=&\left(\pm (\cos\varphi\pm i \sin\varphi) 
\frac{\partial }{\partial \theta}+i\cot\theta (\cos\varphi\pm i\sin\varphi) 
 \frac{\partial }{\partial \varphi}\right)\nonumber\\
&=&\left(\pm e^{\pm i\varphi} 
\frac{\partial }{\partial \theta}+i e^{\pm i\varphi}\cot\theta 
 \frac{\partial }{\partial \varphi}\right)\nonumber\\
&=&e^{\pm i\varphi}\left(\pm 
\frac{\partial }{\partial \theta}+i \cot\theta 
 \frac{\partial }{\partial \varphi}\right).
\end{eqnarray}
Ahora, note que 
\begin{eqnarray}
\frac{\partial }{\partial \theta}\left( \left(\sin\theta\right)^{\mp k}f(\theta)\right)&=&
\mp k\left( \sin\theta\right) ^{\mp k -1} (\cos\theta) f(\theta)+\left(\sin\theta\right)^{\mp k} 
\frac{\partial f(\theta)}{\partial \theta}\nonumber\\
&=&\left(\sin\theta\right)^{\mp k}\left(\frac{\partial f(\theta)}{\partial \theta}
\mp k(\cot\theta) f(\theta)\right),\nonumber
\end{eqnarray}
as\'i
\begin{eqnarray}
\left(\sin\theta\right)^{\pm k}
\frac{\partial }{\partial \theta}\left( \left(\sin\theta\right)^{\mp k}f(\theta)\right)=\left(\frac{\partial f(\theta)}{\partial \theta}
\mp k(\cot\theta) f(\theta)\right).\nonumber
\end{eqnarray}
Entonces, usando que 
$$(\sin\theta)^{1\pm k}= (\sin\theta)^{\pm(k\pm 1)}$$
y el cambio de variable  $u=\cos\theta,$
se llega a
\begin{eqnarray}
-\left(\sin\theta\right) ^{\pm(k\pm 1)}\frac{\partial }{\partial u}\left( \left(\sin\theta\right)^{\mp k}f(\theta)\right)=
\left(\frac{\partial f(\theta)}{\partial \theta}
\mp k(\cot\theta) f(\theta)\right). \nonumber
\end{eqnarray}
Ocupando este resultado se obtiene 
\begin{eqnarray}
L_{\pm}\left(f(\theta)e^{ik\varphi}\right)&=&e^{\pm i\varphi}\left(\pm \frac{ \partial }{\partial \theta}
+i (\cot\theta) \frac{ \partial }{\partial \varphi}\right)\left(f(\theta)e^{ik\varphi}\right)\nonumber\\
&=&e^{i(k\pm 1)\varphi}
\left( \pm \frac{\partial f(\theta) }{\partial \theta}- k (\cot\theta) f(\theta)\right)\nonumber\\
&=&\pm e^{i(k\pm 1)\varphi}
\left( \frac{\partial f(\theta) }{\partial \theta}\mp k (\cot\theta) f(\theta)\right)\nonumber\\
&=&\mp  e^{i(k\pm 1)\varphi}\left(\sin\theta \right)^{\pm(k\pm 1)}\frac{\partial }{\partial u}
\left(\left(\sin\theta\right)^{\mp k}f(\theta)\right).
\nonumber
\end{eqnarray}
Adicionalmente, utilizando esta igualdad y definiendo 
\begin{eqnarray}
k^{\prime}=k\pm 1,\qquad g(\theta)=\left(\sin\theta\right)^{\pm(k\pm 1)}\frac{\partial }{\partial u}\left(
\left(\sin\theta\right)^{\mp k}f(\theta)\right),\nonumber
\end{eqnarray}
se encuentra
\begin{eqnarray}
& L^{2}_{\pm}&\left(f(\theta)e^{ik\varphi}\right)=(\mp) L_{\pm}\left(e^{ik^{\prime}\varphi}g(\theta)\right)\nonumber\\
& &=(\mp)^{2}  e^{i(k^{\prime} \pm 1)\varphi}\left(\sin\theta\right)^{\pm(k^{\prime}\pm 1)}
\frac{\partial }{\partial u}\left(\left(\sin\theta\right)^{\mp k^{\prime}} g(\theta)\right)\nonumber\\
& &=(\mp)^{2}  e^{i(k \pm 2)\varphi}\left(\sin\theta\right)^{\pm(k\pm 2)}\nonumber\\
& &\frac{\partial }{\partial u}\left(\left(\sin\theta\right)^{\mp (k\pm 1)} \left(\sin\theta\right)^{\pm(k\pm 1)}\frac{\partial }{\partial u}\left(\left(\sin\theta\right)^{\mp k}f(\theta)\right) \right)\nonumber\\
& &=(\mp)^{2}  e^{i(k \pm 2)\varphi}\left(\sin\theta\right)^{\pm(k\pm 2)}
\frac{\partial^{2} }{\partial u^{2}}\left(\left(\sin\theta\right)^{\mp k} f(\theta)\right),\nonumber
\end{eqnarray}
en general se tiene
\begin{eqnarray}
(L_{\pm})^{n}\left(f(\theta)e^{ik\varphi}\right)=(\mp)^{n}  e^{i(k \pm n)\varphi}\left(\sin\theta\right)^{\pm(k\pm n)}
 \frac{\partial^{n} }{\partial u^{n}}\left(\left(\sin\theta\right)^{\mp k} f(\theta)\right). \qquad 
\label{eq:l-recurr}
\end{eqnarray}
donde 
\begin{eqnarray}
u=\cos\theta .\nonumber
\end{eqnarray}

\section{El arm\'onico esf\'erico $Y_{ll}(\theta,\varphi)$}

En esta secci\'on obtendremos el arm\'onico esf\'erico $Y_{ll}(\theta,\varphi)$
con el cual podemos construir el resto de los arm\'onico esf\'ericos. De  Eq. (\ref{eq:armo-esfe-sol}) sabemos que 
\begin{eqnarray}
Y_{ll}(\theta,\varphi)=\frac{\alpha_{ll}e^{il\varphi} \Theta^{l}_{l}(\theta)}{\sqrt{2\pi}}
\end{eqnarray}
y  de Eq. (\ref{eq:escalera-norm}) tenemos que se debe cumplir $L_{+}Y_{ll}(\theta,\varphi)=0,$ 
es decir
\begin{eqnarray}
L_{+}Y_{ll}(\theta,\varphi)
=e^{i\varphi}\left( 
\frac{\partial }{\partial \theta}+i \cot\theta 
 \frac{\partial }{\partial \varphi}\right) \frac{\alpha_{ll}e^{il\varphi} \Theta^{l}_{l}(\theta)}{\sqrt{2\pi}}=0.
\end{eqnarray}
Por lo que, $\Theta^{l}_{l}(\theta)$ debe satisfacer
\begin{eqnarray}
\left(\frac{\partial }{\partial \theta}-l \cot\theta\right)  \Theta^{l}_{l}(\theta)=0,
\end{eqnarray}
as\'{\i}
\begin{eqnarray}
\Theta^{l}_{l}(\theta)=\alpha \sin^{l}\theta,\qquad \alpha={\rm constante}.
\end{eqnarray}
Esto quiere decir que 
\begin{eqnarray}
Y_{ll}(\theta,\varphi)=\frac{\alpha_{ll}e^{il\varphi} \sin^{l}\theta}{\sqrt{2\pi}}.
\end{eqnarray}
La constante  $\alpha_{ll}$ se determina  pidiendo la condici\'on $<Y_{ll}|Y_{ll}>=1,$ que es 
\begin{eqnarray}
\int d\Omega Y^{*}_{ll}(\theta,\varphi)Y_{ll}(\theta,\varphi)&=&\int_{0}^{2\pi}d\varphi\int_{0}^{\pi} d\theta \sin\theta
\left(\frac{\alpha_{ll}e^{il\varphi} \sin^{l}\theta}{\sqrt{2\pi}}\right)^{*}
\frac{\alpha_{ll}e^{il\varphi} \sin^{l}\theta}{\sqrt{2\pi}}\nonumber\\
&=&\frac{|\alpha_{ll}|^{2}}{2\pi}\int_{0}^{2\pi}d\varphi\int_{0}^{\pi} d\theta \sin^{2l+1}\theta
e^{i(l-l)\varphi}\nonumber\\
&=&|\alpha_{ll}|^{2} \int_{0}^{\pi} d\theta \sin^{2l+1}\theta=1,\nonumber
\end{eqnarray}
entonces
\begin{eqnarray}
|\alpha_{ll}|^{2}= \frac{1}{\int_{0}^{\pi} d\theta \sin^{2l+1}\theta}.
\label{eq:norma-const-armo}
\end{eqnarray}
Para obtener esta constante, notemos que 
\begin{eqnarray}
\frac{d}{d\theta}\left(\sin^{n-1}\theta\cos\theta\right)&=&(n-1)\sen^{n-2}\theta \cos^{2}\theta-\sin^{n-1}\theta\sin\theta\nonumber\\
&=&(n-1)\sen^{n-2}\theta\left(1- \sin^{2}\theta\right)-\sin^{n}\theta, \nonumber
\end{eqnarray}
de donde 
\begin{eqnarray}
\sin^{n}\theta=\frac{n-1}{n}\sen^{n-2}\theta- \frac{1}{n}\frac{d}{d\theta}\left(\sin^{n-1}\theta\cos\theta\right).
\end{eqnarray}
As\'i,
\begin{eqnarray}
\int_{0}^{\pi}d\theta \sin^{n}\theta=\frac{n-1}{n}\int_{0}^{\pi} d\theta \sen^{n-2}\theta.
\end{eqnarray}
En particular 
\begin{eqnarray}
\int_{0}^{\pi}d\theta \sin^{2l+1}\theta&=&\frac{2l}{2l+1}\int_{0}^{\pi} d\theta \sen^{2l-1}\theta=\frac{2l}{2l+1}
\int_{0}^{\pi}d\theta \sen^{2(l-1)+1}\theta\nonumber\\
&=&\frac{(2l)(2(l-1))}{(2l+1)(2(l-1)+1)}\int_{0}^{\pi}d\theta  \sen^{2(l-2)+1}\theta\nonumber\\
&=& \frac{(2l)(2(l-1))(2(l-2))}{(2l+1)(2(l-1)+1)(2(l-2)+1)}\int_{0}^{\pi} d\theta \sen^{2(l-3)+1}\theta\nonumber\\
&\vdots& \nonumber\\
&=& \frac{(2l)(2(l-1))(2(l-2))\cdots (2(l-i))}{(2l+1)(2(l-1)+1)(2(l-2)+1)\cdots (2(l-i)+1)}\nonumber\\
& &\left(\int_{0}^{\pi} d\theta \sen^{2(l-(i+1))+1}\theta\right).\nonumber
\end{eqnarray}
Este proceso se puede hacer hasta que $2(l-(i+1))+1=1,$ que implica $l=i+1,$ es decir $i=l-1.$ En este caso
\begin{eqnarray}
\int_{0}^{\pi}d\theta \sin^{2l+1}\theta&=&\frac{(2l)(2(l-1))(2(l-2))\cdots 2}{(2l+1)(2l-1)(2l-3)\cdots 3}\int_{0}^{\pi}d\theta \sen\theta\nonumber\\
&=&\frac{2^{l}(l)(l-1)(l-2)\cdots 1}{(2l+1)(2l-1)(2l-3)\cdots 3}2\nonumber\\
 &=&\frac{2^{l+1}l!}{(2l+1)(2l-1)(2l-3)\cdots 3},\nonumber
\end{eqnarray}
considerando  
\begin{eqnarray}
(2l+1)(2l-1)(2l-3)\cdots 3&=&\frac{(2l+1)(2l)(2l-1)(2(l-1))(2l-3)\cdots 3\cdot 2}{(2l)(2(l-1))\cdots 2}\nonumber\\
&=&\frac{(2l+1)!}{2^{l}l!}
\end{eqnarray}
se llega a 
\begin{eqnarray}
\int_{0}^{\pi} d\theta \sin^{2l+1}\theta=\frac{2^{2l+1}(l!)^{2}}{(2l+1)!}.
\end{eqnarray}
Tomando en cuenta este resultado en Eq. (\ref{eq:norma-const-armo}), tenemos que
\begin{eqnarray}
\alpha_{ll}=\sqrt{\frac{(2l+1)!}{2}}\frac{1}{2^{l}l!},
\end{eqnarray}
entonces
\begin{eqnarray}
Y_{ll}(\theta,\varphi)=\sqrt{\frac{(2l+1)!}{4\pi}} \frac{e^{il\varphi}}{2^{l}l!} \sin^{l}\theta.
\label{eq:arm+alto}
\end{eqnarray}

\section{Forma expl\'icita de los arm\'onicos esf\'ericos}

Una vez obtenido un arm\'onico esf\'erico, ocupando el operado escalera, $L_{\pm},$
podemos obtener los dem\'as. Por ejemplo, supongamos que tenemos el arm\'onico
esf\'erico $Y_{ll}.$ Entonces, usando reiteradamente Eq. (\ref{eq:escalera-norm}) se tiene
\begin{eqnarray}
L_{-}Y_{ll}&=&\sqrt{2l}Y_{ll-1},\nonumber\\
 L_{-}^{2}Y_{ll}&=&\sqrt{2l}L_{-}Y_{ll-1}=
\sqrt{(2l)(2l-1)2}Y_{ll-2}=\sqrt{\frac{(2l)!2!}{(2l-2)!}}Y_{ll-2},\nonumber\\
 L_{-}^{3}Y_{ll}&=&\sqrt{\frac{(2l)!2!}{(2l-2)!} }L_{-}Y_{ll-2}= \sqrt{\frac{(2l)!2!}{(2l-2)!}}\sqrt{(2l-2)3}Y_{ll-3}
 \nonumber\\
& &=\sqrt{\frac{(2l)!3!}{(2l-3)!}}Y_{ll-3},\nonumber\\
& &\vdots \nonumber\\
L_{-}^{n}Y_{ll}&=&\sqrt{\frac{(2l)!n!}{(2l-n)!}}Y_{ll-n}.
\end{eqnarray}
Por lo tanto, si $n=l-m$ se encuentra 
\begin{eqnarray}
Y_{lm}(\theta,\varphi)=\sqrt{\frac{(l+m)!}{(2l)!(l-m)!}}\left(L_{-}\right)^{l-m} Y_{ll}(\theta,\varphi).
\label{eq:lm+l}
\end{eqnarray}
Se puede observar que considerando los resultados previos  (\ref{eq:arm+alto}) y (\ref{eq:l-recurr})
para $k=l$ y $n=l-m$ se tiene
\begin{eqnarray}
Y_{lm}(\theta,\varphi)&=& \sqrt{\frac{(l+m)!}{(2l)!(l-m)!}}\left(L_{-}\right)^{l-m}\left( 
\sqrt{\frac{(2l+1)!}{4\pi}} \frac{e^{il\varphi}}{2^{l}l!} \sin^{l}\theta\right)\nonumber\\
&=&\sqrt{\frac{(2l+1)(l+m)!}{4\pi (l-m)!}} \frac{1}{2^{l}l!}\left(L_{-}\right)^{l-m} 
\left(e^{il\varphi} \sin^{l}\theta\right)\nonumber\\
&=&\sqrt{\frac{(2l+1)(l+m)!}{4\pi (l-m)!}} \frac{1}{2^{l}l!}(+)^{l-m}  e^{i(l-(l-m))\varphi}\nonumber\\
& &\sin^{-[l-(l-m)]}\theta
\frac{\partial^{l-m} }{\partial u^{l-m}}\left(\sin^{l}\theta \sin^{l}\theta\right)\nonumber\\
&=&  \sqrt{\frac{(2l+1)(l+m)!}{4\pi (l-m)!}} \frac{e^{im\varphi}}{2^{l}l!} 
\sin^{-m}\theta \frac{\partial^{l-m} \left(\sin\theta\right)^{2l}}{\partial u^{l-m}}\nonumber.
\end{eqnarray}
Note que ninguna propiedad de estas funciones cambia si las multiplicamos por un factor de norma uno, es decir 
por una potencia de $\pm i$ o de $\pm 1.$ En la lite\-ratura existen diferentes elecciones de este factor, 
por conveniencia tomaremos $(-)^{l}.$ Entonces, los arm\'onicos esf\'ericos son
\begin{eqnarray}
Y_{lm}(\theta,\varphi)=(-)^{l}\sqrt{\frac{(2l+1)(l+m)!}{4\pi (l-m)!}} \frac{e^{im\varphi}}{2^{l}l!} 
\left(\sin\theta\right)^{-m} \frac{\partial^{l-m} \left(\sin^{2}\theta\right)^{l}}{\partial u^{l-m}},
\label{eq:expre-final}\\
 u=\cos\theta,\qquad -l\leq m\leq l. \nonumber
\end{eqnarray}
Esta expresi\'on de los arm\'onicos esf\'ericos es com\'un en mec\'anica cu\'antica, en la pr\'oxima secci\'on veremos
otra que es m\'as usual en electrost\'atica.

\section{Polinomios de Legendre y polinomios asociados de Legendre}

El caso $m=0$ es de particular importancia para diferentes aplicaciones, veamos este caso.
Tomando  $m=0$ en Eq. (\ref{eq:expre-final}), se tiene
\begin{eqnarray}
Y_{l0}(\theta,\varphi)&=&\sqrt{\frac{2l+1}{4\pi }} \frac{(-)^{l}}{2^{l}l!} 
\frac{d^{l} }{d u^{l}}\left(\sin^{2l}\theta\right)= 
\sqrt{\frac{2l+1}{4\pi }} \frac{(-)^{l}}{2^{l}l!} 
\frac{d^{l} }{d u^{l}}\left(1-u^{2}\right)^{l}\nonumber\\
&=&
\sqrt{\frac{2l+1}{4\pi }} \frac{1}{2^{l}l!} 
\frac{d^{l} }{d u^{l}}\left(u^{2}-1\right)^{l}= 
\sqrt{\frac{2l+1}{4\pi }}P_{l}(u).
\end{eqnarray}
Donde 
\begin{eqnarray}
 P_{l}(u)= \frac{1}{2^{l}l!} 
\frac{d^{l} }{d u^{l}}\left(u^{2}-1\right)^{l},
\end{eqnarray}
a esta expresi\'on se le llama f\'ormula de Rodrigues de los polinomios de Legendre de grado $l.$ Por lo que el arm\'onico esf\'erico de orden $m=0$ es 
\begin{eqnarray}
Y_{l0}(\theta,\varphi)=
\sqrt{\frac{2l+1}{4\pi }}P_{l}(\cos\theta).
\end{eqnarray}
Con este arm\'onico esf\'erico se pueden obtener el resto. En efecto, de Eq. (\ref{eq:escalera-norm})
se encuentra 
\begin{eqnarray}
L_{\pm}Y_{l0}&=&\sqrt{l(l+1)}Y_{l\pm 1}=\sqrt{\frac{(l+1)!}{(l-1)!}}Y_{l\pm 1},\nonumber \\
\left(L_{\pm}\right)^{2}Y_{l0}&=&\sqrt{\frac{(l+1)!}{(l-1)!}}L_{\pm} Y_{l\pm 1}= \sqrt{\frac{(l+1)!}{(l-1)!}}\sqrt{(l-1)(l+2)} Y_{l\pm 2}\nonumber\\
&=& \sqrt{\frac{(l+2)!}{(l-2)!}}Y_{l\pm2}, \nonumber \\
& & \vdots \nonumber\\
\left(L_{\pm}\right)^{m}Y_{l0}&=&\sqrt{\frac{(l+m)!}{(l-m)!}}Y_{l(\pm m)},\qquad m\geq 0.
\end{eqnarray}
Entonces, tomando en cuenta Eq. (\ref{eq:l-recurr}) se llega a
\begin{eqnarray}
Y_{l(\pm m)}(\theta,\varphi)&=& \sqrt{\frac{(l-m)!}{(l+m)!}}\left(L_{\pm }\right)^{m}Y_{l0}(\theta,\varphi),\nonumber\\
&=&\sqrt{\frac{(l-m)!}{(l+m)!}}(\mp)^{m}e^{\pm im\varphi} \sin^{m}\theta \frac{d^{m}}{du^{m}}
\left(\sqrt{\frac{2l+1}{4\pi }}P_{l}(\cos\theta)\right)\nonumber\\
&=&\sqrt{\frac{(2l+1)(l-m)!}{4\pi (l+m)!}}(\pm )^{m}e^{\pm im\varphi}\left[(-)^{m} (1-u^{2})^{\frac{m}{2}} \frac{d^{m}}{du^{m}}
P_{l}(u)\right].\nonumber
\end{eqnarray}
Definiremos los polinomios asociados de Legendre de grado positivo como
\begin{eqnarray}
P_{l}^{m}(u)= (-)^{m} (1-u^{2})^{\frac{m}{2}} \frac{d^{m}}{du^{m}}P_{l}(u),
\label{eq:polinomios-asociados-legendre}
\end{eqnarray}
as\'{\i}
\begin{eqnarray}
 Y_{l(\pm m)}(\theta,\varphi)=\sqrt{\frac{(2l+1)(l-m)!}{4\pi (l+m)!}}e^{\pm im\varphi} (\pm)^{m} P_{l}^{m}(\cos \theta), \quad m\geq 0,
\label{eq:armonicos}
\end{eqnarray}
es decir, si $m\geq 0,$
\begin{eqnarray}
Y_{lm}(\theta,\varphi)&=&\sqrt{\frac{(2l+1)(l-m)!}{4\pi (l+m)!}}e^{im\varphi}  P_{l}^{m}(\cos \theta),
\label{eq:armonicos-1}\\
Y_{l(-m)}(\theta,\varphi)&=&\sqrt{\frac{(2l+1)(l-m)!}{4\pi (l+m)!}}e^{- im\varphi} (-)^{m} P_{l}^{m}(\cos \theta).
\label{eq:armonicos-2}
\end{eqnarray}
Note que 
\begin{eqnarray}
Y_{lm}^{*}(\theta,\varphi)=(-)^{m} Y_{l(-m)}(\theta,\varphi).
\end{eqnarray}
Ahora, definiendo los polinomios asociados de Legendre de grado negativo como
\begin{eqnarray}
P_{l}^{-m}(u)= (-)^{m}\frac{(l-m)!}{(l+m)!} P^{m}_{l}(u),
\end{eqnarray}
se tiene
\begin{eqnarray}
Y_{l(-m)}(\theta,\varphi)=\sqrt{\frac{(2l+1)(l+m)!}{4\pi (l-m)!}}e^{-im\varphi} P^{-m}_{l}(\cos\theta).
\end{eqnarray}
Se puede observar que si definimos $k=-m$ esta expresi\'on se escribe como
\begin{eqnarray}
Y_{lk}(\theta,\varphi)=\sqrt{\frac{(2l+1)(l-k)!}{4\pi (l+k)!}}e^{ik\varphi} P^{k}_{l}(\cos\theta),
\end{eqnarray}
que tiene la forma de Eq. (\ref{eq:armonicos-1}). Por lo tanto, cualquier arm\'onico esf\'erico se escribe como
\begin{eqnarray}
Y_{lm}(\theta,\varphi)=\sqrt{\frac{(2l+1)(l-m)!}{4\pi (l+m)!}}e^{im\varphi}P_{l}^{m}(\cos \theta), 
\quad -l\leq m\leq l.
\label{eq:armonicos-final}
\end{eqnarray}
Adem\'as, de las relaciones de ortonormalidad Eq. (\ref{eq:p80}) se tiene 
\begin{eqnarray}
& &\delta_{l^{\prime}l}= < Y_{l^{\prime}m}|Y_{lm} >=\int d\Omega Y_{l^{\prime}m}^{*}(\theta,\varphi)Y_{lm}(\theta,\varphi)\nonumber\\
& &=
\int_{0}^{2\pi}d\varphi\int_{0}^{2\pi}d\theta \sin\theta 
\left(\sqrt{\frac{(2l^{\prime}+1)(l^{\prime}-m)!}{4\pi (l^{\prime}+m)!}}e^{im\varphi}P_{l^{\prime}}^{m}(\cos\theta)\right)^{*}\nonumber\\
& &\left(\sqrt{\frac{(2l+1)(l-m)!}{4\pi (l+m)!}}e^{im\varphi}P_{l}^{m}(\cos\theta)\right)\nonumber\\
& =&2\pi\left(\frac{(2l+1)(l-m)!}{4\pi (l+m)!}\right)\int_{0}^{\pi} d\theta \sin\theta P_{l^{\prime}}^{m}(\cos\theta) 
P_{l}^{m}(\cos \theta),\nonumber
\end{eqnarray}
es decir 
\begin{eqnarray}
\int_{0}^{\pi} d\theta \sin\theta P_{l^{\prime}}^{m}(\cos\theta) 
P_{l}^{m}(\cos \theta)= \left(\frac{2 (l+m)!}{(2l+1)(l-m)!}\right)\delta_{l^{\prime}l}.
\end{eqnarray}
Tomando el cambio de variable $u=\cos\theta$ se encuentra
\begin{eqnarray}
\int_{-1}^{1} du P_{l^{\prime}}^{m}(u) 
P_{l}^{m}(u)=\frac{2}{2l+1}\frac{(l+m)!}{(l-m)!} \delta_{ll^{\prime}}.
\end{eqnarray}
Estas son las relaciones de ortonormalidad de los polinomios asociados de Legendre.
En particular, para los polinomios de Legendre, $m=0,$ se llega a
\begin{eqnarray}
\int_{-1}^{1} du P_{l^{\prime}}(u) 
P_{l}(u)=\frac{2}{2l+1} \delta_{ll^{\prime}},
\end{eqnarray}
que son las relaciones de ortonormalidad de los polinomios de Legendre.
\section{Propiedades de los polinomios de Legendre}

Ahora veremos algunas propiedades  de los Polinomios de Legendre. 
Primero recordemos que se cumple la llamada f\'ormula de Rodrigues 
\begin{eqnarray}
 P_{l}(u)= \frac{1}{2^{l}l!} 
\frac{d^{l} }{d u^{l}}\left(u^{2}-1\right)^{l},
\end{eqnarray}
de esta f\'ormula se puede ver que los primeros polinomios de Legendre
son
\begin{eqnarray}
P_{0}(u)&=&1.\\
P_{1}(u)&=&u,\\
P_{2}(u)&=&\frac{1}{2}\left(3u^{2}-1\right),\\
P_{3}(u)&=& \frac{1}{2}\left(5u^{3}-3u\right),\\
P_{4}(u)&=&\frac{1}{8}\left(35u^{4}-30u^{2}+3\right), \\
P_{5}(u)&=& \frac{1}{8}\left(63u^{5}-70u^{3}+15\right). 
\end{eqnarray}
Para obtener la expresi\'on general de los polinomios de Legendre notemos que
si $n$ y $m$ son dos naturales tales que $n\leq m,$ se tiene
\begin{eqnarray}
\frac{d^{n} u^{m}}{du^{n}}&=& \frac{d^{n-1} }{du^{n-1}} \frac{d u^{m}}{du}
=m \frac{d^{n-1} u^{m-1}}{du^{n-1}}=\frac{m!}{(m-1)!} \frac{d^{n-1} u^{m-1}}{du^{n-1}}\nonumber\\
&=&\frac{m!}{(m-1)!}(m-1)\frac{d^{n-2} u^{m-2}}{du^{n-2}}= \frac{m!}{(m-2)!}\frac{d^{n-2} u^{m-2}}{du^{n-2}}=\cdots 
\nonumber\\
&=&
\frac{m!}{(m-n)!}\frac{d^{n-n} u^{m-n}}{du^{n-n}}\nonumber\\
&=& \frac{m!}{(m-n)!} u^{m-n}\nonumber,
\end{eqnarray}
mientras que para el caso  $n> m,$ se encuentra
\begin{eqnarray}
\frac{d^{n} u^{m}}{du^{n}}=0.
\end{eqnarray}
De donde 
\begin{eqnarray}
\frac{d^{n} u^{m}}{du^{n}}&=&\frac{m!}{(m-n)!}u^{m-n}\theta(m-n),\qquad \theta(z)=\left\{
\begin{array}{ll}
1& z\geq 0\\
0& z<0
\end{array} \right.
.\label{eq:recurr}
\end{eqnarray}
En particular, si $l$ y $k$ son naturales se llega a 
\begin{eqnarray}
\frac{d^{l} u^{2(l-k)}}{du^{l}}&=&\frac{(2l-2k)!}{(l-2k)!}u^{l-2k}\theta(l-2k).
\label{eq:recurr}
\end{eqnarray}
As\'{\i}, Eq. (\ref{eq:recurr}) es diferente de cero s\'olo si $l-2k\leq 0.$
Esto quiere decir que el m\'aximo valor que puede tomar $k$ es $l/2.$
Si $l$ es par, es decir $l=2r$ entonces el m\'aximo valor que puede
tomar $k$ es $r.$ De donde, $k$ puede tomar los valores 
$(0,1,2,3,\cdots, r).$ Si $l$ es impar $l=2r+1,$ 
entonces el m\'aximo valor que puede
tomar $k$ es $r+1/2,$ que no es un natural. 
Como $k$ es natural, en este caso $k$ s\'olo puede tomar los valores 
$(0,1,2,3,\cdots, r).$ Definiremos $[l/2]$
como el m\'aximo entero menor o igual a $l/2.$
Entonces, en ambos casos,  $k$ puede tomar los valores 
$(0,1,2,3,\cdots,[l/2]).$ Considerando esta definici\'on, 
Eq. (\ref{eq:recurr}) se puede escribir como
\begin{eqnarray}
\frac{d^{l} u^{2(l-k)}}{du^{l}}&=&\frac{(2l-2k)!}{(l-2k)!}u^{l-2k}
\theta\left(\left[\frac{l}{2}\right]-k\right).
\label{eq:recurr2}
\end{eqnarray}
Tambi\'en recordemos el binomio de Newton
\begin{eqnarray}
(A+B)^{n}=\sum_{k=0}^{n}C_{k}^{n}A^{n-k} B^{k},\qquad  C_{k}^{n}=\frac{n!}{k!(n-k)!},
\label{eq:binomio-newton}
\end{eqnarray}
que implica
\begin{eqnarray}
(u^{2}-1 )^{l}=
 \sum_{k=0}^{l}\frac{(-1)^{k}l!}{k!(l-k)!} u^{2(l-k)}.
\label{eq:bin}
\end{eqnarray}
Por lo que, de  Eq. (\ref{eq:bin}) y Eq. (\ref{eq:recurr2}) se llega a
\begin{eqnarray}
P_{l}(u)&=&\frac{1}{2^{l}l!}
\frac{d^{l}(u^{2}-1)^{l}}{du^{l}}=
\sum_{k=0}^{l}\frac{(-1)^{k}l!}{2^{l}l!k!(l-k)!} 
\frac{d^{l} u^{2l-2k}}{du^{l}}\nonumber\\
&=&\sum_{k=0}^{l}\frac{(-1)^{k} (2l-2k)!}{2^{l}k!(l-k)!(l-2k)!} 
u^{l-2k}\theta\left(\left[\frac{l}{2}\right]-k\right).
\end{eqnarray}
As\'i, la expresi\'on general para los polinomios de Legendre es
\begin{eqnarray}
P_{l}(u)=
\sum_{k=0}^{[l/2]}\frac{(-1)^{k} (2l-2k)!}{2^{l}k!(l-k)!(l-2k)!} 
u^{l-2k}. \label{eq:guera1}
\end{eqnarray}
De esta expresi\'on se puede ver que
\begin{eqnarray}
P_{l}(-u)=(-)^{l}P_{l}(u).
\end{eqnarray}
Por lo tanto, si $l$ es par, $P_{l}(u)$ es par y si $l$ es impar, $P_{l}(u)$
es impar.

\subsection{Funci\'on generadora}

Ahora veremos la funci\'on generadora de los polinomios de Legendre. Probaremos que se cumple
\begin{eqnarray}
\frac{1}{\sqrt{1-2zu+z^{2}}}=\sum_{l\geq 0}z^{l}P_{l}(u), \qquad z<1.
\label{eq:gen-leg}
\end{eqnarray}
Primero note que $u=\cos\theta\leq 1$ y si $0<z<1,$ entonces $z=1-\epsilon,$ con $0<\epsilon<1.$ 
Adem\'as recordemos que si $|\alpha|<1,$ entonces se cumplen las series
\begin{eqnarray}
\frac{1}{1-\alpha}&=&\sum_{n\geq 0}\alpha ^{n},\label{eq:serie1}\\
\frac{1}{\sqrt{1-\alpha}}&=&\sum_{n\geq 0}\frac{(2n)!}{ 2^{2n}(n!)^{2}} \alpha^{n}.
\label{eq:seri2}
\end{eqnarray}
Entonces, 
\begin{eqnarray}
2u<2+ \sum_{n\geq 2} \epsilon^{n}&=& 1-\epsilon+ 1+\epsilon+ \sum_{n\geq 2} \epsilon^{n}=
z+\sum_{n\geq 0} \epsilon^{n}=z+\frac{1}{1-\epsilon}\nonumber\\
&=& z+\frac{1}{z},
\end{eqnarray}
de donde,
\begin{eqnarray}
2uz -z^{2}=z(2u-z)<1.
\end{eqnarray}
Por lo tanto, considerando Eq. (\ref{eq:seri2}) y el binomio de Newton Eq. (\ref{eq:binomio-newton})
se tiene
\begin{eqnarray}
\frac{1}{\sqrt{1-2zu+z^{2}}}&=& \frac{1}{\sqrt{1-z(2u-z)}}=
\sum_{n\geq 0} \frac{(2n)!}{ 2^{2n}(n!)^{2}} z^{n}\left(2u-z\right)^{n}\nonumber\\
&=& \sum_{n\geq 0} \frac{(2n)!}{ 2^{2n}(n!)^{2}} z^{n}\sum_{k=0}^{n}C_{k}^{n}(2u)^{n-k}(-z)^{k}\nonumber\\
& =& \sum_{n\geq 0} \sum_{k=0}^{n} \frac{(2n)!(-)^{k} 2^{n-k} C_{k}^{n}u^{n-k}}{ 2^{2n}(n!)^{2}} z^{n+k} .
\end{eqnarray}
Para simplificar los c\'alculos, definamos $l=n+k,$ entonces $n=l-k.$ Como $n$ es el m\'aximo valor que
puede tener $k,$ se cumple 
$$k\leq n=(l-k),$$ 
esto implica $2k\leq l.$ De donde, el m\'aximo valor que puede tomar
$k$ es el mayor entero menor o igual a $l/2,$ que es $[l/2].$ Con este cambio de variable se encuentra
\begin{eqnarray}
\frac{1}{\sqrt{1-2zu+z^{2}}}&=& 
\sum_{l\geq 0} \sum_{k=0}^{[l/2]} \frac{[2(l-k)]!(-)^{k} 2^{l-2k} C_{k}^{l-k}u^{l-2k}}{ 2^{2(l-k)}[(l-k)!]^{2}} z^{l}
\nonumber\\
&=& 
\sum_{l\geq 0}z^{l} \sum_{k=0}^{[l/2]} \frac{[2(l-k)]!(-)^{k} 2^{l-2k} (l-k)!}{ k!(l-2k)!2^{2(l-k)}[(l-k)!]^{2}} u^{l-2k}
\nonumber\\
&=& 
\sum_{l\geq 0}z^{l}\left( \sum_{k=0}^{[l/2]} 
\frac{[2(l-k)]!(-)^{k}}{ 2^{l}(l-k)!k!(l-2k)!}u^{l-2k}\right).
\end{eqnarray}
Por lo tanto, tomando en cuenta Eq. (\ref{eq:guera1}) se llega a la igualdad Eq. (\ref{eq:gen-leg}).\\

La igualdad  (\ref{eq:gen-leg})  es importante pues permite probar varias
propiedades de los Polinomios de Legendre. Por ejemplo, si
$u=0,$ ocupando la serie Eq. (\ref{eq:seri2}),  se tiene
\begin{eqnarray}
\frac{1}{\sqrt{1+z^{2}}}=\sum_{l\geq 0} 
\frac{(-)^{l}(2l)!}{2^{2l}l!^{2}}z^{2l} =\sum_{l\geq 0}z^{l}P_{l}(0),
\end{eqnarray}
por lo que
\begin{eqnarray}
P_{2l}(0)=\frac{(-)^{l}(2l)!}{2^{2l}l!^{2}}= 
(-1)^{l}\frac{(2l-1)!!}{(2l)!!} \qquad P_{2l+1}(0)=0. 
\end{eqnarray}
Si $u=\pm 1,$ utilizando la serie Eq. (\ref{eq:serie1}), se consigue
\begin{eqnarray}
\frac{1}{\sqrt{1\mp 2z +z^{2}}}= \frac{1}{1\mp z}=
\sum_{l\geq 0} (\pm 1)^{l} z^{l} =\sum_{l\geq 0}z^{l}P_{l}(\pm 1),
\end{eqnarray}
de donde
\begin{eqnarray}
P_{l}(\pm1)=(\pm 1)^{l}. 
\end{eqnarray}

\subsection{Relaciones de recurrencia}

Los polinomios de Legendre  cumplen las siguientes reglas de recurrencia
\begin{eqnarray}
(l+1)P_{l+1}(u)-(2l+1)uP_{l}(u)+lP_{l-1}(u)=0,\label{eq:recurr-1-lengendre} \\
\frac{dP_{l+1}(u)}{du}  -\left(2u\frac{dP_{l}(u)}{du} + P_{l}(u)\right)+ 
\frac{dP_{l-1}(u)}{du}=0 \label{eq:recurr-2-lengendre},\\
\frac{dP_{l+1}(u)}{du}  -u\frac{dP_{l}(u)}{du} -(l+1) P_{l}(u)=0,\label{eq:recurr-3-lengendre} \\
\frac{d P_{l+1}(u)}{du}-\frac{dP_{l-1}(u)}{du}-(2l+1)P_{l}(u)=0,\label{eq:recurr-4-lengendre} \\
(u^{2}-1)\frac{dP_{l}(u)}{du}-luP_{l}(u)+lP_{l-1}(u)=0.\label{eq:recurr-5-lengendre}
\end{eqnarray}

Para probar la identidad (\ref{eq:recurr-1-lengendre})  definamos
\begin{eqnarray}
W(z,u)=\frac{1}{\sqrt{ 1-2uz+z^{2}}} =\sum_{l\geq 0}z^{l}P_{l}(u),
\end{eqnarray}
de donde 
\begin{eqnarray}
 \frac{\partial W(z,u)}{\partial z}=\frac{-(-2u+2z)}{2( 1-2uz+z^{2})^{3/2} } =\frac{(u-z) W(u,z)}{ ( 1-2uz+z^{2})},
 \nonumber
\end{eqnarray}
es decir,
\begin{eqnarray}
( 1-2uz+z^{2})\frac{\partial W(z,u)}{\partial z}=(u-z) W(u,z).
\label{eq:recu-legendre1}
\end{eqnarray}
Adem\'as,
\begin{eqnarray}
 \frac{\partial W(z,u)}{\partial z}=\sum_{l\geq 0} lz^{l-1}P_{l}(u),
 \nonumber
\end{eqnarray}
usando este resultado en Eq. (\ref{eq:recu-legendre1}) se obtiene 
\begin{eqnarray}
& &( 1-2uz+z^{2})\frac{\partial W(z,u)}{\partial z}-(u-z) W(u,z)\nonumber\\
& & =( 1-2uz+z^{2})\sum_{l\geq 0} lz^{l-1}P_{l}(u)-(u-z)\sum_{l\geq 0}z^{l}P_{l}(u)\nonumber\\ 
& &=\sum_{l\geq 0} \left( lP_{l}(u)z^{l-1}-2luP_{l}(u)z^{l}+lP_{l}(u)z^{l+1}-uP_{l}(u)z^{l}+P_{l}(u)z^{l+1}\right)\nonumber\\
& &= \sum_{l\geq 0} \left( lP_{l}(u)z^{l-1}-(2l+1)uP_{l}(u)z^{l}+(l+1)P_{l}(u)z^{l+1}\right)=0.\label{eq:suma-recur-legendre}
\end{eqnarray}
Ahora, note que 
\begin{eqnarray}
\sum_{l\geq 0} lP_{l}(u)z^{l-1}&=& \sum_{l\geq 1} lP_{l}(u)z^{l-1}=\sum_{l\geq 0} (l+1)P_{l+1}(u)z^{l}\nonumber\\
& =&P_{1}(u)+\sum_{l\geq 1} (l+1)P_{l+1}(u)z^{l}
,\nonumber\\
\sum_{l\geq 0} (l+1)P_{l}(u)z^{l+1}&=&\sum_{l\geq 1} lP_{l-1}(u)z^{l}.
\end{eqnarray}
Por lo que, introduciendo estos resultados en (\ref{eq:suma-recur-legendre}), se encuentra
\begin{eqnarray}
P_{1}(u)-uP_{0}(u)+\sum_{l\geq 1}\left[ (l+1)P_{l+1}(u)-(2l+1)uP_{l}(u)+lP_{l-1}(u)\right]z^{l}=0.\nonumber
\end{eqnarray}
Claramente esta igualdad implica Eq. (\ref{eq:recurr-1-lengendre}).\\

Para probar la identidad Eq. (\ref{eq:recurr-2-lengendre})  derivaremos $W(z,u)$ con respecto a $u,$ en ese caso se tiene 
\begin{eqnarray}
\frac{\partial W(z,u)}{\partial u}=\frac{-(-2z)}{2( 1-2uz+z^{2})^{3/2} } =\frac{z W(u,z)}{ ( 1-2uz+z^{2})},
 \nonumber
\end{eqnarray}
es decir,
\begin{eqnarray}
( 1-2uz+z^{2})\frac{\partial W(z,u)}{\partial u}-zW(u,z)=0.
\label{eq:recu-legendre2}
\end{eqnarray}
Adem\'as
\begin{eqnarray}
 \frac{\partial W(z,u)}{\partial u}=\sum_{l\geq 0} z^{l}\frac{dP_{l}(u)}{du},
 \nonumber
\end{eqnarray}
por lo que
\begin{eqnarray}
& &( 1-2uz+z^{2}) \frac{\partial W(z,u)}{\partial u}-zW(u,z)\nonumber\\
& &=
( 1-2uz+z^{2})\sum_{l\geq 0} z^{l}\frac{dP_{l}(u)}{du}-z\sum_{l\geq 0} z^{l} P_{l}(u)\nonumber\\
& &= \sum_{l\geq 0}\left( \frac{dP_{l}(u)}{du} z^{l} -\left(2u\frac{dP_{l}(u)}{du} + P_{l}(u)\right)z^{l+1}+ 
\frac{dP_{l}(u)}{du} z^{l+2}\right)=0.\label{eq:recu-legendre3}
\end{eqnarray}
Tomando en cuenta que $P_{0}(u)=1$ y $P_{1}(u)=u$ se encuentra
\begin{eqnarray}
 \sum_{l\geq 0}\frac{dP_{l}(u)}{du} z^{l}=\sum_{l\geq 1} \frac{dP_{l}(u)}{du} z^{l}=
\sum_{l\geq 0} \frac{dP_{l+1}(u)}{du} z^{l+1}= z+\sum_{l\geq 1} \frac{dP_{l+1}(u)}{du} z^{l+1} ,\nonumber
\end{eqnarray}
adem\'as 
\begin{eqnarray}
 \sum_{l\geq 0}\frac{dP_{l}(u)}{du} z^{l+2}=\sum_{l\geq 1} \frac{dP_{l-1}(u)}{du} z^{l+1}.
\end{eqnarray}
Sustituyendo esto dos resultados en Eq. (\ref{eq:recu-legendre3}) se llega a
\begin{eqnarray}
(1-1)z+\sum_{l\geq 1}\left[ \frac{dP_{l+1}(u)}{du}  -\left(2u\frac{dP_{l}(u)}{du} + P_{l}(u)\right)+ 
\frac{dP_{l-1}(u)}{du}\right] z^{l+1}=0,\nonumber
\end{eqnarray}
que implica Eq. (\ref{eq:recurr-2-lengendre}).\\

Para probar la tercera identidad Eq. (\ref{eq:recurr-3-lengendre}) derivaremos con respecto a $u$ a Eq. (\ref{eq:recurr-1-lengendre}),
que induce  
\begin{eqnarray}
 (l+1)\frac{dP_{l+1}(u)}{du}-(2l+1)P_{l}(u)-(2l+1)u\frac{dP_{l}(u)}{du} +l\frac{dP_{l-1}(u)}{du}=0. \qquad 
 \label{eq:recurr3-lengendre}
\end{eqnarray}
Multiplicando por $l$ a Eq. (\ref{eq:recurr-2-lengendre}) se encuentra
\begin{eqnarray}
l\frac{dP_{l+1}(u)}{du}  -l\left(2u\frac{dP_{l}(u)}{du} + P_{l}(u)\right)+ 
l\frac{dP_{l-1}(u)}{du}=0.\label{eq:recurr4-lengendre}
\end{eqnarray}
Adem\'as, restando Eq. (\ref{eq:recurr3-lengendre}) con Eq. (\ref{eq:recurr4-lengendre}), se consigue 
Eq. (\ref{eq:recurr-3-lengendre}).\\

Ahora, restando a Eq. (\ref{eq:recurr3-lengendre}) el producto de $(l+1)$ con Eq. (\ref{eq:recurr-2-lengendre}), 
se obtiene
\begin{eqnarray}
u\frac{dP_{l}(u)}{du}-lP_{l}(u)-\frac{dP_{l-1}(u)}{du}=0.
\label{eq:recurr0-lengendre}
\end{eqnarray}
Sumando este resultado con Eq. (\ref{eq:recurr-3-lengendre}) se llega a la identidad Eq. (\ref{eq:recurr-4-lengendre}).\\

Si en Eq. (\ref{eq:recurr-3-lengendre}) cambiamos  $l$ por $l-1$ se encuentra
\begin{eqnarray}
\frac{dP_{l}(u)}{du}  -u\frac{dP_{l-1}(u)}{du} -l P_{l-1}(u)=0
.\label{eq:recurr8-lengendre}
\end{eqnarray}
Adem\'as,  de Eq. (\ref{eq:recurr0-lengendre}) tenemos
\begin{eqnarray}
\frac{dP_{l-1}(u)}{du}=u\frac{dP_{l}(u)}{du} -l P_{l}(u).
\end{eqnarray}
Sustituyendo este resultado en Eq. (\ref{eq:recurr8-lengendre}) se obtiene la identidad
 Eq. (\ref{eq:recurr-5-lengendre}).\\

Todas estas identidades son de gran utilidad para resolver diversos problemas de electromagnetismo,
en cap\'itulos posteriores las ocuparemos.

\section{Relaci\'on de completez de los arm\'onicos esf\'ericos}

Hemos demostrado que las funciones propias de los
operadores $L_{z}$ y $L^{2}$ son los arm\'onicos esf\'ericos
$Y_{lm}(\theta,\varphi)$ y que  estas funciones  son  una base ortonormal.
Por lo tanto cualquier otra funci\'on $F(\theta,\varphi)$
se puede escribir como combinaci\'on lineal de esa base, es decir
\begin{eqnarray}
F(\theta, \varphi)=\sum_{l\geq 0} \sum_{m=-l}^{m=l}C_{lm}
Y_{lm}(\theta, \varphi).
\label{eq:serie-arm}
\end{eqnarray}
Ocupando las relaciones de ortonormalidad Eq. (\ref{eq:p80}) se encuentra
\begin{eqnarray}
C_{lm}=\int d\Omega Y^{*}_{lm}(\theta, \varphi) F(\theta, \varphi).
\end{eqnarray}
Note que sustituyendo $C_{lm}$ en Eq. (\ref{eq:serie-arm}) y haciendo
el cambio de variable $u^{\prime}=\cos\theta^{\prime},u=\cos\theta$ 
se llega a 
\begin{eqnarray}
F(\theta, \varphi)&=&\sum_{l\geq 0} \sum_{m=-l}^{m=l}\left(
\int d\Omega^{\prime} Y^{*}_{lm}(\theta^{\prime}, \varphi^{\prime}) 
F(\theta^{\prime}, \varphi^{\prime})\right) Y_{lm}(\theta, \varphi)\nonumber\\
&=&\int d\Omega^{\prime}F(\theta^{\prime}, \varphi^{\prime})
\left(\sum_{l\geq 0} \sum_{m=-l}^{m=l} 
Y^{*}_{lm}(\theta^{\prime}, \varphi^{\prime})Y_{lm}(\theta, \varphi)\right)
\nonumber\\
&=&\int_{0}^{2\pi}d\varphi^{\prime} \int_{0}^{\pi}d\theta^{\prime} \sin\theta^{\prime}F(\theta^{\prime},\varphi^{\prime})
 \left(\sum_{l\geq 0} \sum_{m=-l}^{m=l} 
Y^{*}_{lm}(\theta^{\prime}, \varphi^{\prime})Y_{lm}(\theta, \varphi)\right)
\nonumber\\
&=&\int_{0}^{2\pi} d\varphi\int_{-1}^{1}du^{\prime} 
F(u^{\prime}, \varphi^{\prime}) 
\left(\sum_{l\geq 0} \sum_{m=-l}^{m=l} 
Y^{*}_{lm}(u^{\prime}, \varphi^{\prime})Y_{lm}(u, \varphi)\right).\nonumber
\end{eqnarray}
Por lo tanto, lo que est\'a dentro del par\'entesis debe ser igual a
$\delta(\varphi -\varphi^{\prime})\delta(u -u^{\prime}),$  es decir
\begin{eqnarray}
\sum_{l\geq 0}\sum_{m=-l}^{m=l}Y^{*}_{lm}(\theta^{\prime},\phi^{\prime})
Y_{lm}(\theta,\phi)=\delta(\phi-\phi^{\prime}) 
\delta(\cos\theta-\cos\theta^{\prime} ). 
\label{eq:completez-arm}
\end{eqnarray}
A esta igualdad se le llama relaci\'on de completez.

\section{Teorema de adici\'on de los arm\'onicos esf\'ericos}

Ahora veremos el teorema de adici\'on de los arm\'onicos esf\'ericos, el cual tiene diferentes 
aplicaciones. Este teorema nos dice que si se tiene un vector en los \'angulos $(\theta, \varphi)$
y otro en los \'angulos $(\theta^{\prime}, \varphi^{\prime})$ y adem\'as $\alpha$ es el \'angulo
entre estos dos vectores, entonces se cumple
\begin{eqnarray}
P_{l}(\cos\alpha)=\sum_{m=-l}^{m=l}
\frac{4\pi}{2l+1}Y^{*}_{lm}(\theta^{\prime}, \varphi^{\prime})Y_{lm}(\theta, \varphi).
\label{eq:teorema-de-adicion}
\end{eqnarray}

Para mostrar este teorema primero recordemos que  las funciones rotan con el operador $U(\vec \alpha)=e^{i\vec \alpha\cdot \vec L},$ donde  
$\vec \alpha$ es un vector constante. Por lo que si tenemos 
una funci\'on $f(\vec r),$ la funci\'on rotada es
$$f(\vec r^{\prime})=U(\vec\alpha)f(\vec r).$$ 
Ahora, si tenemos el operador lineal $\hat A$
tal que 
\begin{eqnarray}
\hat Af(\vec r)=g(\vec r),
\end{eqnarray}
con $g(\vec r)$ una funci\'on, como 
\begin{eqnarray}
f(\vec r)&=&U(-\vec \alpha)U(\vec \alpha)f(\vec r)=U(-\vec \alpha)f(\vec r^{\prime}),\nonumber\\
g(\vec r)&=&U(-\vec \alpha)U(\vec \alpha)g(\vec r)=U(-\vec \alpha)g(\vec r^{\prime}),\nonumber
\end{eqnarray}
se encuentra
\begin{eqnarray}
\hat Af(\vec r)= \hat AU(-\vec \alpha)f(\vec r^{\prime})&=&U(-\vec \alpha)g(\vec r^{\prime})
\end{eqnarray}
que implica
\begin{eqnarray}
\hat A^{\prime}f(\vec r^{\prime})=g(\vec r^{\prime}),\qquad \hat A^{\prime}= U(\vec \alpha)\hat AU(-\vec \alpha),
\end{eqnarray}
al operador $\hat A^{\prime}$ le llamaremos operador rotado.\\

Considerando que $L^{2}$ conmuta con $\vec L,$ se tiene 
\begin{eqnarray}
L^{\prime 2}&=&U(\vec \alpha )L^{2}U(-\vec\alpha)= U(\vec \alpha )U(-\vec\alpha)L^{2}=L^{2},\nonumber\\
L^{\prime}_{z}&=&U(\vec \alpha )L_{z}U(-\vec\alpha).\nonumber
\end{eqnarray}

Es claro que en t\'erminos de los \'angulos una rotaci\'on hace la transformaci\'on 
\begin{eqnarray}
(\theta,\varphi)\quad \longrightarrow\quad   (\theta^{\prime},\varphi^{\prime}). 
\end{eqnarray}
En particular para los arm\'onicos esf\'ericos se tiene 
$Y_{lm}(\theta^{\prime},\varphi^{\prime})=U(\vec \alpha)Y_{lm}(\theta,\varphi).$
Adem\'as una rotaci\'on no cambia las reglas de ortonormalidad en el sistema de referencia de las variables
$(\theta^{\prime},\varphi^{\prime}),$ pues ocupando las propiedades del producto escalar
y las reglas de ortonormalidad Eq. (\ref{eq:p80}) se tiene
\begin{eqnarray} 
& &<Y_{lm}(\theta^{\prime},\varphi^{\prime})|Y_{l^{\prime}m^{\prime}}(\theta^{\prime},\varphi^{\prime})>
=
<U(\vec \alpha)Y_{lm}(\theta,\varphi)|U(\vec \alpha)Y_{l^{\prime}m^{\prime}}(\theta,\varphi)>\nonumber\\
&=&
<Y_{lm}(\theta,\varphi)|U^{\dagger}(\vec \alpha)U(\vec \alpha)Y_{l^{\prime}m^{\prime}}(\theta,\varphi)>\nonumber\\
&=& <Y_{lm}(\theta,\varphi)|U(-\vec \alpha)U(\vec \alpha)Y_{l^{\prime}m^{\prime}}(\theta,\varphi)>\nonumber\\
&=&<Y_{lm}(\theta,\varphi)|Y_{l^{\prime}m^{\prime}}(\theta,\varphi)>=\delta_{mm^{\prime}}\delta_{ll^{\prime}}.
\end{eqnarray}
Adicionalmete, se cumple que   
\begin{eqnarray} 
L^{\prime 2}Y_{lm}(\theta^{\prime},\varphi^{\prime})&=&U(\vec \alpha)L^{2}U(-\vec \alpha)U(\vec \alpha)Y_{lm}(\theta,\varphi)
=U(\vec \alpha)L^{2}Y_{lm}(\theta,\varphi)\nonumber\\
&=&U(\vec \alpha)l(l+1)Y_{lm}(\theta,\varphi)
=l(l+1)Y_{lm}(\theta^{\prime},\varphi^{\prime}),\label{eq:valores-primas}\\
L^{\prime }_{z}Y_{lm}(\theta^{\prime},\varphi^{\prime})&=&U(\vec \alpha)L_{z}U(\vec \alpha)U(-\vec\alpha)
Y_{lm}(\theta,\varphi)=U(\vec \alpha)L_{z}Y_{lm}(\theta,\varphi)\nonumber\\
&=&mU(\vec \alpha)Y_{lm}(\theta,\varphi)=m Y_{lm}(\theta^{\prime},\varphi^{\prime}).\nonumber
\end{eqnarray}
Por lo tanto, el conjunto de funciones $Y_{lm}(\theta^{\prime},\varphi^{\prime})$ forman una base de funciones ortonormales 
y cualquier funci\'on $G(\theta^{\prime},\varphi^{\prime})$ se puede escribir en t\'erminos de ellas
\begin{eqnarray}
G(\theta^{\prime}, \varphi^{\prime})=\sum_{l\geq 0} \sum_{m=-l}^{m=l}C_{lm}
Y_{lm}(\theta^{\prime}, \varphi^{\prime}).
\label{eq:serie-arm-2}
\end{eqnarray}
Tomando en cuenta que $\vec \alpha$ es un vector constante, la funci\'on
$$Y_{lm}(\theta^{\prime},\varphi^{\prime})=U(\vec\alpha)Y_{lm}(\theta,\varphi)$$
se puede
ver como una funci\'on que depende de las variables $(\theta,\varphi),$ por lo que se puede
expresar como una serie de arm\'onicos esf\'ericos que dependen de $(\theta,\varphi)$
\begin{eqnarray}
Y_{lm}(\theta^{\prime}, \varphi^{\prime})= \sum_{l^{\prime}\geq 0}\sum_{m^{\prime}=-l^{\prime}}^{m^{\prime}=
l^{\prime}}C_{lml^{\prime}m^{\prime}}
Y_{l^{\prime}m^{\prime}}(\theta, \varphi).
\end{eqnarray}
Pero como se debe cumplir Eq. (\ref{eq:valores-primas}) en esta serie s\'olo contribuyen los t\'erminos que tienen $l^{\prime}=l,$
por lo que
\begin{eqnarray}
Y_{lm}(\theta^{\prime}, \varphi^{\prime})= \sum_{m^{\prime}=-l}^{m^{\prime}=l}C_{lmm^{\prime}}
Y_{lm^{\prime}}(\theta, \varphi),
\label{eq:serie-arm-3}
\end{eqnarray}
con 
\begin{eqnarray}
C_{lmm^{\prime}}=<Y_{lm^{\prime}}(\theta, \varphi)| Y_{lm}(\theta^{\prime}, \varphi^{\prime})>=
\int d\Omega Y_{lm^{\prime}}(\theta, \varphi)U(\vec \alpha)Y_{lm}(\theta, \varphi) .\quad 
\label{eq:def-c}
\end{eqnarray}
De forma analoga, como $\vec \alpha$ es un vector constante, la funci\'on
$$Y_{lm}(\theta,\varphi)=U(-\vec \alpha)Y_{lm}(\theta^{\prime},\varphi^{\prime})$$
se puede ver como una funci\'on que depende de las variables $(\theta^{\prime},\varphi^{\prime})$ por lo que se 
puede expresar como una serie de arm\'onicos esf\'ericos que dependen de $(\theta^{\prime},\varphi^{\prime}).$
Considerando que se debe cumplir 
$$L^{2}Y_{lm}(\theta,\varphi)=l(l+1)Y_{lm}(\theta,\varphi)$$ 
se tiene

\begin{eqnarray}
Y_{lm}(\theta, \varphi)= \sum_{m^{\prime}=-l}^{m^{\prime}=l}D_{lmm^{\prime}}
Y_{lm^{\prime}}(\theta^{\prime}, \varphi^{\prime}),
\label{eq:serie-arm-4}
\end{eqnarray}
con 
\begin{eqnarray}
D_{lmm^{\prime}}=<Y_{lm^{\prime}}(\theta^{\prime}, \varphi^{\prime})| Y_{lm}(\theta, \varphi)>.
\end{eqnarray}
Apelando a las propiedades del producto escalar se encuentra
\begin{eqnarray}
C_{lmm^{\prime}}^{*}&=&\left(<Y_{lm^{\prime}}(\theta, \varphi)| Y_{lm}(\theta^{\prime}, \varphi^{\prime})>\right)^{*}
= <Y_{lm}(\theta^{\prime}, \varphi^{\prime})|Y_{lm^{\prime}}(\theta, \varphi)>\nonumber\\
&=&D_{lm^{\prime}m},
\end{eqnarray}
que se puede escribir como
\begin{eqnarray}
D_{lmm^{\prime}}=C_{lm^{\prime}m}^{*}.
\end{eqnarray}
Por lo tanto,
\begin{eqnarray}
Y_{lm}(\theta, \varphi)= \sum_{m^{\prime}=-l}^{m^{\prime}=l}C_{lm^{\prime}m}^{*}
Y_{lm^{\prime}}(\theta^{\prime}, \varphi^{\prime}).
\label{eq:serie-arm-5}
\end{eqnarray}
Ahora, note que por su definici\'on (\ref{eq:def-c}) las constantes $C_{lm^{\prime}m}$  no pueden depender de $(\theta,\varphi),$ solamente  pueden depender del \'angulo $\vec \alpha.$ Por esta raz\'on, tomaremos el caso m\'as simple. Supongamos que  $\vec r$ est\'a en el eje $z,$ es decir $\theta=0,$   y que  se hace una rotaci\'on con el \'angulo $\vec \alpha=(\theta^{\prime},\varphi^{\prime}),$  claramente el \'angulo final es mismo \'angulo $\vec \alpha.$\\

Adem\'as, considerando Eq. (\ref{eq:armonicos-final}) se encuentra que 
\begin{eqnarray}
Y_{lm}(\theta=0,\varphi)=\sqrt{\frac{2l+1}{4\pi}} \delta_{m0}.
\end{eqnarray}
Por lo que, tomando  $\theta=0$ en Eq. (\ref{eq:serie-arm-3}), se obtiene
\begin{eqnarray}
Y_{lm}(\vec \alpha)&=&Y_{lm}(\theta^{\prime}, \varphi^{\prime})= \sum_{m^{\prime}=-l}^{m^{\prime}=l}C_{lmm^{\prime}}
Y_{lm^{\prime}}(\theta=0, \varphi)=\sum_{m^{\prime}=-l}^{m^{\prime}=l} C_{lmm^{\prime}}\sqrt{\frac{2l+1}{4\pi}} 
\delta_{m^{\prime}0}\nonumber\\
&=&C_{lm0} \sqrt{\frac{2l+1}{4\pi}}\nonumber 
\end{eqnarray}
es decir
\begin{eqnarray}
C_{lm0}=\sqrt{\frac{4\pi}{2l+1}}Y_{lm}(\vec \alpha).\label{eq:coef-arm}
\end{eqnarray}
Esta relaci\'on debe ser cierta para cualquier otros \'angulos $(\theta,\varphi)$ y $(\theta^{\prime}, \varphi^{\prime}).$\\

Ahora, considerando el caso $m=0$ en Eq. (\ref{eq:serie-arm-5}), se llega a
\begin{eqnarray}
Y_{l0}(\theta, \varphi)&=& \sqrt{\frac{2l+1}{4\pi}}P_{l}(\cos\theta) = 
\sum_{m^{\prime}=-l}^{m^{\prime}=l}C_{lm^{\prime}0}^{*}
Y_{lm^{\prime}}(\theta^{\prime}, \varphi^{\prime})\nonumber\\
&=&\sum_{m^{\prime}=-l}^{m^{\prime}=l}
\sqrt{\frac{4\pi}{2l+1}}Y^{*}_{lm^{\prime}}(\vec \alpha)Y_{lm^{\prime}}(\theta^{\prime}, \varphi^{\prime}).
\end{eqnarray}
De donde
\begin{eqnarray}
P_{l}(\cos\theta)=\sum_{m=-l}^{m=l}
\frac{4\pi}{2l+1}Y^{*}_{lm}(\vec \alpha)Y_{lm}(\theta^{\prime}, \varphi^{\prime}).
\label{eq:teorema-deadicion1}
\end{eqnarray}
Note que pasar del  vector de \'angulos $(\theta,\varphi)$ a otro de \'angulos $(\theta^{\prime},\varphi^{\prime})$ se est\'a
haciendo una rotaci\'on con el  \'angulo $\vec \alpha$ que hacen los dos vectores. En el resultado (\ref{eq:teorema-deadicion1}) hemos ocupado tres vectores: el eje $z,$ el vector de \'angulos $P_{1}:(\theta,\varphi)$ y el  vector
de \'angulos $P_{2}:(\theta^{\prime},\varphi^{\prime}).$ Claramente se est\'a suponiendo que esos tres  vectores est\'an en un sistema $S.$\\

Veamos el resultado (\ref{eq:teorema-deadicion1}) en otro sistema de referencia. Consideremos el sistema $\tilde S$ el cual tiene como eje $\tilde z$ el 
vector que est\'a en $P_{1}.$ En este sistema definiremos al vector $\tilde P_{1} $ como el vector $P_{2},$ el cual tiene
\'angulos $\vec \alpha$ con el eje $\tilde z=P_{1}.$ Mientras que definiremos como $\tilde P_{2}$ al eje $z$ que est\'a en los \'angulos 
$(\theta,\varphi)$ del eje $\tilde z.$ Aqu\'i el \'angulo que hacen los  vectores $\tilde P_{1}$ y $\tilde P_{2}$ es $(\theta^{\prime},\varphi^{\prime}).$ En este sistema de referencia el resultado    
Eq. (\ref{eq:teorema-deadicion1})  toma la forma del  teorema de adici\'on de los arm\'onicos
esf\'ericos (\ref{eq:teorema-de-adicion}).\\

El \'angulo $\alpha$ puede ser bastante complicado, por ejemplo supongamos que tenemos los vectores 
\begin{eqnarray}
\vec r_{1}&=&r_{1}(\cos\varphi_{1} \sin\theta_{1}, \sin\varphi_{1} 
\sin\theta_{1}, \cos\theta_{1}), \nonumber\\
\vec r_{2}&=&r_{2}(\cos\varphi_{2} \sin\theta_{2}, \sin\varphi_{2} 
\sin\theta_{2}, \cos\theta_{2}).
\end{eqnarray}
Entonces el \'angulo que forman est\'a dado por  
\begin{eqnarray}
\hat r_{1}\cdot \hat r_{2}= \cos\alpha= 
\sin\theta_{1}\sin\theta_{2}\cos(\varphi_{1}-\varphi_{2})+\cos\theta_{1}
\cos\theta_{2}. \label{eq:ang-arm}
\end{eqnarray}
De donde, el teorema de adici\'on de los arm\'onicos esf\'ericos nos dice que se cumple
\begin{eqnarray}
P_{l}(\cos\alpha)= \frac{4\pi}{2l+1}\sum_{m=-l}^{m=l} 
Y^{*}_{lm} (\theta_{2}, \varphi_{2})Y_{lm}(\theta_{1}, \varphi_{1}).
\label{eq:arm-adi}
\end{eqnarray}

\subsection{Implicaciones del teorema de adici\'on}

El teorema de adici\'on de los arm\'onicos esf\'ericos tiene bastantes aplicaciones,
por ejemplo si $\theta=\theta_{1}=\theta_{2}$ y 
$\varphi=\varphi_{1}=\varphi_{2},$ entonces la ecuaci\'on (\ref{eq:arm-adi}) toma la forma  
\begin{eqnarray}
\sum_{m=-l}^{m=l} |Y_{lm}(\theta, \varphi)|^{2}=
\frac{2l+1}{4\pi} 
\end{eqnarray}
que  se le llama regla de suma  de los arm\'onicos
esf\'ericos.\\

Note que introduciendo Eq. (\ref{eq:arm-adi}) en la relaci\'on de completez
se encuentra
\begin{eqnarray}
\delta(\varphi - \varphi^{\prime}) \delta(\cos\theta- \cos\theta^{\prime})
=\sum_{l\geq 0} \frac{4\pi}{2l+1} P_{l}(\cos\alpha).
\end{eqnarray}
Adem\'as, considerando que la delta de Dirac en coordenadas esf\'ericas
es
\begin{eqnarray}
\delta(\vec r - \vec r^{\prime})= \frac{1}{r^{2}}\delta(r-r^{\prime})
\delta(\varphi - \varphi^{\prime}) \delta(\cos\theta- \cos\theta^{\prime}),
\end{eqnarray}
se llega al resultado
\begin{eqnarray}
\delta(\vec r - \vec r^{\prime})=\frac{4\pi}{2l+1}\sum_{l\geq 0}
\frac{\delta(r-r^{\prime})}{r^{2}} P_{l}(\hat r\cdot\hat r^{\prime} ).
\end{eqnarray}
Para ver otra aplicaci\'on consideremos la funci\'on 
\begin{eqnarray}
\frac{1}{|\vec r_{1}- \vec r_{2}|}=
\frac{1}{\sqrt{r_{1}^{2}-2r_{1}r_{2}
\cos\alpha+r_{2}^{2}}} 
\end{eqnarray}
con $\alpha$ el \'angulo entre $\vec r_{1}$ y $\vec r_{2}^{\prime}$ 
que satisface (\ref{eq:ang-arm}). Ahora, si $r_{1}\not =r_{2}$ definamos 
$r_{<}={\rm min}\{r_{1},r_{2}\}$ y $r_{>}={\rm max}\{r_{1},r_{2}\},$ 
es claro que 
\begin{eqnarray}
\left(\frac{r_{<}}{r_{>}}\right)<1.
\end{eqnarray}
Entonces, usando estas definiciones 
y la funci\'on generatriz Eq. (\ref{eq:gen-leg})
con 
\begin{eqnarray}
z=\frac{r_{<}}{r_{>}},  \qquad u=\cos\alpha, 
\end{eqnarray}
se tiene 
\begin{eqnarray}
\frac{1}{|\vec r_{1}- \vec r_{2}|}=
\frac{1}{r_{>}\sqrt{1-2\left(\frac{r_{<}}{r_{>}}\right)
\cos\alpha+\left(\frac{r_{<}}{r_{>}}\right)^{2}}}
= \frac{1}{r_{>}}\sum_{l \geq 0} 
\left(\frac{r_{<}}{r_{>}}\right)^{l}P_{l}(\cos\alpha). \nonumber
\end{eqnarray}
As\'{\i}, utilizando  el teorema de adici\'on de los arm\'onicos
esf\'ericos Eq. (\ref{eq:arm-adi}), tenemos 
\begin{eqnarray}
\frac{1}{|\vec r_{1}- \vec r_{2}|}=
\sum_{l \geq 0} \sum_{m=-l}^{m=l}
\frac{4\pi}{2l+1} \left(\frac{r_{<}^{l}}{r_{>}^{l+1}}\right)
Y_{lm}^{*}(\theta_{2},\varphi_{2})Y_{lm}(\theta_{1},\varphi_{1}),
\end{eqnarray}
que es la funci\'on de Green de la ecuaci\'on de Poisson en t\'erminos
de los arm\'onicos esf\'ericos.

\section{$L^{2}$ y el Laplaciano}

El espectro de $L^{2}$ es de vital importancia para
la mec\'anica cu\'antica y la electrost\'atica, pues este operador
est\'a relacionado con el Laplaciano. En efecto, ocupando las propiedades del 
tensor de Levi-Civita se tiene 
\begin{eqnarray}
L^{2}&=&L_{i}L_{i}=\left(\epsilon_{ijk}x_{j}p_{k}\right)
\left(\epsilon_{ilm}x_{l}p_{m}\right)=
\epsilon_{ijk}\epsilon_{ilm}x_{j}p_{k}x_{l}p_{m}\nonumber\\
&=&\epsilon_{jki}\epsilon_{ilm}x_{j}\left(x_{l}p_{k}-i\delta_{lk}\right)p_{m}
=\left(\delta_{jl}\delta_{km}-\delta_{jm}\delta_{kl}\right)
\left(x_{j}x_{l}p_{k}p_{m}-ix_{j}\delta_{lk}p_{m}\right)\nonumber\\
&=&\delta_{jl}\delta_{km}x_{j}x_{l}p_{k}p_{m}-
\delta_{jm}\delta_{kl}x_{j}x_{l}p_{k}p_{m}
-i\delta_{jl}\delta_{km}x_{j}\delta_{lk}p_{m}
+\delta_{jm}\delta_{kl}x_{j}\delta_{lk}p_{m}\nonumber\\
&=&x_{j}x_{j}p_{k}p_{k}-x_{m}x_{l}p_{l}p_{m}-ix_{k}p_{k}+i\delta_{kk}x_{l}p_{l}
\nonumber\\
&=&\left(\vec r\right)^{2}\left(\vec p\right)^{2}-x_{m}\vec r\cdot\vec p p_{m} 
-i\vec r\cdot\vec p+3i\vec r\cdot\vec p.
\end{eqnarray}
Adem\'as, como 
\begin{eqnarray}
x_{m}\vec r\cdot\vec p p_{m}&=&x_{m}x_{l}p_{l}p_{m}=x_{m}x_{l}p_{m}p_{l}
=x_{m}\left(p_{m}x_{l}+i\delta_{ml}\right)p_{l}\quad \nonumber\\
&=&x_{m}p_{m}x_{l}p_{l}+ix_{m}\delta_{ml}p_{l}=
\left(\vec r\cdot \vec p\right)\left(\vec r\cdot \vec p\right)+
i\left(\vec r\cdot \vec p\right),\nonumber
\end{eqnarray}
se encuentra
\begin{eqnarray}
L^{2}=\left(\vec r\right)^{2}\left(\vec p\right)^{2}-
\left(\vec r\cdot\vec p\right)^{2}+i\vec r\cdot\vec p,
\end{eqnarray}
de donde 
\begin{eqnarray}
\left(\vec p\right)^{2}&=&\frac{1}{r^{2}}\left( \left(\vec r\cdot\vec p\right)^{2}-i\vec r\cdot\vec p+L^{2}\right).
\end{eqnarray}
Considerando que $\vec p=-i\vec \nabla,$ se tiene
\begin{eqnarray}
\left(\vec p\right)^{2}&=&-\nabla^{2}=\frac{1}{r^{2}}
\left( \left(\vec r\cdot\vec p\right)^{2}-i\vec r\cdot\vec p+L^{2}\right)
\nonumber\\
&=&\frac{1}{r^{2}}\left(\left(-i \vec r\cdot \vec \nabla\right)^{2}-i\left(-i \vec r\cdot \vec \nabla\right)  +L^{2}\right),
\end{eqnarray}
es decir
\begin{eqnarray}
\nabla^{2}=\frac{1}{r^{2}}
\left(\left(\vec r\cdot \vec \nabla\right)^{2}+\left(\vec r\cdot \vec \nabla\right)-L^{2}\right).
\end{eqnarray}
En particular, tomando $\vec \nabla$ en coordenadas esf\'ericas, se encuentra
\begin{eqnarray}
\frac{1}{r^{2}}
\left(\left(\vec r\cdot \vec \nabla\right)^{2}+
\left(\vec r\cdot \vec \nabla\right)\right)
=\frac{1}{r^{2}} \left(
\left(r\frac{\partial}{\partial r}\right)^{2}+r\frac{\partial}{\partial r}
\right)=\frac{1}{r^{2}}\frac{\partial}{\partial r}
\left(r^{2}\frac{\partial}{\partial r}\right).\nonumber
\end{eqnarray}
Por lo tanto, 
\begin{eqnarray}
\nabla^{2}=\frac{1}{r^{2}}\frac{\partial}{\partial r}
\left(r^{2}\frac{\partial}{\partial r}\right)-\frac{1}{r^{2}}L^{2}.\label{eq:laplace-armonicos}
\end{eqnarray}
Posteriormente ocuparemos este resultado para atacar problemas de electrost\'atica y mec\'anica cu\'antica.

\section{Paridad}

La transformaci\'on de paridad est\'a definida por 
\begin{eqnarray}
(x,y,z)\to (-x,-y-z). 
\end{eqnarray}
Ahora, en coordenadas esf\'ericas se tiene
\begin{eqnarray}
-x&=&-r\sin\theta\cos\varphi=r\sin\left(\pi-\theta\right)\cos\left(\pi+\varphi\right),\nonumber\\
-y&=& -r\sin\theta\sin\varphi=r\sin\left(\pi-\theta\right)\sin\left(\pi+\varphi\right),\nonumber\\
-z&=&-r\cos\theta=\cos(\pi-\theta)\nonumber.
\end{eqnarray}
Por lo tanto, la transformaci\'on de paridad en coordenas esf\'ericas  toma la forma
\begin{eqnarray}
(r,\theta,\varphi)\quad \to \quad (r, \pi-\theta, \pi+\varphi). 
\end{eqnarray}
Ahora, veamos como transforman los polinomios asociados de Legendre bajo paridad. 
Primero notemos que bajo paridad se tiene
\begin{eqnarray}
u=\cos\theta \quad \to \quad \cos(\pi-\theta)=-u.
\end{eqnarray} 
Tambi\'en se tiene
\begin{eqnarray}
P_{l}(\cos(\pi-\theta))&=&P_{l}(-\cos\theta)=(-)^{l}P_{l}(\cos\theta)\nonumber\\
P_{l}^{m}(\cos(\pi-\theta))
& =& (-)^{m}\left(1-(-u)^{2}\right)^{\frac{m}{2}}\frac{d^{m}}{d(-u)^{m}}P_{l}(-u)\nonumber\\
& =& (-)^{l+m} (-)^{m} \left(1-u^{2}\right)^{\frac{m}{2}}\frac{d^{m}}{du^{m}}P_{l}(u) \nonumber\\
& =&(-)^{l+m} P_{l}^{m}(\cos\theta).
\end{eqnarray}
Otra identidad de utilidad es 
\begin{eqnarray}
e^{i(\pi+\varphi)m}=e^{i\pi m} e^{i\varphi m}= \left(\cos\pi+i\sin\pi\right)^{m}e^{i\varphi m}= (-)^{m}e^{i\varphi}.
\end{eqnarray}
Por lo tanto, ocupando la definici\'on de los arm\'onicos esf\'ericos (\ref{eq:armonicos-final}) se encuentra
\begin{eqnarray}
Y_{lm}(r, \pi-\theta, \pi+\varphi)= (-)^{l} Y_{lm}(r, \theta, \varphi). 
\end{eqnarray}

\chapter{Ecuaci\'on de Laplace en Coordenadas esf\'ericas}

En este cap\'itulo estudiaremos la ecuaci\'on de Laplace y resolveremos varios problemas de electrost\'atica y
magnetost\'atica.

\section{Soluci\'on general}

Como vimos en el cap\'itulo anterior la ecuaci\'on de Laplace en coordenadas
esf\'ericas Eq. (\ref{eq:laplace-armonicos}) tiene la forma
\begin{eqnarray}
\nabla^{2}\phi=\frac{1}{r^{2}}\frac{\partial}{\partial r}\left(r^{2} \frac{\partial \phi}{\partial r}\right) -\frac{L^{2}\phi}{r^{2}}=0.
\end{eqnarray}
Propondremos como soluci\'on a $\phi( r,\theta,\varphi)=R(r)Y_{lm}(\theta,\varphi),$
de donde
\begin{eqnarray}
\nabla^{2}\phi(r,\theta,\varphi)&=&\frac{1}{r^{2}}\frac{\partial}{\partial r}\left(r^{2} Y_{lm}(\theta,\varphi)\frac{\partial R(r)}{\partial r}\right) -\frac{R(r)L^{2}Y_{lm}(\theta,\varphi) }{r^{2}}\nonumber\\
&=&Y_{lm}(\theta,\varphi) \frac{1}{r^{2}}\frac{\partial}{\partial r}\left(r^{2}\frac{\partial R(r)}{\partial r}\right) -\frac{R(r)l(l+1)Y_{lm}(\theta,\varphi) }{r^{2}}\nonumber\\
&=&\frac{Y_{lm}(\theta,\varphi)}{r^{2}} \left(\frac{\partial}{\partial r}\left(r^{2}\frac{\partial R(r)}{\partial r}\right) -R(r)l(l+1)\right)=0,
\end{eqnarray}
por lo que $R(r)$ debe satisfacer
\begin{eqnarray}
\frac{d}{d r}\left(r^{2}\frac{d R(r)}{d r}\right) -R(r)l(l+1)=0.
\label{eq:radial-laplace}
\end{eqnarray}
Para resolver esta ecuaci\'on haremos la propuesta  $R(r)=a_{\alpha} r^{\alpha},$ con $\alpha$ una constante, entonces
\begin{eqnarray}
\frac{d}{d r}\left(r^{2}\frac{d r^{\alpha}}{d r}\right)=\alpha \frac{d}{d r}\left(r^{2}r^{\alpha-1}\right)= 
\alpha \frac{d}{d r}\left(r^{\alpha+1}\right)= \alpha (\alpha+1) r^{\alpha},
\end{eqnarray}
sustituyendo este resultado en Eq. (\ref{eq:radial-laplace}) se encuentra 
\begin{eqnarray}
r^{\alpha}(\alpha (\alpha+1)-l(l+1))=0, 
\end{eqnarray}
es decir
\begin{eqnarray}
\alpha (\alpha+1)-l(l+1)&=&\alpha^{2}-l^{2}+\alpha-l=(\alpha+l)(\alpha-l)+(\alpha-l)\nonumber\\
&=&(\alpha+l+1)(\alpha-l)=0, 
\end{eqnarray}
entonces, las soluciones para $R(r)$ son $r^{l}$ y $r^{-(l+1)}.$ En  general las soluciones radiales son  
\begin{eqnarray}
R(r)=A_{lm}r^{l}+\frac{B_{lm}}{r^{l+1}},\qquad  A_{lm},B_{lm}={\rm constante}.
\end{eqnarray}
As\'i, las soluciones completas son de la forma
\begin{eqnarray}
\phi_{lm}(r,\theta,\varphi)=\left(A_{lm}r^{l}+\frac{B_{lm}}{r^{l+1}}\right) Y_{lm}(\theta,\varphi) 
\end{eqnarray}
y la soluci\'on general a la ecuaci\'on de Laplace en coordenas esf\'ericas es 
\begin{eqnarray}
\phi(r,\theta,\varphi)=\sum_{l\geq 0}\sum_{m=-l}^{l}\left(A_{lm}r^{l}+\frac{B_{lm}}{r^{l+1}}\right) Y_{lm}(\theta,\varphi). 
\label{eq:solucion-poisson-esfericas}
\end{eqnarray}
Este resultado tiene varias aplicaciones. Por ejemplo, las leyes de la electrost\'atica nos dicen que el campo el\'ectrico, $\vec E,$ satisface las leyes
\begin{eqnarray}
\vec \nabla \cdot \vec E(\vec r)&=&4\pi \rho(\vec r), \nonumber\\ 
\vec \nabla \times \vec E(\vec r)&=&0.
\end{eqnarray}
La primera ley es la llamada ley de Gauss y establece la relaci\'on entre el campo
el\'ectrico y la densidad de carga $\rho.$ La segunda ley establece que existe una funci\'on $\phi$ tal que  
$\vec E=-\vec \nabla \phi,$ por lo que la ley de Gauss toma la forma 
\begin{eqnarray}
\nabla^{2} \phi(\vec r)=-4\pi \rho(\vec r),
\label{eq:katy-poisson}
\end{eqnarray}
que  es la llamada ecuaci\'on de Poisson. Note que $\rho(\vec r)$ solo es diferente de cero donde hay carga, fuera 
de la regi\'on donde hay carga se tiene $\rho(\vec r)=0.$ Por lo tanto la ecuaci\'on de Poisson Eq. (\ref{eq:katy-poisson})
se convierte en la ecuaci\'on de Laplace
\begin{eqnarray}
 \nabla^{2} \phi(\vec r)=0,
\end{eqnarray}
cuya soluci\'on en coordenadas esf\'ericas es Eq. (\ref{eq:solucion-poisson-esfericas}).

\subsection{Problema de la esfera}

Supongamos que tenemos un sistema que consta de una esfera de radio $R$ que est\'a al  potencial $V(\theta,\varphi)$ en su fronterea y que el potencial
es finito en cualquier punto del espacio. El problema consiste en calcular el potencial en todo el espacio, tanto dentro como fuera de la esfera.\\


En el interior de la esfera el potencial Eq. (\ref{eq:solucion-poisson-esfericas}) debe ser finito, esto implica que si $r<R$ los coeficientes $B_{lm}$ deber ser nulos, de lo contrario el potencial diverge en el origen. Si $r>R$ los coeficientes $A_{lm}$ debe ser nulos, de lo contrario el potencial diverge en infinito. Dividiremos el potencial en dos partes, en el interior, $r<R,$
\begin{eqnarray}
\phi_{int}(r,\theta,\varphi)&=&\sum_{l\geq 0}\sum_{m=-l}^{l}A_{lm}\left(\frac{r}{R}\right)^{l} Y_{lm}(\theta,\varphi),
\nonumber
\end{eqnarray}
y en el exterior, $r>R,$
\begin{eqnarray} 
\phi_{ext}(r,\theta,\varphi)&=&\sum_{l\geq 0}\sum_{m=-l}^{l}B_{lm}\left(\frac{R}{r}\right)^{l+1} Y_{lm}(\theta,\varphi).\nonumber
\end{eqnarray}
En la frontera se debe cumplir que $\phi_{int}(R,\theta,\varphi)=\phi_{ext}(R,\theta,\varphi)=V(\theta,\varphi),$ de donde
\begin{eqnarray}
V(\theta,\varphi)=\sum_{l\geq 0}\sum_{m=-l}^{l}A_{lm} Y_{lm}(\theta,\varphi)=\sum_{l\geq 0}\sum_{m=-l}^{l}B_{lm}Y_{lm}(\theta,\varphi),
\end{eqnarray}
es decir $A_{lm}=B_{lm}$ y 
\begin{eqnarray}
V(\theta,\varphi)=\sum_{l\geq 0}\sum_{m=-l}^{l}A_{lm} Y_{lm}(\theta,\varphi),
\end{eqnarray}
por lo que
\begin{eqnarray}
A_{lm}=<Y_{lm}(\theta,\varphi)|V(\theta,\varphi)>=\int d\Omega Y^{*}_{lm}(\theta,\varphi)V(\theta,\varphi).
\label{eq:coeficiente-esfera-laplace}
\end{eqnarray}
As\'i, el potencial del problema es
\begin{eqnarray}
\phi_{int}(r,\theta,\varphi) &=& \sum_{l\geq 0} \sum_{m=-l}^{l} \left( \frac{r}{ R}\right)^{l} A_{lm} Y_{lm}(\theta ,\varphi) \label{eq:poisson-esfera-3-0} \\
\phi_{ext}(r,\theta,\varphi)&=&\sum_{l\geq 0}\sum_{m=-l}^{l}A_{lm}\left(\frac{R}{r}\right)^{l+1} Y_{lm}(\theta,\varphi).
\label{eq:poisson-esfera-3-1}
\end{eqnarray}
con $A_{lm}$ dado por (\ref{eq:coeficiente-esfera-laplace}).

\subsection{F\'ormula de Poisson}

Por otros m\'etodos se puede mostrar que la soluci\'on de la ecuaci\'on de Laplace en coordenadas esf\'ericas con la condici\'on de borde
\begin{eqnarray}
\phi(r=R,\theta,\varphi)&=&V(\theta,\varphi),
\label{eq:poisson-esfera-00-1} 
\end{eqnarray}
es 
\begin{eqnarray}
\phi(r,\theta,\varphi)&=&\mp \frac{R(R^{2}-r^{2})}{4\pi }  \int d \Omega^{\prime} \frac{V(\theta^{\prime},\varphi^{\prime} )}{\left(r^{2}+R^{2}-2Rr\cos\alpha \right)^{\frac{3}{2}}}
\label{eq:poisson-esfera-00} 
\end{eqnarray}
esta es la llamada f\'ormula de Poisson en tres dimensiones. Donde $\alpha$ es el \'angulo entre los vectores $\vec r$ y $\vec r^{\prime}=R\hat e_{r^{\prime}},$ adem\'as
\begin{eqnarray}
 \int d \Omega^{\prime}&=&\int_{0}^{2\pi} d\phi^{\prime}\int_{0}^{\pi}d\theta^{\prime}\sin\theta^{\prime},\nonumber\\
 \cos\alpha&=& \sin\theta^{\prime}\sin\theta \cos(\varphi-\varphi^{\prime})+\cos\theta^{\prime}\cos\theta \nonumber,
\end{eqnarray}
el signo superior $(-)$ corresponde al caso $r>R$ y el signo $(+)$ corresponde al caso $r<R.$\\

Mostraremos que esta soluci\'on es consistente con las soluciones Eq. (\ref{eq:poisson-esfera-3-0}) y  Eq. (\ref{eq:poisson-esfera-3-1}). \\

Note que 
\begin{eqnarray}
& & \phi(r,\theta,\varphi)= \mp \frac{R(R^{2}-r^{2})}{4\pi }  \int d \Omega^{\prime} \frac{V(\theta^{\prime},\varphi^{\prime} )}{\left(r^{2}+R^{2}-2Rr\cos\alpha \right)^{\frac{3}{2}}}\nonumber\\
& & =(\mp) \frac{R}{4\pi } \int d \Omega^{\prime} V(\theta^{\prime},\varphi^{\prime} ) \frac{R(R- r\cos\alpha) - r(r-R\cos\alpha)}{\left(r^{2}+R^{2}-2Rr\cos\alpha \right)^{\frac{3}{2}}} \nonumber\\
& & =(\mp) \frac{R}{4\pi }  \int d \Omega^{\prime} V(\theta^{\prime},\varphi^{\prime} ) 
\Bigg( \frac{R(R- r\cos\alpha)} {\left(r^{2}+R^{2} -2 Rr\cos\alpha \right)^{\frac{3}{2}}}  -\frac{r(r-R\cos\alpha)}{\left(r^{2}+R^{2} -2Rr\cos\alpha \right)^{\frac{3}{2}}}\Bigg)\nonumber\\
& & =(\mp) \frac{R}{4\pi }  \int d \Omega^{\prime} V(\theta^{\prime},\varphi^{\prime} ) (-)
\Bigg( R\frac{\partial }{\partial R}\frac{1}{\sqrt{r^{2}+R^{2}-2Rr\cos\alpha}} \nonumber\\ 
 & & -r\frac{\partial }{\partial r}\frac{1}{\sqrt{r^{2}+R^{2}-2Rr\cos\alpha}}\Bigg)\nonumber\\
  & &= (\pm) \frac{R}{4\pi }  \int d \Omega^{\prime} V(\theta^{\prime},\varphi^{\prime} ) 
  \left( R\frac{\partial }{\partial R}-r\frac{\partial }{\partial r}\right)\left(\frac{1}{\sqrt{r^{2}+R^{2}-2Rr\cos\alpha}}\right)\nonumber\\
 & &= (\pm) \frac{R}{4\pi }  \left( R\frac{\partial }{\partial R}-r\frac{\partial }{\partial r}\right)
  \int d \Omega^{\prime} \frac{ V(\theta^{\prime},\varphi^{\prime} ) } {\sqrt{r^{2}+R^{2}-2Rr\cos\alpha}}\nonumber.
\end{eqnarray}
Definiendo $r_{>}=$menor$\{r,R\},r_{<}=$mayor$\{r,R\}$ y considerando la funci\'on ge\-neratriz de los polinomios de Legendre, junto con
el teorema de adici\'on de los arm\'onico esf\'ericos se llega a   
\begin{eqnarray}
& &\phi(r,\theta,\varphi) = (\pm) \frac{R}{4\pi }  \left( R\frac{\partial }{\partial R}-r\frac{\partial }{\partial r}\right)
  \int d \Omega^{\prime} \frac{ V(\theta^{\prime},\varphi^{\prime} ) } {\sqrt{r_{<}^{2}+r_{>}^{2}-2r_{<}r_{>} \cos\alpha}}\nonumber\\
  &=& (\pm) \frac{R}{4\pi }  \left( R\frac{\partial }{\partial R}-r\frac{\partial }{\partial r}\right)
  \int d \Omega^{\prime} \frac{ V(\theta^{\prime},\varphi^{\prime} ) } {r_{>} \sqrt{1+\left(\frac{r_{<}}{ r_{>}}\right)^{2}-2 \left(\frac{r_{<}}{ r_{>}}\right) \cos\alpha}}\nonumber\\
& =& (\pm) \frac{R}{4\pi }  \left( R\frac{\partial }{\partial R}-r\frac{\partial }{\partial r}\right)
  \int d \Omega^{\prime}  V(\theta^{\prime},\varphi^{\prime} )\frac{1} {r_{>}} \sum_{l\geq 0}  
  \left(\frac{r_{<}}{ r_{>}}\right)^{l}P_{l}(\cos\alpha)\nonumber\\
& =&(\pm) \frac{R}{4\pi }  \left( R\frac{\partial }{\partial R}-r\frac{\partial }{\partial r}\right)
  \int d \Omega^{\prime}  V(\theta^{\prime},\varphi^{\prime} ) \nonumber\\
  & &  \sum_{l\geq 0} \sum_{m=-l}^{l} \frac{4\pi}{2l+1}
\frac{1}{r_{>}} \left( \frac{r_{<}}{ r_{>}}\right)^{l} Y^{*}_{lm}(\theta^{\prime} ,\varphi^{\prime}) Y_{lm}(\theta ,\varphi) \nonumber\\
&= &  \sum_{l\geq 0} \sum_{m=-l}^{l}  \frac{ (\pm)R}{2l+1} \left( R\frac{\partial }{\partial R}-r\frac{\partial }{\partial r}\right)
\left(\frac{1}{r_{>}} \left( \frac{r_{<}}{ r_{>}}\right)^{l}\right) Y_{lm}(\theta ,\varphi) \nonumber\\
& &\int d \Omega^{\prime} Y^{*}_{lm}(\theta^{\prime} ,\varphi^{\prime}) V(\theta^{\prime},\varphi^{\prime} )\nonumber\\
& =& \sum_{l\geq 0} \sum_{m=-l}^{l}  \frac{(\pm)R}{2l+1} \left( R\frac{\partial }{\partial R}-r\frac{\partial }{\partial r}\right) \left(\frac{1}{r_{>}} \left( \frac{r_{<}}{ r_{>}}\right)^{l}\right) A_{lm} Y_{lm}(\theta ,\varphi), 
\label{eq:poisson-esfera-3}
\end{eqnarray}
con 
\begin{eqnarray}
A_{lm}=\int d\Omega Y^{*}_{lm}(\theta,\varphi)V(\theta,\varphi).\nonumber
\end{eqnarray}
Ahora, si  $r<R$ se debe tomar el signo $(-)$ en Eq. (\ref{eq:poisson-esfera-3}), en ese caso
\begin{eqnarray}
\frac{(-)R}{2l+1} \left( R\frac{\partial }{\partial R}-r\frac{\partial }{\partial r}\right)\left(\frac{1}{r_{>}} \left( \frac{r_{<}}{ r_{>}}\right)^{l}\right)&=& \frac{(-)R} {2l+1}\left( R\frac{\partial }{\partial R}-r\frac{\partial }{\partial r}\right)
  \left( \frac{r^{l}}{ R^{l+1}}\right)\nonumber\\
& =&\frac{(-)R}{2l+1}\frac{(-)(2l+1)r^{l}}{R^{l+1}}= \left( \frac{r}{R}\right)^{l},\nonumber
\end{eqnarray}
sustituyendo este resultado en Eq. (\ref{eq:poisson-esfera-3}) se llega a
\begin{eqnarray}
\phi_{int}(r,\theta,\varphi) &=& \sum_{l\geq 0} \sum_{m=-l}^{l} \left( \frac{r}{ R}\right)^{l} A_{lm} Y_{lm}(\theta ,\varphi) \nonumber
\end{eqnarray}
que coincide con  Eq. (\ref{eq:poisson-esfera-3-0}).\\

Para  $R<r$  se debe tomar el signo $(+)$ en Eq. (\ref{eq:poisson-esfera-3}), en cuyo caso
\begin{eqnarray}
\frac{R}{2l+1} \left( R\frac{\partial }{\partial R}-r\frac{\partial }{\partial r}\right)\left(\frac{1}{r_{>}} \left( \frac{r_{<}}{ r_{>}}\right)^{l}\right)&=& \frac{R} {2l+1}\left( R\frac{\partial }{\partial R}-r\frac{\partial }{\partial r}\right)
  \left( \frac{R^{l}}{ r^{l+1}}\right)\nonumber\\
& =&\frac{R}{2l+1}\frac{(2l+1)R^{l}}{r^{l+1}}= \left( \frac{R}{r}\right)^{l+1},\nonumber
\end{eqnarray}
usando este resultado  en Eq. (\ref{eq:poisson-esfera-3}) se obtiene
\begin{eqnarray}
\phi_{ext}(r,\theta,\varphi) &=& \sum_{l\geq 0} \sum_{m=-l}^{l} \left( \frac{r}{ R}\right)^{l+1} A_{lm} Y_{lm}(\theta ,\varphi) \nonumber
\end{eqnarray}
que coincide con  Eq. (\ref{eq:poisson-esfera-3-1}). \\

Por lo tanto, la f\'ormula de Poisson (\ref{eq:poisson-esfera-00}) es la soluci\'on de la ecuaci\'on de Laplace 
con la condici\'on de borde  (\ref{eq:poisson-esfera-00-1})

\subsection{Esfera partida}

Supongamos que tenemos una esfera de radio $R$ cuyo potencial en su frontera es 
\begin{eqnarray}
V(\theta)=\left\{
\begin{array}{ll}
V_{0}&   0\leq \theta \leq \frac{\pi}{2} \\
-V_{0}&   \frac{\pi}{2}< \theta \leq \pi
\end{array} \right.
\label{eq:recurr}
\end{eqnarray}
y es finito en todo el espacio. Determinaremos el potencial en todo el espacio.\\
 
En este caso los potenciales est\'an dados por  Eq. (\ref{eq:poisson-esfera-3-0}) y 
Eq. (\ref{eq:poisson-esfera-3-1}), solo basta determinar los coeficientes $A_{lm}.$
\\
 
Claramente, el sistema es invariantes bajo rotaciones sobre el eje $z,$ por lo que el potencial no puede depender 
de $\varphi.$ Esto implica que si $m\not =0$ los coeficientes $A_{lm}$ son nulos. Los coeficientes  no nulos
son 
\begin{eqnarray}
A_{l0}=A_{l}&=&<Y_{l0}(\theta,\varphi)|V(\theta)>=\int d\Omega Y^{*}_{l0}(\theta,\varphi)V(\theta)\nonumber\\
&=&\int_{0}^{2\pi}d\varphi \int_{0}^{2\pi}  d\theta \sin\theta Y^{*}_{l0}(\theta,\varphi)V(\theta)\nonumber\\
&=& \int_{0}^{2\pi}d\varphi \int_{0}^{2\pi}  d\theta \sin\theta \sqrt{ \frac{2l+1}{4\pi} }P_{l}(\cos\theta) V(\theta)\nonumber\\
&=&\sqrt{ \frac{2l+1}{4\pi}}2\pi  \int_{0}^{2\pi}  d\theta \sin\theta P_{l}(\cos\theta) V(\theta).
\end{eqnarray}
Con el cambio de variable $u=\cos\theta$  se tiene 
\begin{eqnarray}
V(u)=\left\{
\begin{array}{ll}
V_{0}&   0 \leq u \leq 1 \\
-V_{0}&   -1< u \leq 0
\end{array} \right.
\label{eq:recurr}
\end{eqnarray}
y 
\begin{eqnarray}
A_{l}&=&\sqrt{\pi (2l+1)}  \int_{-1}^{1}  du P_{l}(u) V(u).
\end{eqnarray}
Como $V(u)$ es impar, $A_{2l}=0.$ Para el caso impar se tiene
\begin{eqnarray}
A_{2l+1}&=&\sqrt{\pi (2(2l+1)+1)}  2\int_{0}^{1}  du P_{2l+1}(u) V(u)\nonumber\\
 &=& 2V_{0}\sqrt{\pi (4l+3)}\int_{0}^{1}  du P_{2l+1}(u).\nonumber
\end{eqnarray}
Considerando la identidad 
\begin{eqnarray}
P_{2l+1}(u)=\frac{1}{4l+3} \frac{d}{du}\left(P_{2(l+1)}(u) -P_{2l}(u)  \right)
\end{eqnarray}
y  los valores de los polinomios de Legendre en $\pm 1$ y en cero se llega a
\begin{eqnarray}
\int_{0}^{1}  du P_{2l+1}(u)&=& \frac{1}{4l+3}\left(P_{2(l+1)}(u) -P_{2l}(u)  \right)\Bigg|_{0}^{1}\nonumber\\
&= &  
\frac{-1}{4l+3}\left(P_{2(l+1)}(0) -P_{2l}(0)  \right)\nonumber\\
&=& \frac{-1}{4l+3}\left( \frac{(-)^{l+1}(2(l+1))! }{2^{2(l+1)}((l+1)!)^{2}} -\frac{(-)^{l}(2l)!}{2^{2l}(l!)^{2}}\right)\nonumber\\
&=& \frac{(-1)^{l}}{4l+3}\left( \frac{(2l+2)(2l+1)}{2^{2}(l+1)^{2}} \frac{(2l)!}{2^{2l}(l!)^{2}}+ \frac{(2l)!}{2^{2l}(l!)^{2}}  \right)\nonumber\\
&=& \frac{(-1)^{l}}{4l+3}\frac{(2l)!}{2^{2l}(l!)^{2}}\left( \frac{(2l+2)(2l+1)}{2^{2}(l+1)^{2}} + 1\right)\nonumber\\
&=& \frac{(-1)^{l}}{4l+3}\frac{(2l)!}{2^{2l}(l!)^{2}}\left( \frac{2(l+1)(2l+1)}{2^{2}(l+1)(l+1)} + 1\right)\nonumber\\
&=& \frac{(-1)^{l}}{4l+3}\frac{(2l)!}{2^{2l}(l!)^{2}}\left( \frac{4l+3}{2(l+1)}\right)\nonumber\\
&=& \frac{(-1)^{l}(2l)!}{2^{2l}(l!)^{2} 2(l+1)}.
\end{eqnarray}
Entonces 
\begin{eqnarray}
A_{2l+1}&=& \frac{ V_{0}\sqrt{\pi (4l+3)} (-1)^{l}(2l)!}{2^{2l}(l!)^{2}(l+1)},
\end{eqnarray}
que implica

\begin{eqnarray}
V(\theta)&=&\sum_{l\geq 0}A_{2l+1} Y_{2l+1 0}(\theta,\varphi)\nonumber\\
&=&\sum_{l\geq 0} 
\frac{ V_{0} \sqrt{\pi (4l+3)} (-1)^{l}(2l)!} {2^{2l}(l!)^{2}(l+1)} \sqrt{\frac{(4l+3)}{4\pi} } P_{l}(\cos\theta)\nonumber\\
&=&V_{0}\sum_{l\geq 0} \frac{ (-1)^{l}(4l+3)(2l)!} {2^{2l+1}(l!)^{2}(l+1)} P_{2l+1}(\cos\theta), 
\end{eqnarray}
de donde
\begin{eqnarray}
\phi_{int}(r,\theta)&=&V_{0}\sum_{l\geq 0} \frac{ (-1)^{l}(4l+3)(2l)!} {2^{2l+1}(l!)^{2}(l+1)} \left(\frac{r}{R}\right)^{2l+1}P_{2l+1}(\cos\theta),\nonumber\\
\phi_{ext}(r,\theta)&=&V_{0}\sum_{l\geq 0} \frac{ (-1)^{l}(4l+3)(2l)!} {2^{2l+1}(l!)^{2}(l+1)} \left(\frac{R}{r}\right)^{2(l+1)}P_{2l+1}(\cos\theta),
\end{eqnarray}
es el potencial en todo el espacio.

\section{Esfera a potencial cero}

Supongamos que tenemos un esfera de radio $R$ cuya superficie  est\'a a potencial cero y queremos calcular el potencial el\'ectrico en todo el espacio. 
Note que debemos buscar las soluciones de la ecuaci\'on de Laplace que satisfacen
$\phi(R,\theta,\varphi)=0.$ Por conveniencia tomaremos la soluci\'on  general de la forma
\begin{eqnarray}
\phi(r,\theta,\varphi)=\sum_{l\geq 0}\sum_{m=-l}^{l}\left(a_{lm}\left(\frac{r}{R}\right)^{l}+b_{lm}\left(\frac{R}{r}\right)^{l+1}\right) Y_{lm}(\theta,\varphi). 
\end{eqnarray}
Por lo que 
\begin{eqnarray}
\phi(R,\theta,\varphi)=\sum_{l\geq 0}\sum_{m=-l}^{l}\left(a_{lm}+b_{lm}\right) Y_{lm}(\theta,\varphi)=0,\nonumber 
\end{eqnarray}
como los arm\'onicos esf\'ericos son un conjunto de funciones ortonormales,
se debe cumplir 
\begin{eqnarray}
a_{lm}=-b_{lm}. 
\end{eqnarray}
Por lo tanto, la soluci\'on general a este problema es 
\begin{eqnarray}
\phi(r,\theta,\varphi)=\sum_{l\geq 0}\sum_{m=-l}^{l}a_{lm}\left(\left(\frac{r}{R}\right)^{l}-\left(\frac{R}{r}\right)^{l+1}\right) Y_{lm}(\theta,\varphi). \label{eq:esfera-nula}
\end{eqnarray}
\subsection{Plano con protuberancia esf\'erica}

Como una aplicaci\'on del problema anterior, supongamos que tenemos un plano infinito que tiene una protuberancia esf\'erica de radio $R$ y que todo
el sistema est\'a a potencial cero, adem\'as si $r>>R$  el potencia tiene la forma $\phi_{\infty}(r)=-E_{0}z=-E_{0}r\cos\theta.$ 
Para este sistema queremos encontrar el potencial el\'ectrico.\\

Por simplicidad pondremos el plano en el plano $x-y$ de tal forma que el casquete de la protuberancia 
est\'e sobre 
el eje $z.$ Claramente este sistema es invariante bajo rotaciones del eje $z,$ por lo que el potencial no depende de $\varphi.$ Adem\'as, 
como el potencial se debe anular sobre el casquete esf\'erico, \'este debe ser de la forma Eq. (\ref{eq:esfera-nula}) donde lo t\'erminos con
$m\not =0$ no contribuyen. As\'i el potencial debe ser de la forma
\begin{eqnarray}
\phi(r,\theta)=\sum_{l\geq 0}a_{l}\left(\left(\frac{r}{R}\right)^{l}-\left(\frac{R}{r}\right)^{l+1}\right) P_{l}(\cos\theta). 
\end{eqnarray}
Ahora, el potencial se debe anular sobre el plano $x-y$ para cualquier $r.$ Esto  equivale a pedir que el potencial se anule en $\theta=\frac{\pi}{2},$ es decir en $\cos\frac{\pi}{2}=0$ para cualquier $r.$ Entonces, 
\begin{eqnarray}
\phi\left(r,\theta=\frac{\pi}{2}\right)=\sum_{l\geq 0}a_{l}\left(\left(\frac{r}{R}\right)^{l}-\left(\frac{R}{r}\right)^{l+1}\right) P_{l}(0)=0. 
\end{eqnarray}
Considerando que $P_{2l+1}(0)=0,$ se encuentra $a_{2l}=0,$ que implica  
\begin{eqnarray}
\phi(r,\theta)=\sum_{l\geq 0}a_{2l+1}\left(\left(\frac{r}{R}\right)^{2l+1}-\left(\frac{R}{r}\right)^{2(l+1)}\right) P_{2l+1}(\cos\theta).
\end{eqnarray}
Adem\'as, para el caso $r>>R,$ se debe cumplir 
\begin{eqnarray}
\phi_{\infty}(r)=-E_{0}r\cos\theta=\phi(r>>R,\theta)=\sum_{l\geq 0}a_{2l+1}\left(\frac{r}{R}\right)^{2l+1} P_{2l+1}(\cos\theta),\nonumber
\end{eqnarray}
como $P_{1}(\cos\theta)=\cos\theta,$ se llega  $a_{2l+1}=0$ si $l>0$ y  
\begin{eqnarray}
-E_{0}r\cos\theta=a_{1}\frac{r}{R} \cos\theta,
\end{eqnarray}
es decir 
\begin{eqnarray}
a_{1}=-E_{0}R.
\end{eqnarray}
En consecuecia la soluci\'on al problema es
\begin{eqnarray}
\phi(r,\theta)=-E_{0}r\left(1-\left(\frac{R}{r}\right)^{3}\right) \cos\theta.
\end{eqnarray}

\section{Problemas con simetr\'ia azimutal}

La soluci\'on general de la ecuaci\'on de Laplace en coordenadas esf\'ericas 
de problemas que tienen simetr\'{\i}a rotacional 
ante el eje $z,$ es decir que no dependen de $\varphi,$
es
\begin{eqnarray}
\phi(r,\theta)=\sum_{l=0}^{\infty}
\left(A_{l}r^{l}+B_{l}r^{-(l+1)}\right)P_{l}(\cos\theta).
\label{eq:solesf}
\end{eqnarray}
En principio se deben calcular cada uno de los coeficientes
$A_{l}$  y $B_{l},$ sin embargo en algunos casos se pueden ocupar las simetr\'ias del problema
para obtenerlos.

\subsection{Esfera con condiciones especiales} 

Supongamos que tenemos una esfera de radio $R$ cuyo potencial en su superficie
es
\begin{eqnarray}
V(\theta)=V_{0}\left(1+\cos\theta+2\cos\theta+ \cos^{2}\theta\right),\qquad V_{0}={\rm constante}
\label{eq:solesf-especial}
\end{eqnarray}
y el potencial es finito en todo el espacio. Este problema no depende de $\varphi,$ por lo que debemos 
buscar soluciones de la ecuaci\'on de Laplace de la forma Eq. (\ref{eq:solesf}) que satisfagan la condici\'on de borde 
Eq. (\ref{eq:solesf-especial}). Como el potencial es finito en todo el espacio, si $r<R$ el potencial debe ser de la forma
\begin{eqnarray}
\phi_{int}(r,\theta)=\sum_{l\geq 0} a_{l}\left(\frac{r}{R}\right)^{l}P_{l}(\cos\theta).
\end{eqnarray}
Por la misma raz\'on, si $r>R$ el potencial debe escribirse como
\begin{eqnarray}
\phi_{ext}(r,\theta)=\sum_{l\geq 0} b_{l}\left(\frac{R}{r}\right)^{l+1}P_{l}(\cos\theta).
\end{eqnarray}
Antes de continuar escribamos el potencial Eq. (\ref{eq:solesf-especial}) de una forma m\'as sugerente, 
como $\cos2\theta=\cos^{2}\theta-\sen^{2}\theta=2\cos^{2}\theta-1,$ entonces 
\begin{eqnarray}
V(\theta)=V_{0}\left(\cos\theta+3\cos^{2}\theta\right).
\end{eqnarray}
Considerando que 
\begin{eqnarray}
P_{0}(u)=1,\qquad P_{1}(u)=u,\qquad  P_{2}(u)=\frac{1}{2}(3u^{2}-1),
\end{eqnarray}
se encuentra
\begin{eqnarray}
u^{2}=\frac{2}{3}P_{2}(u)+\frac{1}{3}P_{1}(u), 
\end{eqnarray}
as\'i,
\begin{eqnarray}
V(\theta)=V_{0}\left( P_{0}(\cos\theta)+P_{1}(\cos\theta)+ 2P_{2}(\cos\theta) \right).
\end{eqnarray}
Entonces, como $V(\theta)=\phi_{int}(R,\theta)=\phi_{ext}(R,\theta)$ y los polinomios de Legendre son linealmente independientes,
los \'unicos coeficientes diferentes de cero son
\begin{eqnarray}
a_{0}=V_{0},\qquad a_{1}=V_{0},\qquad a_{2}=2V_{0}.
\end{eqnarray}
De donde la soluci\'on al problema es 
\begin{eqnarray}
\phi_{int}(r,\theta)&=&V_{0}\left( P_{0}(\cos\theta)+\frac{r}{R}P_{1}(\cos\theta)+ 2\left(\frac{r}{R}\right)^{2}P_{2}(\cos\theta) \right),\nonumber\\
\phi_{ext}(r,\theta)&=&V_{0}\left( \frac{R}{r} P_{0}(\cos\theta)+\left(\frac{R}{r}\right)^{2}P_{1}(\cos\theta)+ 2\left(\frac{R}{r}\right)^{3}P_{2}(\cos\theta) \right).\nonumber
\end{eqnarray}

\subsection{Potencial de un anillo circular}

Supongamos que tenemos un anillo circular de radio $R$ con
densidad de carga  constante $\lambda,$  paralelo al 
plano $x-y$ y que est\'a a una altura $h$ del origen. Adem\'as
el origen del anillo est\'a sobre el eje $z$.\\

Este sistema es invariante bajo rotaciones del eje $z,$ por lo que
no depende de $\varphi$ y su potencial debe ser de la forma 
Eq. (\ref{eq:solesf}). Note que los coeficientes $A_{l}$ y $B_{l}$  no depende de la posici\'on,
por lo que si los calculamos en un punto o sobre un eje 
los habremos calculado para todo el espacio.\\

Primero calculemos el
potencial para un punto que est\'e sobre el eje $z,$ 
es decir $\theta=0,$ note que esto implica que $r=z.$
Ahora, un elemento de carga del anillo,
$dq=\lambda Rd\varphi$ que est\'a en la posici\'on
$\vec r_{q}=(R\cos\varphi,R \sin\varphi,h)$
contribuye con el potencial
\begin{eqnarray}
d\phi(r=z,\theta=0)=\frac{dq}{|z\hat k-\vec r_{q}|}
=\frac{\lambda Rd\varphi}{\sqrt{R^{2}+(z-h)^{2}}}=
\frac{\lambda Rd\varphi}{\sqrt{R^{2}+h^{2} +z^{2}-2hz}}\nonumber
\end{eqnarray}
de donde
\begin{eqnarray}
\phi(r=z,\theta=0)=
\frac{2\pi\lambda R}{\sqrt{R^{2}+h^{2} +z^{2}-2hz}}.\nonumber
\end{eqnarray}
Este potencial se puede poner de forma m\'as sugerente. En efecto, si $\alpha$ es el \'angulo que hace el eje $z$ con
un  vector que une el origen del sistema coordenado y un punto del anillo,
entonces $R=c\sin\alpha$ y $h=c\cos\alpha,$ con 
$c=\sqrt{R^{2}+h^{2}}.$ Por lo que,
\begin{eqnarray}
\phi(r=z,\theta=0)=
\frac{2\pi\lambda R}{\sqrt{c^{2} +z^{2}-2cz\cos\alpha}}.\nonumber
\end{eqnarray}
Si definimos a $r_{<}=$menor$\{r,c\}$ y  $r_{>}=$mayor$\{r,c\}$ se llega a
\begin{eqnarray}
\phi(r=z,\theta=0)&=&
\frac{2\pi\lambda R}{r_{>}\sqrt{1 +
\left(\frac{r_{<}}{r_{>}}\right)^{2}-2 
\left(\frac{r_{<}}{r_{>}}\right)\cos\alpha}}\nonumber\\
&=&  \frac{2\pi\lambda R}{r_{>}} \sum_{l\geq 0}\left( \frac{r_{<}} {r_{>}}\right) ^{l}P_{l}(\cos\alpha)\nonumber\\
&=& \sum_{l\geq 0} 2\pi\lambda R P_{l}(\cos\alpha) \frac{r_{<}^{l}} {r_{>}^{l+1} }.\label{eq:solan}
\end{eqnarray}
Si $r<c,$ se tiene 
\begin{eqnarray}
\phi_{int}(r=z,\theta=0)=\sum_{l\geq 0} 2\pi\lambda R P_{l}(\cos\alpha) \frac{r^{l}} {c^{l+1} },\label{eq:solan1}
\end{eqnarray}
para el caso $r>c$ se llega a
\begin{eqnarray}
\phi_{ext}(r=z,\theta=0)=\sum_{l\geq 0} 2\pi\lambda R P_{l}(\cos\alpha) \frac{c^{l}} {r^{l+1} }.\label{eq:solan2}
\end{eqnarray}
Para obtener el potencial en todo el espacio consideremos la soluci\'on general Eq. (\ref{eq:solesf}). Si $r\to \infty,$ entonces
$r^{l}$ diverge por lo que en ese caso, que corresponde a $r>c,$ 
se debe cumplir que $A_{l}=0.$ En particular sobre el eje $z,$
es decir $\theta=0,$ como $P_{l}(1)=1,$ se tiene
\begin{eqnarray}
\phi_{ext}(r=z,\theta=0)=\sum_{l\geq 0}B_{l}r^{-(l+1)}.
\end{eqnarray}
Igualando esta soluci\'on con Eq. (\ref{eq:solan2}) se llega a
$B_{l}=2\pi\lambda Rc^{l}P_{l}(\cos\alpha).$
Por lo que, si $r>c,$ la soluci\'on es
\begin{eqnarray}
\phi_{ext}(r,\theta)=\sum_{l\geq 0} \frac{2\pi\lambda Rc^{l} P_{l}(\cos\alpha) }{r^{l+1}} P_{l}(\cos\theta).
\end{eqnarray}
Para el caso $r<c,$ si $r\to 0,$ entonces
$r^{-(l+1)}$ diverge y    
se debe tomar $B_{l}=0.$ En particular sobre el eje $z,$
es decir $\theta=0,$ se consigue
\begin{eqnarray}
\phi_{int}(r=z,\theta=0)=\sum_{l\geq 0}A_{l}r^{l}.
\end{eqnarray}
Igualando esta soluci\'on con Eq. (\ref{eq:solan1}) se encuentra
$A_{l}=\frac{2\pi\lambda R}{c^{l+1}} P_{l}(\cos\alpha).$
De donde, si $r>c,$ la soluci\'on es
\begin{eqnarray}
\phi_{int}(r,\theta)=\sum_{l\geq 0} \frac{2\pi\lambda R P_{l}(\cos\alpha) }{c^{l+1}} r^{l}P_{l}(\cos\theta).
\end{eqnarray}
Claramente la soluci\'on en todo el espacio  es
\begin{eqnarray}
\phi(r,\theta)=2\pi\lambda R\sum_{l\geq 0}
\frac{r_{<}^{l}}{r_{>}^{l+1}}
P_{l}(\cos\alpha)P_{l}(\cos\theta).
\end{eqnarray}

\subsection{Esfera con hueco}

Veamos otro problema que se puede resolver con el m\'etodo
empleado en el ejemplo anterior. Supongamos que tenemos una esfera conductora 
de radio $R$ con un hueco definido por el cono $\theta=\alpha$ y 
que tiene densidad superficial de carga constante $\sigma.$ Queremos calcular 
el potencial el\'ectrico del sistema.\\

Pondremos el origen de coordenadas en el centro 
de la esfera de tal forma que el
eje del cono coincida con el eje $z.$
Este problema no depende de $\varphi,$ por lo que
el potencial es de la forma Eq. (\ref{eq:solesf}). Como en el caso
del anillo, primero calcularemos el potencial sobre el eje $z.$
Un elemento de carga est\'a dado por
$$dq=\sigma da=\sigma R^{2}\sin\theta^{\prime}
d\theta^{\prime}d\varphi^{\prime}$$
y si  tiene la posici\'on
$\vec r^{\prime}=R\hat e_{r^{\prime}},$ entonces contribuye con el potencial
\begin{eqnarray}
d\phi(r=z,\theta=0)&=&\frac{dq}{|z\hat k-R\hat e_{r^{\prime}}|}
=\frac{\sigma R^{2}\sin\theta^{\prime}d\theta^{\prime}d\varphi^{\prime}}
{\sqrt{z^{2}+R^{2}-2Rz\hat k\cdot \hat e_{r^{\prime}}}}\nonumber\\
&=&\frac{\sigma R^{2}\sin\theta^{\prime}d\theta^{\prime}d\varphi^{\prime}}
{\sqrt{z^{2}+R^{2}-2Rz\cos\theta^{\prime}}}
=\frac{\sigma R^{2}\sin\theta^{\prime}d\theta^{\prime}d\varphi^{\prime}}{r_{>} 
\sqrt{1 +\left(\frac{r_{<}}{r_{>}}\right)^{2}-
\left(\frac{r_{<}}{r_{>}}\right)\cos\theta^{\prime}}},\nonumber
\end{eqnarray}
donde $r_{>}=$mayor$\{ r, R\}$ y $r_{<}=$menor$\{ r, R\}.$ 
Por lo que
\begin{eqnarray}
d\phi(r=z,\theta=0)=
\sum_{l\geq 0} 
\frac{r_{<}^{l}}{r_{>}^{l+1}}
P_{l}(\cos\theta^{\prime})\sigma R^{2}
\sin\theta^{\prime}d\theta^{\prime}d\varphi^{\prime},
\end{eqnarray}
as\'{\i} el potencial total sobre un punto en el eje $z$ es
\begin{eqnarray}
\phi(r=z,\theta=0)&=&
\int_{0}^{2\pi}\int_{\alpha}^{\pi}
d\phi(r=z,\theta=0)\nonumber\\
&=&
\sigma R^{2}
\sum_{l\geq 0}
\frac{r_{<}^{l}}{r_{>}^{l+1}}
\int_{0}^{2\pi}d\varphi^{\prime}\int_{\alpha}^{\pi}
P_{l}(\cos\theta^{\prime})
\sin\theta^{\prime}d\theta^{\prime}\nonumber\\
&=& 2\pi \sigma R^{2}
\sum_{l\geq 0}
\frac{r_{<}^{l}}{r_{>}^{l+1}} \int_{\alpha}^{\pi}
P_{l}(\cos\theta^{\prime})
\sin\theta^{\prime}d\theta^{\prime}\nonumber
.
\end{eqnarray}
Con el cambio de variable $x=\cos\theta^{\prime}$ 
y considerando que se cumple
\begin{eqnarray}
P_{l}(x)=\frac{1}{2l+1}\frac{d}{dx}\left(P_{l+1}(x)-P_{l-1}(x)\right),
\qquad P_{l}(-1)=(-1)^{l},
\end{eqnarray}
tenemos 
\begin{eqnarray}
\int_{\alpha}^{\pi}
P_{l}(\cos\theta^{\prime})
\sin\theta^{\prime}d\theta^{\prime}&=&-
\int_{\cos\alpha}^{-1}
P_{l}(x)dx\nonumber\\
&=&\int^{\cos\alpha}_{-1}
\frac{1}{2l+1}\frac{d}{dx}\left(P_{l+1}(x)-P_{l-1}(x)\right)dx\nonumber\\
&=&\frac{1}{2l+1}(P_{l+1}(\cos\alpha)-P_{l-1}(\cos\alpha)), 
\nonumber
\end{eqnarray}
de donde
\begin{eqnarray}
\phi(r=z,\theta=0)=
\sum_{l\geq 0}
\frac{2\pi\sigma R^{2}}{2l+1}
\frac{r_{<}^{l}}{r_{>}^{l+1}}
\left(P_{l+1}(\cos\alpha)-P_{l-1}(\cos\alpha)\right)
\end{eqnarray}
Ahora, tomando en cuenta
la soluci\'on general Eq. (\ref{eq:solesf}), 
el potencial en todo el espacio es
\begin{eqnarray}
\phi(r,\theta)=
\sum_{l\geq 0} 
\frac{2\pi\sigma R^{2}}{2l+1}
\frac{r_{<}^{l}}{r_{>}^{l+1}}
\left(P_{l+1}(\cos\alpha)-P_{l-1}(\cos\alpha)\right)P_{l}(\cos\theta).
\end{eqnarray}

\section{Disco a potencial constante}

Supongamos que tenemos un disco de radio $R$ que est\'a en el plano
$x-y$ centrado en el origen y que tiene potencial constante $V.$ Con la informaci\'on de
que en el disco el campo el\'ectrico  es proporcional a 
\begin{eqnarray}
\frac{4\pi}{\sqrt{R^{2}-\rho^{2}}}
\end{eqnarray}
calcular el potencial si $R<r.$\\

Como el sistema es invariante bajo rotaciones en el eje $z,$ el potencial debe ser de la forma
\begin{eqnarray}
\phi_{ext} (r,\theta)=\sum_{l\geq 0} A_{l} \left(\frac{R}{r}\right)^{l+1}P_{l}(\cos\theta).
\label{eq:disco-f-general}
\end{eqnarray}
Antes de calcular $A_{l}$ note que,  como el potencial es constante en el disco, el campo el\'ectrico, $\vec E=-\vec \nabla\phi,$
debe ser perpendicular a este. Ahora, si $\alpha$ es una constante de proporcionalidad, ocupando la funci\'on generatriz de los
polinomios de Legendre se llega a que el campo el\'ectrico en el disco es 
\begin{eqnarray}
E=\frac{ 4\pi \alpha}{\sqrt{R^{2}-\rho^{2}}}= \frac{4\pi \alpha}{R\sqrt{1-\left(\frac{\rho}{R}\right)^{2}}}=
\frac{4\pi \alpha}{R}\sum_{l\geq 0} (-)^{l}\left(\frac{\rho}{R}\right)^{2l} P_{2l}(0). 
\end{eqnarray}
En el borde del disco, se tiene 
\begin{eqnarray}
\vec E=\frac{4\pi \alpha}{R}\sum_{l\geq 0} (-)^{l} P_{2l}(0) \hat e_{\rho},
\qquad \hat e_{\rho}=(\cos\varphi,\sin\varphi,0), \label{eq:campo-borde-disco}
\end{eqnarray}
con $\hat e_{\rho}$ el vector normal al borde del disco.\\

Ocupando de la funci\'on generatriz de los
polinomios de Legendre Eq. (\ref{eq:gen-leg}) se tiene 
\begin{eqnarray}
\frac{1}{\sqrt{1-\beta^{2}}}=\sum_{l\geq 0} (-)^{l}\beta^{2l} P_{2l}(0). 
\end{eqnarray}
Adem\'as es claro que con el cambio de variable $\beta=\cos\gamma$ se encuentra
\begin{eqnarray}
\int_{0}^{1}\frac{d\beta}{\sqrt{1-\beta^{2}}}=\int_{0}^{\frac{\pi}{2}} d\gamma=\frac{\pi}{2}, 
\end{eqnarray}
por lo que 
\begin{eqnarray}
\int_{0}^{1} d\beta\frac{1}{\sqrt{1-\beta^{2}}}=\int_{0}^{1} d\beta \sum_{l\geq 0} (-)^{l}\beta^{2l} P_{2l}(0) =
\sum_{l\geq 0} \frac{(-)^{l}}{2l+1} P_{2l}(0)=\frac{\pi}{2}. 
\label{eq:integral-dicos-conductor}
\end{eqnarray}
Ahora,  de Eq.   (\ref{eq:disco-f-general}) se encuentra
\begin{eqnarray}
\vec E(r,\theta)&=&-\vec \nabla \phi(r,\theta)=- \sum_{l\geq 0}\Bigg( \frac{-A_{l}(l+1) R^{l+1} }{r^{l+2}} P_{l}(\cos\theta)\hat e_{r} \nonumber\\
& &+ 
\frac{1}{r}\left(\frac{R}{r}\right)^{l+1} A_{l}\frac{dP_{l}(\cos\theta)}{d\theta} \hat e_{\theta}\Bigg).\nonumber
\end{eqnarray}
Con el cambio de variable $u=\cos\theta$ se tiene
\begin{eqnarray}
\frac{dP_{l}(\cos\theta)}{d\theta}=-\sin\theta  \frac{dP_{l}(u)}{du}=-\sin\theta\left( u \frac{dP_{l-1}(u)}{du}+l P_{l-1}(u)\right),
\end{eqnarray}
de donde
\begin{eqnarray}
\vec E(r,\theta)&=& \sum_{l\geq 0}  \frac{A_{l}}{r}\left(\frac{R}{r}\right)^{l+1} 
\Bigg[(l+1) P_{l}(\cos\theta)\hat e_{r} \nonumber\\
& &+
\sin\theta\left( u \frac{dP_{l-1}(u)}{du}+l P_{l-1}(u)\right) \hat e_{\theta}\Bigg]. \label{eq:campo-general-disco}
\end{eqnarray}
En el plano $x-y,$ es decir $\theta=\frac{\pi}{2}$ 
se tiene  
\begin{eqnarray}
\hat e_{r}|_{\frac{\pi}{2}}&=& (\cos\varphi,\sin\varphi,0)=\hat e_{\rho},\nonumber\\
\hat e_{\theta}|_{\frac{\pi}{2}}&=& (0,0,-1)=-\hat k.\nonumber
\end{eqnarray}
Note que el borde del disco se encuentra en $\theta=\frac{\pi}{2}, r=R$ y en ese caso se debe cumplir  Eq. (\ref{eq:campo-borde-disco}).
Entonces,  evaluado en $\left(\theta=\frac{\pi}{2}, r=R\right)$ a Eq. (\ref{eq:campo-general-disco}) e igualando 
el resultado con Eq. (\ref{eq:campo-borde-disco}), se encuentra 
\begin{eqnarray}
\vec E\left(R,\theta=\frac{\pi}{2}\right)&=& \sum_{l\geq 0}  \frac{A_{l}}{R} \left[(l+1) P_{l}(0)\hat e_{\rho} -l P_{l-1}(0) \hat k\right]\nonumber\\
&=&\frac{4\pi \alpha}{R}\sum_{l\geq 0} (-)^{l} P_{2l}(0) \hat e_{\rho}
,\label{eq:disco-conductor}
\end{eqnarray}
que implica 
\begin{eqnarray}
\sum_{l\geq 0} -\frac{A_{l}}{R} l P_{l-1}(0) \hat k=0,
\end{eqnarray}
de donde 
\begin{eqnarray}
A_{2l+1}=0.
\end{eqnarray}
Entonces la igualdad Eq. (\ref{eq:disco-conductor}) toma la forma 
\begin{eqnarray}
\vec E\left(R,\theta=\frac{\pi}{2}\right)= \sum_{l\geq 0}  \frac{A_{2l}}{R} (2l+1) P_{2l}(0)\hat e_{\rho}
=
\frac{4\pi \alpha}{R}\sum_{l\geq 0} (-)^{l} P_{2l}(0) \hat e_{\rho}
,\nonumber
\end{eqnarray}
que induce 
\begin{eqnarray}
A_{2l}=\frac{(-)^{l} 4\pi \alpha}{2l+1} .
\end{eqnarray}
As\'{\i} el potencial exterior tiene la forma 

\begin{eqnarray}
\phi_{ext}(r,\theta)=4\pi\alpha \sum_{l\geq 0} \frac{(-)^{l}}{2l+1} \left(\frac{R}{r}\right)^{2l+1}P_{2l}(\cos\theta),
\end{eqnarray}
este potencial debe satisfacer la condici\'on de borde 
\begin{eqnarray}
\phi\left(R,\theta=\frac{\pi}{2}\right)=4\pi\alpha \sum_{l\geq 0} \frac{(-)^{l}}{2l+1} P_{2l}(0)=V,
\end{eqnarray}
usando Eq. (\ref{eq:integral-dicos-conductor}) se llega a
\begin{eqnarray}
\alpha=\frac{V}{2\pi^{2}}.
\end{eqnarray}
Por lo tanto, el potencial es
\begin{eqnarray}
\phi_{ext}(r,\theta)=\frac{2V}{\pi} \sum_{l\geq 0} \frac{(-)^{l}}{2l+1} \left(\frac{R}{r}\right)^{2l+1}P_{2l}(\cos\theta).
\end{eqnarray}

\section{Distribuci\'on de carga continua}

Supongamos que se tiene una part\'icula puntual de carga $q$ en la posici\'on  $\vec r^{\prime}$, entonces
seg\'un las leyes de la electrost\'atica el potencial el\'ectrico en el punto $\vec r$ est\'a dado por 
\begin{eqnarray}
\phi(\vec r)=\frac{q}{|\vec r-\vec r^{\prime}|}.
\end{eqnarray}
Ahora, si se tiene una densidad de carga $\rho(\vec r)$ en un volumen, $V,$ cada elemento de carga $dq(\vec r^{\prime})=\rho(\vec r)d\vec r$ 
contribuye al potencial en el punto $\vec r$ con
\begin{eqnarray}
d\phi(\vec r)=\frac{dq(\vec r^{\prime})}{|\vec r-\vec r^{\prime}|}= \frac{\rho(\vec r^{\prime})d\vec r^{\prime} }{|\vec r-\vec r^{\prime}|},
\end{eqnarray}
por lo tanto, el potencial total en el punto $\vec r $ es
\begin{eqnarray}
\phi(\vec r)=\int d\vec r^{\prime}   \frac{\rho(\vec r^{\prime}) }{|\vec r-\vec r^{\prime}|}.
\end{eqnarray}
Sea $\alpha$ el \'angulo entre los vectores
\begin{eqnarray}
\vec r=r(\sin\theta\cos\varphi, \sin\theta\sin\varphi,\cos\theta),\qquad
\vec r^{\prime}=r^{\prime} (\sin\theta^{\prime}\cos\varphi^{\prime}, \sin\theta^{\prime}\sin\varphi^{\prime},\cos\theta^{\prime})\nonumber
\end{eqnarray}
tambi\'en  definamos $r_{<}=$menor$\{r,r^{\prime}\},r_{>}=$mayor$\{r,r^{\prime}\},$ entonces 
\begin{eqnarray}
\frac{1}{|\vec r-\vec r^{\prime}|}&=&\frac{1}{\sqrt{r^{2}+r^{\prime 2} -2rr^{\prime}\cos\alpha}}=\frac{1}{\sqrt{r_{>}^{2}+r_{<}^{2} -2r_{<}r_{>}\cos\alpha}}\nonumber\\
&=&\frac{1}{r_{<}\sqrt{1+\left(\frac{r_{<}}{r_{>}}\right)^{2}-2\left(\frac{r_{<}}{r_{>}}\right)\cos\alpha}}=
\frac{1}{r_{<}}\sum_{l\geq 0} \left(\frac{r_{<}}{r_{>}}\right)^{l}P_{l}(\cos\alpha) \nonumber\\
&=&\sum_{l\geq 0}\sum_{m=-l}^{l} \frac{4\pi}{2l+1}\frac{1}{r_{<}}
\left(\frac{r_{<}}{r_{>}}\right)^{l} Y^{*}_{lm}(\theta^{\prime},\varphi^{\prime})  Y_{lm}(\theta,\varphi).\nonumber
\end{eqnarray}
Por lo que
\begin{eqnarray}
\phi(\vec r)&=&\int d\vec r^{\prime}   \frac{\rho(\vec r^{\prime}) }{|\vec r-\vec r^{\prime}|}\nonumber\\
&=&\int d\vec r^{\prime}  \rho(\vec r^{\prime})\left(\sum_{l\geq 0}\sum_{m=-l}^{l} \frac{4\pi}{2l+1}\frac{1}{r_{<}}
\left(\frac{r_{<}}{r_{>}}\right)^{l} Y^{*}_{lm}(\theta^{\prime},\varphi^{\prime})  Y_{lm}(\theta,\varphi)\right)\nonumber\\
&=&\sum_{l\geq 0}\sum_{m=-l}^{l} \frac{4\pi Y_{lm}(\theta,\varphi)  }{2l+1}    \int d\vec r^{\prime} \frac{1}{r_{<}}
\left(\frac{r_{<}}{r_{>}}\right)^{l} Y^{*}_{lm}(\theta^{\prime},\varphi^{\prime})  \rho(\vec r^{\prime}).
\label{eq:desarrollo-dipolar}
\end{eqnarray}
En particular para el potencial exterior, $R<r,$ se tiene 
\begin{eqnarray}
\phi_{ext}(\vec r)=\sum_{l\geq 0}\sum_{m=-l}^{l} \frac{4\pi}{2l+1}  \frac{Q_{lm}}{r^{l+1}} Y_{lm}(\theta,\varphi),\nonumber
\end{eqnarray}
con 
\begin{eqnarray}
Q_{lm}=  \int d \vec r^{\prime}  r^{\prime l} Y^{*}_{lm}(\theta^{\prime},\varphi^{\prime})\rho(\vec r^{\prime}) .
\label{eq:desarrollo-dipolar-1}
\end{eqnarray}
Si $r<r^{\prime},$ se llega a 
\begin{eqnarray}
\phi_{ext}(\vec r)=\sum_{l\geq 0}\sum_{m=-l}^{l} \frac{4\pi}{2l+1} r^{l} q_{lm} Y_{lm}(\theta,\varphi),\nonumber
\end{eqnarray}
con 
\begin{eqnarray}
q_{lm}=  \int d \vec r^{\prime} \frac{1}{ r^{\prime l+1}} Y^{*}_{lm}(\theta^{\prime},\varphi^{\prime})\rho(\vec r^{\prime}) \label{eq:desarrollo-dipolar-2}.
\end{eqnarray}
Entonces, el potencial de cualquier distribuci\'on de carga continua se puede expresar en t\'erminos de 
arm\'onicos esf\'ericos.

\subsection{Esfera cargada}

Ahora veamos un problema sencillo que involucre calcular los cofientes $Q_{lm}$ y $q_{lm}.$ \\

En cuentre el potencial el\'ectrico de una esfera de radio $R$ que en la superficie tiene la distribuci\'on de carga
\begin{eqnarray}
\sigma(\theta)=\sigma_{0}\cos\theta,\qquad \sigma_{0}={\rm constante}. 
\end{eqnarray}

Como solo hay carga en el radio $R,$ las distruci\'on de carga se puede escribir como
\begin{eqnarray}
\rho(\vec r)=\delta(r- R)\sigma_{0}\cos\theta. 
\end{eqnarray}
Adem\'as considerando 
\begin{eqnarray}
Y_{10}(\theta,\varphi)=\sqrt{\frac{3}{4\pi}}P_{1}(\cos\theta)= \sqrt{\frac{3}{4\pi}}\cos\theta 
\end{eqnarray}
se encuentra
\begin{eqnarray}
\rho(\vec r)=\delta(r- R)\sigma_{0}\sqrt{\frac{4\pi}{3} } Y_{10}(\theta,\varphi). 
\end{eqnarray}
Introduciendo este resultado en Eq. (\ref{eq:desarrollo-dipolar-1}) y Eq. (\ref{eq:desarrollo-dipolar-2}), se encuentra
\begin{eqnarray}
Q_{lm}&=&  \int d \vec r^{\prime}  r^{\prime l} Y^{*}_{lm}(\theta^{\prime},\varphi^{\prime})
\delta(r^{\prime}- R)\sigma_{0}\sqrt{\frac{4\pi}{3} } Y_{10}(\theta^{\prime},\varphi^{\prime} )\nonumber\\
&=& \sigma_{0}\sqrt{\frac{4\pi}{3} }\int dr^{\prime} r^{\prime 2} r^{\prime l}\delta(r^{\prime}- R)
\int d\Omega^{\prime} Y^{*}_{lm}(\theta^{\prime},\varphi^{\prime}) Y_{10}(\theta,\varphi)\nonumber\\
&=&\sqrt{\frac{4\pi}{3} } R^{l+2}\sigma_{0}\delta_{l1}\delta_{m0} = \sqrt{\frac{4\pi}{3} } R^{3}\sigma_{0}\delta_{l1}\delta_{m0}, \nonumber\\
q_{lm}&=&  \int d \vec r^{\prime} \frac{1}{ r^{\prime l+1}} Y^{*}_{lm}(\theta^{\prime},\varphi^{\prime})\rho(\vec r^{\prime}) \nonumber\\
&=&\int dr^{\prime} r^{\prime 2} \int d\Omega^{\prime}  \frac{1}{ r^{\prime l+1}} Y^{*}_{lm}(\theta^{\prime},\varphi^{\prime})
 \delta(r^{\prime}- R)\sigma_{0}\sqrt{\frac{4\pi}{3} } Y_{10}(\theta^{\prime},\varphi^{\prime} )\nonumber\\
&=&\sqrt{\frac{4\pi}{3} } \frac{1}{ R^{l-1}} \sigma_{0}\delta_{l1}\delta_{m0} 
 =\sqrt{\frac{4\pi}{3} } \sigma_{0}\delta_{l1}\delta_{m0}\nonumber.
\end{eqnarray}
Es decir, los \'unicos coeficientes $Q_{lm}$ y $q_{lm}$ diferentes de cero son 
\begin{eqnarray}
Q_{10}=\sqrt{\frac{4\pi}{3} } R^{3}\sigma_{0}, \qquad
q_{10}=\sqrt{\frac{4\pi}{3} } \sigma_{0}\nonumber.
\end{eqnarray}
Por lo que los potenciales son
\begin{eqnarray}
\phi_{ext}(r,\theta)=\frac{4\pi}{3}\frac{Q_{10}Y_{10}(\theta,\varphi) }{r^{2}},\qquad
\phi_{int}(r,\theta)=\frac{4\pi}{3} rq_{10}Y_{10}(\theta,\varphi), \nonumber
\end{eqnarray}
es decir 
\begin{eqnarray}
\phi_{ext}(r,\theta)&=&\frac{4\pi R^{3} \sigma_{0}}{3}\frac{\cos\theta }{r^{2}},\nonumber\\
\phi_{int}(r,\theta)&=&\frac{4\pi}{3}\sigma_{0} r\cos\theta=\frac{4\pi}{3}\sigma_{0} z.
\end{eqnarray}

\section{Problemas en magnetismo}

Las leyes de la magnetost\'atica nos dicen que si tenemos una densidad de corriente, $\vec J,$ se produce
un campo magn\'etico $\vec B$ que satisface
\begin{eqnarray}
\vec \nabla \cdot \vec B&=&0,\\
 \vec \nabla \times \vec B&=&\frac{4\pi}{c} \vec J.
\end{eqnarray}
La densidad de corrientes se define como $\vec J=\rho(\vec r) \vec V(\vec r).$ Donde $\rho(\vec r)$ es la densidad de carga y $\vec V(\vec r)$
es la velocidad de las part\'iculas cargadas.\\

La  ecuaci\'on $\vec \nabla \cdot \vec B=0$ implica que existe una funci\'on $\vec A$ tal que $\vec B=\vec \nabla \times \vec A.$ Se puede mostrar que en t\'erminos de la corriente $\vec A$ est\'a dada por \cite{jackson:gnus}
\begin{eqnarray}
\vec A(\vec r)=\frac{1}{c} \int d\vec r^{\prime} \frac{\vec J(\vec r^{\prime})}{|\vec r-\vec r^{\prime}|}.
\end{eqnarray}
En la siguientes secciones veremos dos ejercicios relacionados con este tema.

\subsection{Esfera rotante} 

Supongamos que tenemos una esfera de radio $R$ con densidad de carga superficial constante $\sigma_{0}.$
Si la esfera rota con velocidad constante $\omega,$ encuentre el potencial vectorial y el campo magn\'etico.\\

Como solo hay carga en la superficie de la esfera, la densidad de carga es
\begin{eqnarray}
\rho(\vec r)=\sigma_{0} \delta(r-R).
\end{eqnarray}
Supongamos que el eje de rotaci\'on est\'a en el eje $z,$ entonces la velocidad de un punto de la esfera es
\begin{eqnarray}
\vec V=R\sin\theta \omega \hat e_{\varphi}.
\end{eqnarray}
Adem\'as, sabemos que cualquier vector en dos dimensiones $(x,y)$ se puede escribir como el n\'umero complejo $z=x+iy,$
en particular $\hat e_{\varphi}=(-\sin\varphi,\cos\varphi, 0)$ se puede escribir como
\begin{eqnarray}
\hat e_{\varphi}=-\sin\varphi+i\cos\varphi=i(\cos\varphi+i\sin\varphi)=ie^{i\varphi},
\end{eqnarray}
entonces 
\begin{eqnarray}
\vec V=iR\omega \sin\theta e^{i\varphi}.
\end{eqnarray}
Ahora, de la definici\'on de los arm\'onicos esf\'ericos se tiene 
\begin{eqnarray}
Y_{11}(\theta,\varphi)=(-)\sqrt{\frac{3}{8\pi}} \sin\theta e^{i\varphi},
\end{eqnarray}
de donde 
\begin{eqnarray}
\vec V=-iR\omega \sqrt{\frac{8\pi}{3}}Y_{11}(\theta,\varphi)  .
\end{eqnarray}
As\'i, la densidad de corriente es
\begin{eqnarray}
\vec J=\rho \vec V=-iR\omega \sigma_{0} \delta(r-R)\sqrt{\frac{8\pi}{3}}Y_{11}(\theta,\varphi), 
\end{eqnarray}
entonces, si $r_{>}=$mayor$\{R,r\},r_{<}=$menor$\{R,r\},$  el potencial vectorial magn\'etico es
\begin{eqnarray}
& &\vec A(\vec r)=\frac{1}{c} \int d\vec r^{\prime} \frac{\vec J(\vec r^{\prime})}{|\vec r-\vec r^{\prime}|}
=-\frac{iR\omega \sigma_{0}}{c}
\sqrt{\frac{8\pi}{3}} \int dr^{\prime} r^{\prime 2}\int d\Omega^{\prime}  \delta(r^{\prime}-R)\frac{Y_{11}(\theta^{\prime},\varphi^{\prime})  }{|\vec r-\vec r^{\prime}|}\nonumber\\
&=&  -\frac{iR^{3}\omega \sigma_{0}}{c}
\sqrt{\frac{8\pi}{3}}\int d\Omega^{\prime} Y_{11}(\theta^{\prime},\varphi^{\prime}) \sum_{l\geq 0}\sum_{m=-l}^{l} \frac{4\pi}{2l+1} \frac{r_{<}^{l}}{r_{>}^{l+1}} Y^{*}_{lm}(\theta^{\prime},\varphi^{\prime}) Y_{lm}(\theta,\varphi)\nonumber\\
&= &  -\frac{iR^{3}\omega \sigma_{0}}{c}
\sqrt{\frac{8\pi}{3}} \sum_{l\geq 0}\sum_{m=-l}^{l} \frac{4\pi}{2l+1} \frac{r_{<}^{l}}{r_{>}^{l+1}}  Y_{lm}(\theta,\varphi)
\int d\Omega^{\prime} Y^{*}_{lm}(\theta^{\prime},\varphi^{\prime}) Y_{11}(\theta^{\prime},\varphi^{\prime})\nonumber \\
&= &  -\frac{iR^{3}\omega \sigma_{0}}{c}
\sqrt{\frac{8\pi}{3}} \sum_{l\geq 0}\sum_{m=-l}^{l} \frac{4\pi}{2l+1} \frac{r_{<}^{l}}{r_{>}^{l+1}}  Y_{lm}(\theta,\varphi)
\delta_{l1}\delta_{m1}\nonumber \\
& =&-\frac{iR^{3}\omega \sigma_{0}}{c}
\sqrt{\frac{8\pi}{3}} \frac{4\pi}{3} \frac{r_{<}}{r_{>}^{2}}  Y_{11}(\theta,\varphi) =-\frac{iR^{3}\omega \sigma_{0}}{c}
\sqrt{\frac{8\pi}{3}} \frac{4\pi}{3} \frac{r_{<}}{r_{>}^{2}}  (-)\sqrt{\frac{3}{8\pi}} \sin\theta e^{i\varphi}\nonumber \\
& =&\frac{R^{3}\omega \sigma_{0}}{c} \frac{4\pi}{3} \frac{r_{<}}{r_{>}^{2}}  \sin\theta ie^{i\varphi}=\frac{4\pi \sigma_{0} R^{3}\omega}{3c} \frac{r_{<}}{r_{>}^{2}}  \sin\theta \hat e_{\varphi},
\end{eqnarray}
de donde
\begin{eqnarray}
\vec A(\vec r)=A_{\varphi}\hat e_{\varphi},\qquad A_{\varphi}=\frac{4\pi \sigma_{0} R^{3}\omega}{3c} \frac{r_{<}}{r_{>}^{2}}  \sin\theta.
\end{eqnarray}
Ocupando la expresi\'on del rotacional en coordenadas esf\'ericas Eq. (\ref{eq:rota-esfe}),
se obtiene  el campo magn\'etico 
\begin{eqnarray}
B_{r}&=&\frac{1}{\sin\theta}\frac{\partial }{\partial \theta}\left(\sin\theta A_{\varphi}\right),\nonumber\\
B_{\theta}&=&-\frac{1}{r}\frac{\partial }{\partial r}\left(r A_{\varphi}\right),\nonumber\\
B_{\varphi}&=&0.
\end{eqnarray}
Por lo tanto,
\begin{eqnarray}
B_{r}= \frac{8\pi \sigma_{0} R^{3}\omega}{3c}  \frac{r_{<}}{r_{>}^{2}}  \cos\theta
\end{eqnarray}
Si $r<R,$ se tiene  
\begin{eqnarray}
B_{\theta (int)}=-\frac{8\pi \sigma_{0} R\omega}{3c} \sin\theta,
\end{eqnarray}
mientras que si $r>R,$ se llega a 
\begin{eqnarray}
B_{\theta (ext)}=\frac{4\pi \sigma_{0} R^{4}\omega}{3cr^{3}} \sin\theta.
\end{eqnarray}
Note que 
\begin{eqnarray}
\vec B_{int}= B_{r (int)} \hat e_{r}+ B_{\theta(int)}\hat e_{\theta}= \frac{8\pi \sigma_{0} R\omega}{3c} \hat k,
\end{eqnarray}
es decir el campo magn\'etico en el interior de la esfera es constante y apunta en la direcci\'on del eje de la rotaci\'on de la 
esfera.\\

Cabe se\~nalar que el n\'ucleo de la tierra es met\'alico y tiene cierta carga, por lo que al girar produce un campo magn\'etico. En este
sentido la esfera cargada rotante es un modelo simplificado para explicar el campo magn\'etico terrestre.

\subsection{Anillo de corriente I}

%
%

Ahora veamos el problema  de un anillo circular de radio $R$ por el cual circula
una corriente constante $I$. Sin perdida de generalidad podemos suponer
que el anillo est\'a en el plano $x-y$ y que  su centro est\'a en el origen del sistema coordenado.
Para este sistema la densidad de corriente es 
\begin{eqnarray}
\vec J(\vec r)=\frac{I}{R}\delta(r-R)
\delta\left(\theta-\frac{\pi}{2}\right)\hat e_{\varphi}.
\end{eqnarray}
Ahora, recordemos que  cualquier vector en dos dimensiones, $(x,y),$ se puede escribir como un n\'umero complejo, $z=x+iy.$ En particular 
$\hat e_{\varphi}=(-\sin\varphi,\cos\varphi, 0)$ se puede reprensentar como $\hat e_{\varphi}=-\sin\varphi+i\cos\varphi=
i( \cos\varphi+i\sin\varphi)=ie^{i\varphi},$ por lo que
\begin{eqnarray}
\vec J(\vec r)=\frac{I}{R}\delta(r-R)
\delta\left(\theta-\frac{\pi}{2}\right)i e^{i\varphi}.
\end{eqnarray}
Definamos $r_{>}=$mayor$\{R,r\},r_{<}=$menor$\{R,r\},$  entonces el potencial vectorial  es
\begin{eqnarray}
\vec A(\vec r)&=&\frac{iI}{cR}\int 
\frac{\delta(r^{\prime}-R) \delta\left(\theta^{\prime}-\frac{\pi}{2}\right) 
 e^{i\varphi^{\prime}}} 
{|\vec r-r^{\prime}|}r^{\prime 2}dr^{\prime}d\Omega^{\prime}\nonumber\\
&=&\frac{iI}{cR}
\int\Bigg(  \sum_{l\geq 0}\sum_{m=-l}^{l}\frac{4\pi}{2l+1}
\frac{r_{<}^{l}}{r_{>}^{l+1}} Y_{lm}^{*}
(\theta^{\prime},\varphi^{\prime})Y_{lm}
(\theta,\varphi)\nonumber\\
& &\delta(r^{\prime}-R) \delta\left(\theta^{\prime}-\frac{\pi}{2}\right) 
 e^{i\varphi^{\prime}}
r^{\prime2}dr^{\prime}
d\Omega^{\prime}\Bigg)
\nonumber\\
&=&\frac{iI}{cR}
 \sum_{l\geq 0}\sum_{m=-l}^{l}\frac{4\pi}{2l+1}
\frac{r_{<}^{l}}{r_{>}^{l+1}} R^{2} Y_{lm}
(\theta,\varphi) \int Y_{lm}^{*}
(\theta^{\prime},\varphi^{\prime}) 
\delta\left(\theta^{\prime}-\frac{\pi}{2}\right) e^{i\varphi^{\prime}}
d\Omega^{\prime}.\nonumber
\end{eqnarray}
Ocupando la definici\'on de elemento de \'angulo s\'olido y
de los arm\'onicos esf\'ericos  se tiene 
\begin{eqnarray}
& &\int Y_{lm}^{*}
(\theta^{\prime},\varphi^{\prime}) 
\delta\left(\theta^{\prime}-\frac{\pi}{2}\right) e^{i\varphi^{\prime}}
d\Omega^{\prime}=\int_{0}^{2\pi}d\varphi^{\prime} 
Y^{*}_{lm}\left(\frac{\pi}{2},\varphi^{\prime}\right) 
e^{i\varphi^{\prime}}\nonumber\\
&=&\sqrt{\frac{(2l+1)(l-m)!}{4\pi(l+m)!}} P_{l}^{m}
\left(\cos\frac{\pi}{2}\right)
\int_{0}^{2\pi}d\varphi^{\prime}e^{-im\varphi^{\prime}}e^{i\varphi^{\prime}}
\nonumber\\
&=&\sqrt{\frac{(2l+1)(l-m)!}{4\pi(l+m)!}} P_{l}^{m}\left(0\right)
2\pi \delta_{m1}\nonumber\\
&=&\sqrt{\frac{(2l+1)(l-m)!}{4\pi(l+m)!}} P_{l}^{m}\left(0\right) 
 2\pi \delta_{m1}\nonumber\\
&=&Y^{*}_{lm}\left(\frac{\pi}{2},0\right)2\pi \delta_{m1}.
\end{eqnarray}
Como $m$ debe cumplir que $m \leq l,$ entonces $l\geq 1.$ As\'i, ocupando de nuevo las definiciones 
de los arm\'onicos esf\'ericos Eq. (\ref{eq:armonicos-final}) y de los polinomios asociados de Legendre Eq. (\ref{eq:polinomios-asociados-legendre}), se encuentra
\begin{eqnarray}
\vec A(\vec r)&=&\frac{8\pi^{2} iIR}{c}
 \sum_{l\geq 1}\frac{1}{2l+1}
\frac{r_{<}^{l}}{r_{>}^{l+1}} Y_{l1}
(\theta,\varphi) Y_{l1}^{*}
\left(\frac{\pi}{2},0\right)\nonumber\\
&=&\frac{8\pi^{2} IR}{c}
 \sum_{l=1}^{\infty}\frac{1}{2l+1}
\frac{r_{<}^{l}}{r_{>}^{l+1}} \left(\sqrt{\frac{(2l+1)(l-1)!}{4\pi(l+1)!}}\right)^{2}
P^{1}_{l}(\cos\theta)  P^{1}_{l}(0)ie^{i\varphi} \nonumber\\
&=& \frac{2\pi IR}{c} \sum_{l\geq 1} \frac{(l-1)!}{(l+1)!}
\frac{r_{<}^{l}}{r_{>}^{l+1}} P^{1}_{l}(\cos\theta)  P^{1}_{l}(0)\hat e_{\varphi} \nonumber\\
& =&\frac{2\pi IR}{c}\sum_{l\geq 1} \frac{(2l)!}{(2l+2)!}
\frac{r_{<}^{2l}}{r_{>}^{2l+2}} P^{21+1}_{l}(\cos \theta)   \frac{(-)^{l+1} (2l+1)!}{ 2^{2l}(l!)^{2}} \hat e_{\varphi} \nonumber\\
&=&\frac{2\pi IR}{c}\sum_{l\geq 1} \frac{ (-)^{l+1} (2l)! }{2^{2l+1} (l+1)!l! }
\frac{r_{<}^{2l}}{r_{>}^{2l+2}} P^{21+1}_{l}(\cos\theta)  \hat e_{\varphi}. 
\label{eq:maganillo}
\end{eqnarray}
Por lo tanto, el potencial vectorial solo tiene direcci\'on
$\hat e_{\varphi},$  cuya componente es
\begin{eqnarray}
A_{\varphi}=\frac{\pi IR}{c}\sum_{l\geq 1} \frac{ (-)^{l+1} (2l)! }{2^{2l} (l+1)!l! }
\frac{r_{<}^{2l}}{r_{>}^{2l+2}} P^{21+1}_{l}(\cos\theta) .  
\label{eq:maganillo-1}
\end{eqnarray}
Ocupando el rotacional en coordenadas esf\'ericas Eq. (\ref{eq:rota-esfe}), se encuentra que  el campo magn\'etico es
\begin{eqnarray}
B_{r}&=&\frac{1}{\sin\theta}\frac{\partial }{\partial \theta}\left(\sin\theta A_{\varphi}\right),\nonumber\\
B_{\theta}&=&-\frac{1}{r}\frac{\partial }{\partial r}\left(r A_{\varphi}\right),\nonumber\\
B_{\varphi}&=&0.
\end{eqnarray}
Con el cambio de variable $u=\cos\theta$ se tiene 
\begin{eqnarray}
\frac{\partial }{\partial \theta}=-\sin\theta \frac{\partial }{\partial u},\qquad
\frac{1}{\sin\theta} \frac{\partial A_{\varphi} }{\partial \theta}&=&- \frac{\partial}{\partial u} ((1-u^{2})^{\frac{1}{2}} A_{\varphi})
\end{eqnarray}
Adem\'as, como 
\begin{eqnarray}
P_{l}^{1}(u)=-(1-u^{2})^{\frac{1}{2}}\frac{d P_{l}(u)}{du},\qquad \frac{d}{du}\left((1-u^{2})\frac{d P_{l}(u)}{du}\right)+l(l+1)P_{l}(u)=0,
\nonumber
\end{eqnarray}
se llega a 
\begin{eqnarray}
B_{r}&=&\frac{1}{\sin\theta}\frac{\partial }{\partial \theta}\left(\sin\theta A_{\varphi}\right),\nonumber\\
&=& \frac{\pi IR}{c}\sum_{l\geq 1} \frac{ (-)^{l+1} (2l)! }{2^{2l} (l+1)!l! }
\frac{r_{<}^{2l}}{r_{>}^{2l+2}} (2l+1)(2l+2)P^{21+1}(\cos\theta)\nonumber\\
&=& \frac{2\pi IR}{c}\sum_{l\geq 1} \frac{ (-)^{l} (2l+1)! }{2^{2l} (l!)^{2} }
\frac{r_{<}^{2l}}{r_{>}^{2l+2}} P^{21+1}(\cos\theta).\label{eq:campo-anillo-radial}
\end{eqnarray}
Para la componente $B_{\theta},$ tenemos que si $R<r,$
\begin{eqnarray}
B_{\theta(ext)}
&=&\frac{\pi I}{cr}\sum_{l\geq 1} \frac{ (-)^{l+1} (2l+1)! }{2^{2l} (l+1)!l! }
\left(\frac{R}{r}\right) ^{2l+2} P^{21+1}_{l}(\cos\theta),\label{eq:campo-anillo-angulo-ext}
\end{eqnarray}
mientras que si  $r<R,$

\begin{eqnarray}
B_{\theta(int)}&=&\frac{2\pi I}{cr}\sum_{l\geq 1} \frac{ (-)^{l} (2l)! }{2^{2l} (l!)^{2} }
\left(\frac{r}{R}\right) ^{2l+1} P^{21+1}_{l}(\cos\theta).\label{eq:campo-anillo-angulo-int}
\end{eqnarray}
\subsection{Anillo de corriente II}

Ahora veremos otra forma de resolver el problema de la secci\'on anterior.\\

Fuera de la regi\'on donde hay corriente el\'ectrica, las ecuaciones de la magnetost\'atica son
\begin{eqnarray}
\vec \nabla \cdot \vec B=0,\qquad \vec \nabla \times \vec B=0.
\end{eqnarray}
De la segunda ecuaci\'on se deduce que existe una funci\'on, que llamaremos potencial escalar magn\'etico, $\phi_{m}(\vec r),$ tal que 
\begin{eqnarray}
\vec B=- \vec \nabla \phi_{m}(\vec r).
\end{eqnarray}
Al introducir esta ecuaci\'on en $\vec \nabla \cdot \vec B=0$ se llega a la ecuaci\'on de Laplace
\begin{eqnarray}
 \nabla^{2} \phi_{m}(\vec r)=0,
\end{eqnarray}
cuya soluci\'on en coordenadas esf\'ericas es Eq. (\ref{eq:solucion-poisson-esfericas}). \\

Adem\'as, se puede mostrar que si la corriente circula por una curva cerrada el 
potencial escalar magn\'etico es \cite{landau-medios:gnus}
\begin{eqnarray}
\phi_{m}(\vec r)=\frac{I}{c}\oint_{a}\frac {\hat n^{\prime}\cdot \left(\vec r-\vec r^{\prime}\right)}{ | \vec r-\vec r^{\prime}|^{3} } da^{\prime}.
\label{eq:escalar-magnetico-anillo}
\end{eqnarray}
Donde $a$ es el \'area de la superficie que encierra la curva de corriente y  $\hat n$ la normal a esta superficie.\\

En algunos caso es m\'as conveniente ocupar el potencial escalar 
magn\'etico que el potencial vectorial. Por ejemplo, supongamos que tenemos un anillo de radio $R$ 
por el cual circula una corriente constante $I$.\\

Para resolver este problema pondremos al anillo en el plano $x-y$ y el centro del anillo en
el origen del sistema de referencia coordenado. Claramente el sistema es invariante bajo rotaciones en 
el eje $z,$ por lo que el potencial no depende de $\varphi,$ es decir es de la forma
\begin{eqnarray}
\phi_{m}(r,\theta )=\sum_{l\geq 0}\left( A_{l}\left(\frac{r}{R}\right)^{l}+B_{l}\left(\frac{R}{r}\right)^{l+1}\right)P_{l}(\cos\theta).
\end{eqnarray}
Adicionalmente, los puntos del disco  que encierra el anillo de corriente son de la forma
\begin{eqnarray}
\vec r^{\prime}=r^{\prime}(\cos\varphi^{\prime}, \sin\varphi^{\prime},0)
\end{eqnarray}
mientras que el vector normal es $\hat n^{\prime}=\hat k$
y  el elemento de \'area es 
$da^{\prime}=r^{\prime}dr^{\prime}d\varphi^{\prime}.$
De donde, para un punto 
$\vec r=(x,y,z)=r(\sin\theta\cos\varphi,\sin\theta\sin\varphi,\cos\theta),$
el potencial escalar magn\'etico  Eq. (\ref{eq:escalar-magnetico-anillo}) es
\begin{eqnarray}
\phi_{m}(\vec r)&=&\frac{I}{c}\int_{0}^{2\pi}d\varphi^{\prime}
\int_{0}^{R} r^{\prime}dr^{\prime} \frac{\hat k\cdot
\left((x,y,z)-r^{\prime}(\cos\varphi^{\prime}, \sin\varphi,0) \right)}
{\left(\sqrt{r^{2}+r^{\prime 2}-2rr^{\prime}\sin{\theta}
\cos(\varphi-\varphi^{\prime})}\right)^{3}},\nonumber\\
&=&\frac{I}{c}\int_{0}^{2\pi}d\varphi^{\prime}
\int_{0}^{R}dr^{\prime} \frac{zr^{\prime}}
{\left(\sqrt{r^{2}+r^{\prime 2}-2rr^{\prime}\sin{\theta}
\cos(\varphi-\varphi^{\prime})}\right)^{3}}.\nonumber
\end{eqnarray}
Si $\theta=0,$ es decir, si el punto de observaci\'on est\'a
sobre el eje $z,$ la integral se simplifica notablemente y se encuentra
\begin{eqnarray}
\phi_{m}(0,0,z=r)&=&\frac{I}{c}\int_{0}^{2\pi}d\varphi^{\prime}
\int_{0}^{R}dr^{\prime} \frac{rr^{\prime}}
{\left(\sqrt{r^{2}+r^{\prime 2}}\right)^{3}}\nonumber\\
&=&\frac{2\pi Ir}{c}\int_{0}^{R}dr^{\prime} \frac{r^{\prime}}
{\left(\sqrt{r^{2}+r^{\prime 2}}\right)^{3}}
=\frac{2\pi Ir}{c}\int_{0}^{R}dr^{\prime}\frac{d}{dr^{\prime}}\left(
\frac{-1}{\sqrt{r^{2}+r^{\prime 2}}}\right)\nonumber\\
&=&\frac{2\pi Ir}{c}\left(\frac{-1}{\sqrt{r^{2}+r^{\prime 2}}}\right)
\Big|_{0}^{R}=\frac{-2\pi Ir}{c}
\left(\frac{1}{\sqrt{r^{2}+R^{2}}}-\frac{1}{r}\right)\nonumber\\
&=&\frac{2\pi I}{c}+ \frac{-2\pi I}{c}\frac{r}{\sqrt{r^{2}+R^{2}}}.
\end{eqnarray}
Ahora, definamos $r_{<}=$menor$\{r,R\}$ y $r_{>}=$mayor$\{r,R\},$ entonces ocupando las
propiedades de los polinomios de Legendre, se encuentra
\begin{eqnarray}
\frac{r}{\sqrt{r^{2}+R^{2}}}
=\frac{r}{r_{>}} 
\frac{1}{\sqrt{1+\left(\frac{r<}{r_{>}}\right)^{2}}}
=\frac{r}{r_{>}} \sum_{l\geq 0}
\left(\frac{r_{<}}{r_{>}}\right)^{l}P_{l}(0)=\frac{r}{r_{>}} \sum_{l\geq 0}
\left(\frac{r_{<}}{r_{>}}\right)^{2l}P_{2l}(0).\nonumber
\end{eqnarray}
Por lo tanto,
\begin{eqnarray}
\phi_{m}(0,0,r)
&=&\frac{2\pi I}{c}-\sum_{l\geq 0}\frac{2\pi I}{c}\frac{r}{r_{>}}
\left(\frac{r_{<}}{r_{>}}\right)^{2l}P_{2l}(0).
\label{eq:pmag0}
\end{eqnarray}
Para el caso $r<R$ se tiene
\begin{eqnarray}
\phi_{m(int)}(0,0,r)&=&\frac{2\pi I}{c}-\sum_{l\geq 0}\frac{2\pi I}{c}\frac{r}{R}
\left(\frac{r}{R}\right)^{2l}P_{2l}(0)+\frac{2\pi I}{c}\nonumber\\
&=& \frac{2\pi I}{c}+\sum_{l\geq 0}\left(-\frac{2\pi I}{cR^{2l+1} }P_{2l}(0)\right)r^{2l+1}.\label{eq:pmag2}
\end{eqnarray}
Mientras que si $R<r$ se tiene
\begin{eqnarray}
\phi_{m(ext)}(0,0,r)&=&\frac{2\pi I}{c}
-\sum_{l\geq 0}\frac{2\pi I}{c}\left(\frac{R}{r}\right)^{2l}P_{2l}(0)\nonumber\\
&=&\frac{2\pi I}{c} -\frac{2\pi I}{c}+ \sum_{l\geq 1 }\left(-\frac{2\pi IR^{2l}}{c}P_{2l}(0)\right)
\frac{1}{r^{2l}} \nonumber\\
&=&\sum_{l\geq 1}\left(-\frac{2\pi IR^{2l}}{c}P_{2l}(0)\right)
\frac{1}{r^{2l}}.\label{eq:pmag3}
\end{eqnarray}
Ahora, seg\'un la soluci\'on general de la ecuaci\'on de Laplace
en coordenadas esf\'ericas, si $r<R$ el potencial debe ser de 
la forma
\begin{eqnarray}
\phi_{m(int)}(r,\theta)= \sum_{l\geq 0}A_{l}r^{l}P_{l}(\cos\theta)
\label{eq:pmag-1}.
\end{eqnarray}
Cuando $\theta=0$ esta \'ultima  ecuaci\'on   debe ser igual a Eq. (\ref{eq:pmag2}), por lo tanto
considerando que $P_{l}(1)=1,$ se llega a
\begin{eqnarray}
\phi_{m}(r,\theta=0)&=& \sum_{l\geq 0}A_{l}r^{l}P_{l}(1)= \sum_{l\geq 0}A_{2l}r^{2l}P_{2l}(1)+ \sum_{l\geq 0}A_{2l+1}r^{2l+1}P_{2l+1}(1)\nonumber\\
&=&\frac{2\pi I}{c}+
\sum_{l\geq 0}\left(-\frac{2\pi I}{cR^{2l+1}}P_{2l}(0)\right)r^{2l+1}.
\end{eqnarray}
As\'i, los \'unicos coeficientes  diferentes de cero son
\begin{eqnarray}
A_{0}=\frac{2\pi I}{c} ,\qquad A_{2l+1}=-\frac{2\pi I}{cR^{2l+1}}P_{2l}(0).
\end{eqnarray}
entonces, si $r<R$ se tiene el potencial
\begin{eqnarray}
\phi_{m(int)}(r,\theta)=
\frac{2\pi I}{c}-\frac{2\pi I}{c} \sum_{l\geq 0}P_{2l}(0) 
\left(\frac{r}{R}\right)^{2l+1}P_{2l+1}(\cos\theta).
\end{eqnarray}
Ahora, seg\'un la soluci\'on general de la ecuaci\'on de Laplace
en coordenadas esf\'ericas, si $R<r$ el potencial debe ser de 
la forma
\begin{eqnarray}
\phi_{m(ext)}(r,\theta)= \sum_{l\geq 0}\frac{C_{l}}{r^{l+1}}P_{l}(\cos\theta)
\label{eq:pmag-3}.
\end{eqnarray}
Cuando $\theta=0$ el potencial  Eq. (\ref{eq:pmag3}) debe ser igual a Eq. (\ref{eq:pmag-3}), 
es decir,
\begin{eqnarray}
\phi_{m(ext)}(r,\theta=0)&=& \sum_{l\geq 0}\frac{C_{l}}{r^{l+1}}P_{l}(1)
=\sum_{l\geq 0}\frac{C_{2l}}{r^{2l+1}}+ 
\sum_{l\geq 1}\frac{C_{2l-1}}{r^{2l}}\nonumber\\
&=&\sum_{l\geq 1}\left(-\frac{2\pi IR^{2l}}{c}P_{2l}(0)\right)
\frac{1}{r^{2l}}.
\end{eqnarray}
Por lo tanto,
\begin{eqnarray}
C_{2l}=0,\qquad C_{2l-1}=-\frac{2\pi IR^{2l}}{c}P_{2l}(0) .
\end{eqnarray}
Entonces, si $R<r$ se tiene el potencial
\begin{eqnarray}
\phi_{m(ext)}(r,\theta)&=&-\frac{2\pi I}{c}\sum_{l\geq 1}P_{2l}(0)
\left(\frac{R}{r}\right)^{2l}P_{2l-1}(\cos\theta)\nonumber\\
& =&-\frac{2\pi I}{c}\sum_{l\geq 0}P_{2(l+1)}(0)
\left(\frac{R}{r}\right)^{2(l+1)}P_{2l+1}(\cos\theta).
\end{eqnarray}
Adem\'as, el campo magn\'etico es 
\begin{eqnarray}
\vec B=-\vec \nabla \phi_{m}( r,\theta)=-\left( \frac{\partial \phi_{m} (r,\theta)}{\partial r}\hat e_{r} + \frac{1}{r} \frac{\partial \phi_{m} (r,\theta)}{\partial \theta}\hat e_{\theta}\right),
\end{eqnarray}
es decir
\begin{eqnarray}
B_{r}=- \frac{\partial \phi_{m} (r,\theta)}{\partial r},\qquad  B_{\theta}= -\frac{1}{r} \frac{\partial \phi_{m} (r,\theta)}{\partial \theta},\qquad B_{\varphi}=0.
\end{eqnarray}
De donde, considerando el valor de $P_{2l}(0)$ y rearreglando t\'erminos,  se llega a
\begin{eqnarray}
B_{r(int)}&=&- \frac{\partial \phi_{m(int)} (r,\theta)}{\partial r}=\frac{2\pi I}{c} \sum_{l\geq 0}P_{2l}(0)(2l+1) 
\left(\frac{r}{R}\right)^{2l+1}P_{2l+1}(\cos\theta)\nonumber\\
& =&\frac{2\pi IR}{cr} \sum_{l\geq 0}\frac{(-)^{l}(2l+1)!}{2^{2l}(l!)^{2}} \frac{r^{2l+1}}{R^{2(l+1)}} P_{2l+1}(\cos\theta),\nonumber
\end{eqnarray}
mientras que 
\begin{eqnarray}
B_{r(ext)}&=&- \frac{\partial \phi_{m(ext)} (r,\theta)}{\partial r}=\frac{2\pi I}{c}\sum_{l\geq 0}P_{2(l+1)}(0)(-)2(l+1)
\frac{R^{2l+2}}{r^{2l+3} }  P_{2l+1}(\cos\theta) \nonumber\\
&=& \frac{2\pi IR}{cr} \sum_{l\geq 0} \frac{(-)^{l+1}(2l+2)!}{2^{2l+2}(l+1)!^{2}} \frac{R^{2l+1}}{r^{2l+2} } P_{2l+1}(\cos\theta) \nonumber\\
& =& \frac{2\pi IR}{cr} \sum_{l\geq 0} \frac{(-)^{l+1}(2l+2)!(-)2(l+1)}{2^{2l+2}(l+1)!^{2}} \frac{R^{2l+1}}{r^{2(l+1)} } P_{2l+1}(\cos\theta)\nonumber\\
& =& \frac{2\pi IR}{cr} \sum_{l\geq 0} \frac{(-)^{l}(2l+1)!2(l+1)2(l+1) }{2^{2l+2}(l+1)!^{2}} \frac{R^{2l+1}}{r^{2(l+1)} } P_{2l+1}(\cos\theta)\nonumber\\
& =& \frac{2\pi IR}{cr} \sum_{l\geq 0} \frac{(-)^{l}(2l+1)! }{2^{2l}(l)!^{2}} \frac{R^{2l+1}}{r^{2(l+1)} } P_{2l+1}(\cos\theta).\nonumber
\end{eqnarray}
Note que ambas componentes se pueden escribir como
\begin{eqnarray}
B_{r} =\frac{2\pi IR}{cr} \sum_{l\geq 0} \frac{(-)^{l}(2l+1)! }{2^{2l}(l)!^{2}} \frac{r_{<}^{2l+1}}{r_{>} ^{2(l+1)} } P_{2l+1}(\cos\theta),\nonumber
\end{eqnarray}
que coincide con Eq. (\ref{eq:campo-anillo-radial}).\\

Adicionalmente, consideremos el cambio de variable $u=\cos\theta$ en la ecuaci\'on 
\begin{eqnarray}
\frac{dP_{l}(\cos\theta)}{d\theta}=-\sin\theta  \frac{dP_{l}(u)}{du}=-(1-u^{2})^{\frac{1}{2}}
\frac{dP_{l}(u)}{du}=P_{l}^{1}(u)=P_{l}^{1}(\cos\theta),
\nonumber
\end{eqnarray}
por lo que se encuentra
\begin{eqnarray}
B_{\theta(int)}&=&- \frac{1}{r} \frac{\partial \phi_{m(int)} (r,\theta)}{\partial \theta}= 
\frac{2\pi I}{cr}\sum_{l\geq 0}P_{2l}(0)
\left(\frac{R}{r}\right)^{2l+1}P^{1}_{2l+1}(\cos\theta)\nonumber\\
&=& \frac{2\pi I}{cr}\sum_{l\geq 0}\frac{(-)^{l}(2l)!}{2^{2l}(l!)^{2}} 
\left(\frac{R}{r}\right)^{2l+1}P^{1}_{2l+1}(\cos\theta)
\end{eqnarray}
y 
\begin{eqnarray}
B_{\theta(ext)}&=&- \frac{1}{r} \frac{\partial \phi_{m(ext)} (r,\theta)}{\partial \theta}= 
\frac{2\pi I}{cr}\sum_{l\geq 0}P_{2l+2}(0)
\left(\frac{R}{r}\right)^{2l+2}P^{1}_{2l+2}(\cos\theta)\nonumber\\
&=& \frac{2\pi I}{cr}\sum_{l\geq 0}\frac{(-)^{l+1}(2l+2)!}{2^{2l+2}(l+1)!^{2}} 
\left(\frac{R}{r}\right)^{2(l+1)}P^{1}_{2l+1}(\cos\theta)\nonumber\\
&=& \frac{2\pi I}{cr}\sum_{l\geq 0}\frac{(-)^{l+1}(2l+1)!2(l+1)}{2^{2l+2}(l+1)!(l+1)!} 
\left(\frac{R}{r}\right)^{2(l+1)}P^{1}_{2l+1}(\cos\theta)\nonumber\\
&=& \frac{\pi I}{cr}\sum_{l\geq 0}\frac{(-)^{l+1}(2l+1)!}{2^{2l}(l+1)!l!} 
\left(\frac{R}{r}\right)^{2(l+1)}P^{1}_{2l+1}(\cos\theta).
\end{eqnarray}
Estos resultados coinciden con los obtenidos en la secci\'on anterior, es decir con   Eqs. (\ref{eq:campo-anillo-angulo-ext})-(\ref{eq:campo-anillo-angulo-int})

\chapter{Los Polinomio de Laguerre y el \'atomo de hidr\'ogeno}

En este cap\'itulo estudiaremos el \'atomo de hidr\'ogeno. Para este problema son 
importantes los resultados de cap\'itulo 10.

\section{\'Atomo de hidr\'ogeno}

La ecuaci\'on de  Schr$\rm \ddot o$dinger para una part\'icula en un campo central
es
\begin{eqnarray}
\hat H\psi=E\psi,\qquad \hat H=\frac{1}{2m}\hat P^{2} +V(r), \quad \hat P=-i\hbar \vec \nabla ,
\end{eqnarray}
es decir
\begin{eqnarray}
\left(-\frac{\hbar^{2}}{2m}\nabla^{2}+V(r)\right)\psi=E\psi.\label{eq:Schrodinger-radial}
\end{eqnarray}
Anteriormente vimos que las soluciones de la ecuaci\'on de Schr$\rm \ddot o$dinger con sentido f\'isico forman un conjunto ortonormal.
Es decir, si 
\begin{eqnarray}
\hat H\psi_{a}=E_{a}\psi_{a}, \qquad {\rm y}\qquad \hat H\psi_{b}=E_{b}\psi_{b},
\end{eqnarray}
entonces  
\begin{eqnarray}
<\psi_{a}|\psi_{b}>=\delta_{ab},
\end{eqnarray}
esta propiedad es importante para el desarrollo que haremos posteriormente. \\

En coordenadas esf\'ericas el operador Laplaciano es Eq. (\ref{eq:laplace-armonicos})
\begin{eqnarray}
\nabla^{2}&=&\frac{1}{r^{2}} \frac{\partial }{\partial r}\left( r^{2} \frac{\partial }{\partial r}\right) -\frac{L^{2}}{r^{2}},
\end{eqnarray}
con
\begin{eqnarray}
L^{2}&=&-\left(\frac{1}{\sin\theta} \frac{\partial }{\partial \theta }\left(\sin\theta  \frac{\partial }{\partial \theta }\right)+\frac{1}{\sin^{2}\theta} \frac{\partial^{2} }{\partial \varphi^{2} }\right),
\end{eqnarray}
donde se cumple 
\begin{eqnarray}
L^{2}Y_{lm}(\theta,\phi)=l(l+1)Y_{lm}(\theta,\phi).
\end{eqnarray}
Por lo que, para resolver Eq. (\ref{eq:Schrodinger-radial}) propondremos $\psi(r,\theta,\varphi)=R(r)Y_{lm}(\theta,\phi).$ Como la funci\'on $R(r)$ solo depende de $r,$ mientras que $L^{2}$ y $Y_{lm}(\theta,\phi)$ solo dependen de los \'angulos, se encuentra
\begin{eqnarray}
\nabla^{2}\psi(r,\theta,\varphi)&=&\frac{Y_{lm}(\theta,\phi) }{r^{2}}\frac{\partial }{\partial r}\left(r^{2} \frac{\partial R(r)}{\partial r}\right) -  R(r)\frac{L^{2}Y_{lm}(\theta,\phi) }{r^{2}} \nonumber\\
&=&\frac{Y_{lm}(\theta,\phi) }{r^{2}}\frac{\partial }{\partial r}\left(r^{2} \frac{\partial R(r)}{\partial r}\right) -\frac{R(r)l(l+1)Y_{lm}(\theta,\phi) }{r^{2}} \nonumber\\
&=&Y_{lm}(\theta,\varphi)\left(\frac{1}{r^{2}}\frac{\partial }{\partial r}\left(r^{2} \frac{\partial R(r)}{\partial r}\right) -\frac{l(l+1)R(r) }{r^{2}}\right),\nonumber
\end{eqnarray}
entonces
\begin{eqnarray}
& & \hat H\psi(r,\theta,\varphi)=-\frac{\hbar^{2}}{2m}\nabla^{2} \psi(r,\theta,\varphi) +V(r) \psi(r,\theta,\varphi)\nonumber\\
&=&Y_{lm}(\theta,\varphi)\left(-\frac{ \hbar^{2}}{2m}\left[\frac{1}{r^{2}}\frac{\partial }{\partial r}\left(r^{2} \frac{\partial R(r)}{\partial r}\right) -\frac{l(l+1)R(r) }{r^{2}}\right]   +V(r)R(r)\right)\nonumber\\
&=&E\psi(r,\theta,\varphi)= Y_{lm}(\theta,\phi) E R(r).\nonumber
\end{eqnarray}
As\'i, la ecuaci\'on a resolver es 
\begin{eqnarray}
-\frac{ \hbar^{2}}{2m}\left(\frac{1}{r^{2}}\frac{\partial }{\partial r}\left(r^{2} \frac{\partial R(r)}{\partial r}\right) -\frac{l(l+1)R(r) }{r^{2}}\right)    +V(r)R(r)= E R(r), 
\end{eqnarray}
como solo hay dependencia en la variable $r$ tenemos
\begin{eqnarray}
\frac{1}{r^{2}}\frac{d }{d r}\left(r^{2} \frac{d R(r)}{d r}\right) -\frac{l(l+1)}{r^{2}}R(r) -\frac{2m}{\hbar^{2}} V(r)R(r)+\frac{2m E}{\hbar^{2}} R(r)=0.\nonumber
\end{eqnarray}
Para facilitar los c\'alculos haremos el cambio de variable
\begin{eqnarray}
\rho=\alpha r,\qquad \frac{d }{d r}=\alpha \frac{d }{\partial \rho},
\qquad \alpha=2 \sqrt{\frac{-2mE}{\hbar^{2}} },
\end{eqnarray}
de donde 
\begin{eqnarray}
\frac{1}{\rho^{2}}\frac{d }{d \rho}\left(\rho^{2} \frac{d R(\rho)}{d \rho}\right) -\frac{l(l+1) }{\rho^{2}}R(\rho) -\frac{2m}{\hbar^{2}\alpha^{2}} V\left(\rho\right)R(\rho)-\frac{1}{4} R(\rho)=0.\label{eq:radial}
\end{eqnarray}
Supongamos que tenemos un potencial que cerca del origen a lo m\'as diverge como $V(\rho)\approx \rho^{-1}$ y  en infinito tiende a cero.
Entonces cerca del origen Eq. (\ref{eq:radial})  tiene la forma asint\'otica
\begin{eqnarray}
\frac{1}{\rho^{2}}\frac{d }{d \rho}\left(\rho^{2} \frac{d R(\rho)}{d \rho}\right) -\frac{l(l+1) }{\rho^{2}}R(\rho) 
\approx 0.\label{eq:asintotica-atomo}
\end{eqnarray}
Para resolver esta ecuaci\'on proponemos $R(\rho)=A\rho^{\gamma},$ con $A$ una constante. Al sustituir esta propuesta en Eq. (\ref{eq:asintotica-atomo}) se encuentra
\begin{eqnarray}
\left(\gamma(\gamma+1)-l(l+1)\right) \rho^{\gamma}=0,\nonumber
\end{eqnarray}
lo cual implica 
\begin{eqnarray}
\left(\gamma(\gamma+1)-l(l+1)\right)&=&\gamma^{2}-l^{2}+\gamma-l=\left(\gamma+l\right) \left(\gamma-l\right)+\gamma-l\nonumber\\
&=&\left(\gamma-l\right)\left(\gamma+l+1\right)=0.
\end{eqnarray}
As\'i, los posibles valores de $\gamma$ son $l$ y $-(l+1),$ es decir 
\begin{eqnarray}
R(\rho)\approx A\rho^{l}\qquad {\rm o}\qquad R(\rho)\approx A\rho^{-(l+1)}.
\end{eqnarray}
La funci\'on de onda debe ser finita, de lo contrario no puede ser normalizada, por lo que si $r\ll 1,$ la
\'unica soluci\'on aceptable es
\begin{eqnarray}
R(\rho)\approx A\rho^{l},\qquad \rho\ll 1.
\end{eqnarray}
Si $V(\rho)$ es tal que $V(\rho)\to 0$ cuando $\rho\to \infty,$
en esta caso la ecuaci\'on asint\'otica es
\begin{eqnarray}
\frac{1}{\rho^{2}}\frac{d}{d \rho}\left(\rho^{2} \frac{dR(\rho)}{d \rho}\right) -\frac{1}{4}R(\rho)&=&
\frac{d ^{2} R(\rho) }{d \rho^{2}}+\frac{2}{\rho} \frac{d  R(\rho) }{d \rho} -\frac{1}{4}R(\rho)\nonumber\\
&\approx&
\frac{d ^{2} R(\rho) }{d \rho^{2}} -\frac{1}{4}R(\rho)\approx 0,\nonumber
\end{eqnarray}
que tiene las soluciones,
\begin{eqnarray}
R(\rho)\approx Be^{-\frac{\rho}{2}} , \qquad  R(\rho)\approx Ce^{\frac{\rho}{2}},\quad B,C={\rm constante}.\nonumber
\end{eqnarray}
Debido a que la función de onda debe ser normalizada, la \'unica soluci\'on con sentido es
\begin{eqnarray}
R(\rho)\approx Be^{-\frac{\rho}{2}} , \qquad \rho\gg 1.
\end{eqnarray}
Entonces propondremos como soluci\'on 
\begin{eqnarray}
R(\rho)=\rho^{l} e^{-\frac{\rho}{2}} u(\rho).
\label{eq:propuesta-atomo}
\end{eqnarray}
Note que $u(\rho)$ debe ser una funci\'on tal que $R(\rho \ll 1)\approx A \rho^{l},$ esto implica que $u(0)=a_{0}\not =0.$ 
Adem\'as, si $\rho \to \infty$ debe ocurrir que $ e^{-\frac{\rho}{2}} u(\rho)\to 0,$ por lo tanto en infinito la funci\'on $u(\rho)$
debe ser dominada por $ e^{-\frac{\rho}{2}}.$\\

Para el caso particular del potencial de Coulomb
\begin{eqnarray}
V(r)=-\frac{e^{2}Z}{r}=-\frac{e^{2}Z\alpha}{\rho},
\end{eqnarray}
definiendo 
\begin{eqnarray}
\lambda=\frac{2me^{2}Z}{\hbar^{2}\alpha}=\frac{e^{2}Z}{\hbar}\sqrt{ \frac{-m}{2E}}\label{eq:lambda-atomo} 
\end{eqnarray}
Eq. (\ref{eq:radial}) toma la forma
\begin{eqnarray}
\frac{d^{2} R(\rho)}{d \rho^{2}}+\frac{2}{\rho}\frac{d R(\rho)}{d \rho} +
\left( \frac{\lambda}{\rho}-\frac{l(l+1)}{\rho^{2}} -\frac{1}{4}\right) R(\rho)=0.
\label{eq:radial-atomo}
\end{eqnarray}
Antes de ocupar la propuesta de soluci\'on Eq. (\ref{eq:propuesta-atomo}) notemos que 
\begin{eqnarray}
\frac{d}{d\rho}\left(\rho^{l}e^{-\frac{\rho}{2}}\right)=l\rho^{l-1}e^{-\frac{\rho}{2}}-\frac{1}{2}\rho^{l}e^{-\frac{\rho}{2}}=
\rho^{l}e^{-\frac{\rho}{2}}\left(\frac{l}{\rho}-\frac{1}{2}\right),\nonumber
\end{eqnarray}
de donde 
\begin{eqnarray}
\frac{d^{2}}{d\rho^{2}}\left(\rho^{l}e^{-\frac{\rho}{2}}\right)&=&\frac{d}{d\rho}\left(
\rho^{l}e^{-\frac{\rho}{2}}\left(\frac{l}{\rho}-\frac{1}{2}\right)\right)\nonumber\\
&=&
\left[\frac{d}{d\rho}\left(\rho^{l}e^{-\frac{\rho}{2}}\right)\right]\left(\frac{l}{\rho}-\frac{1}{2}\right)+
\rho^{l}e^{-\frac{\rho}{2}}\left(-\frac{l}{\rho^{2}}\right)\nonumber\\
&=&\rho^{l}e^{-\frac{\rho}{2}}\left(\frac{l}{\rho}-\frac{1}{2}\right)^{2}- \rho^{l}e^{-\frac{\rho}{2}}\left(\frac{l}{\rho^{2}}\right)\nonumber\\
&=&\rho^{l}e^{-\frac{\rho}{2}}\left( \left(\frac{l}{\rho}-\frac{1}{2}\right)^{2}-\frac{l}{\rho^{2}}\right)\nonumber\\
&=&\rho^{l}e^{-\frac{\rho}{2}} \left(\frac{l^{2}-l}{\rho^{2}}-\frac{l}{\rho}+\frac{1}{4}\right).\nonumber
\end{eqnarray}
Entonces, considerando que 
$(fg)^{\prime}=f^{\prime}g+fg^{\prime}$  y   $(fg)^{\prime\prime}=f^{\prime\prime}g+2f^{\prime}g^{\prime}+fg^{\prime\prime}$
se llega a
\begin{eqnarray}
& &\frac{d R(\rho)}{d\rho}=\frac{d}{d\rho}\left( \left(\rho^{l}e^{-\frac{\rho}{2}}\right)u(\rho)\right)=\rho^{l}e^{-\frac{\rho}{2}}\frac{du(\rho)}{d\rho}+
 \frac{d}{d\rho} \left(\rho^{l}e^{-\frac{\rho}{2}}\right)u(\rho) \nonumber\\
&=&\rho^{l}e^{-\frac{\rho}{2}}\frac{du(\rho)}{d\rho}+
\rho^{l}e^{-\frac{\rho}{2}}\left(\frac{l}{\rho}-\frac{1}{2}\right)u(\rho)=\rho^{l}e^{-\frac{\rho}{2}}
\left( \frac{du(\rho)}{d\rho}+\left(\frac{l}{\rho}-\frac{1}{2}\right)u(\rho)\right)\nonumber
\end{eqnarray}
y 
\begin{eqnarray}
& & \frac{d^{2} R(\rho)}{d\rho^{2}}=\frac{d^{2}\left(\rho^{l}e^{-\frac{\rho}{2}}\right) }{d\rho^{2}} u(\rho)+
2\frac{d \left(\rho^{l}e^{-\frac{\rho}{2}}\right)}{d\rho}\frac{d u(\rho)}{d\rho}+ 
\rho^{l}e^{-\frac{\rho}{2}}\frac{d^{2} u(\rho)}{d\rho^{2}}\nonumber\\
&=&\rho^{l}e^{-\frac{\rho}{2}} \left(\frac{l^{2}-l}{\rho^{2}}-\frac{l}{\rho}+\frac{1}{4}\right) u(\rho)+ 2\rho^{l}e^{-\frac{\rho}{2}}\left(\frac{l}{\rho}-\frac{1}{2}\right)\frac{du(\rho)}{d\rho}\nonumber\\
& & + \rho^{l}e^{-\frac{\rho}{2}}\frac{d^{2} u(\rho)}{d\rho^{2}}
\nonumber\\
&=&\rho^{l}e^{-\frac{\rho}{2}}\left(\frac{d^{2} u(\rho)}{d\rho^{2}}+2\left(\frac{l}{\rho}-\frac{1}{2}\right)\frac{du(\rho)}{d\rho} + 
  \left(\frac{l^{2}-l}{\rho^{2}}-\frac{l}{\rho}+\frac{1}{4}\right) u(\rho)  \right)\nonumber.
\end{eqnarray}
Introduciendo estos resultados en Eq. (\ref{eq:radial-atomo}) se obtiene
\begin{eqnarray}
& &\rho^{l}e^{-\frac{\rho}{2}}\left(\frac{d^{2} u(\rho)}{d\rho^{2}}+2\left(\frac{l}{\rho}-\frac{1}{2}\right)\frac{du(\rho)}{d\rho} + 
  \left(\frac{l^{2}-l}{\rho^{2}}-\frac{l}{\rho}+\frac{1}{4}\right) u(\rho)  \right)\nonumber\\
& & +\rho^{l}e^{-\frac{\rho}{2}} \frac{2}{\rho}  \left( \frac{du(\rho)}{d\rho}+\left(\frac{l}{\rho}-\frac{1}{2}\right)u(\rho)\right)  \nonumber\\
& &+\rho^{l}e^{-\frac{\rho}{2}}\left( \frac{\lambda}{\rho}-\frac{l(l+1)}{\rho^{2}} -\frac{1}{4}\right) u(\rho)=0,\nonumber
\end{eqnarray}
por lo que
\begin{eqnarray}
& & \frac{d^{2} u(\rho)}{d\rho^{2}}+2\left(\frac{l+1}{\rho}-\frac{1}{2}\right)\frac{du(\rho)}{d\rho} \nonumber\\
& &+ 
  \left(\frac{l^{2}+l}{\rho^{2}}- \frac{l(l+1)}{\rho^{2}}+\frac{\lambda-(l+1)}{\rho}\right) u(\rho)=0,\nonumber
\end{eqnarray}
es decir la ecuaci\'on que debe satisfacer $u(\rho)$ es
\begin{eqnarray}
\frac{d^{2} u(\rho)}{d\rho^{2}}+\left(\frac{2(l+1)}{\rho}-1\right)\frac{du(\rho)}{d\rho} + 
  \left(\frac{\lambda-(l+1)}{\rho}\right) u(\rho)=0.\label{eq:radial-atomo-2}
\end{eqnarray}
Para resolver esta ecuaci\'on propondremos la serie
\begin{eqnarray}
u(\rho)=\sum_{n\geq 0} a_{n}\rho^{n}, \qquad u(0)=a_{0}\not =0,\nonumber
\end{eqnarray}
que debe cumplir  $u(0)=a_{0}\not =0$ y si $\rho \to \infty,$ entonces $u(\rho)e^{-\frac{\rho}{2}}\to 0.$\\

Para la propuesta de soluci\'on  se tiene 
\begin{eqnarray}
\frac{u(\rho)}{\rho}&=&\sum_{n\geq 0} a_{n}\rho^{n-1}, \qquad \frac{du(\rho)}{d\rho}=\sum_{n\geq 0} a_{n}n\rho^{n-1},\nonumber\\ 
\frac{d^{2}u(\rho)}{d^{2}\rho}&=&\sum_{n\geq 0} a_{n}n(n-1)\rho^{n-2},\nonumber
\end{eqnarray}
entonces
\begin{eqnarray}
\frac{d^{2} u(\rho)}{d\rho^{2}}+\frac{2(l+1)}{\rho}\frac{du(\rho)}{d\rho}&=&\sum_{n\geq 0} a_{n}n\left(n-1+2l+2 \right)\rho^{n-2}\nonumber\\
&=&\sum_{n\geq 1} a_{n}n\left(n+2l+1 \right)\rho^{n-2}\nonumber\\
&=& \sum_{n\geq 0} a_{n+1}(n+1)\left(n+2l+2 \right)\rho^{n-1},\nonumber \\
-\frac{du(\rho)}{d\rho}+\left(\frac{\lambda-(l+1)}{\rho}\right) u(\rho)&=&\sum_{n\geq 0} a_{n}(\lambda-(n+l+1))\rho^{n-1}. \nonumber
\end{eqnarray}
Considerando estos resultados en Eq. (\ref{eq:radial-atomo-2}) se encuentra
\begin{eqnarray}
\sum_{n\geq 0} \Big( a_{n+1}(n+1)\left(n+2l+2 \right) + a_{n}(\lambda-(n+l+1))\Big) \rho^{n-1}=0, \nonumber
\end{eqnarray}
as\'i 
\begin{eqnarray}
 a_{n+1}(n+1)\left(n+2l+2 \right) + a_{n}(\lambda-(n+l+1))=0, \nonumber
\end{eqnarray}
que implica
\begin{eqnarray}
 \frac{a_{n+1}}{a_{n}}=\frac{n+l+1-\lambda}{(n+1)\left(n+2l+2 \right)}. \label{eq:relacion-atomo}
\end{eqnarray}
Si $\rho \to \infty$ los t\'erminos que contribuyen son los m\'as grandes, es decir los que cumplen $n\gg 1,$ en este caso 
se encuentra 
\begin{eqnarray}
 \frac{a_{n+1}}{a_{n}}\approx \frac{n}{n n}=\frac{1}{n}.\nonumber 
\end{eqnarray}
Note que cuando $n$ es muy grande la serie exponencial
$$e^{\rho}=\sum_{n\geq 0}b_{n}\rho^{n},    \qquad b_{n}=\frac{1}{n!},$$
tiene un comportamiento similar 
\begin{eqnarray}
 \frac{b_{n+1}}{b_{n}}= \frac{1}{ n+1}\approx\frac{1}{n}. \nonumber
\end{eqnarray}
Por lo que si $\rho \to \infty,$ entonces $ u(\rho)\to e^{\rho}.$ Este comportamiento no es permitido
pues si $\rho \to \infty,$ entonces $ u(\rho)e^{-\frac{\rho}{2}} \to \infty.$ Para que $u(\rho)$ no tenga ese comportamiento, no dejaremos que su serie pueda tomar valores  grandes de $n$. Esto quiere decir que $u(\rho)$ no puede ser una serie, sino 
un polinomio. As\'i, despu\'es de cierto n\'umero todos los coeficientes
de la serie deben ser nulos. De la relaci\'on Eq. (\ref{eq:relacion-atomo}) es claro que esto se logra si despu\'es de cierto n\'umero $a_{n+1}=0,$ 
que se cumple solo si
\begin{eqnarray}
 n+l+1-\lambda=0. \label{eq:lambda-atomo-1}
\end{eqnarray}
Considerando la definici\'on de $\lambda$ dada en Eq. (\ref{eq:lambda-atomo}), se encuentra
\begin{eqnarray}
 E_{nl}=-\left(\frac{Zme^{2}}{2\hbar^{2}}\right) \frac{Ze^{2}}{(n+l+1)^{2}}.
\end{eqnarray}
Adem\'as, introduciendo Eq. (\ref{eq:lambda-atomo-1}) en Eq. (\ref{eq:radial-atomo-2}) se llega a
\begin{eqnarray}
\frac{d^{2} u(\rho)}{d\rho^{2}}+\left(\frac{2(l+1)}{\rho}-1\right)\frac{du(\rho)}{d\rho} + 
  \frac{n}{\rho} u(\rho)=0.\label{eq:Laguerre0}
\end{eqnarray}
Definamos $\rho=x,\beta=2l+1,u(x)=L_{n}^{\beta}(x),$ entonces Eq. (\ref{eq:Laguerre0}) toma la forma
\begin{eqnarray}
\frac{d^{2} L_{n}^{\beta}(x) }{dx^{2}}+\left(\frac{\beta+1}{x}-1\right)\frac{dL_{n}^{\beta}(x)}{dx} + 
  \frac{n}{x} L_{n}^{\beta}(x)=0,\label{eq:Laguerre}
\end{eqnarray}
esta es la llamada ecuaci\'on de Laguerre, que tambi\'en se puede escribir como
\begin{eqnarray}
x\frac{d^{2} L_{n}^{\beta}(x) }{dx^{2}}+\left(\beta+1-x\right)\frac{dL_{n}^{\beta}(x)}{dx} + 
  n L_{n}^{\beta}(x)=0.\label{eq:Laguerre-1}
\end{eqnarray}
Supongamos que tenemos una soluci\'on, $L_{n}^{\beta}(x),$ de Eq. (\ref{eq:Laguerre}) y definamos la funci\'on
\begin{eqnarray}
f_{n}^{\beta}(x)=\frac{1}{n+1}\left(x\frac{ dL_{n}^{\beta}(x) }{dx}+\left(n+\beta+1-x\right) L_{n}^{\beta}(x) \right).
\label{eq:Laguerre1}
\end{eqnarray}
Considerando Eq. (\ref{eq:Laguerre-1}) se tiene 
\begin{eqnarray}
\frac{df_{n}^{\beta}(x)}{dx}&=&\frac{1}{n+1}\left[x\frac{d^{2}L_{n}^{\beta}(x) }{dx^{2}}+\frac{ dL_{n}^{\beta}(x) }{dx}- L_{n}^{\beta}(x)+\left(n+\beta+1-x\right) \frac{ dL_{n}^{\beta}(x) }{dx} \right]\nonumber\\
&=&  \frac{1}{n+1}\left[x\frac{d^{2}L_{n}^{\beta}(x) }{dx^{2}}+\left(\beta+1-x\right) \frac{ dL_{n}^{\beta}(x) }{dx}+
(n+1)\frac{ dL_{n}^{\beta}(x) }{dx}- L_{n}^{\beta}(x) \right]\nonumber\\
&=&\frac{1}{n+1}\left[(n+1) \frac{ dL_{n}^{\beta}(x) }{dx}- (n+1)  L_{n}^{\beta}(x) \right]=\frac{ dL_{n}^{\beta}(x) }{dx}-  L_{n}^{\beta}(x),\nonumber\\
\frac{d^{2}f_{n}^{\beta}(x)}{dx^{2}}&=&\frac{ d^{2}L_{n}^{\beta}(x) }{dx^{2}}-  \frac{ dL_{n}^{\beta}(x) }{dx}.\nonumber
\end{eqnarray}
Por lo que 
\begin{eqnarray}
& &x\frac{d^{2} f_{n}^{\beta}(x) }{dx^{2}}+\left(\beta+1-x\right)\frac{df_{n}^{\beta}(x)}{dx}\nonumber\\
& & =x
\frac{ d^{2}L_{n}^{\beta}(x) }{dx^{2}}-  x\frac{ dL_{n}^{\beta}(x) }{dx}+ \left(\beta+1-x\right) 
\left(\frac{ dL_{n}^{\beta}(x) }{dx}-  L_{n}^{\beta}(x)\right)\nonumber\\
& & =x\frac{ d^{2}L_{n}^{\beta}(x) }{dx^{2}}+
\left( \beta+1-x\right)\frac{ dL_{n}^{\beta} (x)}{dx}- x\frac{ dL_{n}^{\beta}(x) }{dx}- 
\left(\beta+1-x\right)L_{n}^{\beta}(x)\nonumber\\
& &=-nL_{n}^{\beta}(x)- x\frac{ dL_{n}^{\beta}(x) }{dx}- 
\left(\beta+1-x\right)L_{n}^{\beta}(x)\nonumber\\
& &=-\left( x\frac{ dL_{n}^{\beta}(x) }{dx}+
\left(\beta+n+1-x\right)L_{n}^{\beta}(x)\right)\nonumber\\
& &=-(n+1)f_{n}^{\beta}(x),\nonumber
\end{eqnarray}
es decir 
\begin{eqnarray}
\frac{d^{2} f_{n}^{\beta}(x) }{dx^{2}}+\left(\beta+1-x\right)\frac{df_{n}^{\beta}(x)}{dx}+(n+1)f_{n}^{\beta}(x)=0.\nonumber
\end{eqnarray}
Esto nos indica que si $L_{n}^{\beta}(x)$ es soluci\'on de la ecuaci\'on de Laguerre, entonces  $f_{n}^{\beta}(x)$
tambi\'en es soluci\'on de la ecuaci\'on de Laguerre donde se ha cambiado $n$ por $n+1.$ As\'i podemos llamar 
$L_{n+1}^{\beta}(x)$ a $f_{n}^{\beta}(x),$ es decir
\begin{eqnarray}
L_{n+1}^{\beta}(x)=\frac{1}{n+1}\left(x\frac{ dL_{n}^{\beta}(x) }{dx}+\left(n+\beta+1-x\right) L_{n}^{\beta}(x) \right).
\label{eq:Laguerre-recurrencia}
\end{eqnarray}
Esta relaci\'on es muy \'util, pues nos dice que basta resolver un caso para obtener todos los dem\'as.\\

Claramente el caso m\'as sencillo de la ecuaci\'on de Laguerre es cuando $n=0,$ donde Eq. (\ref{eq:Laguerre}) toma la forma
\begin{eqnarray}
\frac{d^{2} L_{0}^{\beta}(x) }{dx^{2}}+\left(\frac{\beta+1}{x}-1\right)\frac{dL_{0}^{\beta}(x)}{dx}=0.
\end{eqnarray}
Definiendo $U(x)= \frac{d L_{0}^{\beta}(x)}{dx}$ se tiene la ecuaci\'on 
\begin{eqnarray}
\frac{d U(x) }{dx}+\left(\frac{\beta+1}{x}-1\right)U(x)=0,
\end{eqnarray}
cuya soluci\'on es 
\begin{eqnarray}
 U(x)=A\frac{e^{x}}{x^{\beta+1}}, \qquad A={\rm constante}.
\end{eqnarray}
Si $A\not =0,$ esta soluci\'on no es un polinomio, por lo que no nos ayuda a resolver el problema del \'atomo de Hidr\'ogeno. 
As\'i el \'unico caso que podemos tomar es $U(x)=0,$ que implica $L_{0}^{\beta}(x)={\rm constante},$ tomaremos 
\begin{eqnarray}
 L_{0}^{\beta}(x)=1.
\end{eqnarray}
En general los polinomios de Laguerre tienen la forma 
\begin{eqnarray}
 L_{n}^{\beta}(x)=\frac{e^{x}x^{-\beta}}{n!}\frac{d^{n}}{dx^{n}}\left(e^{-x}x^{n+\beta}\right).
 \label{eq:laguerre-rodriguez}
\end{eqnarray}
Probaremos esta afirmaci\'on por inducci\'on,
es claro que se cumple la igualdad para el caso $n=0.$ Para el paso inductivo 
supondremos Eq. (\ref{eq:laguerre-rodriguez}) y mostraremos que se cumple 
\begin{eqnarray}
\frac{e^{x}x^{-\beta}}{(n+1)!}\frac{d^{n+1}}{dx^{n+1}}\left(e^{-x}x^{(n+1)+\beta}\right)= 
L_{n+1}^{\beta}(x), \label{eq:laguerre-rodriguez-1}
\end{eqnarray}
en esta demostraci\'on ocuparemos la igualdad
\begin{eqnarray}
 \frac{d^{n}}{dx^{n}}\left(AB\right)=\sum_{k=0}^{n}C_{k}^{n} \left(\frac{d^{n-k}A}{dx^{n-k}}\right)\left(\frac{d^{k}B}{dx^{k}}\right), \qquad
 C_{k}^{n}=\frac{n!}{k!(n-k)!}.
 \label{eq:newton-derivadas}
\end{eqnarray}
Tomando en cuenta Eq. (\ref{eq:newton-derivadas}) y Eq. (\ref{eq:Laguerre-recurrencia}) se encuentra
\begin{eqnarray}
 & & \frac{ e^{x} x^{-\beta} }{(n+1)!} \frac{d^{n+1}}{dx^{n+1}} \left( e^{-x} x^{n+1+\beta} \right)=
\frac{ e^{x}x^{-\beta}} {(n+1)n!} \frac{d^{n+1}}{dx^{n+1}} \left(xe^{-x}x^{n+\beta } \right) \nonumber\\
&&=\frac{e^{x} x^{-\beta} }{(n+1)n!} \left( x \frac{d^{n+1} } {dx^{n+1}} \left(e^{-x} x^{n+\beta } \right)C_{0}^{n+1}+ 
\frac{d^{n}} {dx^{n}}\left( e^{-x}x^{n+\beta } \right) C_{1}^{n+1} \right)
\nonumber\\
& & = \frac{e^{x}x^{-\beta}}{(n+1)}x \frac{d}{dx} \left(e^{-x}x^{\beta} \frac{e^{x}x^{-\beta}}{n!} 
\frac{d^{n}}{dx^{n}}\left(e^{-x}x^{n+\beta}\right)  \right) + \nonumber\\
 & & \frac{1}{n+1} \frac{e^{x}x^{-\beta}}{n!} 
\frac{d^{n}}{dx^{n}}\left(e^{-x}x^{n+\beta}\right) \frac{(n+1)!}{n!}\nonumber\\
& &= \frac{e^{x}x^{-\beta}}{(n+1)}x \frac{d}{dx}\left(e^{-x}x^{\beta}L_{n}^{\beta}(x)\right)+ L_{n}^{\beta}(x)\nonumber\\
& &=\frac{e^{x}x^{-\beta}}{(n+1)}x \left(   (-) e^{-x}x^{\beta}L_{n}^{\beta}(x)+ \beta x^{\beta-1} e^{-x}L_{n}^{\beta}(x)+
x^{\beta} e^{-x}\frac{dL_{n}^{\beta}(x)}{dx} \right)\nonumber\\
& &+L_{n}^{\beta}(x)\nonumber\\
& &=\frac{1}{n+1}\left(x \frac{dL_{n}^{\beta}(x)}{dx}+(n+1+\beta-x)L_{n}^{\beta}(x)\right)\nonumber\\
& &=L^{\beta}_{n+1}(x),\nonumber
\end{eqnarray}
en el \'ultimo paso se us\'o  la igualdad Eq. (\ref{eq:Laguerre-recurrencia}). Por lo tanto, Eq.  
(\ref{eq:laguerre-rodriguez}) se cumple para cualquier $n.$\\

Ahora veamos la forma expl\'icita de los polinomio de Laguerre, para ello recordemos que, si $n<m,$
\begin{eqnarray}
\frac{d^{n}x^{m}}{dx^{n}}=\frac{m!}{(m-n)!}x^{m-n}.
\end{eqnarray}
Ocupando esta igualdad y Eq. (\ref{eq:newton-derivadas}) se llega a 
\begin{eqnarray}
 L_{n}^{\beta}(x)&=&\frac{e^{x}x^{-\beta}}{n!}\frac{d^{n}}{dx^{n}}\left(e^{-x}x^{n+\beta}\right)=
\frac{e^{x}x^{-\beta}}{n!}\sum_{k=0}^{n}C_{k}^{n} \left(\frac{d^{n-k}x^{n+\beta}}{dx^{n-k}}\right)\left(\frac{d^{k}e^{-x}}{dx^{k}}\right)\nonumber\\
&=&\frac{e^{x}x^{-\beta}}{n!}\sum_{k=0}^{n}C_{k}^{n}\frac{(n+\beta)!}{\left[(n+\beta)-(n-k)\right]!} (-)^{k}x^{n+\beta-(n-k)}e^{-x}\nonumber\\
&=& \sum_{k=0}^{n}\frac{(n+\beta)!}{k!(n-k)!(k+\beta)!} (-x)^{k}, 
\end{eqnarray}
es decir,
\begin{eqnarray}
 L_{n}^{\beta}(x)= \sum_{k=0}^{n}\frac{(n+\beta)!}{k!(n-k)!(k+\beta)!} (-x)^{k}.
\end{eqnarray}

\section{Funci\'on de onda}

Reuniendo los resultados de la secci\'on anterior, encontramos que la funci\'on de onda
del \'atomo de hidr\'ogeno es 
\begin{eqnarray}
\psi_{nlm}(\rho,\theta,\varphi)&=&A\rho^{l} L_{n}^{2l+1}(\rho)e^{-\frac{\rho}{2}}Y_{lm}(\theta,\varphi),\nonumber\\ 
   E_{nl}&=&-\frac{Ze^{2}}{a_{RB}(n+l+1)^{2}},\qquad a_{RB}=\frac{2\hbar^{2}}{Zme^{2}},
\end{eqnarray}
con $A$ una constante de normalizaci\'on. Definiendo  $N=n+l+1,$  la funci\'on de onda toma la  forma  
\begin{eqnarray}
\psi_{Nlm}(\rho,\theta,\varphi)&=&A\rho^{l} L_{N-(l+1)}^{2l+1}(\rho)e^{-\frac{\rho}{2}}Y_{lm}(\theta,\varphi),\quad A={\rm constante}, \nonumber\\
E_{N}&=& -\frac{Ze^{2}}{a_{RB}N^{2}},\qquad N=n+l+1.\nonumber
\end{eqnarray}
Note que el valor m\'inimo que puede tomar $l$ es cero y el m\'aximo valor que puede tomar es $N-1.$ Ahora, como por cada valor de $l$ hay 
$(2l+1)$ valores de $m$ que dan el mismo valor propio $l,$ el n\'umero de funciones propias que dan el mismo valor $N$ es 
\begin{eqnarray}
\sum_{l=0}^{N-1}(2l+1)=N^{2}.
\end{eqnarray}
Este es el grado de degeneraci\'on del \'atomo de hidr\'ogeno.

\chapter{ Ecuaci\'on de Helmholtz}

En este cap\'itulo estudiaremos la ecuaci\'on de Helmholtz y sus soluciones en coordenadas esf\'ericas.

\section{El origen de la ecuaci\'on Helmholtz}

La ecuaci\'on de Helmholtz es 
\begin{eqnarray}
\left(\nabla ^{2} +k^{2}\right)\phi(r,\theta,\varphi)=0,
\label{eq:helmholtz}
\end{eqnarray}
con $k$ una constante real. Esta ecuaci\'on surge en diferentes contextos, por ejemplo la ecuaci\'on de Schr$\rm \ddot o$dinger libre es 
\begin{eqnarray}
-\frac{\hbar^{2}}{2m} \nabla^{2}\psi(x,y,z)=E\psi(x,y,z),
\end{eqnarray}
que se puede escribir como
\begin{eqnarray}
\left( \nabla^{2}+ k^{2}\right) \psi(x,y,z)=0,\qquad k^{2}=\frac{2mE}{\hbar^{2}}.\label{eq:Schrodinger-2d-polares}
\end{eqnarray}
Adem\'as, la ecuaci\'on de onda  es 
\begin{eqnarray}
\left( \nabla^{2}-\frac{1}{c^{2} }\frac{\partial ^{2}}{\partial t^{2}} \right) \Psi(x,y,z,t)=0,
\end{eqnarray}
si proponemos soluciones de la forma $\Psi(x,y,t)=e^{i\omega t}\psi(x,y),$ se obtiene
\begin{eqnarray}
\left( \nabla^{2}+k^{2} \right) \psi(x,y,z)=0, \qquad k^{2}=\frac{\omega^{2}}{c^{2}}. \label{eq:onda-2d-polares}
\end{eqnarray}
Primero estudiaremos el caso de dos dimensiones en coordenas polares y  posteriormente estudiaremos el caso en tres dimensiones con coordenadas
esf\'ericas.

\section{Ecuaci\'on de Helmholtz en dos dimensiones} 

La ecuaci\'on de Helmholtz en dos dimensiones en coordenadas polares es
\begin{eqnarray}
\left(\nabla_{2D}^{2}+k^{2}\right)\phi(r,\theta)=\frac{1}{r}\frac{\partial }{\partial r} \left(r\frac{\partial\psi(r,\theta) }{\partial r}\right)
+\frac{1}{r^{2}}\frac{\partial ^{2}\psi(r,\theta) }{\partial \theta^{2}}   +k^{2} \psi(r,\theta)=0.
\label{eq:polares-helmholtz-2D}
\end{eqnarray}
Para resolver esta ecuaci\'on supongamos que  $\psi(r,\theta)$ es de la forma 
$$\psi(r,\theta)=R(r)\Theta(\theta),$$ 
de donde
\begin{eqnarray}
\left(\nabla_{2D}^{2}+k^{2}\right)\phi(r,\theta)=\frac{\Theta(\theta)}{r}\frac{\partial }{\partial r} \left(r\frac{\partial R(r) }{\partial r}\right)
+\frac{R(r)}{r^{2}}\frac{\partial ^{2}\Theta(\theta) }{\partial \theta^{2}}   +k^{2} R(r)\Theta(\theta)=0,\nonumber
\end{eqnarray}
por lo que 
\begin{eqnarray}
\frac{r^{2} \left(\nabla_{2D}^{2}+k^{2}\right)\phi(r,\theta)}{R(r)\Theta(\theta)}=\frac{r^{2}}{rR(r)}\frac{\partial }{\partial r} \left(r\frac{\partial R(r) }{\partial r}\right)
+\frac{1}{\Theta(\theta)}\frac{\partial ^{2}\Theta(\theta) }{\partial \theta^{2}}   +k^{2} r^{2}=0.\quad \label{eq:separacion0-helmnoltz-2D}
\end{eqnarray}
As\'i, 
\begin{eqnarray}
\frac{\partial}{\partial \theta}\left(\frac{r^{2} \left(\nabla_{2D}^{2}+k^{2}\right)\phi(r,\theta)}{R(r)\Theta(\theta)}\right)
=\frac{\partial}{\partial \theta}\left(\frac{1}{\Theta(\theta)}\frac{\partial ^{2}\Theta(\theta) }{\partial \theta^{2}}\right)=0,\nonumber
\end{eqnarray}
entonces se debe cumplir la ecuaci\'on 
\begin{eqnarray}
\frac{1}{\Theta(\theta)}\frac{\partial ^{2}\Theta(\theta) }{\partial \theta^{2}}=-\nu^{2},\qquad \nu={\rm constante},\label{eq:separacion-helmnoltz-2D}
\end{eqnarray}
es decir 
\begin{eqnarray}
\frac{\partial ^{2}\Theta(\theta) }{\partial \theta^{2}}=-\nu^{2}\Theta(\theta),
\end{eqnarray}
cuya soluci\'on es 
\begin{eqnarray}
\Theta(\theta)= a_{\nu} e^{i\nu \theta } +b_{\nu} e^{-i\nu\theta}.
\end{eqnarray}
Ahora, sustituyendo Eq. (\ref{eq:separacion-helmnoltz-2D}) en Eq. (\ref{eq:separacion0-helmnoltz-2D}) se encuentra
\begin{eqnarray}
\frac{r^{2}}{rR(r)}\frac{\partial }{\partial r} \left(r\frac{\partial R(r) }{\partial r}\right)-\nu^{2}  +k^{2} r^{2}=0,
\end{eqnarray}
que se puede escribir como
\begin{eqnarray}
\frac{1}{r}\frac{\partial }{\partial r} \left(r\frac{\partial R(r) }{\partial r}\right)+\left(k^{2}-\frac{\nu^{2}}{r^{2}}\right)R(r)=0.
\end{eqnarray}
Con el cambio de variable $\zeta=kr$ se obtiene
\begin{eqnarray}
\frac{\partial^{2} R(\zeta) }{\partial \zeta^{2}}+  \frac{1}{\zeta}\frac{\partial R(\zeta) }{\partial \zeta}+
\left(1-\frac{\nu^{2}}{\zeta^{2}}\right)R(\zeta)=0,
\end{eqnarray}
que es la ecuaci\'on de Bessel, as\'i, 
\begin{eqnarray}
R(r)=A_{\nu k} J_{\nu}(kr)+B_{\nu k}J_{-\nu}(kr).
\end{eqnarray}
Por lo tanto, las soluciones de la ecuaci\'on de Helmholtz en dos dimensiones son de la forma
\begin{eqnarray}
\psi(r,\theta)=\left(A_{\nu k} J_{\nu}(kr)+B_{\nu k}J_{-\nu}(kr)\right) \left( a_{\nu} e^{i\nu \theta } +b_{\nu} e^{-i\nu\theta}\right).
\end{eqnarray}
Los coeficientes $A_{\nu k} , B_{\nu k}, a_{\nu}, b_{\nu}$ dependen de las condiciones de borde del problema. Por ejemplo, si $\psi(r,\theta)$
debe ser finito en el $r=0,$ se tiene que $B_{\nu k}=0.$ En ese caso las soluciones toman la forma
\begin{eqnarray}
\psi(r,\theta)= J_{\nu}(kr) \left( C_{\nu k} e^{i\nu \theta } +D_{\nu k} e^{-i\nu\theta}\right).
\end{eqnarray}
Para muchos problemas es importante que $\psi(r,\theta)$ sea una funci\'on univaluada. 
As\'i, como $(r,\theta)$ y $(r,\theta+2\pi)$ representan el mismo
punto, se debe cumplir
\begin{eqnarray}
\psi(r,\theta+2\pi)= \psi(r,\theta),
\end{eqnarray}
de donde 
\begin{eqnarray}
\Theta(\theta+2\pi)=A_{\nu}e^{i\nu (\theta +2\pi)}+ B_{\nu}e^{-i\nu (\theta+2\pi) }
=\Theta(\theta)=A_{\nu}e^{i\nu \theta }+ B_{\nu}e^{-i\nu \theta },
\end{eqnarray}
que induce 
\begin{eqnarray}
e^{i2\pi\nu}=1.
\end{eqnarray}
Por lo tanto, $\nu$ debe ser un n\'umero natural $n.$ Por lo que, en este caso las soluciones son de la forma
\begin{eqnarray}
\psi_{kn}(r,\theta)= J_{n}(kr) \left( C_{n k} e^{in \theta } +D_{n k} e^{-in\theta}\right).
\end{eqnarray}
Si el sistema est\'a restringido a un disco de radio $\tilde R,$ se deben poner la condici\'on de borde
\begin{eqnarray}
\psi(\tilde R,\theta)=0,
\end{eqnarray}
que implica
\begin{eqnarray}
J_{n}(k\tilde R) =0.
\end{eqnarray}
As\'i,  $k\tilde R$ debe ser una ra\'iz de Bessel, $\lambda_{nm}= k\tilde R,$ es decir,
\begin{eqnarray}
k=\frac{\lambda_{nm}}{\tilde R} \label{eq:espectro-helmnoltz-2D}
\end{eqnarray}
y  las soluciones son de la forma
\begin{eqnarray}
\psi_{nm}(r,\theta)= J_{n}\left(\frac{\lambda_{nm} r}{\tilde R}\right) \left( C_{n m} e^{in \theta } +D_{n m} e^{-in\theta}\right).
\end{eqnarray}
La soluci\'on m\'as general es
\begin{eqnarray}
\psi(r,\theta)=\sum_{n\geq 0}\sum_{m\geq 0} J_{n}\left(\frac{\lambda_{nm} r}{\tilde R}\right) \left( C_{n m} e^{in \theta } +D_{n m} e^{-in\theta}\right).
\end{eqnarray}
Note que para la ecuaci\'on de onda en dos dimensiones Eq. (\ref{eq:onda-2d-polares}) la restricci\'on  Eq. (\ref{eq:espectro-helmnoltz-2D}) implica 
que las \'unicas  frecuencias permitidas son
\begin{eqnarray}
\omega_{nm}=\frac{c\lambda_{nm}}{\tilde R}.
\end{eqnarray}
Mientras que para la ecuaci\'on de Schr$\rm \ddot o$dinger en dos dimensiones Eq. (\ref{eq:Schrodinger-2d-polares}) la restricci\'on  Eq. (\ref{eq:espectro-helmnoltz-2D}) implica 
que las \'unicas  energ\'ias permitidas son
\begin{eqnarray}
E_{nm}=\frac{\hbar^{2}}{2m} \left(\frac{\lambda_{nm}}{\tilde R}\right)^{2}.
\end{eqnarray}

\section{Ecuaci\'on de Helmholtz en tres dimensiones} 

Ocupando coordenadas esf\'ericas la ecuaci\'on de Helmholtz Eq. (\ref{eq:helmholtz}) toma la forma
\begin{eqnarray}
& &\left(\frac{1}{r^{2}} \frac{\partial  }{\partial r}\left( r^{2} \frac{\partial \phi(r,\theta,\varphi)}{\partial r}\right) -\frac{L^{2}\phi(r,\theta,\varphi)}{r^{2}}+k^{2}\phi(r,\theta,\varphi)\right)=0 ,\nonumber \\ 
& &L^{2}=-\left(\frac{1}{\sin\theta} \frac{\partial }{\partial \theta }\left(\sin\theta  \frac{\partial }{\partial \theta }\right)+\frac{1}{\sin^{2}\theta} \frac{\partial^{2} }{\partial \varphi^{2} }\right),\nonumber
\end{eqnarray}
con $L^{2}Y_{lm}(\theta,\phi)=l(l+1)Y_{lm}(\theta,\phi).$
Por lo que, para resolver Eq. (\ref{eq:helmholtz}) propondremos la funci\'on $\phi(r,\theta,\varphi)=R(r)Y_{lm}(\theta,\phi).$ Como 
la funci\'on $R(r)$ solo depende de $r$ mientras que $L^{2}$ y $Y_{lm}(\theta,\phi)$ solo dependen de los \'angulos, la ecuaci\'on de Helmholtz toma la forma 
\begin{eqnarray}
& &\left(\nabla^{2}+k^{2}\right)\phi(r,\theta,\varphi)\nonumber\\
&=&Y_{lm}(\theta,\phi) 
\left( \frac{1}{r^{2}} \frac{\partial }{\partial r}\left(r^{2} \frac{\partial R(r)}{\partial r}\right) -  
\frac{l(l+1)R(r) }{r^{2}} +k^{2} R(r)\right)=0,\nonumber
\end{eqnarray}
es decir se debe resolver la ecuaci\'on 
\begin{eqnarray}
\left( \frac{1}{r^{2}} \frac{\partial }{\partial r}\left(r^{2} \frac{\partial R(r)}{\partial r}\right) -  
\frac{l(l+1)R(r) }{r^{2}} +k^{2} R(r)\right)=0.\label{eq:helmholtz-radial}
\end{eqnarray}
Tomaremos la propuesta $R(r)=\frac{u(r)}{\sqrt{r}},$ de donde 
\begin{eqnarray}
\frac{dR(r)}{dr}&=&\frac{d}{dr}\left(\frac{u(r)}{\sqrt{r}}\right)=
\frac{1}{\sqrt{r}} \frac{du(r)}{dr}-\frac{1}{2r \sqrt{r}}u(r)\nonumber\\
&=& \frac{1}{\sqrt{r}}\left( \frac{du(r)}{dr}-\frac{1}{2r}u(r)\right),
\nonumber \\
 r^{2}\frac{d R(r)}{dr}&=& r^{2}\frac{d}{dr}\left(\frac{u(r)}{\sqrt{r}}\right)=\frac{1}{\sqrt{r}}\left(r^{2} \frac{du(r)}{dr}-\frac{r}{2}u(r)\right), \nonumber \\
\frac{d}{dr}\left( r^{2}\frac{dR(r)}{dr}\right)&=&
\frac{d}{dr}\left( r^{2}\frac{d}{dr}\left(\frac{u(r)}{\sqrt{r}}\right)\right)\nonumber\\
&=&
\frac{1}{\sqrt{r}}\left( r^{2} \frac{d^{2}u(r)}{dr^{2}}+r\frac{du(r)}{dr} -\frac{1}{4}u(r)\right).\nonumber
\end{eqnarray}
Considerando estos resultados en Eq. (\ref{eq:helmholtz-radial}) se llega a 
\begin{eqnarray}
\frac{1}{r^{2} \sqrt{r}}\left( r^{2} \frac{d^{2}u(r)}{dr^{2}}+r\frac{du(r)}{dr} -\frac{1}{4}u(r)\right)
-\frac{1}{r^{2} \sqrt{r}}l(l+1)u(r) +\frac{1}{\sqrt{r}} k^{2}u(r)=0,\nonumber
\end{eqnarray}
que se puede escribir como
\begin{eqnarray}
 r^{2} \frac{d^{2}u(r)}{dr^{2}}+r\frac{du(r)}{dr}+\left(k^{2}r^{2} -\left(l+\frac{1}{2}\right)^{2}\right)u(r)=0\nonumber
\end{eqnarray}
con los cambios de variable $z=kr$ y $\nu=l+\frac{1}{2}$ se tiene  
\begin{eqnarray}
z^{2}\frac{d^{2} u(z)}{dz^{2}}+z\frac{du(z)}{dz}+\left(z^{2}-\nu^{2}\right)u(z)=0,\quad \nu=l+\frac{1}{2}
\end{eqnarray}
que es la ecuaci\'on de Bessel Eq. (\ref{eq:bessel-1}) de orden $\nu=l+\frac{1}{2}.$ 
Por lo que las soluciones son combinaciones lineales de las funciones
\begin{eqnarray}
 J_{l+\frac{1}{2}}(kr), \qquad  J_{-\left(l+\frac{1}{2}\right)}(kr). 
\end{eqnarray}
En lugar de $J_{-\left(l+\frac{1}{2}\right)}(kr)$ podemos tomar las funciones de Nuemman 
$N_{l+\frac{1}{2}}(kr).$ Adem\'as, usando la definici\'on de $R(r)$ y de las funciones esf\'ericas de Bessel 
Eqs. (\ref{eq:esfericas-bessel-1})- (\ref{eq:esfericas-bessel-3})
se tiene 
\begin{eqnarray}
R_{l}(r)= a_{l} j_{l}(kr) +b_{l} n_{l}(kr). 
\end{eqnarray}
Por lo tanto, las soluciones de la ecuaci\'on de Helmholtz son de la forma 
\begin{eqnarray}
\phi_{lm}(r,\theta,\varphi)= \left(a_{lm} j_{l}(kr) +b_{lm} n_{l}(kr)\right)Y_{lm}(\theta,\varphi)
\end{eqnarray}
 y la soluci\'on general es 
\begin{eqnarray}
\phi(r,\theta,\varphi)=\sum_{l\geq 0} \sum_{m=-l}^{m=l} \left(a_{lm} j_{l}(kr) +b_{lm} n_{l}(kr)\right)Y_{lm}(\theta,\varphi).
\end{eqnarray}

\section{Aplicaciones}

Una part\'icula cu\'antica est\'a encerrada en una esfera de radio $R,$ encuentre la funci\'on de onda del sistema.\\

En en este caso la funci\'on de onda es 
\begin{eqnarray}
-\frac{\hbar^{2}}{2m} \nabla^{2}\psi=E\psi, 
\end{eqnarray}
de donde 

\begin{eqnarray}
 \left(\nabla^{2}+k^{2}\right) \psi=0,\qquad k^{2}=\frac{2mE}{\hbar^{2}}, 
\end{eqnarray}
que es la ecuaci\'on de Helmholtz. Como la probabilidad de encontrar la part\'icula dentro de la esfera debe ser finita y las funciones de Neumman divergen en el origen, las soluciones aceptables son de la forma $j_{l}(kr).$ Adem\'as, la funci\'on de onda se debe anular en la frontera, por lo que  
\begin{eqnarray}
j_{l}(kR)=0, 
\end{eqnarray}
de donde  $kR=\lambda_{ln}, $ con $\lambda_{ln}$ una ra\'iz de la funci\'on esf\'erica  de Bessel de orden $l$. Considerando la definici\'on de $k,$ las energ\'ias permitidas son 
\begin{eqnarray}
E_{ln}=\frac{\hbar^{2} \lambda_{ln}^{2}}{2mR^{2}}, 
\end{eqnarray}
mientras que las funciones de onda son
\begin{eqnarray}
\psi_{lmn}(r,\theta,\varphi)=j_{l}\left(\frac{\lambda_{ln}r}{R}\right)Y_{lm}(\theta,\varphi). 
\end{eqnarray}

\section{Desarrollo en ondas parciales}

Claramente la funci\'on $\phi(\vec r)=e^{i\vec k \cdot \vec r}$ es soluci\'on a la ecuaci\'on de Helmholtz
Eq. (\ref{eq:helmholtz}). Por lo que esta funci\'on se debe poder expresar en t\'erminos de las funciones 
esf\'ericas de Bessel y los arm\'onicos esf\'ericos. Para probar esta afirmaci\'on el resultado fundamental 
es la integral
\begin{eqnarray}
\int_{-1}^{1} dz e^{i\alpha z}P_{l}(z)=\left( \frac{2\pi}{\alpha}\right)^{\frac{1}{2}} (i)^{l}J_{l+\frac{1}{2}}(\alpha).
\label{eq:integral-ondas-parciales}
\end{eqnarray}
Antes de mostrar esta igualdad es conveniente considerar las identidades 
\begin{eqnarray}
\frac{dJ_{\nu}(dz)}{z}&=&\frac{1}{2}\left( J_{\nu-1}(z)- J_{\nu+1}(z)\right),\label{eq:esfe-bessel-1}\\
\frac{J_{\nu}(z)}{z}&=&\frac{1}{2\nu }\left( J_{\nu-1}(z)+ J_{\nu+1}(z)\right) \label{eq:esfe-bessel-2}
\end{eqnarray}
las cuales se deducen de Eqs. (\ref{eq:iden-iden-1})-(\ref{eq:iden-iden-2}).
Otro resultado de utilidad es
\begin{eqnarray}
(2l+1) \left( \frac{ J_{l+\frac{1}{2}}(\alpha)   }{2\alpha} -
  \frac{\partial J_{l+\frac{1}{2}}(\alpha) }{\partial \alpha} \right)=  
\left((l+1)J_{l+\frac{3}{2}}(\alpha) -l J_{l-\frac{1}{2}}(\alpha) \right),
\label{eq:identidad-para-ondas-p}
\end{eqnarray}
que se obtiene de Eqs.  (\ref{eq:esfe-bessel-1})-(\ref{eq:esfe-bessel-2}), en efecto 
\begin{eqnarray}
& &(2l+1) \left( \frac{ J_{l+\frac{1}{2}}(\alpha)   }{2\alpha} -
  \frac{\partial J_{l+\frac{1}{2}}(\alpha) }{\partial \alpha} \right)\nonumber\\
 & &=( 2l+1)\left( \frac{1}{2\cdot 2 \left(l+\frac{1}{2}\right)}\left( J_{l-\frac{1}{2}}(\alpha)
+ J_{l+\frac{3}{2}}(\alpha)\right)-
\frac{1}{2}\left( J_{l-\frac{1}{2}}(\alpha)- J_{l+\frac{3}{2}}(\alpha) \right)  \right)\nonumber\\
& &=\frac{(2l+1)}{2}\Bigg[ \frac{1}{ \left(2l+1\right)}\left( J_{l-\frac{1}{2}}(\alpha)
+ J_{l+\frac{3}{2}}(\alpha)\right)-
\left( J_{l-\frac{1}{2}}(\alpha)- J_{l+\frac{3}{2}}(\alpha) \right)  \Bigg]\nonumber\\
& &=\frac{(2l+1)}{2(2l+1)}\left[ \left( J_{l-\frac{1}{2}}(\alpha)
+ J_{l+\frac{3}{2}}(\alpha)\right)-(2l+1)
\left( J_{l-\frac{1}{2}}(\alpha)- J_{l+\frac{3}{2}}(\alpha) \right)  \right] \nonumber\\
& &=\frac{1}{2}\left( 2(l+1)J_{l+\frac{3}{2}}(\alpha) -2l J_{l-\frac{1}{2}}(\alpha) \right)=  
\left((l+1)J_{l+\frac{3}{2}}(\alpha) -l J_{l-\frac{1}{2}}(\alpha) \right).\nonumber
\end{eqnarray}
Primero, mostremos la igualdad Eq. (\ref{eq:integral-ondas-parciales}) para $l=0,$ 
\begin{eqnarray}
\int_{-1}^{1} dz e^{i\alpha z}P_{0}(z)&=&\int_{-1}^{1} dz e^{i\alpha z}=\frac{1}{i\alpha}e^{i\alpha}\Bigg|_{-1}^{1}=
\frac{e^{i\alpha}-  e^{-i\alpha}}{i\alpha}\nonumber\\
&=& \left( \frac{2\pi}{\alpha}\right)^{\frac{1}{2}} \left( \frac{\alpha}{2\pi} \right)^{\frac{1}{2}}\frac{2}{\alpha}  
\frac{e^{i\alpha}-  e^{-i\alpha}}{2i}\nonumber\\
&=& \left( \frac{2\pi}{\alpha}\right)^{\frac{1}{2}}   
\left( \frac{2}{\pi \alpha} \right)^{\frac{1}{2}}\sin\alpha\nonumber\\
& =& \left( \frac{2\pi}{\alpha}\right)^{\frac{1}{2}} J_{\frac{1}{2}}(\alpha), 
\end{eqnarray}
es decir 
\begin{eqnarray}
\int_{-1}^{1} dz e^{i\alpha z}P_{0}(z)= \left( \frac{2\pi}{\alpha}\right)^{\frac{1}{2}} J_{\frac{1}{2}}(\alpha).
\end{eqnarray}
Para el caso $l=1$ tenemos 
\begin{eqnarray}
\int_{-1}^{1} dz e^{i\alpha z}P_{1}(z)&=& \int_{-1}^{1} dz e^{i\alpha z}z=(-i)\frac{\partial }{\partial \alpha}
\left(\int_{-1}^{1} dz e^{i\alpha z}\right) \nonumber\\
&=& (-i)\frac{\partial }{\partial \alpha} \left(
\left( \frac{2\pi}{\alpha}\right)^{\frac{1}{2}} J_{\frac{1}{2}}(\alpha)\right)\nonumber\\
& =& (-i)\left[ \left( 2\pi\right)^{\frac{1}{2}}\frac{(-)}{2\alpha^{\frac{3}{2}}} J_{\frac{1}{2}}(\alpha)+
\left( \frac{2\pi}{\alpha}\right)^{\frac{1}{2}} \frac{\partial J_{\frac{1}{2}}(\alpha) }{\partial \alpha}\right] \nonumber\\
&=& (-i)\left( \frac{2\pi}{\alpha}\right)^{\frac{1}{2}} \left[ - \frac{ J_{\frac{1}{2}}(\alpha) }{2\alpha } +
 \frac{\partial J_{\frac{1}{2}}(\alpha) }{\partial \alpha}\right],
\end{eqnarray}
considerando Eqs. (\ref{eq:esfe-bessel-1})-(\ref{eq:esfe-bessel-2}) se tiene 
\begin{eqnarray}
\int_{-1}^{1} dz e^{i\alpha z}P_{1}(z)&=& 
(-i)\left( \frac{2\pi}{\alpha}\right)^{\frac{1}{2}} \Bigg[ - \frac{1}{2}\left( J_{-\frac{1}{2}}(\alpha) + J_{\frac{3}{2}}(\alpha) \right)\nonumber\\
& & + \frac{1}{2} \left(J_{-\frac{1}{2}}(\alpha) - J_{\frac{3}{2}}(\alpha) \right)\Bigg]\nonumber\\
 &=&(i)\left( \frac{2\pi}{\alpha}\right)^{\frac{1}{2}}J_{\frac{3}{2}}(\alpha) 
\end{eqnarray}
Para probar que  Eq. (\ref{eq:integral-ondas-parciales}) es v\'alida para cualquier $l$ ocuparemos el principio de inducci\'on fuerte. Es decir, supondremos que la igualdad  Eq. (\ref{eq:integral-ondas-parciales}) es v\'alida para $l$ y tambi\'en para
todos los valores menores que $l,$ entonces probaremos que se cumple 
\begin{eqnarray}
\int_{-1}^{1} dz e^{i\alpha z}P_{l+1}(z)=\left( \frac{2\pi}{\alpha}\right)^{\frac{1}{2}} (i)^{l+1}J_{l+\frac{3}{2}}(\alpha).
\end{eqnarray}
Antes de iniciar la prueba, notemos que la identidad Eq. (\ref{eq:recurr-1-lengendre}) se puede escribir como
\begin{eqnarray}
P_{l+1}(z)&=&\frac{1}{l+1}\left[ (2l+1)zP_{l}(z)-lP_{l-1}(z)\right].
\end{eqnarray}
Entonces, recurriendo a la identidad Eq. (\ref{eq:identidad-para-ondas-p})  y tomando en cuenta que  $-(i)^{l-1}=(i)^{l+1},$ se llega a
\begin{eqnarray}
& &\int_{-1}^{1} dz e^{i\alpha z}P_{l+1}(z)=\frac{1}{l+1} \int_{-1}^{1} dz e^{i\alpha z}\left[(2l+1) P_{l}(z)-lP_{l-l}(z)\right]\nonumber\\
&=&
\frac{1}{l+1} \int_{-1}^{1} dz e^{i\alpha z}\left[(2l+1) zP_{l}(z)-lP_{l-l}(z)\right]\nonumber\\
&=&\frac{1}{l+1} \left[(2l+1) \int_{-1}^{1} dz e^{i\alpha z}zP_{l}(z)-l \int_{-1}^{1} dz e^{i\alpha z}P_{l-l}(z)\right]\nonumber\\
&=&\frac{1}{l+1}\left[(2l+1) (-i)\frac{\partial }{\partial \alpha}  \left(\int_{-1}^{1} dz e^{i\alpha z}P_{l}(z)\right) -l \left( \frac{2\pi}{\alpha}\right)^{\frac{1}{2}} (i)^{l-1}J_{l-\frac{1}{2}}(\alpha)\right]
\nonumber\\
&=& \frac{1}{l+1}\left[(2l+1) (-i)\frac{\partial }{\partial \alpha}  \left( \left( \frac{2\pi}{\alpha}\right)^{\frac{1}{2}} (i)^{l}J_{l+\frac{1}{2}}(\alpha)  \right) +l \left( \frac{2\pi}{\alpha}\right)^{\frac{1}{2}} (i)^{l+1}J_{l-\frac{1}{2}}(\alpha)\right]
\nonumber\\
&=&\frac{(i)^{l+1} }{l+1}\Bigg[(2l+1) (-)\left( \left(2\pi\right)^{\frac{1}{2}} \frac{(-)}{2\alpha^{\frac{3}{2}}} J_{l+\frac{1}{2}}(\alpha) +
 \left( \frac{2\pi}{\alpha}\right)^{\frac{1}{2}}  \frac{\partial J_{l+\frac{1}{2}}(\alpha) }{\partial \alpha} \right) \nonumber\\
& & +l \left( \frac{2\pi}{\alpha}\right)^{\frac{1}{2}} (i)^{l+1}J_{l-\frac{1}{2}}(\alpha)\Bigg]
\nonumber\\
&=& \frac{(i)^{l+1} }{l+1}\Bigg[(2l+1) \left( \left(\frac{2\pi}{\alpha} \right)^{\frac{1}{2}} \frac{ J_{l+\frac{1}{2}}(\alpha) }{2\alpha}  -
 \left( \frac{2\pi}{\alpha}\right)^{\frac{1}{2}}  \frac{\partial J_{l+\frac{1}{2}}(\alpha) }{\partial \alpha} \right) \nonumber\\
& & +l \left( \frac{2\pi}{\alpha}\right)^{\frac{1}{2}} J_{l-\frac{1}{2}}(\alpha)\Bigg]
\nonumber\\
&=& \frac{(i)^{l+1} }{l+1}  \left(\frac{2\pi}{\alpha} \right)^{\frac{1}{2}} \left[(2l+1) \left( \frac{ J_{l+\frac{1}{2}}(\alpha)   }{2\alpha} -
  \frac{\partial J_{l+\frac{1}{2}}(\alpha) }{\partial \alpha} \right) +l J_{l-\frac{1}{2}}(\alpha)\right],
 \nonumber\\
&= &\left( \frac{2\pi}{\alpha}\right)^{\frac{1}{2}} (i)^{l+1}J_{l+\frac{3}{2}}(\alpha),
\end{eqnarray}
que es lo que queriamos demostrar. Por lo tanto, la igualdad Eq. (\ref{eq:integral-ondas-parciales}) es correcta para cualquier $l$ natural. Considerando la definici\'on de las funciones esf\'ericas de Bessel, la ecuaci\'on
Eq. (\ref{eq:integral-ondas-parciales}) toma la forma 
\begin{eqnarray}
\int_{-1}^{1} dz e^{i\alpha z}P_{l}(z)=2(i)^{l}j_{l}(\alpha). \label{eq:integral-ondas-parciales-esfericas}
\end{eqnarray}
Ahora,  como los polinomios de Legendre son un conjunto de funciones ortogonales en el intervalo $(-1,1),$ 
cualquier funci\'on se puede expresar como una serie de Polinomios de Legendre. En particular la funci\'on exponencial, 
$e^{i\alpha z},$ es decir 
\begin{eqnarray}
 e^{i\alpha z}= \sum_{m\geq 0} a_{m} P_{m}(z),
\end{eqnarray}
entonces
\begin{eqnarray}
 \int_{-1}^{1} e^{i\alpha z} P_{l}(z)&=& \int_{-1}^{1} dz P_{l}(z)\left(\sum_{m\geq 0} a_{m} P_{m}(z)\right)\nonumber\\
 &= &  
\sum_{m\geq 0} a_{m}\int_{-1}^{1} dz P_{l}(z) P_{m}(z)\nonumber\\
&=& \sum_{m\geq 0} a_{m}\frac{2\delta_{lm}}{2l+1}= \frac{2 a_{l}}{2l+1}.
\end{eqnarray}
Por lo tanto, tomando en cuenta Eq. (\ref{eq:integral-ondas-parciales-esfericas}), se encuentra
\begin{eqnarray}
a_{l}=\frac{2l+1}{2}\int_{-1}^{1} e^{i\alpha z} P_{l}(z)=(i)^{l}(2l+1)j_{l}(\alpha),
\end{eqnarray}
que implica 
\begin{eqnarray}
 e^{i\alpha z}= \sum_{l\geq 0} (i)^{l}(2l+1)j_{l}(\alpha)  P_{l}(z).
\end{eqnarray}

Supongamos que $\gamma$ es el \'angulo que hay entre los vectores
\begin{eqnarray}
\vec k=k(\sin\theta^{\prime}\cos \varphi^{\prime}, \sin\theta^{\prime}\sin \varphi^{\prime}, \cos\theta^{\prime}),\qquad
\vec r=r(\sin\theta\cos \varphi, \sin\theta\sin \varphi, \cos\theta). \nonumber
\end{eqnarray}
Entonces, como $\cos \gamma\leq 1,$  se encuentra
\begin{eqnarray}
 e^{i\vec k\cdot \vec r}= e^{ikr\cos\gamma}=
\sum_{l\geq 0}(2l+1)(i)^{l} j_{l}(kr) P_{l}(\cos\gamma).
\end{eqnarray}
Ocupando el teorema de adici\'on de los arm\'onicos esf\'ericos
se llega a 
\begin{eqnarray}
 e^{i\vec k\cdot \vec r}=4\pi  
\sum_{l\geq 0}\sum_{m=-l}^{l} (i)^{l} j_{l}(kr) 
Y_{lm}^{*}\left(\theta^{\prime},\varphi^{\prime}\right)
Y_{lm}\left(\theta,\varphi\right).
\end{eqnarray}
A esta expresi\'on se le llama desarrollo en ondas parciales y tiene diferentes aplicaciones en electromagnetismo y \'optica \cite{jackson:gnus}.

\chapter{ Transformada de Fourier}

En este cap\'itulo estudiaremos la transformada integral de Fourier y  veremos algunas de sus aplicaciones

\section{Definici\'on de transforma de Fourier}

Dada una funci\'on $f$ definida en el eje real, definiremos su transformada de Fourier como
\begin{eqnarray}
\tilde f(k)=F[f(x)]=\frac{1}{\sqrt{2\pi}}\int_{-\infty}^{\infty}dx e^{-ikx}f(x).
\label{eq:transformadadefourier}
\end{eqnarray}
Antes de ver algunos ejemplos, recordemos que si $\alpha >0$ se cumple la integral 
\begin{eqnarray}
\int_{-\infty}^{\infty}dz e^{-\alpha z^{2}}=\sqrt{\frac{\pi}{\alpha}},\label{eq:TFourierG}
\end{eqnarray}
note que esta integral implica 
\begin{eqnarray}
\int_{-\infty}^{\infty}dz \sqrt{\frac{\alpha}{\pi}}e^{-\alpha z^{2}}=1,\label{eq:TFourierG1}
\end{eqnarray}
a\'un cuando $\alpha \to 0$ o si $\alpha \to \infty.$\\

\section{Ejemplos}

Veamos algunos ejemplos de transformada de Fourier.\\

\subsection{ Funci\'on Gaussiana}

Para la funci\'on Gaussiana, $e^{-\alpha x^{2}},$ tenemos 
\begin{eqnarray}
F[e^{-\alpha x^{2} }]=\frac{1}{\sqrt{2\pi}}\int_{-\infty}^{\infty}dx e^{-ikx}e^{-\alpha x^{2}}= \frac{1}{\sqrt{2\pi}}\int_{-\infty}^{\infty}dx 
e^{-\left(\alpha x^{2}+ikx\right)},
\end{eqnarray}
como
\begin{eqnarray}
\alpha x^{2}+ikx&=&\alpha\left(x^{2}+2\left(\frac{ik}{2\alpha}\right)x+\left(\frac{ik}{2\alpha}\right)^{2}- \left(\frac{ik}{2\alpha}\right)^{2}\right)
\nonumber\\
 & =& \alpha\left(x+ \left(\frac{ik}{2\alpha}\right)\right)^{2}+ \frac{k^{2}}{4\alpha},
\end{eqnarray}
entonces
\begin{eqnarray}
F[e^{-\alpha x^{2} }]=\frac{ e^{- \frac{k^{2}}{4\alpha}} }{\sqrt{2\pi}}\int_{-\infty}^{\infty}dx e^{ -\alpha\left(x+ \left(\frac{ik}{2\alpha}\right)\right)^{2} }. 
\end{eqnarray}
Por lo tanto, usando la integral (\ref{eq:TFourierG}) y el cambio de variable $u= x+ \frac{ik}{2\alpha}$  se encuentra 
\begin{eqnarray}
F[e^{-\alpha x^{2} }]=\frac{ e^{- \frac{k^{2}}{4\alpha}} }{\sqrt{2\alpha }}.\label{eq:transformada1}
\end{eqnarray}
\subsection{Funci\'on $e^{-\alpha|x|}$}

La transformada  de la funci\'on $e^{-\alpha |x|}$ es
\begin{eqnarray}
F[e^{-\alpha |x| }]&=&\frac{1}{\sqrt{2\pi}}\int_{-\infty}^{\infty}dx e^{-ikx}e^{-\alpha |x|}\nonumber\\
&=&
 \frac{1}{\sqrt{2\pi}}\left(\int_{-\infty}^{0}dx 
e^{x\left(\alpha-ik\right)}+ \int_{0}^{\infty}dx 
e^{-x\left(\alpha+ik\right)}\right),\nonumber\\
& =& \frac{1}{\sqrt{2\pi}}\left(\frac{ 
e^{x\left(\alpha-ik\right)}}{ \left(\alpha-ik\right)} \Bigg|_{-\infty}^{0}-  
\frac{ e^{-x\left(\alpha+ik\right)}}{\left(\alpha+ik\right)} \Bigg|_{0}^{\infty}\right)\nonumber\\
&=& \frac{1}{\sqrt{2\pi}}
\left(\frac{  1}{ \left(\alpha-ik\right)} +  
\frac{ 1}{\left(\alpha+ik\right)} \right),\nonumber
\end{eqnarray}
por lo tanto
\begin{eqnarray}
F[e^{-\alpha |x| }]&=&\sqrt{\frac{2}{\pi}} \frac{\alpha}{\alpha^{2}+k^{2}}. \label{eq:transformada2}
\end{eqnarray}
\section{Teorema de la convoluci\'on}

Definiremos la convoluci\'on entre dos funciones $f$ y $g$ como
\begin{eqnarray}
(f\star g)(x)=\frac{1}{\sqrt{2\pi}} \int_{-\infty}^{\infty} d\zeta f(x-\zeta)g(\zeta).
\end{eqnarray}
La transformada de Fourier de la convoluci\'on de dos funciones es
\begin{eqnarray}
F\left[(f\star g)(x)\right]&=& \frac{1}{\sqrt{2\pi}}\int_{-\infty}^{\infty}dx e^{-ikx}(f\star g)(x)\nonumber\\
&=&\frac{1}{\sqrt{2\pi}}\int_{-\infty}^{\infty}dx e^{-ikx}
  \frac{1}{\sqrt{2\pi}} \int_{-\infty}^{\infty} d\zeta f(x-\zeta)g(\zeta)\nonumber\\
 &=&  \frac{1}{\sqrt{2\pi}} \int_{-\infty}^{\infty} d\zeta e^{-ik\zeta} g(\zeta) \frac{1}{\sqrt{2\pi}}\int_{-\infty}^{\infty}dx e^{-ik(x-\zeta)}
 f(x-\zeta),\nonumber
\end{eqnarray}
con el cambio de variable $u=x-\zeta$ se tiene 
\begin{eqnarray}
F\left[(f\star g)(x)\right]&=&  \frac{1}{\sqrt{2\pi}} \int_{-\infty}^{\infty} d\zeta e^{-ik\zeta} g(\zeta) \frac{1}{\sqrt{2\pi}}\int_{-\infty}^{\infty}du e^{-iku}
 f(u)\nonumber\\
 &=& \tilde g(k) \tilde f(k).\label{eq:convolucion}
\end{eqnarray}
A este resultado se le llama teorema de la convoluci\'on.

\section{Transformada inversa de Fourier}

En esta secci\'on estudiaremos la transformada inversa de Fourier de una funci\'on $f.$ Primero veamos la funci\'on 
\begin{eqnarray}
g(k)&=&F\left[e^{-\epsilon z^{2} } \tilde f(-z)\right]= \frac{1}{\sqrt{2\pi}}\int_{-\infty}^{\infty} dz e^{-ikz} e^{-\epsilon z^{2}} \tilde f(-z),
\end{eqnarray}
usando la definici\'on de transformada de Fourier  Eq. (\ref{eq:transformadadefourier}) es claro que
\begin{eqnarray}
\tilde f(-k)=\frac{1}{\sqrt{2\pi}}\int_{-\infty}^{\infty}dt e^{ikt}f(t),\label{eq:transformadadefourier1}
\end{eqnarray}
de donde 
\begin{eqnarray}
g(k)&=& \frac{1}{\sqrt{2\pi}}\int_{-\infty}^{\infty} dz e^{-ikz} e^{-\epsilon z^{2}} \frac{1}{\sqrt{2\pi}}\int_{-\infty}^{\infty}dt e^{izt}f(t)\nonumber\\
&=& \frac{1}{2\pi}\int_{-\infty}^{\infty}dt f(t) \int_{-\infty}^{\infty} dz e^{-\left[ \epsilon z^{2} +iz(k-t)\right] }   .
\end{eqnarray}
Para  escribir esta integral de forma m\'as sugerente  notemos que 
\begin{eqnarray}
 \epsilon z^{2} +iz(k-t)&=&\epsilon\left[ z^{2}+2\left(\frac{i(k-t)}{2\epsilon}\right)+ \left(\frac{i(k-t)}{2\epsilon}\right)^{2}- \left(\frac{i(k-t)}{2\epsilon}\right)^{2}\right]\nonumber\\
 & =&\epsilon\left( z+ \frac{i(k-t)}{2\epsilon}\right)^{2}+ \frac{(k-t)^{2}}{4 \epsilon},
\end{eqnarray}
adem\'as, con el cambio de variable $ u=z +i\frac{(k-t)}{2\epsilon},$ se tiene 
\begin{eqnarray}
 \epsilon z^{2} +iz(k-t)=\epsilon u^{2}+ \frac{(k-t)^{2}}{4 \epsilon}.
\end{eqnarray}
As\'i, usando el cambio de variable propuesto y el resultado  Eq. (\ref{eq:TFourierG}), llegamos a 
\begin{eqnarray}
g(k)&=& \frac{1}{2\pi}\int_{-\infty}^{\infty}dt f(t) e^{ - \frac{(k-t)^{2}}{4 \epsilon}} \int_{-\infty}^{\infty} du e^{-\epsilon u^{2}}\nonumber\\
&=& \frac{1}{2\pi}\int_{-\infty}^{\infty}dt \sqrt{\frac{\pi}{\epsilon}} f(t) e^{ - \frac{(k-t)^{2}}{4 \epsilon}} =\int_{-\infty}^{\infty}dt    
\frac{ e^{- \frac{(k-t)^{2}}{4\epsilon} }} {\sqrt{4\pi  \epsilon}}f(t). 
\end{eqnarray}
Se puede observar que tomando  $\alpha=1/(4\epsilon)$ en  Eq. (\ref{eq:TFourierG1}) se consigue 
\begin{eqnarray}
\int_{-\infty}^{\infty}dt    
\frac{ e^{- \frac{(k-t)^{2}}{4\epsilon} }} {\sqrt{4\pi  \epsilon}}=1,
\end{eqnarray}
por lo que 
\begin{eqnarray}
g(k)-f(k)&=&  \int_{-\infty}^{\infty}dt    
\frac{ e^{- \frac{(k-t)^{2}}{4\epsilon} }} {\sqrt{4\pi  \epsilon}}f(t)- f(k) \int_{-\infty}^{\infty}dt    
\frac{ e^{- \frac{(k-t)^{2}}{4\epsilon} }} {\sqrt{4\pi  \epsilon}}\nonumber\\
&=& \int_{-\infty}^{\infty}dt   \left(f(t)- f(k)\right) 
\frac{ e^{- \frac{(k-t)^{2}}{4\epsilon} }} {\sqrt{4\pi  \epsilon}},
\end{eqnarray}
esta igualdad implica 
\begin{eqnarray}
|g(k)-f(k)|&=&  \left| \int_{-\infty}^{\infty}dt   \left(f(t)- f(k)\right) 
\frac{ e^{- \frac{(k-t)^{2}}{4\epsilon} }} {\sqrt{4\pi  \epsilon}}\right|\nonumber\\
&\leq&   \int_{-\infty}^{\infty}dt   \left|f(t)- f(k)\right| 
\frac{ e^{- \frac{(k-t)^{2}}{4\epsilon} }} {\sqrt{4\pi  \epsilon}}\nonumber.
\end{eqnarray}
Recordemos que, por el teorema del valor intermedio \cite{courant-calculo:gnus}, podemos asegurar la  existencia  de $\zeta$ tal que 
\begin{eqnarray}
\frac{f(t)-f(k)}{t-k} =f^{\prime}(\zeta),
\end{eqnarray}
con $f^{\prime}$ la primera derivada de $f.$ Por lo tanto,
\begin{eqnarray}
\left|\frac{f(t)-f(k)}{t-k}\right| \leq {\rm max} \left|f^{\prime}(\zeta)\right|,
\end{eqnarray}
es decir
\begin{eqnarray}
\left|f(t)-f(k)\right| \leq  |t-k|{\rm max}  \left|f^{\prime}(\zeta)\right|.
\end{eqnarray}
De donde
\begin{eqnarray}
|g(k)-f(k)| &\leq&   \int_{-\infty}^{\infty}dt |t-k|{\rm max}  \left|f^{\prime}(\zeta)\right|
\frac{ e^{- \frac{(k-t)^{2}}{4\epsilon} }} {\sqrt{4\pi  \epsilon}}\nonumber\\
&=& \frac{ {\rm max}  \left|f^{\prime}(\zeta)\right|}{\sqrt{4\pi  \epsilon}}
\int_{-\infty}^{\infty}dt |t-k|
 e^{- \frac{(k-t)^{2}}{4\epsilon} }\nonumber.
\end{eqnarray}
Usando el cambio de variable $w=\frac{t-k}{\sqrt{4\epsilon}}$ se tiene 
\begin{eqnarray}
\frac{ 1}{\sqrt{4\pi  \epsilon}}
\int_{-\infty}^{\infty}dt |t-k|
 e^{- \frac{(k-t)^{2}}{4\epsilon} }= \frac{ 4\epsilon}{\sqrt{4\pi  \epsilon}}\int_{-\infty}^{\infty}dw |w| e^{- w^{2}}= 2\sqrt{\frac{4\epsilon}{\pi} }
\int_{0}^{\infty}dw w e^{- w^{2}},\nonumber
\end{eqnarray}
as\'i
\begin{eqnarray}
|g(k)-f(k)| &\leq&   {\rm max}  \left|f^{\prime}(\zeta)\right|2\sqrt{\frac{4\epsilon}{\pi} }
\int_{0}^{\infty}dw w e^{- w^{2}}.
\end{eqnarray}
Este resultado induce 
\begin{eqnarray}
\lim_{\epsilon \to 0} |g(k)-f(k)| =0.
\end{eqnarray}
Por lo tanto,
\begin{eqnarray}
f(x)=g(x)=\lim_{\epsilon \to 0} F\left[e^{-\epsilon k^{2} } \tilde f(-k)\right]= F\left[\tilde f(-k)\right]=
\frac{1}{\sqrt{2\pi}}\int_{-\infty}^{\infty}dk e^{-ikx}\tilde f(-k).\nonumber
\end{eqnarray}
usando el cambio de variable $k^{\prime}=-k$ se encuentra
\begin{eqnarray}
f(x)=\frac{1}{\sqrt{2\pi}}\int_{-\infty}^{\infty}dk e^{ikx}\tilde f(k)= \frac{1}{\sqrt{2\pi}}\int_{-\infty}^{\infty}dk e^{ikx} F\left[f(x)\right].\label{eq:transformadainversa}
\end{eqnarray}
Entonces definiremos la transformada inversa de Fourier como
\begin{eqnarray}
F^{-1}\left [\tilde f(k)\right]= \frac{1}{\sqrt{2\pi}}\int_{-\infty}^{\infty}dk e^{ikx}\tilde f(k),
\end{eqnarray}
esta definici\'on es consistente pues
\begin{eqnarray}
f(x)=F^{-1}\left [\tilde f(k)\right]= F^{-1}\left [F\left [f(x) \right] \right].
\end{eqnarray}

\section{Ejemplos de la transformada inversa de Fourier}

Se puede observar que ocupando la integral Eq. (\ref{eq:transformadainversa}) y tomando en cuenta los resultados  
Eqs. (\ref{eq:transformada1})-(\ref{eq:transformada2}) y Eq. (\ref{eq:convolucion}),  se obtiene 
\begin{eqnarray}
e^{-\alpha|x|}&=&F^{-1}\left[ \sqrt{\frac{2}{\pi}} \frac{\alpha}{k^{2}+\alpha^{2}}\right]= \int_{-\infty}^{\infty} dk  e^{ikx}   \frac{1}{\pi} \frac{\alpha}{k^{2}+\alpha^{2}}, \\
e^{-\alpha x ^{2}}&=&F^{-1}\left[\frac{ e^{- \frac{k^{2}}{4\alpha}} }{\sqrt{2\alpha }}      \right]= \int_{-\infty}^{\infty} dk e^{ikx}  \frac{ e^{- \frac{k^{2}}{4\alpha}} }{\sqrt{4\pi \alpha }},  \\
(f\star g)(x)&=&\frac{1}{\sqrt{2\pi}} \int_{-\infty}^{\infty} d\zeta f(x-\zeta)g(\zeta)= F^{-1}\left[ \tilde f(k)  \tilde g(k)\right]\nonumber \\
&=&  \frac{1}{\sqrt{2\pi} } \int_{-\infty}^{\infty} dk  e^{ikx}  \tilde f(k)  \tilde g(k).
\end{eqnarray}
Tambi\'en note que 
\begin{eqnarray}
F^{-1}\left[ e^{-y|k|} \right]&=&\frac{1}{\sqrt{2\pi}} \int_{-\infty}^{\infty} dk  e^{ikx}  e^{-y|k|}.   
\end{eqnarray}
As\'i, con  los cambios de variable $k=-x^{\prime},x=k^{\prime}$ y usando  Eq.  (\ref{eq:transformada2}) se encuentra 
\begin{eqnarray}
F^{-1}\left[ e^{-y|k|} \right]&=&\frac{1}{\sqrt{2\pi}} \int_{-\infty}^{\infty} dx^{\prime}  e^{-ik^{\prime} x^{\prime}}  e^{-y|x^{\prime}|}=
\sqrt{\frac{2}{\pi}} \frac{y}{y^{2}+k^{\prime 2}},
\end{eqnarray}
es decir
\begin{eqnarray}
F^{-1}\left[ e^{-y|k|} \right]=\sqrt{\frac{2}{\pi}} \frac{y}{y^{2}+x^{ 2}}.\label{eq:Fordinaria}
\end{eqnarray}
Posteriormente ocuparemos estos resultados para resolver ecuaciones diferenciales.

\section{Transfomada de Fourier de la derivada}

Para funciones tales que $f(\pm \infty)=0,$ se tiene 
\begin{eqnarray}
F\left[\frac{d f(x)}{dx}\right ]&=& \frac{1}{\sqrt{2\pi}}\int_{-\infty}^{\infty}dx e^{-ikx}\frac{d f(x)}{dx}\nonumber\\
&=& \frac{1}{\sqrt{2\pi}} \int_{-\infty}^{\infty}dx \left( \frac{d} {dx}\left[e^{-ikx}f(x)\right]+ik e^{-ikx}f(x)\right)\nonumber\\
& =& ik\frac{1}{\sqrt{2\pi}}\int_{-\infty}^{\infty}dxe^{-ikx}f(x)\nonumber\\
&=& ik F[ f(x)]=ik \tilde f(k). 
\end{eqnarray}
Por lo tanto, 
\begin{eqnarray}
F\left[\frac{d^{2} f(x)}{dx^{2}}\right ]=
&=& ik F\left[ \frac{ df(x)}{dx}\right]=  (ik)^{2} F[f(x)] =(ik)^{2}  \tilde f(k). 
\end{eqnarray}
En general se cumple
\begin{eqnarray}
F\left[\frac{d^{n} f(x)}{dx^{n}}\right ]=(ik)^{n}  \tilde f(k). 
\end{eqnarray}
Este resultado nos permitir\'a resolver algunas ecuaciones diferenciales empleando 
la transformada de Fourier.\\

\section{Ecuaci\'on de calor y ecuaci\'on de Schr\"odinger libre}

Se quiere resolver la ecuaci\'on de calor 
\begin{eqnarray}
\frac{\partial u(x,t)}{\partial t} =\sigma \frac{\partial^{2} u(x,t)}{\partial x^{2}},\qquad \sigma={\rm constante},
\label{eq:TFcalor}
\end{eqnarray}
con la condici\'on inicial $u(x,0)=f(x).$ \\

De la condici\'on inicial tenemos  $F[u(x,0)]=F[f(x)],$ es decir $\tilde u(k,0)=\tilde f(k).$ Ahora, al hacer la transformada de Fourier en la variable $x$ a la ecuaci\'on de calor Eq. (\ref{eq:TFcalor}) tenemos  
\begin{eqnarray}
F\left[\frac{\partial u(x,t)}{\partial t}\right] =\sigma F\left[\frac{\partial^{2} u(x,t)}{\partial x^{2}}\right],
\end{eqnarray}
que es equivalente a 
\begin{eqnarray}
\frac{\partial \tilde u(k,t)}{\partial t} =-\sigma k^{2} \tilde u(k,t).
\end{eqnarray}
La soluci\'on de esta ecuaci\'on es 
\begin{eqnarray}
\tilde  u(k,t)=A(k)e^{-\sigma k^{2} t},
\end{eqnarray}
considerando la condici\'on inicial tenemos que $A(k)=\tilde f(k).$ As\'i,  
\begin{eqnarray}
\tilde  u(k,t)=\tilde f(k)e^{-\sigma k^{2} t},
\end{eqnarray}
por lo que, usando la transformada inversa de Fourier y la transformada de Fourier de la funci\'on Gaussiana  Eq. (\ref{eq:transformada1}), con 
$\alpha=1/(4\sigma t),$ se obtiene  
\begin{eqnarray}
u(x,t)&=&F^{-1}\left[\tilde  u(k,t)\right]= F^{-1}\left[\tilde f(k)e^{-\sigma k^{2} t}\right]
= \frac{1}{\sqrt{2\sigma t}}  F^{-1}\left[\tilde f(k)e^{-\sigma k^{2} t} \sqrt{2\sigma t}\right]\nonumber \\
&=& \frac{1}{\sqrt{2\sigma t}} \frac{1}{\sqrt{2\pi}} \int_{-\infty}^{\infty} d\zeta f(\zeta) e^{-\frac{(x-\zeta)^{2}}{4\sigma t}},
\nonumber
\end{eqnarray}
es decir
\begin{eqnarray}
u(x,t)= \int_{-\infty}^{\infty} d\zeta f(\zeta) \frac{ e^{-\frac{(x-\zeta)^{2}}{4\sigma t}}}{\sqrt{4\pi \sigma t}}.
\end{eqnarray}
Note que, por la condici\'on inicial, se tiene  
\begin{eqnarray}
f(x)=\lim_{t\to 0}u(x,t)= \int_{-\infty}^{\infty} d\zeta f(\zeta) \lim_{t\to 0} 
\left(\frac{ e^{-\frac{(x-\zeta)^{2}}{4\sigma t}}}{\sqrt{4\pi \sigma t}}\right),
\end{eqnarray}
por lo tanto
\begin{eqnarray}
\delta(x-\zeta)= \lim_{t\to 0} \frac{ e^{-\frac{(x-\zeta)^{2}}{4\sigma t}}}{\sqrt{4\pi \sigma t}}.
\end{eqnarray}

\subsection{Ecuaci\'on de Schr\"odinger libre}

La ecuaci\'on de Schr\"odinger libre es
\begin{eqnarray}
i\hbar \frac{\partial \psi(x,t)}{\partial t} =\frac{-\hbar^{2}}{2m}  \frac{\partial^{2} \psi(x,t)}{\partial x^{2}},
\label{eq:free-sh}
\end{eqnarray}
que se puede escribir como la ecuaci\'on de calor Eq. (\ref{eq:TFcalor})
\begin{eqnarray}
 \frac{\partial \psi(x,t)}{\partial t} =\sigma \frac{\partial^{2} \psi(x,t)}{\partial x^{2}},\quad \sigma=\frac{i\hbar}{2m}.
\end{eqnarray}
Por lo tanto, si se impone la condici\'on inicial  $\psi(x,0)=\phi(x),$ la funci\'on de onda para cualquier tiempo mayor que 
cero es
\begin{eqnarray}
\psi(x,t)= \int_{-\infty}^{\infty} d\zeta G\left(x-\zeta,t\right)    \phi(\zeta) ,
\end{eqnarray}
con
$$
G\left(x-\zeta,t\right)=\sqrt{\frac{m}{2\pi \hbar t}}e^{\frac{im(x-\zeta)^{2}}{2\hbar t}}.
$$
Debido a que esta \'ultima funci\'on nos da informaci\'on de como la part\'icula pasa del estado inicial $\psi(x,0)$ al estado 
$\psi(x,t)$ se le llama propagador.

\section{ Ecuaci\'on ordinaria de segundo orden}

La transformada de Fourier puede ser \'util para resolver ecuaciones diferenciales ordinarias de segundo orden,
por ejemplo
\begin{eqnarray}
-\frac{d ^{2} u(x)}{dx^{2}}+a^{2} u(x)=f(x).
\end{eqnarray}
Para esta ecuaci\'on tenemos 
\begin{eqnarray}
-F\left[\frac{d ^{2} u(x)}{dx^{2}}\right]+a^{2} F\left[u(x)\right]=F\left[f(x)\right],
\end{eqnarray}
es decir
\begin{eqnarray}
k^{2} \tilde u(k)+a^{2} \tilde u(k)=\tilde f(k),
\end{eqnarray}
de donde 
\begin{eqnarray}
\tilde u(k)=\frac{\tilde f(k)}{k^{2}+a^{2}}.
\end{eqnarray}
Por lo tanto, 
\begin{eqnarray}
u(x)=F^{-1}\left[\tilde u(k)\right]=F^{-1}\left[\frac{\tilde f(k)}{k^{2}+a^{2}}\right]=\frac{1}{a}\sqrt{\frac{\pi}{2}} 
F^{-1}\left[ \tilde f(k) \sqrt{\frac{2}{\pi}}\frac{a}{k^{2}+a^{2}}\right],\nonumber
\end{eqnarray}
usando Eq. (\ref{eq:Fordinaria}) y el teorema de la convoluci\'on, se llega a
\begin{eqnarray}
u(x)=\frac{1}{a}\sqrt{\frac{\pi}{2}}\frac{1}{\sqrt{2\pi}} \int_{-\infty}^{\infty} d\zeta f(\zeta) e^{-a|x-\zeta|},
\end{eqnarray}
es decir 
\begin{eqnarray}
u(x)= \int_{-\infty}^{\infty} d\zeta f(\zeta) \frac{e^{-a|x-\zeta|}}{2a}.
\end{eqnarray}

\subsection{Ecuaci\'on de Laplace en dos dimensiones}

Ahora resolveremos la ecuaci\'on de Laplace en dos dimensiones 
\begin{eqnarray}
\left(\frac{\partial ^{2}}{\partial x^{2}}+ \frac{\partial ^{2}}{\partial y^{2}}\right)u(x,y)=0
\end{eqnarray}
con las condiciones de borde $u(x,0)=f(x)$ y $\lim_{y \to \infty}u(x,y)=0.$ \\

Usando la transformada de Fourier en 
la variable $x$ se consigue 
\begin{eqnarray}
F\left[\left(\frac{\partial ^{2}}{\partial x^{2}}+ \frac{\partial ^{2}}{\partial y^{2}}\right)u(x,y)\right]=
-k^{2}\tilde u(k,y)+ \frac{\partial ^{2} \tilde u(k,y)}{\partial y^{2}}=0.\label{eq:Fpoisson}
\end{eqnarray}
Adem\'as  las condiciones de borde toman la forma 
$$F\left[u(x,0)\right]=F\left[f(x)\right],\quad  \lim_{y \to \infty}F\left[u(x,y)\right]=0,$$ es decir 
\begin{eqnarray}
\tilde u(k,0)=\tilde f(k)\qquad  {\rm y} \qquad \lim_{y \to \infty}\tilde u(k,y)=0.
\end{eqnarray}
La soluci\'on de Eq. (\ref{eq:Fpoisson}) es 
\begin{eqnarray}
\tilde u(k,y)=A(k)e^{\pm \sqrt{k^{2}} y},
\end{eqnarray}
tomando en cuenta las condiciones de borde se obtiene 
\begin{eqnarray}
\tilde u(k,y)=\tilde f(k)e^{-|k| y}.
\end{eqnarray}
Por lo tanto, usando Eq. (\ref{eq:Fordinaria}), se llega a 
\begin{eqnarray}
u(x,y)&=&F^{-1}\left[\tilde u(k,y)\right]=F^{-1}\left[\tilde f(k)e^{-|k| y}\right]\nonumber\\
&=&\frac{1}{\sqrt{2\pi}}\int_{-\infty}^{\infty} d\zeta f(\zeta) \sqrt{\frac{2}{\pi}}\frac{ y}{y^{2}+(x-\zeta)^{2}}, 
\end{eqnarray}
entonces 
\begin{eqnarray}
u(x,y)=\int_{-\infty}^{\infty} d\zeta f(\zeta) \frac{1}{\pi}\frac{ y}{y^{2}+(x-\zeta)^{2}}.
\end{eqnarray}
Note que de la condici\'on de borde $u(x,0)=f(x)$ se infiere que 
\begin{eqnarray}
\delta(x-\zeta)=\lim_{y\to 0}\frac{1}{\pi}\frac{ y}{y^{2}+(x-\zeta)^{2}}.
\end{eqnarray}

\section{Ecuaci\'on de Black-Scholes}

Uno de los problemas m\'as  interesantes en Finanzas  es determinar el precio de un contrato el cual  se desea ejercer a un tiempo $T$. Si $\sigma$ es la volatilidad, 
$r$ es  la tasa de inter\'es y $S$ el precio de una acci\'on, el precio del contrato $C(S,t)$  est\'a determinado por la ecuaci\'on de Black-Scholes \cite{black:gnus,merton:gnus}
\begin{eqnarray}
\frac{\partial C(S,t)}{\partial t}=-\frac{\sigma^{2}}{2}S^{2} \frac{\partial^{2} C(S,t)}{\partial S^{2}}-rS \frac{\partial C(S,t)}{\partial S}+ rC(S,t), 
\end{eqnarray}
la cual se debe  resolver con la condici\'on  
\begin{eqnarray}
C(S,T)=(S-K)\theta(S-K). \label{eq:bs-inicial}
\end{eqnarray}
Note que esta condici\'on nos indica que el m\'inimo valor que puede tomar  $S$ es $K,$ de lo contrario no vale la pena ejercer el contrato. M\'as detalles sobre esta ecuaci\'on
se puede ver en \cite{venegas:gnus}.\\

La ecuaci\'on de Black-Scholes es equivalente a la ecuaci\'on de Schr\"odinger libre en una dimensi\'on (\ref{eq:free-sh}). Para probar esta afirmaci\'on primero notemos que 
usando el cambio de variable 
\begin{eqnarray}
S=e^{x}
\end{eqnarray}
se obtiene 
\begin{eqnarray}
\frac{\partial C(x,t)}{\partial t}=-\frac{\sigma^{2}}{2} \frac{\partial^{2} C(x,t)}{\partial x^{2}}+\left(\frac{\sigma^{2}}{2}-r\right) \frac{\partial C(x,t)}{\partial x}+ rC(x,t). 
\end{eqnarray}
Entonces, con la propuesta 
\begin{eqnarray}
C(x,t)=e^{\left[\frac{1}{\sigma^{2}}\left(\frac{\sigma^{2}}{2}-r\right)x+\frac{1}{2\sigma^{2}}\left(\frac{\sigma^{2}}{2}+r\right)^{2}t \right] }\psi(x,t),
\label{eq:precio1}
\end{eqnarray}
se llega a  la ecuaci\'on de onda
\begin{eqnarray}
\frac{\partial \psi(x,t)}{\partial t}=-\frac{\sigma^{2}}{2} \frac{\partial^{2} \psi(x,t)}{\partial x^{2}},
\label{eq:bss}
\end{eqnarray}
la cual tiene  la misma forma que la ecuaci\'on de Schr\"odinger libre en una dimensi\'on (\ref{eq:free-sh}). Esta equivalencia ha dado origen una nueva disciplina, las llamadas finanzas cu\'anticas \cite{baaquie:gnus}.\\

Como el contrato se desea ejercer al tiempo $T,$ para resolver la ecuaci\'on (\ref{eq:bss})  realizaremos  el cambio de variable 
\begin{eqnarray}
\tau=T-t,
\end{eqnarray}
de donde la ecuaci\'on (\ref{eq:bss}) toma la forma
\begin{eqnarray}
\frac{\partial \psi(x,\tau)}{\partial \tau }=\frac{\sigma^{2}}{2} \frac{\partial^{2} \psi(x,\tau)}{\partial x^{2}}. \label{eq:precio3}
\end{eqnarray}
Adem\'as, con la variable $\tau$ el precio (\ref{eq:precio1}) toma la forma 
\begin{eqnarray}
C(x,\tau)=e^{\left[\frac{1}{\sigma^{2}}\left(\frac{\sigma^{2}}{2}-r\right)x+\frac{1}{2\sigma^{2}}\left(\frac{\sigma^{2}}{2}+r\right)^{2}\left(T-\tau\right) \right] }\psi(x,\tau).
\label{eq:precio2}
\end{eqnarray}
Ahora, como el m\'inimo valor que puede tomar $S$ es $K,$ entonces el m\'inimo valor que puede tomar $x$ es $\ln K.$ Por lo tanto,  la condici\'on inicial (\ref{eq:bs-inicial}) toma la forma
\begin{eqnarray}
C(x,0)=\left(e^{x}-K\right) \theta\left(x-\ln K\right).
\end{eqnarray}
Usando esta condici\'on inicial y  la ecuaci\'on (\ref{eq:precio2}), se obtiene la condici\'on inicial para $\psi(x,\tau)$:
\begin{eqnarray}
\psi(x,0)=\psi_{0}(x)= e^{-\left[\frac{1}{\sigma^{2}}\left(\frac{\sigma^{2}}{2}-r\right)x+\frac{1}{2\sigma^{2}}\left(\frac{\sigma^{2}}{2}+r\right)^{2}T \right] }\left(e^{x}-K\right) \theta\left(x-\ln K\right).\label{eq:precio4}
\end{eqnarray}
Por lo tanto, el problema de obtener la soluci\'on de la ecuaci\'on de Black-Scholes se reduce a resolver la ecuaci\'on diferencial (\ref{eq:precio3}) con la condici\'on inicial (\ref{eq:precio4}).\\

De la ecuaci\'on de calor (\ref{eq:TFcalor}) se puede observar que, haciendo el cambio $\sigma \to \sigma^{2}/2,$ se obtienen la soluci\'on   
\begin{eqnarray}
\psi(x,\tau)= \int_{-\infty}^{\infty} dx^{\prime} \psi_{0}(x^{\prime})  \frac{ e^{-\frac{(x-x^{\prime})^{2}}{2\sigma^{2} \tau}}}{\sqrt{2\pi \sigma^{2} \tau}},
\end{eqnarray}
es decir 
\begin{eqnarray}
\psi(x,\tau)=  \frac{ e^{-\frac{ \left( \frac{\sigma^{2} }{2}+r\right)^{2}T}{2\sigma^{2}} }}  {\sqrt{2\pi \sigma^{2} \tau}} 
\int_{\ln K}^{\infty} dx^{\prime} \left( e^{\left(\frac{1}{2}+\frac{r}{\sigma^{2}}\right)x^{\prime} }-  Ke^{\left(-\frac{1}{2}+\frac{r}{\sigma^{2}}\right)x^{\prime}}   \right)  e^{-\frac{(x-x^{\prime})^{2}}{2\sigma^{2} \tau}}.\label{eq:semiprecio1}
\end{eqnarray}
Antes de obtener $\psi(x,\tau),$ estudiemos  la integral 
\begin{eqnarray}
I_{\pm}=\int_{\ln K}^{\infty} dx^{\prime}  e^{\left(\pm \frac{1}{2}+\frac{r}{\sigma^{2}}\right)x^{\prime} }  e^{-\frac{(x-x^{\prime})^{2}}{2\sigma^{2} \tau}},
\end{eqnarray}
la cual, completando cuadrados,   se puede  escribir como 
\begin{eqnarray}
I_{\pm}=  e^{ \left[\left( \frac{r}{\sigma^{2}} \pm \frac{1}{2}\right) x + \frac{\sigma^{2} \tau}{2} \left(\frac{r}{\sigma^{2}} \pm \frac{1}{2}\right)^{2}\right] } \int_{\ln K}^{\infty} dx^{\prime} 
  e^{-\frac{ \left( (x-x^{\prime})+ \sigma^{2} \tau  \left( \frac{r}{\sigma^{2}} \pm \frac{1}{2} \right) \right)^{2} }{2\sigma^{2} \tau}}. \label{eq:semiprecio2}
\end{eqnarray}
Adem\'as con el cambio de variable 
\begin{eqnarray}
u=\frac{  (x-x^{\prime})+ \sigma^{2} \tau  \left( \frac{r}{\sigma^{2}} \pm \frac{1}{2} \right)  }{\sqrt{\sigma^{2} \tau}}
\end{eqnarray}
se obtiene
\begin{eqnarray}
I_{\pm}=   \sqrt{2\pi \sigma^{2} \tau }e^{ \left[\left( \frac{r}{\sigma^{2}} \pm \frac{1}{2}\right) x + \frac{\sigma^{2} \tau}{2} \left(\frac{r}{\sigma^{2}} \pm \frac{1}{2}\right)^{2}\right] }
 N\left( d_{\pm}\right)
\end{eqnarray}
con
\begin{eqnarray}
 N(z)&=&\int_{-\infty}^{z} du \frac{e^{-\frac{u^{2}}{2}}}{\sqrt{2 \pi}},\\
d_{\pm}&=&\frac{ x-\ln K+ \sigma^{2}\tau \left( \frac{r}{\sigma^{2}}\pm \frac{1}{2}\right) }{\sigma\sqrt{\tau} }= \frac{\ln \left(\frac{S}{K}\right)+ \left(T-t\right) \left( r \pm \frac{\sigma^{2} }{2}\right) }{\sigma\sqrt{T-t} }.\nonumber
\end{eqnarray}
Usando  la integral $I_{\pm}$ en  (\ref{eq:semiprecio1})  se encuentra 
\begin{eqnarray}
\psi(x,\tau)&=&   e^{-\frac{ \left( \frac{\sigma^{2} }{2}+r\right)^{2}T}{2\sigma^{2}} }e^{ \left[\left( \frac{r}{\sigma^{2}} + \frac{1}{2}\right) x + \frac{\sigma^{2} \tau}{2} \left(\frac{r}{\sigma^{2}} + \frac{1}{2}\right)^{2}\right] } N\left( d_{+}\right)\nonumber \\
& &-K e^{-\frac{ \left( \frac{\sigma^{2} }{2}+r\right)^{2}T}{2\sigma^{2}} }e^{ \left[\left( \frac{r}{\sigma^{2}}
 - \frac{1}{2}\right) x + \frac{\sigma^{2} \tau}{2} \left(\frac{r}{\sigma^{2}} - \frac{1}{2}\right)^{2}\right] } N\left( d_{-}\right).\nonumber
\end{eqnarray}
Sustituyendo este resultado en (\ref{eq:precio2}) se llega a 
\begin{eqnarray}
C(s,t)= SN(d_{+})-Ke^{-r(T-t)} N(d_{-}).
\end{eqnarray}
Este resultado es la llamada f\'ormula de Black-Scholes, la cual es una de las expresiones matem\'aticas m\'as usadas.

\section{Delta de Dirac}

Se puede observar que al sustituir la definici\'on de la transformada de Fourier en Eq. (\ref{eq:transformadainversa}) se obtiene 
\begin{eqnarray}
f(x)&=&\frac{1}{\sqrt{2\pi}}\int_{-\infty}^{\infty}dk e^{ikx}\frac{1}{\sqrt{2\pi}}\int_{-\infty}^{\infty}dx^{\prime} e^{-ikx^{\prime}} f(x^{\prime}) \nonumber\\
&=&\int_{-\infty}^{\infty}dx^{\prime}  f(x^{\prime})\int_{-\infty}^{\infty}dk \frac{e^{ik(x-x^{\prime})}}{2\pi}.
\end{eqnarray}
Por lo tanto, en este caso la delta de Dirac la definiremos como
\begin{eqnarray}
\delta(x)=\frac{1}{2\pi} \int_{-\infty}^{\infty}dk e^{ikx},\label{eq:defdeltadirac}
\end{eqnarray}
de donde 
\begin{eqnarray}
f(x)&=&\int_{-\infty}^{\infty}dx^{\prime}  f(x^{\prime}) \delta(x-x^{\prime}).\label{eq:propdeltadirac}
\end{eqnarray}
La delta de Dirac tiene varias propiedades interesantes. Primero mostraremos que 
\begin{eqnarray}
\delta(-x)=\delta(x).
\end{eqnarray}
Esta afirmaci\'on  es correcta, pues con el cambio de variable $k^{\prime}=-k$ se encuentra que
\begin{eqnarray}
\delta(-x)&=& \frac{1}{2\pi} \int_{-\infty}^{\infty}dk e^{-ikx}= \frac{-1}{2\pi} \int_{\infty}^{-\infty}dk^{\prime} e^{ik^{\prime} x}=
\frac{1}{2\pi} \int_{-\infty}^{\infty}dk^{\prime} e^{ik^{\prime}x}=\delta(x), \nonumber
\end{eqnarray}
que es lo que se queria demostrar.\\

De (\ref{eq:propdeltadirac}) se puede ver que si $x=0,$ se tiene 
\begin{eqnarray}
f(0)=\int_{-\infty}^{\infty}dx^{\prime}  f(x^{\prime}) \delta(-x^{\prime})= \int_{-\infty}^{\infty}dx^{\prime}  f(x^{\prime}) \delta(x^{\prime})
\label{eq:prop1deltadirac}.
\end{eqnarray}
Por lo tanto, usando la funci\'on constante $f(x)=1$ se tiene 
\begin{eqnarray}
\int_{-\infty}^{\infty}dx   \delta(x)=1.
\end{eqnarray}
Note que la transformada de Fourier de la delta de Dirac es  
\begin{eqnarray}
F\left[\delta(x)\right]=\frac{1}{\sqrt{2\pi}}\int_{-\infty}^{\infty} dx e^{-ikx}\delta(x)=\frac{1}{\sqrt{2\pi}}.
\end{eqnarray}
La delta de Dirac tambi\'en cumple
\begin{eqnarray}
\delta(ax)=\frac{\delta(x)}{|a|}.\label{eq:delta1}
\end{eqnarray}
Para probar esta afirmaci\'on primero tomemos el caso $a>0$. Con el cambio de variable $k^{\prime}=ak$ se obtiene
\begin{eqnarray}
\delta(ax)&=& \frac{1}{2\pi} \int_{-\infty}^{\infty}dk e^{iakx}= \frac{1}{a} \frac{1}{2\pi} \int_{-\infty}^{\infty}dk^{\prime} e^{ik^{\prime} x}=
\frac{\delta(x)}{a}, \nonumber
\end{eqnarray}
por lo que se cumple Eq. (\ref{eq:delta1}). Para  caso $a<0$ tambi\'en usaremos  el cambio de variable $k^{\prime}=ak,$ de donde 
\begin{eqnarray}
\delta(ax)&=& \frac{1}{2\pi} \int_{-\infty}^{\infty}dk e^{iakx}= \frac{1}{a} \frac{1}{2\pi} \int_{\infty}^{-\infty}dk^{\prime} e^{ik^{\prime} x}=
\frac{1}{-a}\frac{1}{2\pi} \int_{-\infty}^{\infty}dk^{\prime} e^{ik^{\prime} x}
=\frac{\delta(x)}{-a}, \nonumber
\end{eqnarray}
as\'i se cumple Eq. (\ref{eq:delta1}). Por lo tanto, la identidad Eq. (\ref{eq:delta1}) es correcta.\\

En estricto sentido la delta de Dirac solo tiene sentido dentro de una integral. Sin embargo es conveniente manipularla 
 por si sola. Usando la analog\'ia de la delta de Kronecker y la propiedad Eq. (\ref{eq:prop1deltadirac}), se dice que $\delta(x)$ es cero si $x\not =0$ y, considerando la definici\'on  Eq. (\ref{eq:defdeltadirac}), se toma el valor $\delta(0)=\infty.$\\
 
Sea   $f(x)$ una funci\'on, veamos que significa $\delta(f(x)).$ Note que si no existe $x$ tal que $f(x)=0,$ entonces 
$\delta(f(x))=0.$ Supongamos que $f(x)$ tiene un solo cero, $x_{0},$  y que la funci\'on es creciente o decreciente. Si $f(x)$ es creciente, se tiene  que  $\frac{df}{dx}>0$ y 
\begin{eqnarray}
\int_{-\infty}^{\infty} dx g(x)\delta\left(f(x)\right)=\int_{-\infty}^{\infty}dY\frac{g\left(f^{-1}\left(Y\right)\right)}{\frac{df}{dx}} \delta\left(Y\right).\nonumber
\end{eqnarray}
Si la funci\'on $f(x)$ es decrenciente, entonces $\frac{df}{dx}<0$ y
\begin{eqnarray}
\int_{-\infty}^{\infty} dx g(x)\delta\left(f(x)\right)&=&\int_{\infty}^{-\infty}dY\frac{g\left(f^{-1}\left(Y\right)\right)}{\frac{df}{dx}}
 \delta\left(Y\right)\nonumber\\
 &=&\int_{-\infty}^{\infty}dY\frac{g\left(f^{-1}\left(Y\right)\right)}{-\frac{df}{dx}}
 \delta\left(Y\right) .\nonumber
\end{eqnarray}
Por lo tanto, 
\begin{eqnarray}
\int_{-\infty}^{\infty} dx g(x)\delta\left(f(x)\right)=\int_{-\infty}^{\infty}dY\frac{g\left(f^{-1}\left(Y\right)\right)}{\left|\frac{df}{dx}\right|} \delta\left(Y\right)= \frac{g\left(f^{-1}\left(Y\right)\right)\Bigg|_{Y=0}}{\left|\frac{df}{dx}\right|_{Y=0}},
\end{eqnarray}
como $Y=0$ s\'olo si $x=x_{0},$ llega a 
\begin{eqnarray}
\int_{-\infty}^{\infty} dx g(x)\delta\left(f(x)\right)= \frac{g\left(x_{0}\right)}{\left|\frac{df}{dx}\right|_{x=x_{0}} }.
\end{eqnarray}
Adem\'as, note que 
\begin{eqnarray}
\int_{-\infty}^{\infty} dx g(x) \frac{\delta\left(x-x_{0}\right)}{\left|\frac{df}{dx}\right|}= \frac{g\left(x_{0}\right)}{\left|\frac{df}{dx}\right|_{x=x_{0}} }, 
\end{eqnarray}
como este resultado es v\'alido para cualquier funci\'on $g(x),$ podemos tomar 
\begin{eqnarray}
\delta\left(f(x)\right)= \frac{\delta \left(x-x_{0}\right)}{\left|\frac{df}{dx}\right|}.
\end{eqnarray}
Ahora, supongamos que $f(x)$ tiene un n\'umero finito de ceros $\left\{x_{i}\right\}_{i=1}^{n}$tales que 
\begin{eqnarray}
f(x_{i})=0,\qquad  \frac{d f(x)}{dx}\Bigg|_{x=x_{i}}\not =0.\label{eq:delta-dirac-fou}
\end{eqnarray}
Entonces definiremos
$n$ vecindades de radio $\epsilon,$ cada una centrada en un cero $x_{i}.$ Estas vecindades, $V_{i\epsilon}(x_{i}),$ las tomaremos de tal forma que si $x\in V_{i\epsilon}(x_{i})$
y $f(x)=0$, entonces $x=x_{i}.$ Note que debido a que se cumple Eq. (\ref{eq:delta-dirac-fou}), en cada cero podemos elejir la vecindad de tal forma que en ella la funci\'on $f$ solo sea creciente \'o decreciente. Sobre estas vecindades definiremos  las funciones
\begin{eqnarray}
 f_{i}(x)=\left\{
\begin{array}{ll}
f(x)& x\in V_{i\epsilon}(x_{i}),\\
h_{i}(x)  & x \not \in V_{i\epsilon}(x_{i}),
\end{array} \right. 
 \label{eq:recurr}
\end{eqnarray}
donde $h_{i}(x)$ es una funci\'on de tal forma que $f_{i}(x)$ es creciente \'o decreciente en todo el eje real. Entonces,
como la delta de Dirac s\'olo es diferente de cero cuando su argumento es cero, se tiene 
\begin{eqnarray}
\delta\left(f(x)\right)= \sum_{i=1}^{n} \delta \left(f_{i} \left(x\right)\right)=\sum_{i=1}^{n} \frac{ \delta(x-x_{i})}{\left|\frac{df_{i}}{dx}\right|},
\end{eqnarray}
es decir 
\begin{eqnarray}
\delta\left(f(x)\right)= \sum_{i=1}^{n} \frac{ \delta(x-x_{i})}{\left|\frac{df}{dx}\right|},\qquad f(x_{i})=0.
\end{eqnarray}
Por ejemplo, 
\begin{eqnarray}
\delta\left(x^{2}-a^{2}\right)=  \frac{ \delta(x-a) +\delta(x+a)}{2|a|}.
\end{eqnarray}

\subsection{ La funci\'on de Heaviside}

La funci\'on de Heaviside se define como
\begin{eqnarray}
 \theta(x)=\left\{
\begin{array}{ll}
1,& x\geq 0\\
0, & x<0.
\end{array} \right. 
\label{eq:recurr}
\end{eqnarray}
Esta funci\'on se puede ver como la primitiva de la delta de Dirac. En efecto, supongamos que $f$ es una funci\'on tal que 
$f(\pm \infty)=0,$ entonces 
\begin{eqnarray}
\int_{-\infty}^{\infty}dx^{\prime} \frac{d \theta\left(x^{\prime}-x\right)}{dx ^{\prime}} f(x^{\prime})&=& 
\int_{-\infty}^{\infty}dx^{\prime} \left[\frac{d }{dx ^{\prime}}  \left[\theta\left(x^{\prime}-x\right)f(x^{\prime})\right]- 
\theta\left(x^{\prime}-x\right)\frac{d f(x^{\prime})}{dx ^{\prime}} \right]\nonumber\\
& =& \theta\left(x^{\prime}-x\right)f(x^{\prime})\Bigg|_{-\infty}^{\infty}- \int_{-\infty}^{\infty}dx^{\prime}\theta\left(x^{\prime}-x\right)\frac{d f(x^{\prime})}{dx ^{\prime}} \nonumber\\
& =& -\int_{x}^{\infty} dx ^{\prime} \frac{d f(x^{\prime})}{dx ^{\prime}}=f(x),
\end{eqnarray}
es decir
\begin{eqnarray}
\int_{-\infty}^{\infty}dx^{\prime} \frac{d \theta\left(x^{\prime}-x\right)}{dx ^{\prime}} f(x^{\prime})=f(x).
\end{eqnarray}
Como este resultado es v\'alido para cualquier funci\'on $f,$ podemos tomar 
\begin{eqnarray}
\frac{d \theta\left(x^{\prime}-x\right)}{dx ^{\prime}} =\delta\left(x^{\prime}-x\right).
\end{eqnarray}
Integrando la definici\'on de delta de Dirac Eq. (\ref{eq:defdeltadirac}), se puede ver que 
\begin{eqnarray}
\theta(x)=\frac{1}{2\pi i} \int_{-\infty}^{\infty}dk \frac{e^{ikx}}{k}.\label{eq:defdeltadirac}
\end{eqnarray}

\section{Norma de una funci\'on}

Ahora, usando la definici\'on de la transformada inversa de Fourier podemos ver que la norma de una funci\'on es 
\begin{eqnarray}
||f(x)||^{2}&=&\int_{-\infty}^{\infty} dx f^{*}(x)f(x)\nonumber\\
&=& \int_{-\infty}^{\infty} dx  
\left( \frac{1}{\sqrt{2\pi}}\int_{-\infty}^{\infty}dk e^{ikx} \tilde f(k) \right)^{*} \frac{1}{\sqrt{2\pi}}\int_{-\infty}^{\infty}dk^{\prime} e^{ik^{\prime} x} \tilde f(k^{\prime})\nonumber\\
&=& \int_{-\infty}^{\infty}  \int_{-\infty}^{\infty}dk dk^{\prime} \tilde f^{*}(k)\tilde f(k^{\prime})
\left( \frac{1}{2\pi} \int_{-\infty}^{\infty}dx e^{i\left(k^{\prime} -k\right)x} \right)\nonumber\\
&=&  \int_{-\infty}^{\infty}  \int_{-\infty}^{\infty}dk dk^{\prime} \tilde f^{*}(k)\tilde f(k^{\prime})\delta\left(k^{\prime} -k\right)\nonumber\\
&=& \int_{-\infty}^{\infty}dk \tilde f^{*}(k)\tilde f(k)=||\tilde f(k)||^{2}.
\end{eqnarray}
Por lo tanto, la norma de una funci\'on $f(x)$ tiene el mismo valor que su transformada de Fourier $\tilde f(k).$

\section{Transformada de Fourier en $d$ dimensiones}

Hasta ahora hemos trabajado con funciones de una sola variable. Si $f$ es una funci\'on real de $d$ 
variables, $\vec x=(x_{1},\cdots x_{d}),$ la transformada de Fourier se define como
\begin{eqnarray}
\tilde f(\vec k)=F\left[f\left(\vec x\right)\right]= \int_{-\infty}^{\infty} 
\frac{dx_{1}}{\sqrt{2\pi}}\cdots  \frac{dx_{d}}{\sqrt{2\pi}} e^{-i\vec k\cdot \vec x}f(\vec x),\quad 
\vec k=(k_{1},\cdots k_{d}).
\end{eqnarray}
De esta definici\'on es claro que la transformada inversa de Fourier es
\begin{eqnarray}
F^{-1}\left[\tilde f\left(\vec k\right)\right]= \int_{-\infty}^{\infty} 
\frac{dk_{1}}{\sqrt{2\pi}}\cdots  \frac{dk_{d}}{\sqrt{2\pi}} e^{i\vec k\cdot \vec x}\tilde f(\vec k).
\end{eqnarray}
La delta de Dirac la denotaremos como
\begin{eqnarray}
\delta^{(d)}\left(\vec x \right)&=&  \delta(x_{1})\cdots \delta(x_{d})= 
\int_{-\infty}^{\infty} \frac{dk_{1} e^{ix_{1}k_{1}}}{\sqrt{2\pi}}\cdots  
\int_{-\infty}^{\infty} \frac{dk_{d} e^{ix_{d}k_{d}}}{\sqrt{2\pi}}\nonumber\\
&=&
\frac{1}{\sqrt{(2\pi)^{d}} } \int_{-\infty}^{\infty} d\vec k e^{i\vec k\cdot \vec x}.
\end{eqnarray}

\section{Funci\'on de Green}

Supongamos que 
$A( \partial_{x})$ es un operador lineal, con este operador se puede plantear una ecuaci\'on homog\'enea
\begin{eqnarray}
A( \partial_{x})\phi_{0}(x)=0
\end{eqnarray}
y  una ecuaci\'on inhomog\'enea
\begin{eqnarray}
A( \partial_{x})\phi(x)=-4\pi \rho(x).\label{eq:inh}
\end{eqnarray}
Para resolver la ecuaci\'on inhomog\'enea es importante la funci\'on de Green.
Se dice que $G\left(x,x^{\prime}\right)$ es una funci\'on de Green de 
$A( \partial_{x})$ si satisface la ecuaci\'on de Green
\begin{eqnarray}
A( \partial_{x})G\left(x,x^{\prime}\right)=-4\pi \delta \left(x-x^{\prime}\right).
\end{eqnarray}
Empleando una funci\'on  de Green $G\left(x,x^{\prime}\right)$ y una soluci\'on a la ecuaci\'on homog\'enea, $ \phi_{0} \left(x\right),$ se puede construir una soluci\'on de la ecuaci\'on inhomog\'enea. En efecto, la funci\'on
\begin{eqnarray}
\phi(x)=\int dx^{\prime} G\left(x,x^{\prime}\right)\rho(x^{\prime}) + \phi_{0} \left(x\right),\label{eq:SGinh}
\end{eqnarray}
satisface 
\begin{eqnarray}
A( \partial_{x})\phi(x)&=&\int dx^{\prime} A( \partial_{x})G\left(x,x^{\prime}\right)\rho(x^{\prime}) + A( \partial_{x})\phi_{0} \left(x\right)\nonumber \\
&=& -4\pi \int dx^{\prime} \delta\left(x-x^{\prime}\right)\rho(x^{\prime})
=-4\pi \rho(x).\nonumber
\end{eqnarray}
Por lo tanto, Eq. (\ref{eq:SGinh}) es soluci\'on de Eq. (\ref{eq:inh}). La funci\'on de Green y la soluci\'on a la ecuaci\'on de 
homog\'enea se elijen dependiendo de las condiciones de borde de la ecuaci\'on inhomog\'enea.

\subsection{ Funci\'on de Green y funciones propias}

Las funciones de Green del operador $A( \partial_{x})$ est\'an relacionadas con sus funciones de propias.
En efecto, supongamos que tenemos las funciones propias de  $A( \partial_{x}),$ es decir
\begin{eqnarray}
A( \partial_{x})\psi_{\lambda}(x)&=&\lambda \psi_{\lambda}(x).
\end{eqnarray}
Tambi\'en supongamos que las funciones propias forman un conjunto $\{\psi_{\lambda}(x)\}$ ortonormal con 
el producto escalar
\begin{eqnarray}
 <\psi_{\lambda}(x)|\psi_{\lambda^{\prime}}(x)>= \int dx \psi^{*}_{\lambda}(x)\psi_{\lambda^{\prime}}(x)=\delta_{\lambda\lambda^{\prime}}.
\end{eqnarray} 
Por lo tanto, para cualquier funci\'on bien portada se puede hacer el desarrollo de Fourier 
\begin{eqnarray}
f(x)=\sum_{\lambda} a_{\lambda} \psi_{\lambda}(x),\label{eq:TFGd}
\end{eqnarray} 
con los coeficientes de Fourier dados por 
\begin{eqnarray}
a_{\lambda}= <\psi_{\lambda}(x)|f(x)>= \int dx \psi^{*}_{\lambda}(x)f(x).
\end{eqnarray} 
Note que substituyendo los coeficientes de Fourier en Eq. (\ref{eq:TFGd}) se llega a
\begin{eqnarray}
f(x)=\sum_{\lambda} \int dx^{\prime} \psi^{*}_{\lambda}(x^{\prime})f(x^{\prime} )\psi_{\lambda}(x)= 
\int dx^{\prime}f(x^{\prime} )\sum_{\lambda} \psi^{*}_{\lambda}(x^{\prime})\psi_{\lambda}(x),
\end{eqnarray} 
como este resultado es v\'alido para cualquier funci\'on, tenemos 
\begin{eqnarray}
\delta(x-x^{\prime})=\sum_{\lambda} \psi^{*}_{\lambda}(x^{\prime})\psi_{\lambda}(x).
\end{eqnarray} 
Por lo tanto, para este caso la funci\'on de Green est\'a dada por
\begin{eqnarray}
G(x-x^{\prime})=-4\pi \sum_{\lambda} \frac{\psi^{*}_{\lambda}(x^{\prime})\psi_{\lambda}(x)}{\lambda}.
\label{eq:TFGd2}
\end{eqnarray} 
Esta afirmaci\'on es correcta pues 
\begin{eqnarray}
A( \partial_{x}) G(x-x^{\prime})&=&-4\pi \sum_{\lambda} \frac{\psi^{*}_{\lambda}(x^{\prime})A( \partial_{x})\psi_{\lambda}(x)}{\lambda}
= -4\pi \sum_{\lambda} \frac{\psi^{*}_{\lambda}(x^{\prime})\lambda \psi_{\lambda}(x)}{\lambda}\nonumber\\
&=& -4\pi \sum_{\lambda} \psi^{*}_{\lambda}(x^{\prime}) \psi_{\lambda}(x)=- 4\pi\delta(x-x^{\prime}).
\end{eqnarray} 
De Eq. (\ref{eq:TFGd2}) se puede ver que las condiciones de borde que satisface la funci\'on de Green son las mismas que satisfacen las funciones propias.

\section{Ecuaci\'on de Laplace en dos dimensiones}

Hora veremos un ejemplo de funci\'on de Green. Calcularemos la funci\'on de Green del Laplaciano en dos dimensiones
\begin{eqnarray}
\left( \frac{\partial^{2}}{\partial x^{2}}+ \frac{\partial^{2}}{\partial y^{2}}\right) G(\vec x-\vec x^{\prime})=-4\pi \delta^{(2)}(\vec x-\vec x^{\prime}),\quad \vec x=(x,y).
\end{eqnarray} 
Pediremos que la funci\'on de Green se anule en $(x,y)=(0,0)$ y en $(x,y)=(L_{1},L_{2}).$ Entonces debemos buscar 
las funciones propias
\begin{eqnarray}
 \left( \frac{\partial^{2}}{\partial x^{2}}+ \frac{\partial^{2}}{\partial y^{2}}\right) \psi( x,y)= \lambda \psi( x,y).
\label{eq:TFGd3}
\end{eqnarray} 
que se anulen en $(x,y)=(0,0)$ y en $(x,y)=(L_{1},L_{2}).$ \\

Propondremos $\psi( x,y)=X(x)Y(y),$ sustituyendo esta propuesta en (\ref{eq:TFGd3}) se obtiene
\begin{eqnarray}
Y(y) \frac{\partial^{2} X(x)}{\partial x^{2} }+ X(x)\frac{\partial^{2} Y(y)}{\partial y^{2}}= \lambda X(x)Y(y),
\end{eqnarray} 
de donde
\begin{eqnarray}
\lambda=\frac{1}{X(x)} \frac{\partial^{2} X(x)}{\partial x^{2} }+ 
\frac{1} {Y(y)}\frac{\partial^{2} Y(y)}{\partial y^{2}}.
\end{eqnarray} 
Derivando esta ecuaci\'on respecto a $x$ y $y$ se encuentra
\begin{eqnarray}
\frac{\partial }{\partial x}\left(\frac{1}{X(x)} \frac{\partial^{2} X(x)}{\partial x^{2} }\right)=0,\qquad  
\frac{\partial }{\partial x}\left(\frac{1}{Y(y)} \frac{\partial^{2} Y(y)}{\partial y^{2} }\right)=0,
\end{eqnarray} 
de donde 
\begin{eqnarray}
\frac{1}{X(x)} \frac{\partial^{2} X(x)}{\partial x^{2} }=-\alpha^{2},\quad  
\frac{1}{Y(y)} \frac{\partial^{2} Y(y)}{\partial y^{2} }=-\beta^{2},\quad \alpha,\beta={\rm constante}. 
\label{eq:TFGL2}
\end{eqnarray} 
Note que esto implica que  $\lambda=-\alpha^{2}-\beta^{2},$ tambi\'en note que las ecuaciones (\ref{eq:TFGL2})
 son equivalentes a
\begin{eqnarray}
\frac{\partial^{2} X(x)}{\partial x^{2} }=-\alpha^{2} X(x),\qquad  
\frac{\partial^{2} Y(y)}{\partial y^{2} }=-\beta^{2} Y(y), \nonumber
\end{eqnarray} 
cuyas soluciones  son
\begin{eqnarray}
 X_{\alpha}(x)=a_{\alpha} \cos \alpha x +b_{\alpha} \sin \alpha x,\quad   Y_{\beta}(x)=A_{\beta} \cos \beta  +B_{\beta} \sin \beta y, \nonumber
\end{eqnarray} 
donde $a_{\alpha},b_{\alpha},A_{\beta},B_{\beta}$ son constantes. Empleando las condiciones de borde 
se llega a 
\begin{eqnarray}
 X_{n}(x)=\sqrt{\frac{2}{L_{1}}} \sin \frac{n\pi}{L_{1}} x,\quad   Y_{m}(x)=\sqrt{\frac{2}{L_{2}}} \sin \frac{m\pi}{L_{1}} y, \quad \lambda =-\left[\left(\frac{n\pi}{L_{2}}\right)^{2}+  \left(\frac{m\pi}{L_{2}}\right)^{2}\right]\nonumber
\end{eqnarray} 
Por lo tanto, las funciones propias y valores propios son
\begin{eqnarray}
 \psi_{nm}(x,y)=\sqrt{\frac{2}{L_{1}}} \sin \frac{n\pi}{L_{1}} x\sqrt{\frac{2}{L_{2}}} \sin \frac{m\pi}{L_{2}} y, \quad \lambda_{nm} =-\left[\left(\frac{n\pi}{L_{1}}\right)^{2}+  \left(\frac{m\pi}{L_{2}}\right)^{2}\right]\nonumber
\end{eqnarray} 
y la funci\'on de Green del sistema es 
\begin{eqnarray}
 G(\vec x-\vec x^{\prime})= \frac{16}{\pi L_{1}L_{2}} \sum_{n\geq 1}  \sum_{m\geq 1} \frac{ \sin \frac{n\pi}{L_{1}} x^{\prime}  \sin \frac{m\pi}{L_{2}} y^{\prime} \sin \frac{n\pi}{L_{1}} x  \sin \frac{m\pi}{L_{2}} y  } {\left(\frac{n}{L_{1}}\right)^{2}+  \left(\frac{m}{L_{2}}\right)^{2}} .
\end{eqnarray} 

\section{Resultados de variable compleja}

Antes de continuar recordemos dos resultados de variable compleja.\\

Si $z_{0}$ es un polo de $f(z)$ entonces el residuo de $f(z)$ se define 
como 
\begin{eqnarray}
{\rm Res}_{z_{0}}f(z)=\lim_{z\to z_{0}}(z-z_{0})f(z)
\end{eqnarray}
El teorema de Cauchy nos dice que si $C$ en una curva cerrada en el
plano complejo entonces 
\begin{eqnarray}
\oint_{C}f(z)dz=2\pi i\sum_{k=1}^{n}{\rm Res}_{a_{k}}f(z).
\end{eqnarray}
Adem\'as, si $g(x)$ es una funci\'on de variable real con un polo simple
en $a$ y $\delta >0,$ entonces el valor principal de Cauchy se define como 
\begin{eqnarray}
P.P\int_{-\infty}^{\infty}g(x)=
\lim_{\delta\to 0}\left[\int_{-\infty}^{a-\delta}g(x)dx+
\int_{a+\delta}^{\infty}g(x)dx\right].
\end{eqnarray}
Otro resultado es que si  $R(z)$ es una funci\'on que no tiene
polos en el eje real ni en el plano complejo superior y que 
$\lim_{z\to \infty} R(z)=0,$ entonces si $a$ es un n\'umero real, 
se cumple 
\begin{eqnarray}
R(a)=\frac{1}{i\pi} P.P\int_{-\infty}^{\infty}\frac{R(x)}{x-a}
=\frac{1}{i\pi}\int_{-\infty}^{\infty}\frac{R(x)}{x-a}.
\end{eqnarray}  
En particular 
\begin{eqnarray}
 P.P\int_{-\infty}^{\infty}\frac{e^{ix}}{x}
=i\pi,
\end{eqnarray}  
que implica 
\begin{eqnarray}
\int_{-\infty}^{\infty}\frac{\sin x}{x}=\pi.\label{eq:TFintsen}
\end{eqnarray}  
Este resultado lo ocuparemos posteriormente.

\section{Ecuaci\'on de Poisson}

Ahora estudiaremos la ecuaci\'on de Poisson
\begin{eqnarray}
\nabla^{2}\phi(\vec x)=-4\pi \rho\left(\vec x\right),
\end{eqnarray}  
que tiene asociada la ecuaci\'on de Green
\begin{eqnarray}
\nabla^{2}G\left(\vec x-\vec x^{\prime}\right)=-4\pi \delta ^{(3)}\left(\vec x-\vec x^{\prime}\right),
\end{eqnarray}
Tomando la transformadas de Fourier de esta ecuaci\'on obtenemos 
\begin{eqnarray}
F\left[\nabla^{2}G\left(\vec x-\vec x^{\prime}\right)\right]=-4\pi 
F\left[\delta ^{(3)}\left(\vec x-\vec x^{\prime}\right)\right],
\end{eqnarray}
es decir 
\begin{eqnarray}
-\vec k^{2}\tilde G\left(\vec k^{2}\right)=-4\pi\frac{1}{\sqrt{(2\pi)^{3}}}, 
\end{eqnarray}
por lo que 
\begin{eqnarray}
\tilde G\left(\vec k^{2}\right)=4\pi\frac{1}{\sqrt{(2\pi)^{3}}}\frac{1}{ \vec k^{2}},
\end{eqnarray}
de donde,
\begin{eqnarray}
G\left(\vec x-\vec x^{\prime}\right)=\frac{4\pi}{(2\pi)^{3}} \int d\vec k \frac{e^{i\vec k\cdot \left(\vec x-\vec x^{\prime}\right)} }{\vec k^{2}}.
\end{eqnarray}
Usando  coordenadas esf\'ericas y definiendo $\vec R=\vec x-\vec x^{\prime},$ se tiene
\begin{eqnarray}
G\left(\vec x-\vec x^{\prime}\right)&=& \frac{1}{2\pi^{2}} \int d\vec k \frac{e^{i\vec k\cdot \vec R} }{\vec k^{2}}\nonumber\\
&=&\frac{1}{2\pi^{2}} \int_{0}^{\infty} dk k^{2} \int_{0}^{2\pi} d\varphi\int_{0}^{\pi}d\theta \sin\theta \frac{e^{ikR\cos\theta}}{k^{2}}\nonumber\\
&=&\frac{2\pi}{2\pi^{2}} \int_{0}^{\infty} dk  \int_{0}^{\pi}d\theta \sin\theta e^{ikR\cos\theta},
\end{eqnarray}
adem\'as con  el cambio de variable $u=\cos\theta$ se encuentra 
\begin{eqnarray}
 \int_{0}^{\pi}d\theta \sin\theta e^{ikR\cos\theta}&=&-\int_{1}^{-1}du e^{ikR u}= \int_{-1}^{1}du e^{ikR u}=
 \frac{1}{iRk} e^{ikR u}\bigg|_{-1}^{1}\nonumber\\
 &=&\frac{e^{ikR}- e^{-ikR}}{iRk}= 2\frac{\sin kR}{kR}.
\end{eqnarray}
Por lo que
\begin{eqnarray}
G\left(\vec x-\vec x^{\prime}\right)&=&\frac{2}{\pi}  \int_{0}^{\infty} dk   \frac{\sin kR}{kR},
\end{eqnarray}
tomamos $w=kR$  y considerando la integral Eq. (\ref{eq:TFintsen}) se llega a 
\begin{eqnarray}
G\left(\vec x-\vec x^{\prime}\right)&=&\frac{2}{R\pi}  \int_{0}^{\infty} dw  \frac{\sin w}{w}=
\frac{1}{R\pi} \int_{-\infty}^{\infty} dw \frac{\sin w}{w}=\frac{\pi}{R\pi}=\frac{1}{R}\nonumber,
\end{eqnarray}
es decir
\begin{eqnarray}
G\left(\vec x-\vec x^{\prime}\right)&=&\frac{1}{|\vec x-\vec x^{\prime}|}.
\end{eqnarray}
Que implica
\begin{eqnarray}
\nabla^{2}\left(\frac{1}{|\vec x-\vec x^{\prime}|}\right)=-4\pi\delta^{3}\left(\vec x-\vec x^{\prime}\right).\label{eq:Gpoisson}
\end{eqnarray}
Por lo tanto, una soluci\'on a la ecuaci\'on de Poisson es 
\begin{eqnarray}
\phi\left(\vec x\right)=\int d\vec x^{\prime}  \frac{\rho\left(\vec x^{\prime}\right)}{|\vec x-\vec x^{\prime}|} + \phi_{0}(\vec x), 
\end{eqnarray}
con $\phi_{0}(\vec x)$ una soluci\'on a la ecuaci\'on de Laplace $\nabla^{2} \phi_{0}(\vec x)=0.$\\

Para muchos sistemas 
f\'isicos se puede tomar $\phi_{0}(\vec x)=0,$  en esos casos la soluci\'on a la ecuaci\'on de Poisson
es 
\begin{eqnarray}
\phi\left(\vec x\right)=\int d\vec x^{\prime}  \frac{\rho\left(\vec x^{\prime}\right)}{|\vec x-\vec x^{\prime}|}.
\end{eqnarray}
Ahora, note que 
\begin{eqnarray}
\frac{1}{|\vec x- \vec x^{\prime}|}=
\frac{1}{\sqrt{x^{2}-2xx^{\prime}
\cos\alpha+x^{\prime 2} }} 
\end{eqnarray}
con $\alpha$ el \'angulo entre $\vec x $ y $\vec x^{\prime}.$ 
 Ahora, si $|\vec x|\not = |\vec x^{\prime}|$ definamos 
$r_{<}={\rm min}\{|\vec x|, |\vec x^{\prime}|\}$ y $r_{>}={\rm max}\{|\vec x|, |\vec x^{\prime}|\},$ 
es claro que 
\begin{eqnarray}
\left(\frac{r_{<}}{r_{>}}\right)<1.
\end{eqnarray}
Entonces, ocupando  estas definiciones 
y la funci\'on generatriz de los polinomios de Legendre Eq. (\ref{eq:gen-leg})
con 
\begin{eqnarray}
z=\frac{r_{<}}{r_{>}},  \qquad u=\cos\alpha, 
\end{eqnarray}
se tiene 
\begin{eqnarray}
\frac{1}{|\vec x- \vec x^{\prime}|}=
\frac{1}{r_{>}\sqrt{1-2\left(\frac{r_{<}}{r_{>}}\right)
\cos\alpha+\left(\frac{r_{<}}{r_{>}}\right)^{2}}}
= \frac{1}{r_{>}}\sum_{l \geq 0} 
\left(\frac{r_{<}}{r_{>}}\right)^{l}P_{l}(\cos\alpha). \nonumber
\end{eqnarray}
As\'{\i}, usando el teorema de adici\'on de los arm\'onicos
esf\'ericos Eq. (\ref{eq:arm-adi}), tenemos 
\begin{eqnarray}
\frac{1}{|\vec x- \vec x^{\prime}|}=
\sum_{l \geq 0} \sum_{m=-l}^{m=l}
\frac{4\pi}{2l+1} \left(\frac{r_{<}^{l}}{r_{>}^{l+1}}\right)
Y_{lm}^{*}(\theta^{\prime},\varphi^{\prime} )Y_{lm}(\theta,\varphi).
\end{eqnarray}
Fuera de la regi\'on donde est\'a definida  la fuente $\rho(\vec x^{\prime})$ se tiene  
$r_{<}=|\vec x^{\prime}|$ y $r_{>}=|\vec x|.$ Por lo tanto, en esta regi\'on se tiene 
\begin{eqnarray}
\phi\left(\vec x\right)=\int d\vec x^{\prime}  \frac{\rho\left(x^{\prime}\right)}{|\vec x-\vec x^{\prime}|} 
=\sum_{l \geq 0} \sum_{m=-l}^{m=l}  \frac{ Y_{lm}(\theta,\varphi)}{r_{>}^{l+1}} \frac{4\pi}{2l+1} 
\int d\vec x^{\prime} r_{<}^{l} Y_{lm}^{*}(\theta^{\prime},\varphi^{\prime} ) \rho\left(\vec x^{\prime}\right).
\nonumber
\end{eqnarray}
Definiremos  los momentos multipolares como 
\begin{eqnarray}
q_{lm}=\frac{4\pi}{2l+1} 
\int d\vec x^{\prime} r_{<}^{l} Y_{lm}^{*}(\theta^{\prime},\varphi^{\prime} ) \rho\left(\vec x^{\prime}\right)
\end{eqnarray}
por lo que 
\begin{eqnarray}
\phi\left(\vec x\right)=\sum_{l \geq 0} \sum_{m=-l}^{m=l}  \frac{ q_{lm} }{r_{>}^{l+1}} Y_{lm}(\theta,\varphi).
\end{eqnarray}

Adem\'as, usando la norma de Coulomb $\vec \nabla \cdot \vec A=0$, las ecuaciones de la magnetost\'atica se reducen a  
\begin{eqnarray}
\nabla^{2} \vec A(\vec x) =-4\pi \vec J\left(\vec x\right),
\end{eqnarray}  
que son tres ecuaciones de Poisson. Por lo tanto, 
\begin{eqnarray}
\vec A\left(\vec x\right)=\int d\vec x^{\prime}  \frac{\vec J\left(\vec x^{\prime}\right)}{|\vec x-\vec x^{\prime}|}. 
\end{eqnarray}
Claramente este potencial tambi\'en se puede escribir en t\'erminos de los arm\'onicos esf\'ericos.

\section{Funci\'on de Green de la ecuaci\'on Helmholtz}

La ecuaci\'on de  Helmholtz inhomog\'enea es 
\begin{eqnarray}
\left(\nabla^{2}+k^{2}\right) \phi(\vec x) =-4\pi \phi\left(\vec x\right),
\end{eqnarray}  
la cual tiene asociada la ecuaci\'on de Green
\begin{eqnarray}
\left(\nabla^{2}+k^{2}\right) G\left(\vec x-\vec x^{\prime} \right) =-4\pi \delta^{3}\left(\vec x-\vec x^{\prime}\right).
\end{eqnarray}  
Para resolver esta \'ultima ecuaci\'on veamos la funci\'on
\begin{eqnarray}
f(r)=\frac{e^{\alpha r}}{r},\quad r=\sqrt{x^{2}+y^{2}+z^{2}},\label{eq:TFh}
\end{eqnarray}  
la cual tiene una divergencia en $r=0.$ Esta divergencia se puede aislar, en efecto
\begin{eqnarray}
f(r)=\frac{e^{\alpha r}-1+1}{r}=\frac{1}{r}+\frac{e^{\alpha r}-1}{r},
\end{eqnarray}  
note que 
\begin{eqnarray}
\lim_{r\to 0}\frac{e^{\alpha r}-1}{r}=\alpha.
\end{eqnarray}  
Por lo tanto, la divergencia de la funci\'on Eq. (\ref{eq:TFh}) est\'a en t\'ermino $1/r.$ Ahora, observemos
\begin{eqnarray}
\nabla^{2}\left(\frac{e^{\alpha r}}{r}\right)=\nabla^{2}\left( \frac{1}{r}+\frac{e^{\alpha r}-1}{r}\right)
=\nabla^{2}\frac{1}{r}+ \nabla^{2}\left(\frac{e^{\alpha r}-1}{r}\right),
\end{eqnarray}  
empleando la ecuaci\'on Eq. (\ref{eq:Gpoisson}) y el Laplaciano en coordenadas esf\'erica 
Eqs. (\ref{eq:lapla-esfe0})-(\ref{eq:lapla-esfe})
se llega a 
\begin{eqnarray}
\nabla^{2}\left(\frac{e^{\alpha r}}{r}\right)=-4\pi \delta^{(3)}\left(\vec r\right) + \frac{1}{r}\frac{\partial^{2}}{\partial r^{2}} \left(r \frac{e^{\alpha r}}{r}\right)=-4\pi \delta^{(3)}\left(\vec r\right)+ \alpha^{2} \frac{e^{\alpha r}}{r}, \nonumber 
\end{eqnarray}  
es decir 
\begin{eqnarray}
\left( \nabla^{2} -\alpha^{2}\right)\frac{e^{\alpha r}}{r}=-4\pi \delta^{(3)}\left(\vec r\right).
\end{eqnarray}  
Este resultado lo hemos obtenido con las coordenadas $x,y,z.$ Claramente el  resultado no cambia si se consideran las coordenas $x-x^{\prime},y-y^{\prime},z-z^{\prime}.$ Por lo tanto,
\begin{eqnarray}
\left( \nabla^{2} -\alpha^{2}\right)\frac{e^{\alpha |\vec x-\vec x^{\prime}|}}{|\vec x-\vec x^{\prime}|}=-4\pi \delta^{(3)}\left(\vec x- x^{\prime}\right).
\end{eqnarray}  
Si tomamos $\alpha=\pm ik,$ entonces se obtiene 
\begin{eqnarray}
\left( \nabla^{2} +k^{2}\right)\frac{e^{\pm ik |\vec x-\vec x^{\prime}|}}{|\vec x-\vec x^{\prime}|}=-4\pi \delta^{(3)}\left(\vec x- x^{\prime}\right),
\end{eqnarray}  
que es la ecuaci\'on de Green para la ecuaci\'on de Helmholtz inhomog\'enea. As\'i, la funci\'on de Green para  la ecuaci\'on de Helmholtz inhomog\'enea es
\begin{eqnarray}
G\left(\vec x- x^{\prime}\right)=\frac{e^{\pm ik |\vec x-\vec x^{\prime}|}}{|\vec x-\vec x^{\prime}|}.
\end{eqnarray}  
Entonces una soluci\'on a la ecuaci\'on de Helmholtz inhomog\'enea es
\begin{eqnarray}
\phi(\vec x) =\phi_{0}(\vec x)+  \int d\vec x^{\prime} \frac{e^{\pm ik |\vec x-\vec x^{\prime}|}}{|\vec x-\vec x^{\prime}|}\rho(\vec x^{\prime}), 
\end{eqnarray}  
con $\phi_{0}(\vec x)$ una soluci\'on a la ecuaci\'on de Helmholtz libre. Estas soluciones se ocupan para estudiar radiaci\'on y  difracci\'on de onda electromagn\'eticas, el signo se elije dependiendo si las ondas que se estudian son ondas entrantes o salientes.

\subsection{Ecuaci\'on de Lippman-Schwinger}

La ecuaci\'on de Schr$\ddot {\rm o}$dinger 
\begin{eqnarray}
\left(-\frac{\hbar^{2}}{2m} \nabla^{2}+V(\vec x)\right) \psi(\vec x) =E \psi(\vec x) \label{eq:TFscho}
\end{eqnarray}
se puede escribir como una  ecuaci\'on de Helmholtz inhomog\'enea. En efecto, definiendo 
\begin{eqnarray}
k^{2}=\frac{2mE}{\hbar},\qquad \rho(\vec x)=\frac{- m}{2\pi \hbar} V(\vec x)\psi(\vec x)\nonumber
\end{eqnarray}
y realizando operaciones elementales la ecuaci\'on  (\ref{eq:TFscho}) toma la forma
\begin{eqnarray}
\left( \nabla^{2}+k^{2} \right) \psi(\vec x) =-4\pi\rho(\vec x).\nonumber
\end{eqnarray}
Por lo tanto,
\begin{eqnarray}
\psi(\vec x) = \psi_{0}(\vec x)-\left(\frac{ m}{2\pi \hbar}\right)  \int d\vec x^{\prime}\frac{e^{ik |\vec x-\vec x^{\prime}|}}{|\vec x-\vec x^{\prime}|} V(\vec x^{\prime})\psi(\vec x^{\prime}),\label{eq:TFls}
\end{eqnarray}
donde  $\psi_{0}(\vec x)$ es una soluci\'on a la ecuaci\'on de Helmholtz homog\'enea.
A  Eq. (\ref{eq:TFls}) se le llama ecuaci\'on de Lippman-Schwinger y 
es muy \'util para estudiar dispersi\'on de part\'iculas en mec\'anica cu\'antica.

\section{Funci\'on de Green de la ecuaci\'on de onda}

Ahora estudiaremos  la ecuaci\'on de onda inhomog\'enea 
\begin{eqnarray}
\left(\nabla^{2}-\frac{1}{c^{2}}\frac{\partial^{2}}{\partial t^{2}}\right)\phi(\vec x,t)= -4\pi \rho(\vec x, t).
\label{eq:TFondai}
\end{eqnarray}
la cual tiene asociada la ecuaci\'on de Green
\begin{eqnarray}
\left(\nabla^{2}-\frac{1}{c^{2}}\frac{\partial^{2}}{\partial t^{2}}\right)G\left(\vec x- \vec x^{\prime},t-t^{\prime}\right)= -4\pi \delta^{(3)}\left(\vec x- \vec x^{\prime}\right)\delta\left(t-t^{\prime}\right) .
\label{eq:TFonda}
\end{eqnarray}
La ecuaci\'on de onda es invariante relativista, para respetar esta  invariancia en el sector temporal  tomaremos las definiciones
\begin{eqnarray}
\tilde g(\omega)&=&\frac{1}{\sqrt{2\pi}} \int_{-\infty}^{\infty}  dt e^{i\omega t} g(t),\\
g(t)&=& \frac{1}{\sqrt{2\pi}} \int_{-\infty}^{\infty}  d\omega e^{-i\omega t} \tilde g(\omega),\\
\delta(t)&=& \frac{1}{2\pi} \int_{-\infty}^{\infty}  d\omega e^{-i\omega t}.
\end{eqnarray}
Por lo que,
\begin{eqnarray}
G\left(\vec x- \vec x^{\prime},t-t^{\prime}\right)&=&\frac{1}{\sqrt{(2\pi )^{4}}} \int d\vec k d\omega 
e^{-i\left(\omega (t-t^{\prime})-\vec k\cdot (\vec x-\vec x^{\prime})\right)}\tilde G\left(\vec k,\omega\right) ,\nonumber\\
\delta^{(4)}(x^{\mu}-x^{\mu \prime})&=&
\delta^{(3)} \left(\vec x- \vec x^{\prime}\right)\delta(t-t^{\prime})=\frac{1}{\sqrt{(2\pi )^{4}}} \int d\vec k d\omega 
e^{-i\left(\omega (t-t^{\prime})-\vec k\cdot (\vec x-\vec x^{\prime})\right)} \nonumber.
\end{eqnarray}
Por lo tanto, al hacer la transformada de Fourier de Eq. (\ref{eq:TFonda}) se encuentra
\begin{eqnarray}
F\left[\left(\nabla^{2}-\frac{1}{c^{2}}\frac{\partial^{2}}{\partial t^{2}}\right)G\left(\vec x- \vec x^{\prime},t-t^{\prime}\right)\right]= -4\pi F\left[\delta^{(3)}\left(\vec x- \vec x^{\prime}\right)\delta\left(t-t^{\prime}\right)\right],\nonumber
\end{eqnarray}
es decir 
\begin{eqnarray}
\left(-\vec k^{2}+\frac{\omega^{2}}{c^{2}} \right) \tilde G\left(\vec k,\omega\right)&=& \frac{-4\pi}{\sqrt{(2\pi )^{4}}},
\nonumber 
\end{eqnarray}
de donde 
\begin{eqnarray}
\tilde G\left(\vec k,\omega\right)&=& 
\frac{4\pi}{\sqrt{(2\pi )^{4}}}\frac{1}{ \vec k^{2}-\frac{\omega^{2}}{c^{2}}}.
\nonumber 
\end{eqnarray}
Por lo tanto,
\begin{eqnarray}
G\left(\vec x- \vec x^{\prime},t-t^{\prime}\right)&=&\frac{1}{4\pi^{3}} 
\int d\vec k d\omega 
\frac{ e^{-i\left(\omega (t-t^{\prime})-\vec k\cdot (\vec x-\vec x^{\prime})\right)} }{ \vec k^{2}-\frac{\omega^{2}}{c^{2}}}.
\end{eqnarray}
Para hacer esta integral definiremos $\tau=t-t^{\prime},\vec R=\vec x-\vec x^{\prime}$ y tomaremos coordenadas esf\'ericas, por lo que 
\begin{eqnarray}
d\vec k=dkd\theta d\varphi k^{2}\sin\theta,\qquad \vec k\cdot  (\vec x-\vec x^{\prime})=kR\cos\theta. 
\end{eqnarray}
Entonces,
\begin{eqnarray}
G(\vec R, \tau)&=&\frac{1}{4\pi^{3}} 
\int_{-\infty}^{\infty} d\omega \int_{0}^{\infty} d k k^{2} 
\int_{0}^{2\pi} d\varphi \int_{0}^{\pi} d\theta \sin\theta 
\frac{ e^{-i\left( \omega \tau-kR\cos\theta \right)}}{ k^{2}-\frac{\omega^{2}}{c^{2}}}\nonumber\\
& =& \frac{2\pi }{4\pi^{3}}\int_{0}^{\infty} d k k^{2} 
 \int_{0}^{\pi} d\theta \sin\theta e^{ ik R \cos\theta}  \int_{-\infty}^{\infty} d\omega
\frac{ e^{-i\omega \tau}}{ k^{2}-\frac{\omega^{2}}{c^{2}}}.
\end{eqnarray}
Adem\'as, con el cambio de variable $u=\cos\theta,$ se tiene 
\begin{eqnarray}
 \int_{0}^{\pi} d\theta \sin\theta e^{ ik R \cos\theta} &=&-\int_{1}^{-1} du e^{iuRk}
 =\int_{-1}^{1} du e^{iuRk} \nonumber\\
&= &\frac{1}{ikR} e^{iuRk}\Bigg|_{-1}^{1}= \frac{1}{ikR} \left(e^{iRk}-e^{-iRk}\right)=
\frac{2}{kR}\sin k R \nonumber,
\end{eqnarray}
entonces 
\begin{eqnarray}
G(\vec R, \tau)= -\frac{ c^{2}}{\pi^{2}}\int_{0}^{\infty} d k k^{2}  \frac{1}{kR}\sin rR  \int_{-\infty}^{\infty} d\omega
\frac{ e^{-i\omega \tau}}{\omega^{2} -k^{2}c^{2}}.\label{eq:TFgop}
\end{eqnarray}
Ahora estudiaremos la integral
\begin{eqnarray}
I(k)&=& \int_{-\infty}^{\infty} d\omega\frac{ e^{-i\omega \tau}}{\omega^{2} -k^{2}c^{2}}=\int_{-\infty}^{\infty} d\omega\frac{ e^{-i\omega \tau}}{(\omega -kc)(\omega+kc)},
\end{eqnarray}
para facilitar algunos c\'alculos definiremos
\begin{eqnarray}
\omega_{a}=c(k-i\epsilon ),\qquad \omega_{b}=-c(k+i\epsilon),
\end{eqnarray}
por lo que, 
\begin{eqnarray}
I(k)&=& \int_{-\infty}^{\infty} d\omega\frac{ e^{-i\omega \tau}}{\omega^{2} -k^{2}c^{2}}=\lim_{\epsilon \to 0 }\int_{-\infty}^{\infty} d\omega\frac{ e^{-i\omega \tau}}{(\omega -\omega_{a})(\omega-\omega_{b})}.
\end{eqnarray}
%



Pasaremos esta integral al plano complejo, donde $\omega=\omega_{1}+i\omega_{2},$ note que
\begin{eqnarray}
 e^{-i\omega \tau}=e^{i\omega_{1} \tau}e^{\omega_{2} \tau}  .
\end{eqnarray}
Si tomamos $\tau>0,$ entonces  $\lim_{\omega\to -\infty }e^{-i\omega \tau}=0.$
Ahora, sea $C$ la trayectoria semicircular de radio $A$ en el semi plano complejo inferior. Dicha trayectoria  inicia en punto $A$ del eje real y t\'ermina en el punto $-A$ del mismo eje. Claramente  esta trayectoria se recorre en sentido de las manecillas del reloj. Ocupando la trayectoria $C,$ se tiene
\begin{eqnarray}
I(k)&=&\lim_{\epsilon \to 0}\int_{-\infty}^{\infty} d\omega\frac{ e^{-i\omega \tau}}{(\omega -\omega_{a})(\omega-\omega_{b})}+
\int_{C} d\omega\frac{ e^{-i\omega \tau}}{(\omega -\omega_{a})(\omega-\omega_{b})}\nonumber\\
& &-
\int_{C}  d\omega\frac{ e^{-i\omega \tau}}{(\omega -\omega_{a})(\omega-\omega_{b})}.
\end{eqnarray}
Note que si $A \to \infty $ las trayectorias de las dos primeras integrales forman una trayectoria cerrada, $\Gamma,$ que se recorre en el sentido de las manecillas del reloj y que encierra los dos polos $\omega_{a},\omega_{b},$ es decir
\begin{eqnarray}
I(k)&=&\lim_{\epsilon \to 0}\lim _{A\to \infty}\left(\oint_{\Gamma} d\omega\frac{ e^{-i\omega \tau}}{(\omega -\omega_{a})(\omega-\omega_{b})}-
\int_{C}  d\omega\frac{ e^{-i\omega \tau}}{(\omega -\omega_{a})(\omega-\omega_{b})}\right).\nonumber
\end{eqnarray}
Podemos ver que si $A$ es muy grande
\begin{eqnarray}
\left|
\int_{C}  d\omega\frac{ e^{-i\omega \tau}}{(\omega -\omega_{a})(\omega-\omega_{b})}\right|\leq \left|
\int_{C}  d\omega\frac{ e^{-i\omega_{1}\tau}e^{\omega_{2}\tau} }{\omega^{2}}\right|A\leq \frac{e^{\omega_{2}\tau} A}{A^{2}}
\end{eqnarray}
Por lo tanto,
\begin{eqnarray}
\lim_{A\to \infty} \left|
\int_{C}  d\omega\frac{ e^{-i\omega \tau}}{(\omega -\omega_{a})(\omega-\omega_{b})}\right|=0,
\end{eqnarray}
de donde 
\begin{eqnarray}
\lim_{A\to \infty} 
\int_{C}  d\omega\frac{ e^{-i\omega \tau}}{(\omega -\omega_{a})(\omega-\omega_{b})}=0.
\end{eqnarray}
Entonces,
\begin{eqnarray}
I(k)&=& \lim_{\epsilon \to 0}\lim _{A\to \infty}\oint_{\Gamma} d\omega\frac{ e^{-i\omega \tau}}{(\omega -\omega_{a})(\omega-\omega_{b})}\nonumber\\
&=&
-2\pi i\lim_{\epsilon \to 0}\left[ 
\lim_{\omega\to \omega_{a}} \frac{(\omega -\omega_{a}) e^{-i\omega \tau} }{(\omega -\omega_{a})(\omega -\omega_{b})} +     
\lim_{\omega\to \omega_{b}} \frac{(\omega -\omega_{b}) e^{-i\omega \tau} }{(\omega -\omega_{a})(\omega -\omega_{b})}  \right]\nonumber\\
&=& -2\pi i\lim_{\epsilon \to 0}\left[ 
 \frac{ e^{-i\omega_{a} \tau} }{\omega_{a} -\omega_{b}} +  \frac{ e^{-i\omega_{b} \tau} }{\omega_{b} -\omega_{a}}   
  \right]\nonumber\\
 &=& -\lim_{\epsilon \to 0} \frac{2\pi i}{ \omega_{a} -\omega_{b}} \left( e^{-i\omega_{a}\tau}- e^{-i\omega_{b}\tau }\right)
=-\frac{2\pi i}{ 2ck} \left( e^{-ick\tau}- e^{ick\tau}\right)\nonumber\\
&=& -\frac{2\pi }{ck} \frac{ \left(e^{ick\tau}- e^{-ick\tau}\right)}{2i}= -2\pi \frac{ \sin ck\tau}{ck}.
\end{eqnarray}
Ocupando este resultado en Eq. (\ref{eq:TFgop}) se obtiene 
\begin{eqnarray}
G(\vec R, \tau)&=& -\frac{ c^{2}}{\pi^{2}}\int_{0}^{\infty} d k \left(k^{2}  \frac{1}{kR}\sin kR\right) 
\left(-2\pi \frac{ \sin ck\tau}{ck}\right)\nonumber\\
&=&\frac{ 2c}{R\pi}\int_{0}^{\infty} d k   \sin kR \sin ck\tau= \frac{ c}{R\pi}\int_{-\infty}^{\infty} d k   \sin kR \sin ck\tau
\nonumber\\
&=& \frac{ c}{R\pi}\int_{-\infty}^{\infty} d k   \frac{1}{2}\left( \cos k\left(R-c\tau\right) - 
\cos k\left(R+c\tau\right) \right)\nonumber\\
&=& \frac{ c}{4R\pi}\int_{-\infty}^{\infty} d k   \left( e^{ ik\left(R-c\tau\right)} + e^{ -ik\left(R-c\tau\right)}-
 e^{ ik\left(R+c\tau\right)} - e^{ -ik\left(R+c\tau\right)} \right)\nonumber\\
&=& \frac{ c}{2R}\Bigg[\frac{1}{2\pi} \int_{-\infty}^{\infty} d k    e^{ ik\left(R-c\tau\right)} + \frac{1}{2\pi} \int_{-\infty}^{\infty} d ke^{ ik\left(R-c\tau\right)}\nonumber\\
& & -
\frac{1}{2\pi} \int_{-\infty}^{\infty} d k
 e^{ ik\left(R+c\tau\right)}\nonumber
  - \frac{1}{2\pi} \int_{-\infty}^{\infty} d ke^{ -ik\left(R+c\tau\right)} \Bigg] \nonumber\\
  &=& \frac{ c}{2R}\left( \delta( R-c\tau)+ \delta( R-c\tau)- \delta( R+c\tau)- \delta( R+c\tau)\right)\nonumber\\
&=& \frac{ c}{R}\left( \delta( R-c\tau)- \delta( R+c\tau)\right).
\end{eqnarray}
Adem\'as, como $\left(R+c\tau\right)>0,$ se encuentra 
\begin{eqnarray}
G(\vec R, \tau)= \frac{ c}{R} \delta( R-c\tau)=\frac{ 1}{R} \delta\left( \frac{R}{c}-\tau\right),
\end{eqnarray}
es decir
\begin{eqnarray}
G\left(\vec x-\vec x^{\prime}, t-t^{\prime}\right)= \frac{\delta\left(  \frac{\left|\vec x-\vec x^{\prime}\right|}{c} -(t-t^{\prime})\right)  }{ \left|\vec x-\vec x^{\prime}\right|} 
=\frac{\delta\left( t^{\prime} -\left( t- \frac{\left|\vec x-\vec x^{\prime}\right|}{c}\right) \right)  }{ \left|\vec x-\vec x^{\prime}\right|} 
.
\end{eqnarray}
Definiendo el tiempo de retardo como
\begin{eqnarray}
t_{Ret}=t- \frac{\left|\vec x-\vec x^{\prime}\right|}{c} ,
\end{eqnarray}
la funci\'on de Green se puede escribir de la forma
\begin{eqnarray}
G\left(\vec x-\vec x^{\prime}, t-t^{\prime}\right)= \frac{\delta\left( t^{\prime} -t_{Ret}\right)  }{ \left|\vec x-\vec x^{\prime}\right|} 
.
\end{eqnarray}
Por lo tanto, si  $\phi_{0}(\vec x,t)$ una soluci\'on a la ecuaci\'on de onda homog\'enea, la soluci\'on a la ecuaci\'on de onda inhomog\'enea  Eq. (\ref{eq:TFondai}) es 
\begin{eqnarray}
\phi(\vec x,t) &=&\int d\vec x^{\prime} dt^{\prime} \rho\left(\vec x^{\prime},t^{\prime}\right) \frac{\delta\left( t^{\prime} -t_{Ret}\right)  }{ \left|\vec x-\vec x^{\prime}\right|} +\phi_{0}(\vec x,t)\nonumber\\
&=& \int dx^{\prime} \frac{ \rho\left(\vec x^{\prime},t_{Ret}\right)  }{ \left|\vec x-\vec x^{\prime}\right|} +\phi_{0}(\vec x,t)\nonumber,
\end{eqnarray}
es decir,
\begin{eqnarray}
\phi(\vec x,t) = \int d\vec x^{\prime} \frac{ \rho\left(\vec x^{\prime},t- \frac{\left|\vec x-\vec x^{\prime}\right|}{c}  \right)  }{ \left|\vec x-\vec x^{\prime}\right|} +\phi_{0}(\vec x,t).
\end{eqnarray}

\end{document}